\documentclass[11pt,a4paper]{article}
\pdfoutput=1
\usepackage{jheppub}
\newcommand{\refeq}[1]{(\ref{#1})}
\usepackage{graphicx}
\usepackage{amsmath}
\usepackage{amsfonts}
\usepackage{xcolor}

\usepackage{subfig}
\usepackage{graphicx}
\usepackage{multirow}

\relax
\renewcommand{\theequation}{\arabic{section}.\arabic{equation}}
\def\be{\begin{equation}}
\def\ee{\end{equation}}

\newcommand{\ha}{{1 \over 2}}

\newcommand{\bear}{\begin{eqnarray}}
\newcommand{\bea}{\begin{eqnarray}}
\newcommand{\eear}{\end{eqnarray}}
\newcommand{\eea}{\end{eqnarray}}
\def\hri#1#2{\href{http://arxiv.org/abs/#1}{[ArXiv:#1]#2}}
\def\hre#1#2{\href{http://arxiv.org/abs/#1/#2}{[ArXiv:#1/#2]}}

\def\hrj#1#2{\href{https://doi.org/#1}{#2}}
\newbox\pippobox

\def\II{\relax{\rm I\kern-.18em I}}

\def\e{\epsilon}
\def\l{\lambda}
\def\m{\mu}
\def\n{\nu}

\def\r{\rho}
\def\g{\gamma}

\def\pa{\partial}

\def\sp{\;\;\;,\;\;\;}

\def\p{\partial}

\def\f{\varphi}

\def\a{\alpha}
\def\b{\beta}

\def\le{\left}
\def\ri{\right}

\def\Vg{\sqrt{-g}} 
\def\ar{{~~~\Rightarrow~~~}}
\def\l{\lambda}
\def\i{\mathcal{I}}

\newcommand{\eql}[2]
{ \begin{equation} \label{#1}
 #2
\end{equation}}

\title{de Sitter versus Anti de Sitter flows and the (super)gravity landscape: Part II }

\author[1,2]{Elias~Kiritsis,}
\author[3]{Sergio~Morales-Tejera,}
\author[2]{Christopher~Rosen}

\affiliation[1]{\href{http://www.apc.univ-paris7.fr}
{APC, Universit\'e Paris 7}, CNRS/IN2P3, CEA/IRFU, Obs. de Paris, Sorbonne Paris Cit\'e, B\^atiment Condorcet, F-75205, Paris Cedex 13, France (UMR du CNRS 7164).}

\affiliation[2]{ \href{http://hep.physics.uoc.gr/}
 {Crete Center for Theoretical Physics}, Institute for Theoretical and Computational Physics, Department of Physics, University of Crete
 71003 Heraklion, Greece.}
 
\affiliation[3]{\href{https://physics.uvt.ro}
 {West University of Timisoara}, Department of Physics, Bd. Vasile Pârvan 4, Timisoara 300223, Romania}

\abstract{
Generic solutions are studied in Einstein-scalar gravity in an ansatz that can interpolate between  de Sitter and Anti-de Sitter  regimes.
The scalar potential is arbitrary.
All solutions are determined by their end-points in the scalar field space. All such end-points are classified. This provides a complete classification and characterization of the full space of regular solutions.
It is  shown that there are no regular (Centaur) solutions that interpolate between an AdS boundary and a dS interior, within our ansatz, when $d>2$. This no-go theorem persists in the presence of multiple scalar fields with a non-trivial field space metric. The Gubser classification of regular solutions is also upgraded to include cases that are not Lorentz invariant and do not contain AdS boundaries.
}
\preprint{
\begin{flushright}
CCTP-2025-10\\ 
 ITCP-IPP-2020/10
\end{flushright} }

\dedicated{\bf Dedicated to the memory of Umut G\"ursoy,\\ a fine scientist and a gentleman.}

\begin{document}

\maketitle 

\section{Introduction}
\label{sec:gen}

It was believed in the 80's and 90's that Quantum Field Theory is less fundamental than string theory
as it appears as the low-energy limit of string theories. It was also believed that string theory is a UV complete theory as it seemed to be finite\footnote{The fact that non-supersymmetric string theories seemed not be finite, \cite{FS}, did not bother string theorists, as at that time, almost all believed in magic.}.
 Moreover, string theory added the sought-after inclusion of quantum gravity to low-energy QFT interactions, and to a large extent, ``unified" gauge theories and gravity.

However, it soon became obvious that (perturbative) string theories could not be UV-complete theories, as they could not answer questions near or beyond the Planck scale.
Non-perturbative dualities could not help in this direction as they keep the Planck scale fixed, \cite{book}.

The holographic correspondence, \cite{Malda,GKP,Witten98}, has turned the tables and introduced some democracy in theory space. It has provided a contrasting view of string theory, and the associated quantum gravity at least for space-times that are asymptotically AdS. In such contexts, it is expected that the full string theory including its non-perturbative aspects can be reproduced by an appropriate QFT.
To date, no similar picture has emerged for asymptotically flat string theory, despite efforts in that direction,\cite{Susskind,petro}.

Given the AdS/CFT correspondence,  it looks plausible that string theories as we know them, and QFTs as we know them, are collections of local patches in a construct that may be bigger than the framework of QFT or string theory.
It is clear that there may be many string theories that have no weak-coupling limit
and because of this, they are unknown\footnote{Some ideas in this direction can be found in \cite{nstring} but they are by no means unique.}.

A reappraisal of the conceptual view of string theory was imposed by the holographic correspondence: the notion of the landscape of string vacua has undergone a reinterpretation. The string theory landscape has looked rather formidable and unwieldy, \cite{schelle}, but holography forced us to contrast it to the QFT landscape, which,  with the help of non-perturbative techniques,  turns out to be as enormous as the stringy one.

The correspondence between the two landscapes may be central in  understanding the emergence of gravity and space-time, \cite{Seiberg,Malda2,smgrav},  from purely QFT-based concepts.
It also seems important to the several deep problems that plague the coexistence of gravity and quantum mechanics, like the black-hole information paradox, \cite{Harlow,polpar,Marolf}, the cosmological constant problem, \cite{Polchinski} and to some extent the hierarchy problem.

In the case of asymptotically AdS space-times, holography provides a rather credible picture of the (structure of the) space of theories and their connections via string theory/supergravity solutions that correspond to QFT RG flows, (\cite{exotic} and references therein). Moreover, there is a concrete framework to understand the mapping from QFT, \cite{Go}-\cite{Umut}.

There is, however, another set of geometries that does not necessarily seem to fall into the asymptotically flat, or asymptotically AdS categories mentioned.
These include space-times of the type studied in cosmology, and which obviously cannot be neglected.
The eventual asymptotics of such space-times involve regular geometries (de Sitter space) or singular cosmological geometries (big bang or big crunch singularities).
However, such geometries and especially dS space,   present several puzzles, when quantum gravity is assumed.
In particular, some involve the  size of the cosmological constant and the fact that dS seems to be dynamically unstable to quantum corrections, \cite{tw,br,Dvali}.

$\bullet$ It was also observed that  weakly-coupled, weakly-curved string theory seems to be at odds with dS solutions, \cite{DaVa}.
This difficulty, has been elevated to a swampland conjecture, \cite{DS}-\cite{tom}.
It currently states that there are no dS extrema without directions in field space that are ``unstable".
Further cosmological swampland conjectures, like the Transplankian Censorship Conjecture   have been also formulated, \cite{TCC}.

The effort of finding controllable dS solutions in string theory is fully active,  both in the context of the KKLT proposal, \cite{KKLT} and otherwise.

$\bullet$ An alternative realization has been proposed, based on the brane-world idea\footnote{A simplified version of this mechanism  was proposed earlier in \cite{dSe}. In this realization, a bulk RG flow is approximated as an abrupt domain wall between the UV and IR CFT extrema. This approximate realization falls into the fine-tuned category of \cite{dSk}.}, \cite{dSk}. This idea is inspired by the self-tuning mechanism of the cosmological constant\footnote{The cosmology of moving branes has had a longer history in string theory and was related to holography, \cite{kraus}-\cite{holo}.} \cite{selftuning}. It states  that the bulk space can be negatively curved, but the geometry on a collection of branes (that must carry, among other things,  the Standard Model) can be both cosmological and accelerating.

It was assumed in \cite{dSk} that the bulk string theory is holographically related to a boundary QFT, at the asymptotically AdS boundary.
If this
(boundary) dual QFT is defined on flat Minkowski space, it is not possible to have a de Sitter geometry on a brane embedded in the bulk.
However,  if the boundary  QFT is defined on de Sitter space, then it is possible to have a de Sitter geometry on a brane embedded in the bulk.

 Moreover, it was shown that interesting hierarchies can appear between the de Sitter scale of the boundary QFT and of the de Sitter scale on the brane-universe.
The work in \cite{dSk} was achieved in a generic bottom-up context. The implementation of this idea in a controllable bulk string theory framework has not been achieved yet.

$\bullet$ There is an alternative possibility of realizing de Sitter space on a brane moving in asymptotically AdS bulk space. It was shown in \cite{For} that if a brane is moving in an approximate bulk AdS geometry, then the geometry on the brane is approximately de Sitter. A crucial ingredient for this is the presence of an induced Einstein term on the brane. Such terms appear on D-branes of bosonic string theories at tree level, but are induced only at loop level in supersymmetric theories, \cite{KTT}.
In particular, such a mechanism can generate early universe inflation if the associated dual QFT has a ``walking regime", \cite{For}. It remains to be seen if such a mechanism can be realised in a controllable
top-down example.

$\bullet$ de Sitter space has two time-like boundaries, ${\cal I}^+$ and ${\cal I}^-$, whose local geometry is very similar to AdS. In particular, they are conformal boundaries.
 Since the advent of AdS/CFT, it has been suggested that there maybe a de Sitter analogue of holography, called dS/(pseudo)CFT correspondence, \cite{SdS}-\cite{Strom}.
There are several ideas on how this correspondence might be realized. At the scale-invariant points, associated to de Sitter space, the bulk gravitational theory is expected to be dual to a (pseudo)CFT. The precise rules for this (pseudo)CFT were spelled out in \cite{MS}.

In the same context, the full cosmological evolution starting and ending in dS, was analyzed from a holographic viewpoint in \cite{afim}, where it was associated to (pseudo)RG flows. As a consequence, Wilsonian ideas were used to classify inflationary theories\footnote{For a recent discussion from a different starting point, see \cite{We}.}. A sharp contrast can be drawn on the standard view of cosmological solutions and their fine-tuning problems on one side and the holographically-dual picture on the other. In particular, it was argued that the holographic view may be crucial in resolving several fine-tuning problems in cosmology.

$\bullet$ A further idea addressing a ``quantum" description of cosmological geometries, influenced by holography, was to find regular geometries that contain an asymptotically AdS boundary, and which delve far in bulk regions when scalar potentials are positive and the local metric  is of the cosmological type.
In such geometries the dual QFT could be potentially used to ``define" the cosmological regime.

Studies of such cosmological solutions in the AdS context have been performed in \cite{hub,Lowe} following earlier work, \cite{Blau:1986cw} that have used specific metric ansatze. The solutions discussed were rather approximate and used the thin-wall approximation. One of the motivations was to go beyond the results of \cite{Farhi} that argued that all such solutions are singular in the past. However, no successful regular solutions were found.
As we show in this paper, such regular solutions do not exist.

A more recent paper, \cite{AnHo}  addressed the latter question in two space-time dimensions, and found interpolating solutions that contain an AdS$_2$ boundary and a dS$_2$ part in the ``IR" geometry.
The authors of \cite{AnHo} have also speculated that such solutions might exist in higher dimensions. This has not been realized so far.

In \cite{exotic}, a program has started, that aims at a systematic study of holographic solutions in the AdS part of the landscape of gravitational theories.
The goal was  to  produce a precise map with similar RG flows on the QFT side.
Several works have studied different aspects of holographic RG flows in the AdS regime, \cite{exotic,dSk}, \cite{curved}-\cite{Raymond}.

The intricate questions associated with the dS regime imply that an extension of this study to the dS regime is important. The first step was undertaken in \cite{KT} where the first ansatz was studied, which  had the ability to interpolate  between AdS and dS.
It was shown in that work that no regular solutions exist that have at the same time an AdS boundary and a cosmological interior.
The tools used were a detailed study of the topology of regular solutions.

In the present paper we shall study possible interpolating solutions, between the AdS and the dS regimes,  in higher than two dimensions, by choosing the  second of the three interpolating ansatze mentioned in \cite{KT}.

\subsection{Results and outlook}
We consider the gravitational theory
\be\label{eq:bulk}
S\le[g,\f\ri]= \int d^{d+1}x \Vg \le(
			 R
			-\ha\p_a\f \p^a\f-V(\f)  \ri)
			\ee
with a single scalar, $\f$ (in section \ref{multi} we argue that our results are valid in the multiscalar case)\footnote{There are possible generalizations of the multiscalar action that respect our ansatz, but such solutions will be analyzed in the future. They include gauge fields $A_{\mu}$ with only $A_t$ turned on, or $(d-1)$-form fields  proportional to the volume form of $S^{d-1}$.}. Within this theory, we focus our attention on solutions that are contained within the ``spherically-sliced black-hole-like ansatz''
\be
ds^2={du^2\over f(u)}+e^{2A(u)}\left[-f(u)dt^2+R^2~d\Omega_{d-1}^2\right], \qquad \varphi = \varphi(u)
\label{bhSanstz}\ee
which allows for solutions which interpolate between AdS and dS solutions. Our main results will be essentially independent of $d$, and are expected to hold for all $d > 2$.

Special solutions with constant scalar exist only at extrema of the potential and correspond to either AdS$_{d+1}$,
dS$_{d+1}$, dS$_2\times$S$^{d-1}$ or Mink$_{d+1}$ solutions. All other solutions involve a nontrivial scalar field $\f(u)$. With some care, one can use the scalar field instead of $u$  to parametrize the solution.
In the case where these solutions exist in regions with $V<0$, they are holographically dual to Renormalization Group flow solutions in the dual Quantum Field Theory (QFT).
By abuse of language, we shall call all regular solutions of this type ``flow solutions''.

The primary results of this work are three-fold.   We first classify all possible {\it flow endpoints} and subsequently all regular solutions to the gravitational equations that start and end at the flow end-points.
 To accomplish this, we exploit a superpotential formulation of the gravitational equations of motion  developed in several papers in the past, \cite{dVV}-\cite{thermo}, \cite{exotic,curved,gur}.

Flow endpoints away from the boundary of field space are shown to necessarily coincide with extrema of the superpotential $W$, and can be organised into five distinct classes:
\begin{itemize}
\item {\it AdS$_{(d+1)}$ and dS$_{(d+1)}$ Boundaries}  locally coincide with the (holographic) boundary of AdS$_{(d+1)}$ and the past/future ($\mathcal{I}^{\pm}$) boundary of dS$_{(d+1)}$, respectively. They are extrema of both the superpotential as well as the scalar potential (i.e. $W'=V' = 0$ there).

\item {\it dS$_2$ Boundaries} are dS$_2 \times$ S$^{(d-1)}$ asymptotic solutions which can also arise at locations in field space where the superpotential and scalar potential are both extremised.

\item {\it Nariai (Extremal) horizons} are again locally dS$_2 \times $S$^{(d-1)}$ endpoint solutions, but unlike the dS$_2$ boundaries they are characterized by a blackening function $f$ that vanishes quadratically. They coincide locally with the so-called ``Nariai limit'' of de Sitter black holes, in which the event and cosmological horizons coincide.

\item {\it Shrinking points} are flow endpoints in which the spatial sphere in (\ref{bhSanstz}) shrinks to zero size, while the geometry remains regular. A familiar example is the center of Anti de Sitter space in global coordinates. They can arise however in regions of field space in which the scalar potential is positive, negative, or zero.

\item {\it Minkowski Boundaries} are endpoint solutions that have vanishing curvature but the size of the spatial sphere diverges. They are therefore locally identified as the spatial boundary of Minkowski space, $\mathbb{R}^{1,d}$.
\end{itemize}
These endpoint solutions are discussed in further detail in section \ref{sec52}.

We shall also have cause to comment on flows that arrive at the boundaries of field space, which are necessarily singular in our setting. In some cases these singular solutions may be of physical interest. A familiar example are singular solutions of lower-dimensional Einstein-Dilaton theories which satisfy a ``Gubser bound'' \cite{Gubser}.  This is to say that they may be acceptable because they are extremal limits of regular black hole solutions. Moreover, some of them can be lifted to  a higher-dimensional regular  solution. When this is the case, we refer to the solution as ``Gubser-regular''. This mechanism of singularity resolution is familiar from the holography of asymptotically AdS domain-wall solutions \cite{thermo,cgkkm,cgkkm1,exotic}.

These singular flow solutions will be discussed in detail in what follows. For the present survey, we note that the local features of the singularities appearing at the boundary of field space in such flows give rise to a three--fold classification scheme. In particular, ``Type I'' and ``Type II'' singularities may be acceptable in the sense of \cite{Gubser,thermo,cgkkm,cgkkm1,exotic}, whereas ``Type 0'' singularities are always unacceptable.
 
Overall in this paper, we shall call  ``bad" singularities, the singularities of the type 0 solutions, but also the singularities that can appear at finite $\f$, or in type I solutions  with $\a>\a_G$. 

The second major output of this work is a list of rules, enumerated in section \ref{sec:global}, that govern the structure of interpolating solutions within our ansatz. The broad stroke content of these rules can be summarized as follows:
\begin{itemize}
\item Interpolating solutions terminate at regular extrema of the superpotential $W$, which is monotonic as a function of the holographic coordinate $u$ along the flow. If, for example, $W\ge 0$ at an endpoint, then boundary endpoints (Minkowski or (anti) de Sitter) and extremal horizons are {\it minima} of $W$, while shrinking endpoints are {\it maxima}.

\item The blackening function $f$ can have at most one extremum along the flow (excluding endpoints), and this extremum must be a maximum. Therefore, interpolating solutions can have at most two horizons whose locations correspond to the roots of $f$. If the flow terminates in a shrinking endpoint, the blackening function is monotonic along the flow.

\item For solutions with {\it either} an AdS or Minkowski boundary {\it or} a shrinking endpoint, the interpolating solution can have at most one horizon. Conversely, if the solution interpolates between an AdS or Minkowski boundary {\it and} a shrinking endpoint, the solution must be horizonless.

\item If the function $T \equiv e^{-2A}/R^2$, which controls the (inverse) size of the sphere, vanishes, it can only  do so either identically or at boundary endpoints. 

\item In interpolating solutions which have an extremal Nariai horizon endpoint, the blackening function satisfies $f<0$ along the flow and vanishes at the endpoint.

\item A useful quantity for characterizing the properties of interpolating solutions is the combination $\rho \equiv f\dot{\varphi}^2/2-V$ which in part controls the curvature invariants of the solution. For solutions which interpolate between AdS or Minkowski boundaries and shrinking endpoints with $V>0$, the quantity $\rho$ must change sign.
\end{itemize}

Taken in concert, these rules and their corollaries can be used to place strong constraints on the allowed structure of interpolating solutions within our ansatz.

The landscape of these regular interpolating solutions, consistent with the flow rules, is the third major output of this work. It is summarized pictorially in figure \ref{fig:flow}. The important lessons are:
\begin{itemize}
\item {\it Flows from an AdS$_{(d+1)}$ boundary} are privileged in that they are the best understood in the context of holographic duality. We find that such a flow may either terminate in an AdS shrinking endpoint or encounter an event horizon before reaching the boundary of field space. Especially noteworthy is the observation that within our ansatz there are {\it no allowed flows from an AdS boundary to a dS shrinking endpoint}.

\item {\it Flows from a dS$_{(d+1)}$ or dS$_2$ boundary} are comparatively diverse. They may terminate in either AdS or dS shrinking endpoints by passing through a cosmological horizon, or reach the boundary of field space (at a bad singularity) after passing through both cosmological and event horizons. The latter possibility describes a hairy dS--Schwarzschild black hole. 

\item {\it Flows from a Minkowski boundary}  can either end in an AdS shrink point, or pass through an event horizon on the way to a bad singularity. These solutions are not generically constrained by scalar no-hair theorems (see review \cite{nohair}), as such theorems apply to special classes of potentials. In contrast, our potentials are general, but must obey $ V=V'=V''=0$ at a specific point in field space,  in order for such a solution to exist.

\item {\it Flows from Nariai (extremal) endpoints} do not exist. We note that not all locally Nariai solutions are endpoints, and that this flow rule is therefore compatible with the existence of dS black-hole solutions which interpolate between a dS boundary and a bad singularity, passing through a Nariai extremal horizon.

\item More generally, there are no regular flows which connect two boundary endpoints, nor can an interpolating solution connect two shrinking endpoints.
\end{itemize}

For flows involving ``Type I'' and/or ``Type II'' singularities, extra care is necessary to enumerate both the approach to the boundary of field space as well as the asymptotic value of the potential obtained as the singularity is reached. In particular, we find
\begin{itemize}

\item {\it Flows from an AdS$_{(d+1)}$ or Minkowski boundary} can terminate in a Type I/II singularity provided that $V\to-\infty$ as $|\varphi|\to\infty$.

\item {\it Flows from a dS$_{(d+1)}$ or dS$_2$ boundary} involving singular endpoints are again more diverse. They may terminate in a Type I/II singularity with $V\to-\infty$ as $|\varphi|\to\infty$, provided that they pass first through a cosmological horizon. Horizonless flows to a Type I/II singularity are possible if instead $V\to+\infty$ as $|\varphi|\to\infty$.

\item {\it Flows from a Type I/II singularity with $V\to0^-$ as $|\varphi|\to\infty$} obey the same flow rules as AdS and Minkowski boundaries. In particular, they may terminate in AdS shrink points, Type I/II singularities in which $V\to-\infty$ as $|\varphi|\to\infty$, or pass through an event horizon on the way to a bad singularity.

\item {\it Flows from a Type I/II singularity with $V\to0^+$ as $|\varphi|\to\infty$} follow the same flow rules as dS$_{(d+1)}$ and dS$_2$ boundaries. As such, they may pass through a cosmological horizon to terminate in AdS shrink points, Type I/II singularities in which $V\to-\infty$ as $|\varphi|\to\infty$, or dS shrink points. Alternatively, they can pass through both a cosmological and an event horizon on the way to a bad singularity. Finally, there are horizonless flows which terminate in Type I/II singularities with $V\to+\infty$ as $|\varphi|\to\infty$.

\end{itemize}

Within the rich landscape of allowed flows in this simple gravitational theory, it is worth emphasising that our results rule out the possibility of solutions between an AdS boundary and a region of dS. This is in contrast to the situation in $d=1$, where AdS$_2$/dS$_2$ domain walls are known to exist---the so-called ``Centaur solutions'' of \cite{AnHo}. Such a solution would be highly desirable, as it would allow one to bring the power of holographic duality to bear on questions of phenomenological interest to a dS universe.

In section \ref{sec:brane} we attempt to better understand this obstruction in our setup by studying the conditions in which AdS and dS vacua can be joined via a thin brane. We find that such a solution is possible providing the theory on the worldvolume of the brane contributes a particular sort of stress energy to the system. This can be accomplished by endowing the brane action with an Einstein-Hilbert term (generically with a cosmological constant), and comment on the feasibility of obtaining such a gravitational system from a string theory in higher dimensions.

Along the way, our analysis also yields several noteworthy by-products. First, by employing a series of discrete symmetries enjoyed by our system of equations, we are immediately able to exploit our main results to characterise the space of allowed flows for solutions in which the spatial sections are taken to be {hyperbolic} as opposed to spherical. The result is summarised in figure \ref{fig:flowH}.

Importantly, the hyperbolic sliced ansatz {\it also} prohibits horizonless flows between an AdS$_{(d+1)}$ boundary and a dS shrink point. Therefore, of the simple interpolating ansatze reviewed in section \ref{inter}, only the so-called ``dS sliced ansatz'' (\ref{w24}) remains uncharted in its entirety. We leave the systematic exploration of the space of flow solutions to (\ref{eq:bulk}) in the dS sliced ansatz to future work.

Additionally, towards understanding the viability of singular endpoints which appear in our flows, we are led to a generalisation of Gubser's criterion for ``acceptable singularities'' \cite{Gubser} to radial flow geometries which break Lorentz invariance in the transverse directions. The criterion applies to singular solutions which arise at the boundary of field space ($\varphi\to\pm\infty$) and therefore provides a sub-classification of the type 0/I/II endpoints. The details of this analysis are relegated to appendix \ref{asymp}. One highlight is that for singularities in which
\begin{equation}
V \sim -V_\infty e^{\alpha \varphi}
\end{equation}
with $\alpha >0$ as $\varphi\to\infty$, type 0 endpoints are always ``bad'' singularities.

Differently, type I endpoints can be acceptable or ``Gubser-regular'' provided that $\alpha < \alpha_G$ where
\begin{equation}
\alpha_G = \sqrt{\frac{2d}{d-1}}
\end{equation}
which is the Gubser bound of \cite{Gubser}. In this sense, type I singularities are analogous to those that arise in holographic RG flows describing the behaviour of theories on Minkowski space.

Type II endpoints, on the other hand, are ``Gubser-regular'' for {\it any} value of $\alpha$ for these asymptotics. Accordingly, they have no analogue in the standard (Poincar\'e invariant) holographic RG flows.

\section{Einstein-dilaton gravity as a proxy for the gravitational theory} \label{s2}

As a gravitational theory we shall study mostly  Einstein-scalar theory, whose action consists of an Einstein-Hilbert term and a minimally coupled scalar field with a potential. This is the most general two-derivative theory of a metric and a single scalar field and it is a proxy for the more general multiscalar theory . Any solution of a multi-scalar theory can be mapped to a solution of a single scalar theory. We comment on the multi-scalar theory in section \ref{multi}.

We define the Einstein-scalar theory in $d+1$ dimensions, with signature $(-, +\ldots +)$. The action we consider throughout is of the form:
\eql{i1}{
			S\le[g,\f\ri]= \int d^{d+1}x \Vg \le(
			 R
			-\ha\p_a\f \p^a\f-V(\f)  \ri)+S_{GHY}.
			}
where $S_{GHY}$ is the Gibbons-Hawking-York term associated to any boundary that might exist. To arrive to this,  from the most general two-derivative action,  a Weyl rescaling of the metric as well as a redefinition of the scalar are necessary.

\vskip 1cm
$\bullet$ {\bf A holographic ansatz}.
\vskip 1cm

We first present a holographic ansatz
in the  so-called domain-wall coordinate system:
\eql{i2}{
		 \f=\f(u),\qquad ds^2=du^2+e^{2A(u)}ds_d^2
		}
where $u$ is  the holographic coordinate. The $d$-dimensional metric $ds_d^2$ of the manifold $M_{d}$ is a constant curvature, Minkowski signature metric with components $\zeta_{\mu\nu}$.
This is a conical metric where the slice of the cone is isomorphic to the manifold $M_d$ and the radial coordinate $u$ is space-like.

In the regime where $V<0$, the metric may have asymptotically AdS boundaries where the boundary condition for the metric is that of $M_{d}$. Such  solutions have a dual QFT interpretation according to the holographic conjecture, \cite{Malda,GKP,Witten98}. In the holographic correspondence, such solutions describe a state of an appropriate  QFT$_d$, defined on the manifold $M_d$.
The connection between the bulk gravitational setup and the boundary QFT is made by mapping  the bulk metric with the stress energy tensor, $T_{\mu\nu}$,  of the boundary theory and the scalar field with a single-trace scalar operator, $O(x)$.

Using \eqref{i2} and by varying the action \eqref{i1}, we arrive at the following gravitational equations of motion:
\be
\label{i3_1} 2(d-1) \ddot{A} + \dot{\f}^2 + \frac{2}{d} e^{-2A} R^{(\zeta)} =0 \, ,
\ee
\be
\label{i3_2} d(d-1) \dot{A}^2 - \frac{1}{2} \dot{\f}^2 + V - e^{-2A} R^{(\zeta)} =0 \, ,
\ee
where a dot stands for the derivative with respect to $u$ and
 the constant curvature of the slice, $M_d$, is given by
\be
\label{Rz}
R^{(\zeta)}_{\mu \nu} = \bar\kappa \zeta_{\mu \nu} \, , \quad R^{(\zeta)} = d \bar\kappa \, , \quad \textrm{with} \quad \bar\kappa =
\left\{ \begin{array}{lll}
\displaystyle  \frac{(d-1)}{\alpha^2} ,&\phantom{aa}& M_d\sim S^d,\\ \\
\displaystyle ~~~~~~0,&\phantom{aa} &M_d\sim Mink_d\\ \\
\displaystyle - \frac{(d-1)}{\alpha^2} ,&\phantom{aa}&M_d\sim  EAdS_d.
\end{array}\right.
\ee
where $\alpha$ is by definition the radius  of M$_d$.

From (\ref{i3_1}), (\ref{i3_2}),  we may also deduce the Klein-Gordon equation for the scalar, which is given by
\eql{i4}{\ddot{\f}+d\dot A\dot{\f}-V'(\f)=0,}

From now on, we shall call the  scalar field regions where $V(\f)<0$ the ``AdS regime", while regions where $V(\f)>0$ the ``dS regime".
In the AdS regime,  we deduce  from \eqref{i3_1}  that $\dot A(u)$  cannot increase. In the holographic RG, this is related to the holographic c-theorem \cite{GPPZ,Freedman}.

We assume that $V(\f)$ is analytic for all finite $\f$,  as this is a standard property of string theory effective potentials, \cite{book,Trigiante,ceresole}.


\vskip 1cm
$\bullet$ {\bf The cosmological ansatz}.
\vskip 1cm

There is another conical ansatz similar to  (\ref{i2}) where the radial coordinate is timelike
\eql{i5}{
		 \f=\f(t),\qquad ds^2=-dt^2+e^{2A(t)}ds_d^2
		}
where $t$ is  now the time of a cosmological solution. The $d$-dimensional metric $ds_d^2$ of the manifold $M_{d}$ is a constant curvature, Euclidean  signature metric.
In such a cosmological ansatz we describe the time evolution of a $(d+1)$-dimensional universe with constant time slices given by the manifold $M_d$.
In the dS regime, $(V>0)$, such solutions contain natural asymptotically de Sitter (time-like) boundaries.

The equations of the cosmological ansatz can be obtained by a simple set of substitutions from those in the holographic ansatz, (\ref{i3_1}), (\ref{i3_2}), \cite{MS,afim}:
\be
u\to t\sp V\to -V\sp R^{(\zeta)}\to -R^{(\zeta)}
\ee

\section{Interpolating Ans{\"a}tze\label{inter}}

One of the  main  purposes of this work is to study solutions that interpolate between (asymptotically) de Sitter and (asymptotically) Anti de Sitter space-times. To do so, we shall focus our attention on a slightly more general ansatz than that of (\ref{i2}) and (\ref{i5}).

 In particular, we introduce an additional dynamical variable, the \emph{blackening function} $f(u)$, such that the general form of the ansatz becomes 
 \be
 ds^2 = \frac{du^2}{f(u)}+e^{2A(u)}\left[-f(u)dt^2+ds^2_{d-1} \right].\label{eq:fanstz}
 \ee
  The motivation for this is as follows: the vanishing of $f(u)$ yields  a horizon, on either side of which $f$ has a different sign. Therefore, a solution that passes through a horizon, exchanges $u$ from space-like to time-like and vice-versa, offering a simple means of interpolating between (\ref{i2}) and (\ref{i5}).

In \cite{KT} three classes of interpolating ans{\"a}tze were introduced, distinguished essentially by the choice of metric $ds^2_{d-1}$ in (\ref{eq:fanstz}). One corresponding to a flat slicing, one corresponding to a spherical slicing, and finally, one corresponding to  dS slicing. We reproduce them here.

\bigskip

$\bullet$ \emph{The ``black-hole" ansatz with a flat slicing.} The corresponding metric is given by

\be
ds^2={du^2\over f(u)}+e^{2A(u)}\left[-f(u)dt^2+dx_{i}dx^{i}\right].
\label{c38}\ee

Note that when $f=1$ and $e^{A}=e^{-{u\over \ell}}$, (\ref{c38}) reduces to AdS space in Poincar\'e coordinates (where $u$ is space-like). With $f=-1$ and  $e^{A}=e^{-{u\over \ell}}$ the metric reduces to dS space in Poincar\'e coordinates (where $u$ is now time-like).
This ansatz and the associated solutions were studied in \cite{KT}. It was shown that, for $d>2$, no regular solutions can interpolate between a part of AdS containing the boundary  and dS.

\bigskip

$\bullet$ \emph{The black-hole ansatz with a spherical slicing.} This is obtained from the  previous ansatz, by the substitution $dx_{i}dx^{i} \rightarrow d\Omega^2_{d-1}$. The corresponding metric is given by

\be
ds^2={du^2\over f(u)}+e^{2A(u)}\left[-f(u)dt^2+R^2~d\Omega_{d-1}^2\right]
\label{c39}\ee

The radius of the sphere $R$ sets the dimensions, but its value is not of significance, as it may be changed at will, by a shift in $A$ and a rescaling of $t$.

To see that AdS space-time can be obtained from this metric, we set
\be
e^A=e^{-{u\over \ell}}\sp f=1+e^{2{u\over \ell}}\sp R=\ell.
\label{w22}\ee

Using the coordinate transformation $r=\ell~e^{-{u\over \ell}}$, this can be mapped to the static patch metric of AdS in  (\ref{a35}).
Also, in this ansatz we can obtain the AdS metric in global coordinates by setting
\be
e^{A}=\sinh{\rho}\sp f=\coth^2\rho\sp \coth\rho d\rho=du
\ee
 as in (\ref{adsg}). 

To obtain a dS space-time in the same ansatz, we must set
\be
e^{A}=e^{Hu}\sp f=-1+e^{-2Hu}\sp R={1\over H}
\label{w23}\ee
 which can be mapped to the static patch metric in (\ref{a27}) by the coordinate transformation $Hr=~e^{{Hu}}$.

\bigskip

$\bullet$ \emph{The dS sliced ansatz.} This ansatz does not contain the blackness function but only a non-trivial scale factor. It is given by the metric
\be
ds^2=du^2+e^{2A(u)}d\Omega^2_{dS}\sp
d\Omega^2_{dS}\equiv \left(-dt^2+{\cosh^2(Ht)\over H^2}d\Omega_{d-1}^2\right).
\label{w24}\ee
Here, $d\Omega^2_{dS}$ is the de Sitter metric in any coordinates. We have chosen global coordinates above but any other coordinates will do.

By choosing $e^A=\sinh{u\over \ell}$ we obtain AdS as in (\ref{a37}). On the other hand,   setting $e^A=\sin{Hu}$, we obtain dS as in (\ref{a52a}).

In the next two sections, we shall study solutions and their properties that arise from the spherically sliced ansatz in (\ref{c39}) and we leave the final ansatz (\ref{w24}) for  future study.

\section{The black-hole-like ansatz with a spherical slicing}\label{sec5}

In this section, we perform a systematic study of solutions interpolating between two finite values of the scalar field\footnote{When the scalar arrives at the boundaries of its space, $\f\to\pm\infty$, then other options are possible. They are treated in section \ref{sec:7}.} $\f$ in the ansatz (\ref{c39}):

\be
ds^2={du^2\over f(u)}+e^{2A(u)}\left[-f(u)dt^2+R^2~d\Omega_{d-1}^2\right].
\label{f22}\ee

\noindent
where $d\Omega^2_{d-1}$ is the metric of the unit radius $(d-1)$-dimensional sphere.
 $R$ is a length scale that is included for dimensional reasons. Its particular value is irrelevant as it can be changed by shifting $A(u)$ by a constant.

The Einstein equations for the ansatz \eqref{f22} are

\begin{subequations}\label{f23}
\begin{align}
	&2(d-1)\ddot{A}(u)+\dot{\f}^2(u)=0\,,\label{f23a}\\
	&\ddot{f}(u)+d\dot{f}(u)\dot{A}(u)+{2(d-2)\over R^2} e^{-2 A(u)}=0\,,\label{f23b}\\
	 &(d-1)\dot{A}(u)\dot{f}(u)+f(u)\left[d(d-1)\dot{A}^2(u)-\frac{\dot{\f}^2}{2}\right]+V(\f)-{(d-1)(d-2)\over R^2} e^{-2 A(u)} =0\,.\label{f23c}
\end{align}
\end{subequations}
\noindent
The first order equation (\ref{f23c}) will be referred to as the Hubble equation in the rest of the paper.

When $f>0$, by a simple change of the radial coordinate
\be
{du\over \sqrt{f(u)}}=dr
\ee
the metric can be written as
\be
ds^2=dr^2-e^{2A_1(r)}dt^2+e^{2A_2(r)}d\Omega_{d-1}^2
\ee
while for $f<0$, the change of coordinates
\be
{du\over \sqrt{|f(u)|}}=d\tau\sp t=\theta
\ee
maps the metric to
\be
ds^2=-d\tau^2+e^{2A_1(T)}d\theta^2+e^{2A_2(T)}d\Omega_{d-1}^2
\ee
Both are $S^1\times S^{d-1}$ conifold metrics with Minkowski signature.

 Equation (\ref{f23a}) implies that $\dot{A}$ is monotonous along the flow. The Klein-Gordon equation is

\be
\ddot\f(u)+\left(d\dot A(u)+{\dot f(u)\over f(u)}\right)\dot\f(u)-{V'(\f)\over f(u)}=0\,.
\label{f23d}\ee

Not all the equations among (\ref{f23a}-\ref{f23d}) are independent. In fact, the radial derivative of (\ref{f23c}) is implied by the remaining three equations.
In view of the above, the system of equations we are solving has 5 integration constants.

Near an asymptotic AdS boundary, these integration constants have a dual QFT interpretation.
The two integration constants hidden in the $\f$ equation correspond to the source (coupling constant) and vev of the scalar operator dual to $\f$. The value of $f$ at the boundary correspond to the metric coefficient $g_{tt}$ in the boundary QFT metric while the subleading integration constant  corresponds to the vev of the energy. Finally the integration constant of $A$ set the curvature of the sphere in the boundary theory.

Equation (\ref{f23c}) can also be written as
\be
{d\over du}\left(f\dot A e^{dA}\right)=\left(-{V~e^{2A}\over d-1}+{(d-2)\over R^2}\right) e^{(d-2) A(u)}\,.
\label{ev1}\ee
 Then, (\ref{f23b}) becomes
\be
{d\over du}\left(\dot f e^{dA}\right)=-{2(d-2)\over R^2}e^{(d-2)A}
\label{ev2}\ee
while (\ref{f23d}) becomes
\be
{d\over du}\left(f\dot \f e^{dA}\right)=-V' e^{dA}
\label{ev3}\ee

We also introduce the energy momentum tensor of the scalar
\be
T_{\m\n}=\pa_{\m}\f\pa_{\n}\f-{1\over 2}g_{\m\n}(\pa \f)^2-g_{\m\n}V
\label{ev3a}\ee
whose non-zero components for our ansatz are
\be
{T^u}_u={1\over 2}f(\dot\f)^2-V\equiv \rho
\label{ev3b}\ee
\be
{T^i}_j=-p{\delta^{i}}_j\sp p\equiv \left[{1\over 2}f(\dot \f)^2+V\right]
\label{ev3c}\ee
In fact, when $f<0$, $u$ is a time-like coordinate and then $\rho$ can be called the energy density and $p$ is the pressure.
By abuse of language we shall always call $\rho$ the energy density.

We can also rewrite the equations as
\begin{equation}\label{ev4}
(d-1)\frac{d}{du}\left({e^{dA}\mathcal{I}}\right) = -\dot{A}e^{dA}~p
\end{equation}
and
\begin{equation}
\dfrac{\dot{A}}{\dot{\varphi}^2}\dot \r + \dfrac{1}{2(d-1)}\r =- \dfrac{d}{2} \i - \dfrac{(d-1)}{R^2}e^{-2A}=-{d\over 2}\left[f\dot A^2+{d-2\over d~R^2}e^{-2A}\right]\,,
\label{ev5}\end{equation}
where $\mathcal{I}$ is defined as
\begin{equation}
\mathcal{I} \equiv  f\dot{A}^2 - \dfrac{1}{R^2}e^{-2A}\,.
\label{ev7}\end{equation}
In Appendix \ref{sect:inv_sphere} it is shown that $\r,p, \i$  control the curvature invariants.

For subsequent purposes we mention that the metric \ref{f22} features a horizon, located at $u_h$, if its temporal component vanishes at $u_h$\footnote{$e^{A}$ cannot vanish while $f$ is finite, as we find in this paper.}: $g_{tt} = -f(u_h)e^{2A(u_h)}=0$. The Hawking temperature, ${\cal T}$, associated with the horizon can be extracted from the surface gravity $\kappa$ as ${\cal T}=\frac{\kappa}{2\pi}$, with

\begin{equation}\label{ev8}
\kappa^2 =\lim_{u\to u_h}\left( -\dfrac{1}{2} \nabla_{\mu}k_{\nu} \nabla^{\mu} k^\nu \right)= \lim_{u\to u_h} \left(\dfrac{e^{-2A}}{4}\left[\partial_u(e^{2A}f)\right]^2\right)
\end{equation}
where $k^{\mu}=\delta^{\mu}_t$ is a time-like Killing vector. Therefore

\begin{equation}
{\cal T} = \lim_{u\to u_h}\left( \dfrac{e^{-A}}{4\pi} |\partial_{u}(e^{2A}f)|\right)
\label{n105}\end{equation}

\subsection{The first-order formalism and the superpotential}

In previous studies of holographic solutions it was convenient to introduce a first order formalism that
is essentially a Hamilton-Jacobi formalism. It has the advantage of separating the equations with non--trivial integration constant from those with trivial integration constants, as explained in \cite{exotic}.

We introduce the superpotential by defining

\be
W(\f) \equiv - 2(d-1) \dot{A}(u)\,.
\label{w53}\ee

Then equation (\ref{f23a}) is solved by

\begin{equation}\label{w53b}
\dot{\f} = W'\,,
\end{equation}

\noindent
where we denote $\partial_u$ with an overdot and $\partial_\f$ with a prime. Moreover, it is useful to define\footnote{Although $T$ is defined to be positive, $R^2$ becomes an integration constant of the first order equations and can therefore have an arbitrary sign. We comment on this at the end of this section.}

\begin{equation}\label{w54a}
T(\f)\equiv \frac{1}{R^2} e^{-2 A(\f)}\geq 0\,,
\end{equation}

\noindent
which, along with (\ref{w53}) and (\ref{w53b}), implies

\begin{equation}\label{eqtt}
T'={W\over (d-1)W'}T\,,\hspace{1cm} \textrm{and} \hspace{1cm} A(\f)=-\frac{1}{2(d-1)} \int_{\f^*}^\f \frac{W(\f')}{W'(\f')} d \f'\,.
\end{equation}

An important relation is
\be
{dW\over du}={dW\over d\f}{d\f\over du}=(W')^2\geq 0
\ee

Taking the previous definitions into account, the equations of motion (\ref{f23b}), (\ref{f23c}) and (\ref{f23d}) become

\be \label{f10_1}
W' \bigg[ W' f''+\bigg( W''-\frac{d}{2(d-1)} W\bigg)f'\bigg]+2(d-2)T
=0,
\ee

\be
\bigg( \frac{d}{4(d-1)} W^2-\frac{W'^2}{2}\bigg) f-\frac{1}{2} W'Wf'+V-(d-1)(d-2)T=0\,,
\label{w55}\ee

\begin{equation}
\label{w55b}
   W'\left[ W'f' + f\left(W''-\frac{d }{2
   (d-1)}W\right)\right]-V' = 0\,,
\end{equation}

\noindent
respectively. Given some potential $V(\f)$, we may solve for $T$ algebraically from equation (\ref{w55})
to obtain
\be
T={1\over (d-1)(d-2)}\left[\bigg( \frac{d}{4(d-1)} W^2-\frac{W'^2}{2}\bigg) f-\frac{1}{2} W'Wf'+V\right]\,,
\label{reda3a}\ee
and substitute it in the other two equations, obtaining a system of two second order equations for $W,f$,
\be\label{reda1a}
 \frac{f}{4} \left(\frac{d W^2}{d-1}-2 (W')^2\right)- \frac{W' }{4}\Big((d+2) W f'-2 (d-1) \left(f' W'\right)'\Big)+V=0
 \ee
 \be
{W'}\left[ W'f' + f\left(W''-\frac{d }{2
   (d-1)}W\right)\right]-V' = 0\,,
\label{reda2a}
\ee
implying four integration constants.
On can use (\ref{reda2a}) to simplify (\ref{reda1a}) which becomes
\begin{align}
&2(d-1)(W'''W'-(W'')^2)+(d-2)(WW''-(W')^2)=\nonumber\\&-{4V\over f} + {2 (W + 2 (d-1) W'')\over
  f W'}V' - 2(d - 1){V''\over f}
\end{align}

Once $W,f$ have been determined by solving (\ref{reda1a}), (\ref{reda2a}), $T$ is obtained from (\ref{reda3a}) and from it the scale factor $e^A$ via (\ref{w54a}).
Then $\f$ is obtained by solving (\ref{w53b}) adding one more extra integration constant. Therefore, we end up again with 5 integration constants.

A scaling symmetry is obvious in the system (\ref{reda1a}), (\ref{reda2a})
\be
f\to {f\over \l^2}\sp W\to \l W
\label{scaling}\ee
This symmetry reflects the symmetry of the original metric in (\ref{f22}) under rescalings
\be
f\to {f\over \l^2}\sp u\to {u\over \l}\sp t\to {t\over \l}
\ee

The two second-order equations for $W,f$ can be manipulated into a single fourth-order non-linear equation for $W$ which is linear in the bulk scalar potential V and its derivatives. It is given in equation (\ref{w56_1b}) of appendix \ref{structure}.

From $W$ we may then compute $f$ from
\be
f=-\frac{2 (d-1) \left(V'' W'-2 V' W''\right)-2W V'+4 V W'}{W' \left(2 (d-1) (W'')^2+(2-d) W W''+(d-2) W'^2-2 (d-1) W^{(3)} W'\right)}
    \label{w56_6a} \ee
and $T$ from
   \be
   T= \frac{ 2 (d-1) W''^2+d W'^2-2 (d-1) W^{(3)} W'- d W  W''}{ (d-2) (d-1)  \left(2 (d-1) (W'')^2+(2-d) W W''+(d-2) W'^2-2 (d-1) W^{(3)} W'\right)}V+
   \label{w56_7a} \ee
 $$
 +\frac{d W^2 W''-4 (d-1) W'^2 W''+W \left(2 (d-1) W''^2-d W'^2+2 (d-1) W^{(3)} W'\right)}{2 (d-2) (d-1) W' \left(2 (d-1) (W'')^2+(2-d) W W''+(d-2) W'^2-2 (d-1) W^{(3)} W'\right)}V'+
 $$
    $$
    +\frac{W'^2-W W''}{(d-2) \left(W'' \left(2 (d-1) W''-(d-2) W\right)+(d-2) W'^2-2 (d-1) W^{(3)} W'\right)}V''.
$$

If the slice curvature vanishes, we must have $T=0$. Imposing this condition on (\ref{reda1a}), (\ref{reda2a})
we obtain
\be
(f'W')'-{d\over 2(d-1)}Wf'=0
\label{r1}\ee
\be
{W'}\left[ W'f' + f\left(W''-\frac{d }{2
   (d-1)}W\right)\right]-V' = 0\,,
\label{r2}
\ee
When $T=0$, equation (\ref{r1}) can be integrated to give
\be
f'W'=e^{{d\over 2(d-1)}\int_{\f_*}^{\f} du {W\over W'}}
\ee
where the arbitrary point $\f_*$ plays the role of the integration constant.

The system of first order equations (\ref{f10_1})-(\ref{w55b}) has in general solutions that have $T\geq 0$ or $T\leq 0$ or $T$ changing sign during the flow.
Only when $T\geq 0$ the solutions of (\ref{f10_1})-(\ref{w55b}) are solutions of the original set of equations (\ref{f23a})-(\ref{f23c}).
Similarly, if we started with a negative curvature slice, then only solutions with $T\geq 0$ should be also solutions of the original equations.
On the contrary solutions of the first order system, where $T$ changes sign along the flow, are not solutions of the second order equations in (\ref{f23a})-(\ref{f23c}).
Such examples are given in appendix \ref{nopot}.
Therefore for our purposes, we must choose only the solutions for which $T\geq 0$.

A detailed study of the differential equations, as well as their singular points and other properties of interest, is presented in the appendices (in particular appendices \ref{exth}, \ref{structure}, and \ref{WcApp}). In the following section, we highlight the main results relevant to a special class of solutions which interpolate between different ``endpoints''---local geometries in which a radial flow can begin or end.

\section{Flow solutions and flow endpoints}\label{sec52}

Our primary interest in this work is the existence of flow solutions which terminate in (A)dS regions of space-time. Towards this end, we now focus our attention on the properties of solutions in the vicinity of flow ``endpoints''. Flow endpoints are defined as points where a solution ``stops". A solution stops, if $\dot\f$ and $\ddot \f$ are both zero at that point, or the geometry ends (a euclidean cycle shrinks to zero size).
In both cases, $W'=0$.
When this happens at finite values of the scalar field $\f$, then the flow ends.
Therefore, this includes all endpoints at finite values of $\f$, which we call ``finite endpoints".

There can also be also flows that end up at $\f=\pm \infty$. All such flows are singular \cite{Bourdier13}.
We shall discuss them here as well, however, as some such solutions may still be acceptable in the context of an effective gravitational theory \cite{Gubser}.

In addition to (A)dS regions, the analysis of appendix \ref{exth} reveals a multitude of local solutions to the flow equations which correspond to finite endpoints. In particular, we identify five distinct classes of {\it finite} endpoints in which a flow solution within our ansatz may begin or end:

\begin{itemize}

\item {\bf dS$_{(d+1)}$ and AdS$_{(d+1)}$ boundaries} appear as minima (maxima) of the superpotential for $W>0$ ($W<0$).  Moreover, they can only appear at extrema of the potential, i.e. $V'=0$. Depending on the sign of the scalar potential near such an extremum, these endpoints can be e.g. the (holographic) boundary of AdS space, or the past/future boundary of dS space.

\item {\bf dS$_2$ boundaries} are characterised, in our ansatz\footnote{It should be noted that if we replace $S^{d-1}$ in our ansatz with EAdS$_{d-1}$, then there are also $AdS_2\times EAdS_{d-1}$ end-points.}, by a local geometry of the form dS$_2\times$S$^{(d-1)}$. They occur at extrema of the potential: $V'=0$, and can appear only in the dS regime ($V>0$) under the assumption that the slice curvature is positive, $T>0$. They always appear as minima (maxima) of the superpotential for $W>0$ ($W<0$).

\item {\bf Nariai (Extremal) horizons} are local solutions in the de Sitter regime in which the blackening function $f$ has a double zero. The local geometry is similar to the ``extremal'' horizon of a Nariai black hole in de Sitter space. Although these horizons are again locally dS$_2 \times$S$^{(d-1)}$ and occur at extrema of the scalar potential, they are distinct from the dS$_2$ boundaries, as will be clarified in detail below. Importantly, not all Nariai horizons can serve as endpoints of the flow. Whether they can serve as endpoints is controlled by the details of the scalar potential, as explained in more detail in Appendix \ref{sho}.
    As shown in section \ref{sec:global}, they appear as endpoints only in flows that are singular.

\item {\bf Shrinking points}, where the spatial sphere smoothly shrinks to zero size. A familiar example of a shrinking point is the center of AdS in global coordinates. These points are maxima (minima) of the superpotential for $W>0$ ($W<0$), and generically have $V'\neq 0 $. They can arise in regions where  $V$ at the endpoint is positive, negative or zero.

\item {\bf Spatial boundaries of Minkowski space-time} always require a vanishing potential at least cubically, at a given point: $V=V'=V''=0$.  In such solutions, the curvature invariants vanish as we approach the endpoint, while the scale factor that controls the size of the sphere S$^{(d-1)}$ diverges. Hence, the geometry is identified as the spatial boundary of Minkowski space-time. Such solutions are possible endpoints of the flow, as $W'=0$. They correspond to minima (maxima) of the superpotential for $W>0$ ($W<0$).

\end{itemize}

Additionally, we catalogue three classes of {\it singular} flow endpoints that appear as $\f\to\pm \infty$. These local solutions are discussed in considerable detail in appendix \ref{asymp}. In brief, we have
\begin{itemize}
\item {\bf Type 0 endpoints}, which are bonafide singularities in the bulk solution---in other words, they can not be resolved by uplifting the solution to that of a higher dimensional gravitational theory. Consequently, they will only arise as acceptable flow endpoints in our analysis when they are hidden behind an event horizon. Near a type 0 solution, the scale factor governing the size of the S$^{(d-1)}$ vanishes. Depending on the details of the gravitational theory, the magnitude of the blackening function $|f(\infty)|$ either diverges or vanishes. These endpoints need not coincide with extrema of the superpotential.

\item {\bf Type I endpoints} may be resolvable via uplift, and can arise in either AdS or dS regimes. They are characterised by a diverging scalar, and a potential $V$ that can either vanish or diverge exponentially in the scalar. When the scalar potential vanishes at the boundary of field space, type I solutions have a diverging scale factor (controlling the size of the sphere). Conversely, when the scalar potential diverges, the scalar factor vanishes. The blackening function $f$ approaches a constant near a type I endpoint. These endpoints need not coincide with extrema of the superpotential. Type I asymptotics may or may not be Gubser-regular, depending on the details of the local solution, as discussed in appendix \ref{app:l1}.

\item {\bf Type II endpoints} may also be resolvable via uplift, and again can arise in both AdS and dS regimes. Like the type I solutions, they appear at boundaries of field space where the scalar potential may either vanish or diverge exponentially in the scalar.  As in type I endpoints, when the scalar potential vanishes at the boundary of field space, type II solutions have a diverging scale factor. When the scalar potential diverges, the scalar factor vanishes.  These endpoints need not coincide with extrema of the superpotential. Type II asymptotics are Gubser-regular.

\end{itemize}

Finally, we note in passing several singular local solutions that, unlike the type 0/I/II endpoints, can appear at finite values of the scalar $\f$ and are {\it never} acceptable as flow endpoints.
These appear when the scale factor vanishes, $e^A\to 0$, and they can be either generic (singular) shrinking endpoints as case 2 in page \pageref{item:IR} or the Special Shrinking Singular Asymptotics that are described as case 4 in page \pageref{sssa}.

Finally, lying strictly outside of our ansatz is another class of flow endpoint, the so called
{\it Extremal Flat Minkowski Horizons}. As in the Minkowski spatial boundaries introduced above,  these endpoints require a scalar potential fine-tuned such that at least $V= V' = V'' = 0$ at the endpoint. They are locally Minkowski as the curvature vanishes at the endpoint, and flat in the sense that the spatial volume diverges there. They are horizons as characterised by the vanishing of $g_{tt}$, and extremal in that the Hawking temperature associated to the horizon vanishes. An explicit example can be found in the discussion below equation \eqref{case2}. Such solutions constitute possible endpoints of the flow because $W'=0$. Additionally, they are minima (maxima) of the superpotential for $W>0$ ($W<0$).

However, as the inverse scale factor $T$ vanishes identically for this class of endpoints, these local solutions can only appear as asymptotic regions of a flow solution in the ansatz with a flat slicing \eqref{c38}. Accordingly, they will not play a prominent role in our investigation, which focuses on flows in the spherically sliced ansatz. That said, they can be useful for comparison between our results and the more familiar flat domain-wall solutions (such as those explored in \cite{KT}).

We now discuss all the above putative {finite} flow endpoints, present in our ansatz, in further detail. For each class of endpoints, the local solution as well as the form of fluctuations around it are reviewed.
\subsection{Locally AdS$_{(d+1)}$ and dS$_{(d+1)}$ boundaries}
These endpoints can be seen to coincide with extrema of the scalar potential. They correspond to constant scalar solutions, in which the dilaton $\f = \f_*$ such that $W'(\f_*) = V'(\f_*) = 0$. They are distinguished by the sign of $V_* \equiv V(\f_*)$, and we shall often find it helpful to parametrise
\begin{equation}\label{eq:VsParam}
V_* = -\frac{d(d-1)}{\ell^2}, \qquad \mathrm{or} \qquad V_* = d(d-1)H^2
\end{equation}
for these endpoints in AdS ($V_* <0$) or dS ($V_*>0$) regimes, respectively.

Solving equation \refeq{f23a} with constant scalar $\f = \f_*$, one finds
\be
A=a u+A_0
\label{f14}\ee
where $a$ and $A_0$ arise as constants of integration. The (A)dS boundaries have non-zero $a$ (we return to the possibility that $a=0$ below). When $a$ is non-zero, the general solution to (\ref{f23b}) is given by
 \begin{equation}
    f=f_0+{1\over (R a)^2}e^{-2A}+\bar C e^{-dA}
\end{equation}
with $\bar C$ and $f_0$ additional constants of integration. From the Hubble equation \refeq{f23c}
\be
d(d-1) a^2 f_0+V_*=0
\label{f15}\ee
one observes that, for these endpoints, the sign of $f_0$ and $V_*$ is correlated. Moreover, from \refeq{f22}, the ansatz is preserved by rescalings of the radial and time coordinates, which in turn can be used to fix $f_0 = -s$ where $s = \mathrm{sgn} (V_*)$.

Therefore, for any $V_*$, the metric can be brought to the form
\begin{equation}
ds^2 = \frac{dr^2}{1-sa^2 r^2+\frac{\mathcal{C}}{r^{d-2}}}-\left(1-sa^2r^2+\frac{\mathcal{C}}{r^{d-2}} \right)dt^2 + r^2d\Omega_{d-1}^2
\end{equation}
by further changing radial coordinate such that
\begin{equation}
r = Re^{a u+A_0},
\end{equation}
rescaling $t\to aRt$, and introducing the convenient constant
\begin{equation}\label{eqr0}
\mathcal{C} = a^2 R^d \bar{C} .
\end{equation}
We next illustrate some explicit examples of this solution.

Using the parametrisation provided by (\ref{eq:VsParam}), when $V_* <0$, we can take without loss of generality $a = -1/\ell$.  The solution becomes
\be
ds^2={dr^2\over {r^2\over \ell^2}+1+{\mathcal{C}\over r^{d-2}}}-\left({r^2\over \ell^2}+1+{\mathcal{C}\over r^{d-2}}\right)dt^2+r^2~d\Omega_{d-1}^2.
\label{f18}\ee

When $\mathcal{C} = 0$ we recover the AdS metric in ``static patch'' coordinates (\ref{a35}), and for $\mathcal{C}<0$, we find the metric of an AdS black hole in these coordinates. The AdS$_{(d+1)}$ boundary is located at $r\to \infty$ in such coordinates.

If $\mathcal{C}>0$, then there is no horizon, but there is a naked singularity at $r=0$ where the Kretschmann scalar diverges. Accordingly, this solution is not acceptable as a putative endpoint as per our admissibility criteria.

Analogously, if $V_* >0$, the parametrisation of (\ref{eq:VsParam}) allows for the choice $a = -H$. In this case the metric becomes
\be
ds^2={dr^2\over 1-{H^2r^2}+{\mathcal{C}\over r^{d-2}}}-\left(1-{H^2r^2}+{\mathcal{C}\over r^{d-2}}\right)dt^2+r^2~d\Omega_{d-1}^2.
\label{f21}\ee

When $\mathcal{C} = 0$, this is the dS$_{(d+1)}$ metric in static patch coordinates, (\ref{a27}). In these coordinates, the future boundary is located at $r\to \infty$.

Note that $g_{tt}$ has at most one extremum, which is a minimum, located at

\begin{equation}\label{ex}
r_{*} =\left(-\dfrac{d-2}{2 H^2}\mathcal{C}\right)^{1\over d} \,.
\end{equation}

Consider first the case $\mathcal{C}>0$. Then the extremum given in \eqref{ex} is complex and lies outside the domain of $r$, $r\in [0,\infty]$. Accordingly, in this case, the temporal component $g_{tt}$ is monotonic and vanishes only once, since it asymptotes to $-\infty$ as $r\to 0$ and to $+\infty$ as $r\to+\infty$. The vanishing of $g_{tt}$ signals the presence of a horizon which, according to the discussion in Appendix \ref{app:J}, is a cosmological horizon. However, the metric has a coordinate singularity at $r=0$, which is not protected by an event horizon. In other words, it is a naked singularity  and is therefore not an acceptable solution.

For $\mathcal{C}<0$, the extremum of $g_{tt}$ in \eqref{f21} lies at some positive value $r_*$ given by \eqref{ex}. In this case, $g_{tt}$ can be non-vanishing everywhere, and the  curvature singularity at $r=0$ therefore is naked, {\it or} it can vanish at two values: $r_c>r_h>0$. The larger root corresponds to a cosmological horizon, whereas the smaller root yields an event horizon (see again the discussion in Appendix \ref{app:J}).

\subsubsection{Fluctuations around AdS$_{(d+1)}$ boundaries}
We have demonstrated that AdS$^{(d+1)}$ boundaries correspond to constant scalar solutions that extremise the scalar potential. In this section, we describe the leading fluctuations away from this endpoint and consistent with the equations of motion in the first-order (superpotential) formalism.

\subsection*{Local maxima in the AdS regime}

We arrange for the extremum in question to occur at $\f=0$, by a shift in $\f$,  unless otherwise stated. To study the form of solutions near such extrema, we assume a regular expansion for the scalar potential, and a Frobenius-like expansion for the superpotential. The analysis of appendix \ref{structure} shows that expansions of this sort are sufficient to capture the leading behaviour of the solutions of interest:

\be
V=-d(d-1)\frac{1}{\ell^2}+{m^2\over 2}\f^2+\sum_{n=3}^{\infty}V_n{\f^n\over n!}
\sp W=\sum_{n=0}^{\infty}{\f^n\over n!}\left(W_n+\hat{W}_{n+\alpha}\f^\alpha \right)
\label{w64}\ee

\noindent
where $\ell$ is a length scale, and $\alpha$ is assumed to be non-integer valued. We next solve the equations of motion perturbatively for small $\f$. It is convenient to parametrize

\be
m^2\ell^2=\Delta(\Delta-d)=- \Delta_+ \Delta_-\,,
\ee

\noindent
equivalently

\begin{equation}
\Delta_{\pm} = \dfrac{1}{2}\left(d \pm \sqrt{d^2+4m^2\ell^2}\right)\,.
\end{equation}

\noindent
At maxima, we have $m^2<0$ and so the BF bound is respected so long as $-\frac{d^2}{4\ell^2}<m^2<0$. This translates into $0<\Delta_-<\frac{d}{2}$ and $\frac{d}{2}<\Delta_+<d$.

Solving \refeq{w56_1b} perturbatively, we find two branches of solutions, $W_\pm$. These two branches are distinguished by the value of the expansion coefficient $W_2$, see \refeq{DEL}. To develop the perturbative solution for the superpotential further, it is efficient to use the series solution \refeq{w64} with $W_2$ = $W_2^\pm$ as a seed in equation \refeq{eqpe}, and compute small fluctuations around this solution as explained around \refeq{sgs}. This procedure supplements the regular terms in the superpotential with additional non-analytic terms, and introduces new constants of integration. In particular,

\begin{itemize}
\item For the $W_-$ branch: The asymptotic form of the fields appearing in the superpotential formalism depend on three integration constants, denoted $C_W$, $C_f$ and $C_T$.
\begin{equation}
\begin{split}
	W_- &=  \dfrac{1}{\ell}\left( 2 (d-1) + \dfrac{\Delta_{-}}{2}\f^2 + \mathcal{O}(\f^3)  \right) + \dfrac{1}{\ell}C_W |\f|^{\frac{d}{\Delta_-}} -\dfrac{C_f}{\ell}\frac{\Delta_- ^3 }{2 d  (d+2 \Delta_- )}|\f|^{\frac{d}{\Delta_-}+2}\\&-\dfrac{C_T}{\ell}\frac{\Delta_- ^2 (-d+\Delta_- +2) }{2 (\Delta_- +1)  (d-2 (\Delta_-+1))}|\f|^{\frac{2}{\Delta_{-}}+2} +\dots
\end{split}
\end{equation}	

\begin{equation}
f_- = 1 +C_f \dfrac{\Delta_-}{d} |\f|^{\frac{d}{\Delta_-}} + C_T|\f|^{\frac{2}{\Delta_-}} + \dots
\end{equation}
	
\begin{equation}
T_- = \dfrac{1}{\ell^2} C_T |\f|^{\frac{2}{\Delta_-}}+\dots
\end{equation}

\noindent
We have normalized $f=1$ at the endpoint\footnote{The sign of $f$ is correlated with the value of $V$ at the endpoint. If $V<0$ then $f>0$ and vice versa, see (\ref{f15}).}, which further fixes $W_0\,\ell=2(d-1)$. The ellipses in these expansions contains subleading contributions.

To better understand the $W_-$ solution we solve the relations (\ref{w53}) and (\ref{w53b}), to obtain
\begin{align}
	&\f_-(u) = C_- e^{u\Delta_-/\ell} + C_-^{\Delta_+/\Delta_-} \dfrac{ C_Wd}{\Delta_-(d-2\Delta_-)}e^{u\Delta_+/\ell}+ \nonumber\\& + C_T \frac{\Delta_-  (-d+\Delta_- +2)}{2 (d-2 (\Delta_- +1))}e^{(\Delta_-+2)u/\ell}+ \dots
\end{align}
\begin{equation}
A_-(u) = -\dfrac{u-u_*}{\ell} - C_-^2\dfrac{1}{8(d-1)} e^{2u \Delta_-/l} + \dots\\
\end{equation}
\begin{equation}
f_-(u) = 1  +C_f \dfrac{\Delta_-}{d} C_-^{\frac{d}{\Delta_-}}e^{ud/\ell} + C_T C_-^{\frac{2}{\Delta_-}} e^{2u/\ell} + \dots
\end{equation}
In the expansion, we have assumed that $\f$ approaches zero when we arrive at the extremum. Consequently, we must take $u\to-\infty$ and therefore the scale factor diverges linearly as $A\to+\infty$ while $f$ approaches  unity. Therefore, the metric asymptotes to the boundary of $AdS_{d+1}$, and we find that the $W_{-}$ branch at AdS maxima correspond to ``UV endpoints'' when viewed holographically.

As written, this solution appears to contain a total of six integration constants: $C_-, C_W,C_f, C_T$, as well as $u_*$ and the asymptotic value of $f$ near the $AdS_{d+1}$ boundary (which we have chosen above to take the value one). These constants are not all independent, however. In particular, either $u_*$ or $C_T$ can be taken to be the constant of integration for the metric function $A$---both control the overall scale of the space-time's volume form.

Taking a holographic perspective, it is often convenient to think of the constant $C_-$ as controlling the source for the scalar operator dual to $\f$, while $C_W$ roughly controls its expectation value. Similarly, the constant $C_f$ is a proxy for the one-point function of the $tt$-component of the dual stress-energy tensor. As these are all dimensionful quantities in the dual conformal field theory, it is natural to quantify them in units related to the size of the spatial sphere on which the theory resides.

The two quantities in (\ref{ev1}) and (\ref{ev2}) behave as
\be
f\dot A e^{dA}=-{1\over \ell}e^{-{d\over \ell}(u-u_*)}+\cdots~~~\to ~~~-\infty
\ee
\be
\dot f e^{dA}=C_f{\Delta_-\over \ell}C_-^{d\over \Delta_-}e^{d u_*\over \ell}+\cdots ~~~\to ~~~{\rm constant}
\ee

\item For the $W_+$ branch: In this case the asymptotic form of the fields appearing in the superpotential formalism depend only on two integration constants, denoted $C_f$ and $C_T$:

\begin{equation}
\begin{split}
W_+ &=  \dfrac{1}{\ell}\left( 2 (d-1) + \dfrac{\Delta_{+}}{2}\f^2 + \mathcal{O}(\f^3)  \right)  -\dfrac{C_f}{\ell}\frac{\Delta_+ ^3 }{2 d  (d+2 \Delta_+ )}|\f|^{\frac{d}{\Delta_+}+2}\\&-\dfrac{C_T}{\ell}\frac{\Delta_+ ^2 (-d+\Delta_+ +2) }{2 (\Delta_+ +1)  (d-2 (\Delta_++1))}|\f|^{\frac{2}{\Delta_{-}}+2} +\dots
\end{split}
\end{equation}

\begin{equation}
f_+ = 1 +C_f \dfrac{\Delta_+}{d} |\f|^{\frac{d}{\Delta_+}} + C_T|\f|^{\frac{2}{\Delta_+}} + \dots
\end{equation}

\begin{equation}
T_+ = \dfrac{1}{\ell^2} C_T |\f|^{\frac{2}{\Delta_+}}+\dots
\end{equation}

\noindent
We have again normalized $f(\f\to 0)=1$. Solving the relations (\ref{w53}) and (\ref{w53b}) we now obtain

\begin{equation}
	\f_+(u) = C_+ e^{u\Delta_+/\ell} +  C_T \frac{\Delta_+  (-d+\Delta_+ +2)}{2 (d-2 (\Delta_+ +1))}e^{(\Delta_++2)u/\ell}+ \dots
\end{equation}

\begin{equation}
A_+(u) = -\dfrac{u-u_*}{\ell} - C_+^2\dfrac{1}{8(d-1)} e^{2u \Delta_+/\ell} + \dots
\end{equation}

\begin{equation}
f_+(u) = 1  +C_f \dfrac{\Delta_+}{d} C_+^{\frac{d}{\Delta_+}}e^{ud/\ell} + C_T C_+^{\frac{2}{\Delta_+}} e^{2u/\ell} + \dots
\end{equation}

In the expansion we have assumed that $\f$ approaches zero. Consequently, we must have $u\to-\infty$ and the scale factor diverges linearly to $A\to+\infty$ while $f$ approaches unity. Therefore, the metric again asymptotes to the boundary of $AdS$. Accordingly, the $W_+$ branch at an AdS maximum also correspond to UV endpoints.

As before, the $AdS$ boundary in this branch of solutions encourages a holographic interpretation of the various constants which appear. Again we have fixed one constant of integration by demanding that $f(\f\to0) = 1$, and one of $u_*$, $C_T$ is redundant---both control the overall scale of the space-time volume form.

This leaves $C_f$, which again governs the one-point function of the $tt$--component of the dual stress tensor, and $C_+$ which controls the leading fall-off of the scalar field near the boundary. Note, however, that in this branch of solutions this fall-off is proportional to the holographic dual of the scalar operator's expectation value. Importantly, this branch of solutions does not allow for holographic deformations by a source for the scalar operator.

\end{itemize}

\subsubsection*{Local minima in the AdS regime}

For minima in $AdS$, we expand the potential and superpotential near the critical point at $\f=0$ as in \refeq{w64}. The difference is that in this case $m^2>0$, so that necessarily we now have $\Delta_-<0$ and $d<\Delta_+$. The allowed ranges for $\Delta$ determine the allowed deformations about the critical point.

The linearised equation \refeq{eqpe} is solved around the $W_2^\pm$ solution. Contrary to the previous case, now only the $W_+$ branch is allowed \cite{curved}. Regularity of the curvature invariants (computed in appendix \ref{sect:inv_sphere}) for the $W_-$ branch implies that there is no regular flow compatible with the spherically foliated ansatz that ends at a minimum of the potential in the AdS regime.

The asymptotic form of the fields appearing in the superpotential formalism for the $W_+$ branch, depend on two integration constants, denoted  $C_f$ and $C_T$:

\begin{equation}
\begin{split}
	W_+ &=  \dfrac{1}{\ell}\left( 2 (d-1) + \dfrac{\Delta_{+}}{2}\f^2 + \mathcal{O}(\f^3)  \right)  -\dfrac{C_f}{\ell}\frac{\Delta_+ ^3 }{2 d  (d+2 \Delta_+ )}|\f|^{\frac{d}{\Delta_+}+2}\\&-\dfrac{C_T}{\ell}\frac{\Delta_+ ^2 (-d+\Delta_+ +2) }{2 (\Delta_+ +1)  (d-2 (\Delta_++1))}|\f|^{\frac{2}{\Delta_{-}}+2} +\dots
\end{split}
\end{equation}

\begin{equation}
f_+ = 1 +C_f \dfrac{\Delta_+}{d} |\f|^{\frac{d}{\Delta_+}} + C_T|\f|^{\frac{2}{\Delta_+}} + \dots
\end{equation}

\begin{equation}
T_+ = \dfrac{1}{\ell^2} C_T |\f|^{\frac{2}{\Delta_+}}+\dots
\end{equation}

\noindent
We have normalized once more $f(\f\to 0)=1$. Solving the relations (\ref{w53}) and (\ref{w53b}) we obtain

\begin{equation}
\f_+(u) = C_+ e^{u\Delta_+/\ell} +  C_T \frac{\Delta_+  (-d+\Delta_+ +2)}{2 (d-2 (\Delta_+ +1))}e^{(\Delta_++2)u/\ell}+ \dots
\end{equation}

\begin{equation}
A_+(u) = -\dfrac{u-u_*}{\ell} - C_+^2\dfrac{1}{8(d-1)} e^{2u \Delta_+/\ell} + \dots
\end{equation}

\begin{equation}
f_+(u) = 1  +C_f \dfrac{\Delta_+}{d} C_+^{\frac{d}{\Delta_+}}e^{ud/\ell} + C_T C_+^{\frac{2}{\Delta_+}} e^{2u/\ell} + \dots
\end{equation}
In the expansion, we have assumed that $\f$ approaches zero. Consequently, we must have $u\to-\infty$ and the scale factor diverges linearly to $A\to+\infty$ while $f$ approaches unity. Once more, we find that the metric asymptotes to the boundary of $AdS$. Therefore, the $W_+$ branch at AdS {\it minima} also correspond to ``UV endpoints''.

Accordingly, we may once again translate the integration constants appearing in this solution holographically. We fix one constant of integration by demanding that $f(\f\to0) = 1$, and again one of $u_*$, $C_T$ is redundant---both control the overall scale of the space-time volume form.

The constant $C_f$ controls the one-point function of the $tt$--component of the dual stress tensor, leaving $C_+$ which governs the leading fall-off of the scalar field near the boundary. In this branch, it is also the case that this fall-off is roughly the holographically dual of the scalar operator's expectation value. Therefore, this branch of solutions also does not allow for holographic deformations by a source for the scalar operator.

\subsubsection{Fluctuations around dS$_{(d+1)}$ boundaries}

As noted above, the dS$_{(d+1)}$ boundary endpoints appear at extrema of  the scalar potential. Indeed, the discussion of fluctuations around these extrema in the dS regime closely mirrors that of the AdS$_{(d+1)}$ case.

\subsubsection*{Local minima in the dS regime}

Around a minimum in dS, which we again position at $\f=0$, we expand

\be
V=d(d-1)H^2+{m^2\over 2}\f^2+\sum_{n=3}^{\infty}V_n{\f^n\over n!}
\sp W=\sum_{n=0}^{\infty}{\f^n\over n!}\left(W_n+\hat{W}_{n+\alpha}\f^\alpha \right).
\label{w64b}\ee

\noindent
We parametrize

\be
\frac{m^2}{H^2}=\Delta(d-\Delta)= \Delta_+ \Delta_-\,,
\ee

\noindent
or equivalently

\begin{equation}\label{dpm}
\Delta_{\pm} = \dfrac{1}{2}\left(d \pm \sqrt{d^2-\frac{4m^2}{H^2}}\right)\,.
\end{equation}

\noindent
At minima, we have $m^2>0$ and the analogous BF bound in de Sitter gives $0<m^2<\frac{d^2H^2}{4}$. This translates into $0<\Delta_-<\frac{d}{2}$ and $\frac{d}{2}<\Delta_+<d$. Note that formally, the results for $dS$ may be obtained from those in $AdS$ by sending $(V,f,T,W)\to(-V,-f,-T,W)$, which leaves the equations of motion (\ref{f10_1}-\ref{w55b}) unchanged.

Consider first the branch of solutions whose leading behaviour near the critical point is described by \refeq{w64b} with $\hat{W}_{n+\alpha}=0$. In this case, equation \refeq{eqpe} is again solved around the $W_2^\pm$ solution given in \refeq{DEL}. As before there are two branches of solutions:

\begin{itemize}
\item $W_-$ branch. The asymptotic form of the fields appearing in the superpotential formalism depend on three integration constants, denoted by $C_W$, $C_f$ and $C_T$.

\begin{equation}
\begin{split}
	W_- &=  H\left( 2 (d-1) + \dfrac{\Delta_{-}}{2}\f^2 + \mathcal{O}(\f^3)  \right) + H C_W |\f|^{\frac{d}{\Delta_-}} +HC_f\frac{\Delta_- ^3 }{2 d  (d+2 \Delta_- )}|\f|^{\frac{d}{\Delta_-}+2}\\&+HC_T\frac{\Delta_- ^2 (-d+\Delta_- +2) }{2 (\Delta_- +1)  (d-2 (\Delta_-+1))}|\f|^{\frac{2}{\Delta_{-}}+2} +\dots
\end{split}
\end{equation}

\begin{equation}
f_- = -1 +C_f \dfrac{\Delta_-}{d} |\f|^{\frac{d}{\Delta_-}} + C_T|\f|^{\frac{2}{\Delta_-}} + \dots
\end{equation}

\begin{equation}
T_- = H^2 C_T |\f|^{\frac{2}{\Delta_-}}+\dots
\end{equation}

\noindent
We have normalized $f=-1$ at the $\f = 0$ endpoint, this further fixes $W_0=2(d-1)H$. The dots in the expansion stand for higher order contributions. To better understand the solution we again solve the relations (\ref{w53}) and (\ref{w53b}), to obtain
\begin{align}
\f_-(u) &= C_- e^{u\Delta_-H} + C_-^{\Delta_+/\Delta_-} \dfrac{ C_Wd}{\Delta_-(d-2\Delta_-)}e^{u\Delta_+ H} -\nonumber \\ & - C_T \frac{\Delta_-  (-d+\Delta_- +2)}{2 (d-2 (\Delta_- +1))}e^{(\Delta_-+2)uH}+ \dots
\end{align}
\begin{equation}
A_-(u) = - H (u-u_*) - C_-^2\dfrac{1}{8(d-1)} e^{2u \Delta_- H} + \dots
\end{equation}
\begin{equation}
f_-(u) = -1  +C_f \dfrac{\Delta_-}{d} C_-^{\frac{d}{\Delta_-}}e^{ud H} + C_T C_-^{\frac{2}{\Delta_-}} e^{2u H} + \dots
\end{equation}
In the expansion, we have assumed that $\f$ approaches zero. Consequently, we must have $u\to-\infty$ and the scale factor diverges linearly to $A\to+\infty$ while $f$ approaches $-1$. Therefore, as $u\to -\infty$, the spatial sections  become large. This is the near boundary behaviour of dS in global coordinates.
If $u$ is taken to be the time coordinate, then this region corresponds to a past dS boundary, while if $-u$ is the time coordinate this is the future dS boundary.

\item $W_+$ branch. The asymptotic form of the fields appearing in the superpotential formalism depend only on two integration constants, denoted $C_f$ and $C_T$.

\begin{equation}
\begin{split}
	W_+ &=  H\left( 2 (d-1) + \dfrac{\Delta_{+}}{2}\f^2 + \mathcal{O}(\f^3)  \right)  +\dfrac{C_f}{\ell}\frac{\Delta_+ ^3 }{2 d  (d+2 \Delta_+ )}|\f|^{\frac{d}{\Delta_+}+2}\\&+\dfrac{C_T}{\ell}\frac{\Delta_+ ^2 (-d+\Delta_+ +2) }{2 (\Delta_+ +1)  (d-2 (\Delta_++1))}|\f|^{\frac{2}{\Delta_{-}}+2} +\dots
\end{split}
\end{equation}

\begin{equation}
f_+ = -1 +C_f \dfrac{\Delta_+}{d} |\f|^{\frac{d}{\Delta_+}} + C_T|\f|^{\frac{2}{\Delta_+}} + \dots
\end{equation}

\begin{equation}
T_+ = \dfrac{1}{\ell^2} C_T |\f|^{\frac{2}{\Delta_+}}+\dots
\end{equation}

\noindent
We have again normalized $f(\f\to 0)=-1$. Solving the relations (\ref{w53}) and (\ref{w53b}) we obtain

\begin{equation}
\f_+(u) = C_+ e^{u\Delta_+H} -  C_T \frac{\Delta_+  (-d+\Delta_+ +2)}{2 (d-2 (\Delta_+ +1))}e^{(\Delta_++2)u H}+ \dots
\end{equation}

\begin{equation}
A_+(u) = -H(u-u_*) - C_+^2\dfrac{1}{8(d-1)} e^{2u \Delta_+ H} + \dots
\end{equation}

\begin{equation}
f_+(u) = -1  +C_f \dfrac{\Delta_+}{d} C_+^{\frac{d}{\Delta_+}}e^{ud H} + C_T C_+^{\frac{2}{\Delta_+}} e^{2u H} + \dots
\end{equation}
In the asymptotic expansion around the minimum we have assumed that $\f$ approaches zero. This is only compatible with an expansion around $u\to-\infty$. The scale factor diverges linearly to $A\to+\infty$ while $f$ approaches $-1$. Therefore, the metric asymptotes to a $dS$ boundary. Like the previous case, this is the past or future boundary of $dS$ depending on the definition of the time coordinate.

Unlike the AdS case, in dS there is no reason to require that $m^2<\frac{d^2H^2}{4}$. When
$m^2>\frac{d^2H^2}{4}$ the asymptotics of the expansion change and the solutions become oscillatory.
However, the form of the expansion remain similar. The superpotential become complex, but the final solutions for $f$ and $A$ can be made real.
\end{itemize}

\subsubsection*{Local maxima in the dS regime}
Similarly, around maxima in dS we can again expand the potential and superpotential near the critical point $\f=0$ as in \refeq{w64b}. In this case, however, $m^2<0$, necessarily implying $\Delta_-<0$ and $d<\Delta_+$. The allowed ranges for $\Delta$ determine the allowed deformations of the critical point.

Equations (\ref{f10_1})-(\ref{w55b}) are once more solved perturbatively. Unlike in the previous case, now only the $W_+$ branch is allowed. In fact, the undeformed $W_-$ solution can be shown to be incompatible with the spherical foliation of our metric ansatz. Additionally, regularity of the curvature invariants (computed in appendix \ref{sect:inv_sphere}) for the $W_-$ branch requires all the deformations to be set to zero.  This implies that no $W_-$ flow can end regularly in a maximum in a dS regime.

The asymptotic form of the fields appearing in the superpotential formalism for the $W_+$ branch depend on two integration constants, denoted $C_f$ and $C_T$:

\begin{equation}
\begin{split}
	W_+ &=  H\left( 2 (d-1) + \dfrac{\Delta_{+}}{2}\f^2 + \mathcal{O}(\f^3)  \right)  +HC_f\frac{\Delta_+ ^3 }{2 d  (d+2 \Delta_+ )}|\f|^{\frac{d}{\Delta_+}+2}\\&+HC_T\frac{\Delta_+ ^2 (-d+\Delta_+ +2) }{2 (\Delta_+ +1)  (d-2 (\Delta_++1))}|\f|^{\frac{2}{\Delta_{-}}+2} +\dots
\end{split}
\end{equation}

\begin{equation}
f_+ = -1 +C_f \dfrac{\Delta_+}{d} |\f|^{\frac{d}{\Delta_+}} + C_T|\f|^{\frac{2}{\Delta_+}} + \dots
\end{equation}

\begin{equation}
T_+ = H^2 C_T |\f|^{\frac{2}{\Delta_+}}+\dots
\end{equation}

\noindent
We have again normalized $f(\f\to 0)=-1$. Solving the relations (\ref{w53}) and (\ref{w53b}) we obtain

\begin{equation}
\f_+(u) = C_+ e^{u\Delta_+ H} -  C_T \frac{\Delta_+  (-d+\Delta_+ +2)}{2 (d-2 (\Delta_+ +1))}e^{(\Delta_++2)u H}+ \dots
\end{equation}

\begin{equation}
A_+(u) = -H(u-u_*) - C_+^2\dfrac{1}{8(d-1)} e^{2u \Delta_+H} + \dots
\end{equation}

\begin{equation}
f_+(u) = -1  +C_f \dfrac{\Delta_+}{d} C_+^{\frac{d}{\Delta_+}}e^{udH} + C_T C_+^{\frac{2}{\Delta_+}} e^{2uH} + \dots
\end{equation}

In the expansion we assumed that $\f$ approaches zero. Consequently, we must have $u\to-\infty$ and the scale factor diverges linearly to $A\to+\infty$ while $f$ approaches $-1$.  This is the near boundary behaviour of dS in global coordinates. It corresponds to the future or past boundary depending on the choice of $u$ or $-u$ as the time coordinate. Accordingly, we find that the $W_+$ branch at a dS maxima also corresponds to a $dS$ boundary.

\subsection{Locally dS$_2$ endpoints}
As explained in the discussion below equation \refeq{ex}, in a dS regime when the integration constant $\mathcal{C} < 0$ in the metric \refeq{f21}, there exist constant scalar endpoint solutions in which the $g_{tt}$ metric component may possess two roots.

In Appendix \ref{app:J} these two roots were shown to give rise to a cosmological and event horizon in the space-time.

 This fact allows for the possibility that through tuning integration constants, the roots may be made to coincide. This will occur when the extremum of $g_{tt}$ precisely corresponds with a root of $g_{tt}$. In such a case, the location of the (extremal) horizon $r_h$ and the integration constant $\mathcal{C}$ are readily found to be

   \begin{equation}\label{dzero}
   r_h=\sqrt{\dfrac{d-2}{dH^2}} \hspace{2cm} \mathcal{C}_* = -r_h^d\frac{2 H^2}{d-2} = -\frac{1}{2d^{d/2}}\left(\frac{d-2}{H^2}\right)^{\frac{d+2}{2}}.
   \end{equation}

In terms of these quantities, it is interesting to consider, in more detail, the limit in which the event and cosmological horizons present in the space-time with metric \refeq{f21} coincide. To this end, we consider this metric with integration constant $\mathcal{C}$ given by
\begin{equation}
\mathcal{C} = \mathcal{C}_*\left( 1-\epsilon^2\right),
\end{equation}
where $\epsilon$ will be taken to parametrise a small deformation from the degenerate horizon solution in \eqref{dzero}. In this case, the extremum of $f(r)$ will be lifted to a small positive value and the cosmological and event horizons no longer coincide. To understand the region of space-time between these horizons in the small $\epsilon$ limit, we introduce a new radial coordinate $\rho$ such that
\begin{equation}
r = r_h + \epsilon\frac{\sqrt{2}}{d H} \rho.
\end{equation}
with $r_h$ given by \eqref{dzero}.

In terms of $\rho$, the blackening function $f$ in \refeq{f21} is
\begin{equation}
f(\rho) = \frac{2}{d}\left(1-\rho^2 \right)\epsilon^2 + \mathcal{O}(\epsilon^3).
\end{equation}
Further introducing a scaled time coordinate $t\to t\sqrt{d/2}\epsilon$, one can now take the $\epsilon \to 0$ limit to arrive at the metric
\begin{equation}
\lim_{\epsilon\to 0}\,ds^2  = -\left(1-\rho^2 \right)dt^2+\tilde{R}^2\frac{d\rho^2}{\left(1-\rho^2 \right)}+(d-2)\tilde{R}^2\, d\Omega^2_{d-1}
\end{equation}
where we have introduced for convenience the length scale
\begin{equation}
\tilde{R}^2 = \frac{1}{dH^2}.
\end{equation}

We therefore find that in this limit, the solution between the event and cosmological horizons at $\rho = \pm 1$
becomes that of $dS_2 \times S^{d-1}$, the Nariai geometry.

In fact $dS_2 \times S^{d-1}$ space-times arise in this gravitational theory not only as limits of other solutions, but also as bona fide constant scalar solutions to the equations of motion. In particular, returning to \refeq{f14} we consider the special case $a=0$ such that
\begin{equation}
A = A_0.
\end{equation}

Turning next to equation \refeq{f23b}, we integrate to obtain
    \begin{equation}
    \label{eq:divergf}
        f = f_0 + \overline{C} u -  (d-2)\dfrac{e^{-2 A_0}}{R^2} u^2\,,
    \end{equation}
    \noindent
 where again $f_0$ and $\overline{C}$ are integration constants. This solution is consistent with the first order equation \refeq{f23c} provided that
    \begin{equation}\label{eqcons}
    R^2 e^{2A_0} V_*=(d-2)(d-1)\,.
    \end{equation}
     The curvature scale of the sphere is real only if $V_*>0$\, in dimension $d>2$ \footnote{Using discrete symmetries of the equations of motion, one may obtain a solution for $V_*<0$ by analytically continuing $R\to i R$. This  is essentially equivalent to modifying our ansatz with a hyperbolic foliation instead of the spherical foliation in \refeq{f22}. We discuss this case in section \ref{hyperbolic}.}.

    Setting $A_0=0$ without loss of generality, the metric \refeq{f22} becomes
    \begin{equation}\label{cds}
        ds^2 = \dfrac{du^2}{f(u)}-f(u)dt^2 + R^2 d\Omega_{d-1}^2\,,
    \end{equation}
    \noindent
    with $f$ as given in \refeq{eq:divergf}. 
    The behavior of the functions $W,f,T$ around such a point is given in equations (\ref{wds2}), (\ref{gsuno}).
    A constant shift $u \to \overline{u} + \frac{R^2 \overline{C}}{2(d-2)} $ can be used to ``complete the square''  and define

    \begin{equation}
		f(\overline{u}) = \left(f_0+\frac{\overline{C}^2 R^2}{4(d-2)}\right) -\frac{d-2}{R^2} \overline{u}^2\equiv\tilde{f}_0 -h^2 \overline{u}^2\,.
    \end{equation}

  Different possibilities now arise, depending on whether $\tilde{f}_0$ is equal to, greater than, or less than zero.

    Consider first $\tilde{f}_0=0$. In this case we can define $\overline{u} = e^{h U}$ so that the metric becomes

    \begin{equation}\label{sdn}
    ds^2 = -dU^2 + h^2 e^{2 h U} dt^2 + R^2 d\Omega_{d-1}^2\,.
\end{equation}
	This is $dS_2\times S^{d-1} $ where the $dS$ factor is in Poincare coordinates \refeq{a26}, in which the future boundary is located at $U\to\infty$.
	
	If instead $\tilde{f}_0>0$, we can rescale both the time and radial coordinates
	
	\begin{equation}
	T = t \sqrt{\tilde{f}_0}\hspace{0.5cm} U = \dfrac{\overline{u}}{\sqrt{\tilde{f}_0}}
\end{equation}		
	so that we have again $dS_2\times S^{d-1} $, now with $dS_2$ in the static patch coordinates \refeq{a27}:
	
	\begin{equation}\label{eq:dS2S3}
	ds^2 = \dfrac{dU^2}{1-h^2 U^2} - (1-h^2 U^2)dT^2 + R^2 d\Omega_{d-1}^2\,.
	\end{equation}	
In these coordinates, the future boundary is located at $U\to\infty$.	

    Finally, if $\tilde{f}_0<0$, we can rescale the radial and time coordinates

    \begin{equation}
	T = t \sqrt{-\tilde{f}_0}\hspace{0.5cm} U = \dfrac{\overline{u}}{\sqrt{-\tilde{f}_0}}
\end{equation}	
and the metric becomes
    \begin{equation}\label{sdt}
	ds^2 = -\dfrac{dU^2}{1+h^2 U^2} +(1+h^2 U^2)dT^2 +R^2 d\Omega_{d-1}^2\,.
	\end{equation}

This metric is globally related to that of \refeq{eq:dS2S3} by double Wick rotation. Locally, near $U\to\infty$, these two solutions are isometric.

\subsubsection{Fluctuations around dS$_2$ endpoints}
The first order formalism can be used to study the deformations of these dS$_2$ endpoints. These solutions arise when the leading form of the superpotential is as in \eqref{w64b} with $W_n =0$. As we show in appendix \ref{D.2.3}, such solutions are characterised by a superpotential with leading power law behaviour whose exponent is determined by the indicial equation
\be\label{idn5}
(\alpha-2)^2(d-1)V_2-(\alpha-3)V_0 = 0
\ee
where $\a$ is introduced in Eq. \eqref{w64b}, and in the language of \eqref{w64b} we identify $V_0 = d(d-1)H^2$ and $V_2 = m^2$.

By analogy with the preceding solutions, we  find it convenient to parametrise the exponent as

\be
\alpha = 2+\frac{1}{\delta_\pm} \qquad \mathrm{where} \qquad \delta_\pm \equiv \frac{1}{2}\pm\sqrt{\frac{1}{4}-(d-1)\frac{V_2}{V_0}} .
\ee
This implies that there are two branches of solutions, $W_\pm$, with
\begin{align}
W_\pm = W_0\f^{2+1/\delta_\pm} \left(1 -\frac{(d-1) (\delta_\pm -1) (2 \delta_\pm +1) V_3 }{2 \delta_\pm ^2 \left(9 \delta_\pm ^2-1\right)
\label{wds2}   V_0}\f + O(\f^2) \right).
\end{align}
By solving $W'=\dot{\f}$, one finds that to leading order
\begin{equation}
\f \simeq \left(-\frac{W_0 (2 \delta_\pm +1) u}{\delta_\pm ^2}\right)^{-\delta_\pm }+\dots
\end{equation}
while the leading behaviour of the blackening function and scale factor are shown to be given by
\begin{equation}\label{gsuno}
f \simeq -\dfrac{V_0}{d-1} u^2+\ldots  \qquad \mathrm{and}\qquad  T \simeq \dfrac{V_0}{(d-1)(d-2)}+\ldots.
\end{equation}

We first consider the  $W_-$ branch around maxima $(m^2 < 0)$ in a dS region. In this case, as we have taken the critical point to be at $\f=0$ and because $\delta_-<0$, the solution requires $u\to 0$. Therefore, from \eqref{gsuno} we observe that $f$ vanishes quadratically in this case, suggesting the appearance of an extremal horizon. Indeed, this solution corresponds to the {\bf Nariai extremal horizon} limit in which the event and cosmological horizons coincide.

Conversely, for  the $W_+$ branch around maxima and both $W_\pm$ branches around minima, the solution is reached asymptotically as $u\to \infty$. Accordingly, from  \refeq{gsuno} one finds that the blackening function diverges quadratically and the scale factor $T$ approaches a constant. In particular, the metric is of the form

\begin{equation}
ds^2 \simeq -(d-1)\dfrac{du^2}{V_0 u^2} + \dfrac{V_0}{d-1}u^2 dt^2  + \dfrac{(d-1)(d-2)}{V_0} d\Omega_{d-1}^2\,.
\end{equation}
As $u\to \infty$, this space-time coincides with the {\bf future boundary} of $dS_2\times S^{(d-1)}$. We note in passing that in the case of a hyperbolic slicing, analogous solutions would arise in which the metric asymptotes to the boundary of $AdS_2 \times H^{(d-1)}$.

\subsection{Shrinking end-points}\label{shr}

The endpoints we have discussed so far all occur at extrema of the scalar potential. This is not the only possibility, however. Indeed, endpoints can also arise away from extrema of the potential when $e^A\to 0$ and $f$ diverges. We now turn our attention to this possibility. A detailed study of the local properties of these solutions can be found in appendix \ref{G.2.2}.

We parametrize the potential and superpotential in the neighborhood of this critical point with $W'=0$ at $\f=\f_0$ as

\begin{equation}
V=\sum_{n=0}^{\infty}\dfrac{V_n}{n!}(\f-\f_0)^n\, \qquad \mathrm{and} \qquad W(\f)=W_0+\sum_{n=2}^{\infty}{W_n\over n!}(\f-\f_0)^n.
\end{equation}
In this section, we shall be interested in both AdS and dS regimes, and therefore we allow $V_0$ to take either sign. Moreover, by assumption $V_1$ is non-vanishing.

In appendix \ref{G.2.2} we solve the set of equations (\ref{f10_1})-(\ref{w55b}) in this ansatz perturbatively. In particular, from equations (\ref{C110}-\ref{C114}) we find:

\begin{equation}\label{eqdiv}
W = W_0 \left(1-\dfrac{1}{2(d-1)}(\f-\f_0)^2  -{2 (d+2) V_0 +d(d-1) V_2 \over ( d-1)^2 (d+2) V_1}(\f-\f_0)^3+ \mathcal{O}((\f-\f_0)^4)\right)\,,
\end{equation}

\begin{equation}\label{soch}
f = \dfrac{2(d-1)^2}{d}\dfrac{V_1}{W_0^2}\dfrac{1}{\f-\f_0} + \dfrac{(d-1)[d(d-1)V_2-2(d+2)V_0]}{d(d+2)W_0^2}+\mathcal{O}(\f-\f_0)\,,
\end{equation}

\begin{equation}\label{sochu}
T = \dfrac{V_1}{2d}\dfrac{1}{\f-\f_0} + \dfrac{d(d-1)V_2+2(d+2)V_0}{4d(d-1)(d+2)}+\mathcal{O}(\f-\f_0)\,.
\end{equation}

Apart from $\f_0$, the position of this critical point in scalar space, the solution contains a single integration constant, $W_0$.  By employing the symmetry of the ansatz highlighted in (\ref{scaling}), near such a point $W_0$ can be rescaled to unity.

Evidently, both $f$ and $T$ diverge as we approach the critical point. However, the geometry is in fact regular, as demonstrated in appendix \ref{G.2.2} using the curvature invariants computed in appendix \ref{sect:inv_sphere}. We shall call such end-points, the {\it ``shrinking end-points"}.

As $T>0$ by assumption, if the potential is decreasing ($V_1<0$) this solution is only possible if we approach the critical point from the left, $\f\to \f_0^-$. Similarly,  if the potential is increasing ($V_1>0$) this solution is only possible if we approach the critical point from the right $\f\to \f_0^+$. Accordingly, we find that in the vicinity of such a singular point, {\it the scalar field $\f$ climbs \textit{up} the potential away from the shrinking end-point}.

Furthermore, the sign of $f$ and $T$ are correlated as we approach this point. Specifically, since $\lim_{\f\to \f_0}T = +\infty$ one must also have $\lim_{\f\to \f_0}f = +\infty$ independent of the sign of $V_0$. This observation plays an important role when discussing the global structure of the solutions.

To describe the corresponding geometry, we consider the metric \refeq{f22} but use $\f$ as a radial variable, with $du\to\frac{1}{W'} d\f$. Then, using the definition of $T$ (\ref{w54a}) we have

\begin{equation}
	\label{eq:G4b}
	\begin{split}
		&ds^2 = \frac{d\f^2}{fW'^2} - \frac{f}{T}R^2 dt^2 + \frac{1}{T}d\Omega_{d-1}^2 \\
	\end{split}
\end{equation}	   	

\noindent
We change variables once more such that $g_{\f\f}d\f^2 = d\rho^2$. Using the asymptotic solution (\ref{eqdiv}) one finds

\begin{equation}
   		\f-\f_0\simeq \frac{V_1}{2d}\rho^2 + V_1 \frac{2(d+2)V_0+3 d (d-1)V_2}{24 (d-1)(d+2)d^2}\rho^4 +\mathcal{O}(\rho^6)\,,
\end{equation}
where $\f\to \f_0$ as $\rho \to 0\,$.  Together, this change of variables along with (\ref{eqdiv}) yields a metric of the form

   \begin{equation}
   		ds^2 = d\rho^2 - \frac{4(d-1)^2 R^2}{W_0^2}(1-\frac{V_0}{d(d-1)}\rho^2)dt^2 + \left(1 - \frac{1}{3}\frac{V_0}{d(d-1)}\rho^2\right)\rho^2~d\Omega_{d-1}^2+\dots
   	\end{equation}
   	which depends on the value (and sign) of the potential at the singular point,  $V_0$. Parametrising
   $$V_0=-{d(d-1)\over \ell^2}$$ in the $AdS$ regime, we obtain the metric at the center of $AdS$ in global coordinates, i.e. the  metric (\ref{adsg}) as $\rho \to 0$. This corresponds to the ``IR'' of the theory in the holographic RG flow interpretation.
   	
   	On the other hand, in a $dS$ regime, we can parametrise the leading behaviour of the potential as
   $$V_0 = d(d-1)H^2\,,$$ which then coincides with the metric (\ref{a31}) as $\rho\to 0$. This limit corresponds to the location of the observer in the static patch coordinates of $dS$.

Finally, we may solve equations (\ref{w53}) and (\ref{w53b}) to obtain:

\begin{equation}
\f = \f_0+\dfrac{1}{2 d} R^2 e^{2A_0}V_1 e^{-\frac{u W_0}{d-1}} + \dots
\end{equation}

\begin{equation}
A = A_0 -\dfrac{1}{2(d-1)} u W_0 + \dots
\end{equation}

\begin{equation}
f = \dfrac{4(d-1)^2 e^{-2A_0}}{R^2 W_0^2}e^{\frac{u W_0}{d-1}} + \dots
\end{equation}

\noindent
Consistency with the small $\f-\f_0$ expansion requires that $uW_0 \to + \infty$. As a result, the scale factor $e^{2 A}$ vanishes, and the size of the sphere shrinks to zero. This is the justification for the name ``\textit{shrinking end-points}". The combination $e^{2A}f$ remains finite.

The two quantities appearing in (\ref{ev1}) and (\ref{ev2}) asymptote as
\be
f\dot A e^{dA}=-{2\over R^2W_0}e^{(d-2)\left(A_0-{uW_0\over 2(d-1)}\right)}+\cdots~~~\to~~~ 0
\ee

\be
\dot f e^{dA}={4(d-1)\over R^2W_0}e^{(d-2)\left(A_0-{uW_0\over 2(d-1)}\right)}+\cdots~~~\to~~~ 0
\ee

Summarizing, we find that near a shrinking endpoint, in the AdS regime, we arrive at the center of $AdS$ in global coordinates. In the dS regime however, such an endpoint coincides with the location of an observer in the static patch coordinates.

\subsection{Spatial boundaries of Minkowski space-time}
Returning to the constant scalar solutions to the equations of motion, we highlight a final solution of interest. For non-vanishing $a$, \refeq{f15} has a non-trivial solution in a ``Minkowski" regime ($V_*=0$) provided $f_0=0$. In this case, the general solution for the blackening function $f$ is given by
\begin{equation}
f = \frac{1}{(Ra)^2}e^{-2A}+\overline{C}e^{-dA}.
\end{equation}
We first focus on the $\overline{C}=0$ solution and change radial coordinate to $r$ such that

\begin{equation}
r = R e^{au + A_0}
\end{equation}
as before. Rescaling the time coordinate, the metric described by this solution is simply
\begin{equation}\label{eq:MnkSol}
ds^2 = -dt^2 + dr^2 + r^2 d\Omega^2,
\end{equation}
which is Minkowski space in $d+1$ dimensions. We note in passing that solutions with non-zero $\overline{C}$ are higher dimensional analogues of the Schwarzschild black hole (for $\overline{C}<0$) or asymptotically flat space-times with a naked singularity (for $\overline{C}>0$).

Our gravitational system allows for endpoints which are locally isometric to the spatial boundary of Minkowski space, which is achieved in the limit $r\to \infty$ in the coordinates of \refeq{eq:MnkSol} above. To understand this in more detail, we recall the analysis of appendix \ref{exth}. There, we show that these endpoints can appear as local solutions (around a point $\f = 0$ ) when
\begin{equation}\label{eq:MnkVW}
V = \sum^\infty_{n=\beta+2} V_n\frac{\f^n}{n!}, \qquad \mathrm{and} \qquad W = W_0 +\sum^\infty_{n = 2} W_n \frac{\f^n}{n!}
\end{equation}
where $\beta \ge 1$. These solutions are characterised by an blackening function and warp factor of the form
\begin{equation}\label{eq:MnkfT}
f = \f^\beta\sum_{n=0}^\infty f_n \frac{\f^n}{n!}, \qquad \mathrm{and} \qquad T = \f^\beta\sum^\infty_{n=0}T_n \frac{\f^n}{n!}
\end{equation}
respectively.  It is understood that $f_0, \, T_0 \ne 0$ in these sums.

That these solutions asymptotically approach a locally flat region of space-time can be verified by computation of the Kretschmann scalar. Inserting the expansions \refeq{eq:MnkVW} and \refeq{eq:MnkfT} into \refeq{w62a} one finds that to leading order
\begin{equation}
K_2 \propto \f^{2\beta + 4}
\end{equation}
and hence this scalar invariant vanishes as $\f \to 0$. Moreover, as these solutions are described by a scale factor that behaves in the limit like
\begin{equation}
e^{2A}= e^{2A_0}\f^{-\beta}+\ldots
\end{equation}
it is clear that the volume of the spatial sphere diverges as $\f\to0$ while the $tt$--component of the metric remains finite. In fact, by choosing a new radial coordinate $r$ such that $r\propto \f^{-\beta/2}$ we observe that the local metric corresponding to this endpoint can be brought to the form \refeq{eq:MnkSol}.

\section{General properties of solutions} \label{sec:global}

\bigskip

In this section we explore the possible flows allowed by the equations of motion. A summary of the results is presented in figure \ref{fig:flow}. We begin in Subsec. \ref{rul} by establishing a set of rules for the flows that follow from the equations of motion. Subsequently, we establish which flow solutions are not possible in the spherically sliced ansatz in Subsec. \ref{fbd}, and explain in Subsec. \ref{pib} some of the generic properties of the flows that do exist.

\subsection{The global flow rules}\label{rul}

We collect here a list of rules, together with their proofs, that will clarify the structure of the possible flows. We are eventually interested in flows from AdS boundaries (located at extrema of $V$) to any endpoint that is in the de Sitter regime.

\subsubsection*{Rules about the superpotential W and the scale factor $e^A$}

We begin this section addressing generic properties of the superpotential.

\begin{enumerate}
	\item[{\bf rule 0}\label{ru0}]:
	 The superpotential $W$ is monotonic as a function of the holographic coordinate $u$.

\textit{Proof}: This is a direct consequence of Eq. \eqref{w53b}, $W'=\dot{\f}$, which implies ${dW\over du}=(W')^2\geq 0$.

	\item[{\bf rule 1}\label{ru1}] The flow terminates at regular extrema of the superpotential $W$ at finite values of $\f$.

\textit{Proof}: We first assume that the flow terminates at finite $\f$, such that the function $f$ is finite. We shall discuss separately the instances where $f$ is not finite. The scalar $\f$ satisfies a second-order equation,  (\ref{f23d}), and therefore,  for the flow to stop at some  $u$, both $\dot{\f}$ and $\ddot{\f}$ must vanish at that point, provided that $f$ is finite\footnote{Eq. \eqref{f23d} also involves $\dot{A}$, which is finite so long as $W$ is finite at the endpoint, by virtue of Eq. \eqref{w53}.}.

In the superpotential formalism, we have $\dot{\f}=W'$ and $\ddot{\f}=W'W''$. Assuming that $W$ is not divergent at a finite $\f$, around any given point it behaves as $W\simeq W_0 + C (\f-\f_0)^\alpha + \dots$. Consequently, $\dot{\f}\sim (\f-\f_0)^{\alpha-1}$ and $\ddot{\f} \sim (\f-\f_0)^{2\alpha-3}\,$.\footnote{The cases $\alpha=1,2$ are treated separately. If $\alpha=1$ then $\dot{\f}\neq 0$ and the flow does not stop. If $\alpha=2$ it is immediate to check that the flow stops.} The flow will stop if and only if both derivatives vanish, and that only happens for $\alpha>3/2\,$. In particular, \textit{the flow stops at extrema of $W\,$}.

The case $\alpha=3/2$ on the other hand, is realized and corresponds to  a bounce, \cite{exotic}, (change of direction for $\f$), as described in the appendix \ref{bounces}.
However, as shown there, at bounces where $W'=0$ but $W''$ diverges\footnote{Such bounce points are neither maxima nor minima of $W$.} the solutions continues regularly.
The only other possibility with $\a\leq {3\over 2}$ allowed by the equations is $\a=1$, but this is incompatible with $W'=0$.

We therefore conclude that flows end at regular extrema of $W$. 

There are two instances in which the function $f$ diverges at regular extrema of $W$: shrinking endpoints and dS$_2$ boundaries. In the former case, the size of the sphere shrinks to zero, and the flow stops because the geometry ends. In the latter case, $f\sim u^2$ as $u\to \infty$ (see Eq. \eqref{gsuno}), and the term $\dot{f}/f$ in Eq. \eqref{f23d} vanishes. As a consequence, it is again true that the flow stops when both $\dot{\f}$ and $\ddot{\f}$ vanish, and the previous discussion applies.

This rule can be violated for some solutions of the first order equations, as shown in appendix \ref{nopot}, but such solutions are not solutions of the second order Einstein equations.

{\bf Corollary}: In the absence of $\f$-bounces, $W$ is monotonic along the whole flow, as a finite flow is delimited by two consecutive points with $W'=0$. If there are $\f$-bounces along the flow, $W'$ changes sign at each $\f$-bounce.

	\item[{\bf rule 2}\label{ru2}] If $W\geq 0$ around a regular extremum of the superpotential, then a minimum of $W$ corresponds to a boundary endpoint (dS$_{d+1}$, AdS$_{d+1}$, dS$_2$ or Minkowski) or extremal horizon endpoint\footnote{Both flat extremal Minkowski horizons and Nariai horizons.}, while a maximum of $W$ correspond to shrinking endpoints. If $W\leq 0$ then minima of $W$ correspond to shrinking endpoints and maxima of $W$ correspond to boundary endpoints or extremal horizons.

	\textit{Proof}: We prove the statement for each case separately. In appendix \ref{smry} we summarise all possible endpoints. Those are shrinking endpoints, $d+1$ boundary endpoints (dS, AdS and Minkowski), dS$_2$ boundary endpoints and two types of extremal horizons.
	
The shrinking endpoints are described in section \ref{shr} and they have $W'=0$ and $W''/W<0$. Therefore, such points are extrema of $W$, and $W''$ has the opposite sign of $W$ proving  the initial statement in this case.

For solutions describing AdS$_{d+1}$ or dS$_{d+1}$ boundaries we have, from appendices \ref{AdSX} and \ref{dSX} respectively, that $W'=0$ and $W''/W\sim \Delta_{\pm}$ and all regular cases satisfy $\Delta_{\pm}>0$. This proves our statement for such extrema.

The dS$_2$ boundary asymptotics are described in Appendix \ref{g53}. From Eq. \eqref{wds2}, the superpotential $W$ vanishes at least quadratically, and it also satisfies our statement.

Minkowski boundaries and flat extremal Minkowski horizons are studied in Appendix \ref{exth}. According to equations \eqref{e2c}, \eqref{e2cb}, \eqref{eqcoe} (and their analogues for each case), we have $W''/W>0$ while $W'=0$. Therefore these are endpoints of the flow which are maxima (minima) of a negative (positive) superpotential $W$. Finally, there are also Minkowski boundaries with $V'=0$ and for which the superpotential vanishes at least to cubic order (see equation \eqref{184} and the discussion below for an example). Therefore, they also satisfy our statement.

The Nariai horizon asymptotics are studied in Appendix \ref{seho2}. The superpotential behaves like $W\sim \varphi^\alpha\sim \varphi^{2+1/\delta_-}$ (c.f. equations \eqref{eahb2} and \eqref{2hu}). According to the discussion at the end of the same appendix, the Nariai horizons  serve as an endpoint for $\delta_-<-2$. Hence, $W$ has an extremum and vanishes as we approach the Nariai horizons, and our statement is also satisfied for such end-points.

		\item[{\bf rule 3}\label{ru3}] For flows which contain at least one (dS, AdS or Minkowski) boundary endpoint, or a Gubser-regular endpoint with vanishing potential, $W$ can be taken to be always positive (by choosing the direction of the flow), and the scale factor $e^A$ is always monotonic.

\textit{Proof}: There are a few instances in which $W$ changes sign. First, the superpotential can cross zero and diverge at the boundary in field space in both directions: $|\f| \to \infty$. Such solutions contain no endpoint at finite $\f$ and hence no boundary, but are singular at both $\f=\pm\infty$. The fact that $W(\pm \infty)\to \infty$ implies that Gubser-regular solutions at the boundary of field space would also have a diverging potential (see Tables \ref{tab_IS} and \ref{tab_IIS}). For this reason, this instance cannot contain Gubser-regular endpoints with vanishing potential.

 Alternatively, the flow can start at a finite extremum of $W$ and run to infinity in field space. Assuming that $W$ crosses zero, the found extremum is either a maximum for $W>0$ or a minimum for $W<0$. According to rule 2 on page \pageref{ru2}, such an extremum is a shrinking endpoint. As a result, this possibility does not involve any dS, AdS or Minkowski boundary, neither does it involve Gubser-regular endpoints with vanishing potential (where the superpotential should also vanish, c.f. Tables \ref{tab_IS} and \ref{tab_IIS}). Finally, the flow can start and end at two extrema of W at finite values of $\f$. Then, the superpotential could cross zero along the flow that interpolates between the  maximum of the superpotential for $W>0$ and the minimum for $W<0$. Again, according to rule 2 on page \pageref{ru2}, this flow interpolates between two shrinking points and involves no dS, AdS or Minkowski boundaries, or Gubser-regular endpoints with vanishing potential.

We conclude, that for flows that contain at least one boundary, $W$ cannot change sign, and it is therefore either positive or negative along the full flow. The same is true for flows involving Gubser-regular endpoints with vanishing potential. Moreover according to rule 0 on page \pageref{ru0}, it is monotonic along the flow.
By inverting the direction of the flow, $u\to -u$, we can always change the sign of $W$.
Therefore, we can always take $W\geq 0$ without loss of generality, in such flows. In particular, we can take $W\geq 0$ in flows from an extremum of the potential to a shrinking  endpoint. From Eq. \eqref{w53}, the function $e^A$ is monotonic if and only if $W$ does not change sign.

\item[{\bf rule 4}\label{ru4}] The inverse scale factor $T$ has isolated zeros only at boundaries. It can also be zero identically. It never changes sign along a flow that contains a dS or AdS or Minkowski boundary. 

    {\it Proof}: We assume that $T\geq 0$. From equation (\ref{eqtt}) we deduce that, since $W$ is always finite at finite $\f$\footnote{In rule 3 on page \pageref{ru3} we have shown that for flows containing one of the aforementioned boundaries, $W$ cannot vanish in the interior of the flow.}, $A$ can diverge, only if $W'$ vanishes at some finite  $\f$. However, by rule 1 on \pageref{ru1}, $W'$ can only vanish at the end-points of a regular flow\footnote{$W'$ also vanishes at $\f$-bounces but in such a case the integrand in (\ref{eqtt}) is integrable.}.

\end{enumerate}

\subsubsection*{Rules about the blackening function $f$}

We now turn our attention to the general behaviour, allowed by the equations of motion, of the blackening function $f$.

\begin{enumerate}

	\item[{\bf rule 5}\label{ru5}]  If $f$ has an extremum along the flow (i.e. excluding the endpoints), then it is always a maximum. Consequently, $f$ has at most one extremum and the geometry develops at most two horizons, where $f$ has zeros.

	\textit{Proof}: From equation \refeq{f23b}, setting $\Dot{f}=0$ gives $\ddot{f}<0$ at that point. Equivalently, for $f$ as a function of $\f$, we obtain from \refeq{f10_1} that $f'=0$ implies $f''<0$.
The case where $T=0$ or $e^{-A}\to 0$ only takes place at the boundary endpoints of the flow, giving an inflexion point for $f$, namely $\dot{f}=\ddot{f}=0=f'=f''$.

		\item[{\bf rule 6}\label{ru6} ]  $f$ is monotonic in solutions involving a shrinking endpoint.

	\textit{Proof}: From equations (\ref{soch}) and (\ref{sochu}) that describe the behavior of $f$ and $T$  near a shrinking endpoint, we observe that the signs of $f$ and $T$ are correlated at the shrinking endpoint and they are both controlled by the first derivative of the potential at that point. Besides, $T>0$ by assumption, so we must have $T_{\rm shrink}\to +\infty$ and accordingly $f_{\rm shrink}\to + \infty$. Extending the flow away from the shrinking endpoint, $f$ will necessarily decrease. From rule 5 on page \pageref{ru5}, we know that $f$ can have at most one maximum, but such a local maximum would be incompatible with $f_{\rm shrink}\to + \infty$. Therefore, $f$ has no extrema and is monotonous along this class of solutions. In particular, this rule forbids flows between two shrinking endpoints.

\end{enumerate}

\subsubsection*{Rules about horizons}

A horizon is identified by the vanishing of the temporal component of the metric $g_{tt}=-fe^{2A}=0$. The scale factor can only diverge or vanish at endpoints of the flow. Therefore, a horizon along the flow (excluding endpoints), is identified by the vanishing of $f$. We now discuss several rules about the presence of horizons along the flow. In Appendix \ref{app:J}, we classify the different types of horizons.

\begin{enumerate}

	\item[{\bf rule 7}\label{ru7}] There is at most one horizon in solutions involving a shrinking endpoint.

\textit{Proof}: This is a direct corollary of rule 6 on page \pageref{ru6}: since $f$ is monotonic, it can vanish at most once.

\item[{\bf rule 8}\label{ru8}] There is at most one horizon in solutions involving either of the following: (i) an AdS boundary endpoint, (ii) a Minkowski boundary endpoint or (iii) a Gubser-regular endpoint with $V\to 0^-$. If such solutions do feature a horizon, this is a black-hole event horizon.

\textit{Proof}: Firstly, note that the function $f$ is positive in the neighbourhood of (i) an AdS boundary endpoint (see Eq. \eqref{C18}), (ii) a Minkowski boundary endpoint (see Eqs. \eqref{e5e} and \eqref{e6e}) or (iii) a Gubser-regular endpoint with $V\to 0^-$, as specified in Tables \ref{tab_IS} and \ref{tab_IIS}. Secondly, note that $f$ has at most one maximum and no minima (rule 5 on page \pageref{ru5}).

There are three distinct possibilities:

(a) If $f$ increases as the solution departs from the endpoint without having a  maximum, then there is no zero of $f$ and therefore no horizon.

(b)  If $f$ increases as the solution departs from the endpoint, and it has a single maximum, it will have a single zero and therefore one horizon.

(c) If $f$ decreases as it departs from the endpoint, it cannot have a minimum and in that case it can vanish only once.

In the case where there is a horizon, this is a black-hole event horizon according to the discussion in Appendix \ref{app:J}.

\item[{\bf rule 9}\label{ru9}] A solution from (i) an AdS boundary, (ii) a Minkowski boundary, or (iii) a Gubser-regular endpoint with $V\to 0^-$, to a shrinking endpoint, cannot have a horizon. Similarly, a solution from (i) an AdS boundary, (ii) a Minkowski boundary, or (iii) a Gubser-regular endpoint with $V\to 0^-$, to a Gubser-regular endpoint with $V\to -\infty$, cannot have a horizon.

\textit{Proof}: Firstly, note that the function $f$ is positive in the neighbourhood of (i) an AdS boundary endpoint (see Eq. \eqref{C18}), (ii) a Minkowski boundary endpoint (see Eqs. \eqref{e5e} and \eqref{e6e}) or (iii) a Gubser-regular endpoint with $V\to 0^-$, as specified in Tables \ref{tab_IS} and \ref{tab_IIS}.

Secondly, we have seen in rule 6 on page \pageref{ru6} that $f$ is monotonous in solutions involving a shrinking endpoint, and it diverges as $f\to +\infty$ as the shrinking endpoint is approached. Combined with the first observation above, we conclude that $f$ does not vanish in flows connecting the aforementioned  (i), (ii) or (iii) endpoints to shrinking endpoints.

We now proof the second part of the statement. At Gubser-regular type II endpoints with $V\to -\infty$, it is also true that $f\to+\infty$ (see Table \ref{tab_IIS}). We apply the same reasoning as for the shrinking endpoint to conclude that solutions from the (i), (ii) or (iii) endpoints to type II endpoints with $V\to -\infty$ cannot feature a horizon. On the other hand, the function $f$ is positive as a Gubser-regular type I endpoint with $V\to -\infty$ is approached (see Table \ref{tab_IS} for $\a\in (\a_C,\a_G)$). Therefore, solutions connecting the (i), (ii) or (iii) endpoints to Gubser-regular type I endpoints with $V\to -\infty$  could only have horizons if $f$ had a local minimum along the flow, which is not possible according to rule 5 on page \pageref{ru5}. We conclude that such flows cannot have horizons.

\item[{\bf rule 10}\label{ru10}] A solution from (i) a dS$_{d+1}$ boundary, (ii) a dS$_2$ boundary, or (iii) a Gubser-regular endpoint with $V\to 0^+$, to a shrinking endpoint, features a cosmological horizon. Similarly, A solution from (i) a dS$_{d+1}$ boundary, (ii) a dS$_2$ boundary, or (iii) a regular endpoint with $V\to0^+$, to a Gubser-regular endpoint with $V\to -\infty$, also feature a cosmological horizon.

\textit{Proof}: We have seen that the function $f$ diverges to positive values as we approach a shrinking endpoint, $f_{\rm shrink}\to +\infty$, regardless of whether we are in dS or AdS regime of the potential. Additionally, the presence of a shrinking endpoint ensures that $f$ is monotonic (rule 6 on page \pageref{ru6}). On the other hand, the function $f$ is negative in a neighbourhood of a dS$_{d+1}$ boundary (see Eq. \eqref{C18a}), a dS$_2$ boundary (see Eqs. \eqref{fds2} and \eqref{tds2}) and a Gubser-regular endpoint with $V\to 0^+$ (see Tables \ref{tab_IS} and \ref{tab_IIS}). All in all, we are led to conclude that $f$ vanishes along flows connecting the aforementioned (i), (ii) or (iii) endpoints to shrinking endpoints. Therefore, such flows feature a horizon, which is cosmological in agreement with the discussion of Appendix \ref{app:J}.

Finally, at the Gubser-regular endpoints with $V\to -\infty$, the function $f$ is positive (see Tables \ref{tab_IS} and \ref{tab_IIS}). Connecting the (i), (ii) or (iii) endpoints with a Gubser-regular endpoint with $V\to -\infty$ necessarily requires that $f$ changes sing along the flow. Therefore, such flows feature a horizon, which is again cosmological.

\item[{\bf rule 11}\label{ru11}] A solution involving a Gubser-regular endpoint with $V\to +\infty$ does not have a horizon along the flow.

\textit{Proof}: We prove the statement separately for the type I and type II endpoints. At type II endpoints with $V\to +\infty$, the blackening function $f$ vanishes from below (see Table \ref{tab_IIS}), $f\to 0^-$, and therefore it decreases as the solution departs from the type II endpoint. The function $f$ does not have local minima (see rule 5 on page \pageref{ru5}), and therefore the function $f$ is monotonically decreasing in such a solution. In particular, $f\leq 0$ in flows involving type II endpoints with $V\to +\infty$. We conclude that $f$ does not vanish along such a flow, proving the statement for the type II endpoints.

At Gubser regular type I endpoints, i.e. type I endpoints with $\a <\a_G$ in appendix \ref{asymp}, with $V\to +\infty$, the function $f$ is negative, and is also decreasing as it departs from such an endpoint (see Eqs. \eqref{m49}, \eqref{m50} and the discussion below). Again, rule 5 on page \pageref{ru5} prevents $f$ from having local minima, which implies that $f<0$ in flows involving Gubser regular type I endpoints with $V\to +\infty$.  We conclude that $f$ does not vanish along such a flow, proving the statement for the Gubser-regular type I endpoints.

\item[{\bf rule 12}\label{ru12}] Cosmological horizons are always located in the dS regime, i.e. at points with $V>0$.

\textit{Proof}: From the analysis of Appendix \ref{app:J}, we know that a cosmological horizon is the outermost horizon in solutions featuring a dS$_{d+1}$ boundary, a dS$_2$ boundary, or a Gubser-regular endpoint with $V\to 0^+$.

By virtue of rule 3 on page \pageref{ru3}, the scale factor is monotonic, $\dot{A}\neq0$, in solutions involving dS boundaries and Gubser-regular endpoints with vanishing potential, and we can divide \eqref{f23c} by $\dot{A}^2$. We evaluate equation \eqref{f23c} at the location of the cosmological horizon mentioned in rule 10 on page \pageref{ru10}, where $f=0$, to obtain

\begin{equation}\label{eqqv}
		\dfrac{V_h}{\dot{A}^2} =\left. \dfrac{(d-1)(d-2)}{R^2 \dot{A}^2}e^{-2A} - (d-1)\dfrac{\partial f}{\partial A}>  - (d-1)\dfrac{\partial f}{\partial A} \right|_h\,.
	\end{equation}

The function $A$ decreases as the solution departs from dS$_{d+1}$ boundaries, dS$_2$ boundaries, or Gubser-regular endpoints with $V\to 0^+$. On the other hand, the function $f$ is negative around the dS boundary endpoints or around Gubser-regular endpoints with $V\to 0^+$. At the cosmological horizon, $f$ vanishes and increases as it moves away from the boundary. Altogether, this implies that  $\left(\partial f/\partial A \right)_h < 0$, and Eq.  \eqref{eqqv} implies that $V_h>0$.

    \item[{\bf rule 13}\label{ru13}] In flows involving a Nariai (extremal) horizon endpoint, we have $f\leq 0$, where the inequality is saturated only at the Nariai endpoint.

    {\it Proof}: Note that the Nariai endpoints can only happen in the dS regime ($V>0$) under the assumption that $T>0$, as dictated by Eq. \eqref{h20}. Accordingly, from Eq. \eqref{h23} it follows that the blackening function $f$ departs from zero to negative values $f\to0^-$. Combined with rule 5 on page \pageref{ru5}, $f$ cannot have local minima, and we learn that $f$ must remain negative along the flow.

\end{enumerate}

\subsubsection*{Rules about the energy density $\rho$}

We conclude this subsection establishing several rules regarding the behaviour of the energy density $\rho$ defined in Eq. \eqref{ev3b}. These properties shall be crucial in the discussion of Sec. \ref{fbd}.

\begin{enumerate}

\item[{\bf rule 14}\label{ru14}] The energy density $\rho$, defined in Eq. \eqref{ev3b}, must change sign in solutions from an AdS$_{d+1}$ boundary to a shrinking endpoint in the dS regime.

{\it Proof}: The energy density is defined as $\rho = f(\dot{\f})^2/2-V$. In Eqs. \eqref{invads1}, it is shown that the energy density is positive at an AdS$_{d+1}$ boundary, while Eq. \eqref{invshr1} implies that the energy density is negative at a shrinking endpoint in the dS regime. As a result, there must exist a point $\f_{\#}$ along such flows at which the energy density vanishes $\rho_{\#}=0$.

\item[{\bf rule 15}\label{ru15}] The energy density $\rho$, defined in Eq. \eqref{ev3b}, must change sign in solutions from a ${(d+1)}$-Minkowski boundary to a shrinking endpoint in the dS regime.

{\it Proof}: The energy density is defined as $\rho = f(\dot{\f})^2/2-V$. The energy density $\rho$ vanishes at ${(d+1)}$-Minkowski boundaries, and increases as we depart from the boundary, so that $\rho>0$ in the vicinity of the Minkowski boundary, see below \eqref{invMink1} in appendix \ref{exth} for an example. Conversely, at a shrinking endpoint in the dS regime, the energy density is negative $\rho<0$, as dictated by Eq. \eqref{invshr1} for $V_0>0$. As a result, there must exist a point $\f_{\#}$ along such flows at which the energy density vanishes $\rho_{\#}=0$.

\item[{\bf rule 16}\label{ru16}] The energy density $\rho$, defined in Eq. \eqref{ev3b}, must change sign in solutions from a Gubser-regular endpoint with $V\to 0^-$, to a shrinking endpoint in the dS regime.

{\it Proof}: Around a Gubser-regular endpoint with $V\to 0^-$, the energy density vanishes form positive values $\rho\to 0^+$ (see Tables \ref{tab_IS} and \ref{tab_IIS}), while at a shrinking endpoint in the de Sitter regime, the energy density is negative by virtue of Eq. \eqref{invshr1}. As a consequence, there must exist a point $\f_{\#}$ along flows connecting such Gubser-regular endpoints to dS shrinking endpoints at which the energy density vanishes: $\rho_{\#}=0$.

\end{enumerate}

\subsection{On the forbidden flows}\label{fbd}

In this section, we employ the rules of Sec. \ref{rul} to prove that several regular solutions connecting various of the endpoints are actually forbidden. At the end of this section, we also comment on the singularity that appears inside black-hole event horizons, as well as bad singularities that are connected to  $A$-bounces.

\begin{enumerate}

\item[{\bf rule 17}\label{ru17}] There are no regular flows connecting any of the following endpoints with each other: (AdS$_{d+1}$, dS$_{d+1}$, dS$_2$, M$_{d+1}$) boundary endpoints, Nariai endpoints or Gubser-regular endpoints with $V\to 0^{\pm}$

\textit{Proof}: In rule 2 on page \pageref{ru2} we have seen that (AdS$_{d+1}$, dS$_{d+1}$, dS$_2$, M$_{d+1}$) boundary endpoints correspond to minima (maxima) of a positive (negative) superpotential. Additionally, the superpotential $W$ vanishes at Gubser-regular endpoints with $V\to 0^{\pm}$ (see Tables \ref{tab_IS} and \ref{tab_IIS}), so such Gubser-regular endpoints are also minima (maxima) of a positive (negative) superpotential. Besides, by virtue of rule 3 on page \pageref{ru3}, we can assume that $W \geq 0$ in solutions involving any of the endpoints mentioned above. Connecting any of the aforementioned endpoints with each other would require that $W$ flows from a local minimum to another local minimum, and therefore it should encounter a local maximum in between. According to rule 1 on page \pageref{ru1}, the flow must stop at such a maximum, contradicting the previous statement. We conclude that a flow connecting any of the following (AdS$_{d+1}$, dS$_{d+1}$, dS$_2$, M$_{d+1}$) boundary endpoints, Nariai endpoints or Gubser-regular endpoints with $V\to 0^{\pm}$, with each other, does not exist.

\item[{\bf rule 18}\label{ru18}] There are no regular flows connecting any of the following endpoints with each other: shrinking endpoints or Gubser-regular endpoints with $V\to \pm \infty$

\textit{Proof}: We shall show that such flows would require $f$ to have a local minimum, contradicting rule 5 on page \pageref{ru5}.

Firstly, recall that $f\to +\infty$ at any shrinking endpoint, as dictated by Eqs. \eqref{soch} and \eqref{sochu}. Trivially, $f$ increases as it approaches a shrinking endpoint.

As for the Gubser-regular endpoints with $V\to \pm \infty$, we have to distinguish between the type I and type II asymptotic structures discussed in Appendix \ref{asymp}. For the Gubser-regular type I asymptotic solutions with $V\to \pm \infty$, Eqs. \eqref{m49} and \eqref{m50} imply that $f$ increases as it approaches such type I endpoint (see discussion after \eqref{m50}). On the other hand, for the type II endpoints with $V\to -\infty$, then $f\to + \infty$ (see Table \ref{tab_IIS}), while for the type II endpoints with $V\to +\infty$, then $f\to0^-$. In any case, we conclude that $f$ is also increasing as it approaches the type II endpoints with $V\to \pm \infty$.

In short, we have shown that $f$ increases as it approaches shrinking endpoints or Gubser-regular endpoints with a diverging potential. Therefore, a flow connecting any shrinking endpoint or Gubser-regular endpoints with a diverging potential to any other shrinking endpoint or Gubser-regular endpoints with a diverging potential would require that $f$ has a local minimum along the flow, contradicting rule 5 on page \pageref{ru5}. We conclude that such flows are not possible.

\item[{\bf rule 19}\label{ru19}]  There are no regular flows connecting any (i) AdS$_{d+1}$ boundaries, (ii) M$_{d+1}$ boundaries or (iii) Gubser-regular endpoints with $V\to 0^-$, to any dS shrinking endpoint.

 {\it Proof}:  According to rule 3 on page \pageref{ru3}, $\dot A$ can be taken to be positive along such flows and $A$ is monotonic. With this convention, the coordinate $u$ is increasing as we go from the dS shrinking endpoint to either of the aforementioned (i), (ii) or (iii) endpoints.

According to rule 9 on page \pageref{ru9}, during the flows we are considering here, $f$ cannot change sign. Since at a shrinking endpoint it diverges via positive values, we conclude that $f>0$ along the whole flow.

We then consider equation (\ref{ev5}) at any point $u_{\#}$ where $\rho$ vanishes. The right hand side of (\ref{ev5}) is always negative as $f>0$. From the left hand side, as $\rho(u_{\#})=0$, we find that
\be
\dot \rho(u_{\#})<0
\label{main}\ee

According to rules 14 on page \pageref{ru14}, 15 on page \pageref{ru15} and 16 on page \pageref{ru16}, $\rho$ varies from a negative  value at the dS shrinking endpoint to a positive value close to the AdS$_{d+1}$ boundary, M$_{d+1}$ boundary, or Gubser-regular endpoint with $V\to 0^-$, respectively. Therefore, in such flows, $\rho$ must cross zero at least once at some point $u_{\#}$: $\rho(u_{\#})=0$.
We now consider the case where $\rho$ is monotonic along the flow and crosses zero at a single point $u_{\#}$. In that case, $\dot \rho(u_{\#})>0 $ and contradicts (\ref{main}).
If it crosses zero $2n+1>1$ times, then  there are exactly $n+1$ points at which  $\dot \rho>0 $ again contradicting (\ref{main}). We conclude that there are no regular flows connecting (i) AdS$_{d+1}$ boundaries, (ii) M$_{d+1}$ boundaries or (iii) Gubser-regular endpoints with $V\to 0^-$, and dS shrinking endpoints\footnote{It is clear from equation (\ref{ev5}) that changing the direction of the flow, $u\to -u$, does not affect the equation.}.

\item[{\bf rule 20}\label{ru20}] There are no regular flows connecting any (i) AdS$_{d+1}$ boundary, (ii) M$_{d+1}$ boundary or (iii) Gubser-regular endpoint with $V\to 0^-$, to any Gubser-regular endpoints with $V\to +\infty$.

{\it Proof}: We shall show that the existence of such solutions would require that the function $f$ has a local minimum along the flow, contradicting rule 5 on page \pageref{ru5}.

We first focus on the behaviour of $f$ at the Gubser-regular endpoints with $V\to +\infty$. At such Gubser-regular endpoints with type I asymptotics, the function $f$ is negative (see Table \ref{tab_IS}) and decreases as the solution departs from such an endpoint, as it follows from Eqs. \eqref{m49} and \eqref{m50} (see discussion below \eqref{m50}). Similarly, at type II endpoints with $V\to +\infty$, the function $f$ vanishes from below (see Table \ref{tab_IIS}), $f\to 0^-$, and as a consequence it is decreasing as the solution departs from such a type II endpoint. We conclude that $f$ is decreasing as it departs from Gubser-regular endpoints with $V\to +\infty$.

We now turn our attention to the behaviour of $f$ at (i) AdS$_{d+1}$ boundaries, (ii) M$_{d+1}$ boundaries or (iii) Gubser-regular endpoints with $V\to 0^-$. At AdS$_{d+1}$ boundary endpoints, the function $f$ takes a positive value (see Eq. \eqref{C18}). At M$_{d+1}$ boundary endpoints, it vanishes (see e.g. \eqref{e48e} and the discussion below). Finally, at Gubser-regular endpoints with $V\to 0^-$, $f$ is positive for the type I asymptotic structure, while it vanishes from above in the type II asymptotic structure.

Overall, we observe that $f$ is decreasing as it departs from any Gubser-regular endpoint with $V\to +\infty$, and connecting such solution to the aforementioned (i), (ii) or (iii) endpoints would require that $f$ stops decreasing and starts increasing in order to attain the vanishing or positive value that it takes at the  (i), (ii) or (iii) endpoints. In order words, $f$ should have a local minimum along the flow, contradicting rule 5 on page \pageref{ru5}. We conclude that such solutions do not exist.

\item[{\bf rule 21}\label{ru21}] There are no regular flows involving Nariai endpoints.

{\it Proof}: In rule 17 on page \pageref{ru17} we have shown that Nariai endpoints cannot be regularly connected to any boundary endpoint, or to Gubser-regular endpoints with vanishing potential, or to themselves. In principle, they could still be connected to shrinking endpoints or to Gubser-regular endpoints with a diverging potential. We shall now show that these possibilities are not allowed either.

Firstly, from rule 13 on page \pageref{ru13}, we know that the function $f$ is negative along flows involving Nariai endpoints, $f\leq 0$, where the inequality is saturated at the Nariai endpoint. As a consequence, $f$ is increasing as it approaches a Nariai endpoint.

Already, the fact that $f\leq 0$ is incompatible with a shrinking endpoint, where $f$ diverges to positive values, as dictated by Eqs. \eqref{soch} and \eqref{sochu}. We conclude that there is no regular flow connecting Nariai endpoints to shrinking endpoints. Furthermore, $f\to +\infty$ also at type II endpoints with $V\to -\infty$  (see table \ref{tab_IIS}), which is also incompatible with a Nariai endpoint.

We now turn our attention to Gubser-regular type I endpoints with $V\to \pm \infty$ and type II endpoints with $V\to -\infty$. At type II endpoints with $V\to -\infty$, the function $f$ vanishes from below, $f\to 0^-$. Consequently, a flow connecting a Nariai endpoint, where $f\to 0$, with a type II endpoints with $V\to -\infty$, where also $f\to 0$, would require that $f$ has a local minimum, contradicting rule 5 on page \pageref{ru5}.

On the other hand, at Gubser-regular type I endpoints with $V\to \pm \infty$, Eqs. \eqref{m49} and \eqref{m50} imply that $f$ increases as it approaches such type I endpoints (see discussion below \eqref{m50}). Since $f$ is also increasing as it approaches a Nariai endpoint, we conclude that a flow connecting any Nariai endpoints to any type I endpoint with $V\to \pm \infty$ would require that $f$ has a local minimum along the flow, contradicting again rule 5 on page \pageref{ru5}.

All in all, we conclude that there are no regular flows involving Nariai endpoints. As a consequence, a flow departing from a Nariai endpoint necessarily runs into a bad singularity.

\end{enumerate}

The rules presented in this section rule out several global flows connecting various endpoints. We conclude this section with two last rules related to bad singularities.

\begin{itemize}
\item[{\bf rule 22}\label{ru22}] Consider any flow departing from (i) an AdS$_{d+1}$ boundary, (ii) a dS$_{d+1}$ boundary, (iii) a dS$_2$ boundary, (iv) a Minkowski boundary or (v) a  Gubser-regular endpoint with vanishing potential. If such flows encounter a black-hole event horizon, then there is a bad singularity in the interior.

\textit{Proof}: By virtue of rule 3 on page \pageref{ru3}, we take $\dot{A}>0$ along such flows, and $A$ is monotonic. With this convention, the coordinate $u$ is decreasing as we depart from the (i)-(v) endpoints mentioned above.

If a flow from either of the (i)-(v) endpoints,  mentioned above, encounters a black-hole event horizon at some point $u_h$, then the blackening function vanishes at that point $f(u_h)=0$. Additionally, the function $f$ is positive in the outer neighbourhood of the black-hole event horizon (see Appendix \ref{app:J}), and negative inside, which implies $\dot{f}(u_h)>0$.

Note also that $f$ cannot have local minima (rule 5 on page \pageref{ru5}), and therefore the function $f$ must be monotonic inside the black-hole event horizon. For this reason, the flow cannot stop regularly at finite $\f$ inside the black-hole event horizon: rules 17 on page \pageref{ru17} and 21 on page \pageref{ru21} imply that the only finite regular endpoint where the flow could end is a shrinking endpoint; yet, $f\to +\infty$ at shrinking endpoints, which is incompatible with the fact that $f<0$ inside the black-hole event horizon. We therefore conclude that the flow cannot end regularly at finite $\f$ inside the black hole. Below we consider the possibility that the flow ends with Gubser-regular asymptotics inside the horizon.

Rule 17 on page \pageref{ru17} still allows for the flow to end at a Gubser-regular endpoint with diverging potential. From tables \ref{tab_IS} and \ref{tab_IIS}, we observe that $f>0$ at Gubser-regular endpoints with $V \to - \infty$, which is incompatible with $f<0$ inside of the event horizon. Conversely, we can consider the case with a Gubser-regular endpoint with $V\to +\infty$. In such a case, we have $\dot{f}<0$ as we approach $\f\to\infty$: for type II endpoints, this follows from the fact that $f\to 0^-$ there (see Table \ref{tab_IIS}), while for Gubser-regular type I endpoints it follows from Eqs. \eqref{m49}, \eqref{m50} and the discussion below them. However, we know that $\dot{f}(u_h)>0$, and therefore $\dot{f}$ should change sign inside the event horizon. Since $f(u_h)=0$, such extremum of $f$ is necessarily a minimum, which contradicts rule 5 on page \pageref{ru5}. We conclude that the flow cannot end at Gubser-regular endpoints inside the event horizon.

Overall, we have seen that a flow from (i) an AdS$_{d+1}$ boundary, (ii) a dS$_{d+1}$ boundary, (iii) a dS$_2$ boundary, (iv) a Minkowski boundary or (v) a Gubser-regular endpoint with vanishing potential, to a black-hole event horizon, cannot end regularly at finite $\f$ or with Gubser-regular asymptotics. As a result, the only possibility is that there is a bad singularity in the black hole interior.

\item[{\bf rule 23}\label{ru23}] A solution with an $A$-bounce necessarily has a bad singularity. Besides, a solution with an $A$-bounce cannot be connected regularly to any endpoint with $e^A\to \infty$, i.e. endpoints with $T=0$.

\textit{Proof}: From Eq. \eqref{w53}, there is an $A$ bounce, i.e. $\dot{A}=0$, if and only if the superpotential vanishes at that point. We consider the following possibilities: (a) the flow contains two endpoints at finite $\f$, (b) the flow goes from an endpoint at finite $\f$ to the boundary of field space or (c) the flow has no endpoint at finite $\f$.

From rule 3 on page \pageref{ru3}, we know that an $A$-bounce is incompatible with AdS, dS or Minkowski boundary endpoints. Additionally, rule 21 on page \pageref{ru21} shows that there are no regular solutions involving Nariai endpoints. Therefore, the only regular finite endpoints that can appear in a regular flow with an $A$-bounce are shrinking endpoints. Rule 18, on page \pageref{ru18} forbids flows between two shrinking endpoints, and we conclude that a flow with an $A$-bounce cannot have two regular endpoints at finite $\f$.

We consider now the case (b). Again, the only possible regular endpoint at finite $\f$ is a shrinking endpoint. On the other hand, at the boundary of field space ($\f\to \infty$) the superpotential cannot vanish, since in such case $W$ should encounter another local extremum between the $A$-bounce and the boundary of field space, and the flow would stop there (rule 1 on page \pageref{ru1}). We conclude that the superpotential diverges as $\f\to \infty$. Additionally, Gubser-regular endpoints with $W\to \infty$ also have $V\to \pm \infty$ (see Tables \ref{tab_IS} and \ref{tab_IIS}). Rule 18 on page \pageref{ru18} forbids flows from shrinking endpoints to Gubser-regular endpoints with $V\to \pm \infty$ and we conclude that solutions with an $A$-bounce and one regular finite endpoint necessarily involve a bad singularity.

Finally, we consider the case where there is an $A$-bounce and the flow runs to the boundary of field space at both ends. In such a scenario, $W$ does not have local extrema, and it diverges at both endpoints. From Tables \ref{tab_IS} and \ref{tab_IIS} we observe that Gubser-regular endpoints with $W\to \infty$ also have $V\to \pm\infty$. Now, rule 18 on page \pageref{ru18} forbids flows between two Gubser-regular endpoints with diverging potential, and we conclude that, at least one of the endpoints is a bad singularity.

In summary, we have shown that any solution containing an $A$-bounce necessarily contains a bad singularity.

Finally, we have also shown that the only possible regular finite endpoints that can be connected to an $A$-bounce are shrinking endpoints, where $T\to +\infty$ (see Eq. \eqref{sochu}). Similarly, the only Gubser-regular endpoints that can be connected to an $A$-bounce have $V\to \pm \infty$, where also $T\to +\infty$ (see Tables \ref{tab_IS} and \ref{tab_IIS}). We conclude that $A$-bounces cannot be regularly connected to endpoints with $T=0$.

\end{itemize}

\subsection{On the allowed flows}\label{pib}

We begin this part of the section by pointing out which horizons are cosmological and which ones are event horizons in solutions involving a dS, AdS or Minkowski boundary, or involving Gubser-regular endpoints with a vanishing potential. A detailed discussion can be found in Appendix \ref{app:J}. The key feature to distinguish between both horizons is the sign of the metric function $f$ in the outermost region, i.e. the region where the scale factor diverges. In general, there are the following possibilities:

\begin{itemize}
\item Solutions involving AdS boundaries, Minkowski boundaries, or Gubser-regular endpoints with $V\to 0^-$, can have at most one horizon (see rule 8 on page \pageref{ru8}). In the outermost region, the blackening function is positive $f>0$. According to the discussion in Appendix \ref{app:J}, such horizon is necessarily a black-hole event horizon.

\item Solutions involving dS boundaries or Gubser-regular endpoints with $V\to 0^+$ and two horizons. The situation is analogous to a dS black hole, in which the outermost (i.e. the one nearest to the time-like boundary) horizon is cosmological and the innermost horizon is an event horizon. The limit in which both horizons approach each other corresponds to the Nariai limit as discussed below Eq. \eqref{dzero}.

\item Solutions involving dS boundaries or Gubser-regular endpoints with $V\to 0^+$ and one non-extremal horizon, i.e. $f_h=0$ but $\dot{f}_h\neq0$, have $f<0$ in the outermost region, and from Appendix \ref{app:J} this is necessarily a cosmological horizon.
\end{itemize}

\begin{figure}[t]
	\centering
	\includegraphics[width=0.99\linewidth]{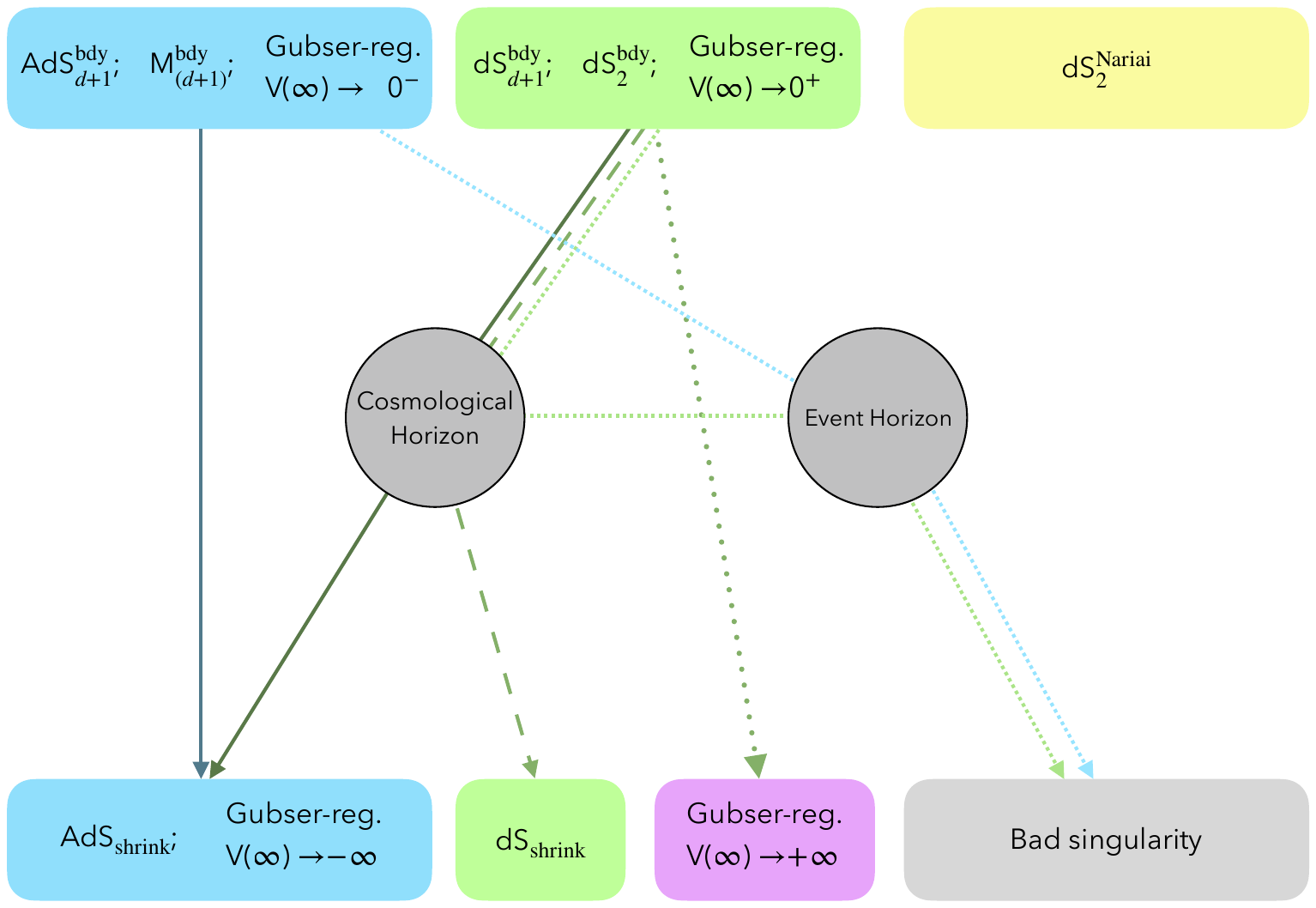}
	\caption{Depiction of the structure of possible flows in the spherically sliced ansatz. All horizons included are regular. The finite endpoints in the upper row are minima (maxima) of a positive (negative) superpotential. The finite endpoints in the lower correspond to maxima (minima) of the positive (negative) superpotential. We have excluded flows with naked singularities, i.e. flows running to a bad singularity that is not covered by a black-hole event horizon. Gubser-reg. stands for Gubser-regular endpoint, extensively discussed in Appendix \ref{asymp}.}\label{fig:flow}
\end{figure}

Now we proceed to discuss possible flows in the spherically sliced ansatz. The classification is based in the rules of Secs \ref{rul} and \ref{fbd}. A graphical summary of the regular flows described is shown in figure \ref{fig:flow}.

\subsubsection*{Flows involving AdS$_{(d+1)}$ boundaries}

The endpoints of the flow lie at extrema of the superpotential $W$ (rule 1 on page \pageref{ru1}). Taking $W>0$ without loss of generality (rule 3 on page \pageref{ru3}), the AdS$_{(d+1)}$ boundary endpoints are always placed at minima of $W$ (rule 2 on page \pageref{ru2}). Therefore, if we start the flow at an AdS$_{(d+1)}$ boundary there are three qualitatively different possibilities: (a) we can run to a maximum of $W$, which is a shrinking endpoint (rule 2 on page \pageref{ru2}) where the flow ends, (b) we can run to an event horizon or (c) we can run to the boundary of field space $|\f|\to \infty$ without crossing a horizon.

The first possibility, (AdS$_{(d+1)}^{\textrm{boundary}} \to$ shrinking endpoint), can never feature a horizon (rule 9 on page \pageref{ru9}). The shrinking endpoint cannot be located in the dS regime (rule 19, on page \pageref{ru19}). Then, the shrinking endpoint is in the AdS regime, and such a solution corresponds to the standard holographic RG-flow on a sphere, which has been extensively studied in the literature.
\be\label{fl1}
 {\rm AdS}_{(d+1)}^{\textrm{boundary}} ~~ \to~~ {\rm AdS ~~shrinking~~ endpoint}.
\ee
The holographically dual {Euclideanised} theory lives on $S^1\times S^{d-1}$, and the solution is what is usually called  an AdS-soliton. We do not show examples of such standard solutions.

The second possibility, where we run to a black hole event horizon, is allowed by rule 8 on page \pageref{ru8}.  Inside the horizon there is a bad singularity, as dictated by rule 22 on page \pageref{ru22}.
\be
 {\rm AdS}_{(d+1)}^{\textrm{boundary}} ~~\to~~ {\rm event~~horizon}~~ \to~~ {\rm bad ~~ singularity}.
\label{extra}
\ee
A familiar example of this kind of solutions are the standard holographic RG-flows at finite temperature. Alternatively, when there is no horizon, the flow hits a naked singularity. Some of these singular solutions may be acceptable holographically as discussed in \cite{Gubser,gkn,thermo,exotic}. These correspond to the Gubser-regular endpoints discussed in Appendix \ref{asymp}. From rules 17 on page \pageref{ru17} and 20 on page \pageref{ru20}, we learn that the only possibility is to connect the AdS boundary with a Gubser-regular endpoint where $V\to -\infty$. From rule 9 on page \pageref{ru9}, such a solution cannot have a horizon.
\be\label{fl3}
 {\rm AdS}_{(d+1)}^{\textrm{boundary}} ~~\to~~{\rm Gubser-regular}\ {\rm endpoint}\ {\rm with}\ (V\to -\infty)\,.
\ee
An example of such a flow is discussed in Section \ref{sec:71}.

The findings above are similar to what Gubser had found for flat-sliced solutions. As we show in appendix \ref{app:l1}, a Gubser-regular singularity, when covered by an infinitesimal horizon becomes a type 0 (bad) singularity as in (\ref{extra}).

\subsubsection*{Flows involving dS$_{(d+1)}$ boundaries}

Similarly to the previous case, we can take $W>0$ without loss of generality (rule 3 on page \pageref{ru3}). Therefore, the dS$_{(d+1)}$ boundary endpoints are located at minima of the superpotential (rule 2 on page \pageref{ru2}). Accordingly, a flow starting from such a boundary, can end either at a maximum of $W$, i.e. a shrinking endpoint, or run to the boundary of field space.

In the first case (dS$_{(d+1)}^{\textrm{boundary}} \to$ shrinking endpoint) the flow necessarily encounters a cosmological horizon between both endpoints (rule 10 on page \pageref{ru10}) located in the dS regime (rule 12 on page \pageref{ru12}). The flow can terminate at a shrinking endpoint regardless of the sign of the potential.

\be\label{fl4}
 {\rm dS}_{(d+1)}^{\textrm{boundary}} ~~\to~~ {\rm cosmological~~ horizon}~~ \to~~ {\rm shrinking~~ endpoint}.
 \ee

Explicit examples where the shrinking endpoint is in the dS or AdS regime are constructed in appendix \ref{app:H}, and described also in section \ref{fine2} (Figs. \ref{fig:H1} and \ref{fig:H2}).

In the second case, the flow necessarily runs into a singularity. The singularity is naked if there is no horizon, or if there is a single horizon (which is necessarily cosmological as explained at the beginning of this section). Conversely, the singularity is shielded, if the flow features two horizons. The outermost horizon is cosmological whereas the innermost is a black-hole event horizon:
\be\label{fl5}
 {\rm dS}_{(d+1)}^{\textrm{boundary}} ~~\to~~ {\rm cosmological~~ horizon}~~ \to~~ {\rm event~~horizon}~~ \to~~ {\rm bad ~ singularity}.
 \ee
The singularity inside the black hole is bad, as it follows from rule 22 on page \pageref{ru22}. One explicit example of this flow is presented in Sec. \ref{sec:dBH}. Such a solution is similar to the Schwarzschild-de Sitter solution. The limit where the cosmological and event horizons approach each other is known as the Nariai limit.  An  example  of the latter is constructed in Appendix \ref{aehor}.

Finally, a solution from a dS$_{d+1}$ can run into a singularity with the Gubser-regular asymptotics while not being covered by a black-hole event horizon. In particular, if the Gubser-regular endpoint has $V\to -\infty$, the flow encounters only a cosmological horizon (rule 10 on page \pageref{ru10}), while if the Gubser-regular endpoint has $V\to +\infty$ such a flow has no horizon (rule 11 on page \pageref{ru11}).

\be\label{fl6}
 {\rm dS}_{(d+1)}^{\textrm{boundary}} ~~\to~~ {\rm cosmological~~ horizon}~~ \to~~ {\rm Gubser-regular}\ {\rm endpoint}\ {\rm with}\ (V\to -\infty)\,,
 \ee

 \be\label{fl7}
 {\rm dS}_{(d+1)}^{\textrm{boundary}} ~~\to~~{\rm Gubser-regular}\ {\rm endpoint}\ {\rm with}\ (V\to +\infty)\,.
 \ee

Explicit examples of these two possibilities are constructed in Appendix \ref{app:71} and discussed in Sec. \ref{sec:71}.

\subsubsection*{Flows involving dS$_2$ boundaries}

In this section, we discuss flows involving dS$_2$ boundaries, where the local geometry is dS$_2\times $S$^{(d-1)}$. Taking $W>0$ without loss of generality (rule 3 on page \pageref{ru3}), such endpoints are located at minima of the superpotential (rule 2 on page \pageref{ru2}). Consequently, the flow terminates either at a neighbouring maximum of the superpotential (rule 1 on page \pageref{ru1}), corresponding to a shrinking endpoint (rule 2 on page \pageref{ru2}), or it runs to the boundary in field space ($\f\to \pm\infty$).

Flows from a dS$_2$ boundary to a shrinking endpoint necessarily encounter a cosmological horizon in between (rule 10 on page \pageref{ru10}) located in the dS regime (rule 12 on page \pageref{ru12}):
\be\label{fl8}
 {\rm dS}_2^{\textrm{boundary}}~~\to~~ {\rm cosmological~~ horizon}~~ \to~~ {\rm shrinking~~ endpoint}.
  \ee
  The shrinking endpoint can be located in the dS or AdS regimes. Explicit examples of these two instances are constructed in Appendix \ref{app:G} and discussed in Sec. \ref{fine2} (Figs. \ref{fig:G3b} and \ref{fig:G1}).

In flows from a dS$_2$ boundary to $\f\to \pm \infty$, we necessarily encounter a singularity at the boundary in field space. If there is no horizon, or only one (necessarily cosmological) horizon, then the singularity is naked. Alternatively, we can have two horizons, where the outermost is cosmological and the innermost is a black-hole event horizon, and the singularity is shielded:
\be\label{fl9}
{\rm dS}_2^{\textrm{boundary}}~~\to~~ {\rm cosmological~~ horizon}~~ \to~~ {\rm event~~ horizon}~~ \to~~ {\rm bad ~ singularity}.
\ee
The singularity inside the black-hole is bad, as it follows from rule 22 on page \pageref{ru22}. An explicit example of this solution is constructed in Appendix \ref{app:dS2BH}, and discussed in Sec. \ref{sec:ds2bh}.

We shall also consider the case where there is a naked singularity, whose asymptotic structure corresponds to the Gubser-regular endpoints found in Appendix \ref{asymp}. In particular, if the Gubser-regular endpoint has $V\to -\infty$, the flow encounters only a cosmological horizon (rule 10 on page \pageref{ru10}), while if the Gubser-regular endpoint has $V\to +\infty$ such a flow has no horizon (rule 11 on page \pageref{ru11}).

\be\label{fl10}
 {\rm dS}_{2}^{\textrm{boundary}} ~~\to~~ {\rm cosmological~~ horizon}~~ \to~~ {\rm Gubser-regular}\ {\rm endpoint}\ {\rm with}\ (V\to -\infty)\,,
 \ee

 \be\label{fl11}
 {\rm dS}_{2}^{\textrm{boundary}} ~~\to~~{\rm Gubser-regular}\ {\rm endpoint}\ {\rm with}\ (V\to +\infty)\,.
 \ee

Explicit examples of these two possibilities are constructed in Appendix \ref{app:74} and discussed in Sec. \ref{sec:74}.

\subsubsection*{Flows involving Minkowski (spatial) boundaries}

In this section, we discuss the viable possibilities to have regular flows involving a boundary of Minkowski space-time. The local structure of the solutions around these endpoints has been addressed in Appendix \ref{exth}. In all of them, the potential must vanish at least cubically ($V = V' =V''=0$)\footnote{This is distinct from a similar situation happening in supergravities emerging from string theory when $\phi\to\pm \infty$. In such cases potentials may vanish as $V\sim e^{-a\f}$ with $a>0$ and $\f\to+\infty$ and therefore, all derivatives of the potential vanish at the boundary of field space. Here, we assume that $V=V'=V''$ happens at finite $\f$. This is therefore a highly-tuned occurrence.}.  Moreover, according to rule 2 on page \pageref{ru2}, they correspond to minima (maxima) of a positive (negative) superpotential. We can take $W>0$ without loss of generality (rule 3 on page \pageref{ru3}). Then, the flow starts at a minimum of $W$. If the flow hits a maximum of the superpotential then this corresponds to a shrinking endpoint and the flow ends there (see rules 1 on page \pageref{ru1} and 2 on page \pageref{ru2}). Such a solution has no horizon (rule 9 on page \pageref{ru9}). Additionally, rule 19 on page \pageref{ru19} shows that the shrinking endpoint cannot be in the dS regime.

\be\label{fl12}
{\rm M}_{d+1}^{\rm boundary}~~ \to ~~{\rm AdS ~~ shrinking ~~ endpoint}.
\ee

Alternatively, the flow can start at a Minkowski boundary and run to the boundary in field space $\f\to \infty$, where it encounters a singularity. One possibility is that the singularity is covered by a black-hole event horizon, inside of which the flow necessarily hits a bad singularity (rule 22 on page \pageref{ru22}):

 \be\label{fl13}
{\rm M}_{d+1}^{\rm boundary}~~ \to ~~{\rm event~~ horizon} \to ~~{\rm bad ~ singularity}.
\ee

An explicit example of such solution is constructed in Appendix \ref{aehor} and discussed in Sec. \ref{sec:dBH}.

Finally, we consider the case where there is a naked singularity with the Gubser-regular asymptotic structure described in Appendix \ref{asymp}. From rules 17 on page \pageref{ru17} and 20 on page \pageref{ru20}, we learn that the only such possibility is to connect the Minkowski boundary with a Gubser-regular endpoint where $V\to -\infty$. From rule 9 on page \pageref{ru9}, such a solution cannot have a horizon:

\be
\label{fl4a}
{\rm M}_{d+1}^{\rm boundary}~~ \to~~{\rm Gubser-regular}\ {\rm endpoint}\ {\rm with}\ (V\to -\infty)\,.
\ee

An explicit example of this kind of flow is constructed in Appendix \ref{app:71} and discussed in Sec. \ref{sec:71}.

\subsubsection*{Flows involving Gubser-regular endpoints with $V \to 0^-$}

In solutions involving a Gubser-regular endpoint with vanishing potential, we can take $W>0$ without loss of generality (rule 3 on page \pageref{ru3}). Therefore, if we start the flow at such a Gubser-regular endpoint, there are two qualitatively different possibilities: (a) we can run to a maximum of $W$, which is a shrinking endpoint (rule 2 on page \pageref{ru2}) where the flow ends, or (b) we can run to the boundary of field space $|\f|\to \infty$.

The first possibility can never feature a horizon (rule 9 on page \pageref{ru9}), and the shrinking endpoint cannot be located in the dS regime (rule 19 on page \pageref{ru19}):
\be\label{fl15}
 {\rm Gubser-regular}\ {\rm endpoint}\ {\rm with}\ (V\to 0^-) ~~ \to~~ {\rm AdS ~~shrinking~~ endpoint}.
\ee
An example of such a solution is constructed in Appendix \ref{app:72} and discussed in Sec. \ref{sec:72}.

In the second possibility, the solution runs to the boundary of field space $|\f|\to \infty$, where we unavoidably encounter a singularity. According to rule 8 on page \pageref{ru8}, it is possible that such a singularity is covered by a black-hole event horizon:
\be\label{fl16}
 {\rm Gubser-regular}\ {\rm endpoint}\ {\rm with}\ (V\to 0^-) ~~\to~~ {\rm event~~horizon}~~ \to~~ {\rm bad ~singularity}.
 \ee
Rule 22, on page \pageref{ru22}, ensures that the singularity inside of the black hole is bad. An example of such a solution is constructed in Appendix \ref{app:75} and discussed in Sec. \ref{sec:75}.

Alternatively, when there is no horizon, the flow hits a naked singularity. We  consider the cases where the naked singularity corresponds to the Gubser-regular asymptotic structure obtained in Appendix \ref{asymp}. From rules 17 on page \pageref{ru17} and 20 on page \pageref{ru20}, we learn that the only possibility is to connect the Gubser-regular endpoint with $V\to 0^-$ with a Gubser-regular endpoint where $V\to -\infty$. From rule 9 on page \pageref{ru9}, such a solution cannot have a horizon.
 \be\label{fl17}
{\rm Gubser-regular}\ {\rm endpoint}\ {\rm with}\ (V\to 0^-) ~\to~{\rm Gubser-regular}\ {\rm endpoint}\ {\rm with}\ (V\to -\infty)
 \ee
An example of such a solution is constructed in Appendix \ref{app:73} and discussed in Sec. \ref{sec:73}.

\subsubsection*{Flows involving Gubser-regular endpoints with $V\to 0^+$ }

In solutions involving a Gubser-regular endpoint with vanishing potential, we can take $W>0$ without loss of generality (rule 3 on page \pageref{ru3}). Therefore, a flow starting from such a Gubser-regular endpoint can end either at a maximum of $W$, i.e. a shrinking endpoint, or run to the boundary of field space.

In the first case, the flow necessarily encounters a cosmological horizon between both endpoints (rule 10 on page \pageref{ru10}) located in the dS regime (rule 12 on page \pageref{ru12}). The flow can terminate at a shrinking endpoint regardless of the sign of the potential.

$$
{\rm Gubser-regular}\ {\rm endpoint}\ {\rm with}\ (V\to 0^+)  ~~\to~~ {\rm cosmological~~ horizon}~~\to
$$
\be\label{fl18}
  \to~~ {\rm shrinking~~ endpoint}.
 \ee
Examples of such a solutions are constructed in Appendix \ref{app:72} and discussed in Sec. \ref{sec:72}.

In the second case, the flow necessarily runs into a singularity. The singularity is naked if there is no horizon, or if there is a single horizon (which is necessarily cosmological). Conversely, the singularity is shielded, if the flow features two horizons, in which case rule 22 on page \pageref{ru22} instructs that the singularity is bad. The outermost horizon is cosmological whereas the innermost is a black-hole event horizon:
$$
{\rm Gubser-regular}\ {\rm endpoint}\ {\rm with}\ (V\to 0^+)  ~\to~ {\rm cosmological~~ horizon}~ \to
 $$
 \be\label{fl19}
 \to~ {\rm event~~horizon}~ \to~~ {\rm bad ~ singularity}.
 \ee
Examples of such a solutions are constructed in Appendix \ref{app:75} and discussed in Sec. \ref{sec:75}.

Finally, a solution departing from a Gubser-regular endpoint with $V\to 0^+$ can run into a singularity with the Gubser-regular asymptotic structure of Sec. \ref{asymp} while not being covered by a black-hole event horizon. In particular, if the Gubser-regular endpoint has $V\to -\infty$, the flow encounters only a cosmological horizon (rule 10 on page \pageref{ru10}), while if the Gubser-regular endpoint has $V\to +\infty$ such a flow has no horizon (rule 11 on page \pageref{ru11}).
$$
{\rm Gubser-regular}\ {\rm endpoint}\ {\rm with}\ (V\to 0^+) ~~\to~~ {\rm cosmological~~ horizon}~~ \to
 $$
 \be\label{fl20}
 \to~~ {\rm Gubser-regular}\ {\rm endpoint}\ {\rm with}\ (V\to -\infty)\,,
 \ee

 \be\label{fl21}
 {\rm Gubser-regular}\ {\rm endpoint}\ {\rm with}\ (V\to 0^+) ~\to~{\rm Gubser-regular}\ {\rm endpoint}\ {\rm with}\ (V\to +\infty)
 \ee

Explicit examples of these two possibilities are constructed in Appendix \ref{app:73} and discussed in Sec. \ref{sec:73}.

\section{Examples of novel flows with finite endpoints}\label{fine2}

In this section, we display flows that explore partially or totally the dS regime of a potential $V$. In particular, we discuss here flows that start and end at finite endpoints of the flow, as well as flows that start at a finite endpoint and end at a singularity that is covered by a black-hole event horizon. For concreteness, we set $d=4$ in the examples of this section.

These have been constructed explicitly in Appendices \ref{app:tuned} and \ref{app:I}. In the former, we start from a suitable superpotential $W$ and, subsequently, solve for the metric functions $e^A$ and $f$ as well as for the potential itself $V$. In the latter, we provide an analytical solution and study particular examples of flows by appropriately choosing the integration constants.

\subsection{Flows from $d+1$ boundary endpoints to shrinking endpoints}

\begin{figure}[h!]
	\centering
	\subfloat{{\includegraphics[width=0.47\linewidth]{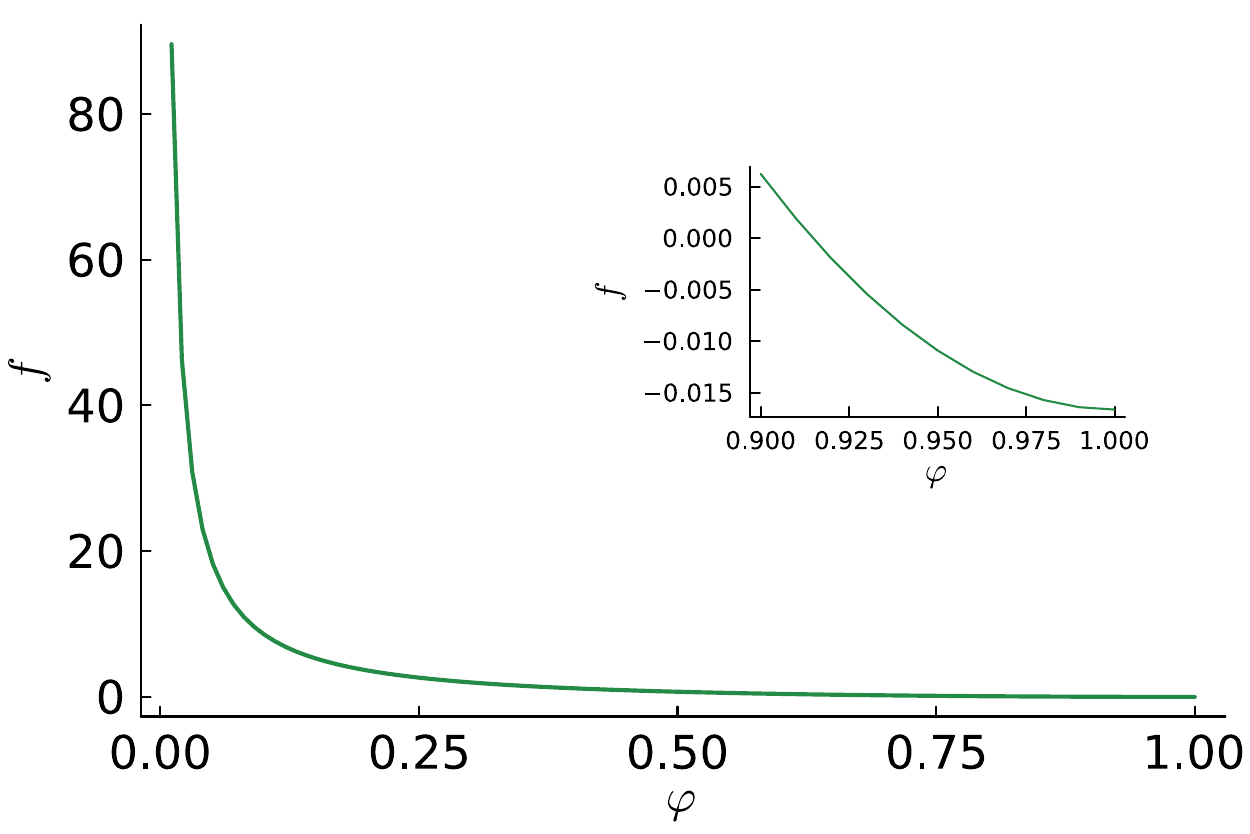}}}
	\qquad
	\subfloat{{\includegraphics[width=0.47\linewidth]{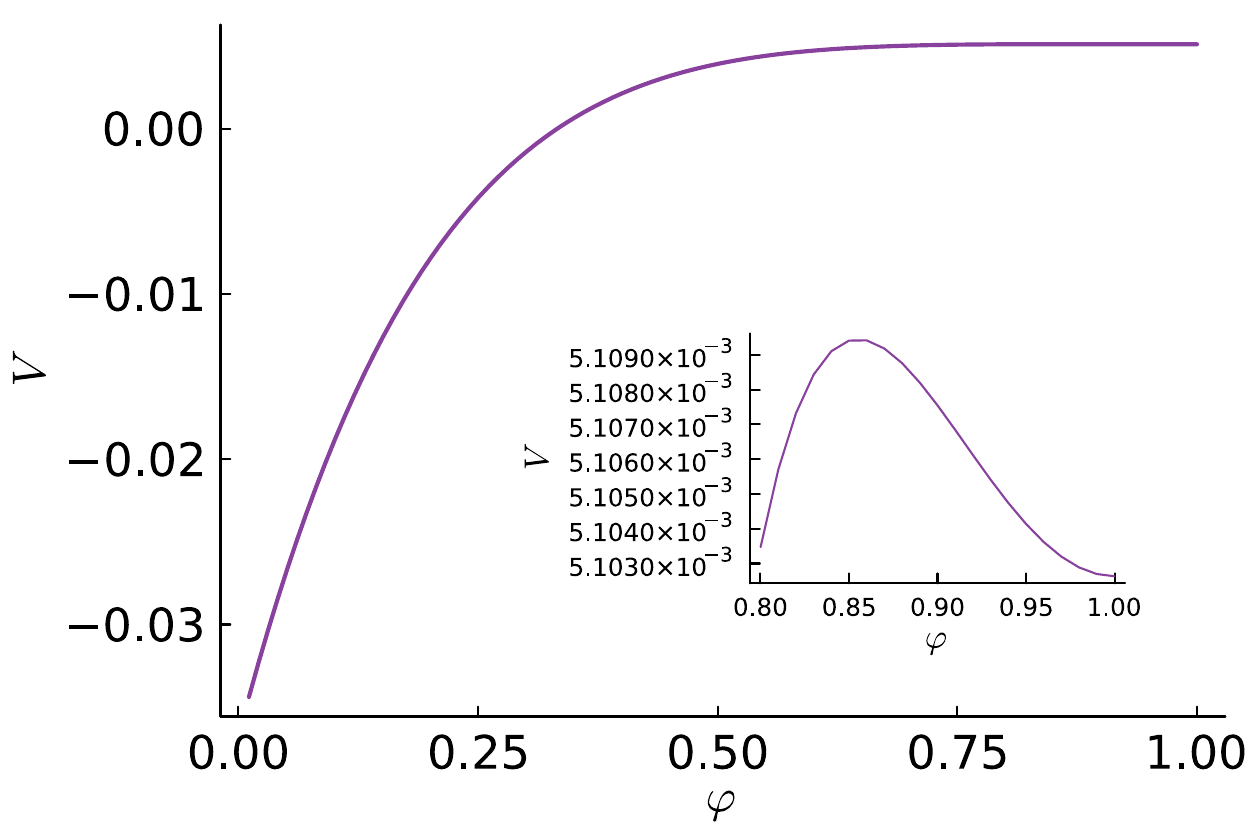}}}
	\qquad
	\subfloat{\includegraphics[width=0.47\linewidth]{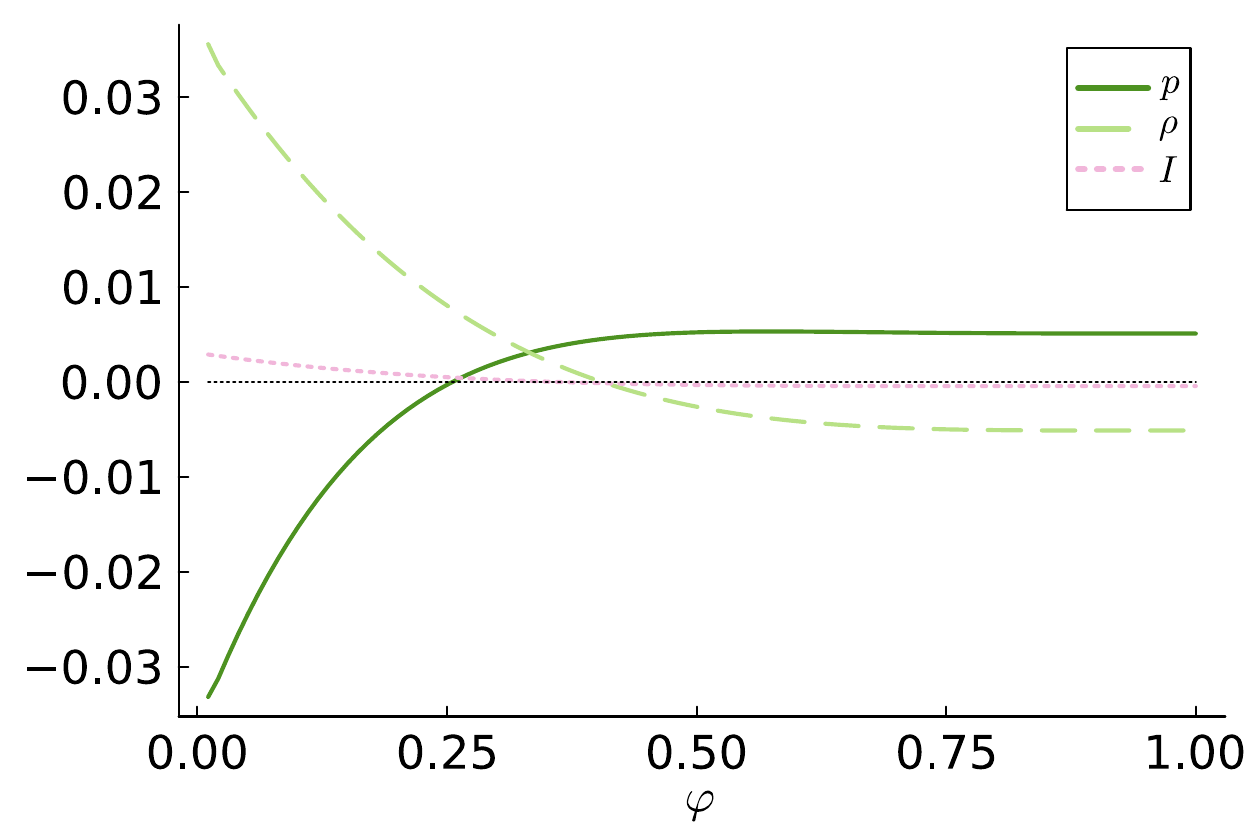}}
	\qquad
	\subfloat{{\includegraphics[width=0.47\linewidth]{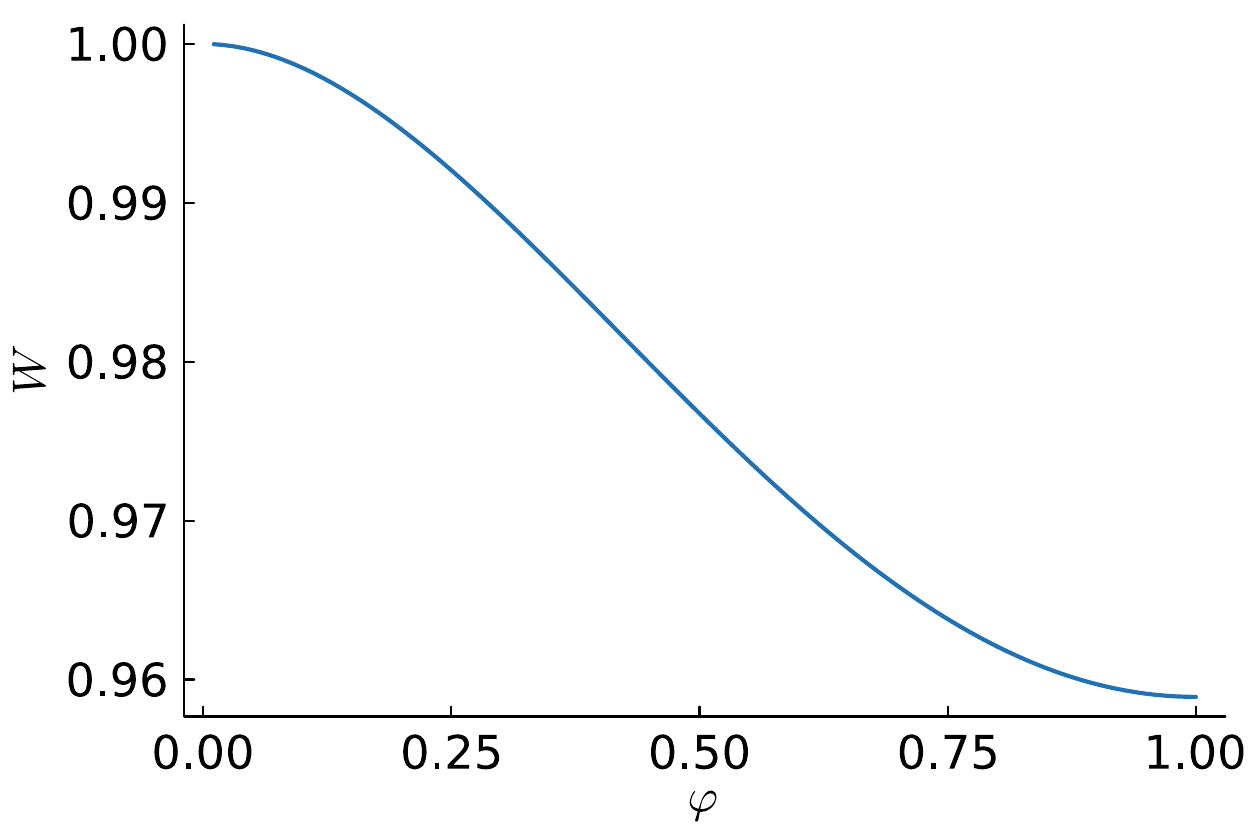}}}
	\caption{Flow from a boundary of dS$_5$ at $\f=1$, to the center of AdS$_5$ at $\f=0$. The solution has a horizon in the dS regime. $T$ and $f$ diverge at $\f=0$ in a correlated manner so that the curvature invariants are finite. We added zoom-in plots for $f$ and $V$ to emphasize the presence of a horizon and of the extremum respectively. In the bottom-left panel, we display the functions that appear in the curvature invariants, defined in Appendix \ref{sect:inv_sphere}. They are regular everywhere along the flow. }
	\label{fig:H1}
\end{figure}

\begin{figure}[h!]
	\centering
	\subfloat{{\includegraphics[width=0.47\linewidth]{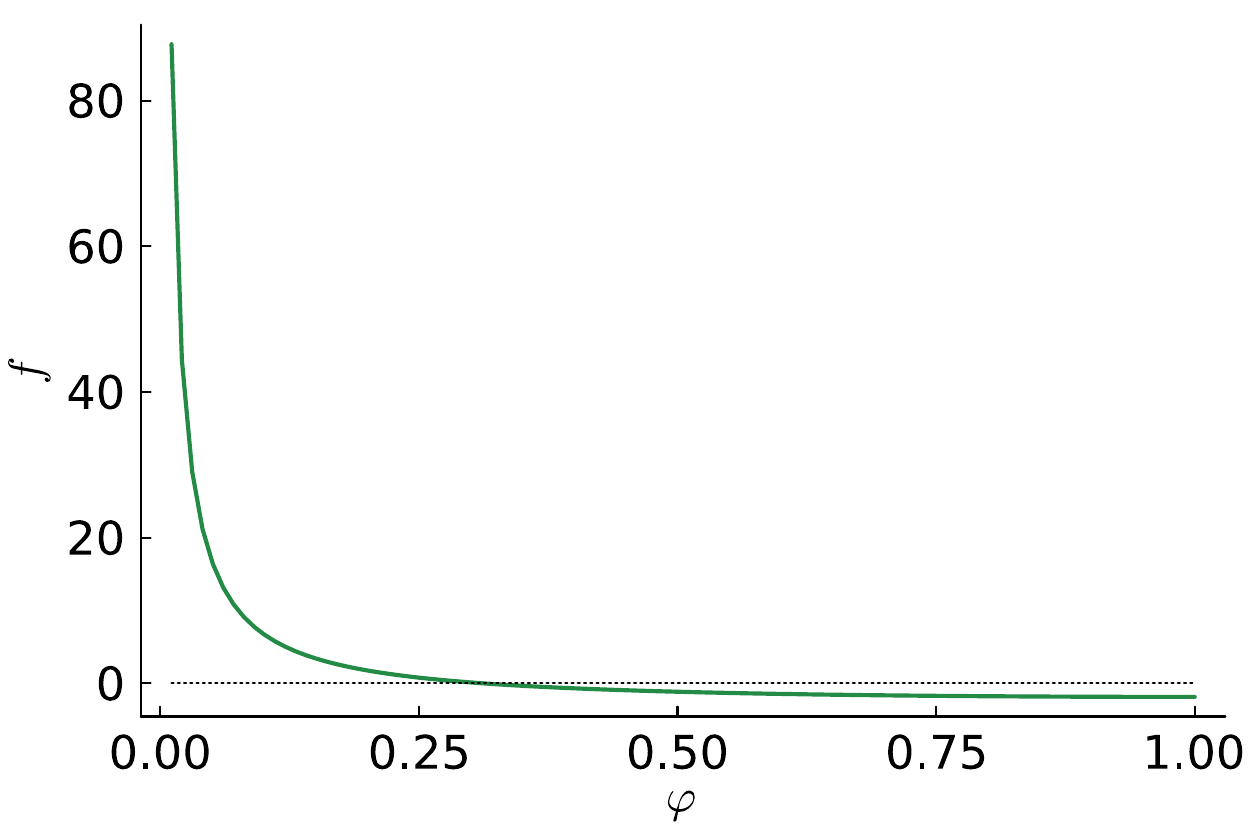}}}
	\qquad
	\subfloat{{\includegraphics[width=0.47\linewidth]{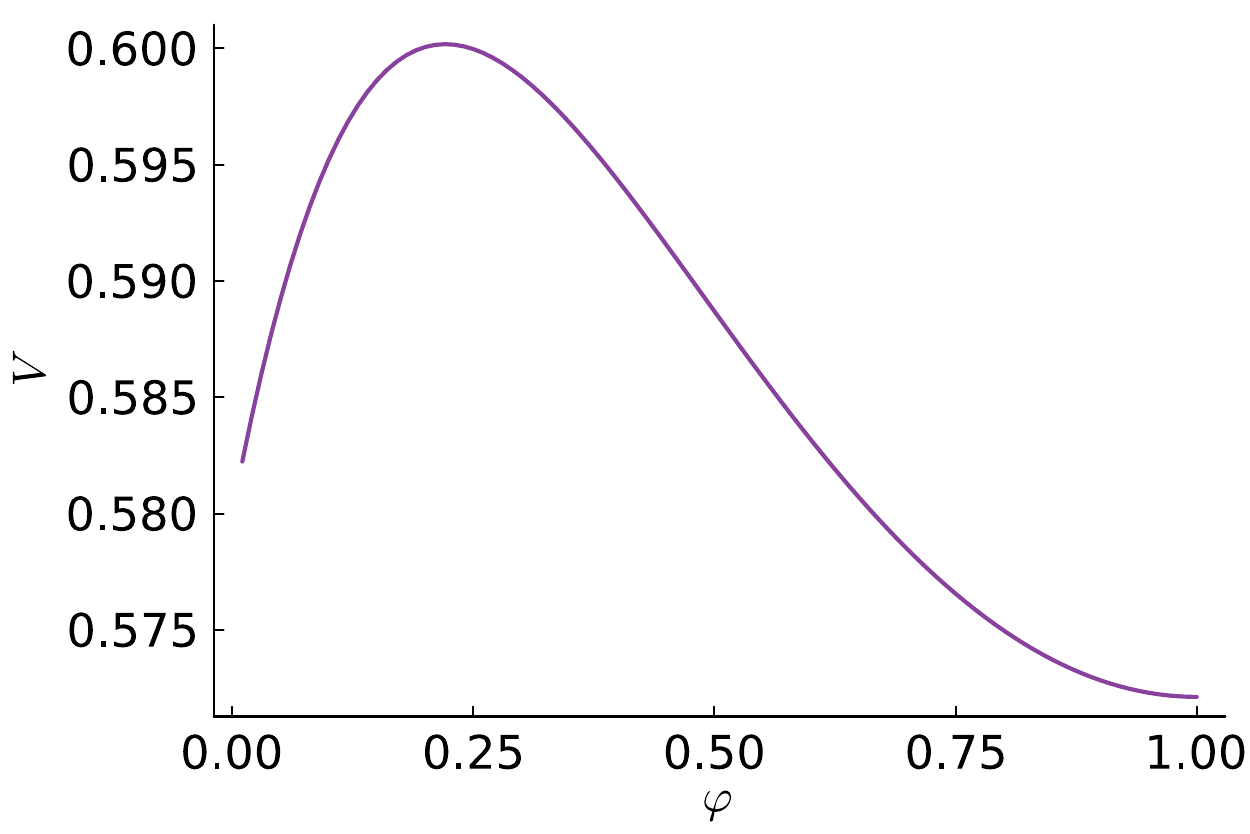}}}
	\qquad
	\subfloat{\includegraphics[width=0.47\linewidth]{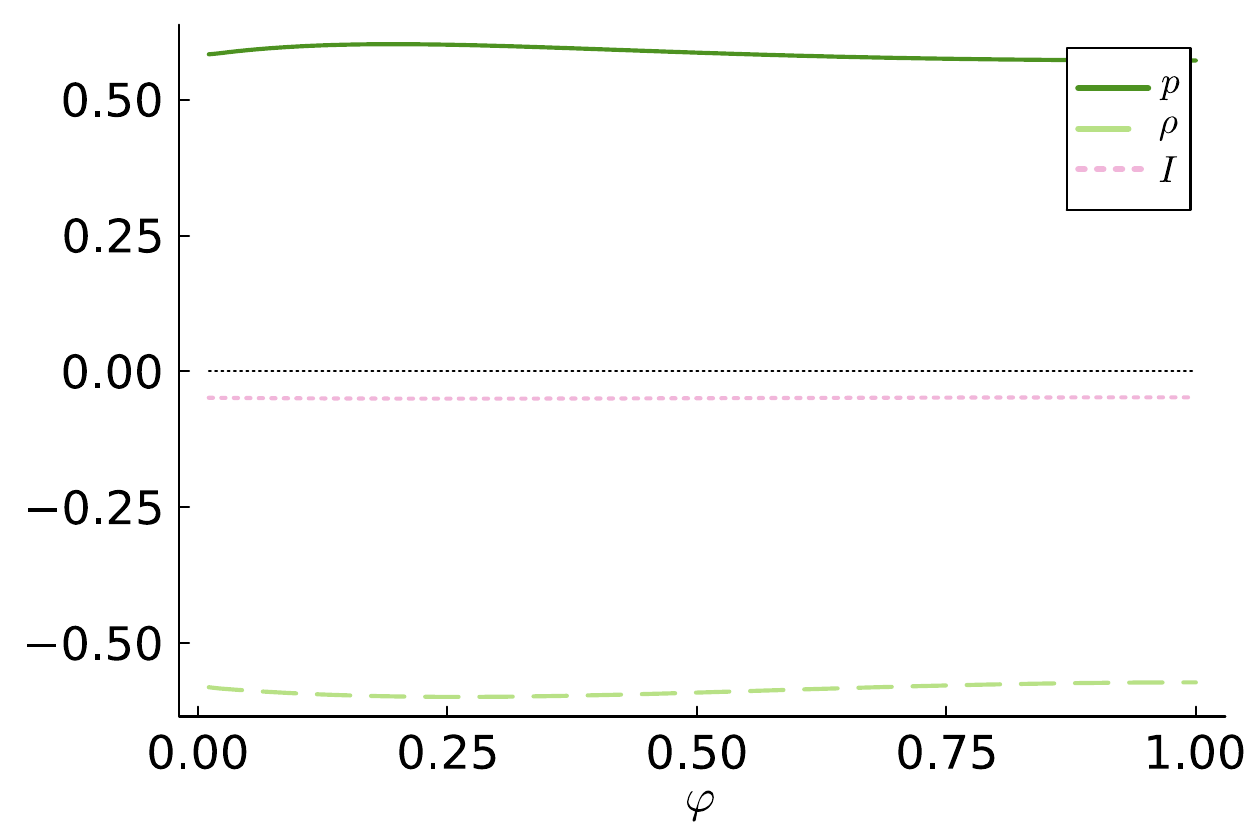}}
	\qquad
	\subfloat{{\includegraphics[width=0.47\linewidth]{dS5-AdS5-tuned-W}}}
	\caption{Flow from a boundary of dS$_5$ at $\f=1$ to the center of dS$_5$ at $\f=0$. The flow develops a horizon in the dS regime. $T$ and $f$ diverge at $\f=0$ in a correlated manner so that the curvature invariants are finite. We display the combinations that appear in the curvature invariants \ref{sect:inv_sphere}. }
	\label{fig:H2}
\end{figure}

\begin{figure}[h!]
	\centering
	\subfloat{{\includegraphics[width=0.47\linewidth]{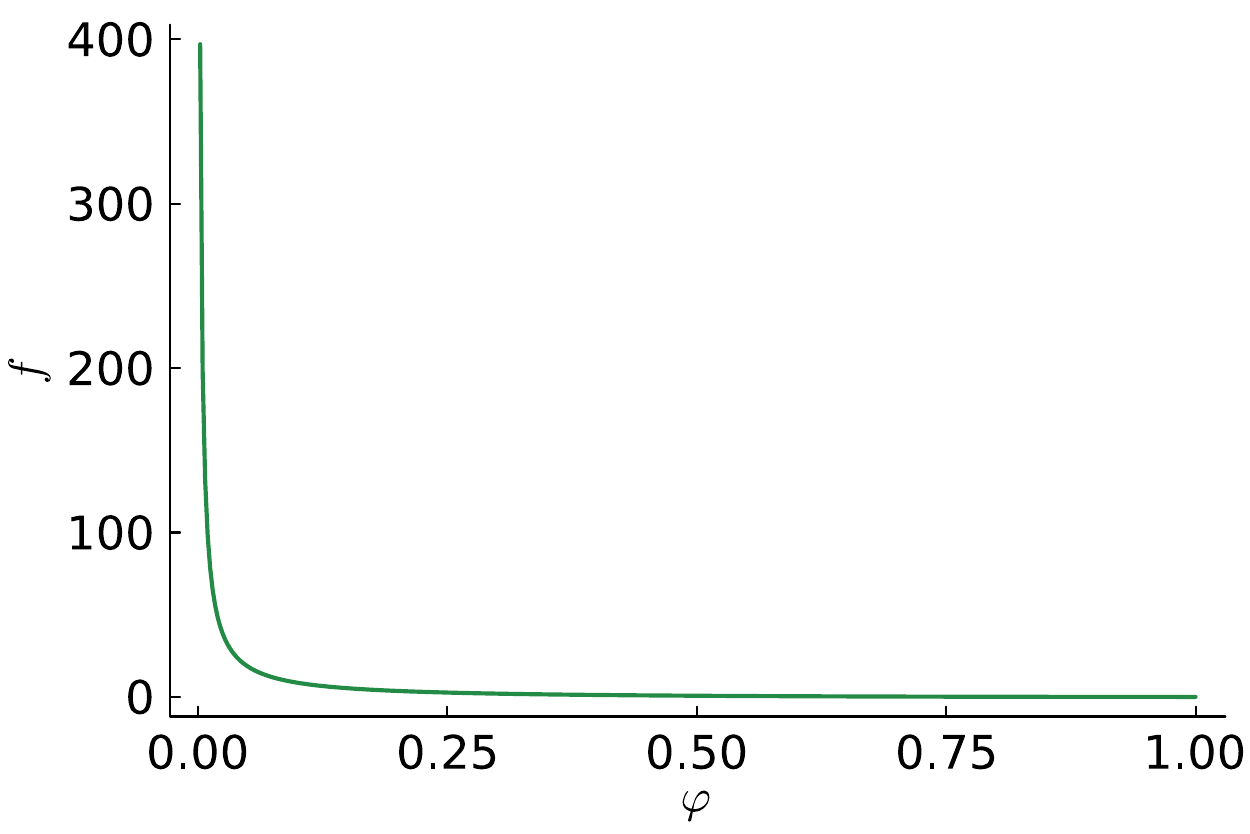}}}
	\qquad
	\subfloat{{\includegraphics[width=0.47\linewidth]{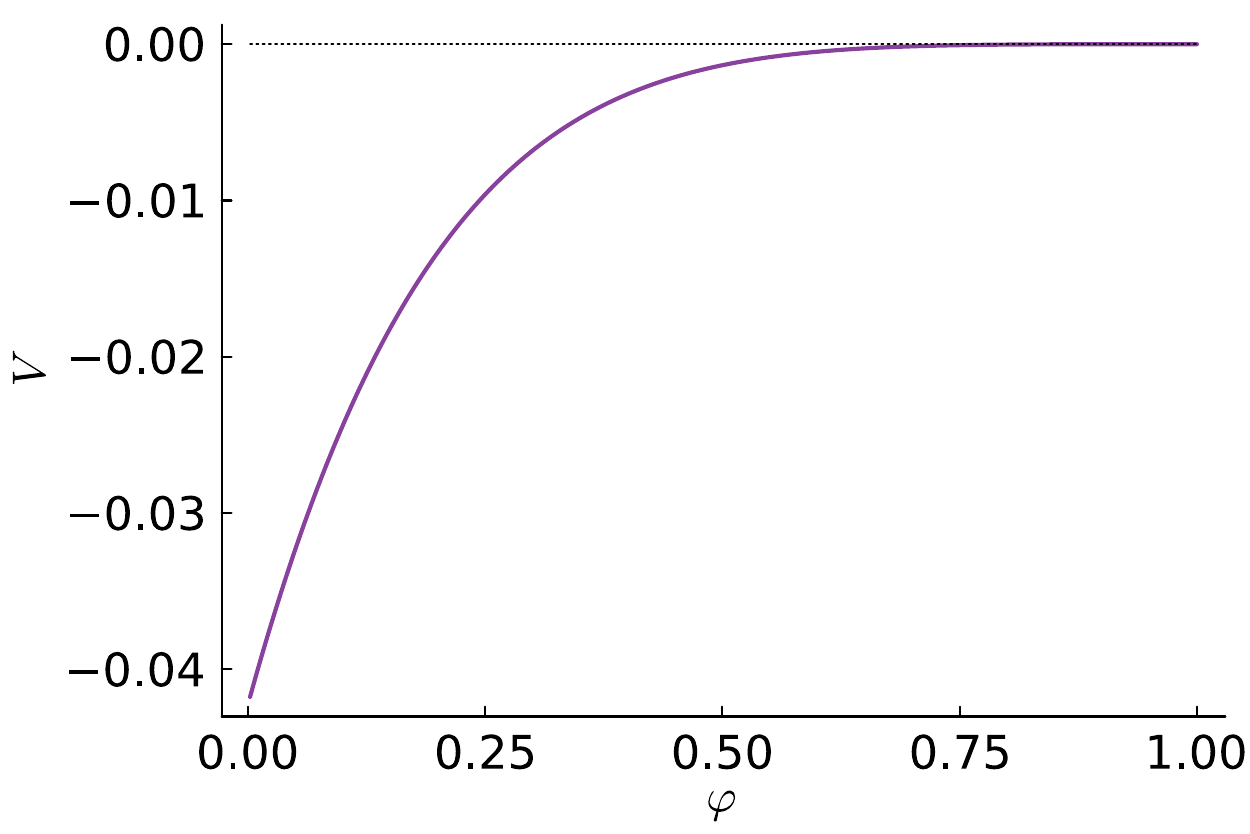}}}
	\qquad
	\subfloat{\includegraphics[width=0.47\linewidth]{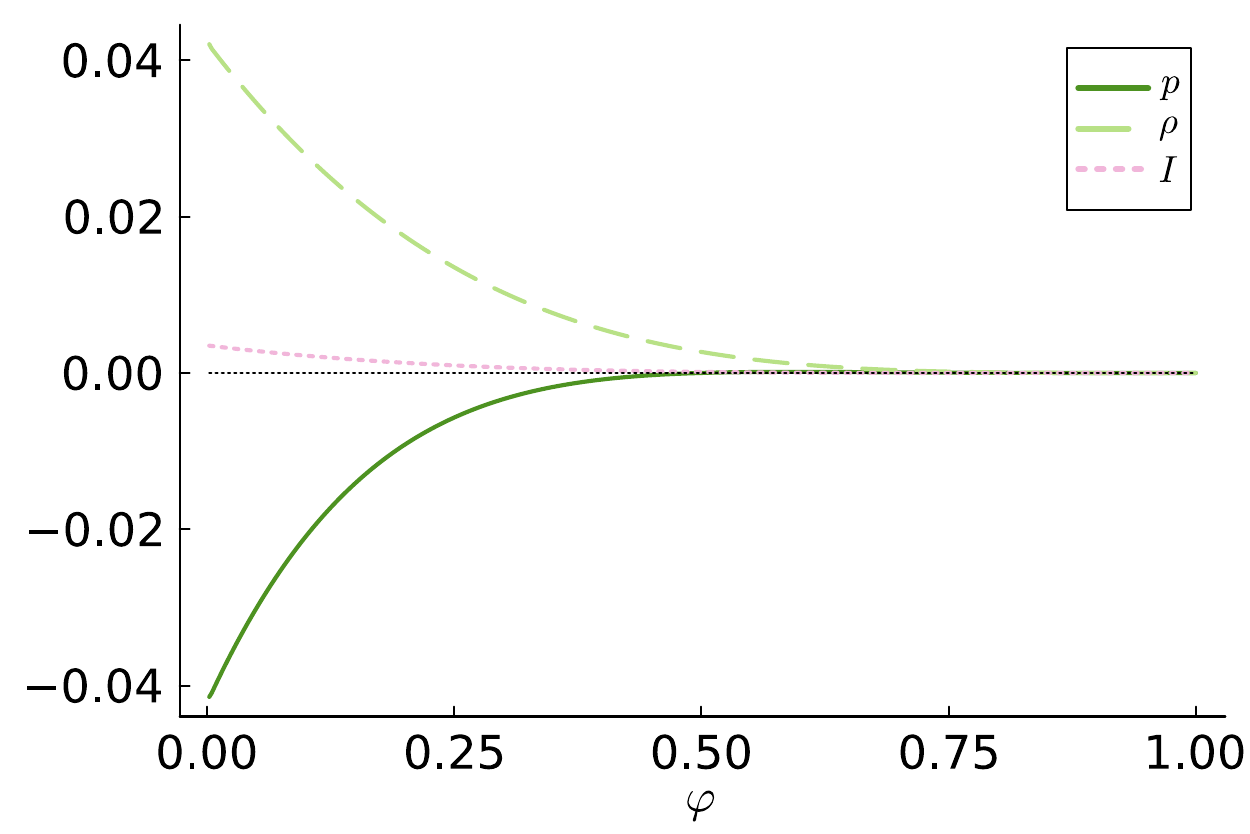}}
	\qquad
	\subfloat{{\includegraphics[width=0.47\linewidth]{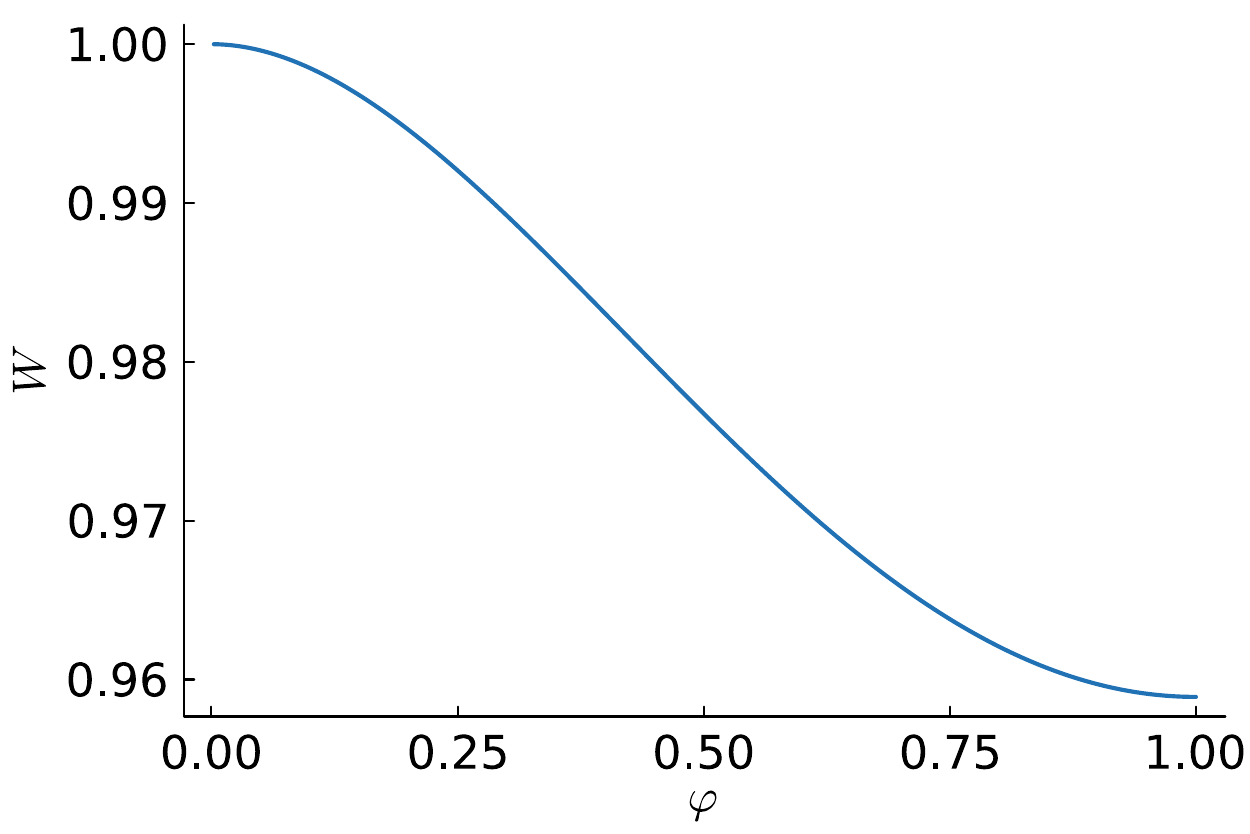}}}
	\caption{Flow from a boundary of M$_5$ at $\f=1$ to the center of AdS$_5$ at $\f=0$. $T$ and $f$ diverge at $\f=0$ in a correlated manner so that the curvature invariants are finite. We display the combinations that appear in the curvature invariants \ref{sect:inv_sphere}. }
	\label{fig:M5}
\end{figure}

We first discuss flows that start at $d+1$ boundary endpoints and end at shrinking endpoints. According to the rules of Sec. \ref{sec:global}, the possible flows connecting such boundary endpoints and shrinking endpoints are
$$
{\rm AdS}_{d+1}^{\rm bdy} \to {\rm AdS}_{d+1}^{\rm shr}\,,\quad {\rm dS}_{d+1}^{\rm bdy} \to {\rm dS}_{d+1}^{\rm shr}\,, \quad {\rm dS}_{d+1}^{\rm bdy} \to {\rm AdS}_{d+1}^{\rm shr}\,,\quad {\rm M}_{d+1}^{\rm bdy} \to {\rm AdS}_{d+1}^{\rm shr}\,.
$$
These solutions are constructed in Appendix \ref{app:H}. The strategy we follow to construct these flows is to start with a conveniently chosen superpotential that features two extrema: one corresponding to a shrinking endpoint, and one corresponding to a boundary endpoint. Subsequently,  equations \eqref{eqtt}, \eqref{f10_1} and \eqref{w55} are solved to obtain the inverse scale factor $T$, the function $f$ and to reconstruct the potential $V$. The nature of the shrinking and boundary endpoints is determined by the sign of the potential at each endpoint, which ultimately depends on the boundary conditions chosen for $T$ and $f$. For the solutions outlined above, the chosen superpotential is
\begin{equation}
W(\f) = 1-\frac{\f ^2}{6}+\frac{37 \f ^3}{219}-\frac{19 \f ^4}{438}\,,
\end{equation}
which has a boundary endpoint of a five-dimensional manifold at $\f=1$ while it has a shrinking endpoint at $\f=0$. If the boundary endpoint is either dS or AdS, then the superpotential implies that $\Delta_-=1$, by virtue of Eqs. \eqref{DEL}, \eqref{dwm1} and \eqref{dwm3}. Conversely, if the boundary endpoint is a Minkowski boundary, then it corresponds to the asymptotic structure of the first relation in Eqs. \eqref{e2cb}.

Depending on the boundary conditions chosen for $f$ we find the following four possibilities:
\begin{itemize}

\item[(a)] A flow without horizon from the boundary of AdS$_5$ to a shrinking endpoint in the AdS regime; This has the standard holographic interpretation, as dual to the ground state of a holographic QFT on $R\times S^{d-1}$. We do not show explicitly this well-known solution.

\item[(b)] A flow from the boundary of dS$_5$ at $\f=1$ to a shrinking endpoint in the AdS regime at $\f=0$, with a cosmological horizon located in the dS regime. The potential, superpotential, and blackening function for this solution is shown in Fig. \ref{fig:H1}, together with the quantities controlling the curvature invariants.

\item[(c)] A flow from the boundary of dS$_5$ at $\f=1$ to a shrinking endpoint in the dS regime at $\f=0$, again with a cosmological horizon in the dS regime. The function $f$, the potential, and the superpotential for this solution is shown in Fig. \ref{fig:H2}, together with the quantities controlling the curvature invariants.

\item[(d)] A flow from the boundary of M$_5$ at $\f=1$ to a shrinking endpoint in the AdS regime at $\f=0$. Again, the potential, superpotential, and blackening function, and the quantities controlling the curvature invariants, are shown in Fig. \ref{fig:M5}.

\end{itemize}

The Penrose diagram of dS$^{\rm bdy}_5\to$ AdS$^{\rm shr}_5$ solutions is similar to the dS$^{\rm bdy}_5\to$ dS$^{\rm shr}_5$ solutions and this is similar to the Penrose diagram of dS space in static coordinates.

\subsection{Flows from $d+1$ boundary endpoints to a black hole}\label{sec:dBH}

\begin{figure}[h!]
	\centering
	\subfloat{{\includegraphics[width=0.47\linewidth]{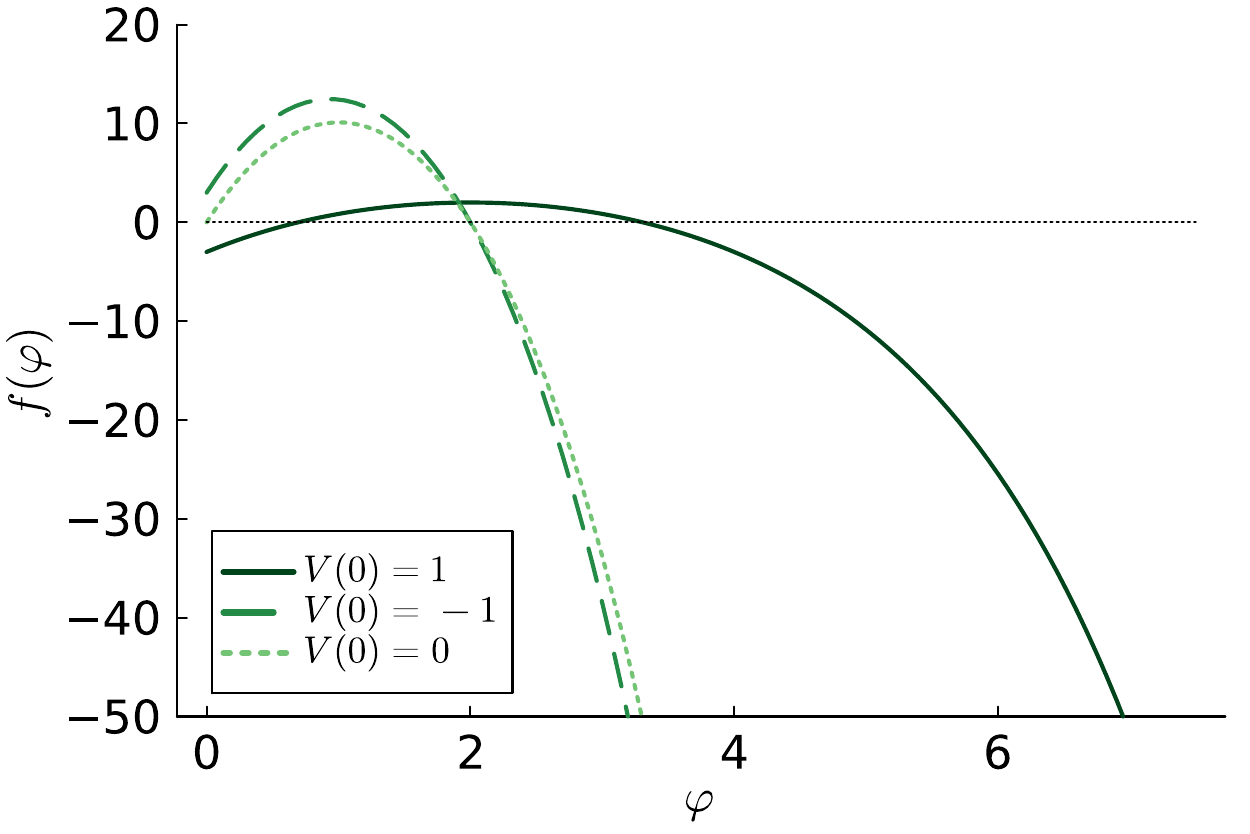}}}
	\qquad
	\subfloat{{\includegraphics[width=0.47\linewidth]{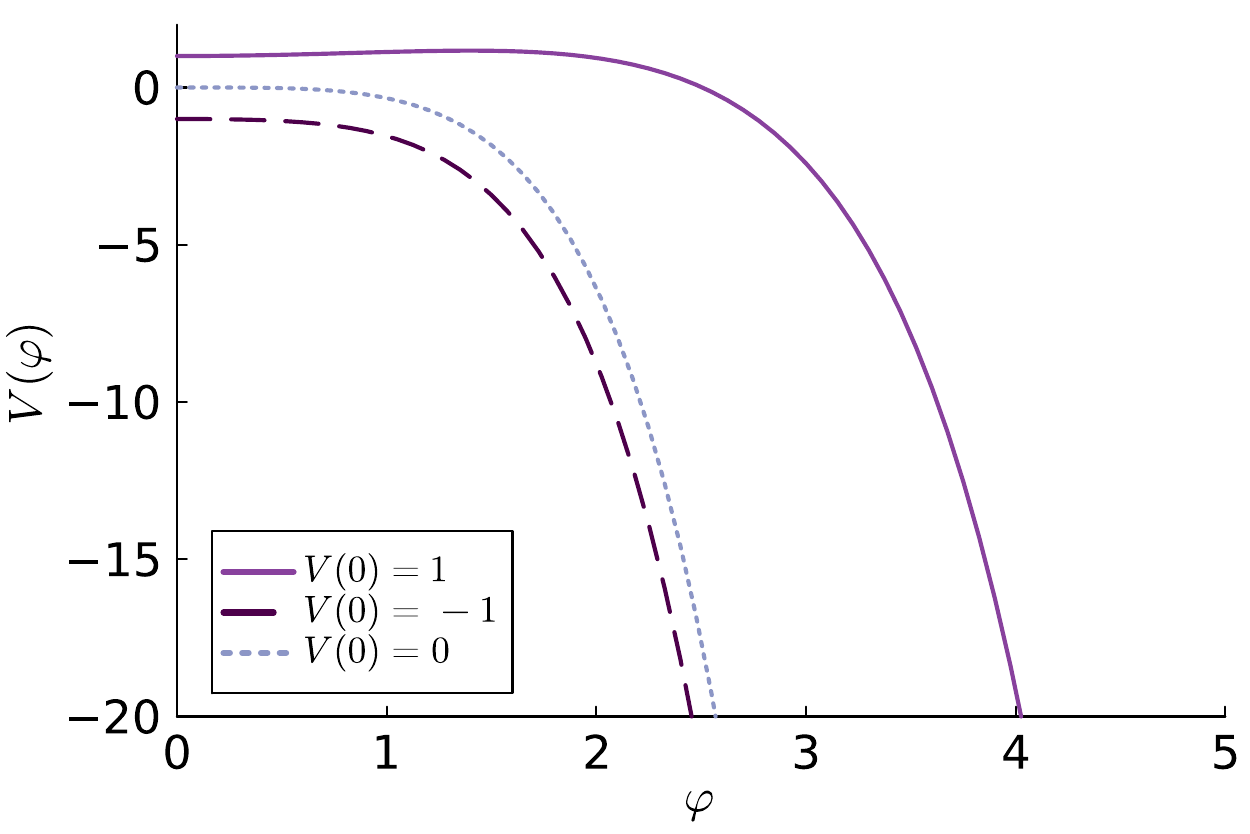}}}
	\qquad
	\subfloat{{\includegraphics[width=0.47\linewidth]{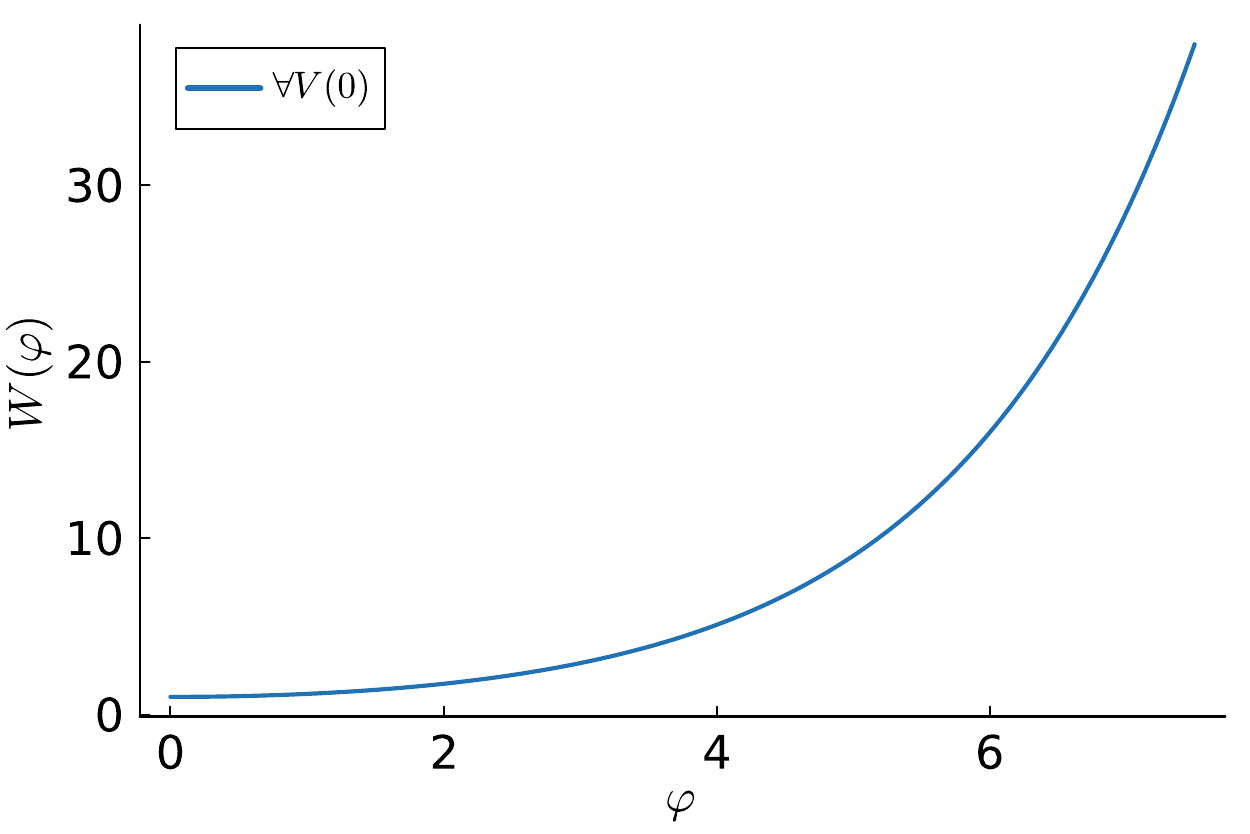}}}
	\caption{Flow from a $d+1$ boundary endpoint at $\f=0$ to a black hole. The sign of the potential at the boundary, $V(0)$, determines whether this is an AdS, dS or Minkowski boundary. The vanishing of $f$ signals the presence of a horizon. All three solutions have a black-hole event horizon, with a singularity in the interior. }
	\label{fig:BH}
\end{figure}

We now discuss solutions that start at a finite endpoint corresponding to the boundary of a $d+1$-dimensional manifold, to a black-hole event horizon. According to the rules discussed in Sec. \ref{sec:global}, we have the following three possibilities:
$$
{\rm AdS}_{d+1}^{\rm bdy} \to {\rm black~~ hole}\,,\quad {\rm dS}_{d+1}^{\rm bdy} \to {\rm black~~ hole}\,,\quad {\rm M}_{d+1}^{\rm bdy} \to {\rm black ~~hole}\,.
$$
In Appendix \ref{app:I} we constructed a family of exact solutions to the equations of motion. A subclass of such solutions contains a flow from $d+1$ boundary endpoints to a black-hole event horizon. The explicit family of solutions is given by

\begin{equation}\label{615}
W = \cosh\left(\dfrac{\varphi}{\sqrt{3}}\right)\,, \qquad T = -C_t \sinh\left(\dfrac{\varphi}{\sqrt{3}}\right)\,,
\end{equation}
\begin{equation}\label{616}
f= f_0 + 36 C_t e^{-\f/\sqrt{3}}-\frac{1}{2}f_1\cosh\left(\dfrac{\varphi}{\sqrt{3}}\right)\,,
\end{equation}
\begin{equation}\label{617}
V=-\dfrac{1}{4}f_0+\left(\frac{1}{6}f_1 -12C_t\right)\cosh\left(\dfrac{\varphi}{\sqrt{3}}\right)-\dfrac{1}{12}f_0\cosh\left(\dfrac{2\varphi}{\sqrt{3}}\right)\,,
\end{equation}
where $f_0$, $f_1$ and $C_t$ are integration constants. At $\f = 0$, the superpotential is positive and has a minimum. Therefore, at $\f=0$ there is a five-dimensional boundary endpoint. The superpotential has no other extremum, so the flow necessarily runs to the boundary of field space. How to set the integration constants is discussed in Appendix \ref{aehor}. Generically, we demand that $C_t<0$ and study flows for $\f>0$. In such a case, the function $T$ is positive as required by the spherically sliced ansatz. We also demand that there is a horizon at some finite location $\f_h$, where $f(\f_h)=0$, and demand that the potential is either positive, negative, or vanishing at the boundary endpoint $(\f=0)$. In this way, we construct examples of the three possible cases:
\begin{itemize}
\item[(a)] $V(0)<0$: Flow from an AdS$_5$ boundary endpoint at $\f=0$ to a black-hole event horizon at $\f_h=2$. This is a standard holographic RG-flow at finite temperature. An example of this solution is shown in Fig. \ref{fig:BH} (dashed lines).
\item[(b)] $V(0)=0$: Flow from a M$_5$ boundary endpoint at $\f=0$ to a black-hole event horizon at $\f_h=2$. This is an example of a black hole with scalar hair in an asymptotically flat space-time.  An example of this solution is shown in Fig. \ref{fig:BH} (dotted lines).
\item[(c)] $V(0)>0$: Flow from a dS$_5$ boundary endpoint at $\f=0$ to a black hole at $\f_h\simeq 3.3$. This solution also has a cosmological horizon at $\f_h\simeq 0.7$.  An example of this solution is shown in Fig. \ref{fig:BH} (solid lines). This solution should be understood as a generalization of a dS black hole with a running dilaton. The cosmological and event horizons can be made coincident, as exemplified in Appendix \ref{aehor}.
\end{itemize}

\subsection{Flow from dS$_2$ boundary endpoints to shrinking endpoints}

\begin{figure}[h!]
	\centering
	\subfloat{\includegraphics[width=0.47\linewidth]{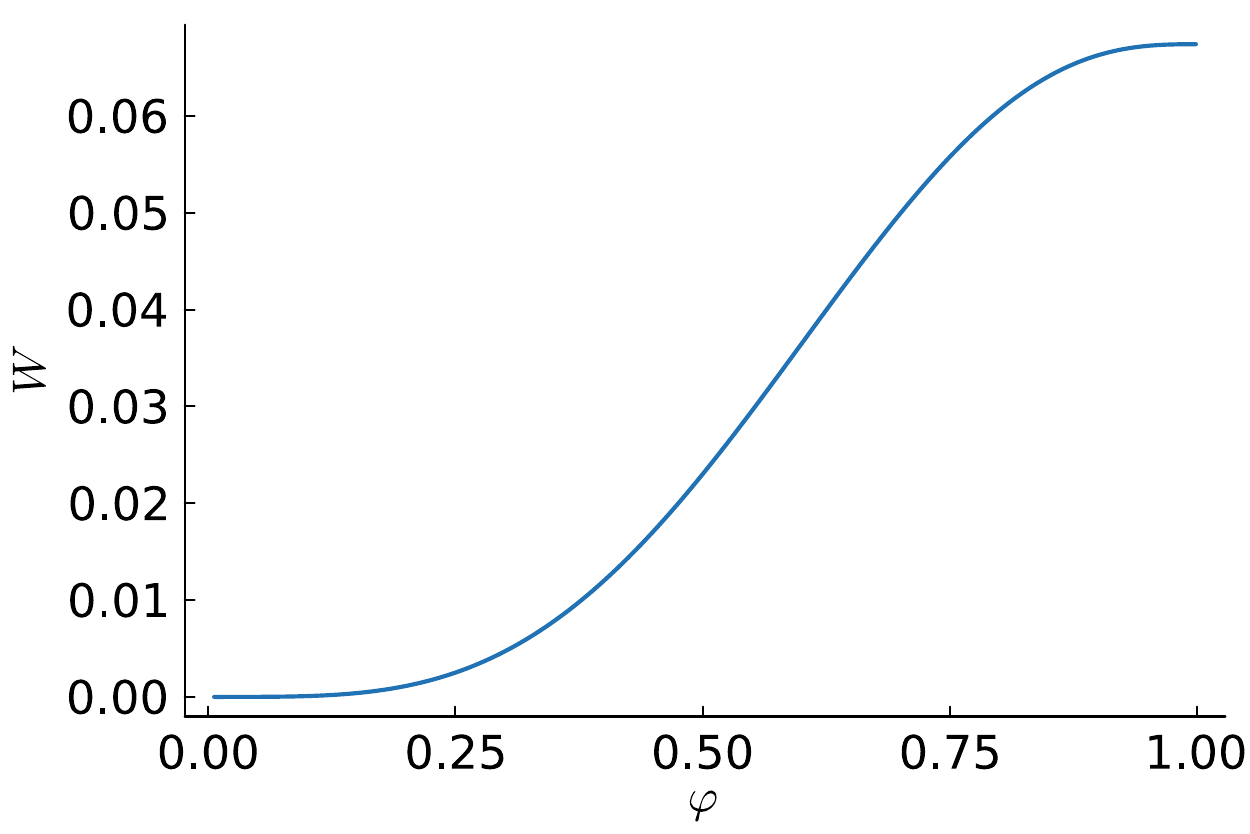}}
	\qquad
	\subfloat{\includegraphics[width=0.47\linewidth]{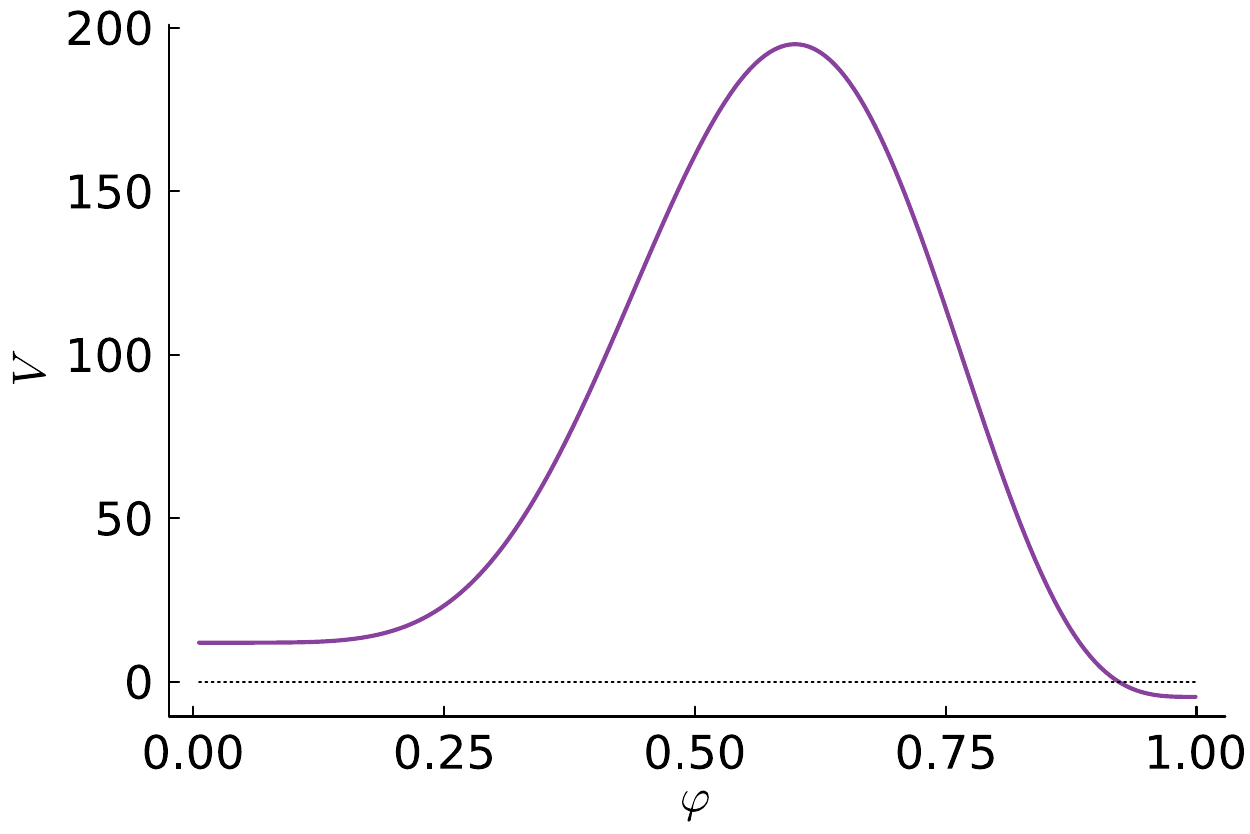}}
	\qquad
	\subfloat{\includegraphics[width=0.47\linewidth]{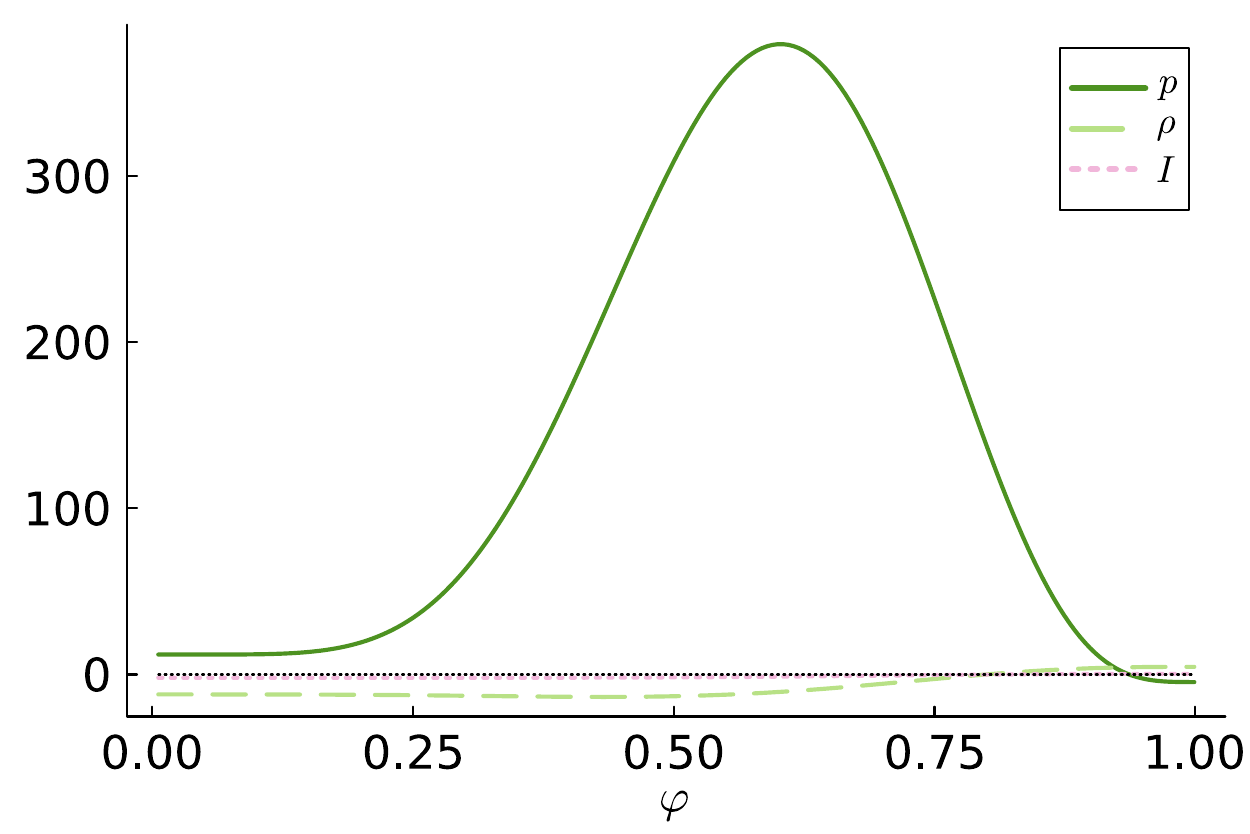}}
	\qquad
	\subfloat{\includegraphics[width=0.47\linewidth]{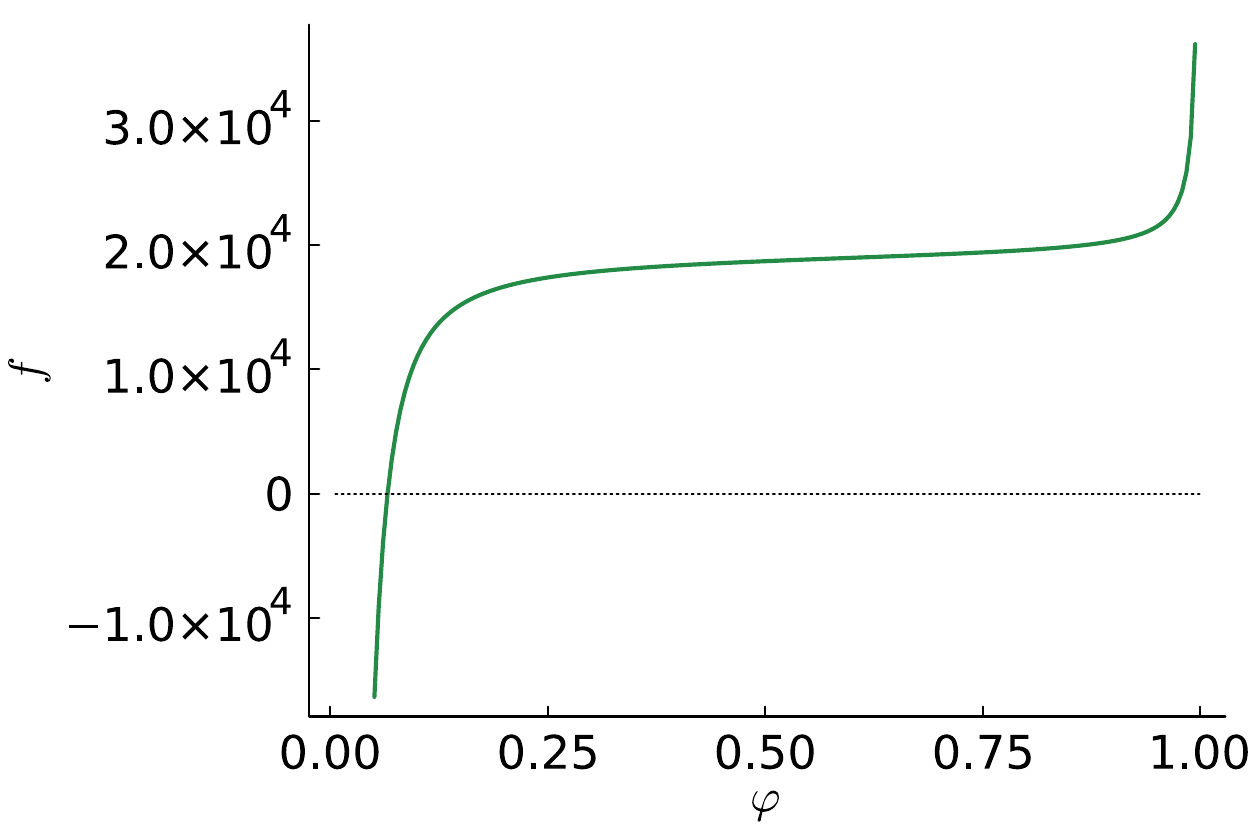}}
	\caption{Flow from the boundary of dS$_2 \times $S$_3$ at $\f=0$ to the interior of AdS$_5$ at $\f=1$. The superpotential vanish as $\f^4$ near $\f=0$, while $f$ diverges at both endpoints and $T$ diverges at $\f=0$. The curvature invariants are regular along the flow, since the pressure, energy density and $\mathcal{I}$ are finite, as shown in the bottom left panel.}
	\label{fig:G3b}
\end{figure}

\begin{figure}[h!]
\centering
\subfloat{\includegraphics[width=0.47\linewidth]{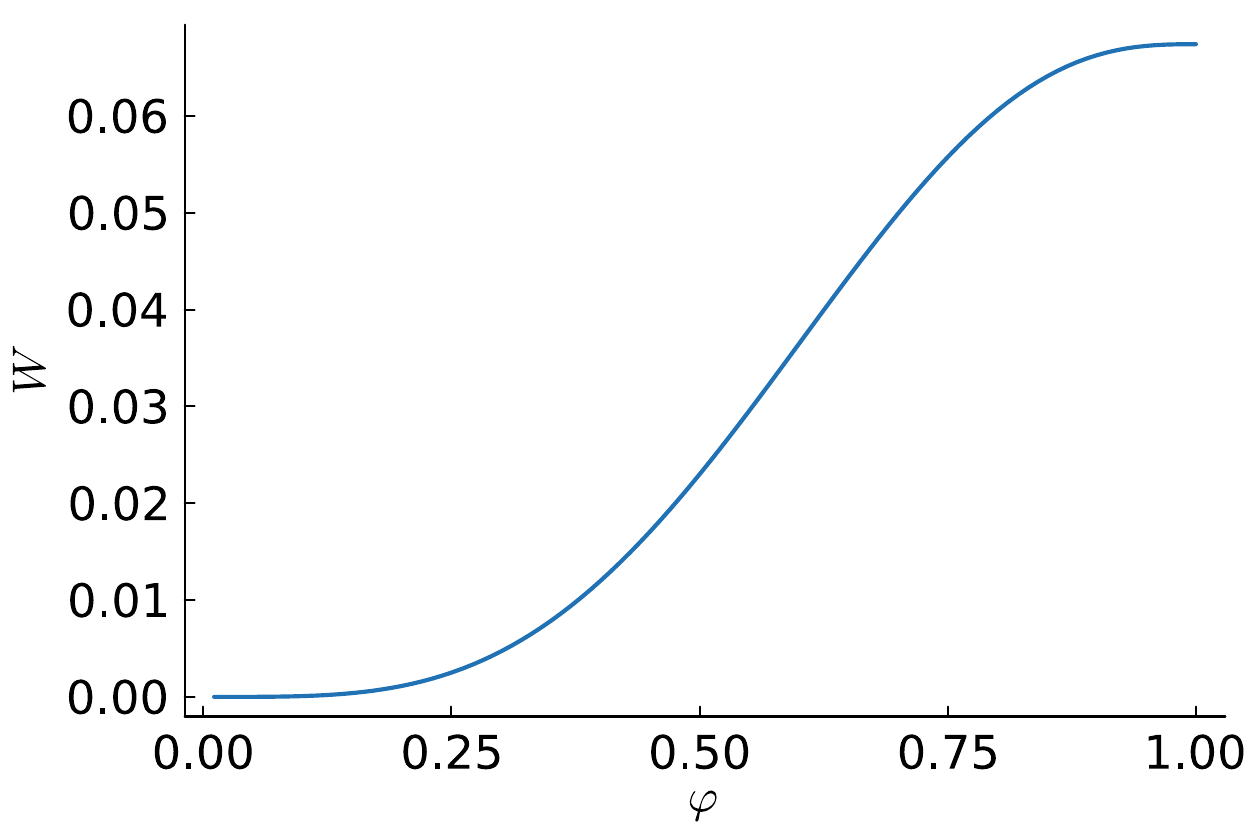}}
	\qquad
	\subfloat{\includegraphics[width=0.47\linewidth]{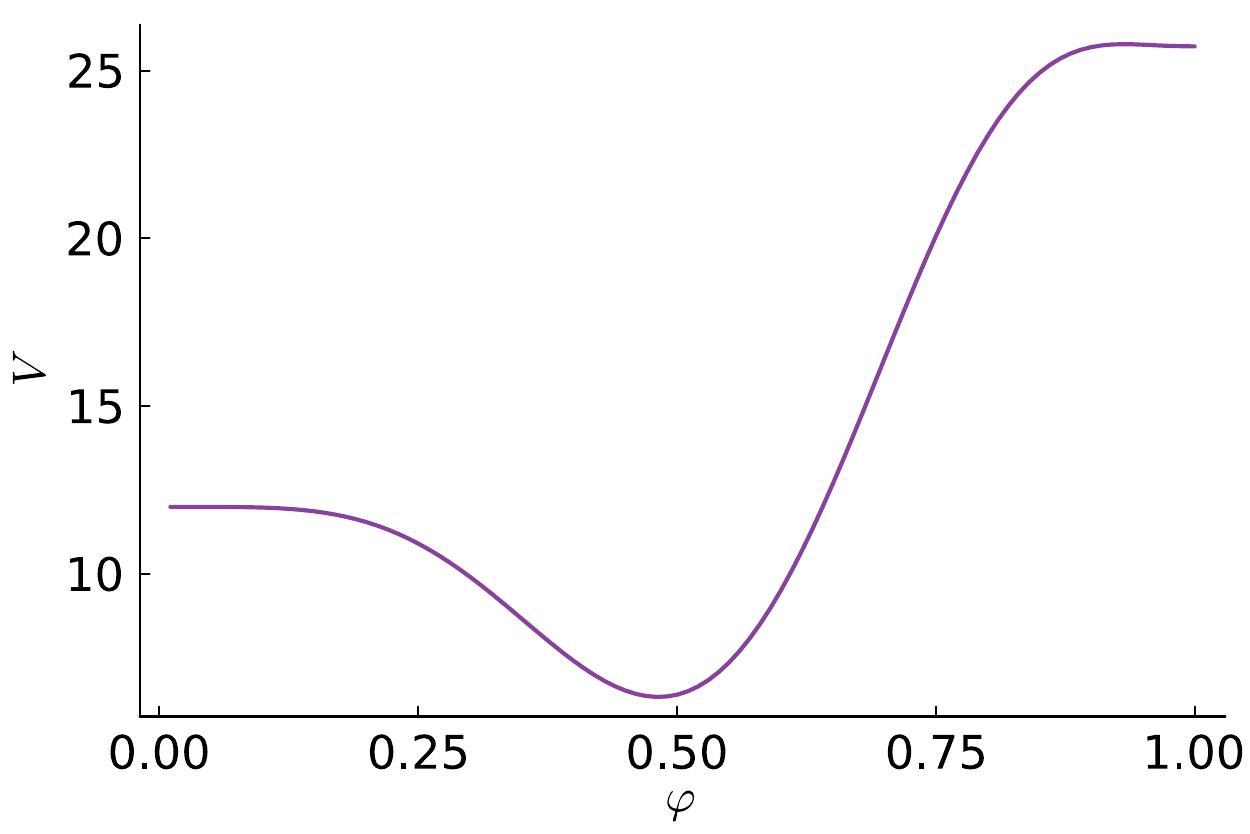}}
	\qquad
	\subfloat{\includegraphics[width=0.47\linewidth]{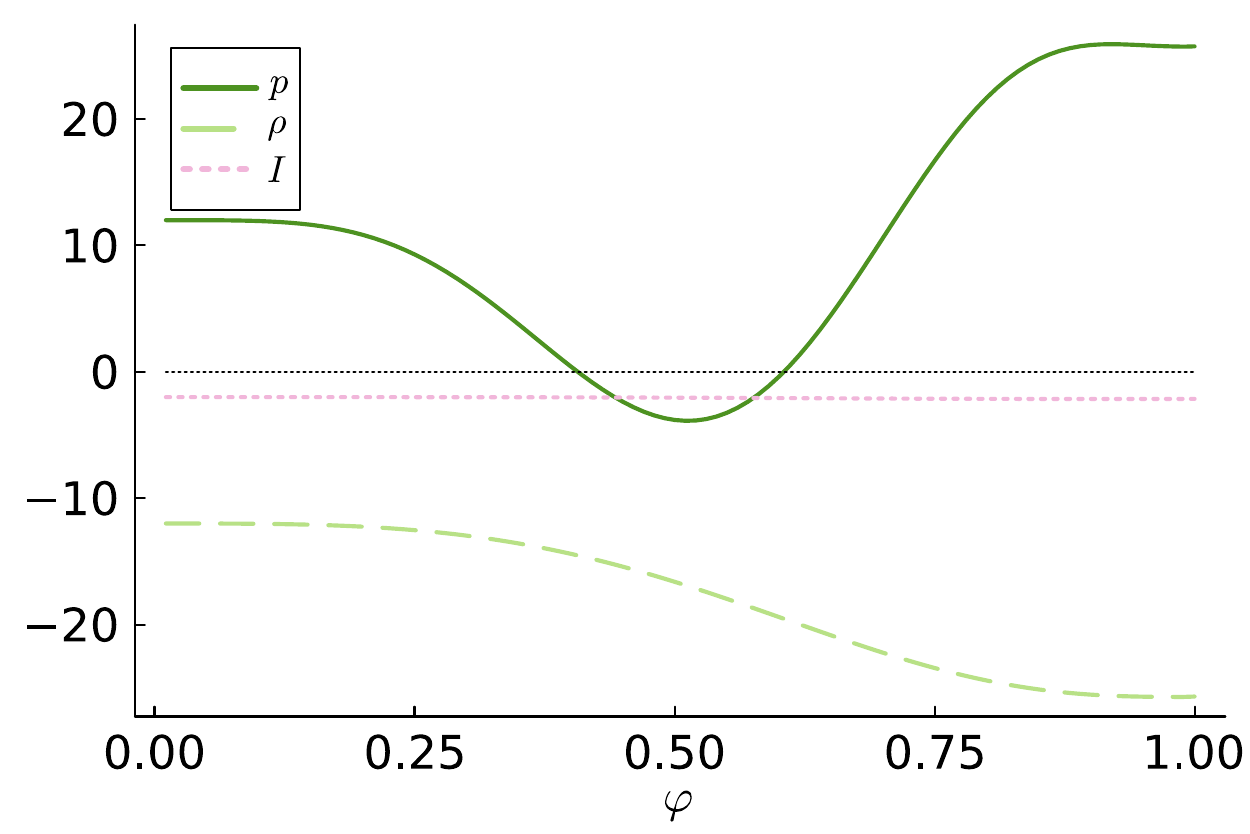}}
	\qquad
	\subfloat{\includegraphics[width=0.47\linewidth]{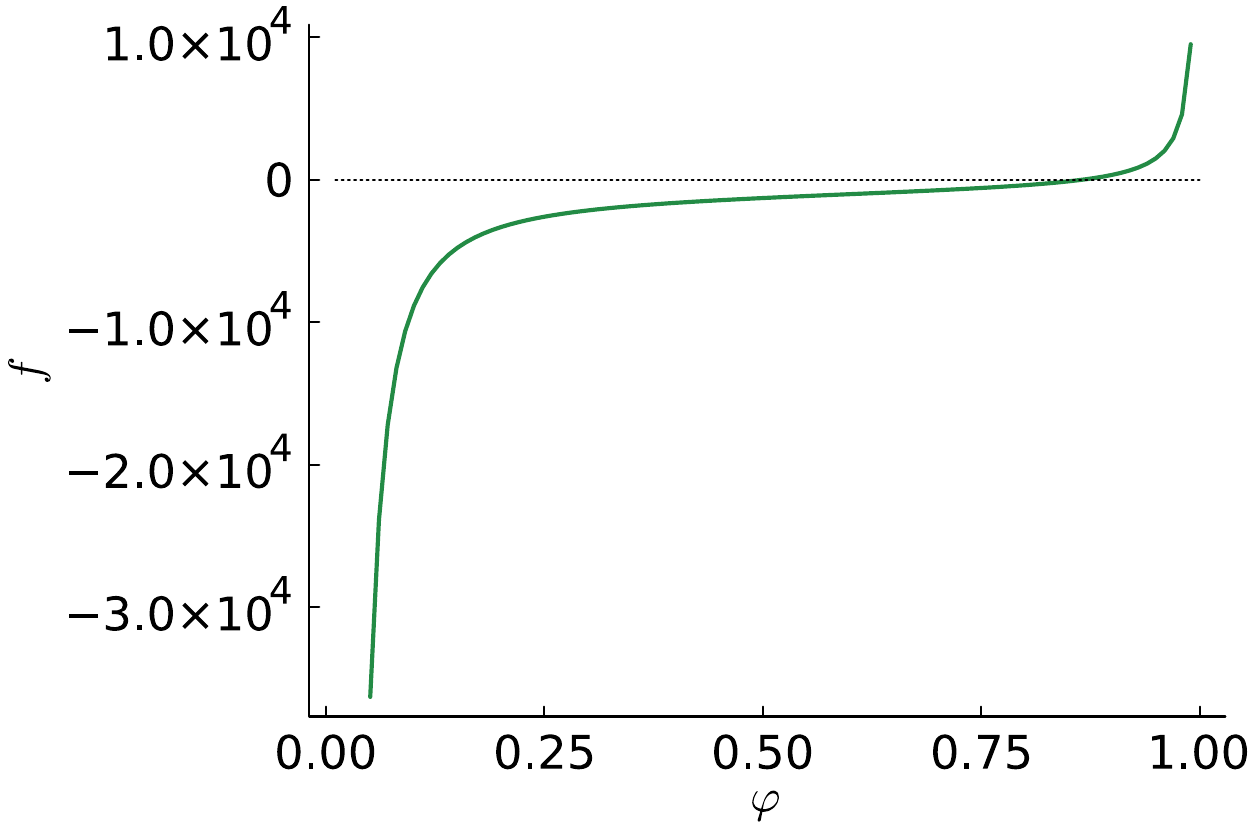}}
\caption{Flow from the boundary of dS$_2 \times $S$^3$ at $\f=0$ to {the location of an observer in dS$_5$ in the static patch coordinates} at $\f=1$. The superpotential vanishes as $\f^4$ near $\f=0$, while $f$ diverges at both endpoints and $T$ diverges at $\f=1$. The curvature invariants are regular along the whole flow including the endpoints, since the quantities controlling the curvature invariants, $(p,\rho,\mathcal{I})$ are finite.
}
\label{fig:G1}
\end{figure}

In order to construct solutions from a dS$_2$ boundary to shrinking endpoints, we  engineer a superpotential $W$ that features both endpoints. We ensure the presence of such endpoints by demanding that $W$ has two extrema whose local form is dictated by the local structure of the solutions derived in Appendix \ref{g53} and \ref{G.2.2} respectively. In particular, we choose

\begin{equation}\label{71a}
W= \f^4 - \dfrac{142}{89} \f^5 + \dfrac{59}{89}\f^6\,.
\end{equation}
\noindent
Note that the dS$_2$ boundary corresponds to the minimum located at $\f=0$ with the parameter defined in Eq. \eqref{alfa} set to $\delta_{\pm}=1/2$. The superpotential has a maximum at $\f=1$ where the shrinking endpoint is located. Given the superpotential in Eq. \eqref{71a}, we proceed to numerically solve the equations of motion \refeq{f10_1}-\refeq{w55b} in the range $\f\in[0,1]$. The nature of the shrinking endpoint, either dS or AdS, is determined by the sign of the potential at such an endpoint, which is ultimately controlled by the integration constants. The detailed construction of the solutions with this superpotential is contained in Appendix \ref{app:G}, and we refer to this appendix for more details. Depending on the choice of integration constants, we find the following two possibilities:

\begin{itemize}
\item[(a)] Flow from a dS$_2$ boundary to an AdS shrinking endpoint.

This solution is presented in figure \ref{fig:G3b}. The potential $V$ has a minimum in the dS regime at $\f=0$, which corresponds to the dS$_2$ boundary. At such point, the blackening function $f$ diverges to negative infinity as $f\sim 1/\f^4$, while the superpotential vanishes as $W\sim \f^4$ (see Eqs. \eqref{71a} and \eqref{j4}). Therefore, the curvature invariants in the appendix \ref{sect:inv_sphere} are finite and the geometry is regular around $\f=0$. Moving away from the dS$_2$ boundary, the potential $V$ grows and finds a maximum. Then $V$ decreases and goes to negative values. At $\f=1$ we find a shrinking endpoint as described in the appendix \ref{G.2.2}. Around this point, both $f$ and $T$ diverge to $+\infty$ in a correlated way so that the curvature invariants also remain finite.

The blackening function vanishes once along the flow, signalling the presence of a horizon, which is cosmological (see Appendix \ref{app:J}). At the horizon, all the functions are finite, so the geometry is also regular there. The horizon is located in the dS regime, in agreement with rule 12 on page \pageref{ru12} in Sec. \ref{rul}.

\item[(b)] Flow from a dS$_2$ boundary to an dS shrinking endpoint.

This solution is presented in figure \ref{fig:G1}. Again, the potential has a minimum at $\f=0$, where the geometry corresponds to a dS$_2$ boundary, as shown in Appendix \ref{app:G}. As we depart from the boundary, the flow skips three extrema of the potential: two maxima and one minimum. At $\f=1$, the flow ends at a shrinking endpoint located in the dS regime. Along the flow, the function $f$ vanishes once, signalling the presence of a cosmological horizon, that is again located in the dS regime.

The geometry is regular at both endpoints, in spite of the apparent divergence of the functions $f$ and $T$. This is explicitly demonstrated in the bottom left panel of Fig. \ref{fig:G1}, where the energy density, pressure, and $\mathcal{I}$ controlling the curvature invariants are shown to be finite.
\end{itemize}

\subsection{Flow from a dS$_2$ boundary to a black hole}\label{sec:ds2bh}

\begin{figure}[h!]
	\centering
	\subfloat{{\includegraphics[width=0.47\linewidth]{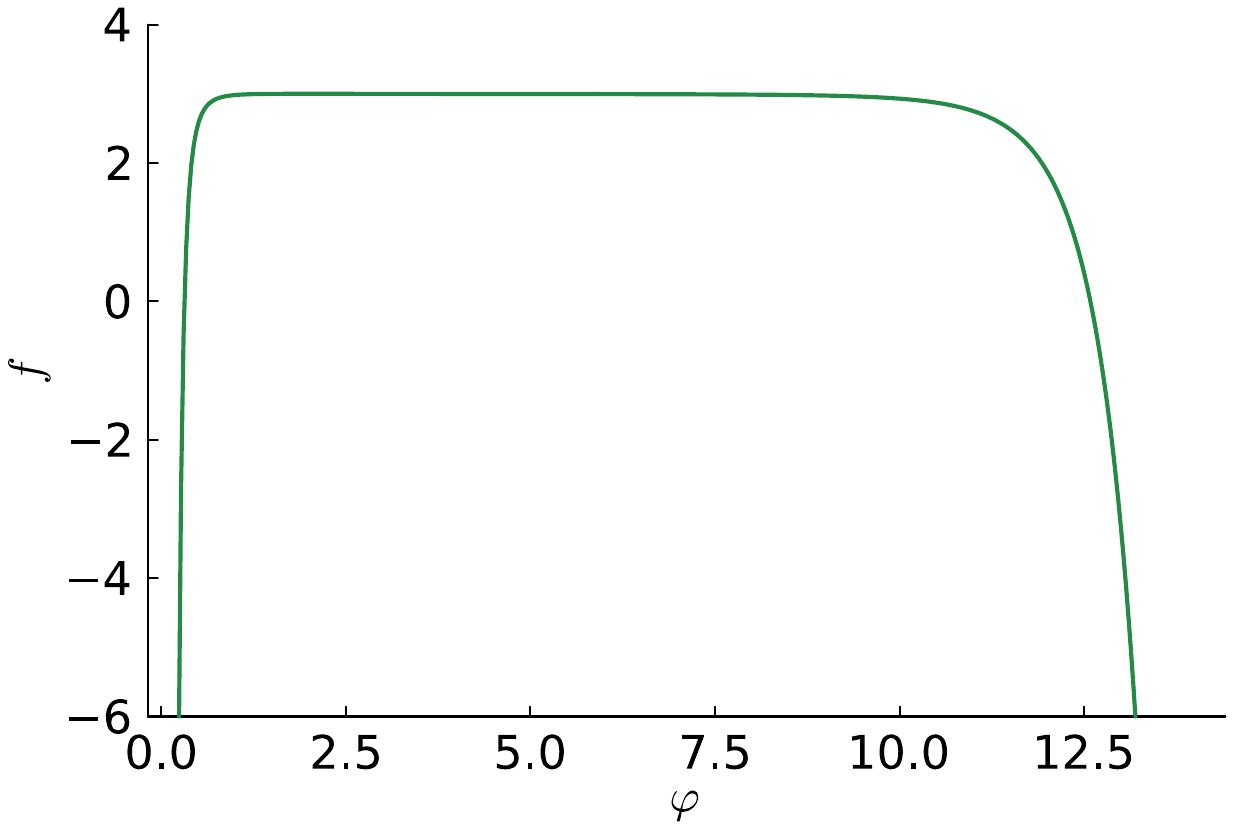}}}
	\qquad
	\subfloat{{\includegraphics[width=0.47\linewidth]{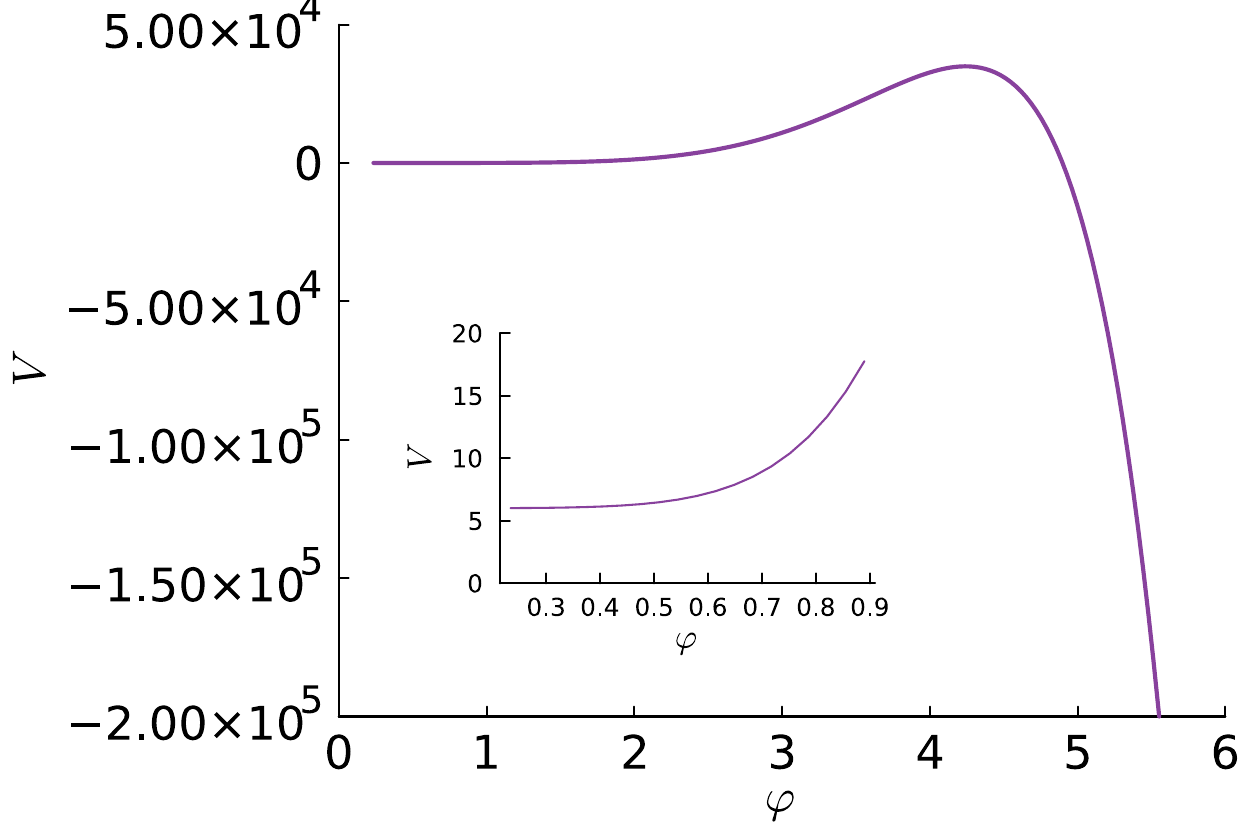}}}
	\qquad
	\subfloat{\includegraphics[width=0.47\linewidth]{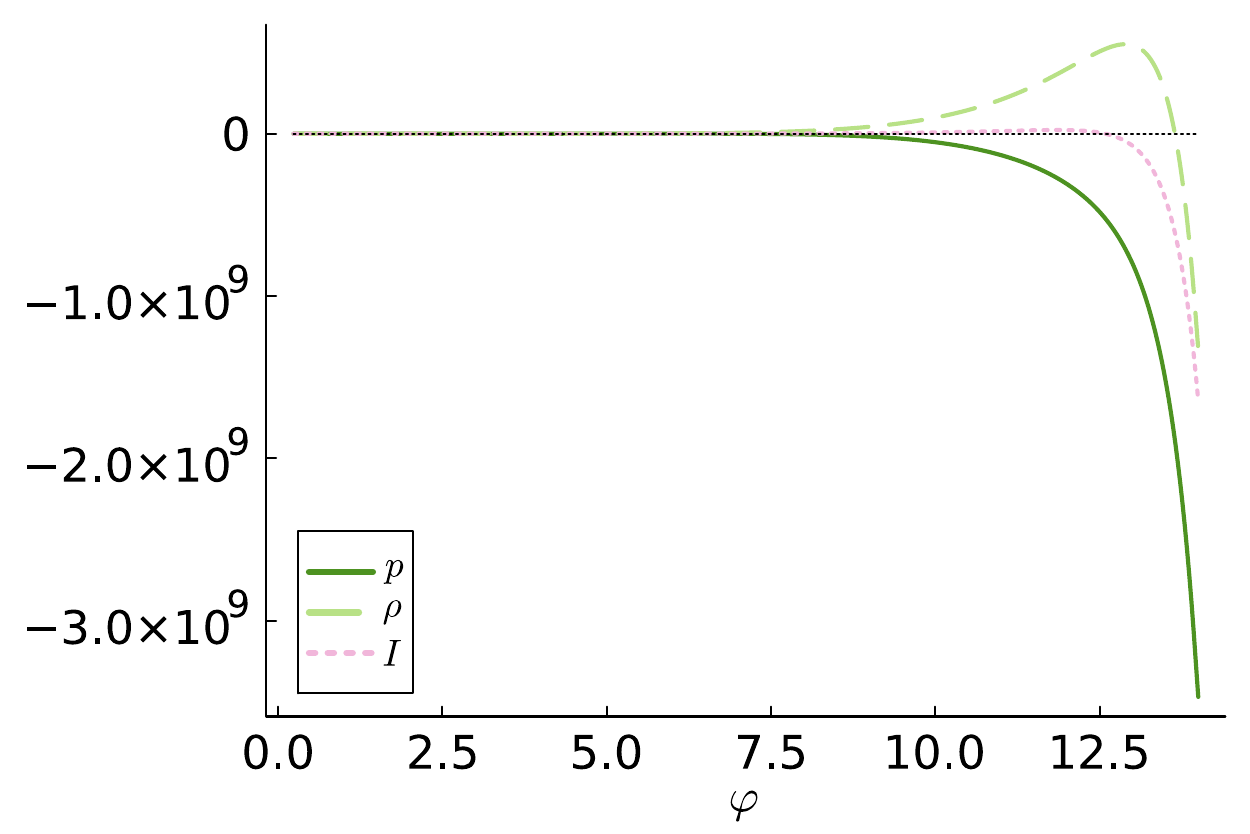}}
	\qquad
	\subfloat{{\includegraphics[width=0.47\linewidth]{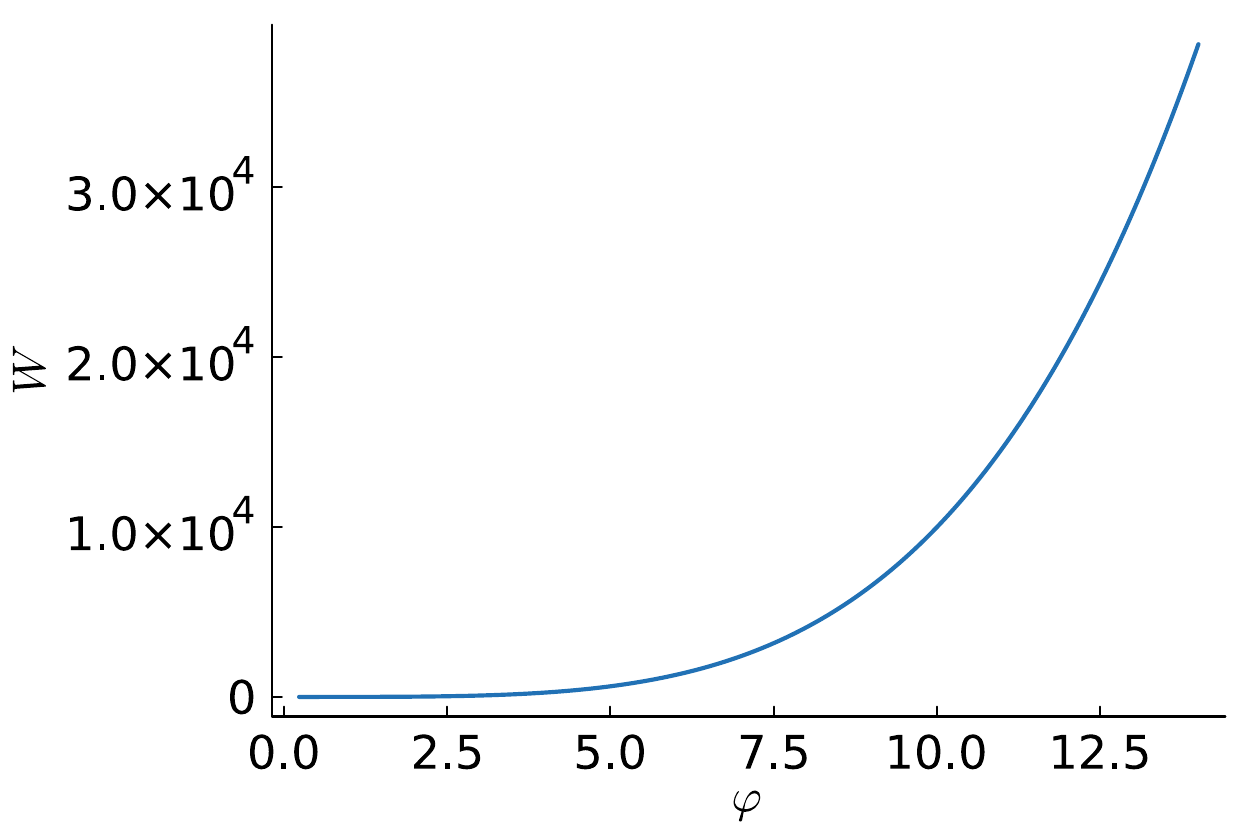}}}
	\caption{Flow from a boundary of dS$_2$ at $\f=0$, to the boundary of field space $\f\to\infty$. The flow develops two horizons, the outermost corresponds to a cosmological one while the inner one is a black-hole event horizon. The function $f$ (top left panel) diverges at $\f=0$ in a correlated manner with the vanishing of the superpotential $W$ (bottom right panel), so that the curvature invariants are finite at $\f=0$, as shown in the bottom left panel. Finally, the top right panel shows the potential $V$. }
	\label{fig:ds2bh}
\end{figure}

We now present a solution from the boundary of dS$_2\times $S$^3$ across two horizons, the outermost being cosmological while the inner one corresponds to a black-hole event horizon. This solution has been constructed in Appendix \ref{app:dS2BH}. Similarly to the previous section, we engineer a superpotential which has the desired properties, and subsequently compute the functions $T$, $f$, and the potential $V$. In this case, we use the following superpotential:
\begin{equation}
W= \f^4\,.
\end{equation}
At $\f=0$, the superpotential vanishes in agreement with the dS$_2$ asymptotic solution, Eq. \eqref{wds2}, for $\delta_{\pm}=1/2$. There are no other extrema of the superpotential and, as a consequence, a flow starting at the dS$_2$ boundary necessarily runs to the boundary of field space $\f\to\infty$, where it encounters a singularity. We construct a flow for $\f>0$ without loss of generality. As explained in Appendix \ref{app:dS2BH}, we choose integration constants such that the singularity is covered by a black hole horizon.

In Fig. \ref{fig:ds2bh} we show a concrete example of the solution described in Appendix \ref{app:dS2BH}.
The solution features a dS$_2$ boundary at $\f=0$, where the potential $V$ is positive and the function $f$ diverges to $-\infty$ as $\f^{-4}$. The curvature invariants remain finite at $\f=0$ despite the apparent divergence of $f$. As the solution departs from the dS$_2$ boundary, the function $f$ vanishes twice. The outermost vanishing signals the presence of a cosmological horizon, while the innermost vanishing is tied to the presence of a black-hole event horizon. Inside the black hole, the solution hits a bad singularity at $\f\to\infty$. The potential increases as the solution departs from the dS$_2$ boundary until it finds a maximum, and later diverges to $-\infty$.

\section{Flows that end at Gubser-regular endpoints ($\f\to \pm\infty$)}\label{sec:7}

In this section, we discuss solutions that run to the boundary of field space, $\f\to\pm\infty$, such that the asymptotic solution can be made regular in the Gubser sense (see Appendix \ref{asymp}).  Generically, we assume that the asymptotic potential behaves exponentially as
\be
V\sim e^{\a\f}\sp \f\to \infty\;,
\label{aasym}\ee
motivated by string theory. The solutions that are Gubser-regular, exist only if
\begin{equation}
\a<\a_G =\sqrt{\dfrac{2d}{d-1}}\,,
\end{equation}
in type I solutions, where $\a_G$ is referred to as the Gubser bound, or for any $\a$ in type II solutions.
There is also the conjectured TCC bound, \cite{TCC},
$$|\a|\geq \a_{TCC}\equiv {2\over \sqrt{d-1}}.$$
For $d>2$, we have the inequality
\be
\a_{C} < \a_{TCC}<\a_G
\ee
where the confinement bound, $\a_C$, was defined in (\ref{ms019}).

According to the analysis of Sec. \ref{asymp}, Gubser-regular asymptotic solutions come in two classes, which we have named type I and type II. We refer collectively to both classes as the Gubser-regular endpoints as $\f\to\pm \infty$. Depending on the sign of $\alpha$, the potential diverges or vanishes as we approach the $\f\to\pm\infty$  endpoints. Moreover, the potential can be either in the dS or AdS regimes. Overall, we have four qualitatively distinct ways to arrive at such  Gubser-regular endpoints. Below, we discuss all the possible scenarios.

\begin{itemize}
\item Gubser-regular endpoint with $V\to0^\pm$ as $\f\to \pm\infty$.

From rule 17 on page \pageref{ru17} in  section \ref{sec:global}, we have deduced that such solutions cannot be connected to any boundary or any Nariai endpoint. Instead, we can connect it to a shrinking endpoint, or to another Gubser-regular endpoint at $\f\to\infty$, where it is necessary that $V\to\pm\infty$.

\begin{itemize}
\item[(a)] If $V\to0^+$, there exist Gubser-regular solutions that are connected to a shrinking endpoint either in the dS or AdS regime. Similarly, there are solutions that connect the $V\to0^+$ regular endpoint with another regular endpoint where the potential $V$ can diverge either to $+\infty$ or to $-\infty$. Explicit examples of all these four cases are presented below, in Secs. \ref{sec:72} and \ref{sec:73} respectively. Finally, the flow can be connected to a bad singularity that is covered by a black-hole event horizon. An explicit example of this is shown in Sec. \ref{sec:75}.

\item[(b)] If $V\to 0^-$, we have shown in Sec. \ref{sec:global} that the flow cannot be connected to a shrinking endpoint in the dS regime, nor can it be connected to another Gubser-regular endpoint with $V\to +\infty$. As a result, a regular endpoint with $V\to 0^-$ can only flow to a shrinking endpoint in the AdS regime, or to another Gubser-regular endpoint at the boundary of field space with $V\to -\infty$. Examples of the two allowed flows are constructed in Secs. \ref{sec:72} and \ref{sec:73} respectively. Finally, the flow can be connected to a bad singularity that is covered by a black-hole event horizon. An explicit example of this is constructed  in Sec. \ref{sec:75}.
\end{itemize}

\item Gubser-regular endpoints with $V\to\pm \infty$ as $\f\to \pm\infty$.

From rule 18 on page \pageref{ru18} in section  \ref{sec:global}, we have deduced that such solutions cannot be connected to any shrinking endpoint. We have also shown that they cannot be connected to any Nariai endpoint. Alternatively, they can be connected to boundary endpoints (AdS$_{d+1}$, dS$_{d+1}$ ,M$_{d+1}$, dS$_2$), or to another Gubser-regular endpoint at $\f\to\pm\infty$, where it is necessary that $V\to 0^{\pm}$. We discussed this last possibility in the previous item. Therefore, we focus on the possibility of flowing to a finite endpoint.

\begin{itemize}
\item[(a)] If $V\to+\infty$, we have shown in Sec. \ref{sec:global} that such an endpoint cannot be connected to any AdS$_{d+1}$ boundaries, neither can they be connected to Minkowski boundaries (rule 20 on page \pageref{ru20}). Alternatively, they can be connected to a dS$_{d+1}$ boundary or to dS$_2$ boundaries. Examples of both cases are presented in  Secs. \ref{sec:71} and \ref{sec:74} respectively.

\item[(b)] If $V\to -\infty$, there exist Gubser-regular solutions that are connected to either of the boundaries. Explicit examples of all these cases are presented below, in sections \ref{sec:71} and \ref{sec:74} respectively.
\end{itemize}

\end{itemize}

Below, we construct examples of solutions that run to the boundary of field space, $\f\to\infty$, and which admit the Gubser-regular asymptotic structure of Appendix \ref{asymp}, i.e. with type I asymptotics and $\a<\a_G$ or with type II asymptotics. In the remainder of the section, we work with $d=4$ space dimensions.

\subsection{From $d+1$-dimensional boundaries to $V(\infty)\to \pm\infty$}\label{sec:71}

\begin{figure}[h!]
	\centering
	\subfloat{{\includegraphics[width=0.47\linewidth]{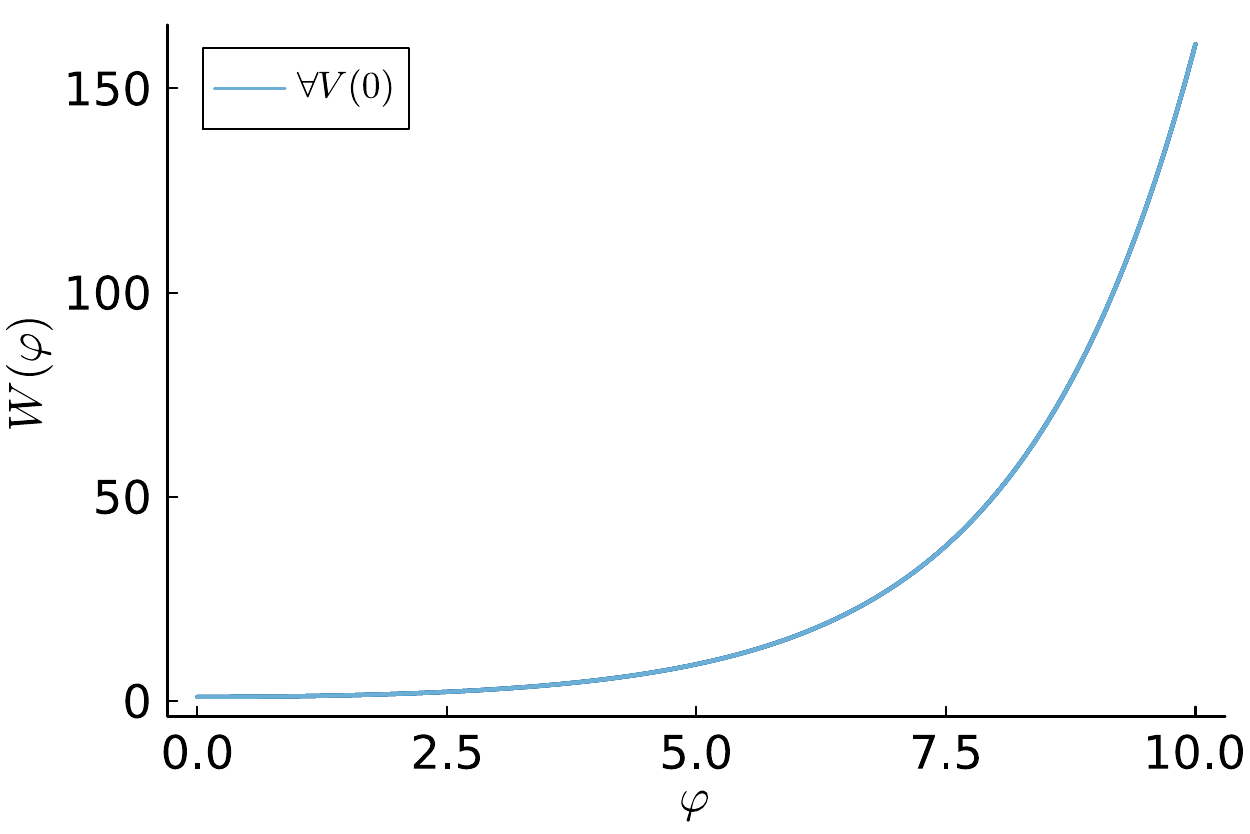}}}
	\qquad
	\subfloat{{\includegraphics[width=0.47\linewidth]{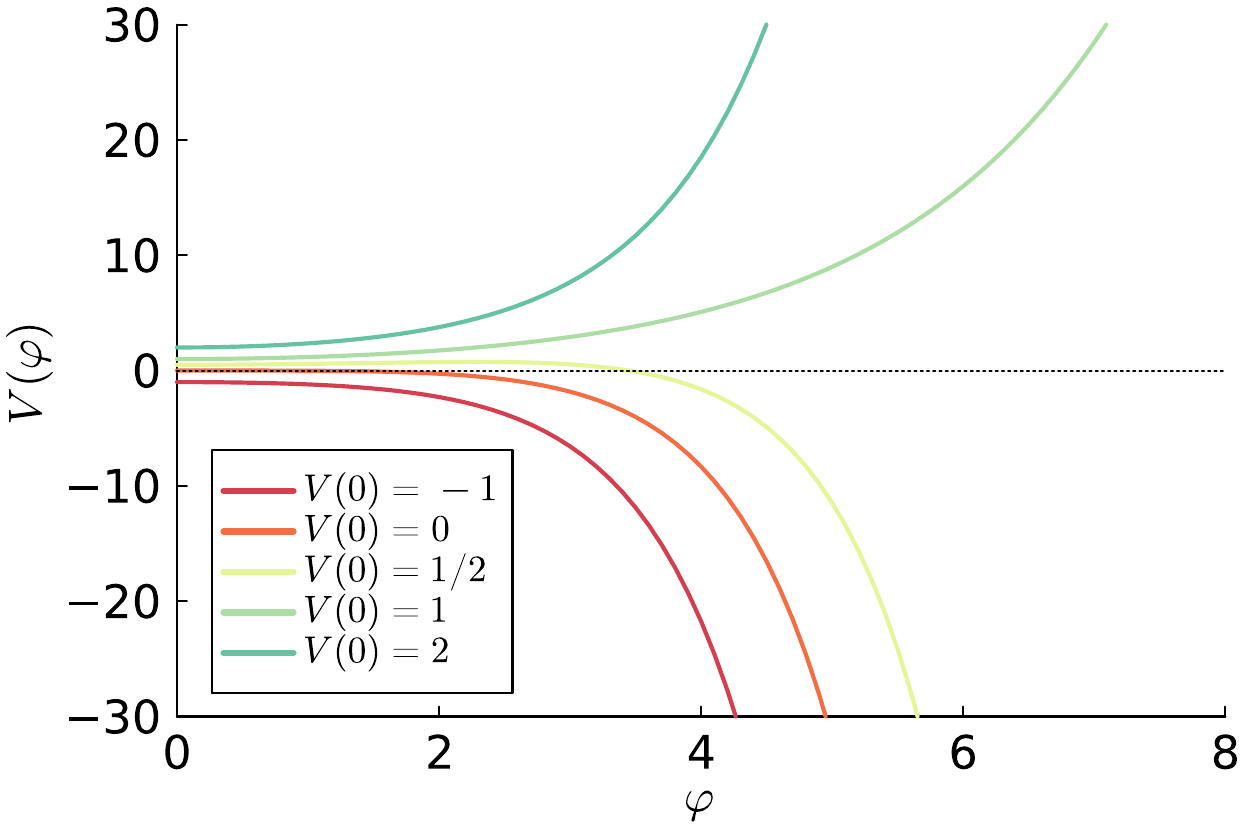}}}
	\qquad
	\subfloat{\includegraphics[width=0.47\linewidth]{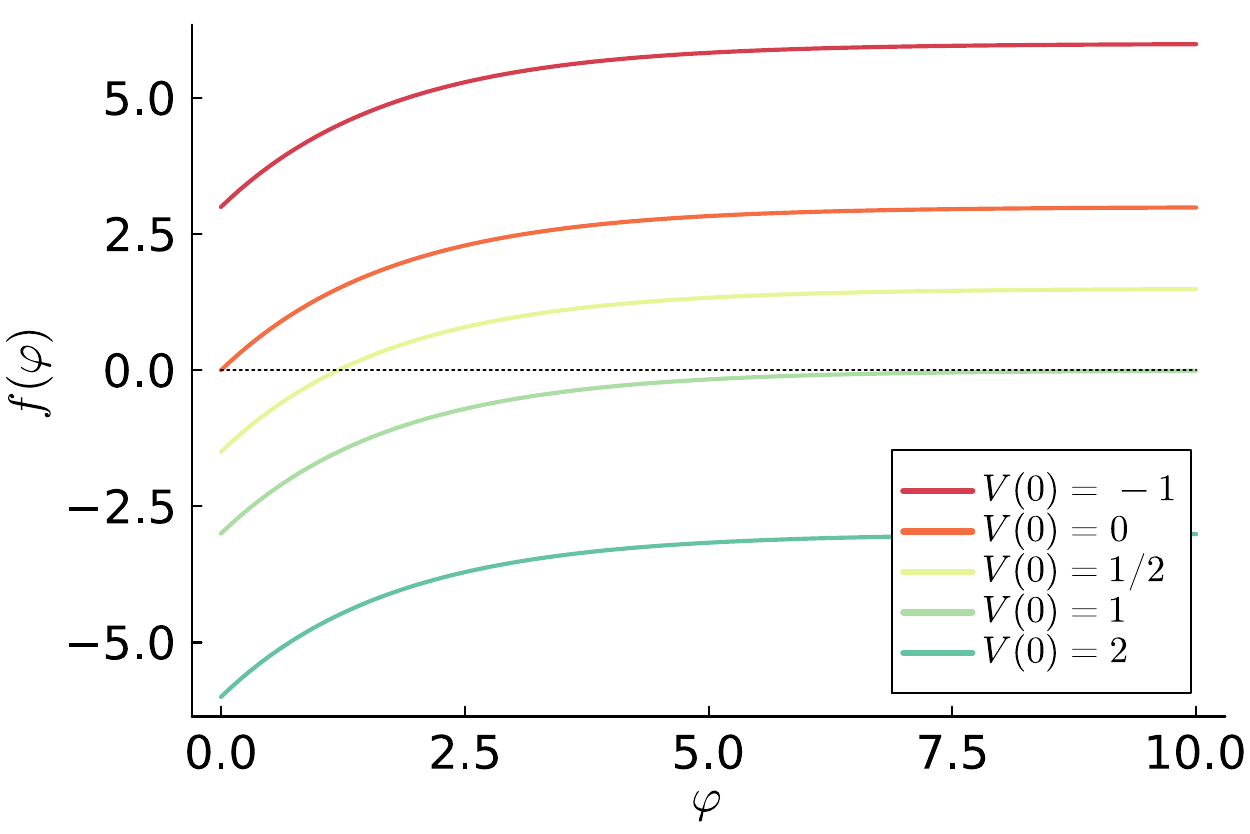}}
	\caption{Solution from a boundary endpoint (AdS$_5$, dS$_5$ or M$_5$) at $\f=0$ to type I or type II endpoints at the boundary of field space $\f\to+\infty$. The solutions are explicitly given in Eqs. \eqref{715}-\eqref{717}, with $f_0 =  3(1-V(0))$. The parameter $V(0)$ controls the value of the potential at $\f=0$.}
	\label{fig:gubx}
\end{figure}

In Appendix \ref{app:71} we have constructed the following exact solution to the equations of motion \eqref{f10_1}-\eqref{w55b}:

\begin{equation}\label{715}
W = \cosh\left(\dfrac{\varphi}{\sqrt{3}}\right)\,, \qquad T = \frac{1}{12} \sinh\left(\dfrac{\varphi}{\sqrt{3}}\right)\,,
\end{equation}
\begin{equation}\label{716}
f= f_0 - 3 e^{-\f/\sqrt{3}}\,,
\end{equation}
\begin{equation}\label{717}
V=-\dfrac{1}{4}f_0+\cosh\left(\dfrac{\varphi}{\sqrt{3}}\right)-\dfrac{1}{12}f_0\cosh\left(\dfrac{2\varphi}{\sqrt{3}}\right)\,,
\end{equation}
where $f_0$ is an integration constant that we parametrise in terms of the value of the potential at $\f=0$:
\begin{equation}
f_0\equiv -3(V(0)-1)\,.
\end{equation}
The superpotential \eqref{715} has a single extremum at $\f=0$, corresponding to a five dimensional boundary (dS$_5$, AdS$_5$ or M$_5$). In the absence of an event horizon, the flows contained in \eqref{715} can only be from $\f=0$ to $\f\to\infty$. The solution is constructed such that it approaches the boundary of field space $\f\to+\infty$ with the Gubser-regular type of asymptotics described in Appendix \ref{asymp}. In particular, the blackening function and potential asymptote to
\begin{equation}\label{coonf}
f(\f\to+\infty) = -3(V(0)-1) + O(e^{-\f/\sqrt{3}})\,, \quad V(\f\to+\infty) = \dfrac{1}{8}(V(0)-1)e^{2\f/\sqrt{3}} + O(e^{\f/\sqrt{3}})\,.
\end{equation}

If $V(0)\neq 1$, the solution approaches $\f\to+\infty$ asymptotically as dictated by the type I endpoint, while for $V(0)=1$ it does so with the type II asymptotic structure. Note that for the type I case, the potential diverges with an exponent, $2/\sqrt{3}$, that lies in between the confinement and the Gubser bounds: $\a_C =\sqrt{2/3}\,$; $\a_G = 2\sqrt{2/3}$. Conversely, in the type II case the potential diverges with the exponent $1/\sqrt{3}$, which is below both bounds.

Different values of $V(0)$ give rise to five qualitatively different possibilities, all of which are shown in Fig. \ref{fig:gubx}:
\begin{itemize}
\item $V(0)<0$: The extremum of the superpotential at $\f=0$ corresponds to an AdS$_5$ boundary, which is connected to a type I endpoint in the AdS regime, $V(\infty)\to -\infty$. The function $f$ remains positive and bounded along the flow, and there is no horizon in this flow.
The particular exponent in (\ref{coonf}) corresponds to a dual confining theory in the context of holography.

\item $V(0)=0$:  At $\f=0$ the potential vanishes as $V= -\f^4/72+ O(\f^6)$, and the extremum of the superpotential at $\f=0$ is identified with an M$_5$ boundary. The solution runs to the boundary of field space and connects to a type I endpoint in the AdS regime. The function $f$ is positive and vanishes at the $M_5$ boundary as $f = \sqrt{3}\f + O(\f^2)$, in agreement with the asymptotic solution of Eq. \eqref{e5e} for $d=4$. This solution does not feature any horizon.

\item $0<V(0)<1$: In this case, $\f=0$ corresponds to a dS$_5$ boundary, that is connected to a type I endpoint in the AdS regime, where $V\to -\infty$. There is a horizon at $\f_h =-\sqrt{3}\log(1-V(0))$, where $f(\f_h)=0$. According to the discussion of Appendix \ref{app:J}, this is a cosmological horizon.

\item $V(0)=1$: The minimum of the superpotential at $\f=0$ corresponds to a dS$_5$ boundary. Interestingly, the potential and superpotential coincide $V=W = \cosh(\f/\sqrt{3})$. At $\f\to \infty$, the potential diverges to $+\infty$ while the function $f$ vanishes. This is a type II endpoint in the dS regime.\footnote{Note that for $V\sim e^{\a\f}\to \pm \infty$ in the type II asymptotics with a spherical slicing, we require (see Eq. \eqref{m54}) that $\a>\a_C$ if $V\to -\infty$ or that $0<\a<\a_C$ if $V\to +\infty$, where $\a_C=\sqrt{2/(d-1)}\simeq 0.81$ is the confinement bound. In our case, we have chosen that the superpotential diverges as $W=e^{\f/\sqrt{3}}$, which in the type II asymptotics, see Eq. \eqref{ms19}, implies that $\a = 1/\sqrt{3}\simeq 0.58$. Therefore, the choice $\beta=1/\sqrt{3}$ for the superpotential \eqref{k15} can only accommodate type II asymptotics with $V\to +\infty$.} The function $f$ is negative along the flow, and vanishes at $\f\to\infty$.

\item $V(0)>1$: At $\f=0$ there is a dS$_5$ boundary, that is connected to a type I endpoint in the dS regime at $\f\to\infty$. In this case, the function $f$ remains negative along the flow, and there is no horizon.
\end{itemize}

In summary, we have described solutions connecting a dS$_5$ boundary with a type I endpoint in the AdS or dS regime, an M$_5$ boundary connected to a type I endpoint in the AdS regime, and an AdS$_5$ boundary connected to a type I endpoint in the AdS regime. Additionally, we described a solution from a dS$_5$ boundary to a type II endpoint in the dS regime.

\subsection{From dS$_2$ boundaries to $V(\infty)\to\pm\infty$}\label{sec:74}

\begin{figure}[h!]
	\centering
	\subfloat{{\includegraphics[width=0.47\linewidth]{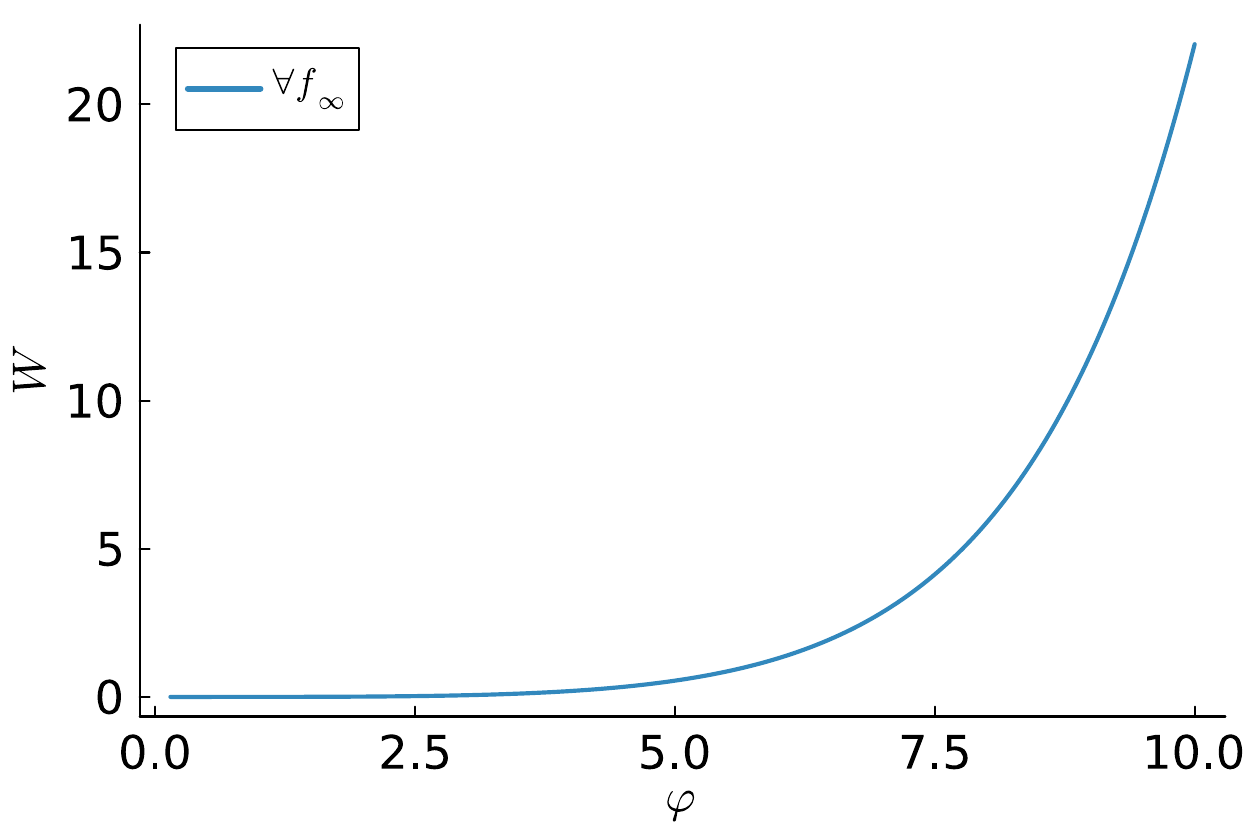}}}
	\qquad
	\subfloat{{\includegraphics[width=0.47\linewidth]{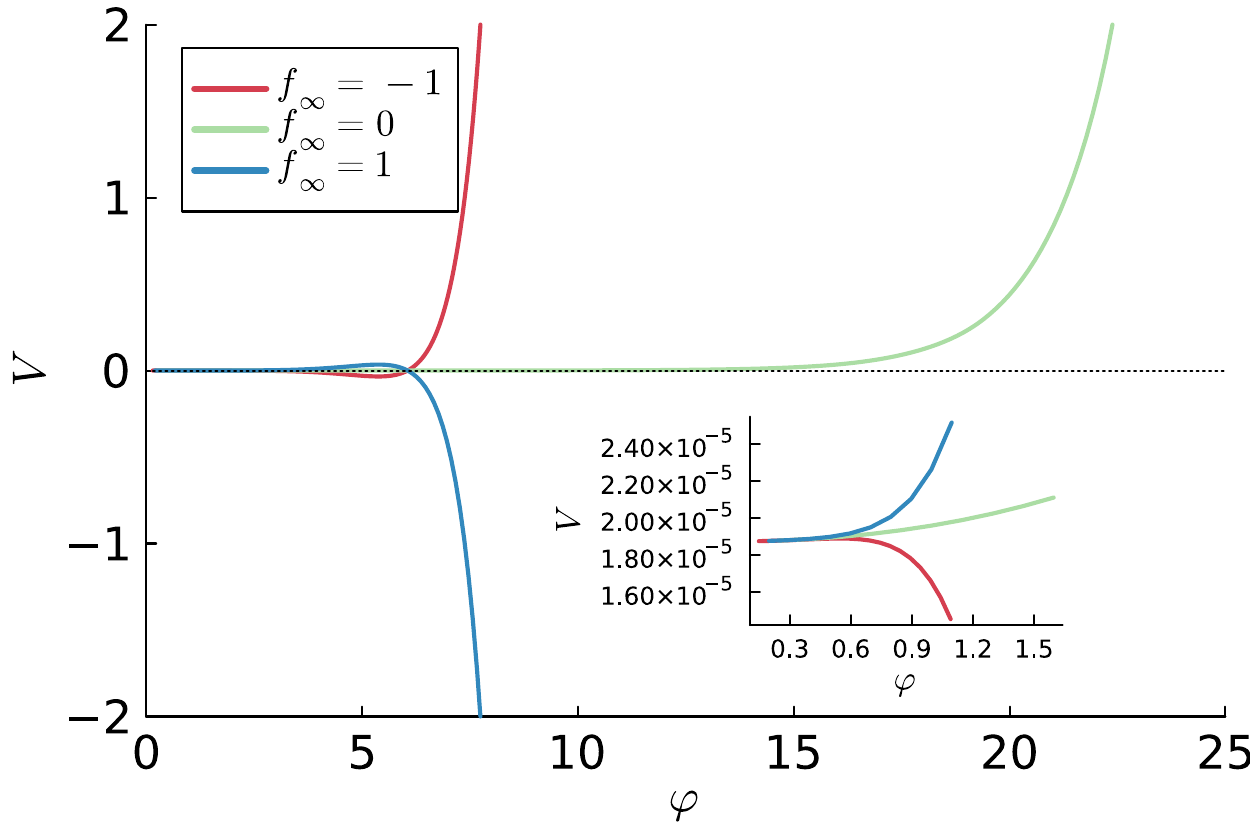}}}
	\qquad
	\subfloat{\includegraphics[width=0.47\linewidth]{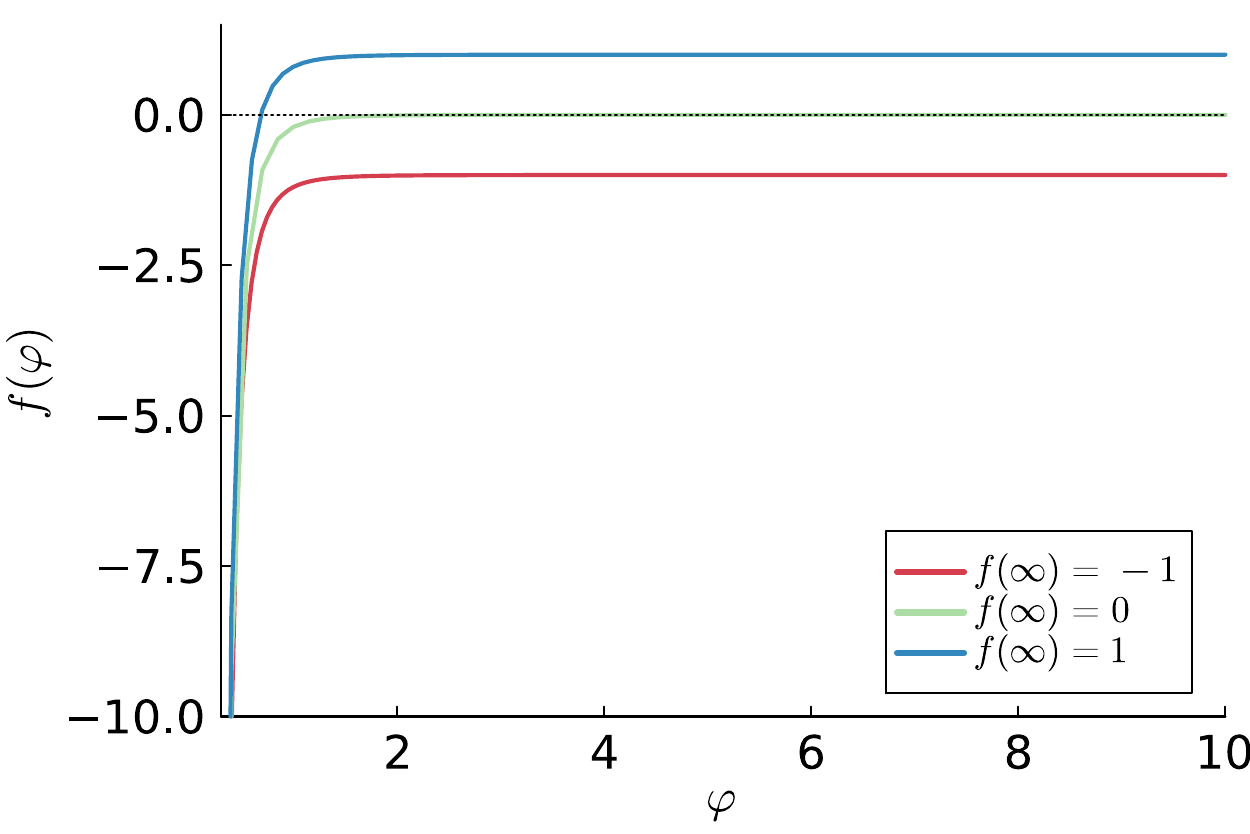}}
	\caption{Solution from a dS$_2$ boundary at $\f=0$ to type I or type II endpoints at the boundary of field space $\f\to+\infty$ where the potential diverges. The superpotential is given in Eq. \eqref{wgds2}, while $f$ and $V$ are obtained through the numerical integration of Eq. \eqref{fpg} with different choices of boundary conditions $f(\infty)$. The constant $C_T$ is set such that $f'=1/\f^5 + O(\f^{-4})$ in Eq. \eqref{fpct}. }
	\label{fig:gubxs}
\end{figure}

In this section, we discuss the solutions that interpolate between dS$_2$ boundaries and Gubser-regular endpoints, where the potential is necessarily divergent. Such solutions have been explicitly constructed in Appendix \ref{app:74}.

We have engineered such a solution by choosing some superpotential with the appropriate behaviour. At a dS$_2$ boundary, the superpotential vanishes as dictated by \eqref{wds22}, while at the boundary of fields space $\f\to+\infty$ we assume that it diverges exponentially. These conditions are satisfied by the following superpotential:
\begin{equation}\label{wgds2}
W = \left[\cosh\left(\frac{1}{2}\b \f\right)-\cosh\left(\frac{1}{2}\b_2\f\right)\right]^2\,.
\end{equation}
We assume that $\b_2<\b$ without loss of generality. The superpotential \eqref{wgds2} vanishes at $\f=0$ as $\f^4$. This corresponds to the asymptotic solution \eqref{wds22} with $\delta_{\pm}=\frac{1}{2}$, and we can identify the point $\f=0$ with a dS$_2$ boundary. The superpotential \eqref{wgds2} has only one regular extremum, at $\f=0$, and, according to rule 1 on page \pageref{ru1}, the flow connects $\f=0$ with $\f\to +\infty$ or with $\f\to -\infty$. We shall restrict ourselves to a flow from a dS$_2$ boundary to a Gubser-regular endpoint at $\f\to+\infty$. An equivalent construction can be made demanding that the regular endpoint is at $\f\to-\infty$.

We choose the following values for $\b$ and $\b_2$:
\begin{equation}
\b = \dfrac{2}{\sqrt{15}}\,,\qquad \b_2 = \dfrac{1}{\sqrt{15}}\,,
\end{equation}
which we have shown in Appendix \ref{app:74} give rise to solutions where the potential diverges for $\f\to\infty$ as $V\sim e^{\a \f}$, such that $\a_C<\a<\a_G$ for the type I asymptotics. As shown in Appendix \ref{app:l1}, these bounds on $\a$ are required for type I solutions in the spherically sliced ansatz that can be accepted \` a la Gubser.

In Appendix \ref{app:74} we discuss how to compute the inverse scale factor $T$, the function $f$ and the potential $V$. $T$ is given in Eq. \eqref{o11}, while $f$ and $V$ require to integrate numerically Eq. \eqref{fpg}. It is shown (see below Eq. \eqref{o16}) that qualitatively different solutions depend on a single parameter, which we choose to be the value of $f$ at the boundary of field space: $f(\infty)$. In Eq. \eqref{o18} we show that the behaviour of the potential $V$ as we approach the dS$_2$ boundary ($\f=0$) and as we approach the type I endpoint ($\f\to+\infty$) is given by
\begin{equation}
V(\f\to 0) = 162 C_T \left(1 + \dfrac{1}{24}\f^2 + O(\f^4)\right)\,, \quad V(\f\to\infty) = -\frac{f(\infty)}{80}e^{\frac{4}{\sqrt{15}}\f} + O\left(e^{\frac{\sqrt{5}}{2\sqrt{3}}}\right)\,,
\end{equation}
in agreement with the dS$_2$ boundary asymptotic solution of Sec. \eqref{g53} and with the type I asymptotic solution of Eq. \ref{ms18}. Note that in the particular case where $f(\infty) =0$, the asymptotic behaviour becomes that of the type II solutions. Additionally, we observe that the sign of the potential as $\f\to\infty$ is anti-correlated with the sign of $f(\infty)$, while at the dS$_2$ boundary the potential is always positive. Therefore, depending on the choice of $f(\infty)$, we encounter three qualitatively different solutions. An example of each case is shown in Fig. \ref{fig:gubxs}.

\begin{itemize}
\item $f(\infty)<0$: The solution connects a dS$_2$ boundary at $\f=0$ with a type I endpoint with $V\to +\infty$ as $\f\to\infty$. The blackening function is always negative and there is no horizon.
\item $f(\infty)=0$: The solution connects a dS$_2$ boundary at $\f=0$ with a type II endpoint with $V\to +\infty$ as $\f\to\infty$. The function $f$ is negative along the flow, and vanishes at the type II endpoint.
\item $f(\infty)>0$: The solution connects a dS$_2$ boundary with a type I endpoint with $V\to -\infty$ as $\f\to\infty$. The function $f$ vanishes once, and the solution has a cosmological horizon.
\end{itemize}

\subsection{From shrinking endpoints to  $V(\infty)\to 0$}\label{sec:72}

\begin{figure}[h!]
	\centering
	\subfloat{{\includegraphics[width=0.47\linewidth]{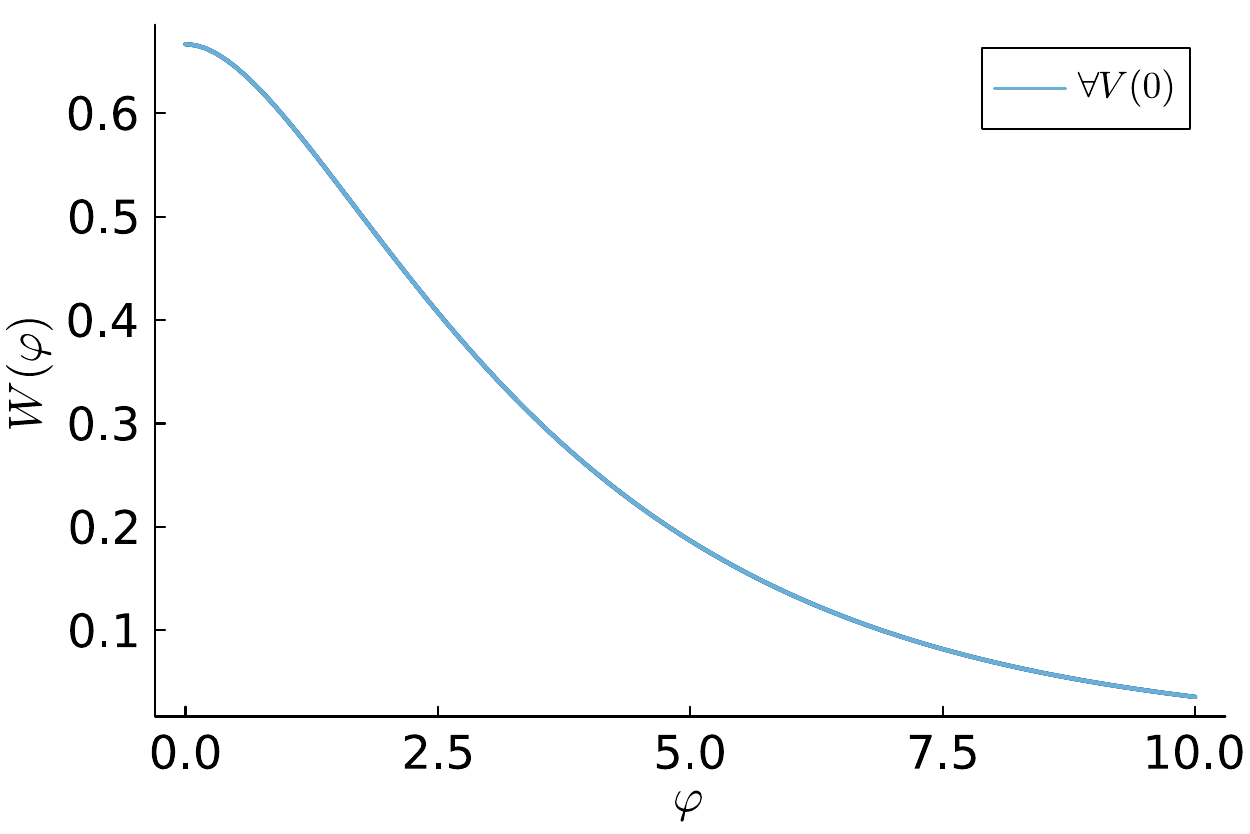}}}
	\qquad
	\subfloat{{\includegraphics[width=0.47\linewidth]{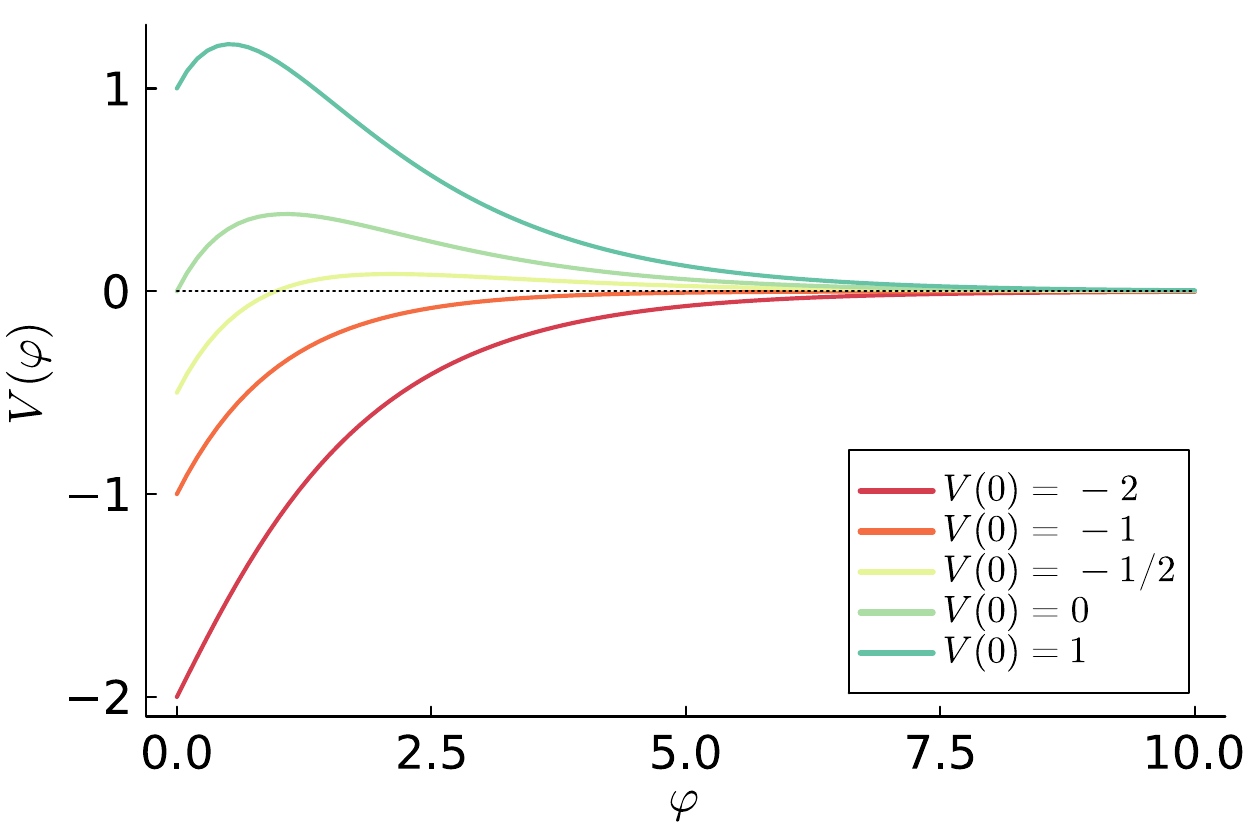}}}
	\qquad
	\subfloat{\includegraphics[width=0.47\linewidth]{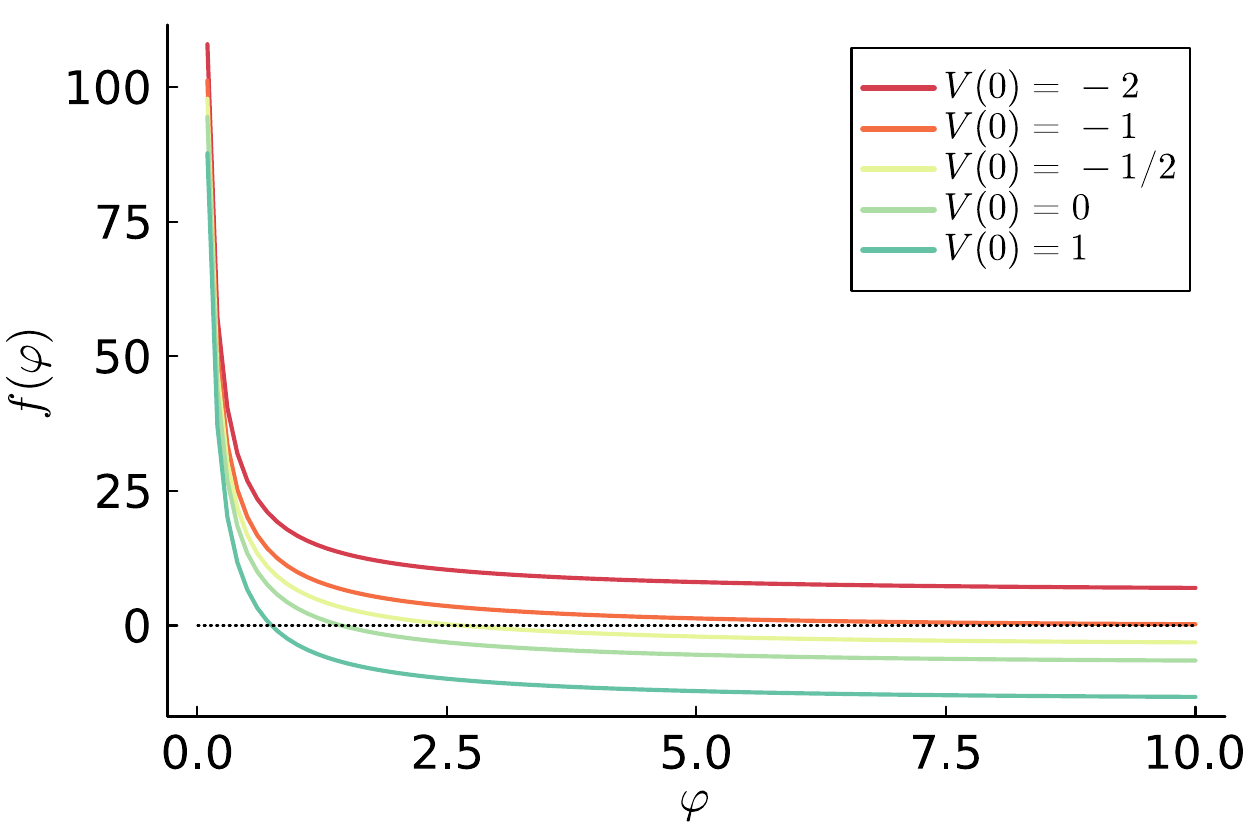}}
	\caption{Solution from a shrinking endpoint at $\f=0$ to type I or type II endpoints at the boundary of field space $\f\to+\infty$ where the potential vanishes. The solutions are explicitly given in Eqs. \eqref{sugu}, \eqref{7122}, and \eqref{7133} with $f_0 =  -27(1+V(0))/4$. The parameter $V(0)$ controls the value of the potential at the shrinking endpoint $\f=0$.}
	\label{fig:gubxv}
\end{figure}

We now describe solutions where the potential vanishes exponentially $V\to 0^{\pm}$ as we approach the boundary in field space $\f\to \infty$, and such that they admit the Gubser-regular asymptotic structure of Appendix \ref{asymp}. In this case, Eqs. \eqref{ms18} and \eqref{ms19} imply that both the potential $V$ and the superpotential $W$ vanish exponentially as $\f\to\infty$. The detailed construction of these solutions is presented in Sec. \ref{app:72}. We assume that the same flow has another endpoint at a finite $\f$, which must be a shrinking endpoint in agreement with rules 17 on page \pageref{ru17} and 18 on page \pageref{ru18} of section \ref{rul}.

We consider the following superpotential:
\begin{equation}\label{sugu}
W = c_1 e^{\beta \f} + c_2 e^{\beta_2 \f}\,,
\end{equation}
with parameters
\begin{equation}
c_1 = 1 \sp c_2 = - \dfrac{1}{3} \sp \beta=-\dfrac{1}{3} \sp \beta_2 = -1 \,.
\end{equation}
As shown in Appendix \ref{app:72}, this choice of parameters ensure that there is a shrinking endpoint at $\f=0$, and no other extremum of $W$ for $\f>0$. Additionally, the superpotential vanishes as $\f\to\infty$. The solutions to the equations of motion \eqref{f10_1}-\eqref{w55b} that admit Gubser-regular asymptotics as $\f\to\infty$ are given by
\begin{equation}\label{7122}
T= \frac{1}{12}\frac{ e^{-\f}}{1-e^{-2
   \varphi/3}} \,,\quad f =  f_0 + \frac{27 }{8\sinh  \frac{\varphi }{3}}\,,
\end{equation}
\begin{equation}\label{7133}
V = -\dfrac{f_0e^{-2\f/3}}{54}\left(15 - 6 e^{-2\f/3}-e^{-4\f/3}\right)- e^{-\f}\,.
\end{equation}
We defer the reader to Appendix \ref{app:72} for further details about the construction of the previous solutions. $f_0$ is an integration constant, that we parametrise in terms of the value of the potential at the shrinking endpoint $V(0)$:
\begin{equation}
 f_0\equiv-\frac{27}{4}(1 + V(0))
\end{equation}
As we approach the boundary of field space $\f\to\infty$, the potential $V$ and function $f$ behave as
\begin{equation}\label{vgub}
V =\frac{1}{8} (1 + V(0))e^{-2\f/3}+O(e^{-\f}) \sp f= -\frac{27}{4}(1 + V(0)) + O(e^{-\f/3})
\end{equation}
If $V(0)\neq -1$, the previous asymptotic behaviour match the type I asymptotic solution of Appendix \ref{asymp} that are regular \`a la Gubser \eqref{ms18}\,, while for $V(0)=-1$ it corresponds to the type II asymptotics. Therefore, such solutions connect a shrinking endpoint at $\f=0$ with an endpoint at $\f\to \infty$, with a Gubser-regular asymptotic structure, where the potential vanishes.

Depending on the value of $V(0)$, we encounter five qualitatively different solutions, shown in Fig. \ref{fig:gubxv}:

\begin{itemize}
\item $V(0)<-1$: The potential is negative at the shrinking endpoint, while $f_0>0$, such that $\lim_{\f \to \infty}V = 0^-$. Therefore, this is a flow from an AdS shrinking endpoint to a type I endpoint with $V\to 0^-$.

\item $V(0)= -1$:  The potential is again negative at the shrinking endpoint but now $f_0=0$. In this case the potential is simply $V=-e^\f$. The asymptotic behaviour of the functions at the boundary of field space matches with the type II asymptotics of Appendix \ref{asymp}. We conclude that this is a flow from an AdS shrinking endpoint to a type II endpoint with $V\to 0^-$.\footnote{Note that for $V\sim e^{\a\f}\to 0$ in the type II asymptotics with a spherical slicing, we require (see Eq. \eqref{m54}) that $\a<-\a_C$ if $V\to 0^-$ or that $-\a_C<\a<0$ if $V\to 0^+$, where $\a_C=\sqrt{2/(d-1)}\simeq 0.81$ is the confinement bound. In our case, we have chosen that the superpotential vanishes as $W=e^{-\f/3}$, which in the type II asymptotics, see Eq. \eqref{ms19}, implies that $\a = -1$. Therefore, the choice $\beta=-1/3$ for the superpotential \eqref{sugu} can only accommodate type II asymptotics with $V\to0^-$.}

\item $-1<V(0)<0$: In this case the shrinking endpoint is again in the AdS regime, while $f_0<0$, and the potential $V$ at $\f\to\infty$ vanishes from above. The flow contains a cosmological horizon. This is a flow from an AdS shrinking endpoint to a type I endpoint with $V\to 0^+$.

\item $V(0)=0$: This case is similar to the previous one, with the difference that the potential vanishes at the shrinking endpoint.

\item $V(0) >0$: The shrinking endpoint is in the dS regime and we have $f_0<0$, such that the potential vanishes from above as $\f\to \infty$. The flow contains a cosmological horizon. This is a flow from a dS shrinking endpoint to a type I endpoint with $V\to 0^+$.

\end{itemize}

In summary, we have described solutions connecting a shrinking endpoint in the dS regime to type I endpoints where $V\to 0^+$, as well as solutions from a shrinking endpoint in the AdS regime with type I endpoints where $V\to 0^\pm$. Additionally we described a solution from a shrinking endpoint in the AdS regime to a type II endpoint with $V\to 0^-$.

\subsection{From $V(\infty)\to \pm \infty$ to $V(\infty)\to 0^{\pm}$}\label{sec:73}

\begin{figure}[h!]
	\centering
	\subfloat{{\includegraphics[width=0.47\linewidth]{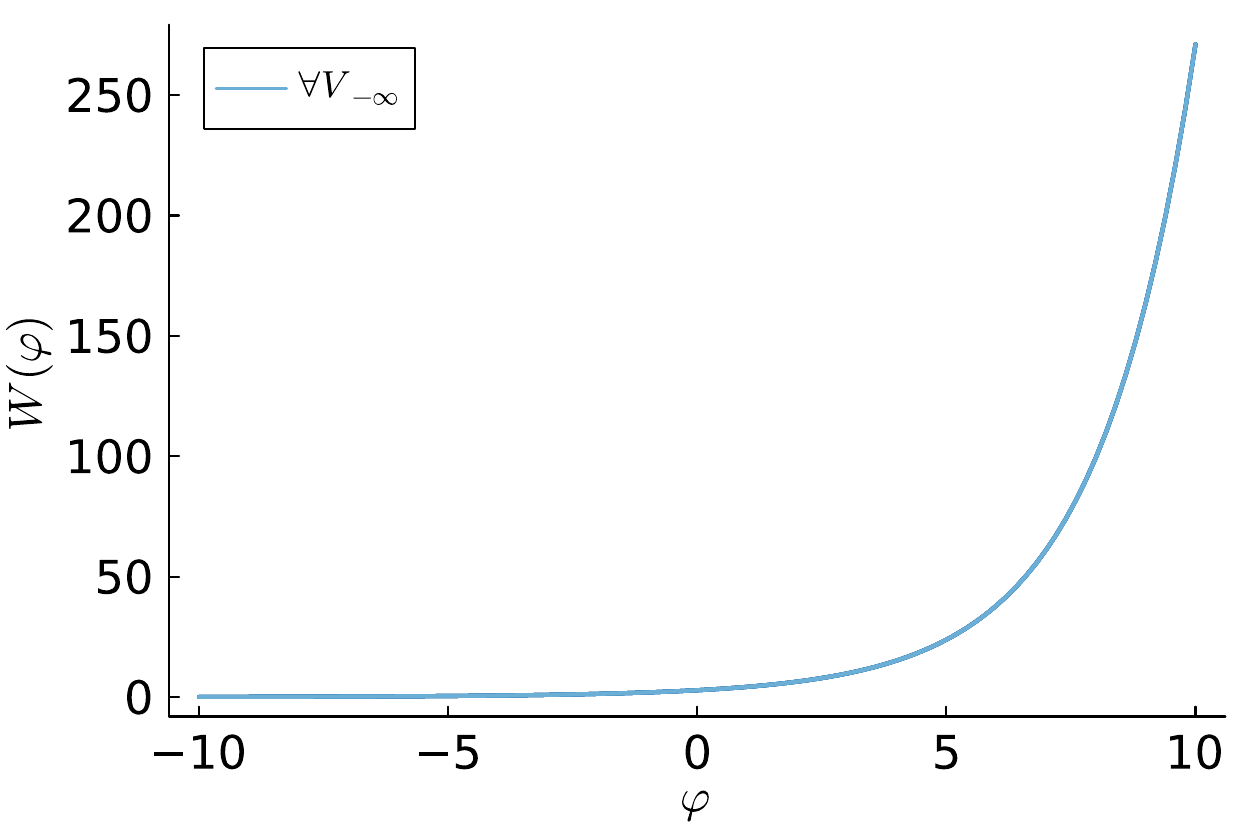}}}
	\qquad
	\subfloat{{\includegraphics[width=0.47\linewidth]{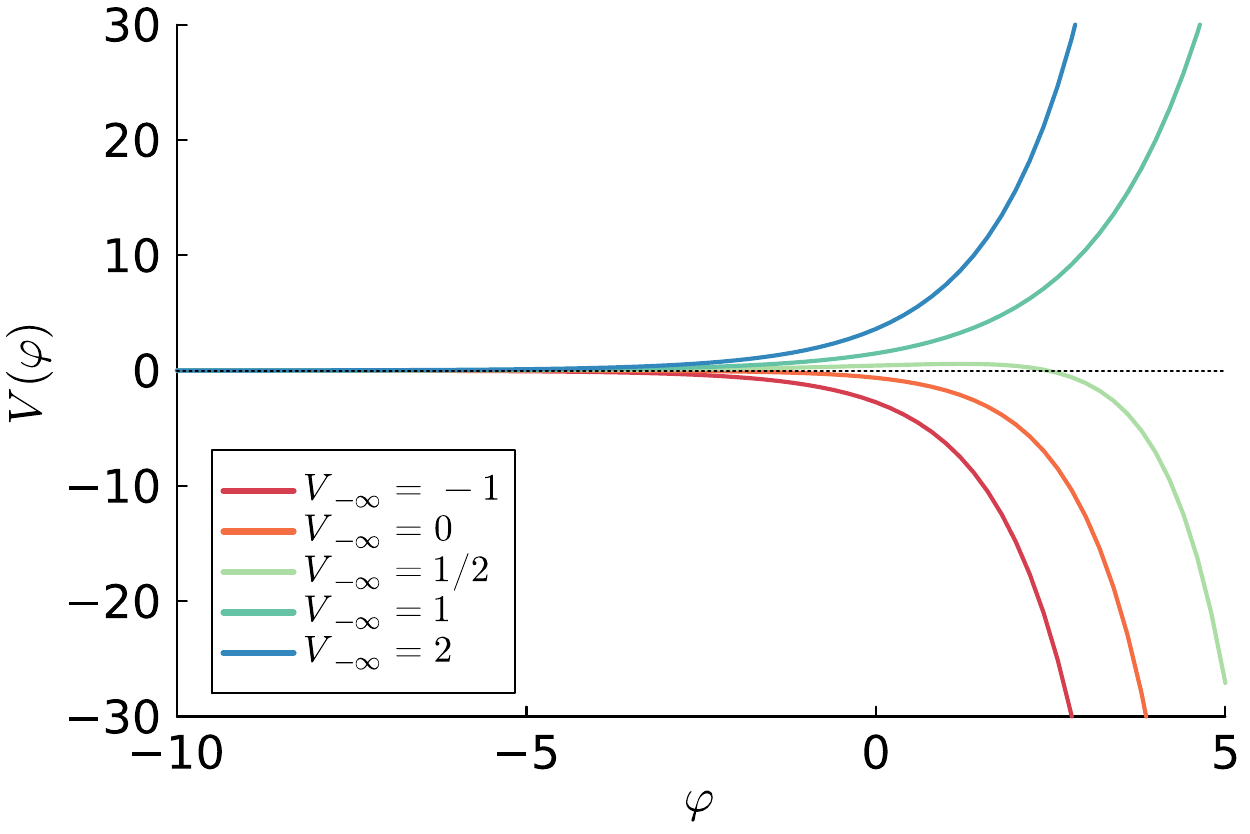}}}
	\qquad
	\subfloat{\includegraphics[width=0.47\linewidth]{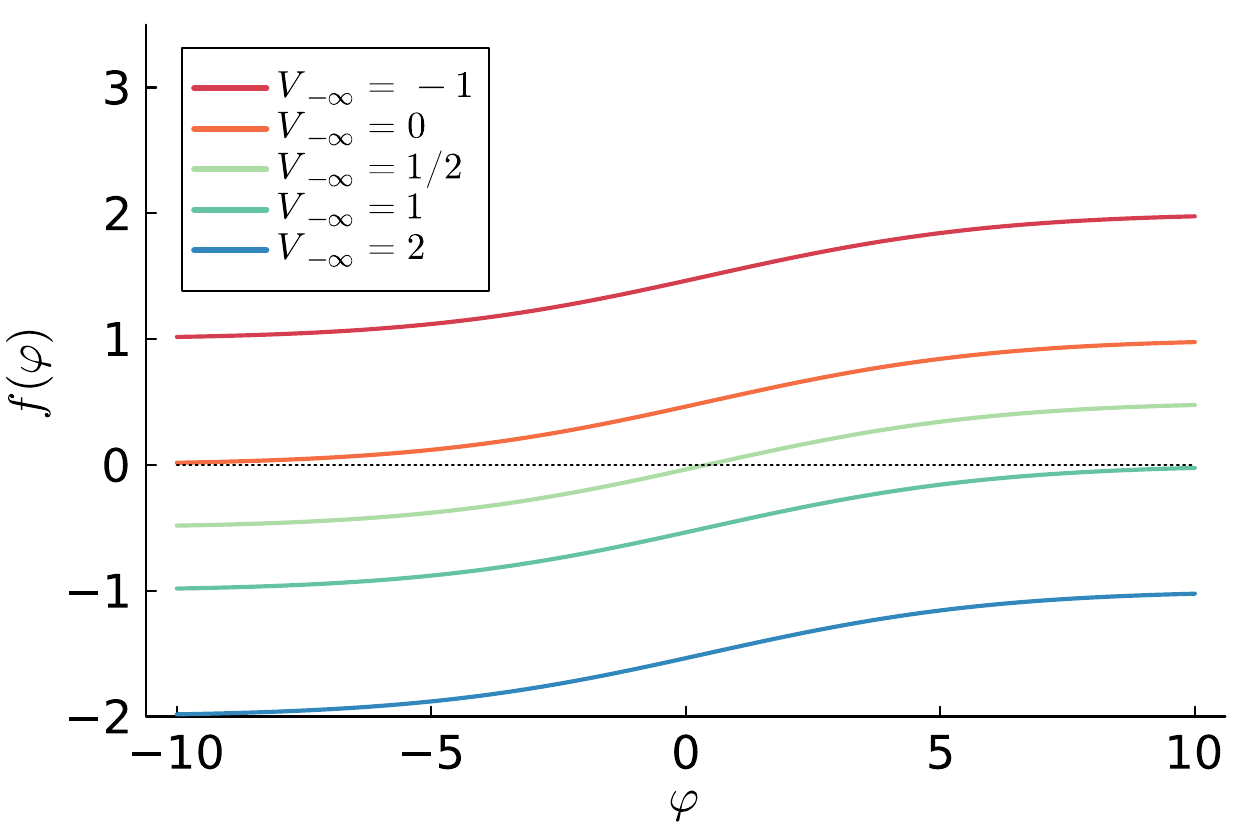}}
	\caption{Flow solutions interpolating between two Gubser-regular endpoints with vanishing potential as $\f\to-\infty$ and diverging potential as $\f\to+\infty$. The analytical solution is given in Eqs. \eqref{sugu2}, \eqref{tgg}, \eqref{fgg} and \eqref{vgg} and it is parametrised in terms of a single parameter $V_{-\infty}$ that controls the behaviour of the potential as $\f\to-\infty$. There are five inequivalent solutions, and we show a representative example of each of them.}
	\label{fig:gubxi}
\end{figure}

In this section we describe flow solutions that run between two Gubser-regular endpoints as $|\f|\to \infty$. From rules 17 on page \pageref{ru17} and 18 on page \pageref{ru18}, it is necessary that the potential vanishes at one of the endpoints, while it diverges at the second one. The detailed construction of these solutions can be found in Appendix \ref{app:73}.

Consider the following superpotential, also given in Eq. \eqref{sugu}:
\begin{equation}\label{sugu2}
W = c_1 e^{\beta \f} + c_2 e^{\b_2 \f}\,,
\end{equation}
now with parameters
\begin{equation}
c_1=1\,, \quad \b= \sqrt[3]{\dfrac{2}{3}}\,, \quad \b_2 = \dfrac{1}{4}\sqrt{\dfrac{3}{2}}\,,
\end{equation}
and $c_2>0$. As we discussed in Appendix \ref{app:73}, this ensures that the superpotential has no local extrema, and therefore can only describe flows from $\f\to-\infty$, where $W\to 0 $, to $\f\to +\infty$, where $W\to + \infty$. Additionally, the chosen values for $\beta$ and $\beta_2$ translate, in the type I solutions, to an asymptotic behaviour of the potential that is compatible with the spherical slicing and respects the Gubser bound at both ends.

The explicit solution for the inverse scale factor $T$, the blackening function $f$, and the potential $V$ is given in Eqs. \eqref{tgg}, \eqref{fgg} and \eqref{vgg} respectively. Similarly to the discussion in the previous sections, qualitatively distinct solutions are distinguished by a single integration constant, which we parametrise in terms of the leading coefficient of the potential $V$ as $\f\to -\infty$, and which we denote $V_{-\infty}$. In particular, the asymptotic behaviour of the function $f$ and of the potential as $\f$ diverges is given by Eqs. \eqref{o36} and \eqref{o37}, which we reproduce here:
\begin{equation}
V|_{\f\to+\infty} = -\frac{5}{27}(1-V_{-\infty}) e^{2\sqrt[3]{\frac{2}{3}}\f} +\dots\,,\qquad V|_{\f\to-\infty} = V_{-\infty}e^{\frac{2}{2}\sqrt{\frac{3}{2}}\f}+\dots
\end{equation}
\begin{equation}
f|_{\f\to+\infty} = (1-V_{-\infty}) +\dots\,,\qquad f|_{\f\to-\infty} = -V_{-\infty}+\dots
\end{equation}
We encounter five distinct cases depending on the asymptotic behaviour of the potential, which is controlled by $V_{-\infty}$. These cases are shown in Fig. \ref{fig:gubxi}.

\begin{itemize}
\item $V_{-\infty}<0$: The potential diverges to $-\infty$ as $\f\to +\infty$ while it vanishes as $\f\to-\infty$ from below. In both cases, we have type I Gubser-regular asymptotics. The function $f$ remains positive along the flow and there is no horizon.
\item $V_{-\infty} = 0$: In this case, the leading contribution to $V$ and $f$ at $\f\to -\infty$ vanishes. Therefore, we encounter the type II asymptotic structure as $\f\to-\infty$, with $V\to0^-$ and $f\to 0^+$. Conversely, at $\f\to\infty$ the asymptotic behaviour of the solution is of type I; the potential diverges to negative values while $f$ attains a constant positive value along the flow. $f$ vanishes at the type II endpoint.
\item $0<V_{-\infty}<1$: This solution connects a type I endpoint at $\f \to -\infty$ with $V\to 0^+$ with a type I endpoint at $\f\to +\infty$ with $V\to -\infty$. The blackening function vanishes once along the flow, signalling the presence of a horizon, which is cosmological.
\item $V_{-\infty}=1$: Now, the leading contribution to the asymptotic behaviour of the solution vanishes as $\f\to +\infty$. This solution connects a type I endpoint at $\f\to-\infty$ with $V\to 0^+$, with a type II endpoint at $\f\to-\infty$ with $V\to +\infty$. The blackening function is negative along the flow, and vanishes only at the type II endpoint.
\item $V_{-\infty}>1$: This solution connects a type I endpoint at $\f\to -\infty$ with $V\to 0^+$ with another type I endpoint at $\f\to +\infty$ with $V \to +\infty$.  The function $f$ remains negative along the flow.
\end{itemize}

\subsection{From $V(\infty)\to 0$ to a black hole}\label{sec:75}

\begin{figure}[h!]
	\centering
	\subfloat{{\includegraphics[width=0.47\linewidth]{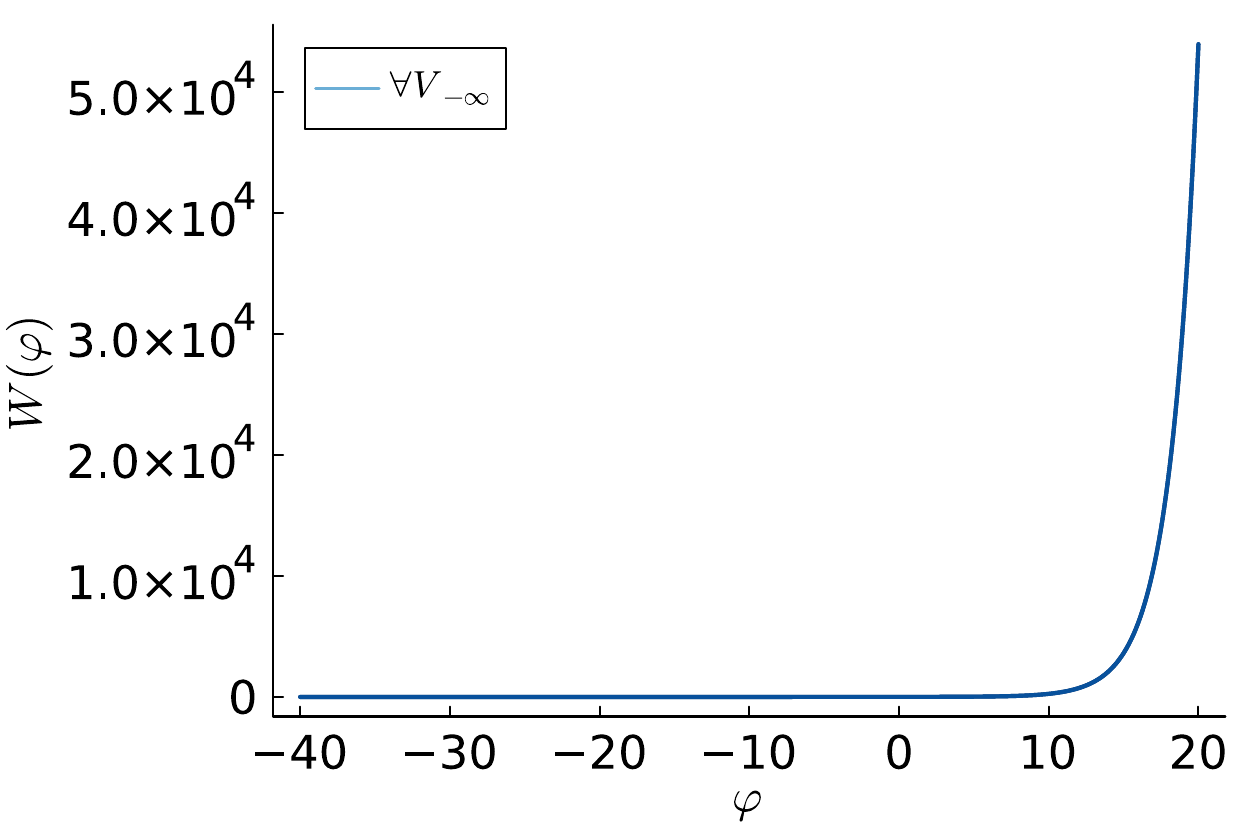}}}
	\qquad
	\subfloat{{\includegraphics[width=0.47\linewidth]{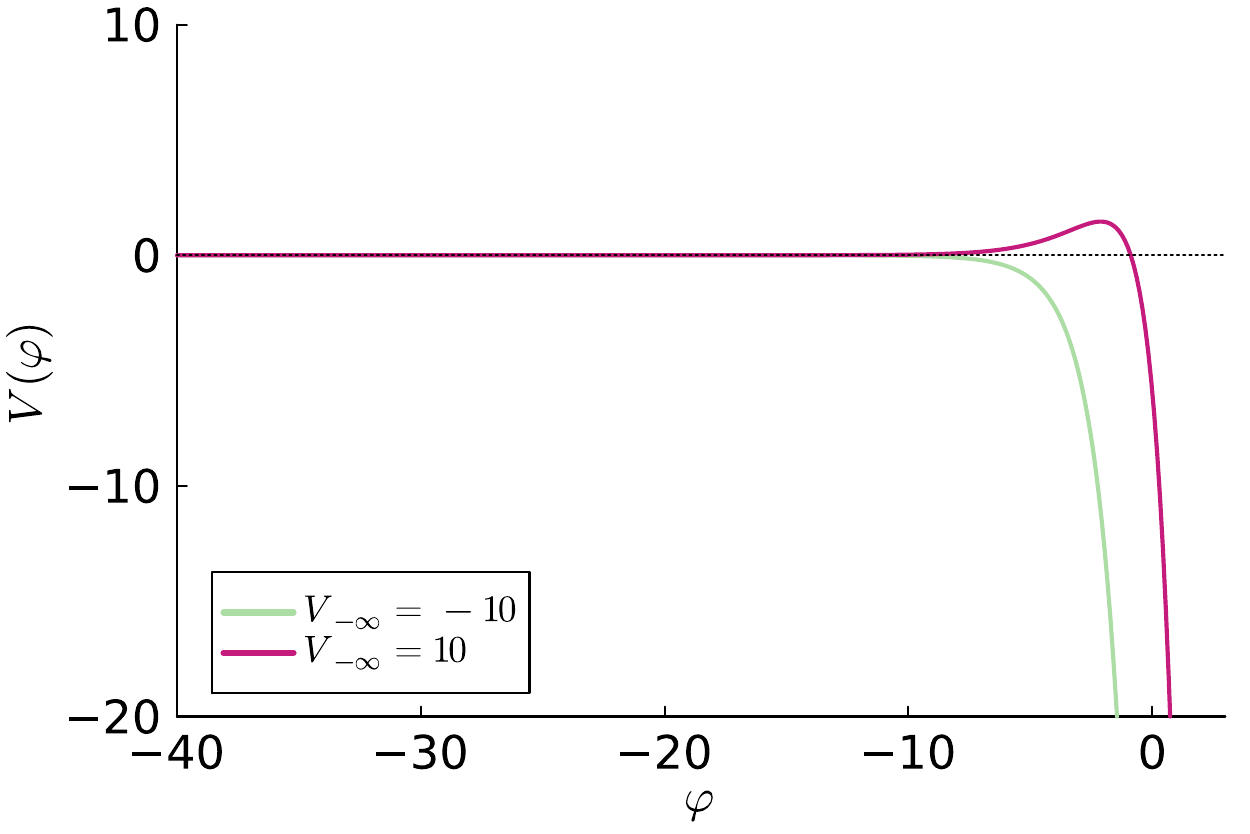}}}
	\qquad
	\subfloat{\includegraphics[width=0.47\linewidth]{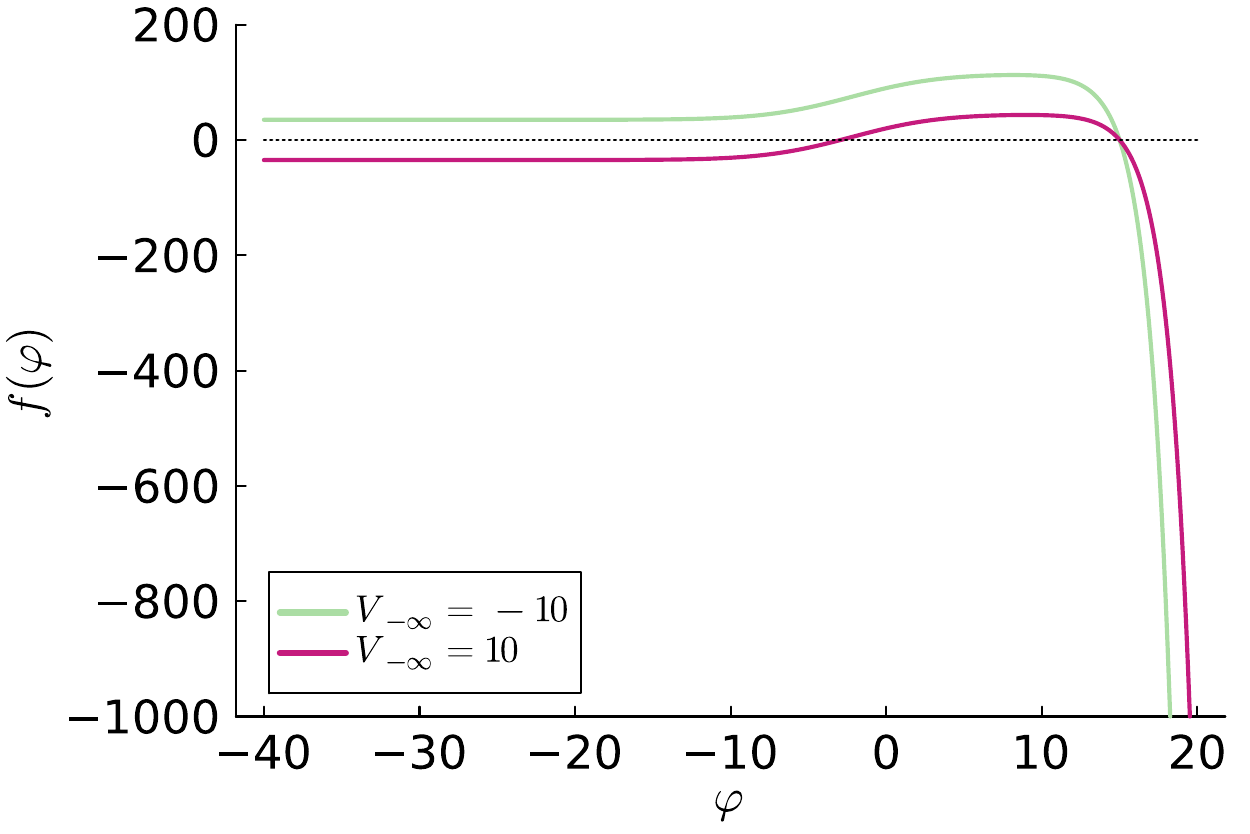}}
	\caption{Flow solutions interpolating between two Gubser-regular endpoints with vanishing potential as $\f\to-\infty$ and a black hole at $\f_h=15$. The analytical solution is given in Eqs. \eqref{asugu2}, \eqref{tgg2}, \eqref{fgg2} and \eqref{vgg2} and it is parametrised in terms of a single parameter $V_{-\infty}$ that controls the sign of the potential as $\f\to-\infty$. }
	\label{fig:gubx5}
\end{figure}

In this section, we discuss solutions that feature a Gubser-regular endpoint where $V\to 0^{\pm}$ together with a black-hole event horizon. According to the classification of horizons in Appendix \ref{app:J}, the presence of a black hole requires that $f$ vanishes once if $V\to 0^-$, or that $f$ vanishes twice if $V\to 0^+$. Inside the black hole, the flow runs to a bad singularity, as shown in rule 22 on page \pageref{ru22}. The details about the construction of this solution are presented in Appendix \ref{app:75}.

We consider again the superpotential of Sec. \ref{sec:73}, which has no endpoint at finite $\f$:
\begin{equation}
W= e^{\sqrt[3]{\frac{2}{3}}\f} + e^{\frac{1}{4}\sqrt{\frac{3}{2}}}\,,
\end{equation}
where we have also set $c_2=1$ for concreteness. At $\f\to -\infty$, the superpotential vanishes. Therefore, the potential $V$ also vanishes at Gubser-regular endpoints as $\f\to -\infty$ (see Tables \ref{tab_IS} and \ref{tab_IIS}). In the previous section, the integration constants are chosen such that the solution admits the Gubser-regular asymptotic structure as $\f\to + \infty$. In this section, we fix the integration constants such that the singularity at $\f\to+\infty$ is covered by a black-hole event horizon. It follows from rule 22 on page \pageref{ru22} that such a singularity is bad.

The explicit solution for the inverse scale factor $T$, the blackening function $f$, and the potential $V$ is given in Eqs. \eqref{tgg2}, \eqref{fgg2} and \eqref{vgg2} respectively.

The solution has two integration constants: $f_0$ and $f_1$. In order to construct solutions running from a Gubser-regular endpoint to a horizon, we must demand that $f$ vanishes at least one, at a location $\f_h$. For concreteness, we set $\f_h=15$. This condition fixes one of the integration constants. The second integration constant can be fixed in terms of the asymptotic behaviour of the potential. As we approach the Gubser-regular endpoint at $\f\to-\infty$, the function $f$ approaches a constant value, while the potential vanishes asymptotically:
\begin{equation}
f(-\infty)= f(-\infty)+\dots\,,\quad
V(-\infty)= -\frac{55}{192} f(-\infty) e^{\frac{1}{2} \sqrt{\frac{3}{2}} \varphi }+\dots \equiv -V_{-\infty}e^{\frac{1}{2} \sqrt{\frac{3}{2}} \varphi }+\dots\,.
\end{equation}
The behaviour of $f$ and $V$ is compatible with the  type I asymptotic solutions of Appendix \ref{asymp} as $\f\to -\infty$. The second integration constant can be fixed in terms of $V_{-\infty}$. Specifically, both $f_0$ and $f_1$ are given by
\begin{equation}
f_0 = 79.6885\, -3.49091 V_{-\infty}\,, \qquad f_1 =-2360.5 + 103.488 V_{-\infty}.
\end{equation}

The value $V_{-\infty}$ distinguishes qualitatively different solutions. In Fig. \ref{fig:gubx5} we present an example of two inequivalent cases:
\begin{itemize}
\item $V_{-\infty}=-10$. The potential vanishes from below at $\f\to-\infty$, while $f$ asymptotes to a positive constant at the regular endpoint. Additionally, $f$ vanishes once, at $\f=15$, signalling the presence of a black-hole event horizon. Inside the black hole there is a bad singularity. This is a solution from a type I endpoint with $V\to0^-$ to a black hole.
\item $V_{-\infty}=10$. The potential vanishes from above at $\f\to-\infty$, while $f$ asymptotes to a negative constant at the regular endpoint. Additionally, $f$ vanishes twice: $\f\simeq -2.91$ is the outermost root of $f$, and is therefore a cosmological horizon - as discussed in Appendix \ref{app:J}; $\f\simeq 15$ is the innermost root, which corresponds to a black-hole event horizon. Inside the black hole there is a bad singularity. This is a solution from a type I endpoint with $V\to 0^+$ to a black hole.
\end{itemize}

The case where $V_{-\infty}=0$ would give a solution from a Gubser-regular endpoint with type II asymptotics and $V\to 0^-$ to a black-hole event horizon.

\section{The multiscalar Case\label{multi}}

So far we have analysed gravity coupled to a single scalar. In this section we shall argue that our no-go results are valid in the presence of an arbitrary number of scalars.

In such a case, the general two derivative gravitational action that generalizes (\ref{i1}) can be written as
\eql{m1}{
			S\le[g,\f^I\ri]= \int d^{d+1}x \Vg \le(
			 R
			-\ha G_{IJ}(\f)\p_a\f^I \p^a\f^J-V(\f^I)  \ri)+S_{GHY}.
			}
where $S_{GHY}$ is the Gibbons-Hawking-York term associated to any boundary that might exist. Again , we arrive to this action,  from the most general two-derivative action,  after a Weyl rescaling of the metric, $g_{\m\n}$,  as well as a redefinition of the scalar fields if necessary.
The metric $G_{IJ}(\f)$ of the scalar manifold is assumed to be positive definite as is the case in effective actions of string theory.

The geometry of the scalar manifold is assumed to be regular for finite values of the scalars. This excludes the case of conifold singularities that are known to appear in string theory, and which appear at finite distances in the scalar space\footnote{Although we exclude conifold singularities, we believe that our results are valid even if they are included.}, \cite{book}. On the other hand, the singularities that appear at the boundaries of moduli space, are of the standard types: decompactications or emergent string singularities, \cite{swamp}. By choosing appropriately an adapted coordinate system near the boundaries, the potential can be again parametrized as in (\ref{aasym}) with $\f$ appropriately defined, while the metric $G_{IJ}$ is regular.

Given again the interpolating metric ansatz (\ref{f22}) the equations become

\be
	2(d-1)\ddot{A}+G_{IJ}\dot{\f}^I\dot\f^J=0\,,\label{f23am}
\ee
\be
	\ddot{f}(u)+d\dot{f}(u)\dot{A}(u)+{2(d-2)\over R^2} e^{-2 A(u)}=0\,,\label{f23bm}
\ee
\be	 (d-1)\dot{A}(u)\dot{f}(u)+f(u)\left[d(d-1)\dot{A}^2(u)-\frac{G_{IJ}\dot{\f}^I\dot{\f}^J}{2}\right]+V(\f^I)-{(d-1)(d-2)\over R^2} e^{-2 A(u)} =0\,.\label{f23cm}
\ee

 Equation (\ref{f23am}) implies again that $\dot{A}$ is monotonous along the flow. The Klein-Gordon equations for the scalars
 \be
 \square \f^I+{\Gamma^I}_{JK}\pa\f^J\pa\f^K-G^{IJ}{\pa V\over \pa \f^J}=0
 \ee
 which for the present ansatz becomes
 { \be
f\ddot\f^I+f\left(d\dot A+{\dot f\over f}\right)\dot\f^I+f{\Gamma^I}_{JK}\dot\f^J\dot\f^K-G^{IJ}{\pa V\over \pa\f^J}=0\,.
\label{f23dm}\ee}

Equation (\ref{f23cm}) can also be written as
\be
{d\over du}\left(f\dot A e^{dA}\right)=\left(-{V~e^{2A}\over d-1}+{(d-2)\over R^2}\right) e^{(d-2) A(u)}
\label{ev1m}\ee
and (\ref{f23bm}) becomes
\be
{d\over du}\left(\dot f e^{dA}\right)=-{2(d-2)\over R^2}e^{(d-2)A}\;.
\label{ev2m}\ee

We also introduce the energy momentum tensor of the scalars as
\be
T^{\f}_{\m\n}=G_{IJ}\pa_{\m}\f^I\pa_{\n}\f^J-{1\over 2}g_{\m\n}G_{IJ}\pa_{\m}\f^I\pa^{\m}\f^J-g_{\m\n}V
\label{ev3am}\ee
whose non-zero components for our ansatz are
\be
{T^u}_u={1\over 2}f~G_{IJ}\dot\f^I\dot \f^J-V\equiv \rho
\label{ev3bm}\ee
\be
{T^i}_j=-p{\delta^{i}}_j\sp p\equiv \left[{1\over 2}f~G_{IJ}\dot\f^I\dot \f^J+V\right]
\label{ev3cm}\ee
In fact, when $f<0$, $u$ is a time-like coordinate and then $\rho$ can be called the energy density and $p$ is the pressure.
By abuse of language we shall always call $\rho$ it the energy density
{
We can also rewrite the equations as
\begin{equation}\label{ev4m}
(d-1)\frac{d}{du}\left({e^{dA}\mathcal{I}}\right) = -\dot{A}e^{dA}~p
\end{equation}
and
\begin{equation}
\dfrac{\dot{A}}{G_{IJ}\dot{\varphi}^I\dot\f^J}\dot \r + \dfrac{1}{2(d-1)}\r =- \dfrac{d}{2} \i - \dfrac{(d-1)}{R^2}e^{-2A}=-{d\over 2}\left[f\dot A^2+{d-2\over d~R^2}e^{-2A}\right]\,,
\label{ev5m}\end{equation}
where $\mathcal{I}$ is defined as
\begin{equation}
\mathcal{I} \equiv  f\dot{A}^2 - \dfrac{1}{R^2}e^{-2A}\,.
\label{ev7m}\end{equation}
Like in the single scalar case, $\r,p, \i$  control the curvature invariants of the geometry.
}

There is an adapted first order formalism in the multi-scalar case that was developed in \cite{papa,exotic} but we shall not need it here. 

Given the above, all of our flow rules in section \ref{rul} are again valid, as the only scalar property we have used in the single scalar case to prove them was the non-negativity of $\dot\f^2$ in the gravitational equations. Since now $\dot\f^2$ is replaced by $G_{IJ}\dot\f^I\dot\f^J$ which is again non-negative, all such properties remain true.

Finally, our asymptotic solutions described in appendix \ref{asymp} in the single scalar case, remain intact provided that near the asymptotic region of the scalar potential, we choose adapted coordinates so that the potential behaves as in (\ref{z1}) where the scalar $\f$ is the one that runs to infinity.
Consequently, the classification of asymptotic solutions is as described in appendix \ref{asymp} for the single scalar case.

\section{Thin wall solutions with (A)dS asymptotics}\label{sec:brane}
A simple setting in which one might hope to construct domain-wall solutions, which interpolate between an AdS boundary and a dS interior is by way of a ``thin wall'' construction. In particular, one may wonder under what conditions a solution within our ansatz exists in which the space-time consists of the vacuum AdS solution  joined along a co-dimension one surface to vacuum dS.

This question was addressed in \cite{hub}, extending the earlier work of \cite{Blau:1986cw} on the dynamics of de Sitter vacua bubbles in asymptotically flat space-times to the case of negative curvature. In particular, the authors of \cite{hub} study domain-wall solutions in the thin-wall limit in which a part of the (Schwarzschild) AdS solution which includes the time-like boundary is joined via a tensionful ``brane'' to a region of dS vacuum. The branes may be either static or dynamical, and the asymptotically AdS side of the junction is allowed to have a non-vanishing black-hole mass. The authors' primary interest is in constructing solutions which contain dS infinity, which they realize in various examples.

In what follows, we depart from this analysis in several ways. Our present interest is in static thin brane solutions in which the solution interpolates between a region of vacuum AdS which includes the timelike boundary, and vacuum dS.  Unlike the analysis of \cite{hub}, we do not restrict the stress-energy of the brane to take the form of a cosmological constant on its worldvolume. Instead, we demand that the desired solution exists, and then determine a brane action on its worldvolume which allows the gravitational equations of motion to be satisfied.

We  now show that such a solution exists provided the co-dimension one ``brane'' is endowed with a very specific form of localized stress-energy\footnote{This differs from analogous solutions in the ansatz with dS slicing (\ref{w24}), in which a ``tensionless'' solution can be found.}. This is in contrast with the Einstein-dilaton theory studied in the previous sections, in which such spherically symmetric domain-wall solutions were shown not to exist (recall figure \ref{fig:flow}). Here, we highlight the results of this thin wall construction---the details are presented in appendix \ref{app:thinW}.

We first partition the space-time into two regions $\mathcal{M}^\pm$ separated by the brane, whose worldvolume is taken to be the hypersurface $\Sigma$. We use conventions in which the unit normal to $\Sigma$ points towards $\mathcal{M}^+$. Introducing the notation
\begin{equation}
\left[T \right] = T\left(\mathcal{M}^+\right)\big |_\Sigma-T\left(\mathcal{M}^-\right)\big |_\Sigma.
\end{equation}
for any tensor $T$ defined on either side of the hypersurface, the junction conditions for the putative gravitational solution are given by
\begin{equation}\label{eq:JCthin}
\left[\gamma_{ij} \right] = 0 \qquad \mathrm{and} \qquad \Big(\left[K_{ij}\right]-\left[K\right]\gamma_{ij} \Big)=-S_{ij}.
\end{equation}
Here $\gamma_{ij}$ is the metric induced on $\Sigma$, $K_{ij}$ is its extrinsic curvature with trace $K$, and  $S_{ij}$ allows for the addition of a brane stress energy tensor, which contributes to the bulk equations of motion like
\begin{equation}
T_D^{\mu\nu}=\delta(s)S^{ij}\theta^\mu_i\theta^\nu_j.
\end{equation}
In this expression we have taken $s$ to measure the proper distance from the hypersurface, and the $\theta^i=\partial_i$ are a set of vectors tangent to $\Sigma$.

Within the spherically sliced ansatz of (\ref{c39}), we can take without loss of generality the metric on $\mathcal{M}^-$ to be
\begin{equation}
\mathrm{d}s^2_- = \frac{\mathrm{d}u^2}{\left(1+e^{2u}\right)} + e^{-2u}\left[-\left(1+e^{2u}\right)\mathrm{d}t^2 +\mathrm{d}\Omega_{d-1}^2\right]
\end{equation}
which is AdS$_{d+1}$ with unit radius.  In these coordinates, the boundary of AdS is attained as $u\to-\infty$. We wish to match this to a dS$_{d+1}$ solution in the region $\mathcal{M}^+$ with metric
\begin{equation}
\mathrm{d}s^2_+ = \frac{\mathrm{d}\mathfrak{u}^2}{\left(e^{2 \frac{H}{\alpha}\mathfrak{u}}-1\right)} + \alpha^2\,e^{-2\frac{H}{\alpha}\mathfrak{u}}\left[-\left(e^{2\frac{H}{\alpha}\mathfrak{u}}-1\right)\mathrm{d}t^2 +\frac{1}{H^2}\mathrm{d}\Omega_{d-1}^2\right].
\end{equation}
In these coordinates, the ``shrinking endpoint'' (which is identified with the location of the observer in this static patch of dS) is located at $\mathfrak{u}\to\infty$. Note that we have introduced the radial coordinate $\mathfrak{u}$ to emphasise the fact that the radial coordinates on $\mathcal{M}^\pm$ need not be the same. Similarly, one can choose distinct time coordinates on either side of the brane. The scaling symmetries of the ansatz allow one to take these time coordinates to be proportional to one another, with proportionality constant $\alpha$.

Direct calculation shows that these two spaces can indeed be joined along a thin brane in a manner consistent with the junction conditions (\ref{eq:JCthin}) provided that
\begin{equation}
\alpha > 1
\end{equation}
and that there is stress-energy on the brane described by the brane stress tensor
\begin{equation}\label{eq:STc1}
S_{tt} = \frac{1}{\alpha}(1-d)\left(\frac{\alpha^2+H^2}{\alpha^2-1} \right)^{1/2}\left(1-\alpha\right)\gamma_{tt}
\end{equation}
and
\begin{equation}\label{eq:STc2}
S_{\sigma\rho} = \left(1-\alpha\right)\left[\frac{1}{\alpha}(1-d)-\left(\frac{1+H^2}{\alpha^2+H^2}\right) \right]\left(\frac{\alpha^2+H^2}{\alpha^2-1} \right)^{1/2}\gamma_{\sigma\rho}
\end{equation}
where $\sigma$, $\rho$ are directions on the $S^{d-1}$.

An immediate question is what brane action gives rise to such a stress tensor. The simplest covariant action on the brane consistent with such a stress tensor is Einstein-Hilbert with a cosmological constant:
\begin{equation}
S_D = -\frac{1}{2\kappa_D^2} \int_\Sigma\mathrm{d}x^d \sqrt{-\gamma}\Big( R[\gamma]+\mu\Big).
\end{equation}
The total bulk plus brane action is therefore
\be
S=S_{bulk}+S_D
\ee
with $S_{bulk}$ given in (\ref{i1}).

Indeed, we find that this brane action reproduces (\ref{eq:STc1}, \ref{eq:STc2}) provided one makes the identification between parameters $(\kappa_D,\mu)$ and $(\alpha,H)$ like
\begin{equation}\label{eq:kD}
 \frac{1}{2\kappa_D^2} = \left(\frac{\alpha-1}{d-2}\right)\sqrt{\frac{\alpha^2-1}{\alpha^2+H^2}}
\end{equation}
and
\begin{equation}\label{eq:muD}
\mu = (1-d)(d-2)\left(\frac{\alpha^2+H^2}{\alpha^2-1} \right)\left[\frac{2}{\alpha}+\frac{1+H^2}{\alpha^2+H^2} \right] .
\end{equation}
Note that, since $\a>1$, the tension of the brane is negative and the brane Planck scale is also negative.

Although this class of thin-wall solutions is constructed within a very simple gravitational model, their utility is spoiled somewhat by the fact that it is not clear how such a theory could arise as a low-energy limit of string or M-theory. In particular, the Einstein-Hilbert term on the brane is known to be absent in the leading $\alpha'$ corrections to the single $D$-branes action in type II string theory. Such brane actions have however been previously studied both in the context of the bosonic string where an Einstein term appears at tree level, \cite{CLR} or in the superstrings where it appears a one loop level \cite{KTT,AMV}, as well as phenomenologically in \cite{DGP}.
Note also that in \cite{CLR} the tree-level brane Planck scale (i.e the overall coefficient of the brane action) is similarly negative.

\section{The black-hole ansatz with hyperbolic slicing}\label{hyperbolic}

Sections \ref{sec5}-\ref{sec:brane} have been devoted to characterizing all possible solutions in the spherically sliced ansatz \eqref{c39}. In this section, we review the main differences in the space of solutions, when one considers a black-hole ansatz with hyperbolic slices\footnote{The related ansatz with flat slices has been analysed earlier in \cite{KT}.}:

\begin{equation}
ds^2={du^2\over f(u)}+e^{2A(u)}\left[-f(u)dt^2+R^2~dH_{d-1}^2\right]\,,
\label{eq:hyp}
\end{equation}
where H$_{d-1}$ is the $d-1$ Euclidean space of constant negative curvature. Einstein's equations for the ansatz \eqref{eq:hyp} are equivalent to the equations in a spherically sliced ansatz, Eqs. \eqref{f23a}-\eqref{f23d}, upon the analytical continuation $R\to i R$. We can similarly define the energy density, pressure, and $\mathcal{I}$ as given in Eqs. \eqref{ev3b}, \eqref{ev3c} and \eqref{ev7}, which again obey the equations \eqref{ev4} and \eqref{ev5}, with $R\to i R$.

In the first order formalism, we define the superpotential as in \eqref{w53}, such that the solution to \eqref{f23a} is \eqref{w53b}. We define the inverse scale factor now as in the spherically sliced ansatz up to a minus sign:
\begin{equation}
T(\f) = -\dfrac{1}{R^2}e^{-2A}\leq 0\,.
\end{equation}
With the previous redefinitions, the equations of motion in the hyperbolically sliced ansatz are identical to Eqs. \eqref{f10_1}, \eqref{w55} and \eqref{w55b}, except that in the hyperbolic ansatz we should demand that $T<0$. A key observation relies on the fact that these equations are invariant under the following reflection symmetry:
\begin{equation}\label{reflex}
	f\to - f\sp V\to -V \sp T\to -T\,.
\end{equation}

As a consequence of this symmetry, the results obtained for the local structure of the solutions, provided in Appendices \ref{structure}-\ref{Marginal} and \ref{asymp}, that do not depend on the sign of $T$, are also valid in the hyperbolically sliced ansatz. We list now the major similarities and differences of the endpoints in the hyperbolic ansatz compared to the spherical ansatz:

\begin{itemize}
\item AdS$_{d+1}$ boundary endpoints, discussed in \ref{AdSX} are qualitatively similar in both ansatze.
\item dS$_{d+1}$ boundary endpoints, discussed in \ref{dSX} are also qualitatively similar in both ansatze.
\item dS$_2 \times$S$^{d-1}$ regions in the spherical ansatz, described in Appendix \ref{g53}, become AdS$_2\times $H$^{d-1}$ regions in the hyperbolically sliced ansatz.
\item Spatial boundaries of Minkowski space-time, uncovered in Appendix \ref{exth} for the spherically sliced ansatz, are also present in the hyperbolic ansatz. In the spherical ansatz, the function $f$ vanishes from above at the boundary, while in the hyperbolic ansatz, $f\to 0^-$ at the boundary.
\item Shrinking endpoints, described in Appendix \ref{G.2.2} for the spherically sliced ansatz, are similarly present in the hyperbolically sliced ansatz, with the main difference that $f\to -\infty$ as the shrinking endpoint is approached in the latter, as it follows from Eqs. \eqref{soch} and \eqref{sochu} together with $T\leq 0$.
\item Gubser-regular endpoints, classified in Appendix \ref{asymp} for the spherically sliced ansatz, appear in the hyperbolic ansatz as well. Their properties in the hyperbolic ansatz are summarised in Tables \ref{tab_IH} and \ref{tab_IIH}.
\end{itemize}

Another important difference concerns the number and nature of the horizons that can appear in the flows with a hyperbolically sliced ansatz:
\begin{itemize}
\item The reflection symmetry \eqref{reflex}, combined with rule 8 on page \pageref{ru8} of Sec. \ref{rul}, reveals that flows involving dS boundaries, Minkowski boundaries, or Gubser-regular endpoints with $V\to 0^+$, can have at most one horizon. Additionally, the function $f$ is negative in the region close to the boundary endpoint, and the analysis of Appendix \ref{app:J} indicates that such a horizon is cosmological.
\item The reflection symmetry \eqref{reflex}, applied to rule 5 on page \pageref{ru5} of Sec. \ref{rul}, indicates that $f$ can only have one local minimum along the flow. As a consequence, solutions from AdS$_{d+1}$ boundaries, AdS$_2$ boundaries or Gubser-regular endpoints with $V\to 0^-$, where $f$ is positive, can develop at most two horizons. Precisely because $f$ is positive around those endpoints, the outermost horizon must be a black-hole event horizon (see discussion of Appendix \ref{app:J}), while the innermost horizon would correspond to a Cauchy horizon. The limit of coincident horizons would give rise to an extremal horizon.
\end{itemize}

\begin{figure}[t]
	\centering
	\includegraphics[width=0.99\linewidth]{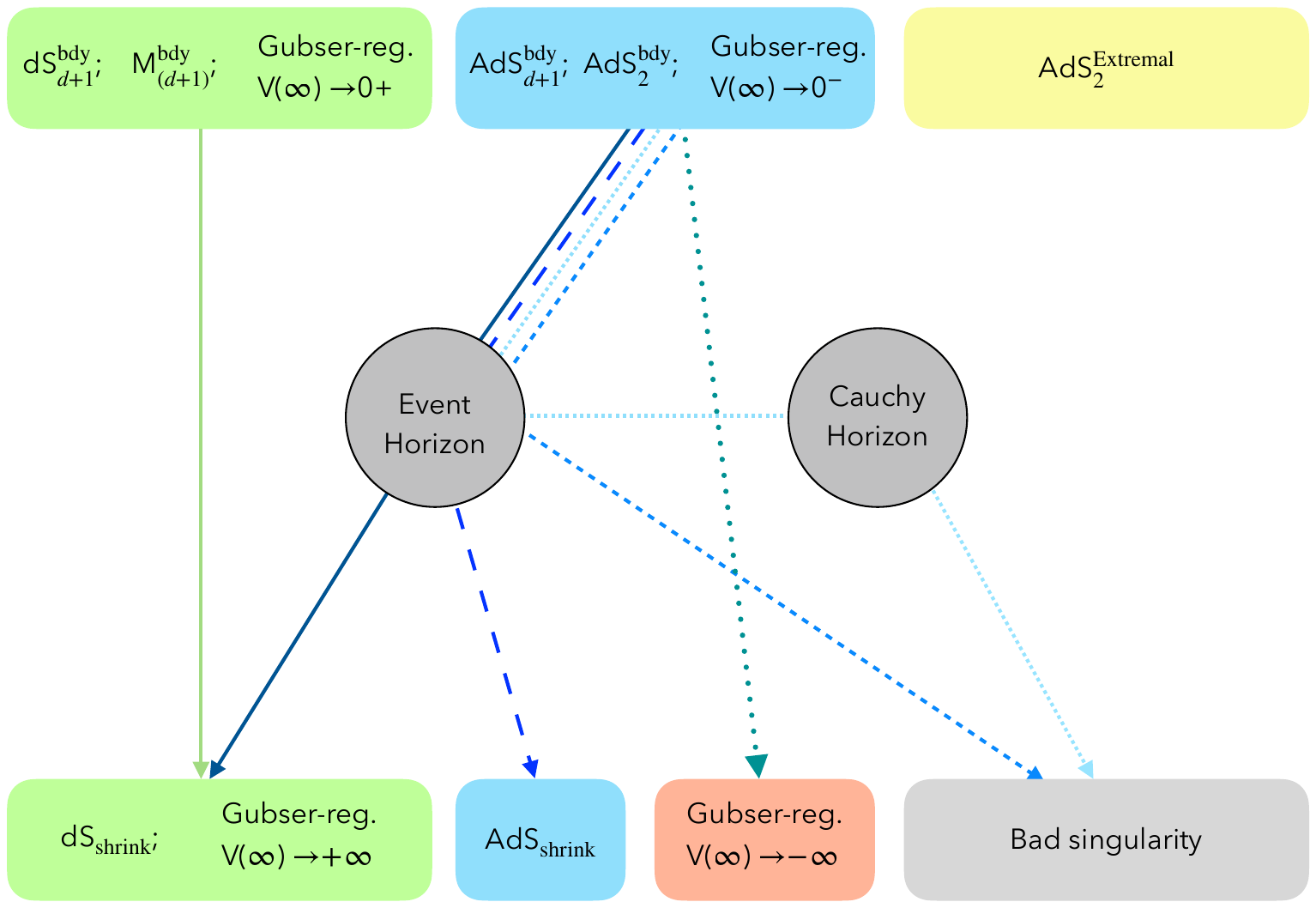}
	\caption{Depiction of the structure of possible flows in the hyperbolically sliced ansatz. All horizons included are regular. The finite endpoints in the upper row are minima (maxima) of a positive (negative) superpotential. The finite endpoints in the lower correspond to maxima (minima) of the positive (negative) superpotential. We have excluded flows with naked singularities, i.e. flows running to a bad singularity that is not covered by a black-hole event horizon. Gubser-reg. stands for Gubser-regular endpoint.}
	\label{fig:flowH}
\end{figure}

As a consequence of the reflection symmetry \eqref{reflex}, any solution that exists in the spherically sliced ansatz can be mapped to a solution in the hyperbolically sliced ansatz with an inverted potential. For instance, a standard solution from an AdS$_{d+1}$ boundary to an AdS shrinking endpoint in the spherically sliced ansatz would be mapped to a solution from a dS$_{d+1}$ boundary to a dS shrinking endpoint in the hyperbolic ansatz. Similarly, any rule that forbids a given flow in the spherical ansatz (see Sec. \ref{fbd}), will give a similar rule for a flow that is forbidden in the hyperbolic ansatz. This allows to characterise automatically all the possible flows in the hyperbolic ansatz.

A summary of the possible regular flows in the hyperbolically sliced ansatz is provided in Fig. \ref{fig:flowH}. We have excluded the flows that are connected to a bad naked singularity. Two comments are in order. Firstly, the flows
\begin{equation}\label{flh1}
\left[ {\rm AdS}_{(d+1)}^{\textrm{bdy}}\,,{ \rm M}_{(d+1)}^{\rm bdy} \,, {\rm Gubser-reg.}\, V(\infty)\to 0^- ~\to~ {\rm event~horizon}~ \to~ {\rm bad~ singularity}\right]^{\rm Sph.}
\end{equation}
that are possible in the spherical ansatz, are mapped to

$$
\left[ {\rm dS}_{(d+1)}^{\textrm{bdy}}\,,{ \rm M}_{(d+1)}^{\rm bdy} \,, {\rm  Gubser-reg.}\, V(\infty)\to 0^+ ~\to~ {\rm cosmological~horizon}~ \to \right.
$$
\begin{equation}\label{flh2}
\left. \to~ {\rm bad ~singularity}\right]^{\rm Hyp.}
\end{equation}
under the reflection symmetry \ref{reflex}. The superscripts Sph. and Hyp. denote spherical and hyperbolic respectively. The newly obtained flows contain a naked singularity and therefore are not included in Fig. \ref{fig:flowH}. On the other hand, there are three flows with a naked singularity in the spherical ansatz that, under the reflection symmetry \eqref{reflex}, are mapped to flows with a horizon-covered singularity. Specifically,

$$
\left[ {\rm dS}_{(d+1)}^{\textrm{bdy}}\,,{ \rm dS}_{2}^{\rm bdy} \,, {\rm Gubser-reg.}\, V(\infty)\to 0^+ ~\to~ {\rm cosmological~horizon}~ \to \right.
$$
\begin{equation}\label{flh3}
 \left. \to~ {\rm bad~ singularity}\right]^{\rm Sph.}
\end{equation}
are mapped to
\begin{equation}\label{flh4}
\left[ {\rm AdS}_{(d+1)}^{\textrm{bdy}}\,,{ \rm AdS}_{2}^{\rm bdy} \,, {\rm Gubser-reg.}\, V(\infty)\to 0^- ~\to~ {\rm event~horizon}~ \to~ {\rm bad ~singularity}\right]^{\rm Hyp.}
\end{equation}
This possibility is included in Fig. \ref{fig:flowH} for the hyperbolic ansatz.

\section*{Acknowledgements}
\addcontentsline{toc}{section}{Acknowledgements}

We would like to thank David Andriot, Dio Anninos, Tarek Anous, Roberto Emparan, Diego Hofman, Carlos Hoyos, Takaki Ishii, Romuald
Janik, Matti Jarvinen, Dieter L\"ust, Javier Mas, Vasilis Niarchos, Francesco Nitti, George Obied, Angel Paredes, Jose-Manuel Penin-Ascariz, Achilleas Porfyriadis,
Edwan Preau, Alfonso Ramallo, Matthew Roberts, Javier Subils, Houri Tarazi, Cumrun Vafa and Timm Wrase.

This work is partially supported by the European MSCA grant HORIZON MSCA-
2022-PF-01-01 ``BlackHoleChaos" No.101105116 and by the H.F.R.I call ``Basic
research Financing (Horizontal support of all Sciences)" under the National Recovery
and Resilience Plan ``Greece 2.0" funded by the European Union  \\
-NextGenerationEU (H.F.R.I. Project Number: 15384.),  the In2p3 grant ``Extreme
Dynamics'' and the ANR grant ``XtremeHolo'' (ANR project n.284452)

This work is also partially funded by the EU’s NextGenerationEU instrument through the National Recovery and Resilience Plan of Romania - Pillar III-C9-I8, managed by the Ministry of Research, Innovation and Digitization, within the project entitled ``Facets of Rotating Quark-Gluon Plasma'' (FORQ), contract no. 760079/23.05.2023 code CF 103/15.11.2022.


\newpage
\appendix
\renewcommand{\theequation}{\thesection.\arabic{equation}}
\addcontentsline{toc}{section}{Appendix}
\section*{Appendices}

\section{Coordinate systems in de Sitter\label{a}}

In this work, we encounter de Sitter space in various coordinate systems. Here we provide a brief overview of some of the more common coordinates on dS.  For further details see, for example \cite{dSlec,anninos}.

The (d+1)-dimensional de Sitter space is defined as the submanifold
\be
-X_0^2+\sum_{i=1}^{d+1}X_i^2={1\over H^2},
\label{a24}\ee
embedded in $\mathbb{R}^{d+2}$. This embedding manifests the $SO(d+1,1)$ isometry group of dS$_{d+1}$.

\subsection{Global coordinates}

We can construct global coordinates for de Sitter by first introducing the spherical coordinates
\be
X^i=r n^i\sp i=1,2,\cdots,d+1\sp n^in^i=1\sp r>0
\label{a39}\ee
where $n^i$ parametrises $S^d$. We next define the coordinates, $w,v$, such that
\be
r=w{\cosh(v)}\sp X^0=w{\sinh(v)}\sp v\in \mathbb{R}\sp w\geq 0
\label{a42}\ee
where $v\in \mathbb{R}$ as $X^0$ can be both positive and negative.

These are global coordinates on dS, defined through the embedding
\be
X^{0}={\sinh(v)\over H}\sp X^i={\cosh(v)\over H} n^i\sp i=1,2,\cdots d+1\sp v\in \mathbb{R}.
\ee
They yield the induced metric
\be
\mathrm{d}s^2={1\over H^2}\left(-\mathrm{d}v^2+\cosh^2(v)\,\mathrm{d}\Omega_{d}^2\right)
\ee
In these coordinates, the past boundary ${\cal I}^-$ is obtained as $v\to-\infty$ and  $r\to \infty$.
The future boundary ${\cal I}^+$ is obtained as $v\to +\infty$ and $r\to \infty$.

One may imagine the de Sitter$_{d+1}$ manifold in global coordinates as a spatial $d$-sphere with variable radius. In the infinite past, the sphere has infinite radius and this describes the ${\cal I}^-$ past boundary. As time increases, the sphere radius decreases, shrinking to a minimum size at $v=0$. As $v>0$  increases, the sphere expands once more eventually becoming infinitely large in the far future. This is the future boundary, ${\cal I}^+$.

The analytic continuation $v\to i v$ brings  de Sitter to a $(d+1)$-dimensional sphere.
In this paper we typically do not distinguish between past and future boundaries, as we are not concerned with the direction time's arrow.

\subsection{dS sliced coordinates}

One may similarly construct an embedding that yields the dS$_d$ sliced metric of dS$_{d+1}$ (\ref{a52}). This embedding is given by
\be
X^{0}={\sin\theta\over H}\sinh(z)\sp X^i={{\sin\theta\over H}\cosh(z)} n^i\sp i=1,2,\cdots d
\label{a53}\ee
\be
 X^{d+1}={\cos\theta\over H}\sp z\in \mathbb{R}\sp \theta\in[0,\pi]
\label{a54}\ee
Here the $n^i$, satisfying $n^i n^i = 1$, are constrained coordinates on $S^{d-1}$.

The induced metric from this embedding is 
\be
ds_{dS}^2={1\over H^2}\left[\mathrm{d}\theta^2+\sin^2\theta\left(-\mathrm{d}z^2+\cosh^2(z)\mathrm{d}\Omega_{d-1}^2\right)\right]
\label{a52}\ee
By defining coordinates $w,t$ with units of length as $w={\theta\over H},t={z\over H}$ this metric  becomes
 \be
\mathrm{d}s_{dS}^2=\mathrm{d}w^2+{\sin^2(wH)}\left(-\mathrm{d}t^2+{\cosh^2(tH)\over H^2}\mathrm{d}\Omega_{d-1}^2\right)
\label{a52a}\ee

Alternatively, one may introduce
\be
{\cal W}=\log\left(\tan{\theta\over 2}\right)\sp \sin\theta={1\over \cosh{\cal W}}\sp \cos\theta={-\tanh{\cal W}}
\ee
such that the metric becomes
\be
\mathrm{d}s^2={H^2\over \cosh^2{\cal W}}\left[\mathrm{d}{\cal W}^2+\left(-\mathrm{d}z^2+\cosh^2(z)\mathrm{d}\Omega_{d-1}^2\right)\right]
\ee

In these coordinates one may obtain ${\cal I}^-$ when $z<0$. Noting that the (square of the) global radial coordinate, $r^2$ is given by
\be
r^2={{\sin^2\theta\over H^2}\cosh^2(z)}+{\cos^2\theta\over H^2}
\ee
we see that this becomes infinite when $z\to -\infty$. This does not yield the full $S^d$ sphere at infinity, but only an $S^{d-1}$ subsphere.
We thus see that the ${\cal I}^{-}$ boundary of the dS$_d$ slice is an $S^{d-1}$ subsphere of the $S^d$ at the  ${\cal I}^{-}$ boundary of dS$_{d+1}$. Similar statements apply to the ${\cal I}^{+}$ boundary.

\subsection{Other coordinate systems}

\begin{itemize}

\item {\bf Poincar\'e coordinates}

The coordinate systems
\be
\mathrm{d}s^2=-dT^2+e^{2HT}\mathrm{d}\vec x^2={1\over (Ht)^2}(-\mathrm{d}t^2+\mathrm{d}\vec x^2)
\label{a26}
\ee
provide Poincar\'e coordinates for dS. Here $T\to\infty$ is the future boundary ${\cal I}^+$,  while $T\to-\infty$ is a single point on the past boundary, ${\cal I}^-$. Here, this is a (past) Poincar\'e horizon. The time $t$ is a conformal time, which takes values between $t=-\infty$ (the Poincar\'e horizon) and $t\to 0$, which is the future boundary.

The point at $T=-\infty$ resembles a big-bang singularity, as space evidently shrinks to a point. However, this is in fact a coordinate singularity similar to the coordinate singularity at the origin of spherical coordinates. Indeed, the dS curvature is everywhere finite and constant.

\item {\bf Static patch coordinates}

These coordinates cover the region a static observer has causal access to in de Sitter space. The metric can be written
\be
\mathrm{d}s^2=-\left(1-H^2r^2\right)\mathrm{d}t^2+{\mathrm{d}r^2\over (1-H^2r^2)}+r^2~\mathrm{d}\Omega_{d-1}^2\;.
\label{a27}\ee
In these coordinates, $r=0$ is the position of the static observer and $r={1\over H}$ is the cosmological horizon. Thus, the region the observer may access is given by $r\in \left[0,{1\over H}\right]$.  The future boundary is obtained as $r\to+\infty$.

Upon further change of radial coordinate to $Hr=\sin(Hu)$ with $u\in\left[0,{\pi\over 2H}\right]$,
The above metric becomes
\be
\mathrm{d}s^2=-\cos^2(Hu)\mathrm{d}t^2+\mathrm{d}u^2+{\sin^2(Hu)\over H^2}~\mathrm{d}\Omega_{d-1}^2
\label{a31}\ee
This is a metric  on  the interior of the cosmological  horizon. For $Hr>1$ we may take
\be
Hr=\cosh(Hu)
\label{a32}\ee
such that the ${\cal I}^{+}$ boundary is obtained as $u\to \infty$.
In this case, the metric is given by
\be
\mathrm{d}s^2=\sinh^2(Hu)\mathrm{d}t^2-\mathrm{d}u^2+{\cosh^2(Hu)\over H^2}\mathrm{d}\Omega_{d-1}^2
\label{a33}\ee

\item {\bf AdS sliced coordinates}

Finally, one may foliate dS$_{d+1}$ by EAdS$_d$ slices: 
\be
\mathrm{d}s^2=-\mathrm{d}t^2+{\sinh^2(Ht)\over H^2}\mathrm{d}H_{d}^2=-\mathrm{d}t^2+{\sinh^2(Ht)\over H^2}\left(\mathrm{d}R^2+\sinh^2 R \mathrm{d}{\Omega}^2_{d-1}\right)
\label{a28}\ee
where $\mathrm{d}H^2_d$ is the metric of unit radius Euclidean AdS$_d$---hyperbolic space. The future boundary ${\cal I}^+$ is obtained as
$t\to\infty$,  while the $t\to 0$ limit takes one to a big-bang like singularity. This is in fact a coordinate singularity.

\end{itemize}

\section{Coordinate systems in anti de Sitter\label{b}}

Here we provide a brief overview of some of the more common coordinates on AdS.  For further details see, for example \cite{adsreview}.

The (d+1)-dimensional anti de Sitter space is defined as the manifold described by the surface :
\be
X_0^2+X_1^2- \sum_{i=2}^{d+1}X_i^2=\ell^2
\label{a30}\ee
embedded in flat space with (2,d) signature. This embedding makes manifest the $SO(d,2)$ isometry group of AdS$_{d+1}$.

\subsection{Global coordinates}
Global coordinates on AdS can be constructed through the embedding
\be
X^0=r_1\cos\theta\sp X^1=r_1\sin\theta\sp X^i=r_2 n^i\sp i=2,3,\cdots,d+1
\label{aa59}\ee
where $n^i n^i = 1$,  $\theta\in[0,2\pi]$ and $r_1,r_2\geq 0$.

Changing coordinates again such that
\be
r_1=\ell\cosh\r\sp r_2=\ell\sinh\r
\label{aa64}\ee
with $\r\geq 0$ so that $r_2>0$, we obtain the metric
\be
\mathrm{d}s^2=\ell^2\left(-\cosh^2\rho~\mathrm{d}\theta^2+\mathrm{d}\r^2+\sinh^2\rho~\mathrm{d}\Omega_{d-1}^2\right)
\label{adsg}\ee
Here $\theta$ is a time-like coordinate, taking values in $[0,2\pi]$.
We can make this time coordinate non-compact by extending the range of $\theta$ to the whole real line. This universal cover of the surface in (\ref{a30}) can be taken as the definition of anti de Sitter (AdS) space with Minkowski signature.

This spacetime possesses a single boundary located at $\rho\to\infty$, with the geometry of $\mathbb{R}\times S^{d-1}$. When $\rho=0$, one obtains the center of AdS$_{d+1}$.

Further introducing the coordinate $\varphi$ such that $\tan\varphi=\sinh\rho$,  the radial coordinate is compactified and the metric becomes
\be
\mathrm{d}s^2={\ell^2\over \cos^2\varphi}\left(-\mathrm{d}\theta^2+\mathrm{d}\varphi^2+\sin^2\varphi \mathrm{d}\Omega_{d-1}^2\right)={\ell^2\over \cos^2\varphi}\left(-\mathrm{d}\theta^2+\mathrm{d}\Omega_{d}^2\right) \sp \varphi\in \left[0,{\pi\over 2}\right)
\label{a34}\ee
This coordinate system makes manifest the fact that the time it takes a null radial geodesic to arrive from the center of the spacetime to the boundary is $\Delta\theta=\pi/2$.

\subsection{Other coordinate systems}

\begin{itemize}

\item {\bf Poincar\'e coordinates}

Much like in the dS case, one can define Poincar\'e coordinates on AdS, like
\be
\mathrm{d}s^2=-\mathrm{d}u^2+e^{-2{u\over \ell}}(-\mathrm{d}t^2+\mathrm{d}\vec x^2)={\ell^2\over z^2}(\mathrm{d}z^2-\mathrm{d}t^2+\mathrm{d}\vec x^2).
\label{a34_1}
\ee
Here when $u\to-\infty$ one arrives at the AdS boundary,  while $u\to \infty$ is the location of the Poincar\'e horizon.

\item {\bf Static patch coordinates}

Are named by analogy (primarily in appearance) to the dS coordinates of the same name. They are more commonly referred to as another presentation of  ``global coordinates'' on AdS:
\be
\mathrm{d}s^2=-\left(1+{r^2\over \ell^2}\right)\mathrm{d}t^2+{\mathrm{d}r^2\over \left(1+{r^2\over \ell^2}\right)}+r^2~\mathrm{d}\Omega_{d-1}^2
\label{a35}\ee
The AdS boundary in these coordinates is located at $r\to \infty$, while the center is at $r=0$.

\item {\bf AdS slice coordinates}

AdS$_{d+1}$ can also be foliated by EAdS$_d$ slices. In particular,
\be
\mathrm{d}s^2=\mathrm{d}u^2+\ell^2\cosh^2{u\over \ell}\mathrm{d}H_{d}^2=\mathrm{d}u^2+\cosh^2{u\over \ell}\left(\mathrm{d}R^2+\sinh^2 R \mathrm{d}{\Omega}^2_{d-1}\right)
\label{a36}\ee
where $\mathrm{d}H^2_d$ is the metric of unit radius Euclidean AdS$_d$. In these coordinates, 
$u\to\pm\infty$ are two pieces of  the boundary that meet along the equator.

\item {\bf dS slice coordinates}

Finally, one may foliate AdS$_{d+1}$ with dS$_d$ slices:
\be
\mathrm{d}s^2=\mathrm{d}w^2+\ell^2\sinh^2{w\over \ell}\mathrm{d}S_{d}^2=\mathrm{d}w^2+\sinh^2{w\over \ell}\left(-\mathrm{d}t^2+{\cosh^2(Ht)\over H^2}\mathrm{d}\Omega_{d-1}^2\right)
\label{a37}\ee
In these coordinates, $w$ takes values in $w\in (-\infty,0]$, while as $w\to-\infty$ one arrives at the AdS boundary.

\end{itemize}

\vskip 1cm

\section{Curvature invariants and regularity in spherically sliced coordinates} \label{sect:inv_sphere}

\vskip 1cm

In this appendix we shall  calculate the invariants for the metric (\ref{f22}) in our ansatz. They are useful in determining the regularity of solutions.

 The scalar curvature is given by

\be
R=\frac{d+1}{d-1} V(\f)+\frac{1}{2}f(u) \dot{\f}^2(u)=\frac{d+1}{d-1} V(\f)+{1\over 2}fW'^2={dp-\rho\over d-1}
\label{w58}\ee
\be
\sp \partial_\mu\f \partial^\mu\f=f(u) \dot{\f}^2(u)=fW'^2=\r+p.
\label{w59}\ee
where $\r$, $p$ were defined in (\ref{ev3b}) and (\ref{ev3c}).

The square of the Ricci tensor is,
\be
R_{\mu \nu}R^{\mu \nu}=\frac{d+1}{(d-1)^2} V^2+\frac{1}{d-1}V~ f\dot{\f}^2+\frac{1}{4}f^2\dot{\f}^4=
\label{w60}\ee
$$
=\frac{d+1}{(d-1)^2} V^2+\frac{1}{d-1}f V W'^2+\frac{1}{4}f^2W'^4
$$

The Kretchmann invariant is
\be
K_2\equiv R_{\mu \nu \kappa \lambda}R^{\mu \nu \kappa \lambda}={{d^2-4d+7\over (d-1)^2}V^2}+(d-2)(d-1)^2d \left(f\dot A^2-{e^{-2A}\over R^2}\right)^2+
\label{w61}\ee
$$
+{(d-1)(d-2)\left(2V-f\dot\f^2\right)\left(f\dot A^2-{e^{-2A}\over R^2}\right)}-{{d-3\over d-1}V~f~\dot\f^2}+{{d+1\over 4(d-1)}f^2\dot\f^4}
$$
\be
={{d^2-4d+7\over (d-1)^2}V^2}+(d-2)(d-1)^2d \left({fW^2\over 4(d-1)^2}-T\right)^2+
\label{w62}\ee
$$
+{(d-1)(d-2)\left(2V-fW'^2\right)\left({fW^2\over 4(d-1)^2}-T\right)}-{{d-3\over d-1}V~f~W'^2}+{{d+1\over 4(d-1)}f^2 W'^4}
$$

We may now conclude from (\ref{w58}) (\ref{w59})  and (\ref{w60}) that for any point in $\f$-space where the potential is regular, regularity of (generalized) curvature  invariants implies that
\begin{equation}\label{mathi1}
\r+p \equiv f\dot\f^2 =f W'^2
\end{equation}
is finite.
Regularity of the  Kretschmann invariant in (\ref{w61}), (\ref{w62}) then implies that
\be
\mathcal{I} \equiv f\dot A^2-{e^{-2A}\over R^2}={fW^2\over 4(d-1)^2}-T~~~~{\rm is}~~{\rm also}~~{\rm finite}
\label{w63}\ee

In the case where the scalar is constant, $\dot \f=0$, the only non-trivial condition to satisfy is (\ref{w63}). This is the condition that is relevant at the center of AdS (or dS) where $T\to\infty$ and therefore $f\to\infty$ in a correlated fashion as dictated by (\ref{w63}).

\vskip 1cm
\section{Perturbative solutions I: general considerations and solutions around an ordinary point. } \label{structure}
\vskip 1cm

In this appendix we take a slightly different approach to understanding the local behaviour of the solutions to equations (\ref{f10_1}), (\ref{w55}) and \refeq{w55b}. In particular, these equations may be used to derive a single fourth-order equation satisfied by $W$. We write this equation as
\be
b_0 V+b_1 V'+b_2V''+b_3V^{(3)}=0
   \label{w56_1b}\ee

\noindent
where the functions $b_i$ are given by
\be
b_0=-2 W' \left[4 (d-1)^2 \left(W''^3- W W''^2+ W^{(4)} W'^2+ W W^{(3)} W'\right)-\right.
 \label{w56_2}\ee
$$
\left.-8 (d-1)^2 W^{(3)} W' W''+d(d-2) (W^2 W''- W W'^2)\right],
$$

\be
b_1=-4 (d-1)^3 W''^4+(d-1) (d^2-2d-4) W^2 W''^2-d(d-1)(d-2) W'^4+
   \label{w56_3}\ee
$$
+4(2d-3)(d-1)^2 W^{(3)} W'^3-12 (d-1)^3 (W^{(3)})^2 W'^2-4(2d-3)(d-1)^2 W'^2 W''^2+
$$
$$
8 (d-1)^3 W^{(4)} W'^2 W''+8 (d-1)^3 W^{(3)} W'W''^2-8 (d-1)^2 W
   W''^3+d(d-2) W^3 W''-
$$
$$
-d(d-2) W^2 W'^2+
4 (d-1)^2 W W^{(4)} W'^2+4 (d-1)^2 W^2 W^{(3)} W'-
$$
$$
-
4(d-1)(d-2) W W'^2 W''+4 (d-1)^2 W W^{(3)} W' W'',
$$

   \be
   b_2=-(d-1) W' \left[-4 (d-1)^2 W''^3+(d^2-4)( W^2 W''- W W'^2)+4 (d-1)^2 W^{(4)} W'^2-\right.
   \label{w56_4}\ee
$$-
 \left.4 (d-1)(d-2) W'^2 W''-8 (d-1) W W''^2+4 d(d-1) W W^{(3)}
   W' \right],
$$

and

   \be
   b_3=-2 (d-1)^2 W'^2 \left[2 (d-1) W''^2+(2-d) W W''+(d-2) W'^2-2 (d-1) W^{(3)} W'\right].
\label{w56_5}\ee

It will often be useful to examine the local behaviour of the functions $f$ and $T$ as well. For a given $V$, these can be written in terms of $W$ and its derivatives:
\be
f=-\frac{2 (d-1) \left(V'' W'-2 V' W''\right)-2W V'+4 V W'}{W' \left(2 (d-1) (W'')^2+(2-d) W W''+(d-2) W'^2-2 (d-1) W^{(3)} W'\right)}
    \label{w56_6} \ee
and
   \be
   T= \frac{ 2 (d-1) W''^2+d W'^2-2 (d-1) W^{(3)} W'- d W  W''}{ (d-2) (d-1)  \left(2 (d-1) (W'')^2+(2-d) W W''+(d-2) W'^2-2 (d-1) W^{(3)} W'\right)}V+
   \label{w56_7} \ee
 $$
 +\frac{d W^2 W''-4 (d-1) W'^2 W''+W \left(2 (d-1) W''^2-d W'^2+2 (d-1) W^{(3)} W'\right)}{2 (d-2) (d-1) W' \left(2 (d-1) (W'')^2+(2-d) W W''+(d-2) W'^2-2 (d-1) W^{(3)} W'\right)}V'+
 $$
    $$
    +\frac{W'^2-W W''}{(d-2) \left(W'' \left(2 (d-1) W''-(d-2) W\right)+(d-2) W'^2-2 (d-1) W^{(3)} W'\right)}V''.
$$
The previous approach is valid so long as the denominator of Eqs. \eqref{w56_6} and \eqref{w56_7} does not vanish. The exceptional case in which such a denominator vanishes is extensively discussed in Appendix \ref{app:I}.

To investigate the solutions to equation \refeq{w56_1b}, we must study their properties both in the vicinity of  ``ordinary points'' in scalar field space, as well as near ``singular points''. For a specified scalar potential $V$, \refeq{w56_1b} is a fourth order non-linear differential equation for the superpotential $W$. We shall call ``ordinary points" those around which the superpotential is an analytic function of $\f$.

``Singular points'' of non-linear differential equations can be movable or fixed. Movable singularities are sensitive to the boundary conditions imposed on the solution to the differential equation, whereas the location of fixed singular points are determined by the equation alone.

In particular, the singular points of the superpotential equation (\ref{w56_1b})  correspond to points where the coefficient of the leading derivative $W^{(4)}$ vanishes.
This coefficient is
\be
W'^2((d-1) V''W'-2(d-1) V' W''-W V'+2 V W').\label{WEQsp}
\ee
We observe  that one class of singular points are given by the extrema of $W$---those points at which
\be
W'=0.\label{WXp}
\ee
Below, we  show that singular points in this class may correspond to:
\begin{itemize}
\item Extrema of $V$.

\item Points where the spatial sphere foliating the geometry shrinks smoothly to a point (``shrinking endpoints'').

\item Points around which the scalar trajectory changes direction in field space--i.e. ``bouncing points'' where $\dot\f=0,\ddot\f\neq0$.
\end{itemize}
The properties of solutions in this class of singular points are explored in appendix \ref{WcApp}.

Another class of singular points are those points for which
\be
(d-1) V''W'-2(d-1) V' W''-W V'+2 V W'=0
\ee
From (\ref{w56_6}) we observe that at such points $f$ vanishes, and we shall later deduce that these singular points correspond to:

\begin{itemize}
\item Horizons. This class of solutions is examined in appendix \ref{sho}
\item Points where $V=V'=V''=0$. These ``Minkowski points''---named in reference to the vanishing potential---are described in detail in Appendix \ref{exth} and shall not be discussed further here.
\end{itemize}

In what follows, we shall first study the solutions to \refeq{w56_1b} around a generic, i.e. ordinary, point and then turn our attention to the properties of solutions near the singular points introduced above. In doing so, we assume that the scalar potential $V$ is analytic in $\f$, (for finite values of $\f$), and can therefore be expanded about {\it any} point (which we take without loss of generality to be $\f=0$ via a shift of $\f$)  as a Taylor series in $\f$:
\be
V(\f)=\sum_{n=0}^{\infty}{V_n\over n!}\f^n.\label{genVX}
\ee
This assumption is motivated by the form of the scalar potentials that arise in supergravity theories which descend from string/M-theory in higher dimensions.

Determining the generic local behaviour of the superpotentials that solve \refeq{w56_1b} is a complicated task which we shall not presently attempt. For the solutions discussed in this work, we find that an expansion of the form
\be
W(\f)=\sum_{n=0}^{\infty}{\f^n\over n!}\left(W_{n} +\hat{W}_{n+\alpha}\f^\alpha+\tilde{W}_{n}\f^\alpha\log \f\right)+\ldots\label{genWX}
\ee
can be used to describe the {\it leading} behaviour of the solution near both ordinary and the singular points of interest\footnote{There is single exception to this parametrization and this involves nearly-marginal cases and is  discussed in more detail in appendix \ref{Marginal}.}.
At higher orders, generically one expects terms non-linear in the functions of $\f$ multiplying $W_n$, $\hat{W}_{n+\alpha}$ and $\tilde{W}_n$ to enter the expansion. All such terms, together with other sub-leading non-analyticities, are included in the ellipses in \refeq{genWX}. Generically speaking these are resurgent expansions.

The exponent $\alpha$ appearing in \refeq{genWX} allows for the possibility of non-analytic behaviour in the leading form of the superpotential near a singular point, as a putative solution to  the indicial equation in the method of Frobenius.

In general, a solution $W$ to the master equation \eqref{w56_1b} has four integration constants. It is possible that the expansion \eqref{genWX} does not capture all of them, in which case the missing integration constants come along with the non-analyticities in the ellipses in \eqref{genWX}. Since the constants are arbitrary, we can assume that they are small and linearize \eqref{w56_1b} around a given solution $W_b$. In particular, we take $W= W_b + \delta W$ and obtain $\delta W$ from the linearized version of \eqref{w56_1b}:

\begin{equation}\label{eqpe}
c_0 \delta W + c_1 \delta W' + c_2 \delta W'' + c_3 \delta W^{(3)} + c_4 \delta W^{(4)}=0\,,
\end{equation}
where the coefficients $c_i$ depend on the background solution $W_b$ as well as on the potential $V$. We shall provide the explicit form of \eqref{eqpe} in the particular cases where it is needed. We shall similarly write $f=f_b +\delta f$ and $T = T_b+\delta T$, where the first contributions in both of them are obtained from Eqs. \eqref{w56_6} and \eqref{w56_7} evaluated with $W_b$, while the corrections $\delta f$ and $\delta T$ are computed from the same equations linearised around $\delta W$.

\vskip 1cm

\subsection{Solutions around an ordinary point}\label{ord}

\vskip 1cm

We begin our analysis by studying the solution around an ordinary point, which we place at $\f=0$ with a shift of $\f$. For this, we make the ansatz $\alpha = \tilde{W}_n = 0$ in \refeq{genWX}, such that

\be
W(\f)=\sum_{n=0}^{\infty}{W_n\over n!}\f^n\,.
\label{C1b}\ee

Now $W_0,W_1,W_2,W_3$ are constants of integration in the solution to the fourth order equation (\ref{w56_1b}), and the first coefficient in the expansion of the superpotential that is fixed by the equation is
\be
W_4={d(d-2)(W_0^2W_2-W_0W_1^2)+4(d-1)^2(W_2^3-W_0^2W_2+W_0W_1W_3-2W_1W_2W_3)\over 4(d-1)^2W_1^2( 2 V_0 W_1 -V_1 W_0 - 2 (d-1) V_1 W_2+ (d-1) V_2 W_1 )}V_0+
\label{C9}\ee
$$
+{d(d-2)(W_0^3W_2-W_0^2W_1^2)-d(d-1)(d-2)W_1^4-4(d-1)(d-2)W_0W_1W_2\over 4(d-1)^2W_1^2( 2 V_0 W_1 -V_1 W_0 - 2 (d-1) V_1 W_2+ (d-1) V_2 W_1 )}V_1+
$$
$$
+{ (d^2-2d-4) W_0^2 W_2^2 -4 (d-1) (2d-3)W_1^2W_2^2-8(d-1)W_0W_2^3-4(d-1)^2 W_2^4
\over 4(d-1)W_1^2( 2 V_0 W_1 -V_1 W_0 - 2 (d-1) V_1 W_2+ (d-1) V_2 W_1 )}V_1+
$$
$$
+{4(d-1)^2 W^2_0W_1W_3+4(2d-3)(d-1)^2W_1^3W_3+4(d-1)^2 W_0W_1W_2W_3
\over 4(d-1)^2W_1^2( 2 V_0 W_1 -V_1 W_0 - 2 (d-1) V_1 W_2+ (d-1) V_2 W_1 )}V_1+
$$
$$+
{8(d-1)^3 W_1W_2^2W_3-12(d-1)^3 W_1^2W_3^2
\over 4(d-1)^2W_1^2( 2 V_0 W_1 -V_1 W_0 - 2 (d-1) V_1 W_2+ (d-1) V_2 W_1 )}V_1+
$$
$$
{(d^2-4)(W_0W_1^2-W_0^2W_2)+4(d-1)(d-2)W_1^2W_2+8(d-1)W_0W_2^2+4(d-1)^2W_2^3
\over 4(d-1)W_1( 2 V_0 W_1 -V_1 W_0 - 2 (d-1) V_1 W_2+ (d-1) V_2 W_1 )}V_2
$$
$$-{dW_0W_3\over ( 2 V_0 W_1 -V_1 W_0 - 2 (d-1) V_1 W_2+ (d-1) V_2 W_1 )}V_2-
$$
$$
-{(d-2) W_1^2 -(d-2) W_0W_2 + 2 (d-1) W_2^2 - 2 (d-1) W_1 W_3
\over 2( 2 V_0 W_1 -V_1 W_0 - 2 (d-1) V_1 W_2+ (d-1) V_2 W_1 )}V_3
$$

Using (\ref{w56_6}) and (\ref{w56_7}) together with the expansion of $W$ around an ordinary point, we can also compute the expansion coefficients of $f$ and $T$ locally. Expanding
\be
f=\sum_{n=0}^{\infty}f_n~\f^n\sp T=\sum_{n=0}^{\infty}T_n~\f^n\;.
\label{C100}\ee
we find
\be
f_0=2{V_1 W_0 - (2 V_0 + (d-1) V_2) W_1 + 2(d-1) V_1 W_2\over W_1((d-2)W_1^2 - (d-2)W_0 W_2 + 2(d-1) (W_2^2 -
  W_1 W_3)}
\label{C10}\ee

\be
f_1={(4(d-1)W_1W_2-2dW_0W_1)V_0\over (d-1)W_1^2((d-2)(W_1^2 - W_0 W_2) + 2(d-1)( W_2^2 -  W_1 W_3))}
\label{C11}\ee
$$
+{dW_0^2+(d-1)(d-2)W_1^2+d(d-1)W_0W_2-2(d-1)^2(W_2^2+W_1W_3)
\over (d-1)W_1^2((d-2)(W_1^2 - W_0 W_2) + 2(d-1)( W_2^2 -  W_1 W_3))}V_1
$$
$$
+{2(d-1)^2W_1W_2-d(d-1)W_0W_1
\over (d-1)W_1^2((d-2)(W_1^2 - W_0 W_2) + 2(d-1)( W_2^2 -  W_1 W_3))}V_2
$$

\noindent
in the expansion of $f$, while for $T$ we have
\be
T_0={e^{-2A_0}\over R^2}=-{  (d(W_0 W_2-W_1^2) + 2(d-1)(W_1W_3- W_2^2)
\over (d-1)(d-2) \left[(d-2)(W_1^2 - W_0 W_2) + 2(d-1)( W_2^2 - W_1 W_3)\right]}V_0+
\label{C12}\ee

$$+
{  W_0^2 W_2 -4(d-1) W_1^2 W_2 +
      W_0 (-d W_1^2 + 2(d-1) W_2^2 + 2(d-1) W_1 W_3)\over 2(d-1)(d-2)W_1 \left[(d-2)(W_1^2 - W_0 W_2) + 2(d-1)( W_2^2 - W_1 W_3)\right]}V_1
$$
$$+
{ W_1^2 - W_0 W_2\over (d-2)\left[(d-2)(W_1^2 - W_0 W_2) + 2(d-1)( W_2^2 - W_1 W_3)\right]}V_2
$$
Note that relation (\ref{C12}) can be used to express the integration constant $W_3$
in terms of the value of $T$ at the expansion point, $T_0$ as
 \be
 W_3={(d-2)\left[(d-2)W_1^2-(d-2)W_0W_2+2(d-1)W_2^2\right]\over (V_1 W_0 +2(d-1)(d-2) T_0 W_1 - 2 V_0 W_1)}T_0-
 \label{C101}\ee
 $$
 -{dW_1^2-dW_0W_2+2(d-1)W_2^2\over (d-1) (V_1 W_0 +2(d-1)(d-2) T_0 W_1 - 2 V_0 W_1)}V_0+
$$
$$
+{dW_0W_1^2-dW_0^2W_2+4(d-1)W_1^2W_2-2(d-1)W_0W_2^2\over 2 (d-1) W_1 (V_1 W_0 +2(d-1)(d-2) T_0 W_1 - 2 V_0 W_1)}V_1-
$$
$$
-{W_1^2-W_0W_2\over  (V_1 W_0 +2(d-1)(d-2) T_0 W_1 - 2 V_0 W_1)}V_2
$$

One can further solve for $W_{1,2}$ as a function of $f_{0,1}$ and $T_0$, leading to two branches of solution:

$$
W^{\pm}_1= - \frac{f_1 W_0}{2f_0}\mp
$$
\be
 \frac{\mp\sqrt{(d-1) \left(-8 (d-1) f_0 \left(\left(d-1)(d-2\right) T_0-V_0\right)+2 d f_0^2 W_0^2+(d-1) f_1^2 W_0^2\right)}}{2 (d-1) f_0}
\label{C102}\ee
and
\be
W_2^\pm = \frac{2 (d-2) (d-1) T_0+f_0 (W_1^\pm)^2-2
   V_0}{f_0 W_0}+\frac{V_1}{f_0 W_1^\pm}
\ee

\vskip 1cm

\section{A Fr\"obenious approach to local solutions}\label{exth}

\vskip 1cm

In this appendix we shall  study  the local properties of solutions to (\ref{reda3a})--(\ref{reda2a}) by analyzing a general Fr\"obenius-like ansatz for the system of differential equations.
We expand around an arbitrary {\it finite} point $\f_0$ in scalar field space, that by a shift, we can set to be $\f_0=0$. We then assume the following expansions near $\f=0$:

\be\label{eahb}
V = \sum_{n=0}^\infty V_n \dfrac{\f^n}{n!}\qquad W = \f^\alpha \sum_{n=0}^\infty W_{n} \dfrac{\f^n}{n!} \qquad f = \f^\beta \sum_{n=0}^\infty f_{n} \dfrac{\f^n}{n!} \qquad T = \f^\gamma \sum_{n=0}^\infty T_{n} \dfrac{\f^n}{n!}
\ee
where by definition  $f_{0},W_{0},T_0\not=0$.

To proceed, we must solve equations (\ref{reda1a}), (\ref{reda2a}) for $W,f$ which we reproduce here,
\be\label{reda1}
 \frac{f}{4} \left(\frac{d W^2}{d-1}-2 (W')^2\right)- \frac{W' }{4}\Big((d+2) W f'-2 (d-1) \left(f' W'\right)'\Big)+V=0
 \ee
 \be
{W'}\left[ W'f' + f\left(W''-\frac{d }{2
   (d-1)}W\right)\right]-V' = 0\,,
\label{reda2}
\ee
and then determine $T$ from (\ref{reda3a}):
\be
T={1\over (d-1)(d-2)}\left[\bigg( \frac{d}{4(d-1)} W^2-\frac{W'^2}{2}\bigg) f-\frac{1}{2} W'Wf'+V\right]\,.
\label{reda3aa}\ee
It must be non-negative everywhere for the ansatz with spherical slicing \eqref{c39} that we study in this work.

Some of the solutions appear as particular cases of those described in appendices \ref{structure}, \ref{WcApp} and \ref{sho}. For solutions that are not described elsewhere, we complete the analysis by looking for perturbations around the given solution, in order to complete all possible integration constants. Generically, a solution will contain pieces beyond the ones assumed in equation (\ref{eahb}).
 Writing $W\to W + \delta W\,, f\to f + \delta f$, where $W,f$ are the series in  (\ref{eahb}), substituting into Eqs. \eqref{reda1} and \eqref{reda2} and assuming that the perturbations are small, we find the
 linearized equations obeyed by the perturbations:

$$
\frac{d \left(W \left(f \delta W'+\delta f
   W'\right)+\delta W f W'\right)}{2-2 d}+2
   \delta W' f' W'+W'' \left(f \delta W'+\delta f W'\right)+
$$
\begin{equation}\label{pert1}
+f \delta W'' W'+\delta f' W'^2 = 0\,,
\end{equation}
\begin{align}\label{pert2}
&-W' \left((d+2) \left(\delta W f'+W \delta f' \right)-2 (d-1) \left(\delta W 'f''+\delta W'' f'+\delta f' W''+\delta f'' W'\right)\right)\\&-\delta W' \left((d+2) W f'-2
   (d-1) \left(f'' W'+f' W''\right)\right)+f
   \left(\frac{2 d \delta W W}{d-1}-4 \delta W'
   W'\right)\nonumber \\&+\delta f \left(\frac{d W^2}{d-1}-2 W'^2\right) = 0\nonumber\,.
\end{align}

As mentioned in Sec. \ref{sec5}, the equations above are invariant under the rescaling
\be
W\to \l W\sp f\to {f\over \l^2}\,,
\label{scaling2}\ee
with $T$ invariant under this scaling. This rescaling can be related to the rescaling in the metric (\ref{f22})
\be
u\to {u\over \l}\sp t\to \l t\sp f\to {f\over \l^2}
\ee
that leaves the metric invariant. This scaling parameter $\l$ is always one of the integration constants of the solution and it will be identified with $W_0$, see equation (\ref{eahb}).

\subsection{The associated geometry\label{fgeometry}}

We shall now calculate the geometry of solutions using the expansions
 (\ref{eahb}) near $\f\to 0$,  with $f_0\not=0,W_0\not= 0$ without loss of generality. We use (\ref{w53}) and (\ref{w53b}) to write

\be\label{f77}
{dA\over d\f}=-{1\over 2(d-1)}{W\over W'}.
\ee
Substituting the expansion for $W$ we obtain
\be
{dA\over d\f}=-{1\over 2(d-1)\a}\f+\cdots  \ar A=A_0-{1\over 4(d-1)\a}\f^2 + \dots
\ee
which indicates that as long as $\a\not=0$, the scale factor $e^{A}$ is regular and finite at $\f=0$.

For $\a=0$ we have instead
\be
{dA\over d\f}=-{1\over 2(d-1)}{W_0\over W_1}+\cdots
\ee
where we have assumed that  $W_1\not=0$.
We obtain
\be
A=A_0-{1\over 2(d-1)}{W_0\over W_1}\f+\cdots
\ee
and therefore again  the scale factor $e^A$ is finite at $\f=0$.

If, on the other hand, $\a=0$ and $W_1=0,W_2\not=0$ then,
\be
e^{A}=e^{A_0}\f^{-{W_0\over 2(d-1)W_2}}+\cdots
\ee
and $e^A\to 0$ if $W_0W_2<0$ while $e^A\to \infty$ if $W_0W_2>0$. We also have
 \be
  -(d-1)(d-2)T =   - (V_0 + V_1 \f +{\cal O}(\f^2) ) +\f^{2\a+\b-2}\left[{\a (\a + \b) f_0 W_0^2\over 2} +\right.\ee
    $$\left. + {W_0 (\a (\a + \b+1) f_1 W_0 + (\b+ 2 \a(1+\a+\b))  f_0 W_1)\over
 	2}\f+{\cal O}(\f^2)\right] \,.
$$
 $T$ either asymptotes to a constant or diverges depending on the sign of $2\a+\b-2$.
 When $\a=\b=0$ then

	\be
	-(d-1)(d-2)T=\dfrac{1}{4}\left(-4V_0 - {d f_0 W_0^2\over  d-1} + 2 W_1 (f_1 W_0 + f_0 W_1)\right)\nonumber
	\ee
and is therefore finite.

The other scale factor that controls the size of the time direction is
\be
g_{tt}=-e^{2A}f
\ee

$\bullet$ When $\a\not=0$, or $\a=0$ and $W_1\not=0$, then we found that $e^A$ is finite and therefore $g_{tt}$ is controlled by the exponent $\b$. If $\beta >, =,<0$ then $g_{tt}$ respectively vanishes, is finite, or diverges at $\f=0$.

$\bullet$ When $\a=0$ and $W_1=0$, then
\be
g_{tt}=-e^{2A_0}f_0~\f^{\b-{W_0\over (d-1)W_2}}+\cdots
\ee

To  calculate the Kretchmann scalar from (\ref{w62})
\be
K_2={{d^2-4d+7\over (d-1)^2}V^2}+(d-2)(d-1)^2d \left({fW^2\over 4(d-1)^2}-T\right)^2+
\label{w62a}\ee
$$
+{(d-1)(d-2)\left(2V-fW'^2\right)\left({fW^2\over 4(d-1)^2}-T\right)}-{{d-3\over d-1}V~f~W'^2}+{{d+1\over 4(d-1)}f^2 W'^4}
$$
we need
\be
\left({fW^2\over 4(d-1)^2}-T\right)={\f^{2\a+\b-2}\over 2(d-1)(d-2)}\Big[\a(\a+\b)f_0W_0)^2+
\ee
$$
+
W_0 \Big(\a (\a + \b+1) f_1 W_0 + (\b + 2 \a ( \a + \b+1)) f_0 W_1\Big)\f+{\cal O}(\f^2)\Big]
$$
$$
- \dfrac{1}{(d-1)(d-2)} (-V_0 -V_1 \f +{\cal O}(\f^2))
$$

\be
fW'^2=\f^{2\a+\b-2}\Big[\a^2f_0W_0^2+\a W_0 \Big(\a f_1 W_0 + 2 (\a+1) f_0 W_1\Big)\f+{\cal O}(\f^2)\Big]
\ee

\begin{align}
	&\left({fW^2\over 4(d-1)^2}-T\right)^2 ={\f^{4\a+2\b-4}\over 4(d-1)^2(d-2)^2}\Big[\a^2(\a+\b)^2 f_0^2W_0^4+\nonumber \\
	&+2\a (\a + \b) f_0 W_0^3 \Big(\a ( \a + \b+1) f_1 W_0 + (\b + 2 \a ( \a + \b+1)) f_0 W_1\Big)\f+{\cal O}(\f^2)\Big]\nonumber\\ & \dfrac{1}{(d-2)^2(d-1)^2}\left( V_0^2 + 2 V_0 V_1 \f + \mathcal{O}(\f ^2) \right) + \dfrac{\f^{2\a + \b -2}}{(d-2)^2(d-1)^2}\left[-\a (\a + \b) f_0 V_0 W_0^2 + \right.\nonumber \\ &\left.W_0 \left(\alpha  W_0 \left(f_0 V_1 (-\alpha -\beta )-V_0 (\alpha +\beta +1)
   f_1\right)-f_0 V_0 (2 \alpha  (\alpha +\beta +1)+\beta ) W_1\right) \f  + \mathcal{O}(\f^2)\right]
\end{align}

To leading order, $K_2$ is

 \begin{align}
 	& K_2 = \frac{2 (3 d-5) V_0^2}{(d-2) (d-1)^2} +\f^{2\a + \b -2}\left(-\frac{2 \alpha  f_0 V_0 W_0^2 (\alpha +\beta  (d-1))}{(d-2) (d-1)} + \mathcal{O}(\f)\right) \nonumber\\
 	& + \f ^{4\a + 2\b -4} \left( \frac{\alpha ^2 f_0^2 W_0^4 \left(\alpha ^2 (4 d-6)+4 \alpha  \beta  (d-1)+\beta ^2
   (d-1) d\right)}{4 (d-2) (d-1)} + \mathcal{O}(\f)\right)
 \end{align}
 When $\a=\b=0$ the leading contribution to $K_2$ is

 \begin{align}
 & K_2 =\frac{8 (d-1)
   f_0 V_0 W_0^2+d f_0^2 W_0^4+8 (3 d-5) V_0^2}{4 (d-2) (d-1)^2} - \frac{W_0 f_1 W_1 \left(d f_0 W_0^2+4
   (d-1) V_0\right)}{2 (d-2) (d-1)} + \nonumber\\ & \frac{(1-d) W_1^2 \left(8 f_0 V_0-W_0^2 \left((d-1) d f_1^2-4
   f_0^2\right)\right)}{4 (d-2) (d-1)^2} +\frac{f_0 W_0 f_1 W_1^3}{d-2}+\frac{(2 d-3)
   f_0^2 W_1^4}{2 (d-2) (d-1)} + O(\f)
 \end{align}

\subsection{Classification of the  solutions of the first order equations}

Substituting the expansions \eqref{eahb} into Eq. \eqref{reda1}, \eqref{reda2} we obtain,

\be\label{redb}
\f ^{2 \alpha +\beta } \left[{p_0\over \f^4}+{p_1\over \f^3}+{p_2\over \f^2}+{\cal O}(\f^{-1})\right]+V_0+V_1\f+{\cal O}(\f^2) = 0
\ee
\be\label{redb1}
\f ^{2 \alpha +\beta } \left[{q_0\over \f^3}+{q_1\over \f^2}+{\cal O}(\f^{-1})\right]-V_1+V_2\f+{\cal O}(\f^2) = 0
\ee
with
\be
p_0=\frac{(d-1) \alpha ^2 \beta  (\alpha +\beta -2) f_{0}
   W_{0}^2}{2}
   \ee
   \be
     p_1={(d-1)\a W_0\over 2}\Big[\a(\b+1)(\a+\b-1)f_1W_0+\b(\a+1)(2\a+2\b-3)f_0W_1\Big]
   \ee
   \be
   p_2=-{\a\over 4}(2\a + (d + 2)\b) f_0W_0^2+{(d-1)\b(\a+1)^2(\a+\b-1)\over 2}f_0W_1^2+
   \ee
   $$
   +{(d-1)\a ( \a+2) \b ( \a + \b-1) \over 2}f_0W_0W_2+{(d-1)\a (\a+1) (\b+1) (2\a+2\b-1)\over 2}f_1W_0W_1+
   $$
   $$
   +{(d-1)\a^2 (\b+2) (\a + \b)\over 4}f_2 W_0^2
   $$
\be
q_0={\alpha ^2 (\alpha +\beta -1) f_0 W_0^2}\sp q_1=\a W_0\left[\a (\a + \b) f_1 W_0+(\a+1) (2 \a + 2 \b-1) f_0 W_1\right]
\ee

It is evident that if $2\a+\b\notin {\mathbb Z}$ then the potential must vanish, $V=0$. As we are interested in non-trivial potentials, we henceforth turn our attention to $2\a+\b\in {\mathbb Z}$ .

Below, we consider the possible cases:

\vskip 1cm
$\bullet$  $2\a+\b < -1$.
\vskip 1cm

Setting $p_0=p_1=p_2=p_3=p_4=q_0=q_1=q_2=q_3=0$ solves the equations \eqref{redb} and \eqref{redb1} before the potential kicks in. We obtain two possible solutions
\be
\a=0\sp \b=-n\sp W_1=0\sp W_2={W_0\over \b(d-1)}\sp n> 1
\ee
and
\be
\a=0\sp \b=-n\sp W_1=0\sp W_2={dW_0\over 2\b(d-1)}\sp n> 1
\ee
However, checking the higher order terms we find  that there are no solutions with $W_0\not=0$.

\vskip 1cm
$\bullet$  $2\a+\b = -1$.
\vskip 1cm

 In this case, setting $p_0=p_1=p_2=p_3=p_4=q_0=q_1=q_2=q_3=0$ solves the equations \eqref{redb} and \eqref{redb1} before the potential kicks in, and we obtain two possible solutions. The first one has

\be\label{shr2}
\a=0\sp \b=-1\,,
\ee
and the first few coefficients are given by

\be
 W_1=0\sp W_2=-{W_0\over (d-1)}
\sp W_3 =-{(2 (d+2) V_0 + d(d-1) V_2) W_0\over (d-1)^2 (d+2) V_1}\,,
\ee

\be
f_0 = {2 (d-1)^2 V_1\over d W_0^2}\sp
 f_1 = {(d-1) (-2 (d+2) V_0 + d(d-1)V_2))\over d (d+2) W_0^2}\,.
 \ee
It has a single integration constant: $W_0$.
From Eq. \eqref{reda3aa} we also have
\be
\g=-1\sp T_0={V_1\over 2 d}\sp  T_1 = {2 (d + 2) V_0 + d (d - 1) V_2\over
 4 d(d-1) (d+2)}
 \ee
which indicates that the scale factor shrinks to zero in this solution, while the sign of the slice curvature is the same as that of $V_1$.
Therefore in our case we must have $\f V_1>0$ which implies that the solution exists to the right or the left of that point, depending on the sign of $V_1$.
From Eq. \eqref{f77} we find

\begin{equation}
A = \dfrac{1}{2}\log \f + A_0 + O(\f)\,,
\end{equation}
where $A_0$ is an integration constant. The previous equation implies that the radius of the thermal circle approaches a constant value:

\begin{equation}
g_{tt} = -f e^{2A} = - \dfrac{f_0 }{\f} e^{2A_0}\f + \dots = - f_0 e^{2A_0} + O(\f)
\end{equation}

The Kretschmann invariant for this solution approaches a constant value:

\begin{equation}
K_2 = \frac{2 (d+1) V_0^2}{(d-1)^2 d} +\frac{4 (d (d+2)-1) V_1 V_0  }{(d-1)^2 d^2}\f +\mathcal{O}(\f^2)\,.
\end{equation}

Finally, since $W_1 = 0$, the previous solution corresponds to an endpoint of the flow. This is the known \textbf{shrinking endpoint} where the sphere smoothly shrinks to zero size (see  Appendix \ref{G.2.2}).

The other solution is given by the following first few coefficients,
\be
\a=0\sp \b=-1\sp W_1=0\sp W_2=-{dW_0\over 2(d-1)}\sp W_3 =- {d ((d-1)V_2 +2 d  V_0) W_0\over 4 (d-1)^2 V_1}
\ee
\be
f_0= {4 (d-1)^2 V_1\over d^2 W_0^2}\sp
 f_1= -{(d-1) (2 d V_0 -(d-1) V_2)\over d^2 W_0^2}
 \ee
It has a single integration constant, $W_0$.
For this solution $T$ vanishes  order by order.
When $T=0$, then the other functions satisfy
\be
f'W'=e^{{d\over 2(d-1)}\int_{\f_0}^{\f} {W\over W'}d\f}
\ee
This is compatible with the solution above with
\be
\f_0={2(d-1)V_1\over dW_0}
\ee

This solution has a diverging Kretschmann scalar:

\begin{equation}
K_2 =\frac{(d-2) d f_0^2 W_0^4}{16 (d-1)^2 \f ^2} + O(\f^{-1})\,.
\end{equation}

This solution was encountered in the last item of Sec. \ref{G.2.2} and it is concluded that the solution is singular and therefore not acceptable.

\vskip 1cm
$\bullet$ $2\a+\b= 0$
\vskip 1cm

In this case, setting $p_0=p_1=p_2=p_3=q_0=q_1=q_2=0$  solves the equations \eqref{redb} and \eqref{redb1} up to $O(\f^{-1})$ before the potential kicks in, we obtain only one solution, the trivial one:
\be\label{ab0}
\a=\b=0\,.
\ee

Here, the scale factor approaches a constant value:
\be
\gamma=0\sp T_0={4(d-1)V_0+df_0W_0^2-2(d-1)W_1(f_1W_0+f_0W_1)\over 4(d-1)^2(d-2)}\,.
\ee

This is a particular case of the ansatz for the expansion of the superpotential given in Eq. \eqref{genWX}. It has four integration constants: $f_0,f_1,W_0,W_1$. From Eq. \eqref{f77} we find

\begin{equation}
A = A_0 + \dfrac{W_0}{2(d-1)W_1}\f +O(\f^2)\,,
\end{equation}
where we have temporarily assumed that $W_1\neq 0$. Consequently, the $g_{tt}$ factor of the metric approaches a constant value

\begin{equation}
g_{tt} =  -e^{2A_0}f_0\,.
\end{equation}

The solution is regular and the Kretschmann scalar approaches a constant value:

\begin{align}
 & K_2 =\frac{8 (d-1)
   f_0 V_0 W_0^2+d f_0^2 W_0^4+8 (3 d-5) V_0^2}{4 (d-2) (d-1)^2} - \frac{W_0 f_1 W_1 \left(d f_0 W_0^2+4
   (d-1) V_0\right)}{2 (d-2) (d-1)} + \nonumber\\ & \frac{(1-d) W_1^2 \left(8 f_0 V_0-W_0^2 \left((d-1) d f_1^2-4
   f_0^2\right)\right)}{4 (d-2) (d-1)^2} +\frac{f_0 W_0 f_1 W_1^3}{d-2}++\frac{(2 d-3)
   f_0^2 W_1^4}{2 (d-2) (d-1)} + O(\f)
 \end{align}

This case corresponds to the expansion about a \textbf{regular point}. Finally, for $W_1=0$ the solution matches the expansion around a singular point studied in  Appendix \ref{structure} corresponding to \textbf{dS$^{(d+1)}$ boundaries} and \textbf{AdS$^{(d+1)}$ boundaries}.

\vskip 1cm
$\bullet$ $2\a+\b= 1$
\vskip 1cm

In this case, setting $p_0=p_1=p_2=q_0=q_1=0$  solves the equations \eqref{redb} and \eqref{redb1} up to $O(\f^{-1})$ before the potential kicks in, and we obtain three branches of solutions with

\be \label{e48e}
\a=0\sp \b=1\,.
\ee

\textbf{Case 1:} $W_1=0$.

These solutions exist for $V_0=V_1=V_2=0$. There are two subcases:

\begin{equation}\label{e2c}
 W_2=\dfrac{W_0}{d-1} \ \ \textrm{and} \ \ W_2=\dfrac{dW_0}{2(d-1)}\,.
\end{equation}

\textbf{Case 1. Branch 1:} $W_2=\dfrac{W_0}{d-1}$.
The few subleading coefficients in the expansion are found to be\footnote{Notably, there is an exception for $d=4$ dimensions. In such a case, we need require $V_3=0$ as well. The first coefficient of the solution is now given by
\begin{equation}
W_3=\frac{3 V_4-2 W_0^2 f_1}{9 f_0 W_0}\,,
\end{equation}
while $f_0, f_1$ and $W_0$ are integration constants. We further find $\gamma=1$ with
\begin{equation}\label{51e}
T_0 = \dfrac{1}{36} f_0 W_0^2\,.
\end{equation}
}

\begin{equation}\label{e5e}
f_0 = -\frac{(d-1)^2 V_3}{(d-4) W_0^2} \sp W_3 = -\frac{(d-4) V_4 W_0}{6 (d-1) V_3}\, \sp f_1 = -\frac{(d-1)^2 V_4}{12 W_0^2}
\end{equation}
while the scale factor vanishes to leading order as

\begin{equation}\label{e6e}
\g = 1\sp T_0=-\dfrac{V_3}{4(d-4)}
\end{equation}
{We must have $\f V_3<0$, which implies that the solution always exists to the left or the right depending on the sign of $V_3$.}
This branch of solutions has two integration constants: $W_0$ and $f_1$. From equation \eqref{f77} we find

\begin{equation}
A = -\dfrac{W_0}{2(d-1)W_2}\log \f + A_0 + O(\f) = -\dfrac{1}{2}\log \f +A_0 + O(\f)\Rightarrow e^{2A} = e^{2A_0}\f^{-1} + O(\f)
\end{equation}
The fact that $f$ vanishes as we approach this solution is compensated by the divergence of the scale factor so that the temporal component of the metric asymptotes to a constant value:

\begin{equation}\label{5e5}
g_{tt} = - e^{2A}f = -e^{2A_0}f_0 + O(\f)
\end{equation}

The geometry is regular, since all the functions appearing in the curvature invariants \eqref{w58}-\eqref{w62} are finite. In particular, the invariants vanish to leading order, e.g. the Kretschmann invariant to leading order is given by

\begin{equation}
K_2 = \frac{\left(d^4-7 d^3+5 d^2-23 d+474\right) V_3^2 }{18 (d-6)^2 (d-4)^2 (d-1)^2}\f ^6 + O(\f^7)
\end{equation}
The pressure, energy density and quantity $\mathcal{I}$ controlling the curvature invariants (see Appendix \ref{sect:inv_sphere}) evaluate to \footnote{
Again, there is an exception for $d=4$. In such case, the quantities controlling the curvature invariants are given, to leading order, by
\begin{equation}\label{invMink14}
\rho = \frac{1}{18} f_0 W_0^2 \varphi ^3 + \dots \sp p = \frac{1}{18} f_0 W_0^2 \varphi ^3+ \dots
\end{equation}
\begin{equation}\label{invMink24}
\mathcal{I} = \frac{1}{216}  \left(4 f_1 W_0^2+3 V_4\right)\varphi ^2+ \dots
\end{equation}
}
\begin{equation}\label{invMink1}
\rho = -\frac{(d-1) V_3}{6 (d-4)}\varphi ^3 + \dots \sp p = \frac{(d-7) V_3}{6 (d-4)}\varphi ^3 + \dots
\end{equation}
\begin{equation}\label{invMink2}
\mathcal{I} = -\frac{(d-7) V_3 }{6 (d-6) (d-4) (d-1)}\varphi ^3 + \dots
\end{equation}
Note that the leading contribution to the energy density $\rho$ is proportional to $T_0$ in Eq. \eqref{e6e}. Therefore, for the spherical slicing ($T>0$), the energy density $\rho$ increases from zero as we depart from this solution.

We look for deformations around the given solution by solving Eqs. \eqref{pert1} and \eqref{pert2}. To leading order in $\f$ we find

\begin{equation}
\delta W = C_0 + C_1\f ^{\frac{1}{4} \left(-\sqrt{d^2-20 d+68}+d-2\right)+1}+C_2\f ^{\frac{1}{4}
   \left(\sqrt{d^2-20 d+68}+d-2\right)+1}+C_3\f ^{\frac{d}{2}+1}+\dots
\end{equation}
We set $C_0=0$ because the constant term is not subleading with respect to the unperturbed solution. The term proportional to $C_3$ is subleading provided that $d>2$. On the other hand,  whether the terms proportional to $C_1$ and $C_2$ are subleading with respect to the unperturbed solution depends on $d$. If $d < 4$ then only $C_2$ is subleading and we have to set $C_1=0$. If $4< d <6$ both $C_1=C_2=0$, while if $6<d$ both deformations are allowed. Whenever the exponents are complex, a real solution can be constructed by appropriately combining $C_1$ and $C_2$ (for an example, see equations \eqref{im1} and \eqref{im2}). All in all, this branch of solutions can have up to four integration constants: $W_0,C_1,C_2,C_3$.

Finally, the perturbation for $f$ is found to be
\begin{align}
&\delta f = C_1 \frac{(d-1)^3 \left(\sqrt{d^2-20 d+68}+2 d-10\right) V_3 \f ^{\frac{1}{4}
   \left(-\sqrt{d^2-20 d+68}+d-2\right)}}{(d-4)^2 W_0^3} +C_3 \frac{3 (d-1)^3 V_3 \f ^{d/2}}{(d-4) W_0^3} \nonumber\\ & -C_2\frac{(d-1)^3 \left(\sqrt{d^2-20 d+68}-2 d+10\right) V_3 \f ^{\frac{1}{4}
   \left(\sqrt{d^2-20 d+68}+d-2\right)}}{(d-4)^2 W_0^3}+ \dots
\end{align}
while from \eqref{reda3aa} we obtain

\begin{align}\label{e58}
&\delta T = C_1 \frac{(d-1) \left(\sqrt{d^2-20 d+68}+2 d-10\right) V_3 \f ^{\frac{1}{4}
   \left(d-\sqrt{d^2-20 d+68}\right)-\frac{1}{2}}}{4 (d-4)^2 W_0}  +C_3\frac{(d-1) V_3 \f ^{d/2}}{2 (d-4) (d-2) W_0} \nonumber \\
   &-C_2 \frac{(d-1) \left(\sqrt{d^2-20 d+68}-2 d+10\right) V_3 \f ^{\frac{1}{4}
   \left(\sqrt{d^2-20 d+68}+d\right)-\frac{1}{2}}}{4 (d-4)^2 W_0}+\dots
\end{align}

The solution is asymptotically flat. On top of that, the $S^{(d-1)}$ asymptotes to infinite size and we can identify this asymptotics with the spatial boundary of Minkowski space-time. Since $W_1=0$, this constitutes a possible endpoint of the flow.
\vskip 1cm

\textbf{Case 1. Branch 2:} $W_2=\dfrac{dW_0}{2(d-1)}$.
The next few coefficients of the expansion are given by

\begin{equation}
f_0 = \frac{2 (d-1)^2 V_3}{d^2 W_0^2} \sp W_3 =\frac{d V_4 W_0}{6 (d-1) V_3}\sp f_1 = -\frac{(d-1)^2 V_4}{3 d^2 W_0^2}
\end{equation}
The inverse scale factor $T$ obtained from \eqref{reda3aa} vanishes order by order in the expansion. The solution has one integration constant: $W_0$. From equation \eqref{f77}, we compute the metric function $A$ for this solution:

\begin{equation}\label{60a}
A = -\frac{W_0}{2(d-1)W_2}\log \f + A_0 +O(\f)= -\frac{1}{d}\log \f + A_0+O(\f)\ \Rightarrow e^{2A} = e^{2A_0}\f^{-2/d}+\dots \,.
\end{equation}

Contrary to the solution found in the previous branch, in this case the temporal component of the metric vanishes (for $d>2$), signalling the presence of a horizon:

\begin{equation}
g_{tt} = -f e^{2A} = -f_0 e^{2A_0}\f^{1-2/d}+\dots
\end{equation}

The solution is regular and the Kretschmann invariant is given by

\begin{equation}
K_2 = \frac{(d-2) (d-1)^2 V_3^2 }{4 d^3} \f ^2+ O(\f^3)
\end{equation}
and vanishes as $\f\to 0$.

We now solve equations \eqref{pert1} and \eqref{pert2} that give us the deformations around the given solution. To leading order, we obtain the perturbation for the superpotential $\delta W$

\begin{equation}
\delta W = C_0 + C_1\f^{1+2/d} + C_2 \f^{1-\sqrt{2}} + C_3\f^{1+\sqrt{2}} + \dots
\end{equation}

The terms proportional to $C_0$ and $C_2$ are not subleading with respect to the unperturbed solution, and the same is true for $C_1$ provided that $d\geq 2$. Therefore, we set $C_0=C_1=C_2=0$ for consistency. Conversely, the deformation proportional to $C_3$ is always subleading and hence allowed. The deformations for $f$ and $T$ are obtained again from equations \eqref{pert1}, \eqref{pert2} combined with \eqref{reda3aa}:

\begin{equation}
\delta f = -\frac{4 \left(3+2 \sqrt{2}\right) C_3 (d-1)^3 V_3 \f^{\sqrt{2}}}{d^3 W_0^3}+\dots
\end{equation}

\begin{equation}\label{65dt}
\delta T = 0
\end{equation}

The fact that $T$ vanishes identically means that this solution only appears for the flat sliced ansatz \eqref{c38}.

We conclude that there are two integration constant for this solution: $W_0,C_3$. The geometry of the solution in this limit is that of a horizon with infinite size, as the volume of spatial slices diverges. Again, this solution has $W_1=0$ while $W_2$ is finite, so this is a possible endpoint of the flow.

\vskip 1cm

\textbf{Case 2:} $W_1\neq0$
\be\label{neh}
 W_2 = -{ W_0\over 2(d-1)}+{  (2 V_0  +(d-1) V_2) W_1\over 2(d-1)V_1}
 \ee
\be
f_0 = {V_1\over W_1^2}\sp
 f_1= {(d+3) V_1 W_0 - (6 V_0 +(d-1) V_2) W_1\over 4 (d-1) W_1^3}
 \ee
while the scale factor approaches a constant value
\be
\g=0\sp T_0={V_0 - {V_1 W_0\over 2 W_1}\over (d-1)(d-2)}
\ee
We must require that $T_0>0$ which implies

\begin{equation}
\dfrac{W_0}{V_1 W_1} < 2 V_0\,.
\end{equation}

This solution has two integration constants, $W_0,W_1$ and does not exist at an extremum of the potential where $V_1=0$.
According to Eq. \eqref{f77}:

\begin{equation}
A = A_0 - \dfrac{W_0}{2(d-1)W_1}\f +O(\f^2)
\end{equation}
and the $tt$ component of the metric vanishes linearly in $\f$:

\begin{equation}
g_{tt} = -e^{2A}f = -e^{2 A_0}f_0 \f\,,
\end{equation}
signalling the presence of a horizon.
The geometry is regular around this solution and the Kretschmann scalar approaches a constant value:

\begin{equation}
K_2 = \dfrac{1}{4(d-2)}\left(\frac{d V_1^2 W_0^2}{W_1^2}+\frac{8 (3 d-5) V_0^2}{(d-1)^2}-\frac{8 V_1 V_0
   W_0}{W_1}\right) +\mathcal{O}(\f)
\end{equation}
This solution matches the \textbf{non-extremal horizons} described in Appendix \ref{sho}.

\vskip 1cm
$\bullet$ $2\a+\b= 2$.
\vskip 1cm

In this case, setting $p_0=p_1=q_0=0$ solves the equations \eqref{redb} and \eqref{redb1} up to $O(\f^{-1})$ before the potential kicks in, and we obtain two  solutions for $\a$ and $\b$ and several different possibilities within one of them. The first solution is

\begin{equation}
\a=0\sp \b=2
\end{equation}

There are three non-equivalent possibilities, all of which require an extremum of the potential $V_1=0$ as well as the following fine-tuning:

\be
2V_0+(d-1)V_2=0\;.
\ee

\textbf{Case 1:} $W_1=0$

In this case, the fine-tuning condition for the potential is more strict, and the solution appears only for $V_0 =V_1= V_2 =V_3= 0$. There are two subcases:

\begin{equation}\label{e2cb}
 W_2=\dfrac{W_0}{2(d-1)} \ \ \textrm{and} \ \ W_2=\dfrac{dW_0}{4(d-1)}\,.
\end{equation}

\textbf{Case 1. Branch 1:} $W_2 = \dfrac{W_0}{2 ( d-1)}$

The following coefficients are found to be

\begin{equation}\label{e76}
f_0 = -\frac{2 (d-1)^2 V_4}{3 (d-3) W_0^2} \sp W_3 = -\frac{(d-3) V_5 W_0}{4 (d-1)^2 V_4}\sp f_1 =-\frac{(d-1) V_5}{3 W_0^2}\,.
\end{equation}
From Eq. \eqref{reda3aa} we find that the inverse scale factor $T$ vanishes to leading order as

\begin{equation}\label{j83}
\g = 2 \sp T_0 = -\frac{V_4}{6(d-3)}
\end{equation}
We must have $V_4<0$.
There is a single integration constant, $W_0$.

In order to understand the geometry, we compute the metric components $e^{2A}$ and $g_{tt}$. From equation \eqref{f77} we have

$$
A = -\dfrac{W_0}{2(d-1)W_2}\log\f + A_0 + O(\f) = -\log\f + A_0 + O(\f) \Rightarrow
$$
\begin{equation}
 \Rightarrow e^{2A} = e^{2A_0}\f^{-2} + O(\f^{-2})\,.
\end{equation}
As a consequence, $g_{tt}$ asymptotes to a constant value in spite of the vanishing of $f$:

\begin{equation}
g_{tt} = -f e^{2A} = -e^{2A_0}f_0 + O(\f)
\end{equation}
The geometry is regular and the Kretschmann scalar vanishes as

\begin{equation}
K_2= \frac{\left(d^4-9 d^3+39 d^2-103 d+144\right) V_4^2 }{288 (d-4)^2 (d-3)^2
   (d-1)^2}\f^8 +O(\f^9)
\end{equation}
The pressure, energy density and quantity $\mathcal{I}$ controlling the curvature invariant also vanish:
\begin{equation}\label{invMink21}
\rho = -\frac{(d-1) V_4} {24 (d-3)}\varphi ^4 + \dots \sp p = \frac{(d-5) V_4}{24 (d-3)}\varphi ^4 + \dots
\end{equation}
\begin{equation}\label{invMink22}
\mathcal{I} =-\frac{(d-5) V_4}{24 (d-4) (d-3) (d-1)}\varphi^4+ \dots
\end{equation}
Note that the leading contribution to the energy density $\rho$ is proportional to $T_0$ in Eq. \eqref{j83}. Therefore, for the spherical slicing ($T>0$), the energy density $\rho$ increases from zero as we depart from this solution.

The solution is asymptotically flat. Since the sphere factor of the metric $S^{(d-1)}$ asymptotes to infinite size, we can identify this asymptotics with the spatial boundary of Minkowski space-time. In addition, this is a possible endpoint of the flow because $W_1=0$ while $W_2$ is finite.

We now look for perturbations around the previous solution by solving equation \eqref{pert1} and \eqref{pert2}. To leading order we find for $\delta W$:

\begin{equation}
\delta W =C_0 + C_1 \f ^{\frac{1}{2} \left(-\sqrt{d^2-16 d+40}+d-2\right)+1}+C_2 \f
   ^{\frac{1}{2} \left(\sqrt{d^2-16 d+40}+d-2\right)+1}+C_3 \f ^d +\dots
\end{equation}
The constant deformation $C_0$ is not subleading with respect to the unperturbed solution and therefore we set $C_0=0$ for consistency. Conversely, the deformation proportional to $C_3$ is always subleading for $d>2$, so $C_3$ is an integration constant of the full solution. As for $C_1$ and $C_2$, they are leading or subleading depending on the dimension. It is easy to check that for $d=2$, $C_2$ is subleading while $C_1$ is not. For $d=3$ and $d=4$, none of those two is subleading, while for $d>4$ both are subleading. In addition, for $3<d<13$ the exponents are complex numbers and a real solution can be constructed by appropriately combining the integration constants (for an example see equations \eqref{im1} and \eqref{im2}). All in all, the full solution has up to four integration constants: $W_0,C_1,C_2,C_3$.

Again from \eqref{pert1} and \eqref{pert2} we extract the perturbation for the blackening function $\delta f$ and from \eqref{reda3aa} the perturbation for the inverse scale factor $\delta T$:

\begin{align}
\delta f &= \frac{4 C_1 (d-1)^3 \left(\sqrt{d^2-16 d+40}+3 d-10\right) V_4 \f ^{\frac{1}{2}
   \left(d-\sqrt{d^2-16 d+40}\right)}}{3 (d-3)^2 W_0^3}\nonumber\\&+\frac{4 C_2(d-1)^3
   \left(-\sqrt{d^2-16 d+40}+3 d-10\right) V_4 \f ^{\frac{1}{2} \left(d+\sqrt{d^2-16
   d+40}\right)}}{3 (d-3)^2 W_0^3}\nonumber\\&+\frac{16 C_3(d-1)^3 d V_4
   \f ^{d}}{3 (d-3) W_0^3} + \dots
\end{align}

\begin{align}
\delta T &= C_1 \frac{(d-1) \left(d^3-12 d^2+\left(-d^2+4 d-9\right) \sqrt{d^2-16 d+40}+29 d\right) V_4
   \f ^{\frac{1}{2} \left(d-\sqrt{d^2-16 d+40}\right)}}{6 (d-3)^2 (d-2) W_0} \nonumber \\ & C_2 \frac{(d-1) \left(d^3-12 d^2-\left(-d^2+4 d-9\right) \sqrt{d^2-16 d+40}+29 d\right) V_4
   \f ^{\frac{1}{2} \left(d+\sqrt{d^2-16 d+40}\right)}}{6 (d-3)^2 (d-2) W_0} \nonumber\\& \frac{(d-1)^2 d (d+1) V_4}{3 (d-3)^2 (d-2) W_0}\f^d + \dots
\end{align}

\textbf{Case 1. Branch 2:} $W_2  = \dfrac{d W_0}{4(d-1)}$
\vskip 0.5cm

The next few coefficients for $W$ are given by

\begin{equation}
f_0=\frac{8 (d-1)^2 V_4}{3 d^2 W_0^2}\sp W_3 =\frac{d V_5 W_0}{8 (d-1) V_4}\sp f_1 = -\frac{4 (d-1)^2 V_5}{3 d^2 W_0^2}\,.
\end{equation}

The inverse scale factor $T$ obtained from \eqref{reda3aa} vanishes order by order. There is single integration constant: $W_0$. Similarly to the previous case, we compute the metric functions $e^{2A}$ and $g_{tt}$ in order to understand the geometry. From \eqref{f77} we obtain

\begin{equation}
A  = - \dfrac{W_0}{2(d-1)W_2}\log \f + A_0 + O(\f) = -\dfrac{2}{d}\log\f + A_0 + O(\f)\Rightarrow e^{2A} = e^{2A_0} \f^{-4/d}\,,
\label{e83}\end{equation}
confirming that the scale factor diverges. Conversely, the $g_{tt}$ factor vanishes for $d>2$:

\begin{equation}
g_{tt} = -fe^{2A} = -f_0 e^{2A_0} \f^{2(1-2/d)}\,,
\end{equation}
signalling the presence of a horizon. The geometry is regular and the Kretschmann scalar vanishes as

\begin{equation}
K_2 = \frac{4 (d-2) (d-1)^2 V_4^2 }{9 d^3}\f^4 + O(\f^5)
\end{equation}

The solution is asymptotically flat. We now look for deformations around the previous solution by solving equations \eqref{pert1} and \eqref{pert2}. To leading order  we find, for $\delta W$,

\begin{equation}
\delta W = C_0 + C_1 \f^{1+\sqrt{3}} + C_2 \f^{1-\sqrt{3}} + C_3 \f^{4/d}+\dots
\label{e86}\end{equation}
The only term that is subleading with respect to the original solution (for $d>2$) is the one proportional to $C_1$. Therefore, for consistency we set $C_0 = C_2 = C_3 = 0$.

The perturbation for the blackening function, $\delta f$, and for the inverse scale factor $\delta T$, are obtained again from equations \eqref{pert1}, \eqref{pert2}, and \eqref{reda3aa} respectively. We quote the result to leading order:

\begin{equation}
\delta f=-\frac{64 \left(2+\sqrt{3}\right) C_1 (d-1)^3 V_4 \varphi ^{1+\sqrt{3}}}{3 d^3
   W_0^3}   + \dots
\end{equation}

\begin{equation}\label{e91}
\delta T= 0
\end{equation}

The fact that $T$ vanishes identically means that this solution only appears for the flat sliced ansatz \eqref{c38}. In that case, this is a possible endpoint of the flow, because $W_1=0$ while $W_2$ is finite.

We conclude that there are two integration constant for this solution: $W_0,C_1$. The geometry of the solution in this limit is that of horizon with infinite size, as the volume of spatial slices diverges.

\vskip 1cm

\textbf{Case 2:} $W_1 \neq 0$

In that case
\be\label{89}
W_2=-{2 V_0 W_0 + (d-1)^2 V_3 W_1\over 4 (d-1) V_0}
\ee
\be
f_0 = -{V_0\over (d-1) W_1^2}\sp  f_1 = -{ (d+5) V_0 W_0 \over 6 (d-1)^2W_1^3}-{V_3\over 4W_1^2}
   \ee
It has two integration constants $W_0,W_1$.
In addition, the scale factor approaches a constant value:
\be
\gamma=0\sp T_0 = {V_0\over (d-1)(d-2)}\,,
\ee
and we must have $V_0>0$.

According to \eqref{f77}, the metric function $A$ also approaches a constant value

\begin{equation}
A =  A_0 - \dfrac{W_0}{2(d-1)W_1}\f + O(\f^2)\,,
\end{equation}
and consequently the $g_{tt}$ component of the metric vanishes as

\begin{equation}
g_{tt}=-f e^{2A} = - e^{2 A_0} f_0 \f^2  + O(\f^3)\,,
\end{equation}
signalling the presence of a horizon. The geometry is regular and the Kretschmann scalar approaches a constant value:

\begin{equation}
K_2 = \frac{2 (3 d-5) V_0^2}{(d-2) (d-1)^2} + \mathcal{O}(\f)
\end{equation}

The above solution matches the first family of extremal horizons  found in Appendix \ref{seho}, (see below Eq. \eqref{eahb3}).
The horizon is extremal since $\dot f=f'W'\to 0$ at this particular point.

The second solution for $\a$ and $\b$ is

\be \label{95} ~~~\a=1\sp \b=0\ee
\be
W_2 = {4 V_0 W_0 + 2 (d-1) V_2 W_0 + (d-2) f_0 W_0^3 - 8 (d-1) V_1 W_1 +
 8 (d-1) f_0 W_0 W_1^2)\over 6 (d-1) f_0 W_0^2}
  \ee
\be
f_1 = -{-V_1 + 2 f_0 W_0 W_1\over W_0^2}\sp  f_2 = -{(
  2 V_0 W_0 - f_0 W_0^3 + 2 (d-1) V_1 W_1 - 4(d-1) f_0 W_0 W_1^2)\over (d-1) W_0^3}
   \ee
There are three integration constants $W_0,W_1,f_0$.
The scale factor approaches a constant value:
\be
\g=0\sp
T_0 = {2 V_0 - f_0 W_0^2\over 2 (d-1)(d-2)}\,.
\ee
and we must impose that $T_0>0$.
From Eq. \eqref{f77} we obtain the behaviour of the metric function $A$ as we reach $\f \to 0$:

\begin{equation}
A = A_0 - \dfrac{1}{4(d-1)}\f + O(\f^2)\,.
\end{equation}
As a consequence, the $tt$ component of the metric becomes

\begin{equation}
g_{tt}= -e^{2 A_0} f_0 +O(\f)\,.
\end{equation}

The geometry is regular and the Kretschmann scalar approaches a constant value

\begin{equation}
K_2 = \frac{\left(2 d^2-5 d+3\right) f_0^2 W_0^4-4 (d-1) f_0 V_0 W_0^2+4 (3 d-5)
   V_0^2}{2 (d-2) (d-1)^2}+ \mathcal{O}(\f)
\end{equation}

This solution is a particular case of the expansion around a generic point (obtained in Eq. \eqref{ab0} and extensively discussed in Appendix \ref{ord}) where the leading coefficient of the superpotential $W_0$ is set to zero.

\vskip 1cm
$\bullet$ $2\a+\b= 3$
\vskip 1cm

In this case, setting $p_0=0$ solves the equations \eqref{redb} and \eqref{redb1} up to $O(\f^{-1})$ before the potential kicks in, and we obtain three possibilities.\\

\textbf{Case 1.}
\be
\a=0\sp \b=3\,,
\ee
\be
W_1=V_0=V_1=V_2=V_3=V_4 = 0\,.
  \ee
and we find two branches for the coefficient $W_2$:

\begin{equation}\label{eqcoe}
W_2=\frac{W_0}{3(d-1)} \qquad W_2=d\frac{W_0}{6(d-1)}
\end{equation}

{\bf Case 1. Branch 1}: $W_2=\frac{W_0}{3(d-1)}$

The following few leading coefficients are found to be

\begin{equation}
	f_0=-\frac{3 (d-1)^2 V_5}{4 (3 d-8) W_0^2}\,,\qquad f_1 = -\frac{3 (d-1)^2 V_6}{40 (d-2) W_0^2}\,, \qquad W_3=-\frac{(3 d-8) V_6 W_0}{45 (d-2) (d-1) V_5}\,.
\label{case1}\end{equation}

The inverse scale factor $T$ vanishes as
\be
\g=3\sp
T_0 = - 3\frac{V_5}{16(3d-8 )}\,,
\label{e106}\ee
and we must have $\f V_5>0$, which implies that the solution exists to the left or to the right of that point, depending on the sign of $V_5$.

In principle, this branch has one integration constant: $W_0$. From equation \ref{f77} we find

\begin{equation}
A = -\frac{W_0}{2(d-1)W_2}\log \f + A_0 +O(\f)= -\frac{3}{2}\log \f + A_0+O(\f)\ \Rightarrow e^{2A} = e^{2A_0}\f^{-3}+ O(\f^{-2}) \,,
\end{equation}
confirming that the scale factor diverges for this solution. The behavior of $e^A$ is compatible with the leading power of $T$ in (\ref{e106}). In addition, the $g_{tt}$ component of the metric approaches a constant value in spite of the vanishing of $f$:

\begin{equation}
g_{tt} = -e^{2A}f = -e^{2A_0} f_0 +O(\f)
\end{equation}

The geometry is regular and the Kretschmann scalar vanishes as

\begin{equation}
K_2 =\frac{(d (3 d (3 d (9 d-79)+983)-6679)+6810) V_5^2 \f ^{10}}{7200 (8-3 d)^2 (10-3
   d)^2 (d-1)^2}+ \mathcal{O}(\f^{11})
\end{equation}
The pressure, energy density and quantity $\mathcal{I}$ controlling the curvature invariant also vanish:
\begin{equation}\label{invMink31}
\rho = -\frac{(d-1) V_5}{40 (3 d-8)}\varphi ^5 + \dots \sp p = \frac{(3 d-13) V_5}{120 (3 d-8)}\varphi ^5 + \dots
\end{equation}
\begin{equation}\label{invMink32}
\mathcal{I} = -\frac{(3 d-13) V_5}{40 (d-1) (3 d-10) (3 d-8)}\varphi^5+ \dots
\end{equation}
Note that the leading contribution to the energy density $\rho$ is proportional to $T_0$ in Eq. \eqref{e106}. Therefore, for the spherical slicing ($T>0$), the energy density $\rho$ increases from zero as we depart from this solution.

This solution is asymptotically flat. Since the $S^{3}$ asymptotes to infinite size, we can identify this asymptotics with the spatial boundary of Minkowski space-time.  Since $W_1=0$ and $W_2$ is finite, this is a possible endpoint of the flow.

We further look for perturbations around this solution by solving Eqs. \eqref{pert1} and \eqref{pert2}, which for this particular solution become, to leading order,

$$
\frac{(d-1) (3 d-14) V_5 \f ^3 \delta W'}{8 (3 d-8) W_0}-\frac{(d-1)
   V_5 \f ^4 \delta W''}{4 (3 d-8) W_0}+\frac{d V_5 \f ^4
   \delta W}{8 (3 d-8) W_0}+
$$
\begin{equation}
+\frac{W_0^2 \f ^2 \delta f'}{9 (d-1)^2}-\frac{(3 d-2) W_0^2 \f  \delta f}{18
   (d-1)^2} = 0
\end{equation}

$$
\frac{3 (d-1)^2 (3 d-4) V_5 \f ^3 \delta W'}{16 (3 d-8) W_0}-\frac{3
   (d-1)^2 V_5 \f ^4 \delta W''}{8 (3 d-8) W_0}-\frac{(3 d+4) W_0^2
   \f ^2 \delta f'}{36 (d-1)}+
$$
\begin{equation}
+\frac{W_0^2 \f ^3 \delta f''}{18 (d-1)}+\frac{d W_0^2 \f  \delta f}{4 (d-1)}=0
\end{equation}

The solution to the previous system of equations is given by

\begin{equation}\label{im1}
\delta W =  C_1 + C_2 \f ^{\frac{3 d}{2}-1} + C_3 \f^{(3d-2)\over 4} \cos\left(\frac{1}{4}\sqrt{-9 d^2+132 d-292}\log \f + C_4 \right) +\dots\,,
\end{equation}

\begin{align}\label{im2}
\delta f &= \frac{9 C_1 (d-1)^2 d V_5 \f ^3}{4 (8-3 d)^2 W_0^3} -\frac{45 C_2 (d-1)^2 \left(39 d^2-46 d+16\right) V_5 \f ^{3 d/2}}{8 (8-3 d)^2
   W_0^3} \nonumber\\
   & \frac{9 C_3 (d-1)^3 V_5 \f ^{\frac{3 d}{4}+\frac{1}{2}} \left(\sqrt{-9 d^2+132
   d-292} \sin \left(C_4+\frac{1}{4} \sqrt{-9 d^2+132 d-292} \log (\f )\right)\right)}{4 (8-3 d)^2 W_0^3}
   \nonumber\\
   & \frac{9 C_3 (d-1)^3 V_5 \f ^{\frac{3 d}{4}+\frac{1}{2}} \left(2
   (6 d-17) \cos \left(C_4+\frac{1}{4} \sqrt{-9 d^2+132 d-292} \log (\f
   )\right)\right)}{4 (8-3 d)^2 W_0^3}+\dots\,,
\end{align}
where we have assumed that $2<d<12$ so that $-9 d^2+132 d-292>0$. The contribution coming from $C_1$ is not subleading with respect to the unperturbed solution and therefore one has to set $C_1=0$. The solution proportional to $C_2$ is subleading so long as $d>2$, while the solution coming from $C_3$ and $C_4$ is subleading only when $d\geq 4$. Therefore this branch of solutions has four integration constants: $W_0, C_2, C_3, C_4$. Finally, writing $T\to T+\delta T$ in \eqref{reda3aa} we find

\begin{align}
\delta T &= \frac{9 C_2 (d-1) (3 d-2) V_5 \f ^{3 d/2}}{16 (d-2) (3 d-8) W_0} \nonumber \\ &-\frac{9 C_3 (d-1) (6 d-17) V_5 \f ^{\frac{3 d}{4}+\frac{1}{2}} \cos
   \left(C_4+\frac{1}{4} \sqrt{-9 d^2+132 d-292} \log (\f )\right)}{8 (8-3 d)^2
   W_0}\nonumber\\& -\frac{9 C_3 (d-1) \sqrt{-9 d^2+132 d-292} V_5 \f ^{\frac{3 d}{4}+\frac{1}{2}}
   \sin \left(C_4+\frac{1}{4} \sqrt{-9 d^2+132 d-292} \log (\f )\right)}{16 (8-3
   d)^2 W_0}+\dots
\end{align}

{\bf Case 1. Branch 2}: $ W_2=\frac{d W_0}{6(d-1)}$

The second branch requires one extra fine-tuning: $V_6=0$. The first few coefficients are found to be

\begin{equation}
	f_0 = \frac{3 (d-1)^2 V_5}{2 d^2 W_0^2}\qquad W_3 =-\frac{2 d^3 W_0^3 f_1}{27 (d-1)^3 V_5},
\label{case2}\end{equation}
while the inverse scale factor $T$ extracted from \eqref{reda3aa} vanishes order by order. Note however that $T$ may vanish with non-integer powers of $\f$. This branch has two integration constants:$f_1$ and $W_0$. From equation \eqref{f77} we learn that the metric function $A$ diverges as we approach this solution

\begin{equation}
A = -\frac{W_0}{2(d-1)W_2}\log \f + A_0 +O(\f)= -\frac{3}{d}\log \f + A_0+O(\f)\ \Rightarrow e^{2A} = e^{2A_0}\f^{-6/d}+\dots \,.
\label{e116}\end{equation}
Accordingly, the $tt$ component of the metric vanishes (provided that $d>2$) as

\begin{equation}
g_{tt} = -f e^{2A} = -f_0 e^{2A_0} \f^{3\left(1-2/d\right)}+\dots\,,
\end{equation}
signalling the presence of a horizon. The geometry is regular and the Kretschmann scalar vanishes as

\begin{equation}
K_2 =\frac{9 (d-2) (d-1)^2 V_5^2 \f ^6}{64 d^3}+ \mathcal{O}(\f^7)
\end{equation}

The solution is asymptotically flat. Since $W_1=0$ and $W_2$ is finite, this is a possible endpoint of the flow. We look for perturbations around this solution by solving the equations of motion for the perturbations given in \eqref{pert1} and \eqref{pert2}. To leading order we find

\begin{equation}
\delta W = C_1 + \frac{C_2 }{\f} + C_3 \f^3 + C_4 \f^{-1 + {6\over d}}+\dots
\end{equation}
The contributions coming from $C_1$, $C_2$ and $C_4$ are not subleading with respect to the unperturbed solution for $d>2$, and one has to set $C_1=C_2=C_4=0$. Taking this into account, the perturbation for the function $f$ is found to be

\begin{equation}
\delta f = -\frac{81 C_3 (d-1)^3 V_5 \f ^4}{d^3 W_0^3}+\dots
\end{equation}
The contribution coming from $C_3$ is always subleading with respect to the unperturbed solution. However, it can be reabsorbed into $f_1$. The solution has only two integration constants: $W_0,f_1$. Finally, we solve for the perturbation in $T$:

\begin{equation}
\delta T = 0
\end{equation}

The fact that $T$ vanishes identically means that this solution only appears for the flat sliced ansatz \eqref{c38}.

We conclude that there are two integration constant for this solution: $W_0,f_1$. The geometry of the solution in this limit is that of a horizon with infinite size, as the volume of spatial slices diverges. This is a possible endpoint of the flow.

\textbf{Case 2.}

\be\a=1\sp \b=1 \ee
\be\label{neh2}
W_1 = {(2 V_0 + (d-1) V_2) W_0\over 4 (d-1) V_1}
\ee
\be
W_2={(12 V_0^2 +
   4 (d-1) V_0 V_2 + (d-1) (2 d V_1^2 -(d-1)V_2^2 +
      4 (d-1) V_1 V_3)) W_0)\over 36 (d-1)^2 V_1^2}
\ee
\be
f_0 = {V_1\over W_0^2}\sp  f_1 = -{6 V_0 +(d-1) V_2\over 4 (d-1) W_0^2}
\ee
\be
  f_2 = {
 84 V_0^2 +
  56 (d-1) V_0 V_2 + (d-1) (7(d-1) V_2^2 - 4(d-1) V_1 V_3 +
     4d  V_1^2 )\over 36 (d-1)^2 V_1 W_0^2}
\ee

There is only one integration constant, $W_0$.
The inverse scale factor approaches a constant value:
\be
\gamma=0\sp T_0 = {V_0\over (d-1)(d-2)}\,.
\ee
Clearly such a solution exists only in the dS regime, $V_0>0$.

According to Eq. \eqref{f77},

\begin{equation}
A = A_0 - \dfrac{1}{4(d-1)}\f^2 +O(\f^3)\ \Rightarrow e^{2 A} = e^{2A_0} +O(\f^2)\,,
\end{equation}
and the $g_{tt}$ factor in the metric vanishes linearly in $\f$:

\begin{equation}
g_{tt} = -f e^{2A} = -f_0 e^{2 A_0} \f +O(\f^2)
\end{equation}
signalling the presence of a horizon. The geometry is regular and the Kretschmann scalar approaches a constant value

\begin{equation}
K_2 =\frac{2 (3 d-5) V_0^2}{(d-2) (d-1)^2}+ \mathcal{O}(\f)\,.
\end{equation}

The previous solution matches the standard horizon found for $2\alpha+\beta=1$, also described in Appendix \ref{sho}. However here the superpotential $W$ vanishes at the horizon. This implies that $\dot A$ changes sign there, and the monotonicity of $A$ is compromised.
Nonetheless, this solution is a special case of the previous one.

\vskip 1cm

\textbf{Case 3.}

\be\label{bcp}
\a={3\over 2}\sp \b=0
\ee
In this case the solution is
\be
W_1 = {3 (6 V_0 + (d-1) V_2) W_0\over
 20 (d-1) V_1}
 \ee
 \be
   W_2= {348 V_0^2 +
    4 (d-1) V_0 V_2 + (d-1) (32 d V_1^2 - 9(d-1) V_2^2  +
       24 (d-1) V_1 V_3)\over 336 (d-1)^2 V_1^2}W_0
       \ee
\be
f_0 = {8 V_1\over 9 W_0^2}\sp  f_1 = -{16 V_0\over 9 (d-1) W_0^2}\sp  f_2 =
{ 16 V_0 (6 V_0 + (d-1) V_2)\over 27 (d-1)^2 V_1 W_0^2}
\ee
There is a single integration constant: $W_0 $. The inverse scale factor approaches a constant value,
\be
\gamma=0\sp T_0 = {V_0\over (d-1)(d-2)}\,,
\ee
as well as the $g_{tt}$ component of the metric:
{We must also have $V_0>0$}

\begin{equation}
g_{tt} =  -f e^{2A} = f_0 e^{2A_0}+O(\f)\,,
\end{equation}
where we have used that

\begin{equation}
A = A_0 - \dfrac{1}{6(d-1)}\f^2 + O(\f^3)\,,
\end{equation}
as it follows from \eqref{f77}.

The geometry is regular and the Kretschmann scalar approaches a constant value

\begin{equation}
K_2 =\frac{2 (3 d-5) V_0^2}{(d-2) (d-1)^2}+ \mathcal{O}(\f)\,.
\end{equation}

Finally, from $W' =\dot{\f}$ we obtain $\f = \dfrac{3}{4}W_0^2 u^2 + O(u^3)$ so that the previous solution describes a Bounce point (where $W'=0$ without the flow stopping). The Bounce solutions described in Appendix \ref{bounces}, reduces to this one once the leading coefficient in $W$ there is set to $0$.
They exist only in the de Sitter regime $V_0>0$.

\vskip 1cm
$\bullet$ $2\a+\b= 4$.
\vskip 1cm

Here we find two possibilities, with two branches each.

{\bf Case 1.} $\a=0\,, \beta=4$

This solution exists only when $V_0=V_1=V_2=V_3=V_4=V_5=0$.

There are two branches of solutions

{\bf Case 1. Branch 1}:  $W_2 = {W_0\over 4 (d-1)}$.

The first few coefficients are given by
\be\label{e141}
W_1=0\sp W_2 = {W_0\over 4 (d-1)}\sp W_3 = -{(2d-5) V_7\over 12 (13 - 19 d + 6 d^2) V_6}W_0
\ee
\be
f_0 = -{2 (d-1)^2 V_6\over 15 (2d-5) W_0^2}\sp f_1= -{4 (d-1)^2 V_7\over 45 (6d-13) W_0^2}
\ee

There is a single integration constant: $W_0$. The inverse scale factor vanishes according to

\be\label{j153}
\gamma=4\sp T_0=\dfrac{V_6}{150-60 d}\,,
\ee
{and we must demand that $V_6>0$.}

On the other hand, solving \eqref{f77} we observe that the metric function $A$ diverges according to

\begin{equation}
A =-2\log \f +A_0+O(\f)\,, \Rightarrow e^{2A} = e^{2 A_0} \f^{-4}+O(\f^{-3})\,.
\end{equation}
The $tt$ component of the metric approaches a constant value in spite of the vanishing of $f\simeq f_0\f^4$:

\begin{equation}
g_{tt}= -f e^{2A}=-f_0e^{2 A_0} +O(\f)\,.
\end{equation}

Finally, the Kretschmann scalar vanishes, to leading order, as

\be
K={(123 + d (-139 + d (67 + d (-17 + 2 d)))) V_6^2 \over 129600 (2d-5)^2 (d-3)^2 (d-1)^2}\f^{12}+\cdots
\ee
and the geometry is regular. The pressure, energy density and quantity $\mathcal{I}$ controlling the curvature invariant also vanish:
\begin{equation}\label{invMink41}
\rho = \frac{(d-1) V_6}{12 (150-60 d)}\varphi ^6 + \dots \sp p = -\frac{(d-4) V_6}{12 (150-60 d)}\varphi ^6+ \dots
\end{equation}
\begin{equation}\label{invMink42}
\mathcal{I} = -\frac{(d-4) V_6 }{360 (d-3) (d-1) (2 d-5)}\varphi^6+ \dots
\end{equation}
Note that the leading contribution to the energy density $\rho$ is proportional to $T_0$ in Eq. \eqref{j153}. Therefore, for the spherical slicing ($T>0$), the energy density $\rho$ increases from zero as we depart from this solution.

The solution is asymptotically flat. Since $W_1=0$ and $W_2$ is finite, this is a possible endpoint of the flow.
We look for deformations around the previous solution by dissecting Eqs. \eqref{pert1} and \eqref{pert2}. To leading order we find

\begin{equation}
\delta W = C_1+C_2 \f ^{2 d-2}+C_3 \f ^{d-1} \cos \left(\sqrt{-d^2+14 d-29}
   \log (\f )\right)+
\end{equation}
$$
+C_4 \f ^{d-1} \sin \left(\sqrt{-d^2+14 d-29} \log (\f
   )\right)+\dots
   $$
where it is assumed that $2<d<12$ in order to have a positive radicand. We further set $C_1=0$ for consistency with the unperturbed solution. In addition, we have

\begin{align}
&\delta f  =  \frac{32 (d-1)^3 V_6 \f ^d \left(C_3 \left(2 d^2+3 d-21\right)+C_4
   \sqrt{-d^2+14 d-29} (2 d-7)\right) }{15 (5-2 d)^2 W_0^3}\times \nonumber\\
&\sin \left(\sqrt{-d^2+14 d-29} \log (\f
   )\right)+C_2\frac{64 (d-1)^3 d \left(2 d^2-7 d+5\right) V_6 \f ^{2 d-1}}{5 (5-2 d)^2 W_0^3} +  \nonumber\\
   &+\frac{32 (d-1)^3 V_6 \f ^d \left(C_3 \sqrt{-d^2+14 d-29} (7-2 d)+C_4
   \left(2 d^2+3 d-21\right)\right) }{15 (5-2 d)^2 W_0^3}\times\nonumber\\& \times \cos \left(\sqrt{-d^2+14 d-29} \log (\f
   )\right)\,,
\end{align}
to leading order. The term proportional to $C_2$ is allowed for $d>3$ whereas the terms accompanying $C_3$ and $C_4$ are permitted if $d>4$. Therefore, the full solution has up to $4$ integration constants: $W_0,C_2,C_3,C_4$. Finally, we find the perturbation in $T$ by solving Eq. \eqref{reda3aa}:

\begin{align}
&\delta T = \frac{4 (d-1) V_6 \f ^d \left(C_3 \sqrt{-d^2+14 d-29} (10 d-21)+C_4
   \left(38 d^2-177 d+203\right)\right) }{15 (5-2 d)^2 (d-2) W_0}\times\nonumber\\
   &\sin \left(\sqrt{-d^2+14 d-29} \log (\f
   )\right) +\frac{8 C_2 (d-1)^2 d V_6 \f ^{2 d-1}}{5 (d-2) (2 d-5) W_0}+ \cos \left(\sqrt{-d^2+14 d-29} \log (\f
   )\right) \times \nonumber\\
   & \frac{4 (d-1) V_6 \f ^d \left(C_3 \left(38 d^2-177 d+203\right)+C_4
   \sqrt{-d^2+14 d-29} (21-10 d)\right) }{15 (5-2 d)^2 (d-2) W_0}+\dots
\end{align}

This solution is asymptotically flat. Since the $S^{3}$ asymptotes to infinite size, we can identify this asymptotics with the spatial boundary of Minkowski space-time.

\vskip 1cm

{\bf Case 1. Branch 2}: $W_2 = {d W_0\over 8 (d-1)}$.

The first few coefficients are given by

\be
W_1=0\sp W_2 = {d W_0\over 8 (d-1)}
\ee
\be
f_0 = {8 (d-1)^2 V_6\over 15 d^2 W_0^2} \qquad f_1 =\frac{16 (d-1)^2 V_7}{45 d^2 W_0^2} \qquad W_3 = \frac{d V_7 W_0}{24(1-d) V_6}
\ee
There is a single integration constant $W_0$. The inverse scale factor $T$ vanishes order by order. From the solution to \eqref{f77}

\begin{equation}
A = - \dfrac{4}{d}\log \f + A_0 + O(\f) \ \Rightarrow e^{2A} = e^{2A_0} \f^{-8/d}+\dots
\end{equation}
we observe that the function $A$ diverges. Conversely, the $g_{tt}$ factor of the metric vanishes for $d>2$, signalling the presence of a horizon:

\begin{equation}
g_{tt} = -e^{2A} f= -e^{2A_0}f_0 \f^{4\left(1-2/d\right)}+\dots
\end{equation}

The geometry is regular and the Kretschmann invariant vanishes as

\begin{equation}
K_2 = \frac{4 (d-2) (d-1)^2 V_6^2 \f ^8}{225 d^3} + \mathcal{O}(\f^9)
\end{equation}

The solution is asymptotically flat. Since $W_1=0$ and $W''$ is finite, this is a possible endpoint of the flow.

We look now for deformations around this solution by analyzing Eqs. \eqref{pert1} and \eqref{pert2}. The solution is given by

\begin{equation}
\delta W  = C_1+C_2 \f ^{1-\sqrt{5}}+C_3 \f ^{1+\sqrt{5}}+C_4 \f
   ^{\frac{8}{d}-2}+\dots\,
\end{equation}
The terms proportional to $C_1$ and $C_2$ are not subleading and therefore these constants are set to zero by consistency. Taking that into account, we find $\delta f$ to be

\begin{equation}
\delta f =-\frac{128 (d-1)^3 V_6 \left(\left(3+\sqrt{5}\right) C_3 d (3 d-8) \f
   ^{3+\sqrt{5}}+2 C_4 \left(d^2-16\right) \f ^{8/d}\right)}{15 d^4 (3 d-8)
   W_0^3} + \dots
\end{equation}
Again we find that the term with $C_3$ is subleading whereas the term proportional to $C_4$ is leading for $d>2$ and we require $C_4=0$.

Therefore, this solution has two integration constants: $W_0, C_3$. Finally, we solve \eqref{reda3aa} to find

\begin{equation}
\delta T = 0
\end{equation}

The fact that $T$ vanishes identically means that this solution only appears for the flat sliced ansatz \eqref{c38}.

We conclude that there are two integration constant for this solution: $W_0,C_3$. This solution describes a flat horizon in Minkowski space-time.

\vskip 1cm

{\bf Case 2}

In this case, the equations only demand that $V_1=0$ and the exponents and first coefficients are found to satisfy

\begin{equation}\label{156}
	 \alpha_\pm = 2+\dfrac{1}{\delta_{\pm}} \qquad \beta_\pm = -\frac{2}{\delta_{\pm}} \qquad f_0^\pm (W_0^\pm)^2 = -\frac{\delta_\pm^4 V_0}{(d-1) (2 \delta_{\pm} +1)^2}\,,
\end{equation}

where we have defined

\begin{equation}\label{157}
	\delta_{\pm}= \dfrac{1}{2} \left(1\pm\sqrt{1-\dfrac{4(d-1)V_2}{V_0}}\right)\in \left[\frac{1}{2},\pm \infty \right)\,,
\end{equation}
or equivalently

\begin{equation}
	V_2 = \dfrac{V_0}{d-1} \delta(1-\delta)\,.
\end{equation}
This solution has a single integration constant: $W_0$. Again, Eq. \eqref{reda3aa} implies $\gamma=0$ and

\begin{equation}
	T_0 = \frac{V_0}{(d-1)(d-2)}\,.
\end{equation}
{So that $V_0>0$.}

Depending on the ranges of $\delta_{\pm}$ the above solutions describe generic extremal horizons (see Appendix \ref{seho}) or the boundary of dS$_2\times$ S$^{(d-1)}$ (see Appendix \ref{g53}).
 Note that $f\sim \f^{-2/\delta_{\pm}}$, and therefore, the sign of $\delta$ plays a crucial role. Whenever it is negative, $f$ vanishes and we obtain a horizon, whereas if it is positive, $f$ diverges and we obtain the boundary of $dS_2$.

In both cases the geometry is regular and the Kretshcmann scalar approaches a constant value:

\begin{equation}
K_2 = \frac{2 (3 d-5) V_0^2}{(d-2) (d-1)^2} + \mathcal{O}(\f)
\end{equation}

\vskip 1cm
$\bullet$ $2\a+\b= 5$.
\vskip 1cm

In this case the equations demand that $V_0=V_1=V_2=0$ and three possible values of $\a$:  $\a=0,5/2,3$.

{\bf Case 1}. $\a=0,\b=5$.

The $\a=0,\b=5$ solution requires $V_0=V_1=V_2=V_3=V_4=V_5=V_6=0$.
Then  we have two branches for this solution.

{\bf Case 1. Branch 1.} $W_2 = {W_0\over 5 (d-1)}$.

\be\label{e163}
W_1=0\sp   W_2 = {W_0\over 5 (d-1)}\sp W_3 = -{(5d-12) V_8 W_0\over 70 (d-1) (5d-11) V_7}
\ee
\be
f_0= -{5 (d-1)^2 V_7\over 72 (5d-12) W_0^2}\sp f_1 = -{25 (d-1)^2 V_8\over 2016 (5d-11) W_0^2}
\ee
There is a single integration constant: $W_0$. The inverse scale factor vanishes according to
\be\label{j177}
\g=5\sp T_0=-{5 V_7\over 288 (5d-12)}
\ee
Again, the assumption that $T>0$ translates into  $V_7\f>0$, which means that the solution exists either to the left or to the right of $\f=0$ depending on the sign of $V_7$. Solving now \eqref{f77} we find the behaviour of the function $A$:

\begin{equation}
A = -\dfrac{5}{2}\log \f + A_0 +\dots\ \Rightarrow e^{2A} = e^{2A_0}\f^{-5}+\dots
\end{equation}
As a consequence, the $g_{tt}$ component of the metric approaches a constant value

\begin{equation}
g_{tt} = -f e^{2A}=-f_0 e^{2A_0}+O(\f)
\end{equation}

The solution is regular and the Kretchsmann invariant vanishes as

\begin{equation}
K_2 = \frac{\left(625 d^4-5175 d^3+19605 d^2-38407 d+31290\right) V_7^2 \f ^{14}}{12700800
   (12-5 d)^2 (14-5 d)^2 (d-1)^2}+\mathcal{O}(\f^{15})
\end{equation}
The pressure, energy density and quantity $\mathcal{I}$ controlling the curvature invariant also vanish:
\begin{equation}\label{invMink51}
\rho =-\frac{(d-1) V_7}{1008 (5 d-12)} \varphi ^7 + \dots \sp p = \frac{(5 d-19) V_7}{5040 (5 d-12)} \varphi ^7+ \dots
\end{equation}
\begin{equation}\label{invMink52}
\mathcal{I} = -\frac{(5 d-19) V_7 \phi ^7}{1008 (d-1) (5 d-14) (5 d-12)}\varphi^7+ \dots
\end{equation}
Note that the leading contribution to the energy density $\rho$ is proportional to $T_0$ in Eq. \eqref{j177}. Therefore, for the spherical slicing ($T>0$), the energy density $\rho$ increases from zero as we depart from this solution.

The solution is asymptotically flat. It corresponds to the spatial boundary of Minkowski space. Since $W_1=0$ and $W''$ is finite, this is a possible endpoint of the flow, provided that the potential vanishes to the given order: $V=V_7 \f^7 +\dots$. It is similar to the one in (\ref{case1}).

We look for perturbations about this solution by solving Eqs. \eqref{pert1} and \eqref{pert2}. To leading order we extract:

\begin{equation}
\delta W = C_1+C_2 \f ^{\frac{5 d}{2}-3}+C_4 \f ^{\frac{1}{4} (5 d-10)+1}
   \cos \left(C_4+\frac{1}{4} \sqrt{-25 d^2+340 d-676} \log (\f )\right)+\dots
\end{equation}
Note that $C_1$ is not subleading with respect to the unperturbed solution. Accordingly, we set $C_1=0$. The perturbation for $f$ is found to be

\begin{align}
&\delta f = C_2\frac{175 (d-1)^3 (5 d-6) V_7 \f ^{5 d/2}}{144 (5 d-12) W_0^3} \nonumber\\
& +\frac{25 C_3 (d-1)^3 \sqrt{-25 d^2+340 d-676} V_7 \f ^{\frac{5
   d}{4}+\frac{3}{2}} \sin \left(C_4+\frac{1}{4} \sqrt{-25 d^2+340 d-676} \log
   (\f )\right)}{72 (12-5 d)^2 W_0^3}\nonumber\\
   & +\frac{25 C_4 (d-1)^3 (15 d-37) V_7 \f ^{\frac{5 d}{4}+\frac{3}{2}} \cos
   \left(C_4+\frac{1}{4} \sqrt{-25 d^2+340 d-676} \log (\f )\right)}{36 (12-5
   d)^2 W_0^3}+\dots
\end{align}
The terms involving $C_2$, $C_3$ and $C_4$ are subleading, hence allowed, so long as $d>2$. Therefore, the full solution has four integration constants: $ W_0, C_2, C_3, C_4$. Finally, we solve Eq. \eqref{reda3aa} to find:

\begin{align}
&\delta T = \frac{25 C_2 (d-1) (5 d-6) V_7 \f ^{5 d/2}}{288 (d-2) (5 d-12) W_0}\nonumber\\
&+\frac{25 C_4 (d-1) \sqrt{-25 d^2+340 d-676} V_7 \f ^{\frac{5 d}{4}+\frac{3}{2}}
   \sin \left(C_4+\frac{1}{4} \sqrt{-25 d^2+340 d-676} \log (\f )\right)}{288
   (12-5 d)^2 W_0}\nonumber\\
   &+ \frac{25 C_3 (d-1) (15 d-37) V_7 \f ^{\frac{5 d}{4}+\frac{3}{2}} \cos
   \left(C_4+\frac{1}{4} \sqrt{-25 d^2+340 d-676} \log (\f )\right)}{144 (12-5
   d)^2 W_0}+\dots
\end{align}

\vskip 1cm

{\bf Case 1. Branch 2.} $ W_2 = {d W_0\over 10 (d-1)}$
\be
W_1 = 0\sp  W_2 = {d W_0\over 10 (d-1)}\sp  W_3 = -{d V_8 W_0\over 70 (d - 1) V_7}
\ee
\be
f_0 = {5 (d-1)^2 V_7\over 36 d^2 W_0^2}\sp  f_1= {25 (d-1)^2 V_8\over
	504 d^2 W_0^2}
\ee
There is a single integration constant: $W_0$. The inverse scale factor $T$ is order by order zero in this solution, while the metric function $A$ can be obtained from \eqref{f77}:

\begin{equation}
A = A_0 - \dfrac{5}{d} \log \f + O(\f)\ar e^{A}\sim \f^{-{5\over d}}\,.
\end{equation}

The $g_{tt}$ component of the metric vanishes for $d>2$:

\begin{equation}
g_{tt} = -fe^{2A}= -e^{2A_0}f_0 \f^{5(1-2/d)}+\dots\,,
\end{equation}
signalling the presence of an extremal  horizon.
The solution is regular and the Kretschmann scalar vanishes as

\begin{equation}
K_2 = \frac{25 (d-2) (d-1)^2 V_7^2 \f ^{10}}{20736 d^3} + \mathcal{O}(\f^{11});
\end{equation}

The solution is asymptotically flat. Since $W_1=0$, this is a possible endpoint of the flow provided the potential $V$ vanishes like $\f^7$.
We look for deformations about the previous solution by solving \eqref{pert1} and \eqref{pert2}. To leading order we have

\begin{equation}
\delta W  = C_1+C_2 \f ^{1-\sqrt{6}}+C_3 \f ^{1+\sqrt{6}}+C_4 \f
   ^{\frac{10}{d}-3}+\dots
\end{equation}
Only the terms proportional to $C_3$ and $C_4$ are subleading with respect to the unperturbed solution. Therefore we set $C_1=C_2=0$. Then, the perturbation for $f$ becomes

\begin{equation}
\delta f = -\frac{25 (d-1)^3 V_7 \left(\left(7+2 \sqrt{6}\right) C_3 d (2 d-5) \f
   ^{4+\sqrt{6}}+C_4 \left(3 d^2+5 d-50\right) \f ^{10/d}\right)}{18 d^4 (2 d-5)
   W_0^3}+\dots
\end{equation}
We note that the term proportional to $C_4$ is not subleading for $d>2$ and consequently we are forced to set $C_4=0$ for consistency. The full solution has two integration constants: $W_0, C_3$. Finally, we solve for \eqref{reda3aa} to find that the perturbation in $T$ vanishes order by order:

\begin{equation}
\delta T = 0
\end{equation}
The fact that $T$ vanishes identically means that this solution only appears for the flat sliced ansatz \eqref{c38}.

The geometry of the solution in this limit is that of horizon with infinite size, as the volume of spatial slices diverges. This solution is similar to (\ref{case2}).

\vskip 1cm

{\bf Case 2.} $\a=5/2,\b=0$.

For the solution with $\a={5\over 2},\b=0$ we have
  \be
  f= {4 V_3\over 75 W_0^2}
  \ee
  to all orders, while the first coefficients for $W$ are
  \be
  W_1 ={5 V_4 W_0\over
 56 V_3}\sp  W_2 = {128d V_3^2 - 25(d-1) V_4^2 + 80(d-1) V_3 V_5\over 2880 (d-1) V_3^2}W_0
    \ee
There is a single integration constant: $W_0$. We also have $T=0$ to all orders.
When $T=0$ then $f$ and $W$ satisfy
\be
f'W'=Ce^{{d\over 2(d-1)}\int {W\over W'}d\f}\,,
\ee
and we observe that $f'=0$ compatible with $C=0$ in the relation above. The behaviour of the metric function $A$ can be extracted from \ref{f77}:

\begin{equation}
A = A_0 -\dfrac{1}{10(d-1)}\f^2+\dots
\end{equation}

Trivially, the $tt$ component of the metric also approaches a constant value. The solution is regular and the Kretschmann scalar vanishes as

\begin{equation}
K_2 = \frac{d V_3^2 \f ^6}{9 (d-1)^2} +\mathcal{O}(\f^7)
\end{equation}

All the metric components asymptote to a constant value, which implies that the metric asymptotes to flat space. In addition, from the flow equation $\dot{\f}=W'$ we find the dependence of $\f$ in the holographic coordinate $u$:

\begin{equation}
\sqrt{\f} = -\dfrac{4}{5uW_0}\,.
\end{equation}
The small $\f$ expansion translates into $u W_0\to -\infty$, and therefore the metric asymptotes to the spatial boundary of Minkowski space.

We look for fluctuations around the previous solution by solving equations \eqref{pert1} and \eqref{pert2}. For the superpotential we find, to leading order

\begin{equation}
\delta W = C_1 + C_2 \phi ^{5/2}+C_3 \phi ^2 + \frac{C_4}{\sqrt{\phi }}\,.
\end{equation}

\noindent
None of the integration constants is subleading with respect to the unperturbed solution. Therefore, we conclude that $\delta W=0$ and the full solution has a single integration constant: $W_0$.

Since $\a>3/2$, this is a possible endpoint of the flow provided that $V_0=V_1=V_2=0$. The fact that $T=0$ to all orders implies that this solution can only appear for a flat slicing of the metric.

\vskip 1cm

{\bf Case 3.} $\a=3,\b=-1$.

For $\a=3,\b=-1$ we have
\be \label{184}
W_1= 0\sp  W_2=
 {1\over 70} \left({5 d\over (d-1)} + {V_5\over
    V_3}\right) W_0
    \ee
    \be
     W_3= {(-5 V_4 (8 d V_3 + 3 (d-1) V_5) +
    21 (d-1) V_3 V_6) W_0\over 2940 (d-1) V_3^2}
\ee
\be
f_0= {V_3\over 18 W_0^2}\sp  f_1 = {V_4\over
 108 W_0^2}\sp  f_2 = {-2 d V_3 + (d-1) V_5\over
 756 (d-1) W_0^2}
 \ee
 \be
   f_3 = {-5 V_4 (8 d V_3 + 3 (d-1) V_5) +
  21 (d-1) V_3 V_6\over 52920 (d-1) V_3 W_0^2}
  \ee
  There is a single integration constant: $W_0$. We also have $T=0$ to all orders. The metric function $A$ approaches a constant value. In particular, solving \eqref{f77}:

  \begin{equation}
  A = A_0 -\dfrac{1}{12(d-1)}\f^2+\dots
  \end{equation}
  The $tt$ component of the metric diverges as

  \begin{equation}
  g_{tt}=-f e^{2A} = -\frac{f_0e^{2A_0}}{\f}+\dots
  \end{equation}
  However, the geometry is regular and the Kretschmann scalar vanishes as

  \begin{equation}
  K_2 = \frac{(d (d+32)-17) V_3^2 \f ^6}{144 (d-1)^2} +\mathcal{O}(\f^7)
  \end{equation}

  \vskip 1cm

Since $\a>3/2$, this is a possible endpoint of the flow.

We look for deformations around the previous solution by solving equations \eqref{pert1} and \eqref{pert2}. In particular, we can combine them to give the following equation for $\delta W$:

\begin{align}
-\frac{d V_3 \delta W}{6 (d-1) W_0 \f }+\frac{V_3 \delta W'}{3 W_0 \f ^2}-\frac{V_4 \delta W''}{9
   W_0}-\frac{V_3 \delta W^{(3)}}{6 W_0}+\frac{V_3 \f
   \delta W^{(4)}}{6 W_0} + \dots=0\,,
\end{align}
where the dots contain higher order terms in $\f$ for each of the coefficients of $\delta W$ and its derivatives. The leading order solution to the previous equation is found to be

\begin{equation}
\delta W = C_0 + C_1\f^3 + C_2\f^{2+\sqrt{2}} + C_3 \f^{2-\sqrt{2}}+\dots
\end{equation}
From the previous solution, only the deformation proportional to $C_2$ is subleading with respect to the unperturbed solution, and therefore we set $C_0=C_1=C_3=0$ for consistency. Again from equations \eqref{pert1} and \eqref{pert2} we obtain the perturbation for the blackening function $\delta f$ and from \eqref{reda3aa} we find that the perturbation for the inverse scale factor $\delta T$ vanish order by order.

\begin{equation}
\delta f = -\frac{\left(3+2 \sqrt{2}\right) C_2 V_3}{54 W_0^3} \f^{\sqrt{2}-2} + \dots
\end{equation}

\begin{equation}\label{dte}
\delta T = 0
\end{equation}

Consequently, the full solution has two integration constants: $W_0$ and $C_2$. The fact that $T=0$ implies that these kind of solutions only exist for the flat sliced ansatz \eqref{c38}.

The metric for this solution is, to leading order

\begin{equation}
ds^2 = \f \dfrac{du^2}{f_0} - f_0e^{2A_0}\frac{dt^2}{\f} +\frac{e^{2A_0}}{\f}dx_idx^i\,,
\end{equation}
From the flow equation $W' = \dot{\f}$ we find $\f$ as a function of the radial coordinate $u$:

\begin{equation}
W' = 3 W_0 \f^2 = \dot{\f} \Rightarrow \f = \dfrac{1}{3 W_0 (u_*-u)}\,,
\end{equation}
where $u_*$ is an integration constant. The small $\f$ expansion translates now into an expansion around $u\to \pm \infty$. We take $u\to-\infty$ without loss of generality. Substituting $\f(u)$ into the above metric we obtain

\begin{equation}
ds^2 = \dfrac{1}{3W_0 f_0 (u_*-u)}du^2 - 3 W_0 f_0 e^{2A_0} (u_*-u)dt^2 + 3W_0  e^{2A_0} (u_*-u) dx_i dx^i\,.
\end{equation}
Finally we can change variables as

\begin{align}
&\dfrac{1}{3W_0 f_0 (u_*-u)}du^2 = 2 e^{2 r } dr^2 \Rightarrow \dfrac{u_*-u}{3 W_0 f_0}=e^{2r} \nonumber\\& 9 W_0^2 f_0^2 e^{2A_0}dt^2 = d\tilde{t}^2\, \qquad dx_i dx^i = 9 W_0^2 e^{2A_0} f_0 d\tilde{x}_id\tilde{x}^i.
\end{align}
The expansion around $u\to -\infty$ translates now into $r\to\infty$. Therefore, the metric becomes

\begin{align}
ds^2 &= e^{2r}(dr^2-d\tilde{t}^2 +d\tilde{x}_i d\tilde{x}^i)=dR^2+R^2(-d\tilde{t}^2 +d\tilde{x}_i d\tilde{x}^i)\sp R\to +\infty
\end{align}
This is an asymptotically flat metric as $R\to+\infty.$.

\vskip 1cm
$\bullet$  $2\a+\b>5$
\vskip 1cm

The structure is similar to the previous examples. Setting $2\alpha+\beta=5+m$ with $m\in \mathbb{N}$, we encounter three distinct cases.

Firstly, $\alpha=0$ $\beta=m+5$ and $V_0=V_1=\dots=V_{m+6}=0$, within which there are two branches distinguished by the value of $W_2$. One branch is identified with the boundary of Minkowski space-time, the inverse scale factor vanishes as
$$T\sim- V_{m+7}\f^{5+m}\;.$$
 The assumption that $T>0$ restricts $ V_{m+7}\f^{5+m}>0$.  The other branch has a horizon with a diverging volume and only exists for the flat sliced ansatz \eqref{c38}.

Secondly, $\alpha=(m+5)/2$, $\beta=0$ and $V_0=\dots=V_{m+2}=0$, while $T$ vanishes order by order. This kind of solution is only possible with a flat sliced \eqref{c38}.

Finally, $\alpha=3+m$, $\beta=-m-1$ and $V_0=\dots=V_{m+2}=0$. One finds again the boundary of Minkowski space-time. The inverse scale factor $T$ vanishes order by order, and so this kind of solutions only exist for the flat sliced ansatz \eqref{c38}.

All three cases have a regular geometry with a vanishing Kretschmann scalar.

\subsection{Summary of classes of  solutions found}\label{smry}

We conclude this section by summarising the distinct solutions encountered in this analysis:

\begin{itemize}
\item {\bf Shrinking endpoints}, where the sphere smoothly shrinks to zero. They appear for $2\a + \b = -1$ (c.f. discussion below equation \eqref{shr2}) and they are studied in more detail in Appendix \ref{G.2.2}. They have $W' = 0$ and therefore, according to the discussion in section \ref{sec:global}, they are a possible endpoint of the flow. In addition, they are maxima (minima) of the superpotential for $W>0$ ($W<0$). In general, these solutions appear for $V'\neq 0 $. The requirement that $T>0$ enforces $(\f-\f_0) V_1(\f_0)>0$, where $\f_0$ is the position of the shrinking endpoint. This implies that the solution exists to the left or the right of that point, depending on the sign of the first derivative of the scalar potential, $V_1(\f_0)$ at that point.

\item {\bf Regular points}. They show up for $\a = \b=0$, see below equation \eqref{ab0}. They are described in more detail in Appendix \ref{ord}. They cannot be endpoints of the flow because $W'\neq 0$ in this type of solutions. They also appear for $\a = 1$ and $\b=0$, see equation \eqref{95} and below, in which case they are a limiting case of the solution in Appendix \ref{ord} in which $\dot{A}$ reverses sign.

\item {\bf dS$^{(d+1)}$ and AdS$^{(d+1)}$ boundaries} similarly appear for $\a = \b=0$ (see below \eqref{ab0}) but they have $W'=0$ and therefore are possible endpoints of the flow. They always appear as minima (maxima) of the superpotential for $W>0$ ($W<0$). Finally, they can only appear at extrema of the potential, i.e. $V'=0$. A detailed discussion of these boundaries is presented in appendices \ref{AdSX} and \ref{dSX}.

\item {\bf dS$_2$ boundaries}. They appear for $2\a + \b = 4$, and the explicit form of $\a$ and $\b$ is given in equation \eqref{156}. The geometry for this class of solutions is dS$_2\times$S$^{(d-1)}$. They require to have an extremum of the potential: $V'=0$. A complete analysis of these endpoints is provided in Appendix \ref{g53}. There it is shown that they are possible endpoints of the flow that appear only in the dS regime of the potential ($V>0$) under the assumption that $T>0$. Finally, they always appear as minima (maxima) of the superpotential for $W>0$ ($W<0$).

\item {\bf Spatial boundaries of Minkowski space-time}. There are two subcases, they always require a vanishing potential at least cubically\footnote{Such conditions are satisfied naturally for potentials that vanish exponentially at the boundaries of field space.}: $V_0=V_1=V_2=0$. The main difference between the two is that one appears for the ansatz with spherical slicing \eqref{c39}, while the other one is present only in the ansatz with a flat slicing \eqref{c38}.

In the first subcase, with the spherical slicing, we have  $\a=0$ and $\b \geq 1$ (see below equation \eqref{case1} for an example). Higher integer values of $\b$ require more coefficients of the potential to vanish: $V_0 = \dots = V_{\b +1}=0$. Moreover, the assumption that $T>0$ further implies $V_{\b+2}\f^\b>0$. Additionally, it implies that the energy density $\rho$ increases from zero as we depart from the boundary, see Eqs. (\ref{invMink1},\ref{invMink21},\ref{invMink31},\ref{invMink41},\ref{invMink51}). The curvature invariants vanish as we approach the solution, while the scale factor that controls the size of the sphere S$^{(d-1)}$ diverges. Hence, the geometry is identified as the spatial boundary of Minkowski space-time. This kind of solution constitutes a possible endpoint of the flow, since $W'=0$. Furthermore, it corresponds to minima (maxima) of the superpotential for $W>0$ ($W<0$).

The second subclass exists for $2\a + \b \equiv m+5 \geq 5$, and in particular $\alpha=3+m$ and $\beta=-m-1$. Now the potential vanishes up to $V_0=\dots=V_{m+2}=0$. The solution exists only in the ansatz with a flat slicing due to the fact that the inverse scale factor $T$ vanishes identically. The metric can be shown to be asymptotically flat. An explicit example (for $m=0$) can be found in equation \eqref{184} and the subsequent discussion.

\item {\bf Extremal Flat Minkowski  Horizons}. These are flat extremal horizons (with infinite volume)  in locally Minkowski space (zero curvature).  Similarly to the previous case, they appear for $\a=0$ and $\b>1$, and require some fine-tuning of the potential: $V_0 = \dots = V_{\b +1}=0$. The inverse scale factor $T$ vanishes identically, which implies that these asymptotics only appear in the ansatz with a flat slicing \eqref{c38}. In this case the scale factor controlling the size of the slices also diverges but the temporal component of the metric $g_{tt}$ vanishes, signalling the presence of a horizon.
    Using (\ref{n105}), we can compute the Hawking temperature of such horizons and it is vanishing. We conclude that these asymptotics correspond to flat extremal horizons.

     For an explicit example, see the discussion below equation \eqref{case2}. Such solutions constitute possible endpoints of the flow because $W'=0$. Additionally, they are minima (maxima) of the superpotential for $W>0$ ($W<0$).

\item {\bf Non-extremal horizons}. They appear in two different incarnations. Firstly, in the discussion below equation \eqref{neh}, we have $\a=0$ and $\b=1$. These are the standard horizons and they match the description given in Appendix \ref{sho}. Secondly, they appear for $\a=1$ and $\b=1$ (see discussion below \ref{neh2}). In this case the superpotential vanishes and this is a special case of the previous example. The monotonicity of $A$ is compromised in the second type. In both cases, we have $W'\neq 0 $ and therefore the flow cannot stop at non-extremal horizons.

\item {\bf Nariai (Extremal) horizons}. Here, the blackening function $f$ has a double zero and this only happens in the de Sitter regime. The local geometry is similar to the extremal horizon of a Nariai black hole in de Sitter space. They have $\b=2$ while $\a=0$, see the discussion below \eqref{89}. Interestingly, the extremal horizons also appear for $2 \a + \b=4$, see equation \eqref{156} and below. Note that these are distinct from the solutions for dS$_2$ boundaries, the distinction arising from the range of values for $\delta_\pm$, which is ultimately controlled by the ratio $V_2/V_0$ (see equation \ref{157}). Both situations are captured and described in detail in appendices \ref{seho} and \ref{seho2}. In both cases, the assumption that $T>0$ implies that $V_0>0$, i.e. these horizons are allowed only in the dS regime.
    The extremal horizons of appendix \ref{seho} do not correspond to end-points of the flow, and we do not consider them further. However, the extremal horizons of appendix \ref{seho2} are end-points of the flow, and they are therefore interesting for our purposes.  The endpoints of appendix  \ref{seho2} appear as minima (maxima) of the superpotential for $W>0$ ($W<0$).

\item {\bf Bounce points}. These appear for $\a = 3/2$ and $\b=0$, see equation \eqref{bcp} and the discussion below.
    As it happens in any bounce point, $\dot{\f}=0$ but $\ddot{\f}\neq 0$. Therefore, the flow reverses direction without stopping. The fact that $W$ vanishes implies a monotonicity change  for $A$.
     Therefore these are points both both $\f$ and $e^A$ change monotonicity.

     In addition, these bounce points appear only in the dS regime ($V_0>0$) under the assumption that $T>0$. Note that the standard bounce points, those described in Appendix \ref{bounces}, include the ones found in this appendix in the particular case in which the leading coefficient of $W$ vanishes. Moreover, the bounce points described in Appendix \ref{bounces} would appear here for $\a=\b=0$ if we allowed for half-integer powers of $\f$ in the ansatz for $W$ in equation \eqref{eahb}.
    Bounce points do not correspond to end-points of the flow.
\end{itemize}

\vskip 1cm
\section{Perturbative solutions II: solutions around a singular point with $W'=0$}\label{WcApp}
\vskip 1cm

As alluded to in Appendix \ref{structure}, singular points of the superpotential equation reveal interesting features in the space of solutions. We next discuss these features in detail, starting with the class characterised by \refeq{WXp}. These solutions will be of the form \refeq{genWX}, with $W_1 = 0$.

\subsection{Extrema of the scalar potential: solutions near $V'=W'=0$} \label{D.2.3}

We begin by studying solutions near an extremum of the potential (that we arrange via shifts in $\f$ to occur at $\f=0$). Accordingly, we assume that the potential around this point is of the form
\be
V(\f) = V_0 + \sum^\infty_{n=2}\frac{V_n}{n!}\f^n\label{VxCP}
\ee
and investigate under what conditions a solution to \refeq{w56_1b} exists. There are various possibilities, which we distinguish by the behaviour of the  {\it leading} term in the expansion around the singular point:

\begin{enumerate}
\item $\hat{W}_n= \tilde{W}_n=0$

These are solutions whose leading behaviour is analytic around the singular point. In this case, equation \eqref{w56_1b} gives to leading order (up to a constant non-vanishing prefactor):

\begin{equation}\label{ind}
W_2   \left[(d-2) W_0-4 (d-1) W_2\right] \left[d V_2 W_0^2+2 d V_0 W_2 (4 W_2-W_0)-8
   V_0 W_2^2\right]\f =0
\end{equation}

Solutions in this class exist for:
\begin{itemize}

\vskip 0.3cm

\item $W_2 = W_0 = 0$

Such solutions either give rise to the trivial solution (which is to say $W = 0$), or they arise as special cases of the endpoints discussed in Appendix \ref{g53}, in which the parameter $\delta_\pm$, defined in Eq. \eqref{alfa}, is of the form $\delta_\pm =1/n$, with $n\in \mathbb{Z}$. These are endpoints with local dS$_2\times $S$^{d-1}$ geometry.

\vskip 0.3cm
\item $W_2 = 0$, $W_0\ne 0$

Only the trivial solution $W = W_0$ exists unless $V_2$ = 0. In this case, the solution is contained in the $W_2^-$ branch of solutions introduced below.

\vskip 0.3cm
\item $W_2^\pm =\frac{\Delta_\pm}{2(d-1)}W_0$

Where we have defined
\be
\Delta_\pm \equiv \frac{d}{2} \pm\sqrt{\frac{d^2}{4}-d(d-1)\frac{V_2}{V_0}}\label{DEL}
\ee
These solutions govern ``endpoints" of the flows, in which the solution terminates at an (A)dS$_{d+1}$ {boundary}.

\vskip 0.3cm
\item $W_2 = \frac{(d-2)}{2(d-1)}W_0$

In this case the solution can be determined iteratively. It can be shown by induction that the solution obtained solves
\be
2 (d-1) (W'')^2+(2-d) W W''+(d-2) W'^2-2 (d-1) W^{(3)} W'=0
\label{C120z}\ee
to all orders. This is precisely the case where the denominator of Eqs. \eqref{w56_6} and \eqref{w56_7} vanishes. Additionally, all the coefficients in Eq. \eqref{w56_1b} vanish identically for this solution. The potential $V$ that gives rise to this solution is determined indirectly through Eq. \eqref{w55}. We can obtain the local behaviour of the metric functions and of the potential from Eqs. \eqref{C120}, \eqref{eqtt}, \eqref{f10_1} and \eqref{w55}:
\begin{equation}
W = W_0 + \dfrac{(d-2)}{4(d-1)}W_0\f^2 + \frac{1}{6}W_3\f^3 + \dots \qquad T =0 \,,
\end{equation}
\begin{equation}
f = -\frac{4(d-1)V_0}{d W_0^2}\,,\qquad V = V_0 + \dfrac{(d-2)V_0}{d(d-1)}\f^2+\dots
\end{equation}
where both $W_0$ and $W_3$ are integration constants. Locally, this solution is equivalent to (A)dS$_{d+1}$ boundary endpoints with $\Delta_+ = d-2$.

The exact solution to Eq. \eqref{C120z} is discussed in detail in appendix \ref{app:I}.

\end{itemize}

\item $\alpha \ne 0,1$, and $\tilde{W}_n =0$

This is a solution which is {\it generically} non-analytic to leading order around the singular point. The relevant indicial equation is
\be\label{idn}
(\alpha-2)^2(d-1)V_2-(\alpha-3)V_0 = 0
\ee
with solutions
\be
\alpha = 2+\frac{1}{\delta_\pm} \qquad \mathrm{where} \qquad \delta_\pm \equiv \frac{1}{2}\pm\sqrt{\frac{1}{4}-(d-1)\frac{V_2}{V_0}} .
\ee
Note the similarity between the $\delta_\pm$ appearing here, and the $\Delta_\pm$ defined in \refeq{DEL} above.

It is important to keep in mind that for special values of $\delta_\pm$, the Frobenius index $\alpha$ may be integer valued. When this happens, the solution's leading behaviour near the singular point is analytic, and therefore the solution may overlap with those itemized above. In any case, solutions in this branch are locally dS$_2 \times$ S$^{d-1}$.
\end{enumerate}

Summarizing, we find that for solutions near a point where $V' = W' = 0$, the local form of the solution is either trivial (the scalar does not run), an (A)dS$_{d+1}$ boundary, or a dS$_2 \times$ S$^{d-1}$ region. We shall find that the latter may correspond to either a boundary or an extremal horizon.

The details of these solutions can depend importantly on the precise form of the scalar potential $V$, in particular, both the magnitude and sign of the coefficients $V_n$ entering \refeq{VxCP}. In what follows, we further explore these solutions for the special case that $V_0 = -{d(d-1)\over \ell^2}<0$, corresponding to what we term an ``AdS region" of space, and for $V_0 = d(d-1)H^2>0$ corresponding to a ``dS region".

\vskip 0.3cm
\subsubsection{Extrema in an AdS region: AdS$_{d+1}$ endpoints} \label{AdSX}

\vskip 0.3cm

Here we consider the solutions introduced above, in the special case where
\be
V_0 = -\frac{d(d-1)}{\ell^2} \qquad \mathrm{and} \qquad V_2 = m^2\,, \qquad \mathrm{where} \qquad \ell^2m^2 = \Delta\left(\Delta-d\right).
\ee
Note that the solutions $\Delta_\pm$ to the quadratic equation above are the same as those appearing in Eq.  \refeq{DEL}:

\begin{equation}\label{dD}
\Delta_{\pm} = \frac{d}{2} \pm\sqrt{\frac{d^2}{4} +   m^2 \ell^2}\,.
\end{equation}

We have seen that the possible non-trivial solutions to the superpotential equation near singular points where $V'=W'=0$ are classified as either (A)dS$_{d+1}$ boundary endpoints or dS$_2 \times$ S$^{d-1}$ regions  (boundaries or extremal horizons). Consider first the AdS$_{d+1}$ boundary endpoint branch of solutions, in which $W_2^\pm = \Delta_\pm W_0/2(d-1)$.  The superpotential can be determined iteratively---for example the first few terms are
\be
{W^{\pm}_3\over W_0}={\ell^2 V_3\over  2(d-1)(3\Delta_{\pm}-d)}
\ee
and
\be
 {W^{\pm}_4\over W_0}={ 3d\Delta_{\pm}^2(3\Delta_{\pm}-d)^2 -6(d-1) \ell^4 V_3^2 +2(d-1)
   (3\Delta_{\pm}-d)^2 \ell^2 V_4\over 4(d-1)^2 (3\Delta_{\pm}-d)^2 (4\Delta_{\pm}-d)}.
\label{C17}\ee
Taking the limit $\f \to 0$ in which we approach the singular point for equations \eqref{w56_6} and \eqref{w56_7} we find that $f$ is constant and $T$ vanishes:

\be
f^{\pm} = {4(d-1)^2\over \ell^2 W_0^2}\sp T^{\pm} = 0.
\label{C18}\ee

The power series solution found in this way, is the same as for superpotentials in an ansatz with flat slicing (c.f. \cite{KT}). The differences will appear in subleading non-analytic contributions, which we find by solving \eqref{eqpe}:

\begin{align}\label{sgs}
 & \frac{d \Delta_{\pm} ^2 W_0^3 \left(d^2-d (3 \Delta_{\pm} +2)+2 \Delta_{\pm}  (\Delta_{\pm} +2)\right)}{4
   (d-1)^3 \ell^2}\f  \delta W
 +\frac{\Delta_{\pm} ^4 W_0^3  (d-\Delta_{\pm} -2) }{4 (d-1)^2 \ell^2}\f^3 \delta W^{(4)}
   \nonumber \\
  &+\frac{\Delta_{\pm} ^2 W_0^3   (d-\Delta_{\pm} -2) \left(d^2+4 d-\Delta_{\pm}  (\Delta_{\pm}+2)\right) }{4 (d-1)^2 \ell^2}\f \delta W'' -\nonumber \\
&-\frac{(\Delta_{\pm} +2) \Delta_{\pm}  W_0^3
   (d-\Delta_{\pm} -2) \left(d^2-\Delta_{\pm} ^2\right) \delta W'}{4 (d-1)^2
   \ell^2}-\nonumber\\
   &-\frac{\Delta_{\pm} ^3 W_0^3  \left(d^2-d (2 \Delta_{\pm} +1)+\Delta_{\pm} ^2+\Delta_{\pm} -2\right)
   }{2 (d-1)^2 \ell^2}\f^2 \delta W^{(3)}
   +\dots = 0
\end{align}
where the dots contain subleading contributions to the coefficients of $\delta W$ and its derivatives.\footnote{We have assumed that $m^2\neq0$, or equivalently that $\Delta_{+}\neq d$ and $\Delta_-\neq0$. For a vanishing mass the operator in the dual field theory is marginal. In such case, the leading order of equation \eqref{sgs} changes and the solution is also different. See \cite{Bourdier13} for an example in the flat sliced ansatz.} The solution to the previous equation is given by

\begin{align}\label{edw}
\delta W^{\pm} =& C_0 + \dfrac{C_W}{\ell} \f^{d/\Delta_{\pm}} -\dfrac{C_T}{\ell}\frac{ \Delta_{\pm} ^2 (-d+\Delta_{\pm} +2) }{2 (\Delta_{\pm} +1)
   (d-2 (\Delta_{\pm} +1))} \f^{2+2/\Delta_{\pm}} \nonumber \\ &- \dfrac{C_f}{\ell} \frac{ \Delta_{\pm} ^3 }{2 d (d+2 \Delta_{\pm}
   )} \f^{2+d/\Delta_{\pm}} + \dots
\end{align}
The integration constants are $C_0$, $C_W$, $C_T$ and $C_f$. The derivation of Eq. \eqref{sgs} requires the knowledge of the leading solution up to $\f^2$. As a result, the consistency of the solution requires that $\delta W$ vanishes faster that $\f^2$, which immediately implies $C_0 = 0$. Whether the remaining three powers in $\f$ are subleading or not depends on the value of $\Delta_+$ ($\Delta_-$ respectively), which in turn depends on the sign of $m^2$. Prior to distinguishing the subcases, we extract the corrections $\delta f$ and $\delta T$ to the leading order solution for $f$ and $T$ in \eqref{C18} from equations \eqref{w56_6} and \eqref{w56_7} . Then

\begin{align}\label{df1}
\delta f^{\pm} =\frac{8 (d-1)^3 }{\ell^3 W_0^3}\left( C_f \frac{ \Delta_{\pm} }{d } \f^{d/\Delta_{\pm} } -C_T\f ^{2/\Delta_{\pm} }\right) + \dots
\end{align}

\begin{align} \label{dt1}
\delta T^{\pm} = \dfrac{2(d-1)}{\ell W_0} \dfrac{C_T}{\ell^2} \f^{2/\Delta_{\pm}} + \dots
\end{align}
The ellipsis contains higher-order terms which only depend on the shown integration constants, in other words $\delta f$ does not depend on $C_W$ while $\delta T$ does not depend on $C_f$ or $C_W$.

The solution is guaranteed to be regular, as long as the terms in $\delta W$ \eqref{edw} are finite. In particular, the Ricci scalar \eqref{w58} is given by

\begin{align}
R^{\pm} &= \left(-\frac{d(d-1)}{\ell^2} + O(\f)\right) + C_W\frac{2 (d-1) d}{\ell^3 W_0}\f^{d/\Delta_{\pm}} -C_f\frac{(d-1) \Delta_{\pm} ^3}{d \ell^3 W_0}\f^{2+d/\Delta_{\pm}}\nonumber \\
 &+ C_T \frac{2 (d-1) \Delta_{\pm} ^2 (d-\Delta_{\pm} -2)}{\ell^3 W_0 (d-2 (\Delta_{\pm} +1))} \f ^{2+2/\Delta_{\pm}} +\dots
\end{align}
Additionally, the quantities controlling the curvature invariants (pressure, energy density and $\mathcal{I}$), introduced in Eqs. (\ref{ev3b},\ref{ev3c},\ref{ev7}), take the following form:
\begin{equation}\label{invads1}
\rho^{\pm} = \dfrac{d(d-1)}{\ell^2} + \dfrac{d\Delta_{\pm}}{2\ell^2}\varphi^2 + \dots \sp p^{\pm} = -\dfrac{d(d-1)}{\ell^2} + \dfrac{\Delta_{\pm}(2\Delta_{\pm}-d)}{2\ell^2}\varphi^2 + \dots
\end{equation}
\begin{equation}\label{invads3}
\mathcal{I}^{\pm} = \dfrac{1}{\ell^2} + \dfrac{\Delta_{\pm}}{2(d-1)\ell^2}\varphi^2 + \dots
\end{equation}
Note that the terms proportional to the integration constants $C_W, C_f, C_T$ appear in subleading contributions, collectively denoted with dots.

We now distinguish under which conditions the contributions to \eqref{edw} are subleading:

\begin{itemize}
\item Maxima in the AdS regime: $m^2<0$. \\
According to the definition of $\Delta_{\pm}$ \eqref{dD}, and assuming that the BF bound $m^2>-d^2/4\ell^2$ is respected, we have $0<\Delta_-<d/2$ and $d/2<\Delta_+<d$. Note that for both the $+$ and $-$ branches, the deformations in \eqref{edw} proportional to $C_T$ and $C_f$ are allowed, since $2+2/\Delta_{\pm}>2$ and $2+d/\Delta_{\pm}>2$. However, the deformation proportional to $C_W$ is only subleading for the $-$ branch, since $d/\Delta_->2$ but $d/\Delta_+<2$. Consequently, we find

\begin{align}\label{dwm1}
W_- &= W_0\left(1 + \dfrac{\Delta_{-}}{4(d-1)}\f^2 + O(\f^3)\right) + \dfrac{C_W}{\ell} \f^{d/\Delta_{-}}\nonumber \\
 &-\dfrac{C_T}{\ell}\frac{ \Delta_{-} ^2 (-d+\Delta_{-} +2) }{2 (\Delta_{-} +1)
   (d-2 (\Delta_{-} +1))} \f^{2+2/\Delta_{-}} - \dfrac{C_f}{\ell} \frac{ \Delta_{-} ^3 }{2 d (d+2 \Delta_{-}
   )} \f^{2+d/\Delta_{-}} + \dots
\end{align}

\begin{align}\label{dwp1}
W_+ &= W_0\left(1 + \dfrac{\Delta_{+}}{4(d-1)}\f^2 + O(\f^3)\right)\nonumber \\
 &-\dfrac{C_T}{\ell}\frac{ \Delta_{+} ^2 (-d+\Delta_{+} +2) }{2 (\Delta_{+} +1)
   (d-2 (\Delta_{+} +1))} \f^{2+2/\Delta_{+}}  - \dfrac{C_f}{\ell} \frac{ \Delta_{+} ^3 }{2 d (d+2 \Delta_{+}
   )} \f^{2+d/\Delta_{+}} + \dots
\end{align}
for each branch respectively. While

\begin{align}\label{dfb}
 f^{\pm} = {4(d-1)^2\over \ell^2 W_0^2} +  \frac{8 (d-1)^3 }{\ell^3 W_0^3}\left( C_f \frac{ \Delta_{\pm} }{d } \f^{d/\Delta_{\pm} } -C_T\f ^{2/\Delta_{\pm} }\right) + \dots
\end{align}

\begin{align} \label{dtb}
T^{\pm} = \dfrac{2(d-1)}{\ell W_0} \dfrac{C_T}{\ell^2} \f^{2/\Delta_{\pm}} + \dots
\end{align}

for both branches.

\item Minima in the AdS regime: $m^2>0$.\\
From the definition \eqref{dD} we find $\Delta_-<0$ and $\Delta_+>d$.

Interestingly, none of the terms in the solution \eqref{edw} are subleading for the $-$ branch because $d/\Delta_-<2$, $2 + d/\Delta_-<2$ and  $2 + 2/\Delta_-<2$. Therefore

\begin{align}\label{dwmc}
W_- &= W_0\left(1 + \dfrac{\Delta_{-}}{4(d-1)}\f^2 + O(\f^3)\right)
\end{align}
\be
f^{-} = {4(d-1)^2\over \ell^2 W_0^2}\sp T^{-} = 0.
\label{C188}\ee

The fact that there are no deformations in the $-$ branch, along with $T=0$ in the leading solution \eqref{C188}, implies that this solution only appears for a flat slicing, and is therefore incompatible with the spherical slicing we study throughout this work.

Conversely, the $+$ branch of the solution does not admit the deformation proportional to $C_W$ but it does admit the other two, since $d/\Delta_+<2$ but $2 + d/\Delta_+>2$ and  $2 + 2/\Delta_+>2$. Therefore, around minima in the AdS regime we obtain

\begin{align} \label{dwp2}
W_+ &= W_0\left(1 + \dfrac{\Delta_{+}}{4(d-1)}\f^2 + O(\f^3)\right) -\dfrac{C_T}{\ell}\frac{ \Delta_{+} ^2 (-d+\Delta_{+} +2) }{2 (\Delta_{+} +1)
   (d-2 (\Delta_{+} +1))} \f^{2+2/\Delta_{+}}\nonumber \\
    - &\dfrac{C_f}{\ell} \frac{ \Delta_{+} ^3 }{2 d (d+2 \Delta_{+}
   )} \f^{2+d/\Delta_{+}} + \dots
\end{align}

\begin{align}\label{dfc}
 f^+ = {4(d-1)^2\over \ell^2 W_0^2} +  \frac{8 (d-1)^3 }{\ell^3 W_0^3}\left( C_f \frac{ \Delta_+ }{d } \f^{d/\Delta_+ } -C_T\f ^{2/\Delta_+ }\right) + \dots
\end{align}

\begin{align} \label{dtc}
T^+ = \dfrac{2(d-1)}{\ell W_0} \dfrac{C_T}{\ell^2} \f^{2/\Delta_+} + \dots
\end{align}

\end{itemize}

These are AdS$_{d+1}$ boundary endpoints. Their interpretation as such is presented in section \ref{sec52}.

\subsubsection{Extrema in a dS region: dS$_{d+1}$ endpoints} \label{dSX}
Here we consider the solutions introduced above in the special case where
\be
V_0 = d(d-1)H^2\qquad \mathrm{and} \qquad V_2 = m^2 \qquad \mathrm{where} \qquad \frac{m^2}{H^2} = \Delta\left(\Delta-d\right).
\ee
Note in particular that the solutions $\Delta_\pm$ to the quadratic equation above are the same as those appearing in \refeq{DEL}:

\begin{equation}\label{dD2}
\Delta_{\pm} = \frac{d}{2} \pm\sqrt{\frac{d^2}{4} -   \frac{m^2}{H^2}}\,.
\end{equation}

Consider the dS$_{d+1}$ boundary endpoint branch of solutions, in which $W_2^\pm = \Delta_\pm W_0/2(d-1)$.  The discussion closely mirrors that of the previous section. Again the superpotential can be determined iteratively---for example the first few terms are
\be
{W^{\pm}_3\over W_0}=-{V_3\over  2(d-1)(3\Delta_{\pm}-d)H^2}
\ee
and
\be
 {W^{\pm}_4\over W_0}={ 3d\Delta_{\pm}^2(3\Delta_{\pm}-d)^2H^4 -6(d-1)  V_3^2 -2(d-1)
   (3\Delta_{\pm}-d)^2 V_4H^2\over 4(d-1)^2H^4 (3\Delta_{\pm}-d)^2 (4\Delta_{\pm}-d)}.
\label{C17a}\ee
 The function
$f$ is again constant but now negative at the singular point:
\be
f^{\pm}=-{4(d-1)^2H^2\over W_0^2}\sp T^{\pm}=0.
\label{C18a}\ee
As was the case in for the AdS$_{d+1}$ endpoints, the solutions found as power series above contain a single integration constant, namely $W_0$. However, we expect the general solution to be determined by up to four arbitrary integration constants, since the master equation \eqref{w56_1b} is fourth order in derivatives. Once again we expect that the ``missing" integration constants in our series solution control subleading non-analytic behaviour in the superpotential. We find them by solving \eqref{eqpe}, which coincides with \eqref{sgs} of the previous section up to an overall constant.
The solution $\delta W$ is given by

\begin{align}\label{edw2}
\delta W^{\pm} =& C_0 + H C_W \f^{d/\Delta_{\pm}} +H C_T\frac{ \Delta_{\pm} ^2 (-d+\Delta_{\pm} +2) }{2 (\Delta_{\pm} +1)
   (d-2 (\Delta_{\pm} +1))} \f^{2+2/\Delta_{\pm}} \nonumber \\ &+ H C_f \frac{ \Delta_{\pm} ^3 }{2 d (d+2 \Delta_{\pm}
   )} \f^{2+d/\Delta_{\pm}} + \dots
\end{align}
We need to set $C_0 = 0$ for the solution to be subleading. The corrections $\delta f$ and $\delta T$ to the leading order solution for $f$ and $T$ in \eqref{C18a} are obtained from equations \eqref{w56_6} and \eqref{w56_7}:

\begin{align}\label{df2}
\delta f^{\pm} =H^3\frac{8 (d-1)^3 }{ W_0^3}\left( C_f \frac{ \Delta_{\pm} }{d } \f^{d/\Delta_{\pm} } -C_T\f ^{2/\Delta_{\pm} }\right) + \dots
\end{align}

\begin{align} \label{dt2}
\delta T^{\pm} = H^3 \dfrac{2(d-1)}{ W_0} C_T \f^{2/\Delta_{\pm}} + \dots
\end{align}

The solutions presented above are regular provided that $\delta W$ remains finite as we approach the extremum of $V$ at $\f\to 0$. The Ricci scalar is now given by

\begin{align}
R^{\pm} &= \left(H^2 d(d-1) + O(\f)\right) - H^3 C_W\frac{2 (d-1) d}{ W_0}\f^{d/\Delta_{\pm}} -C_f\frac{(d-1) \Delta_{\pm} ^3}{d  W_0}H^3\f^{2+d/\Delta_{\pm}}\nonumber \\
 &+ C_TH^3 \frac{2 (d-1) \Delta_{\pm} ^2 (d-\Delta_{\pm} -2)}{ W_0 (d-2 (\Delta_{\pm} +1))} \f ^{2+2/\Delta_{\pm}} +\dots
\end{align}
In addition, the pressure, energy density and $\mathcal{I}$, introduced in Eqs. (\ref{ev3b},\ref{ev3c},\ref{ev7}), that control the curvature invariants, take the following form:
\begin{equation}\label{invds1}
\rho^{\pm} = -d(d-1)H^2 + \dfrac{\Delta_{\pm}(d-2\Delta_{\pm})}{2}H^2\varphi^2 + \dots  \sp  p^{\pm} = d(d-1) H^2 - \dfrac{d\Delta_{\pm}}{2}H^2\varphi^2 + \dots
\end{equation}
\begin{equation}\label{invds2}
\mathcal{I}^{\pm} = - H^2 -  \dfrac{\Delta_{\pm}}{2(d-1)}H^2\varphi^2 + \dots
\end{equation}
Note that the terms proportional to the integration constants $C_W, C_f, C_T$ appear in subleading contributions, collectively denoted with dots.

Similarly to the previous section, we proceed to discuss which of the deformations in \eqref{edw2} is allowed depending on the value of $m^2$.

\begin{itemize}
\item Minima in the dS regime: $m^2>0$. \\
From $\Delta_{\pm}$ \eqref{dD2}, and assuming that the analogous of the BF bound in dS regions $m^2<H^2d^2/4$ is respected, we have $0<\Delta_-<d/2$ and $d/2<\Delta_+<d$. For either of the $\pm$ branch, the deformations in \eqref{edw2} proportional to $C_T$ and $C_f$ are allowed, since $2+2/\Delta_{\pm}>2$ and $2+d/\Delta_{\pm}>2$. However, the deformation proportional to $C_W$ is only subleading for the $-$ branch because $d/\Delta_->2$ but $d/\Delta_+<2$. Consequently, we find

\begin{align}\label{dwm3}
W_- &= W_0\left(1 + \dfrac{\Delta_{-}}{4(d-1)}\f^2 + O(\f^3)\right) +  H C_W \f^{d/\Delta_{-}}\nonumber \\
 &+H C_T \frac{ \Delta_{-} ^2 (-d+\Delta_{-} +2) }{2 (\Delta_{-} +1)
   (d-2 (\Delta_{-} +1))} \f^{2+2/\Delta_{-}} +H C_f \frac{ \Delta_{-} ^3 }{2 d (d+2 \Delta_{-}
   )} \f^{2+d/\Delta_{-}} + \dots
\end{align}

\begin{align}\label{dwp3}
W_+ &= W_0\left(1 + \dfrac{\Delta_{+}}{4(d-1)}\f^2 + O(\f^3)\right)\nonumber \\
 &+H C_T\frac{ \Delta_{+} ^2 (-d+\Delta_{+} +2) }{2 (\Delta_{+} +1)
   (d-2 (\Delta_{+} +1))} \f^{2+2/\Delta_{+}}  +H C_f \frac{ \Delta_{+} ^3 }{2 d (d+2 \Delta_{+}
   )} \f^{2+d/\Delta_{+}} + \dots
\end{align}
for each branch respectively, while both branches have

\begin{align}\label{df2f}
 f^{\pm} =-{4(d-1)^2H^2\over W_0^2} + H^3\frac{8 (d-1)^3 }{ W_0^3}\left( C_f \frac{ \Delta_{\pm} }{d } \f^{d/\Delta_{\pm} } -C_T\f ^{2/\Delta_{\pm} }\right) + \dots
\end{align}

\begin{align} \label{dtf2}
 T^{\pm} = H^3 \dfrac{2(d-1)}{ W_0} C_T \f^{2/\Delta_{\pm}} + \dots
\end{align}

\item Maxima in the dS regime: $m^2<0$.\\
From the definition \eqref{dD2} we find $\Delta_-<0$ and $\Delta_+>d$.

In this case, none of the terms in the ``$-$" branch of the solution \eqref{edw2} are subleading because $d/\Delta_-<2$, $2 + d/\Delta_-<2$ and  $2 + 2/\Delta_-<2$ and we have

\begin{align}\label{dwmd3}
W_- &= W_0\left(1 + \dfrac{\Delta_{-}}{4(d-1)}\f^2 + O(\f^3)\right) \sp f^{-}=-{4(d-1)^2H^2\over W_0^2}\sp T^{-}=0.
\end{align}

The fact that there are no deformations in the ``$-$" branch, along with $T=0$ in the leading solution \eqref{C18a}, implies that this solution only appears for a flat slicing---it is incompatible with the global spherical slicing.

Conversely, the $+$ branch of the solution admits the deformations proportional to $C_T$ and $C_f$ but does not permit the other one because $d/\Delta_+<2$ but $2 + d/\Delta_+>2$ and  $2 + 2/\Delta_+>2$. Therefore, around maxima in the dS regime we have

\begin{align} \label{d2wp2}
W_+ &= W_0\left(1 + \dfrac{\Delta_{+}}{4(d-1)}\f^2 + O(\f^3)\right) +H C_T \frac{ \Delta_{+} ^2 (-d+\Delta_{+} +2) }{2 (\Delta_{+} +1)
   (d-2 (\Delta_{+} +1))} \f^{2+2/\Delta_{+}}\nonumber \\
    + &H C_f \frac{ \Delta_{+} ^3 }{2 d (d+2 \Delta_{+}
   )} \f^{2+d/\Delta_{+}} + \dots
\end{align}

\begin{align}\label{df22f}
 f^{+} =-{4(d-1)^2H^2\over W_0^2} + H^3\frac{8 (d-1)^3 }{ W_0^3}\left( C_f \frac{ \Delta_+ }{d } \f^{d/\Delta_+ } -C_T\f ^{2/\Delta_+ }\right) + \dots
\end{align}

\begin{align} \label{dt2f2}
 T^{+} = H^3 \dfrac{2(d-1)}{ W_0} C_T \f^{2/\Delta_+} + \dots
\end{align}

\end{itemize}

These are dS$_{d+1}$ boundary endpoints. Their interpretation as such is presented in section \ref{sec52}.

\subsubsection{Extrema in a dS region: dS$_{2}\times$S$^{d-1}$ solutions} \label{g53}
The remaining branch of solutions in this class were advertised as dS$_{2}\times$S$^{d-1}$ ``regions'', and in this section we explain in what sense this is true. The analysis will show that such solutions are only consistent when $V_0>0$, which is to say in a dS regime, but we  begin by not imposing any sign on $V_0$.

We have seen that the leading behaviour of such solutions near the singular point corresponds to a superpotential of the form \refeq{genWX} such that $W_n=\tilde W_n = 0\,\,\forall n$.

Solving the indicial equation \eqref{idn} led to\footnote{The solutions $\alpha=0$ or $\alpha=1$ correspond to a Taylor series and are excluded by assumption.}

\begin{equation}\label{alfa}
\alpha_\pm = 2 + \dfrac{1}{\delta_\pm} \, \, \ \ \textrm{with} \ \  \delta_{\pm}= \dfrac{1}{2} \left(1\pm\sqrt{1-\dfrac{4(d-1)V_2}{V_0}}\right)\in \left[\frac{1}{2},\pm \infty \right)\,,
\end{equation}
where $\delta_{\pm}$ can be equivalently defined through
\begin{equation}
V_2 = \dfrac{V_0}{d-1} \delta(1-\delta)\,.
\end{equation}
Note the similarity between $\delta_{\pm}$ and the conformal dimension of the dual operator to a scalar field in AdS$_2$.
Equation \eqref{alfa} states that the leading behaviour of the superpotential is $W\sim \f^{2+1/\delta_{\pm}}$ as $\f\to 0$. Solving Eq. \eqref{w56_1b} iteratively gives

\begin{align}
W_\pm = W_0\f^{2+1/\delta_\pm} \left(1 -\frac{(d-1) (\delta_\pm -1) (2 \delta_\pm +1) V_3 }{2 \delta_\pm ^2 \left(9 \delta_\pm ^2-1\right)
\label{wds22}   V_0}\f + O(\f^2) \right)
\end{align}
The regularity of the solution depends on the range of $\delta_{\pm}$, which is ultimately dictated by the ratio $V_2/V_0$. Previous to discussing the allowed ranges, we complete the solution by computing $f$ and $T$ and studying fluctuations around the given solution.

The blackening function $f$ as well as the inverse scale factor $T$ are obtained by direct substitution of the previous equation \eqref{wds22} onto Eqs. \eqref{w56_6} and \eqref{w56_7}, giving

\begin{equation}\label{fds2}
f^{\pm} = -\frac{\delta_\pm ^2 }{ (d-1) (2 \delta_\pm +1)^2}\frac{\f ^{-2/\delta_\pm }}{W_0^2}\left( \delta_\pm ^4 V_0 + \dfrac{(d-1)V_3}{1-3\delta_{\pm}}\f + \dots \right)\,,
\end{equation}

\begin{equation}\label{tds2}
T^{\pm} = \dfrac{V_0}{(d-1)(d-2)}\left(1  + \dfrac{\delta_{\pm}}{2 (-1 + d) (1 + 2 \delta_{\pm})} \f^2+\dots\right)\,,
\end{equation}
The fact that $T$ is positive by definition together with the solution \eqref{tds2} reveals that such solutions can exist only in the dS regime (i.e. $V_0>0$). Interestingly, had we placed a hyperbolic slicing in our ansatz instead of the spherical slicing of Eq \eqref{c39}, then the solution would exist in the AdS regime. Such solutions have been found in \cite{GKN}.

Note that the previous equations \eqref{wds22}-\eqref{tds2} are not valid for $\delta_{\pm}=1/3$ ($\delta_{\pm}=-1/3$ is incompatible with a regular solution).
In general, there will be exceptions whenever $\delta_{\pm}=1/n$, with $n$ a natural number. Indeed, interpreting $\delta_+$ as the conformal dimension of the perturbing operator in  AdS/CFT correspondence, it corresponds to a scaling dimension of a 1-dimension QFT (associated to the AdS$_2$/dS$_2$ asymptotics). As happens in higher dimensional cases, the n-th power of this operator will have dimension $n\delta_+$ and therefore when $n\delta_+=1$ this multi-trace operator is marginal, and this signals the appearance of logs in the expansion (\ref{wds22}).

This phenomenon has been studied in higher-dimensional cases before, \cite{Ahar}, and the qualitative behavior of flows are similar even in these cases.
We temporarily assume that $\delta_{\pm}\neq 1/n$. We shall later work in detail the exceptions  $\delta_{\pm}= 1/2$ (appearing at $4$-th order in Eq. \eqref{wds22}) and $\delta_{\pm}=1/3$.

The solution \eqref{wds22}-\eqref{tds2} has a single integration constant, $W_0$, out of the four allowed by the fourth order differential equation \eqref{w56_1b}. The remaining integration constants appear as further non-analytical contributions, which we find by perturbing the given solution, i.e. by solving \eqref{eqpe}. To leading order \eqref{eqpe} in this case is given by

\begin{align}
&\delta_{\pm} ^3 \f ^3 \delta W^{(4)}+2 (\delta_{\pm} -3) \delta_{\pm} ^2 \f ^2
   \delta W^{(3)}-\delta_{\pm}  \left(\delta_{\pm} ^2+2 \delta_{\pm} -11\right) \f
   \delta W'' \nonumber \\
   &+(\delta_{\pm} +1) \left(\delta_{\pm} ^2+\delta_{\pm} -6\right)
   \delta W'-\frac{\delta_{\pm}  \f  ((d+2) \delta_{\pm} +d-4) \delta W}{d-1} +\dots = 0
\end{align}
up to an overall non-vanishing constant. The solution to the previous equation around $\f \to 0$ is given by

\begin{equation}\label{pew}
\delta W_{\pm} = C_0 + C_1 \f ^{2+ 1/\delta_{\pm}} + C_2 \f^{2/\delta_{\pm}} + C_3 \f^{2+3/\delta_{\pm}} + \dots
\end{equation}
Requiring that the solution for $\delta W$ is subleading with respect to \eqref{wds22} already imposes $C_0 = C_1 = 0$. Whether the deformations proportional to $C_2$ and $C_3$ are allowed depends on the range of $\delta_{\pm}$ and will be discussed case-wise later in this section. The solution \eqref{pew} induces fluctuations in $f$ and $T$ that we obtain by direct substitution in \eqref{w56_6} and \eqref{w56_7}:

\begin{align}
\delta f^{\pm} = C_2\frac{2 \delta_\pm ^3 V_0  }{(d-1) (2 \delta_\pm +1)^3 W_0^3}\f^{-2-1/\delta_\pm} + C_3 \frac{2 \delta_\pm ^5 (2 \delta_\pm +3) V_0}{(d-1) (\delta_\pm +1) (2 \delta_\pm +1)^3 W_0^3} + \dots
\end{align}

\begin{align}
& \delta T^{\pm} = C_2 \frac{\delta_{\pm} ^2 (2 \delta_{\pm} -1) V_0}{(d-2) (d-1)^2 (2 \delta_{\pm} +1)^2 W_0}\f^{1/\delta_{\pm}}\nonumber \\ & - C_3 \frac{\delta_{\pm} ^2 V_0 }{(d-2) (d-1)^2 (\delta_{\pm} +1) (2 \delta_{\pm} +1)^2 W_0}\f^{2+2/\delta_{\pm}}+\dots
\end{align}

The quantities controlling the curvature invariants, $p$, $\rho$ and $\mathcal{I}$, introduced in Eqs. (\ref{ev3b},\ref{ev3c},\ref{ev7}), take the following form:
\begin{equation}\label{invds21}
\rho^{\pm} = - V_0 -\frac{\delta_\pm  \left(\delta_\pm ^3-\delta_\pm +1\right) V_0 }{2 (d-1)}\varphi ^2+ \dots \sp p^{\pm} =  V_0 -\frac{\delta_\pm  \left(\delta_\pm ^3-\delta_\pm +1\right) V_0 }{2 (d-1)}\varphi ^2 + \dots
\end{equation}
\begin{equation}\label{invds23}
\mathcal{I}^{\pm} = -\frac{V_0}{(d-1)(d-2)}-\frac{\delta_\pm  V_0 }{2 (d-2) (d-1)^2 (2 \delta_\pm +1)}\varphi ^2+ \dots
\end{equation}
where terms proportional to the integration constants $C_2, C_3$ appear in subleading contributions, collectively denoted with dots. We further evaluate the Ricci scalar from \eqref{w58}:

\begin{align}\label{ric}
R = &\left(\dfrac{d+1}{d-1}V_0  + O(\f)\right) +  C_2 \frac{(1-2 \delta_\pm ) \delta_\pm  V_0}{(d-1) (2 \delta_\pm +1) W_0} \f^{1/\delta_{\pm}}\nonumber \\ & -C_3\frac{\delta_\pm ^2 (2 \delta_\pm +3) V_0}{(d-1) (\delta_\pm +1) (2 \delta_\pm +1) W_0}\f^{2 + 2/\delta_{\pm}}+\dots
\end{align}

We now discuss the regularity of the solution, the allowed ranges for $\delta_{\pm}$ and the possible integration constants for the solution depending on whether we have a maximum or a minimum in the dS regime\footnote{Recall that \eqref{tds2} implies that we this solution exists in the dS regime. With a hyperbolic slicing we would have an AdS regime and the same conclusions for dS apply under the replacement maxima $\leftrightarrow$ minima.}.

\begin{itemize}
\item Minima in the dS regime. We parametrise $V_0 = (d-1)H^2 >0$ and $V_2 = m^2 >0$.

From the definition \eqref{alfa} we have that $0<\delta_-<1/2$ and $1/2<\delta_+<1$, where we have assumed that the analogue of the BF bound in dS$_2$ is satisfied: $m^2<H^2/4$.

In the $-$ branch, the deformations associated with $C_2$ and $C_3$ are both subleading, since $2 + 1/\delta_- < 2/\delta_-$ and $2+1/\delta_-<2+3/\delta_-$. In addition, the solution is regular since all the exponents are positive in \eqref{ric} and the Ricci scalar approaches a constant as $\f \to 0$. Therefore, for the $-$ branch we find

\begin{align}
W_- = W_0\f^{2+1/\delta_-} \left(1 + O(\f) \right) + C_2 \f^{2/\delta_{-}} + C_3 \f^{2+3/\delta_{-}} + \dots
\end{align}

\begin{align}
f^{-} &= -\frac{\delta_- ^2 }{ (d-1) (2 \delta_- +1)^2}\frac{\f ^{-2/\delta_- }}{W_0^2}\left( \delta_- ^2 V_0  + O(\f) \right) + C_2\frac{2 \delta_- ^3 V_0  }{(d-1) (2 \delta_- +1)^3 W_0^3}\f^{-2-1/\delta_-}\nonumber \\&+ C_3 \frac{2 \delta_- ^5 (2 \delta_- +3) V_0}{(d-1) (\delta_- +1) (2 \delta_- +1)^3 W_0^3} + \dots
\end{align}

\begin{align}
T_- &= \dfrac{V_0}{(d-1)(d-2)}\left(1  +O(\f^2)\right)+ C_2 \frac{\delta_{-} ^2 (2 \delta_{-} -1) V_0}{(d-2) (d-1)^2 (2 \delta_{-} +1)^2 W_0}\f^{1/\delta_{-}}\nonumber \\ & - C_3 \frac{\delta_{-} ^2 V_0 }{(d-2) (d-1)^2 (\delta_{-} +1) (2 \delta_{-} +1)^2 W_0}\f^{2+2/\delta_{-}}+\dots
\end{align}

As for the $+$ branch, the exponents satisfy $2+1/\delta_+ > 2/\delta_+$ and $2+1/\delta_+<2+3/\delta_+$, which in turns means that we have to set $C_2 = 0$ for consistency, while $C_3$ remains arbitrary. Hence,

\begin{align}
W_+ = W_0\f^{2+1/\delta_-} \left(1 + O(\f) \right) + C_3 \f^{2+3/\delta_{+}} + \dots
\end{align}

\begin{align}
f^{+} &= -\frac{\delta_- ^2 }{ (d-1) (2 \delta_- +1)^2}\frac{\f ^{-2/\delta_- }}{W_0^2}\left( \delta_- ^2 V_0  + O(\f) \right) +
\end{align}
$$
+ C_3 \frac{2 \delta_- ^5 (2 \delta_- +3) V_0}{(d-1) (\delta_- +1) (2 \delta_- +1)^3 W_0^3} + \dots
$$

\begin{align}
T^+ &= \dfrac{V_0}{(d-1)(d-2)}\left(1  +O(\f^2)\right)  - 
\end{align}
$$
-C_3 \frac{\delta_{+} ^2 V_0 }{(d-2) (d-1)^2 (\delta_{+} +1) (2 \delta_{+} +1)^2 W_0}\f^{2+2/\delta_{+}}+\dots
$$

From section \ref{rul}, we know that this is an endpoint of the flow if $W$ vanishes faster than $\f^{3/2}$. Since in both cases $\delta_{\pm}>0$ and $W\sim \f^{2+1/\delta_{\pm}}$, the condition is always satisfied and these solutions are possible endpoints of the flow.

\item Maxima in the dS regime: $V_0 = (d-1)H^2 >0$ and $V_2 = m^2 <0$.

In this case we have $-\infty<\delta_-<0$ and $1<\delta_+<\infty$. From a similar analysis than in the previous section we conclude that none of the deformations are subleading for the $-$ branch of the solutions. Consequently both $C_2=C_3=0$ and the solution is simply the leading one:

\begin{align}\label{f80}
W_- = W_0\f^{2+1/\delta_-} \left(1 + O(\f) \right)
\end{align}

\begin{align}
f^{-} &= -\frac{\delta_- ^2 }{ (d-1) (2 \delta_- +1)^2}\frac{\f ^{-2/\delta_- }}{W_0^2}\left( \delta_- ^2 V_0  + O(\f) \right)
\end{align}

\begin{align}\label{f82}
T_- &= \dfrac{V_0}{(d-1)(d-2)}\left(1  +O(\f^2)\right)
\end{align}

Again, the previous solution is a possible endpoint of the flow if $W$ vanishes faster than $\f^{3/2}$. Then the endpoints satisfy $2+1/\delta_->3/2$, which implies $\delta_-<-2$. If the inequality is saturated, the solution is a bounce point. Alternatively, the flow continues for $2+1/\delta_-=1$, i.e. $\delta_-=-1$. In the complementary range $\delta_-\in(-2,-1)\cup(-1,0)$, the solution is regular but the second derivative of $\f$ diverges: $\ddot{\f} = W'W''\sim \f^{1+2/\delta_-}$.

Finally, for the $+$ branch of the solutions the exponents satisfy $2+1/\delta_+ > 2/\delta_+$ and $2+1/\delta_+<2+3/\delta_+$. Accordingly, we set $C_2=0$ for consistency and the solution is:

\begin{align}
W_+ = W_0\f^{2+1/\delta_-} \left(1 + O(\f) \right) + C_3 \f^{2+3/\delta_{+}} + \dots
\end{align}

\begin{align}
f^{+} &= -\frac{\delta_- ^4 V_0}{ (d-1) (2 \delta_- +1)^2}\frac{\f ^{-2/\delta_- }}{W_0^2}\left( 1  + O(\f) \right) + C_3 \frac{2 \delta_- ^5 (2 \delta_- +3) V_0}{(d-1) (\delta_- +1) (2 \delta_- +1)^3 W_0^3} + \dots
\end{align}

\begin{align}
(d-2)(d-1)T^+ &= V_0\left(1  +O(\f^2)\right)  - C_3 \frac{\delta_{+} ^2 V_0 }{(d-1) (\delta_{+} +1) (2 \delta_{+} +1)^2 W_0}\f^{2+2/\delta_{+}}+\dots
\end{align}

In this case, $W$ vanishes faster than $\f^{3/2}$ and the solution serves as a possible endpoint of the flow.

\end{itemize}

These solutions correspond to dS$_2\times S^{(d-1)}$ regions corresponding to the Nariai limit for the $-$ branch around maxima in the dS regime, or dS$_2\times S^{(d-1)}$ boundaries in the other three cases ($+$ branch around maxima and $\pm$ branches around minima). Their interpretation as such is presented in  appendix \ref{seho} and in section \ref{sec52} respectively. Interestingly, the previous analysis shows that the flow can either stop, cross or bounce in the Nariai horizon limit.

We conclude this section by showing how the leading solution is constructed for some of the exceptional $\delta_{\pm} = 1/n$ found in \eqref{wds22}:

\begin{itemize}

\item  $\delta_{\pm} = 1/2$ (i.e. $V_2 = \frac{V_0}{4(d-1)}$).

In such case, $\delta_+=\delta_-$ so that $2 +\frac{1}{\delta_\pm}=4$ and it is expected that the logarithmic piece in the expansion of the superpotential \eqref{genWX} is non-trivial. Indeed, substituting the expansion \eqref{genWX} into the master equation for the superpotential \eqref{w56_1b} and solving it perturbatively we find

\begin{align}
W &= W_4 \f^4 \left( 1  + \frac{8 (d-1) V_3}{5 V_0}\f +\frac{64 (d-1)^3 V_3^2-(d-2)V_0^2}{24 V_0^2(d-1)}\f^2\right.\nonumber\\
+ &\left. \frac{V_0^2 ((267 d-94) V_3-20 (d-1) V_5)-160 (d-1)^2 V_4 V_3 V_0+6720
   (d-1)^3 V_3^3}{1890 V_0^3}\f^3 \right)\nonumber \\
   +& W_8 \f^8 \left(1 + \frac{ \left(-32 (d-1)^2 V_4 V_0+896 (d-1)^3 V_3^2+(d+4)
   V_0^2\right)}{1024 (d-1) V_0^2}W_4 \log(\f) \right) + O(\f^9)\,,
\end{align}
where the two free integration constants are now $W_4$ and $W_8$\footnote{In order to look for the missing integration constants we write $W +\delta W$ with $\delta W\ll W$ by assumption. To leading order one finds $\delta W\sim \f^k$ with $k \in\{0,4,8\}$. None of them give additional integration constants.}. The logarithmic piece would be absent if we fine tune the potential so that its coefficient vanishes, i.e. for

\begin{equation}
V_4 = \frac{(4 + d) V_0}{32 (d-1)^2} + \frac{28 ( d-1) V_3^2}{V_0}\,.
\end{equation}

We can compute $f$ and $T$ from Eqs. \eqref{w56_6} and \eqref{w56_7}:

\begin{equation}
f = -\frac{V_0}{64 (d-1) W_4^2 \f ^4}+\frac{V_3}{8 W_4^2 \f ^3} + O(\f^{-2})\,,
\end{equation}
\begin{equation}
T = \frac{V_0}{(d-2) (d-1)} +\frac{V_0 }{8 (d-2) (d-1)^2}\f ^2 + O(\f^3)\,.
\end{equation}
In order to find the geometry of the solution we first solve the flow equation $W' = \dot{\f}$. To leading order we find

\begin{equation}
\f_{\pm} = \pm\dfrac{1}{\sqrt{-8 W_4 u}} + \dots
\end{equation}
The plus (minus) sign corresponds to reaching the solution from the right (left) of the critical point. The assumption that $\f\to 0$ at the critical point implies that the solution is approached as $W_4 u\to -\infty$. Accordingly, the blackening function diverges quadratically and $T$ approaches a constant as given in \eqref{tds2} and the metric asymptotes to the boundary of dS$_2 \times$ S$^{(d-1)}$.

\item $\delta_- = 1/3$ (i.e. $V_2 = \frac{2 V_0}{9(d-1)}$).

Then we have $2 + \frac{1}{\delta_-} = 5$ and $2 + \frac{1}{\delta_+} = \frac{7}{2}$. Each exponent provides a class of solutions for the superpotential. We first consider the case where

\begin{equation}\label{eqter}
W = \f^{7/2}\sum_{n=0}^\infty \left(W_n + \tilde{W}_n \log(\f)\right)\dfrac{\f^n}{n!}\,.
\end{equation}
Substituting the previous expansion into \eqref{w56_1b} and solving it perturbatively gives

\begin{align}
	W_1 &= \frac{7 (d-1) V_3 W_0}{24 V_0} \quad  \tilde{W}_0 =\tilde{W}_1=\tilde{W}_2=0 \quad \tilde{W}_3 = -\frac{5 (d-1) V_0 W_0}{39 V_0}\nonumber\\
   W_2 &= \frac{W_0 \left(-168 (d-1)^2 V_4 V_0+441 (d-1)^3 V_4^2+(864-512 d)
   V_0^2\right)}{3520 (d-1) V_0^2}
\end{align}
Both $W_0$ and $W_3$ remain as free integration constants. Now we may look for the missing integration constants by replacing $W \to W + \delta W$, where $\delta W$ is assumed to be small. From the master equation \eqref{w56_1b} and assuming that the leading behaviour of $\delta W$ is $\f^k$ we find

\begin{equation}
1715 V_0 W_0^3 \left(-4 (d-1) k^4+52 (d-1) k^3-211 (d-1) k^2+273 (d-1)
   k \right) \f ^{k+\frac{7}{2}}+\dots=0\,
\end{equation}
from which we learn that $k\in\{0,3,7/2,13/2\}$. The only solution for $k$ which is subleading with respect to \eqref{eqter} is $k=13/2$. Solving \eqref{w56_1b} for the subleading contributions to $\delta W$ yields

\begin{align}
\delta W &= W_{13/2}\f^{\frac{13}{2}}\left(1+ \frac{299 (d-1) V_3 }{120 V_0}\f\right. \nonumber \\
+&\frac{ \left(4103736 (d-1)^2 V_4 V_0+96464823 (d-1)^3 V_3^2-32 (43448
   d-47331) V_0^2\right)}{10472000 (d-1) V_0^2}\frac{\f^2}{2} \nonumber \\ + & \left. -\frac{31 (d-1) V_3 W_0}{399 V_0}\f^3 \log(\f) + \dots \right)
\end{align}
Therefore, this solution has three integration constants, namely $ W_0$, $W_3$ and $W_{13/2}$.

Alternatively, we may construct the solution with $2+1/\delta_-=5$:

\begin{equation}
W = \f^{5}\sum_{n=0}^\infty \left(W_n + \tilde{W}_{n+1} \f \log(\f) + \hat{W}_{n+2} \f^2 \log(\f)^2\right)\dfrac{\f^n}{n!}\,.
\end{equation}

Solving now \eqref{w56_1b} pertubatively reveals

\begin{align}
 &\tilde{W}_{1} = \frac{15 (d-1) V_3 W_0}{2 V_0} \quad \tilde{W}_2 = \frac{27 (d-1) V_3 \left(5 (d-1) V_3 W_0+28 V_0 W_1\right)}{49 V_0^2} \nonumber\\ &\hat{W}_2=\frac{405 (d-1)^2 V_3^2 W_0}{7 V_0^2}\nonumber \\
 & W_2 = \frac{W_0 \left(1470 (d-1)^2 V_4 V_0+62865 d ((d-3) d+3) V_3^2-98 d
   V_0^2-62865 V_3^2\right)}{1372 (d-1) V_0^2}\nonumber \\ &+\frac{36 (d-1) V_3
   W_1}{49 V_0}+\frac{72 W_1^2}{35 W_0}
\end{align}
We have two integration constants: $W_0$ and $W_1$. Similarly to the previous case, we can look for the missing integration constants by linearising around the known perturbative solution: $W\to W + \delta W$. Assuming that to leading order $\delta W\sim \f^k$ give two possible allowed values: $k=6$ and $k=11$. The case $k=6$ is already captured in the fact that $W_1$ is an integration constant. For $k=11$ we find

\begin{equation}
\delta W = W_{11} \f^{11} \left( 1 + \frac{858 (d-1) V_3}{49 V_0}+\frac{319 W_1}{70 W_0} \f + \frac{957 (d-1) V_3}{28 V_0} \f\log(\f) \right)\,,
\end{equation}
with $W_{11}$ an integration constant.

\end{itemize}

\subsection{Shrinking endpoints: solutions where $W'=0\,$, $V'\neq0$}\label{G.2.2}

This class of solutions we find to behave near the singular point (again taken to be at $\f=0$) as in \refeq{genWX} with $\alpha = \tilde{W}=0$. In other words, they generically take the form
\be
W(\f)=W_0+\sum_{n=2}^{\infty}{W_n\over n!}\f^n
\label{C1}\ee

\noindent
to leading order around $\f=0$. We obtain the following possibilities:
\begin{enumerate}

\item $W_2=0$.

This case leads to the trivial solution $W=constant$. It  implies that $\f$ does not run, and this is why it is trivial.

\item \label{item:IR} $W_2=-{W_0\over d-1}$.

This branch of solutions we call \textbf{shrinking endpoint} solutions for reasons that will shortly become apparent.
As we show below the general solution where the scale factor shrinks to zero is singular, however there is a one parameter family of solutions that are regular.

In this case, solving \refeq{w56_1b} pertubatively we obtain for the first few coefficients
    \be
    W_3=-{2 (d+2) V_0 +d(d-1) V_2 \over ( d-1)^2 (d+2) V_1}W_0
\label{C110}\ee
\be
W_4=\frac{2 d(d+2)(d-1)\left[4{V_0} {V_2} -(d+2)(d-1){V_1}
   {V_3}\right]-4 (d+2)^2 (d+4) {V_0}^2}{(d-1)^3 (d+2)^2 (d+4) {V_1}^2}W_0+
   \label{C111}\ee
   $$+
   \frac{ (d-1)^2 d^2 (d+8) {V_2}^2
  -(d-1) (d+2) (5 d^2+14d-8) {V_1}^2}{(d-1)^3 (d+2)^2 (d+4) {V_1}^2}W_0
$$
and so on. A single constant of integration, $W_0$, appears in this series solution.

Shrinking endpoint solutions have diverging $f$ and $T$ functions at the singular point. It is straightforward to show that \refeq{w56_6} and \refeq{w56_7} imply a local behaviour of the form
\be
f={f_{-1}\over \f}+\sum_{n=0}^{\infty}f_n\f^n\sp T={T_{-1}\over \f}+\sum_{n=0}^{\infty}T_n\f^n.
\label{C112}\ee
The first few coefficients appearing in these expansions are
\be
f_{-1}={2(d-1)^2\over d}{V_1\over W_0^2}\sp f_0={(d-1) \left( d(d-1 )V_2-2 (d+2)  V_0\right)\over d (2 + d) W_0^2}
\label{C113}\ee

\begin{equation}
f_1 = -\frac{8 d \left(d^2+d-2\right) V_0 V_2-2 d \left(d^2+d-2\right)^2 V_1
   V_3+(d-1)^2 d^2 (d+8) V_2^2}{6 d (d+2)^2 (d+4) V_1 W_0^2}-
\end{equation}
$$
-\frac{-4 (d+2) (d+4) V_0^2+4 (d-1)  (d
   (d+4)-4) V_1^2}{6 d (d+2) (d+4) V_1 W_0^2}
$$
and
\be
T_{-1}={V_1\over 2d}\sp T_0={2 (d+2) V_0 +d (d-1)V_2\over 4 d(d-1)(d+2)}.
\label{C114}\ee
\begin{equation}
T_{1} = \frac{-8 d \left(d^2+d-2\right) V_0 V_2+2 d \left(d^2+d-2\right)^2 V_1
   V_3-(d-1)^2 d^2 (d+8) V_2^2}{24 d (d+4) \left(d^2+d-2\right)^2 V_1}+
\end{equation}
$$
 +\frac{4 (d+2)^2 (d+4) V_0^2+8 (d-1) (d+2) (d
   (d+4)+2) V_1^2}{24 d (d+4) \left(d^2+d-2\right)^2 V_1}
$$

This solution describes the shrinking of the foliating $S^{d-1}$ to zero size while at the same time $f\to\infty$ so that $f/T$ is finite. This can be seen upon writing the local solution in terms of the radial $u$ coordinate, where it reads
\begin{equation}
\begin{split}
&f(u) =\frac{4(d-1)^2 e^{\frac{u W_0}{d-1}-2 A_0}}{R^2
   W_0^2}+\dots\\
   & A(u) =  A_0-\frac{u
   W_0}{2(d-1)}+\dots\\
   & \f(u) = \frac{1}{2d} R^2 e^{2 A_0-\frac{u W_0}{(d-1)}} V_1+\dots
\end{split}
\end{equation}
Small $\f$ implies that $uW_0 \to \infty$ near the singular point, so that indeed the volume form on the sphere $\mathrm{vol}_{S^{d-1}}\sim e^{(d-1)A}\to 0$ there.

Despite the appearance of singular behaviour in the metric, direct computation shows that in fact the curvature invariants remain finite for these solutions in the vicinity of the singular point. The quantities controlling the curvature invariants, $p,\rho $ and $\mathcal{I}$, introduced in Eqs.  (\ref{ev3b},\ref{ev3c},\ref{ev7}) around a shrinking endpoint are given by
\begin{equation}\label{invshr1}
\rho = - V_0 -\frac{(d-1) V_1 }{d}\varphi+ \dots \sp p =  V_0 +\left(\frac{1}{d}+1\right) V_1 \varphi+ \dots
\end{equation}
\begin{equation}\label{invshr3}
\mathcal{I} = -\frac{V_0}{d(d-1)}  - \frac{V_1 }{2 d(d-1)}\varphi + \dots
\end{equation}
Additionally, the Kretschmann scalar of Eq. (\ref{w61}) is
\be
K_2={2(d+1)\over d(d-1)^2}V_0^2+{\cal O}(\f).
\label{C116}\ee
Parametrizing the value of the potential as $V_0=\pm{d(d-1)\over \ell^2}$ we obtain
\be
K_2={2d(d+1)\over \ell^4}+{\cal O}(\f)
\label{C117}\ee
which is the value of the Kretschmann scalar for (A)dS$_{d+1}$ with radius of curvature $\ell$.

We now proceed to look for perturbations around the previously found solution. First we quote the explicit form of Eq. \eqref{eqpe} for this particular case to leading order in each of the $c_i$ coefficients in \eqref{eqpe}:

\begin{align}
&\frac{(d-2) d V_1 W_0^3
   \delta W}{2 (d-1)^3}-\frac{(d-2) d W_0^3 (2
   (d+2) V_0+(d-1) d V_2) \delta W'}{2 (d-1)^3 (d+2)}\nonumber \\
   &+\frac{\left(d^2-2 d\right) V_1 W_0^3 \delta W''}{2 (d-1)^2}+\frac{2 d V_1
   W_0^3 \f  \delta W^{(3)}}{(d-1)^2}+\frac{2
   V_1 W_0^3 \f ^2 \delta W^{(4)}}{(d-1)^2} + \dots=0\,.
\end{align}
The solution to the previous equation is, again to leading order,

\begin{equation}
\delta W = C_0 + C_1\f + C_2 \f^{(4-d)/2} + + C_3 \f^{(6-d)/2} + \dots
\end{equation}
Note that all the exponents of $\f$ in the previous equation are smaller than $2$ (for $d>2$). Therefore, the solution is not subleading with respect to the unperturbed solution and consistency requires that we set $C_0=C_1=C_2=C_3 = 0$. In other words, this regular solution does not admit deformations that preserve the regularity and there is only one integration constant: $W_0$.
Of course, there is a four parameter family of singular shrinking solutions. The regular solution above is the codimension 3 manifold that does not have a curvature singularity.

The geometry around the shrinking endpoint is discussed below Eq. \eqref{eq:G4b}. In the AdS regime, the shrinking endpoint corresponds to the center of AdS space in global coordinates, while in the dS regime it corresponds to the location of an observer in the static patch coordinates.
Moreover, this is an endpoint of the flow because the sphere $S^{(d-1)}$ shrinks to zero size and the geometry ends there. This is therefore an  IR endpoint of the flow, similar to the situation of flows on S$^d$ studied in detail in \cite{curved} and \cite{F}.

\item \label{item:exact} $W_2={(d-2)\over 2(d-1)}W_0$.

In this case the solution can again be determined iteratively. It can be shown by induction that the solution obtained solves
\be
2 (d-1) (W'')^2+(2-d) W W''+(d-2) W'^2-2 (d-1) W^{(3)} W'=0
\label{C120}\ee
to all orders. This corresponds to the case where the denominator of Eqs. \eqref{w56_6} and \eqref{w56_7} vanishes. Additionally, all the coefficients in Eq. \eqref{w56_1b} vanish identically for this solution. The potential $V$ that gives rise to this solution is determined indirectly through Eq. \eqref{w55}. We can obtain the local behaviour of the metric functions and of the potential from Eqs. \eqref{C120}, \eqref{eqtt}, \eqref{f10_1} and \eqref{w55}:
\begin{equation}
W = W_0 + \dfrac{(d-2)}{4(d-1)}W_0\f^2 + \frac{1}{6}W_3\f^3 + \dots \qquad T =0 \,,
\end{equation}
\begin{equation}
f = -\frac{4(d-1)V_0}{d W_0^2}\,,\qquad V = V_0 + \dfrac{(d-2)V_0}{d(d-1)}\f^2+\dots
\end{equation}
where both $W_0$ and $W_3$ are integration constants. Note that the potential $V$ has a local extremum. In this subsection we study local solutions with $V'\neq 0$, and we shall not discuss this solution further here. A detailed discussion of this local solution can be found below Eq. \eqref{C120z} or in Appendix \ref{app:I}.

\item $W_2=-{d W_0 \over 2(d-1)}$.\label{sssa}

We solve Eq. \eqref{w56_1b} perturbatively and obtain the functions $f$ and $T$ from Eqs. \eqref{w56_6} and \eqref{w56_7}:
\begin{equation}\label{fe10}
W= W_0 -\frac{dW_0}{4(d-1)}\f^2  -\frac{d W_0 (d (2 V_0+V_2)-V_2)}{24 (d-1)^2 V_1}\f^3 + O(\f^4)\,,
\end{equation}
\begin{equation}
f = \frac{4 (d-1)^2 V_1}{d^2 W_0^2 \f }-\frac{(d-1) (2 d V_0-d V_2+V_2)}{d^2 W_0^2} + O(\f)\,, \qquad  T=0\,.
\end{equation}
In order to fully characterize this solution, we study the deformations of it, governed by Eq. \eqref{eqpe}. In this case, the leading contribution to \eqref{eqpe} is given by
\begin{equation}
d \f ^2 \delta W^{(4)}+(d+2)  \f  \delta W ^{(3)}-(d-2)  \delta W''-\frac{2  (d (2 V_0+V_2)-V_2) }{(d-1) V_1}\delta W'-\frac{(d-2) d \delta W}{2
   (d-1)}=0\,.
\end{equation}
We write  $\delta W = \f^k \delta w(\f)$, where $\delta w$ has derivatives at $\f=0$ up to the forth that are finite. The previous relation reduces to
\begin{equation}
\f^k\left(\frac{d  (k-1)^2 k (d (k-3)+2)}{\f ^2}\delta w(0) + O(\f^{-1})\right)=0\,,
\end{equation}
and we find the possible solutions $k=\{0,1,1,3-2/d\}$. The only solution that is subleading with respect to the unperturbed solution \eqref{fe10} is $k=3-2/d$. From Eqs. \eqref{w56_6} and \eqref{w56_7} we obtain the corrections to $f$ and $T$, denoted as $\delta f$ and $\delta T$, that are proportional to $\delta w(0)$:
\begin{equation}
\delta f = \f^{-2/d}\left(\frac{8 (d-1)^3 (3 d-2) \delta w(0) V_1}{d^4 W_0^3} + O(\f)\right)\,, \quad
\end{equation}
\begin{equation}
\delta T = \f^{-2/d}\left(\frac{4 (d-1)^2 (3 d-2) \delta w(0) V_1}{(d-2) d^4 W_0} + O(\f) \right)\,.
\end{equation}
The function $T$ is non-trivial, and therefore this is compatible with the spherically sliced ansatz. Additionally, the function $T$ diverges to $+\infty$ as $\f\to 0$ and, since $T\propto e^{-2A}$, the scale factor $e^{A}$ vanishes for this local solution. This branch of solutions suffers from a naked singularity. In particular, the quantity $\mathcal{I}$, defined in Eq. \eqref{w63}, diverges
\begin{equation*}\label{finitesing}
    \mathcal{I} = \frac{f W^2}{4(d-1)^2}-T = \frac{V_1}{d^2}\frac{1}{\f} + O(\f^0)
\end{equation*}
\noindent
which results in a divergent Kretchmann invariant.
We call this asymptotic, the {\it special singular shrinking asymptotic (SSSA)}.
 As this singular behaviour does not arise at the boundary of field space, we do not anticipate that it can be resolved upon uplift to a higher-dimensional solution
 as in the cases discussed in the literature, \cite{cgkkm1,Ahmad}.  Accordingly, we shall consider such solutions are unacceptable singularities and shall not explore them further in this work.

\end{enumerate}

\subsection{$\f$-Bounces}\label{bounces}

$\f$-Bounces are solutions in which the scalar field reverses direction along its trajectory. They are known to correspond to points where the solution $W$ is regular but its derivatives may be singular. To explore this class of solutions, we insert \refeq{genWX} into \refeq{w56_1b} with $\tilde{W}_n = 0$.

We find a Frobenius type solution in which the indicial equation is satisfied for $\alpha = 3/2$. Indeed, we then immediately observe that at such singular points $\f$ reverses direction, since $\dot \f=W'=0+\dots$ and $\ddot \f=W'W''=\frac{9}{8}\hat{W}_{3/2}^2+\dots$.

We can simplify the discussion by reorganising the expansion near such a singular point such that
\be
W=W_0+\sum_{n=3}^{\infty} ~W_{n/2}~\f^{n\over 2}.
\label{CC5}\ee
Note that this is more general than the standard Frobenius ansatz.

Solving the equation (\ref{w56_1b}) we determine all higher coefficients of $W$ in terms  of  the arbitrary integration constants $W_0,W_{3\over 2}$, and $W_2$. For example,
\be
W_{5\over 2}=\frac{9 (d-1) W_{3\over 2}^2 \left((d-1) V_2+6V_0\right)+V_1 \left((6-d) d W_0^2+36 (d-1) W_2 W_0+52 (d-1)^2 W_2^2\right)}{60 (d-1)^2 V_1 W_{3\over 2}}
\label{C105}\ee
and so on for the higher coefficients.

Moreover, we find that we can iteratively solve for $f$ and $T$ from \refeq{w56_6} and \refeq{w56_7} by employing expansions of the form
\be
f=f_0+\sum_{n=1}^{\infty} ~f_{n/2}~\f^{n\over 2}
\sp
T=T_0+\sum_{n=1}^{\infty} ~T_{n/2}~\f^{n\over 2}.
\label{C106}\ee
In particular, we have
\be
f_0=\frac{8 V_1}{9 W_{3\over 2}^2}\sp f_{1\over 2}=\frac{8 V_1 \left(d W_0-6 (d-1) W_2\right)}{27 (d-1) W_{3\over 2}^3}
\label{C107}\ee
and
\be
T_0=\frac{V_1 W_0 \left(d W_0+6 (d-1) W_2\right)+9 V_0 (d-1) W_{3\over 2}^2}{9 (d-2) (d-1)^2 W_{3\over 2}^2}
\label{C108}\ee
\be
T_{1\over 2}=\frac{4 W_0  \left(V_1 W_0 \left(d W_0+6 (d-1) W_2\right)+9 V_0 (d-1) W_{3\over 2}^2\right)}{27 (d-2) (d-1)^3 W_{3\over 2}^3}
\label{C109}\ee

These solutions can be thought of as the generalization of the bounces of \cite{exotic} to the spherical-sliced ansatz. Despite various divergences appearing in derivatives of the superpotential and metric functions, all curvature invariants are finite at a bounce. The quantities controlling the curvature invariants are $p,\rho$ and $\mathcal{I}$ (see appendix \ref{sect:inv_sphere}), and are given by
\begin{equation}\label{invbnc1}
\rho = - V_0 + \frac{V_1  (d (W_0+2 W_2)-2 W_2)}{3 (d-1) W_{3/2}}\varphi ^{3/2}+ \dots
\end{equation}
\begin{equation}\label{invbnc2}
p =  V_0 +2 V_1 \varphi + \frac{V_1  (d (W_0+2 W_2)-2 W_2)}{3 (d-1) W_{3/2}}\varphi ^{3/2} + \dots
\end{equation}
\begin{equation}\label{invbnc3}
\mathcal{I} = -\frac{\frac{V_1 W_0 \left(6 (d-1)^2 W_2+((d-3) d+4) W_0\right)}{(d-1)^2
   W_{3/2}^2}+9 V_0}{9 (d-2)}+ \dots
\end{equation}
which are indeed finite.

\section{Perturbative solutions III: solutions around a singular point corresponding to a horizon} \label{sho}

The conditions under which a horizon appears in a solution within our ansatz are explored in detail in appendix \ref{app:J}. There we show that solutions with a horizon are characterized by locations in field space, $\f_h$, where the $tt$ component of the metric vanishes $e^{2A(\f_h)}f(\f_h)=0$.
 Generically, the scale factor vanishes only at shrinking endpoints or at the boundaries of field space. However these places do not correspond to horizons but endpoints of the geometry.
  We therefore focus on the case where the scale factor does not vanish, and $f(\f_h)=0$. We shall first assume that $f$ vanishes linearly around $\f_h$ and then turn our attention to alternative behaviours.

\subsection{Non-extremal horizons}

The master equation for the superpotential \eqref{w56_1b} can be solved perturbatively around a horizon. However, in this context it is more natural to work directly with the system of differential equations given in \eqref{f10_1}-\eqref{w55b}.
 We denote the location of the horizon in field space as $\f_h$ and expand the potential $V(\f)$, the superpotential $W(\f)$, the blackening function $f(\f)$ and the scale factor $T(\f)$ in Taylor series:

\begin{equation}\label{eah}
V = \sum_{n=0}^\infty V_n \dfrac{(\f-\f_h)^n}{n!}\qquad W = \sum_{n=0}^\infty W_n \dfrac{(\f-\f_h)^n}{n!}
\end{equation}
\begin{equation}\label{eahg}
	 f = \sum_{n=1}^\infty f_n \dfrac{(\f-\f_h)^n}{n!} \qquad T = \sum_{n=0}^\infty T_n \dfrac{(\f-\f_h)^n}{n!}
\end{equation}%
The presence of the horizon is encoded in the fact that the expansion for $f$ starts at $n=1$. Expanding the equations of motion \eqref{f10_1}-\eqref{w55b} we can solve for the coefficients in the expansion:

\begin{align}\label{heq}
&W_0^\pm = \pm \dfrac{2V_0-2(d-2)(d-1)T_0}{\sqrt{f_1 V_1}} \qquad W_1^\pm =\pm \sqrt{\dfrac{V_1}{f_1}} \qquad W_2^\pm = \pm \dfrac{2(d-2)T_0+V_2}{2\sqrt{f_1 V_1}}\nonumber \\
&T_1^\pm = 2T_0\dfrac{(d-1)(d-2)T_0+ V_0}{V_1(d-1)} \qquad f_2^\pm = \frac{f_1 \left(-2 (d-2)(d+3) T_0+2 \frac{d}{d-1} V_0- V_2\right)}{2
   V_1}  \nonumber \\
&T_2^\pm  = \dfrac{T_0}{d-1}\dfrac{6 ( d-2)^2 T_0^3}{V_1^2} + \frac{T_0 ((d-2) (d-1) T_0 ((d-1) V_2-10 V_0)+V_0 (4 V_0-(d-1)V_2))}{(d-1)^2
   V_1^2}
\end{align}
The ``$-$" branch of the solutions can be mapped into the ``$+$" branch with the change of coordinates $u\to -u$. Around the horizon we have three integration constants, namely $f_1$, $T_0$ and the location of the horizon $\f_h$. The first one sets the temperature of the horizon while the second controls the curvature of the spherical slices. Finally, $\f_h$ sets the value of the coupling at the horizon. The three integration constants provide two dimensionless physically relevant ratios.

The quantities controlling the curvature invariants of the appendix \ref{sect:inv_sphere}, $\rho$, $p$ and $\mathcal{I}$, introduced in Eqs. \eqref{ev3b}, \eqref{ev3c} and \eqref{ev7} are given by
\begin{equation}
\rho  = -V_0 -V_1(\varphi-\varphi_h)+\dots \sp p  = V_0 - \dfrac{3}{2}V_1(\varphi-\varphi_h)+\dots
\end{equation}
\begin{equation}
\mathcal{I} = -T_0 +\frac{ \left((d-4) (d-2) T_0^2+\frac{V_0^2}{(d-1)^2}-2 T_0 V_0\right)}{V_1}(\varphi -\varphi_h) + \dots
\end{equation}
The previous quantities are finite, and therefore the solution is regular across the horizon.

\subsection{Extremal horizons. Part I}\label{seho}

Note that the solution around the horizon Eq. \eqref{heq} is not valid for $f_1=0$, i.e. for extremal horizons. Note that for extremal horizons both $f(u_h)=\dot{f}(u_h)=0$, and according to Eq. \eqref{f23c}

\begin{equation}\label{tttb}
	V_h = \dfrac{(d-1)(d-2)}{R^2}e^{-2 A}\,,
\end{equation}
which can only happen in the de-Sitter regime ($V>0$). We shall solve equations (\ref{reda1a}), (\ref{reda2a}) for $W,f$ which we reproduce here,
\be\label{reda1b}
\frac{f}{4} \left(\frac{d W^2}{d-1}-2 (W')^2\right)- \frac{W' }{4}\Big((d+2) W f'-2 (d-1) \left(f' W'\right)'\Big)+V=0
\ee
\be
{W'}\left[ W'f' + f\left(W''-\frac{d }{2
	(d-1)}W\right)\right]-V' = 0\,,
\label{redab2}
\ee
and then determine $T$ from (\ref{reda3a}) that we reproduce here,
\be
T={1\over (d-1)(d-2)}\left[\bigg( \frac{d}{4(d-1)} W^2-\frac{W'^2}{2}\bigg) f-\frac{1}{2} W'Wf'+V\right]\,,
\label{reda3aab}\ee

In Appendix \ref{exth} we show that two families of extremal horizons appear, where the metric fields admit a Fr\"obenius expansion:

\be\label{eahb2}
V = \sum_{n=0}^\infty V_n \dfrac{(\f-\f_h)^n}{n!}\qquad W = (\f-\f_h)^\alpha \sum_{n=0}^\infty W_{n} \dfrac{(\f-\f_h)^n}{n!}
\ee

\be\label{eahb3}
f = (\f-\f_h)^\beta \sum_{n=0}^\infty f_{n} \dfrac{(\f-\f_h)^n}{n!} \qquad T = (\f-\f_h)^\gamma \sum_{n=0}^\infty T_{n} \dfrac{(\f-\f_h)^n}{n!}
\ee
where $\f_h$ is the location of the horizon in field space. The first family of extremal horizons appears for $\alpha=0$ and $\beta=2$. In this case, equations \eqref{reda1b} and \eqref{redab2} become

\begin{align}
&0=\frac{1}{2} (d-1) f_0 W_1^2+V_0 +\nonumber  \\ &+	\f  \left(\frac{1}{4} (2 (d-1) f_0 W_1 W_2-W_1 ((d+2) f_0 W_0-2 (d-1) (f_1
	W_1+2 f_1 W_2)))+V_1\right)
\end{align}
\begin{equation}
-V_2 +\f  \left(f_0 W_1^2-V_2\right)=0
\end{equation}
whose solution requires an extremum of the potential $V_1=0$ as well as fine-tuning of the potential:

\begin{equation}\label{otra}
	V_2=-\frac{2 V_0}{d-1}\qquad f_0 W_1^2 = -\frac{2 V_0}{(d-1) }
\end{equation}
\begin{equation}\label{una}
	W_2 = -\frac{(d-1)^2 W_1 V_3+2 V_0 W_0}{4 (d-1) V_0}  \qquad  f_1=-\frac{(d+5) V_0 W_0}{(d-1)^2 W_1^3} -\frac{3 V_3}{2 W_1^2}
\end{equation}

From \eqref{reda3aab} we learn that

\begin{eqnarray}
	T_0=\frac{V_0}{(d-1)(d-2)}\,.
\end{eqnarray}
From the flow equation $W'=\dot{\f}$  we observe that $\f$ approaches the previous solution linearly in $u$ and therefore $f$ vanishes quadratically and this is indeed a extremal horizon. Finally, the fact that $W_1\neq0$ implies that the flow does not stop at this kind of extremal horizons. The solution has two integration constants: $W_0,W_1$, as well as the location of the horizon $\f_h$.

In this case, the quantities controlling the curvature invariants of the appendix \ref{sect:inv_sphere}, $\rho$, $p$ and $\mathcal{I}$, evaluate to
\begin{equation}
\rho  = -V_0 \frac{1}{4}  \left(-\frac{2 (d+5) V_0 W_0}{(d-1)^2W_1}-4
   V_1-3 V_3\right)(\varphi-\varphi_h)+\dots
\end{equation}
\begin{equation}
p  = V_0 + \left(V_1-\frac{(d+5) V_0 W_0}{2 (d-1)^2 W_1}-\frac{3V_3}{4}\right)(\varphi-\varphi_h)+\dots \sp \mathcal{I} = -\frac{V_0}{d^2-3 d+2}\dots
\end{equation}
The previous quantities are finite, and therefore the solution is regular across this type of extremal horizon.

\subsection{Extremal horizons. Part II}\label{seho2}

The second family of extremal horizons is achieved when $2\a + \b = 4$. Consequently, Eqs. \eqref{reda1b} and \eqref{redab2} become

\begin{equation}
	(V_0 + (-1 + d) (-2 + \alpha)^2 \alpha^2 f_{0} W_0^2) + O(\f)=0\,,
\end{equation}
\begin{equation}
   -V_1 + (-V_2 - (-3 + \alpha) \alpha^2 f_{0} W_0^2) \f + O(\f^2) = 0\,.
\end{equation}
The solution is possible at extrema of the potential in the $dS$ regime. To leading order we obtain

\begin{equation}\label{2hu}
	V_1 = 0\qquad \alpha_\pm = 2+\dfrac{1}{\delta_{\pm}} \qquad \beta_\pm = -\frac{2}{\delta_{\pm}} \qquad f_0^\pm (W_0^\pm)^2 = -\frac{\delta_\pm^4 V_0}{(d-1) (2 \delta_{\pm} +1)^2}\,,
\end{equation}
where we have defined

\begin{equation}
	\delta_{\pm}= \dfrac{1}{2} \left(1\pm\sqrt{1-\dfrac{4(d-1)V_2}{V_0}}\right)\in \left[\frac{1}{2},\pm \infty \right)\,,
\end{equation}
or equivalently

\begin{equation}
	V_2 = \dfrac{V_0}{d-1} \delta(1-\delta)\,.
\end{equation}
Eq. \eqref{reda3aab} implies $\gamma=0$ and

\begin{equation}\label{h20}
	T_0 = \frac{V_0}{(d-1)(d-2)}\,.
\end{equation}

Note that $f\sim(\f-\f_h)^{-2/\delta_{\pm}}$. Therefore, requiring the presence of a horizon enforces $\delta_{\pm}<0$, which is only true for the $-$ branch of the solutions whenever

\begin{equation}
	\dfrac{V_2}{V_0}<0\,,
\end{equation}
that is, around maxima of the potential in the dS regime.  In fact, because this is an extremum of the potential, this solution has been extensively studied in appendix \ref{D.2.3}. In particular, it is given in equations \eqref{f80}-\eqref{f82} and it has a single integration constant: $W_0$.
The quantities controlling the curvature invariants of appendix \ref{sect:inv_sphere} are evaluated in Eqs. \eqref{invds21} and \eqref{invds23}, and are finite.

Solving for the flow equation $W'=\dot{\f}$ gives to leading order

\begin{equation}
	\f-\f_h \simeq \left(-\frac{W_0 (2 \delta_- +1) u}{\delta_- ^2}\right)^{-\delta_- }\,.
\end{equation}

The horizon is located at $\f_h$ and, since $\delta_-<0$ this requires that $u\to0$. Therefore, the blackening function $f$ in terms of $u$ behaves as

\begin{equation}\label{h23}
	f = -\dfrac{V_0}{d-1}u^2 + O(u^3)\,
\end{equation}
and such horizons are always extremal. In section \ref{sec:global} we show that whenever a solution features two horizons, one of them is cosmological while the inner one is an event horizon. Therefore, the presence of an extremal horizon corresponds to the limit of coinciding event and cosmological horizons, also known as Nariai limit. The geometry asymptotes to dS$_2\times$S$^{(d-1)}$.

From the discussion below \eqref{f82}, we know that this is an endpoint of the flow provided that $\delta_-<-2$ ($V_2\leq -\frac{6V_0}{(d-1)}$). Alternatively, the flow crosses the horizon regularly for the fine-tuned case where $\delta_-=-1$ ($V_2 = \frac{-2V_0}{d-1}$). Finally, for $\delta_-\in(-2,-1)\cup(-1,0)$ the solution is regular at the horizon but the second derivative of $\f$ diverges.

\vskip 1cm
\section{Marginally relevant boundaries\label{Marginal}}
\vskip 1cm

In the previous sections we have found solutions around singular points of the equations where the solution to the superpotential equation \eqref{w56_1b} is assumed to be power-like. In this section we generalize the previous ansatze to include exponentially vanishing behaviour around a singular $\f=0$, while the potential $V$ is expanded in a Taylor series:

\begin{equation}\label{anse}
W = e^{-c \f^{-n}}\f^{m}\sum_{l=0}^\infty W_l\dfrac{\f^l}{l!}\sp V = \sum_{l=0}^\infty V_l \dfrac{\f^l}{l!}\,.
\end{equation}
We shall assume that $n>0$ and $c\f^{-n}>0$, otherwise the exponential admits a Taylor expansion and it falls into the class of solutions studied in the previous sections. Besides,  we assume that $W_0\neq 0$ without loss of generality.

Substituting the ansatz \eqref{anse} into \eqref{w56_1b} gives, to leading order,

\begin{equation}\label{eqse}
e^{-4 c \f^{-n}}\f^{-8-7n+8m}\left[ 4 c^7(d-1)n^7(1+n)W_0^4 V_1 + O(\f,\f^n,\f^{2n})\right] = 0\,.
\end{equation}
The only solution compatible with the assumptions that $c \neq 0$, $n>0$ and $W_0\neq0$ is to have an extremum of the potential, i.e. $V_1=0$. Under this condition, which assumes an extremum of the scalar potential,  \eqref{eqse} vanishes to leading order. The next-to-leading contribution gives

\begin{equation}\label{eqse2}
e^{-4 c \f^{-n}}\f^{-8-7n+8m}\left[4 c^7 ( d -1) n^7 (1 + n) W_0^4 V_2\f + O(\f^2,\f^{1+n},\f^{1+2n}) \right]=0\,.
\end{equation}
A non-trivial solution compatible with our assumptions, requires that we have an inflexion point of the potential: $V_2=0$. In the AdS case, this implies that the scalar $\f$ is dual to a marginal operator, $\Delta=d$.  This is the reason we call such asymptotics the {\em marginally-relevant boundaries} in the title. In such a case, the NNLO\footnote{NNLO stands for next-to-next-to leading order. Similarly, N$^3$LO refers to the next-to-next-to-next-to leading order correction and so on.} contribution to \eqref{w56_1b} is

\begin{equation}\label{eqse3}
2c^6n^6(1+n) W_0^4\left(  c (d-1) n  V_3 \f^2 -
 2   V_0 \f^{1+n}\right) + O(\f^3,\f^{2+n},\f^{1+2n}) = 0
\end{equation}
where we have suppressed the prefactor $e^{-4 c \f^{-n}}\f^{-8-7n+8m}$ for compactness. We shall continue to suppress this factor in the rest of this section. We distinguish three possibilities in Eq. \eqref{eqse3}:

\begin{itemize}
\item[G.1] $n<1$. Then $\f^{1+n}$ is dominant and a solution requires $V_0=0$. Eventually, this possibility leads to $V_l=0\ \forall\ l$  or to the trivial solution with $W_l=0\ \forall\ l$.
    Therefore there is no non-trivial solution in this case.

\item[G.2] $n=1$. In this case both powers of $\f$ in Eq. \eqref{eqse3} are of the same order and the parameter $c$ is determined as

\begin{equation}\label{A}
c =\dfrac{2 V_0}{ (d-1) V_3}\,, \quad c/\f>0\,.
\end{equation}

The N$^3$LO in the main equation  \eqref{w56_1b} becomes

\begin{equation}
\frac{256 V_0^6 W_0^4 \f ^3 \left(3 (d-1) (m-2) V_3^2+2 V_0
   V_4\right)}{3 (d-1)^6 V_3^7} + O(\f^4)=0\,,
\end{equation}
fixing the exponent $m$ in \eqref{anse} to be
\begin{equation}
m = 2-\frac{2 V_0 V_4}{3 (d-1) V_3^2}\,.
\end{equation}
The rest of the solution can be obtained by standard  perturbation theory around this leading solution. The next coefficient $W_1$ is given by

\begin{equation}\label{w1}
W_1 = -\frac{W_0 \left(-6 (d-1) V_0 V_3^2 V_4+54 (d-1)^2 V_3^4+3
   V_0^2 V_3 V_5-4 V_0^2 V_4^2\right)}{18 (d-1) V_0
   V_3^3}\,
\end{equation}
and so on.

We compute $f$ from Eq. \eqref{w56_6}, and we obtain  to leading order a function that, according to Eq. \eqref{A}, diverges as $\f\to 0$:

\begin{equation}
f=-\frac{(d-1)^3 V_3^4 }{16 V_0^3 W_0^2}e^{\frac{4 V_0}{(d-1) V_3 \f }} \f ^{\frac{4 V_0
   V_4}{3 (d-1) V_3^2}+4}+\dots
\end{equation}
The function $T$ can be obtained form \eqref{w56_7} and is constant to leading order

\begin{equation}\label{t10}
T = \frac{V_0}{(d-1)(d-2)} + \dots
\end{equation}

Despite the diverging $f$ function, the geometry is regular and the Ricci scalar and Kretschman invariant attain a finite value:

\begin{equation}
	R= \dfrac{d+1}{d-1}V_0 + \dots \ \ \ \ K_2 = \frac{2 (3 d-5) V_0^2}{(d-2) (d-1)^2} + \dots
\end{equation}

The solution has a single integration constant $W_0$, it constitutes a possible endpoint of the flow provided that $c/\f>0$ as $\f\to0$, because in such a case, the superpotential $W$ vanishes faster that $\f^{3/2}$. This is a sufficient condition to have an endpoint, as discussed in Sec. \ref{sec:global}. In order to determine the geometry of the solution around this endpoint, we write the metric as a function of $\f$:

\begin{equation}
ds^2 = \dfrac{d\f^2}{fW'^2} - e^{2A}f dt^2 + \dfrac{1}{T}d\Omega_{(d-1)}^2\,.
\end{equation}
Where we have used that $\dot\f =  W'$ and the definition of $T$ in Eq. \eqref{w54a}. The fact that $T>0$ along with Eq. \eqref{t10} enforces $V_0>0$. The condition that we have an endpoint ($c/\f>0$) further gives $V_3\f>0$ as we approach $\f\to 0$. Making use of the solution for $W$ and $f$, we compute the $g_{\f\f}$ component of the metric.

\begin{equation}
g_{\f\f} = \dfrac{1}{fW'^2} = -\dfrac{\zeta^2}{\f^4} + \dots \ \ \ \textrm{with} \ \ \ \zeta^2 \equiv \dfrac{4V_0}{(d-1)V_3^2}\,.
\end{equation}
On the other hand, the dependence of $A$ as a function of $\f$ can be obtained from the second equation in \eqref{eqtt}, giving as a result

\begin{equation}
A = A_0 -\frac{V_3}{12 V_0}\f^3 + \dots
\end{equation}
Therefore, the metric takes the asymptotic form

\begin{equation}
ds^2 = -\dfrac{\zeta^2d\f^2}{\f^4} + \dfrac{4e^{2A_0}}{\zeta^6W_0^2V_3^2}\f^{4+\zeta^2V_4/3}e^{\zeta^2V_3\over \f}dt^2 + \dfrac{(d-1)(d-2)}{V_0}d\Omega_{(d-1)}^2 +\dots
\end{equation}
where we used the dilaton as the holographic coordinate.
We now make a change of coordinates given by

\begin{equation}
r^2 = \f^{4+\zeta^2V_4/3}e^{\zeta^2V_3\over \f}\,,
\end{equation}
where $r^2\to\infty$ as we approach the endpoint at $\f\to 0$. From the change of variables, we obtain

\begin{equation}
2r dr = -V_3\zeta^2\dfrac{r^2}{\f^2}d\f +\dots \ \Rightarrow \ \zeta^2\dfrac{d\f^2}{\f^4} = \dfrac{4}{\zeta^2V_3^2}\dfrac{dr^2}{r^2}
\end{equation}
Finally, rescaling the time coordinate as
\begin{equation}
\tilde{t}^2 =  \dfrac{16e^{2A_0}}{\zeta^8W_0^2V_3^4} t^2
\end{equation}
we arrive at
\begin{equation}
ds^2 = -\dfrac{4}{\zeta^2V_3^2}\dfrac{dr^2}{r^2} + \dfrac{\zeta^2V_3^2}{4}r^2d\tilde{t}^2 + \dfrac{(d-1)(d-2)}{V_0}d\Omega_{(d-1)}^2 +\dots
\end{equation}
The $\tilde{t}-r$ part of the metric is identified with the boundary of two dimensional de Sitter space in Static patch coordinates (see Eq. \eqref{a27} for $r\to\infty$). We identify the dS scale with

\begin{equation}
H^2 = \frac{1}{4}\zeta^2V_3^2 = \dfrac{V_0}{d-1}\,.
\end{equation}
We conclude that the geometry around this endpoint is given by a boundary of dS$_2$ times the $d-1$ dimensional sphere $S^{(d-1)}$. Both the radius of the sphere and the dS$_2$ scale are controlled by the value of the potential $V_0$ at the endpoint.

These solutions can be thought of as limits of the dS$_2$ of \ref{g53} as $V_2\to 0$ in equation (\ref{alfa}).
Clearly, although not of interest in this paper, there are similar solutions with asymptotic geometry AdS$_2\times $R$_t\times$ EAdS$_{d-1}$, when the slice manifold (R$_t\times $EAdS$_{d-1}$ here) has constant negative curvature.
The holographic interpretation of such solutions is that the driving operator on the dual one-dimensional QFT is marginally relevant.

 Similar looking solutions for marginally-relevant boundaries were found in the AdS regime with flat slices  in \cite{Bourdier13}. The main difference from the solution above, is that in that case, W asymptotes to a constant and the exponential behavior appears at the vev level.
Here, it controls the leading behavior of the superpotential.

\item[G.3] $n>1$. Then the $\f^2$ term in Eq. \eqref{eqse3} dominates and the solution requires $V_1=V_2=V_3=0$. In this case, the next contributions to \eqref{eqse3} are

\begin{equation}\label{eqse4}
\frac{2}{3} c^6 n^6 (1 + n)  W_0^4 \left(A (d-1) n V_4 \f^3  - 6  V_0 \f^{1+n} \right) + O(\f^4,\f^{2+n},\f^{1+2n})=0
\end{equation}
and we are again confronted with three options, with the same structure as the options considered so far.
\begin{itemize}
\item[G.3.1] $1<n<2$. Then the vanishing of the leading term in Eq. \eqref{eqse4} enforces $V_4=0$. Pursuing this branch of solutions eventually leads to either $W=0$ or $V=0$ to all orders and therefore there is no non-trivial solution.

\item[G.3.2] $n=2$. Now both terms in \eqref{eqse4} are of the same order and a non-trivial solution is obtained in a similar spirit as in Eqs. \eqref{A}-\eqref{w1}. We quote the result:

\begin{equation}
c= \dfrac{3V_0}{(d-1)V_4}\,,\ m = 3- \dfrac{3V_0V_6}{10(d-1)V_4^2} \,, \ W_1=-\dfrac{V_0 W_0V_7}{20(d-1)V_4^2}\,.
\end{equation}

Similarly to the case with $n=1$, this solution is regular and constitutes a possible endpoint of the flow provided that $c>0$. There is a single integration constant: $W_0$.

We compute the blackening function $f$ from \eqref{w56_6}. To leading order in $\f$ as $\f \to 0$ we have

\begin{equation}
f = -\frac{(d-1)^3 V_4^4 e^{\frac{6 V_0}{(d-1) V_4 \phi ^2}} \phi ^{\frac{3
   V_0 V_6}{5 (d-1) V_4^2}+6}}{1296 V_0^3 W_0^2} +\dots
\end{equation}
On the other hand, we compute the inverse scale factor $T$ from Eq. \eqref{w56_7}:

\begin{equation}
T= \dfrac{V_0}{(d-1)(d-2)}+\dots
\end{equation}
Analogously to the case with $n=1$, the Ricci scalar and the Kretschman invariant approach a constant value as $\f\to 0$:

\begin{equation}
	R= \dfrac{d+1}{d-1}V_0 + \dots \ \ \ \ K_2 = \frac{2 (3 d-5) V_0^2}{(d-2) (d-1)^2} + \dots
\end{equation}

Following similar steps to the case $n=1$, we find that the geometry asymptotes to a boundary of dS$_2\times$S$^{(d-1)}$.

\item[G.3.3] $n>2$. The $\f^3$ term in \eqref{eqse4} dominates and a non-trivial solution requires $V_4=0$. The next contribution to Eq. \eqref{w56_1b} is given now by

\begin{equation}
	\frac{1}{6} c^6 n^6 (1 + n)  W_0^4 \left(A (d-1) n V_5 \f^4  - 24  V_0 \f^{1+n} \right) + O(\f^4,\f^{2+n},\f^{1+2n})=0\,.
\end{equation}

The structure of cases repeats, in analogy with the previous discussion.

\end{itemize}

\end{itemize}

Overall, we conclude that there will be non-trivial solutions for $n\in \mathbb{N}$ which require that $V_1=V_2=\dots=V_{n+1}=0$. For each solution to be a possible endpoint of the flow, it is required that $\f^n V_{n+2}>0$ as we approach $\f\to0$. The geometry around the endpoint gives a boundary of dS$_2\times$S$^{(d-1)}$.

All such solutions have a single integration constant, $W_0$. In order to find other integrations constants not captured by the ansatz \eqref{anse}, we linearise the equations of motion around the known solution, as in Eq. \eqref{eqpe}. For concreteness, we study the solution for $n=1$. In this case, Eq. \eqref{eqpe} becomes to leading order

\begin{align}\label{peex}
& \left(\frac{16 V_0^3W_0^3 \f ^7 }{(d-1)^2 V_3^2} + O(\f^8)\right)\delta W^{(4)} -\left(\frac{192 V_0^4W_0^3 \f ^5 }{(d-1)^3 V_3^3}+ O(\f^6)\right)\delta W^{(3)}
 \nonumber\\
&  +\left( \frac{704 V_0^5W_0^3 \f ^3 }{(d-1)^4 V_3^4}+ O(\f^4)\right)\delta W'' -\left(\frac{768 V_0^6W_0^3 \f }{(d-1)^5 V_3^5}+ O(\f^2)\right) \delta W' \nonumber\\ & -\left(\frac{64 (d-4) V_0^5 W_0^3 \f ^3}{(d-1)^5 V_3^4}+ O(\f^4)\right) \delta W=0\,.
\end{align}

In order to find the solution to the previous linear equation, we write an ansatz inspired by the leading exponential solution,

\begin{equation}
\delta W(\f) = e^{-z/\f}\f^y \delta w(\f)
\end{equation}
where $z$ and $y$ are determined by perturbatively solving the differential equation. In particular, substituting the previous ansatz into \eqref{peex} gives to leading order

\be
\frac{16 V_0^3 W_0^3 z \left((d-1) V_3 z \left((d-1)^2
   V_3^2 z^2-12 (d-1) V_3 V_0 z+44 V_0^2\right)-48
   V_0^3\right)}{(d-1)^5 V_3^5~ e^{z\over \f}~\f^{-(y+1)}}  \delta w(0)+\dots = 0
\ee
from which we find

\begin{equation}
z=0 \sp z= \dfrac{2 V_0}{(d-1)V_3} \sp z= \dfrac{4 V_0}{(d-1)V_3} \sp z= \dfrac{6 V_0}{(d-1)V_3}\,.
\end{equation}

The solution $z=0$ is discarded since it is not subleading with respect \eqref{anse}. Similarly,  $z= \frac{2 V_0}{(d-1)V_3}$ gives again the leading solution, see Eq. \eqref{A}, so we can reabsorb this solution into the leading one. Finally, the two possibilities
$$ z_1= \dfrac{4 V_0}{(d-1)V_3}\,, z_2= \dfrac{6 V_0}{(d-1)V_3}$$ are subleading and therefore allowed. Solving Eq. \eqref{peex} to subleading order fixes the value of $y$ in each case:

\begin{align}
 &z_1= \dfrac{4 V_0}{(d-1)V_3}\sp y_1 = - 3 - \dfrac{4V_0V_4}{3(d-1)V_3^2}\,,\nonumber\\
 &z_2= \dfrac{6 V_0}{(d-1)V_3}\sp y_2 = - 2 - \dfrac{2V_0V_4}{3(d-1)V_3^2}\,.
\end{align}

We conclude that the solutions studied in this appendix have three integration constants. In this case, where $V_1=V_2=0\neq V_0,V_3$, the full asymptotic solution for the superpotential reads

\begin{align}
W &= W_0 \f^{2- \frac{2V_0V_4}{3(d-1)V_3^2}} e^{-\frac{2 V_0}{(d-1)V_3 \f}}(1 + O(\f)) + C_1   \f^{-3 - \frac{4V_0V_4}{3(d-1)V_3^2}} e^{-\frac{4 V_0}{(d-1)V_3 \f}}(1 + O(\f))\nonumber \\+& C_2   \f^{-2 - \frac{2V_0V_4}{3(d-1)V_3^2}} e^{-\frac{6 V_0}{(d-1)V_3 \f}}(1 + O(\f))
\end{align}

\vskip 1cm
\section{Novel  solutions}\label{app:tuned}
\vskip 1cm

In this appendix we collect the construction of the novel flows presented in section \ref{fine2}. The strategy to construct such flows, is to choose a superpotential featuring maxima and minima that correspond to the endpoints of interest in each case. Given a superpotential $W$, we solve the first Eq. \eqref{eqtt} to find the inverse scale factor $T$. The blackening function $f$, is later obtained by solving Eq. \eqref{f10_1}. Finally, we reconstruct the potential $V$ by solving algebraically Eq. \eqref{w55}.
In all of this appendix, we have set $d=4$.

\subsection{Solutions from (A)dS$_2$ boundary endpoints to shrinking endpoints}
\label{app:G}

In this subsection, we display specific examples involving extrema where the metric becomes the near-boundary dS$_2\times $S$^3$ or AdS$_2 \times $H$^3$ metric. The local structure of the dS$_2$ solutions has been described in appendix \ref{g53}. According to Eq. \eqref{tds2} the inverse scale factor $T$ is proportional to the value of the potential $V_0$ at the given point. Since we need $T>0$ for our analysis, as we are in the spherical sliced ansatz \eqref{f22}, such endpoints can only happen in the dS regime ($V_0>0$).

The AdS$_2$ endpoints can be obtained instead with a hyperbolic slicing. The equations of motion for the hyperbolic slicing can be obtained from Eqs. \eqref{f23a}-\eqref{f23d} under the analytical continuation $R\to i R$. Therefore, in such slicing $T<0$ and the AdS$_2$ endpoints are obtained from Appendix \ref{g53} with $V_0<0$.

As shown in section \ref{pib}, flows involving dS$_2$ (or AdS$_2$) boundary endpoints can end regularly at shrinking endpoints, described in Sec. \ref{shr}. Such shrinking endpoints correspond to the center of AdS$_5$ in global coordinates, or to the location of an observer in the static patch coordinates of dS$_5$. Which of the two cases is relevant  depends on the sign of the potential $V$ in the relevant region.

We begin by constructing a superpotential featuring two types of extrema: a minimum with dS$_2$ or AdS$_2$ endpoints and a maximum with a shrinking endpoint. The chosen superpotential is:

\begin{equation}
    W(\f) = W_4 \left(\f^4 -\frac{142}{89}\f^5 + \frac{59}{89}\f^6\right)
\label{n100}\end{equation}
We set $W_4=1\,$ without loss of generality by virtue of the scaling symmetry \eqref{scaling}. The endpoints of this flow, $\f=0$ and $\f=1$ are two of  the extrema of this superpotential\footnote{There is a third extremum at $\f={178\over 177}>1$ which is not relevant for this flow.} . We  study flows between $\f=0$ and $\f=1$. At $\f=0$ there is a minimum of $W$ where the metric will become asymptotic to dS$_2 \times $S$^3$ or AdS$_2 \times $H$^3$, whereas at $\f=1$ there is a maximum, where the metric behaves as \eqref{C112}. The behaviour of the superpotential at the minimum matches that of a  (A)dS$_2$ boundary endpoint, Eq. \eqref{wds2}, with $\delta_\pm = \frac{1}{2}$.

Given the superpotential in (\ref{n100}), we first solve equation \eqref{w54a} for the function $T$ and find

\begin{equation}
    T =C_t \frac{ e^{\frac{\f  (177 \f -142)}{6372}} (178-177 \f
   )^{\frac{277235}{281961}}}{1-\f}\,,
\end{equation}

\noindent
where $C_t$ is the  constant of integration. Note that $T$ diverges as $\frac{1}{1-\f}$ as we approach the extremum at $\f_*=1$ as dictated by \eqref{C112}. By virtue of Eqs. \eqref{C113} and \eqref{C114}, the signs of $T$ and $f$ are correlated. For this solution $C_t$ determines the sign of $T$ at $\f_*=1$ and hence the sign of $f$.  Choosing $C_t>0$, then $T>0$ and we land in the center of AdS$_5$ in global coordinates  as shown in Sec. \ref{shr}. At $\f=0$, we have $$T(0)= 178^{277235/281961} C_t\;.$$

We may now solve Eq. \eqref{f10_1} numerically to find  $f$. From Eq. \eqref{fds2} we observe that our superpotential in (\ref{n100}) corresponds to  $\delta_{+}=\delta_{-} =  \frac{1}{2}$. Next, from  \eqref{fds2} we know that $f$ diverges at the (A)dS$_2$ endpoint as $f\sim \f^{-4}$. On the other hand, from Eq. \eqref{C112}, we obtain that $f$ diverges as $f\sim (\f - 1)^{-1}$ near the shrinking endpoint. We emphasize that these solutions give a regular geometry despite the diverging behaviour of $f$ and $T$. For numerical convenience, we redefine $f$ extracting its divergences:

\begin{equation}\label{j4}
    f(\f):= \frac{f_{-1}}{1-\f} + \frac{1}{\f^4}f_s(\f).
\end{equation}

\noindent
Solving Eq. \eqref{f10_1} near $\f=1$ fixes $f_{-1} = 7921 e^{\frac{35}{6372}}C_t\,$. Different values of $f_s(1)$ in Eq. \eqref{j4} label different solutions. Once $f$ is computed, the potential $V$ is extracted from \eqref{w55}. The value of $V(0)$ is also proportional to $C_t$, whereas $f_s(0)\propto -C_t$. We discuss now three particular solutions.

\subsubsection*{From dS$_2 \times$ S$_3$ to dS$_5$}
\label{sec:G1}

We fix $C_t$ so that $T>0$. Specifically, we set $T(0)=2$ without loss of generality, since the equations of motion are invariant under $(f,T,V)\to\lambda (f,T,V)$. Via (\ref{tds2}) this leads to $V(0)=12\,$.
Finally, we demand $f_s(1) = 2\,$. The solution is shown in figure \ref{fig:G1}. Along the flow $V>0$ whereas $f$ vanishes at one point signalling the presence of a horizon. This is the cosmological horizon from the point of view of the observer that is at $\f=1$.

Below we discuss the geometry around both endpoints of the flow.
\begin{itemize}
    \item The metric around $\f=0$.\\
    First, we solve the flow equation $\dot{\f}=W'$ near $\f=0$. This gives
\begin{equation}
\f(u) =\frac{1}{\sqrt{-8u}}+\dots
\end{equation}
and as a consequence the holographic coordinate $u\to -\infty$ as $\f\to 0$.
    We know that the blackening function $f$ is negative and diverges as $f=f_s(0)/\f^4+\dots = 64 f_s(0)u^2 + \dots$, whereas $T$ approaches a constant positive value. Therefore, the ansatz \eqref{f22} becomes

    \begin{equation*}
        ds^2 = -\frac{du^2}{64|f_s(0)|u^2} + u^2 64|f_s(0)|e^{2A_0}dt^2 + ds^2(S^3)
    \end{equation*}
    \begin{equation}\label{eae}
    = - d\rho^2 +
        e^{-16|f_s(0)|\rho} d\tilde{t}^2  + ds^2(S^3)\,,
    \end{equation}

    \noindent
    where we have taken into account that $f_s(0)<0$. In the last step we defined $8\sqrt{|f_s(0)|}\rho = - \log |u|$ and $\tilde{t} = 8\sqrt{|f_s(0)|}e^{A_0}$. As $u\to-\infty$ then $\rho \to -\infty$ and we have the future boundary $\mathcal{I}^+$ of dS$_2$ times the 3-sphere.

    \item The metric around $\chi:=\f-1=0$.\\
	We use the asymptotic solution in \eqref{C112} and work with the metric as a function of $\chi$,
	
	\begin{equation}
	\label{eq:G4}
	\begin{split}
		&ds^2 = \frac{d\chi^2}{fW'^2} - \frac{f}{T}R^2 dt^2 + \frac{1}{T}d\Omega_3^2 \\
		&= \frac{d\chi^2}{V_1^2}\left(\frac{2V_1}{\chi} - \frac{2(V_0+3V_2)}{3}\right)-\frac{36 R^2}{W_0^2}\left(1-\frac{V_0 \chi}{3 V_1}\right)dt^2 + \left(\frac{8 \chi}{V_1}-\frac{8(V_0+V_2)}{3 V_1^2}\chi^2\right)d\Omega_3^2 \\&+\dots.
	\end{split}
    \end{equation}	   	
   	
   	The fact that $T>0$ along with Eqs. \eqref{C112} and \eqref{C114} impose $V_1/\chi>0$. The dots stem for higher orders in $\chi$. We change variables so that $g_{\chi\chi}d\chi^2 = d\rho^2$. This gives
   	\begin{equation}
   		\chi\simeq \frac{V_1}{8}\rho^2 + V_1 \frac{V_0+3V_2}{576}\rho^4 +\mathcal{O}(\rho^6)\,,
   	\end{equation}
   	where $\chi\to0$ translates into $\rho \to 0\,$. Substituting  into the above metric yields
   	
   	\begin{equation}\label{mj8}
   		ds^2 = d\rho^2 - \frac{36 R^2}{W_0^2}(1-\frac{V_0}{12}\rho^2)dt^2 + \frac{12}{V_0}\left(\frac{V_0}{12}\rho^2 - \frac{1}{3}\frac{V_0^2}{12^2}\rho^4\right)d\Omega_3^2+\dots
   	\end{equation}
   	which depends on $V_0$ as anticipated by \eqref{C116}. We may now parametrize  $V_0 = 12 H^2\,$. This gives the metric \eqref{a31} for small radial coordinate, so this corresponds to the location of an observer in  dS$_5$ in the static patch coordinates.
\end{itemize}

\vskip 1cm
\subsubsection*{From AdS$_2 \times$ H$_3$ to AdS$_5$}
\vskip 1cm

\begin{figure}[h!]
\centering
\subfloat{\includegraphics[width=0.47\linewidth]{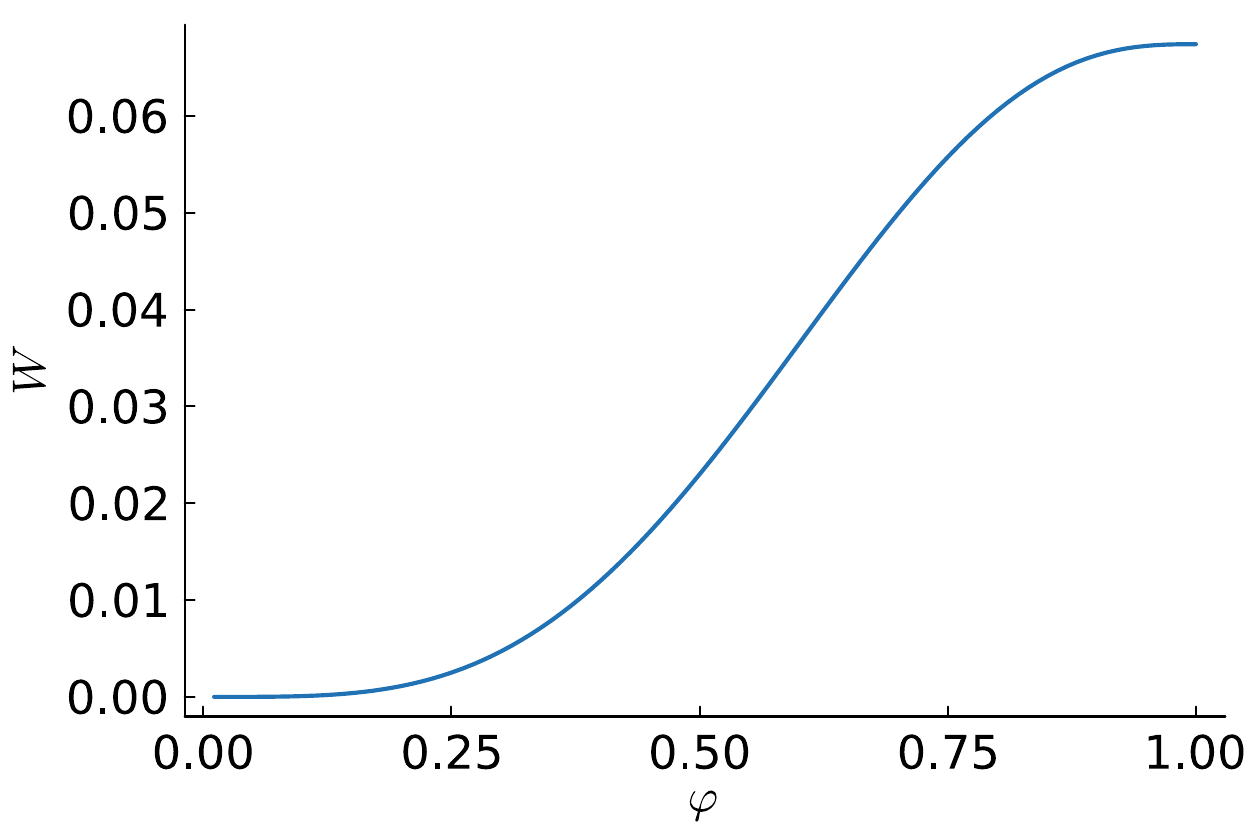}}
	\qquad
	\subfloat{\includegraphics[width=0.47\linewidth]{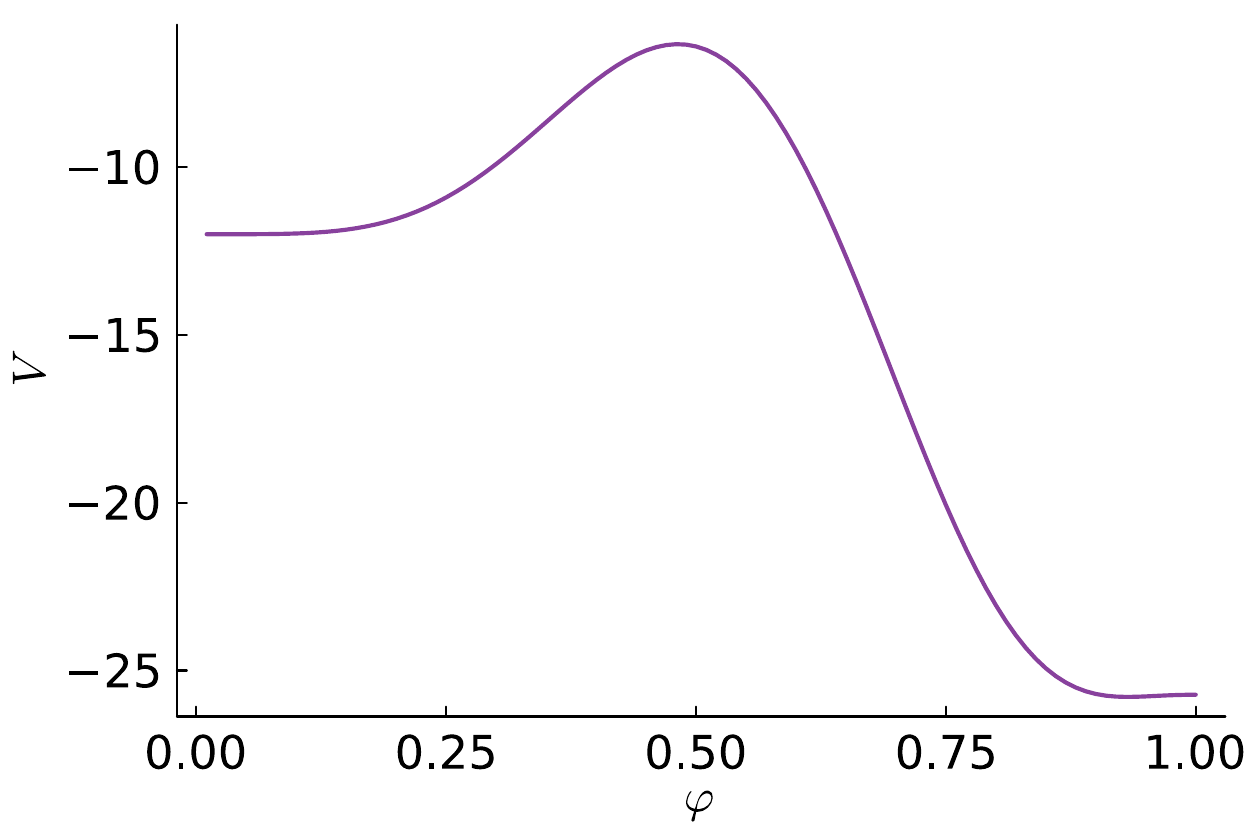}}
	\qquad
	\subfloat{\includegraphics[width=0.47\linewidth]{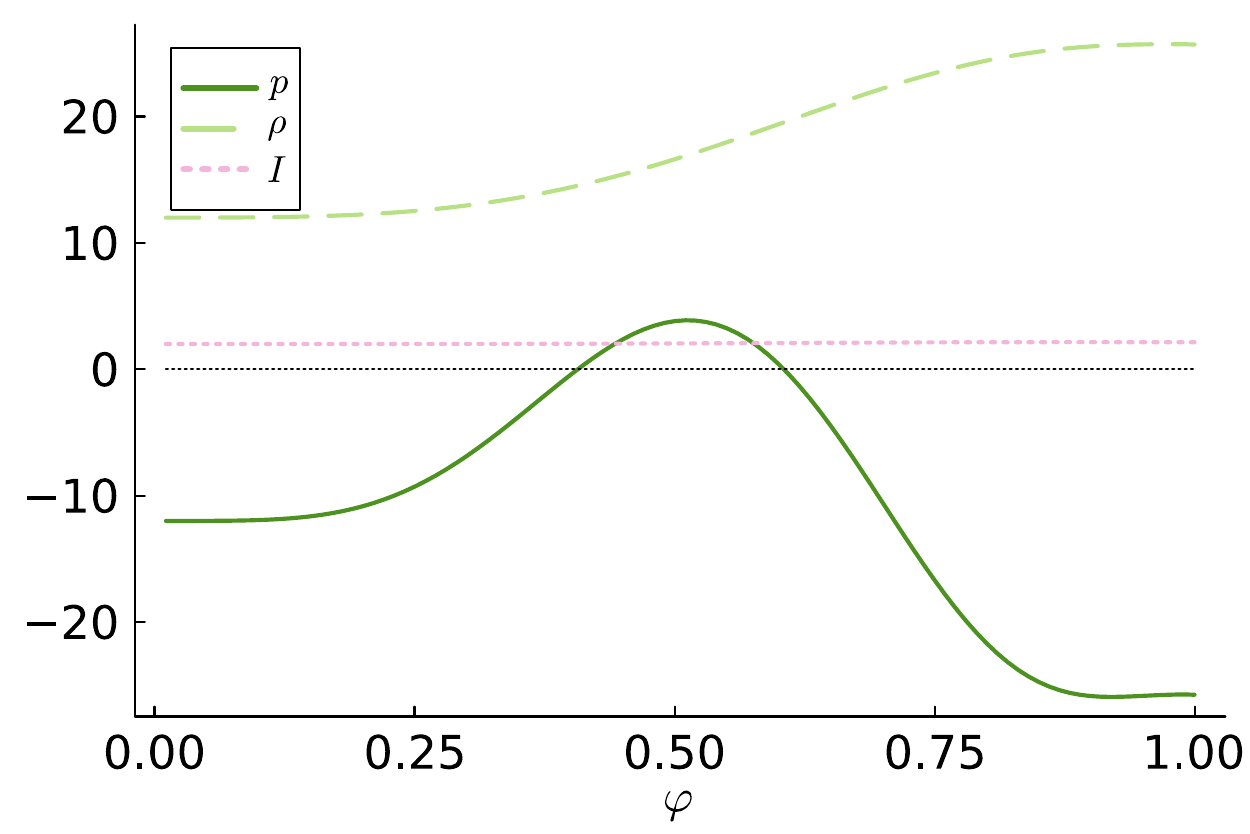}}
	\qquad
	\subfloat{\includegraphics[width=0.47\linewidth]{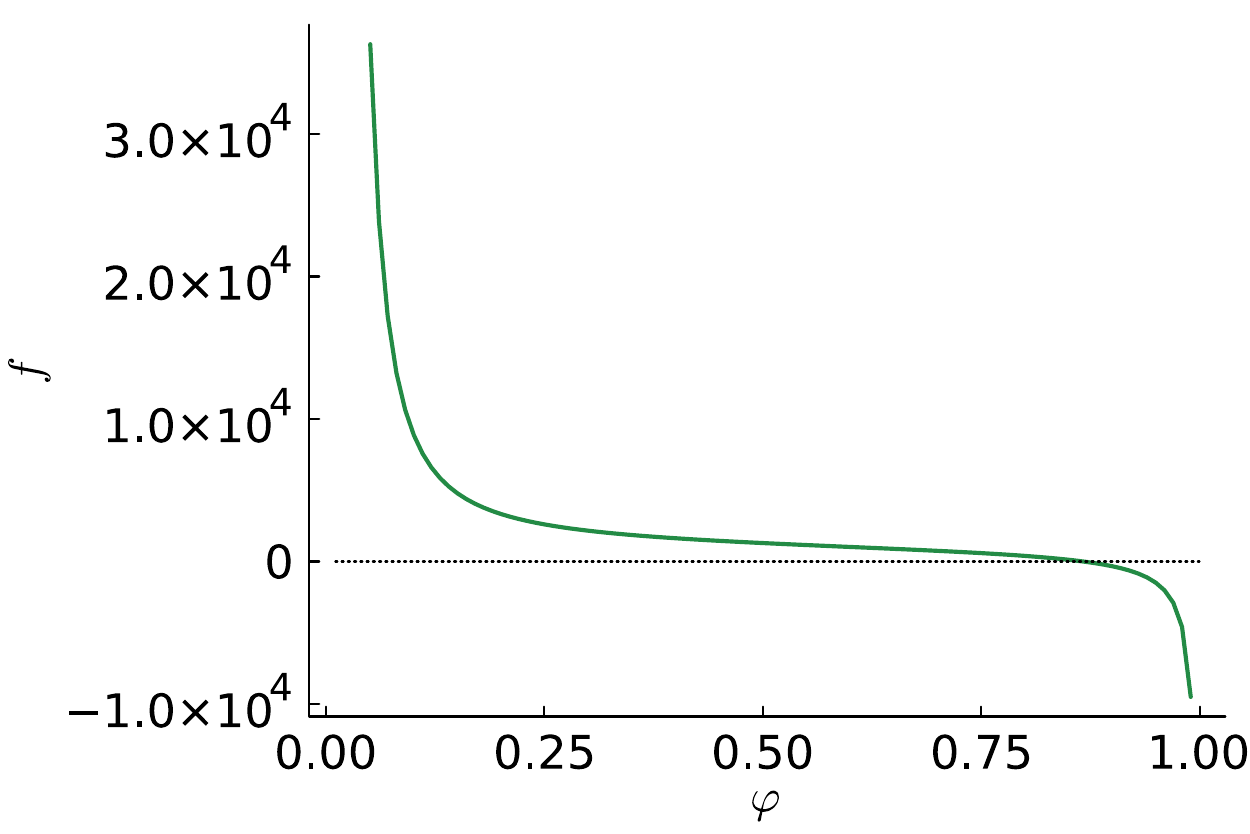}}
\caption{Flow from the boundary of AdS$_2 \times $H$_3$ at $\f=0$ to the center of AdS$_5$ in global coordinates, at $\f=1$. The superpotential vanishes as $\f^4$ near $\f=0$, while $f$ diverges at both endpoints and $T$ diverges at $\f=1$. The curvature invariants are regular along the flow and at the end-points.
}
\label{fig:G2}
\end{figure}

This is the twin solution to the previous one, with $(f,T,V)\to(-f,-T,-V)\,$.
Although, this solution is not relevant to our sphere-sliced ansatz, we present it here as we think it is new.

We fix $C_t$ so that $T(0)=-2$. This time it leads to $V(0)=-12\,$. Finally, we demand $f_s(1) = -2\,$. The solution is shown in figure \ref{fig:G2}. The potential is negative along the flow, $V<0$, whereas the blackening function $f$ vanishes at one point signalling the presence of a horizon.

We add a few more details about this solution.

\begin{itemize}
    \item Metric around $\f=0$.

	First, we solve the flow equation $\dot{\f}=W'$ near $\f=0$. This gives
\begin{equation}
\f(u) =\frac{1}{\sqrt{-8u}}+\dots
\end{equation}
and as a consequence the holographic coordinate $u\to -\infty$ as $\f\to 0$.
    We know that the blackening function $f$ is positive and diverges as $f=f_s(0)/\f^4+\dots = 64 f_s(0)u^2 + \dots$, whereas $T$ approaches a constant positive value. Therefore, the ansatz \eqref{f22} becomes

    \begin{equation*}
        ds^2 = \frac{du^2}{64f_s(0)u^2} - u^2 64f_s(0)e^{2A_0}dt^2 + ds^2(H^3)
    \end{equation*}
    \begin{equation}\label{mj10}
    =  d\rho^2 -
        e^{-16f_s(0)\rho} d\tilde{t}^2  + ds^2(H^3)\,,
    \end{equation}

    \noindent
    where we have taken into account that $f_s(0)>0$. In the last step we defined $8\sqrt{f_s(0)}\rho = - \log |u|$ and $\tilde{t} = 8\sqrt{f_s(0)}e^{A_0}$. As $u\to-\infty$ then $\rho \to -\infty$ and we have the boundary of AdS$_2\times$H$^3$.

    \item Metric around $\chi:=\f-1=0$.

    	We use the asymptotic solution in \eqref{C112} and work with the metric as a function of $\chi$,
	
	\begin{equation}
	\label{eq:G4b2}
	\begin{split}
		&ds^2 = \frac{d\chi^2}{fW'^2} - \frac{f}{T}R^2 dt^2 + \frac{1}{T}dH_3^2 \\
		&=  \frac{d\chi^2}{V_1^2}\left(\frac{2V_1}{\chi} - \frac{2(V_0+3V_2)}{3}\right)-\frac{36 R^2}{W_0^2}\left(1-\frac{V_0 \chi}{3 V_1}\right)dt^2 + \left(\frac{8 \chi}{V_1}-\frac{8(V_0+V_2)}{3 V_1^2}\chi^2\right)dH_3^2 \\&+\dots.
	\end{split}
    \end{equation}	   	
   	
   	The fact that $T<0$ in this solution, along with Eqs. \eqref{C112} and \eqref{C114} impose $V_1/\chi<0$. The dots stem for higher orders in $\chi$. We change variables to
   	\begin{equation}
   		\chi\simeq \frac{V_1}{8}\rho^2 - V_1 \frac{V_0+3V_2}{576}\rho^4 +\mathcal{O}(\rho^6)\,,
   	\end{equation}
   	where $\chi\to0$ translates into $\rho \to 0\,$. Substituting  into the above metric yields
   	
   	\begin{equation}
   		ds^2 = d\rho^2 - \frac{36 R^2}{W_0^2}(1-\frac{V_0}{12}\rho^2)dt^2 + \frac{12}{V_0}\left(\frac{V_0}{12}\rho^2 - \frac{1}{3}\frac{V_0^2}{12^2}\rho^4\right)dH_3^2+\dots
   	\end{equation}
   	which depends on $V_0$ as anticipated by \eqref{C116}. We may now parametrize  $V_0 = -12 /\ell^2\,$. This gives the center AdS$_5$ in global coordinates with hyperbolic foliation, as it can be seen from Eq. \eqref{adsg} as $\rho\to0$.
\end{itemize}

\vskip 1cm
\subsubsection*{From dS$_2 \times$ S$_3$ to AdS$_5$}\label{secJ13}
\vskip 1cm

We fix $C_t$ so that $T(0)=2$. This automatically leads to $V(0)=12\,$. Finally, we demand $f_s(1) = 20000\,$. The solution is shown in figure \ref{fig:G3b}. Along the flow $V$ changes sign. Similarly $f$ vanishes at one point signalling the presence of a horizon, which is again cosmological. The geometry around the endpoints is the same as is \ref{sec:G1} despite the fact that $V$ changes sign:

We now discuss the geometry around both endpoints:

\begin{itemize}
    \item The metric around $\f=0$.
   	The situation is analogous to the one presented above. At $\f=0$ the metric is given again by Eq. \eqref{eae}, describing the future boundary $\mathcal{I}^+$ of dS$_2$ times the 3-sphere.

    \item Metric around $\chi:=\f-1$.
   	We again have \eqref{eq:G4} with $R^2>0$ and $V_0<0$. Therefore this corresponds to the center of AdS$_5$ in global coordinates.
\end{itemize}

\vskip 1cm
\subsubsection*{From AdS$_2 \times$ H$_3$ to dS$_5$}
\vskip 1cm

\begin{figure}[h!]
\centering
\subfloat{{\includegraphics[width=0.47\linewidth]{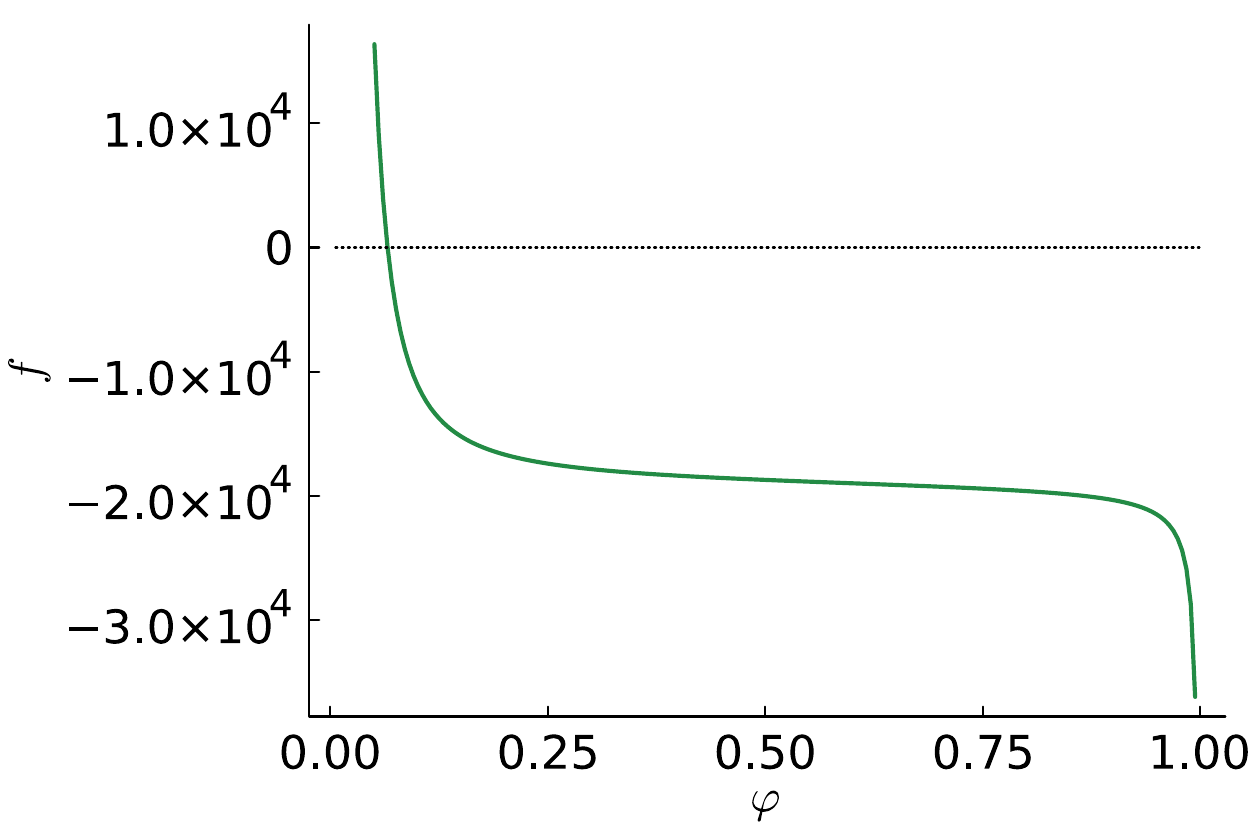}}}
	\qquad
	\subfloat{{\includegraphics[width=0.47\linewidth]{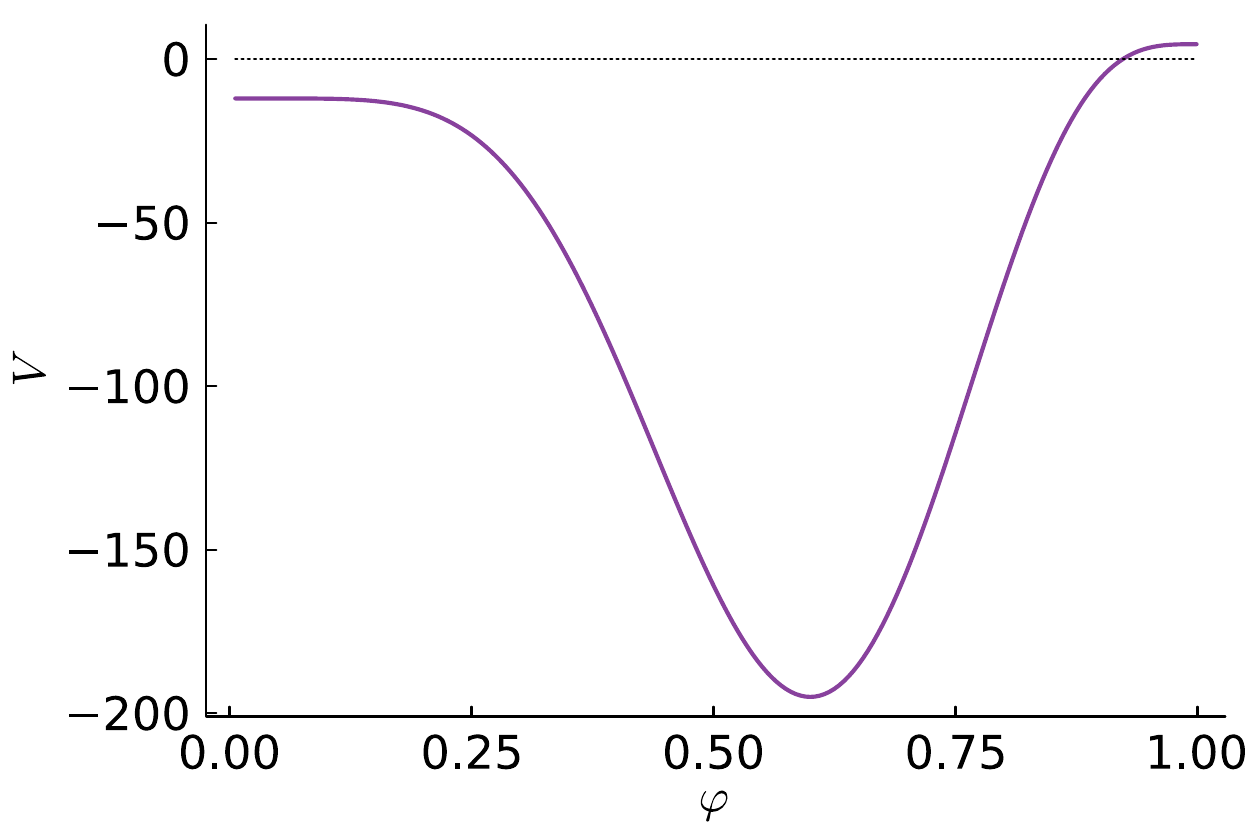}}}
	\qquad
	\subfloat{\includegraphics[width=0.47\linewidth]{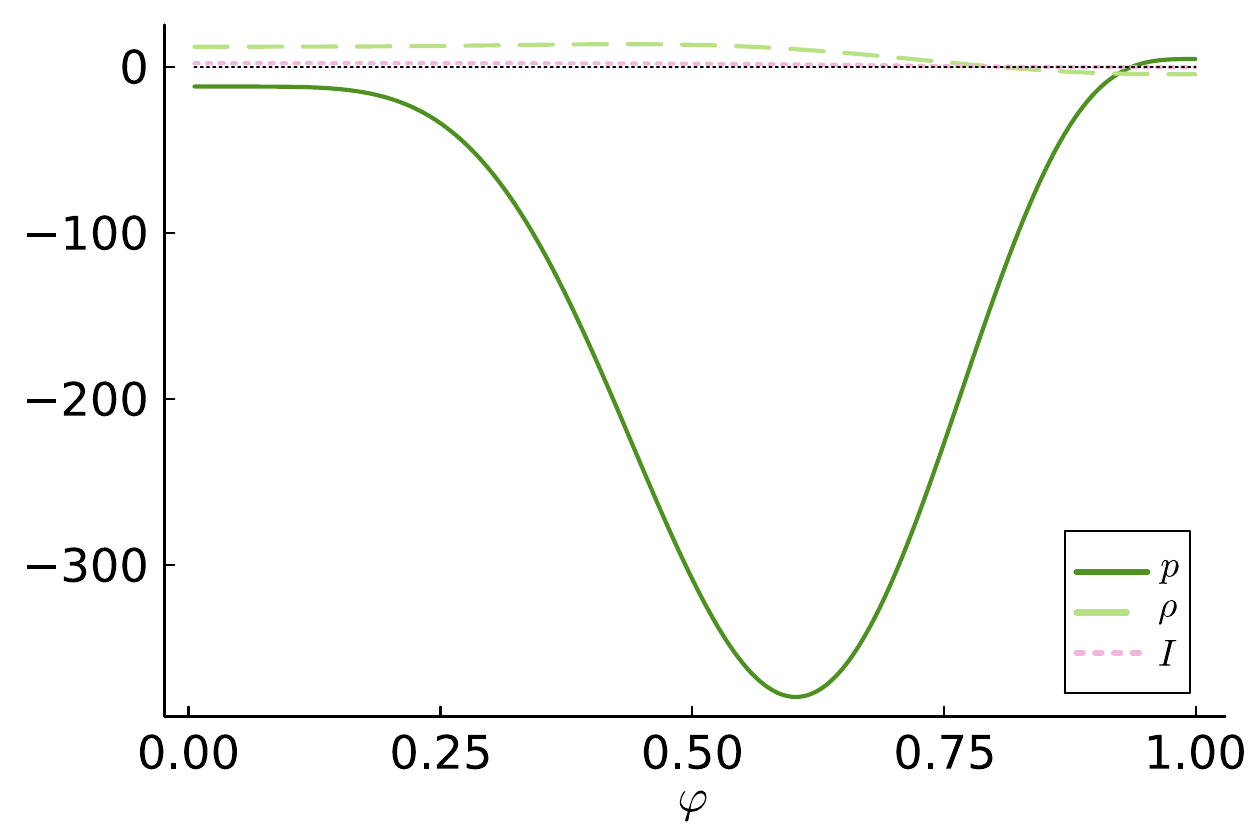}}
	\subfloat{{\includegraphics[width=0.47\linewidth]{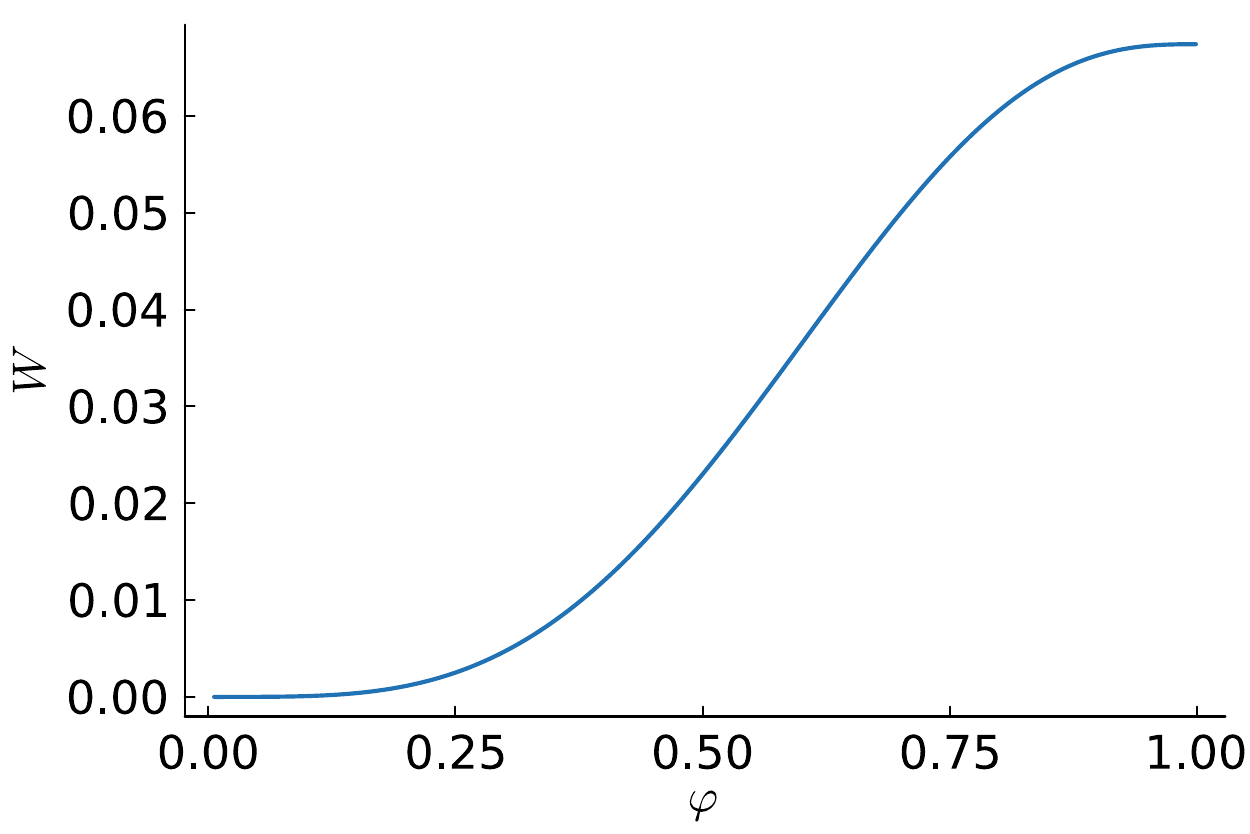}}}
\caption{Flow from the boundary of AdS$_2 \times$ S$_3$ at $\f=0$ to the interior of dS$_5$ at $\f=1$. The superpotential vanish as $\f^4$ near $\f=0$, while $f$ diverges at both endpoints and $T$ diverges at $\f=1$. The curvature invariants are regular along the flow.
}
\label{fig:G4}
\end{figure}

This solution can be trivially obtained from the previous one by the transformation $(f,T,V)\to(-f,-T,-V)$. The result is displayed in figure \ref{fig:G4}. The metric around $\f=0$ is again given by \eqref{mj10}, and it describes boundary of AdS$_2\times$H$^3$. On the other hand, around $\f=1$, we obtain the metric as in \eqref{mj8} with hyperbolic foliation, corresponding to the location of an observer in  dS$_5$. We conclude that the solution describes a flow from the boundary of AdS$_2$ towards a shrinking endpoint in the dS regime. The blackening function $f$ vanishes once along the flow, signalling the presence of a horizon.

\subsection{Solution from a dS$_2$ boundary to a black-hole event horizon.}\label{app:dS2BH}

In this section, we construct a solution from the boundary of dS$_2\times $S$^3$ across two horizons, the outermost being cosmological while the inner one corresponds to a black-hole event horizon. Similarly to the previous section, we engineer a superpotential which has the desired properties, and subsequently find the functions $T$, $f$, and the potential $V$. In this case, we use the following superpotential:
\begin{equation}
W= \f^4\,.
\end{equation}
At $\f=0$, the superpotential vanishes in agreement with the dS$_2$ asymptotic solution, Eq. \eqref{wds2}, for $\delta_{\pm}=1/2$. There are no other extrema of the superpotential and, as a consequence, a flow starting at the dS$_2$ boundary necessarily runs to the boundary of field space $\f\to\infty$, where it encounters a singularity. We construct a flow for $\f>0$ without loss of generality.

We now solve the first relation in Eq. \eqref{eqtt} to obtain the inverse scale factor $T$, and integrate once Eq. \eqref{f10_1} to obtain
\begin{equation}\label{i2b}
T= C_T e^{\frac{1}{24}\f^2}\,,\quad f'= \frac{24 C_T e^{\frac{\varphi ^2}{24}}+e^{\frac{\varphi ^2}{12}} \varphi ^2
   \left(C_T\text{Ei}\left(-\frac{\varphi ^2}{24}\right)+192 f_1\right)}{192 \varphi ^5}\,,
\end{equation}
where Ei$(x)$ is the exponential integral function, while $C_T$ and $f_1$ are integration constants. We set the integration constant $C_T=1$ without loss of generality. Note that, according to the analysis of Appendix \ref{g53}, the blackening function $f$ diverges to $-\infty$ as it approaches the dS$_2$ boundary. Additionally, a solution from a dS$_2$ boundary to a black-hole event horizon requires that $f$ vanishes twice, such that the outer root corresponds to a cosmological horizon and the inner root to the event horizon. We conclude that a necessary and sufficient condition for such a flow to exist is that $f$ has an extremum, $f'(\f_*)=0$, at a point $\f_*$ where $f(\f_*)>0$. The condition $f'(\f_*)=0$ fixes the integration constant $f_1$:
\begin{equation}
f_1 =-\frac{1}{192} \text{Ei}\left(-\frac{\f_*^2}{24}\right)-\frac{e^{-\frac{\f_*^2}{24}}}{8 \f_*^2}
\end{equation}
The function $f$ shall be obtained by numerically integrating Eq. \eqref{i2b}. The boundary condition controls the value of $f$ at $\f_*$. As we discussed earlier, we shall demand $f(\f_*)>0$.

Once $f$ is found, we reconstruct the potential $V$ algebraically solving Eq. \eqref{w55}:
\begin{equation}
V= \frac{1}{96}  e^{\frac{\varphi ^2}{12}} \varphi ^4 \text{Ei}\left(-\frac{\varphi
   ^2}{24}\right)+\frac{1}{4} e^{\frac{\varphi ^2}{24}} \left( \varphi ^2+8 f_1
   e^{\frac{\varphi ^2}{24}} \varphi ^4+24\right)+\left(8 \varphi ^6-\frac{\varphi ^8}{3}\right)
   f(\varphi )\,.
\end{equation}

In Fig. \ref{fig:ds2bh} we show a concrete example of the solution described in this appendix. We demand that the maximum of $f$ is located at $\f_*=2$, and that $f(\f_*)=3$. The solution features a dS$_2$ boundary at $\f=0$, where the potential $V$ is positive and the function $f$ diverges to $-\infty$ as $\f^{-4}$. The curvature invariants remain finite at $\f=0$ despite the apparent divergence of $f$. As the solution departs from the dS$_2$ boundary, the function $f$ vanishes twice. The outermost vanishing signals the presence of a cosmological horizon, while the innermost vanishing is tied to the presence of a black-hole event horizon. Inside the black hole, the solution hits a bad singularity at $\f\to\infty$.

\subsection{Solutions from $d+1$ boundary endpoints to shrinking endpoints}
\label{app:H}

In this subsection we describe a solution that interpolates between the boundary  dS$_5$ endpoint and a shrinking endpoint in the (A)dS regime. This constitutes a proof of existence of such flows. Firstly, we engineer a superpotential featuring one extremum that  generates the boundary of dS$_5$ and another extremum generating  a shrinking endpoint:

\begin{equation}
	W(\f) = W_0\left(1-\frac{\f ^2}{6}+\frac{37 \f ^3}{219}-\frac{19 \f ^4}{438}\right)\,,
\label{n101}\end{equation}

\noindent
and we set $W_0=1$ for by virtue of the scaling symmetry \eqref{scaling}. This superpotential has (1) a maximum at $\f=0$  corresponding to a shrinking endpoint (2) a minimum at $\f=1$ corresponding to a boundary endpoint, with $\Delta_-=1$ according to the relation above Eq. \eqref{DEL}. We now study the flow between these two extrema.

We first solve \ref{w54a} for the function T:

\begin{equation}
	T =C_t \frac{ e^{\frac{1}{456} \f  (19 \f -37)} (\f -1)^2}{(73-38 \f
   )^{18661/17328} \f }\,,
\end{equation}

\noindent
where $C_t$ is a constant of integration. Note that $T$ diverges as $1/\f$ as we approach $\f_*=0$ as dictated by \eqref{C112}, while it vanishes as we approach the extremum $\f_*=1$ . Eq. \eqref{C112} also implies that the function $f$  diverges as $1/\f$ at $\f_*=0$ . For numerical convenience we redefine:

\begin{equation}
     f(\f):=\frac{f_s(\f)}{\f}\,.
\end{equation}

We further fix $C_t=\frac{73}{36} 73^{1333/17328}$, which is tantamount to demanding $f_s(0)=1$. Note that the equations of motion are invariant under $(T,f,V)\to\lambda (T,f,V)$, and therefore the value of $C_t$ is irrelevant. We solve \eqref{f10_1} numerically to find the form of $f$. Then, different values of $f_s'(0)$ label different solutions. Once $f$ is found, we compute $V$ using equation \eqref{w55}. Depending of $f_s'(0)$ four classes of solutions arise:
\begin{itemize}

\item (a) A flow without horizon from the boundary of AdS$_5$ to a shrinking endpoint in the AdS regime; This has the standard holographic interpretation, as dual to the ground state of a holographic QFT on $R\times S^{d-1}$.

\item (b) A flow from the boundary of dS$_5$ to a shrinking endpoint in the AdS regime with a cosmological horizon located in the dS regime.

    \item (c) A flow from the boundary of dS$_5$ to a shrinking endpoint in the dS regime, again with a cosmological horizon in the dS regime.

    \item (d) A flow from the boundary of M$_5$ to a shrinking endpoint in the AdS regime.

        \end{itemize}

   We did not find initial conditions that generate  a flow from the boundary of AdS$_5$ to the center of dS$_5$, in agreement with the discussion of Sec. \ref{sec:global}.

We present examples of solutions to cases (b), (c) and (d) above.

The results of dS$^{bdy}_5\to$ AdS$^{shrink}_5$ (case b) are shown in figure \ref{fig:H1}, the given boundary condition is $f_s'(0)=-2.82055$. The results of dS$^{bdy}_5\to$ dS$^{shrink}_5$  (case c) are shown in figure \ref{fig:H2}, the given boundary condition is $f_s'(0)=-6.52055$. The results of M$^{bdy}_5\to$ AdS$^{shrink}_5$ are shown in Fig. \ref{fig:M5}, the given boundary condition is $f_s(1)=0$.

The Penrose diagram of dS$^{bdy}_5\to$ AdS$^{shrink}_5$ solutions is similar to the dS$^{bdy}_5\to$ dS$^{shrink}_5$ solutions and this is similar to the Penrose diagram of dS space in static coordinates.

\section{Exact flows}
\label{app:I}

In this appendix we discuss the solution obtained that corresponds to the case where the derivation of Eq. \eqref{w56_1b}, used in the analysis of appendices \ref{structure}, \ref{WcApp} and \ref{sho}, does not apply. Specifically, it corresponds to the case where the superpotential satisfies

\be
2 (d-1) (W'')^2+(2-d) W W''+(d-2) W'^2-2 (d-1) W^{(3)} W'=0
\label{C120C}\ee

Then, the general equation (\ref{w56_1b}) is satisfied independent of the potential $V(\f)$, because all factors $b_0,b_1,b_2,b_3$ vanish when (\ref{C120C}) is satisfied\footnote{However, as we show later, the potential is determined indirectly.}, while the denominator of Eqs. \eqref{w56_6} and \eqref{w56_7} vanishes. The general solution to equation \eqref{C120C} is given by

\begin{equation}
\label{eq:d42}
    W = (C_1+C_2 z)z^{-\alpha}
\end{equation}

\noindent
where we have defined

\begin{equation}\label{k2}
	z = {\exp} \left({\frac{1}{\alpha}\sqrt{\frac{\alpha}{1-\alpha}} \sqrt{\frac{d-2}{2(d-1)}} \f }\right)\,.
\end{equation}

\noindent
$C_{1,2}$ and $\alpha$ are the three integration constants of equation (\ref{C120}).\footnote{The variable $\a$ have also been used in other appendices. There is no relation between this variable in different appendices.} 
We focus on the solutions where the three integration constants, $\a$, $C_1$ and $C_2$, are non-trivial.

A real superpotential is achieved only if $\alpha\in(0,1)$. Besides, $\f \to (0,+\infty,-\infty)$ are mapped to $z\to(1,+\infty,0)$ respectively. Note that when $\f$ diverges, the superpotential diverges exponentially. The superpotential has at most one extremum, located at

\begin{equation}\label{k3}
	z_* = \frac{C_1}{C_2}\frac{\alpha}{1-\alpha}\,.
\end{equation}
Besides, $W'_*W''_*=0$, revealing that a flow may start at $z_*$ but it does not stop at a finite value of $\f$.

One may solve $W'=\Dot{\f}$ to obtain $u(\f)$

\begin{equation}\label{k4}
    u -c_\f=\frac{2 (\alpha -1) (d-1) z^{\alpha } \, _2F_1\left(1,\alpha ;\alpha +1;\frac{C_2
    (1-\alpha)}{C_1 \alpha }z\right)}{\alpha  C_1 (d-2)}\,.
\end{equation}

Solving for \eqref{f10_1} gives the explicit form of $T$

\begin{equation}\label{k5}
    T(z)=C_t \left[z^{\alpha } \left(\frac{\alpha  C_1}{z(1-\alpha
   )}-C_2\right)\right]^{\frac{2}{d-2}}\,,
\end{equation}

\noindent
which vanishes at $z_*$. Now, equation \eqref{f10_1} gives $f$:
\begin{equation}
\begin{split}
	&f(z)=f_0 - \frac{T(z)}{C_t} \left[\frac{(\alpha -1)^2 C_2 f_1 z^{2 \alpha }}{2 \alpha  C_1 (\alpha
   (d-1)-1)} \, _2F_1\left(1,\frac{2 \alpha  (d-1)}{d-2};\frac{2 \alpha  (d-1)}{d-2}+\frac{d-4}{d-2};\frac{(1-\alpha) C_2 }{\alpha  C_1}z\right)\right. \\
   &\left. - \frac{(\alpha -1) z^{2 \alpha -1} \left(8 \alpha  C_t (d-1)^2-C_1 C_2
   f_1\right)}{C_1 C_2 (2 \alpha  (d-1)-d)} \times \right. \\& \left. \times  \, 
_2F_1\left(1,\frac{2 \alpha  (d-1)}{d-2}-1;\frac{2 \alpha  (d-1)}{d-2}-\frac{2}{d-2};\frac{(1-\alpha) C_2 }{\alpha  C_1}z\right) \right]
\end{split}
\end{equation}
Eq. \eqref{w55} can now be solved to find $V(\f)$:
\begin{equation}\label{vz}
\begin{split}
& V(z) = z^{-2\alpha} (h_0 + h_1 z + h_2 z^2) + T(z)\left[h_3 + (h_4+h_5z + h_6z^2)\, _2F_1\left(1,a;a+\frac{d-4}{d-2};c z\right)\right. \\
&+\left. h_7 \, _2F_1\left(1,a-1;a-\frac{2}{d-2};c z\right) + (h_8 + h_9 z + h_{10} z^2)\, _2F_1\left(2,a;a+\frac{d-4}{d-2};c z\right) \right. \\
& \left.+ (h_{11}z+h_{12}z^2+h_{13}z^3)\, _2F_1\left(2,a+1;a+\frac{2(d-3)}{d-2};c z\right) \right]
\end{split}
\end{equation}
\noindent
where
$$
a = \frac{2 \alpha  (d-1)}{d-2}\,, \qquad  c = \frac{(1-\alpha) C_2 }{\alpha  C_1} \,,\qquad h_0 = \frac{C_1^2 f_0 (2 \alpha -2 \alpha  d+d)}{4 (\alpha -1) (d-1)}\,, \qquad h_1 = -C_1 C_2 f_0
$$
$$
h_2 = \frac{C_2^2 f_0 (2 \alpha -2 \alpha  d+d-2)}{4 \alpha  (d-1)}\,,\quad h_3 = (d-2) (d-1)\,,\quad h_4 = -\frac{(\alpha -1) C_1 C_2 (d-2) f_1}{8 \alpha  C_t
   (d-1) (\alpha  (d-1)-1)}\,,
$$
$$
h_5 = \frac{2 C_2}{C_1} h_4 \,, \qquad h_6 =\frac{(\alpha -1) C_2^2}{\alpha  C_1^2}h_4\,, \qquad h_7 = \frac{(d-2) \left(8 \alpha  C_t (d-1)^2-C_1 C_2
   f_1\right)}{4 \alpha  \text{Ct} (d-1) ((2 \alpha -1) d-2 \alpha )}\,,
$$
$$
 h_8 = \frac{(\alpha -1) (d-2) (2 \alpha  (d-1)-d+2) \left(C_1 C_2
   f_1-8 \alpha  C_t (d-1)^2\right)}{8 \alpha  \text{Ct} (d-1)
   (\alpha  (d-1)-1) (2 \alpha  (d-1)-d)}\,, \quad h_9 = \frac{(2 \alpha -1) C_2}{\alpha  C_1}h_8\,,
$$
$$
h_{10} =\frac{(\alpha -1) C_2^2}{\alpha  C_1^2} h_8\,, \qquad h_{11} = \frac{(\alpha -1)^2 C_2^2 (d-2) f_1}{4 \alpha C_t (\alpha
   (d-1)-1) (2 \alpha  (d-1)+d-4)} \,,
$$
\begin{equation}
 h_{12} = \frac{(2 \alpha -1) C_2}{\alpha  C_1}h_{11}\,, \qquad h_{13} = \frac{(\alpha -1) C_2^2}{\alpha  C_1^2}h_{11}\,.
\end{equation}

\begin{figure}[h!]
	\centering
	\subfloat{\includegraphics[scale=0.32]{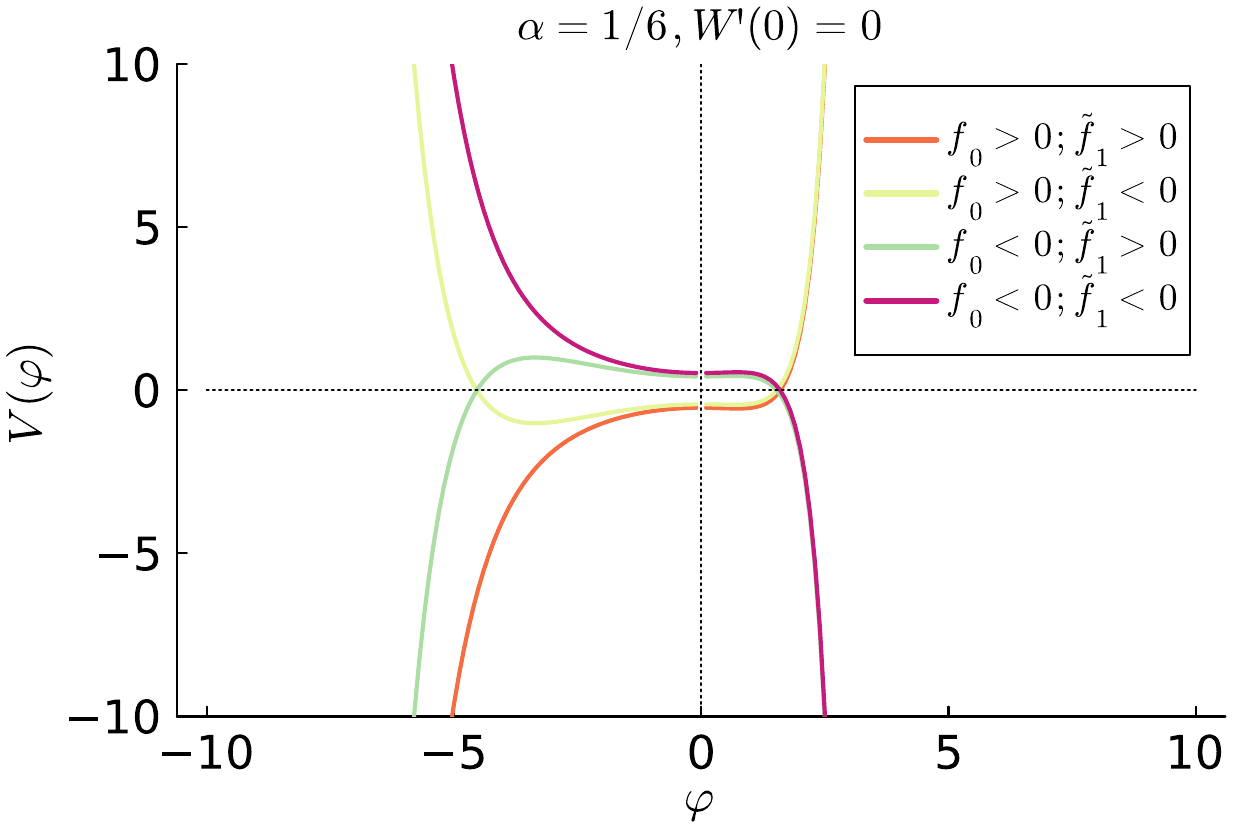}}
	\qquad
	\subfloat{\includegraphics[scale=0.32]{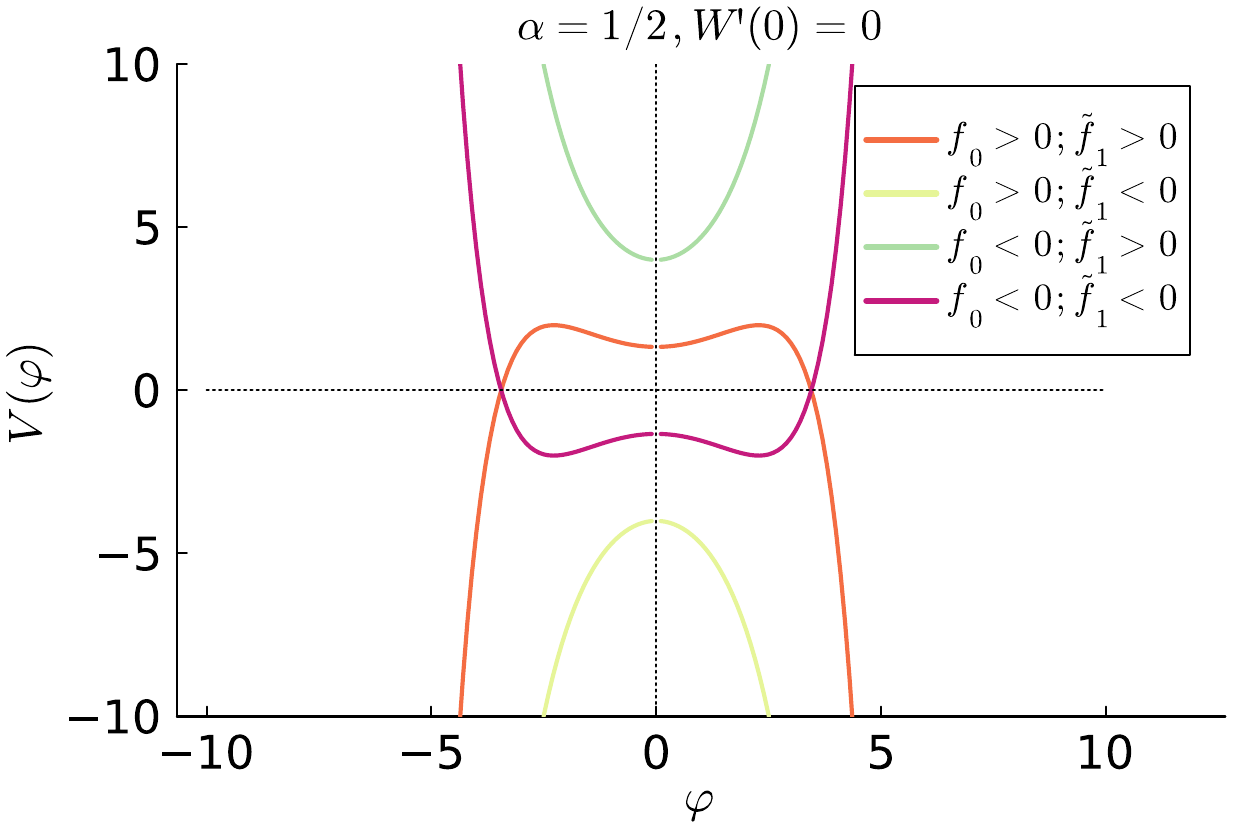}}
	\qquad
	\subfloat{{\includegraphics[scale=0.32]{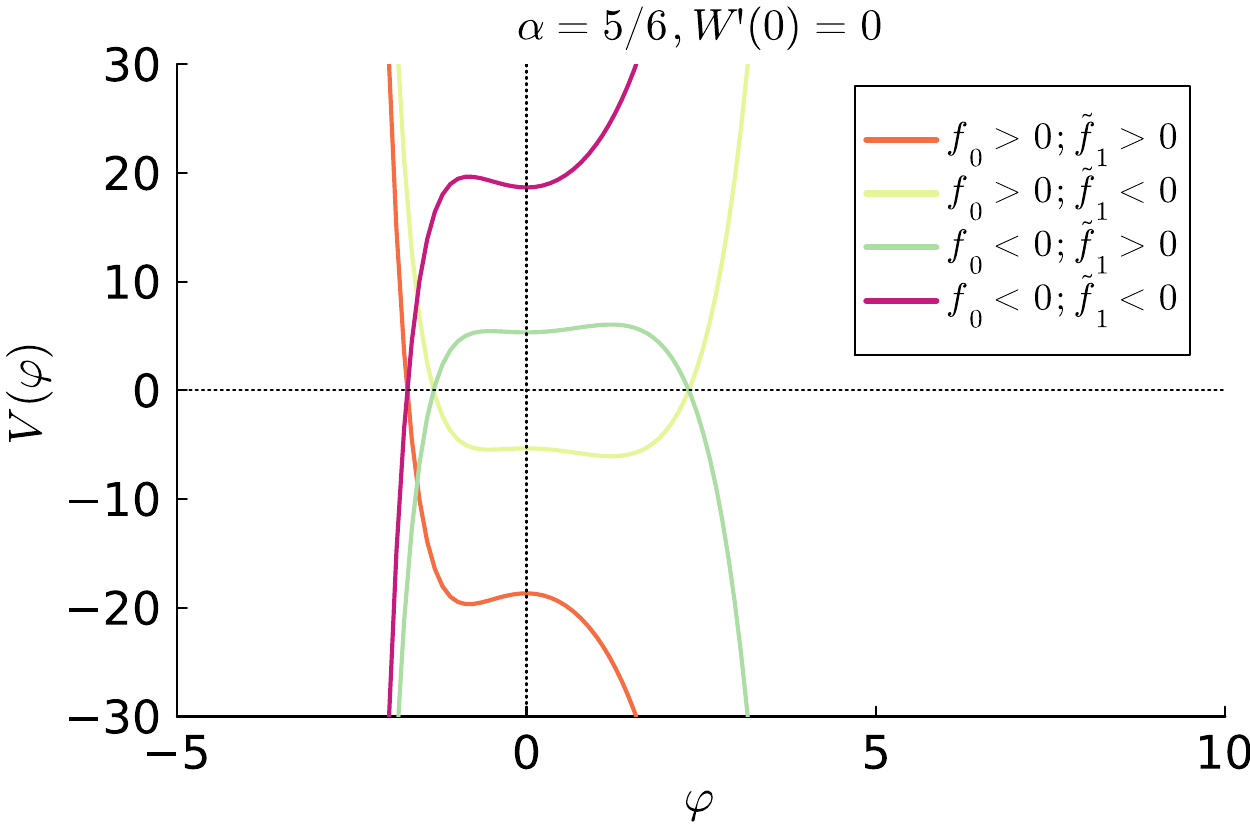}}}
	\caption{Potential $V$ given in Eq. \eqref{vz} for $d=4$ (see also Eq. \eqref{k11}), for $C_1=C_t=1$. We demand that $W$ has an extremum at $\f=0$, which sets $C_2=\a/(1-\a)$. We choose three qualitatively distinct values of $\a$ and vary the integration constants $f_0$ and $\tilde{f}_1$. The latter is defined in Eq. \eqref{ftild}. }
	\label{fig:0K1}
\end{figure}

\begin{figure}[h!]
	\centering
	\subfloat{\includegraphics[scale=0.32]{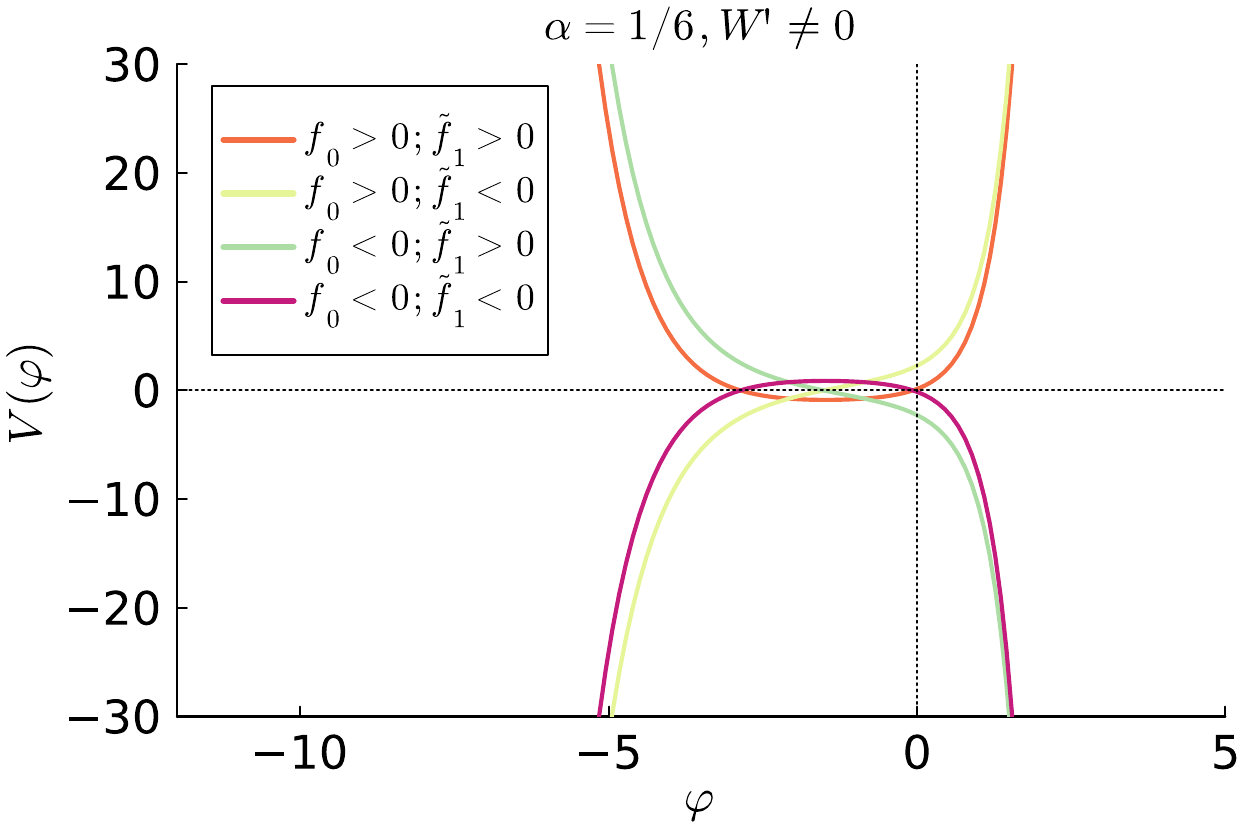}}
	\qquad
	\subfloat{\includegraphics[scale=0.32]{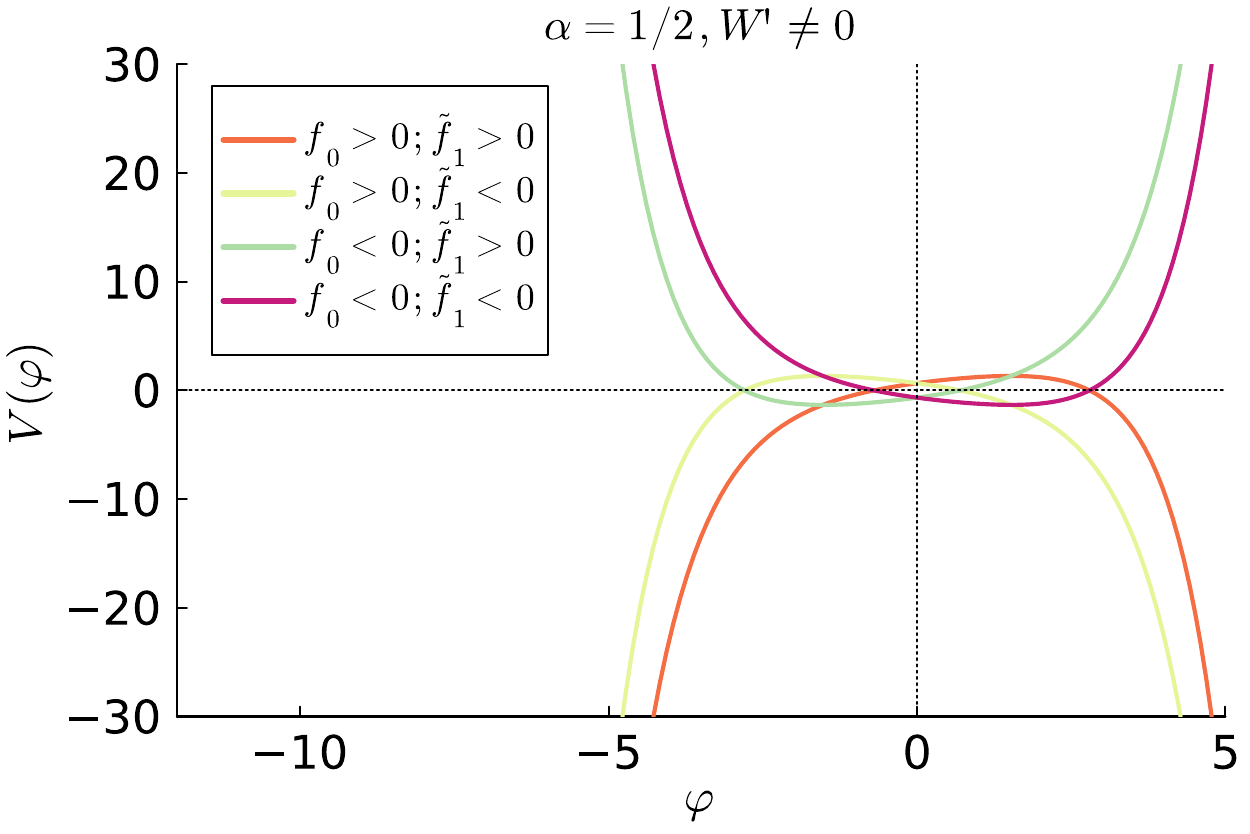}}
	\qquad
	\subfloat{{\includegraphics[scale=0.32]{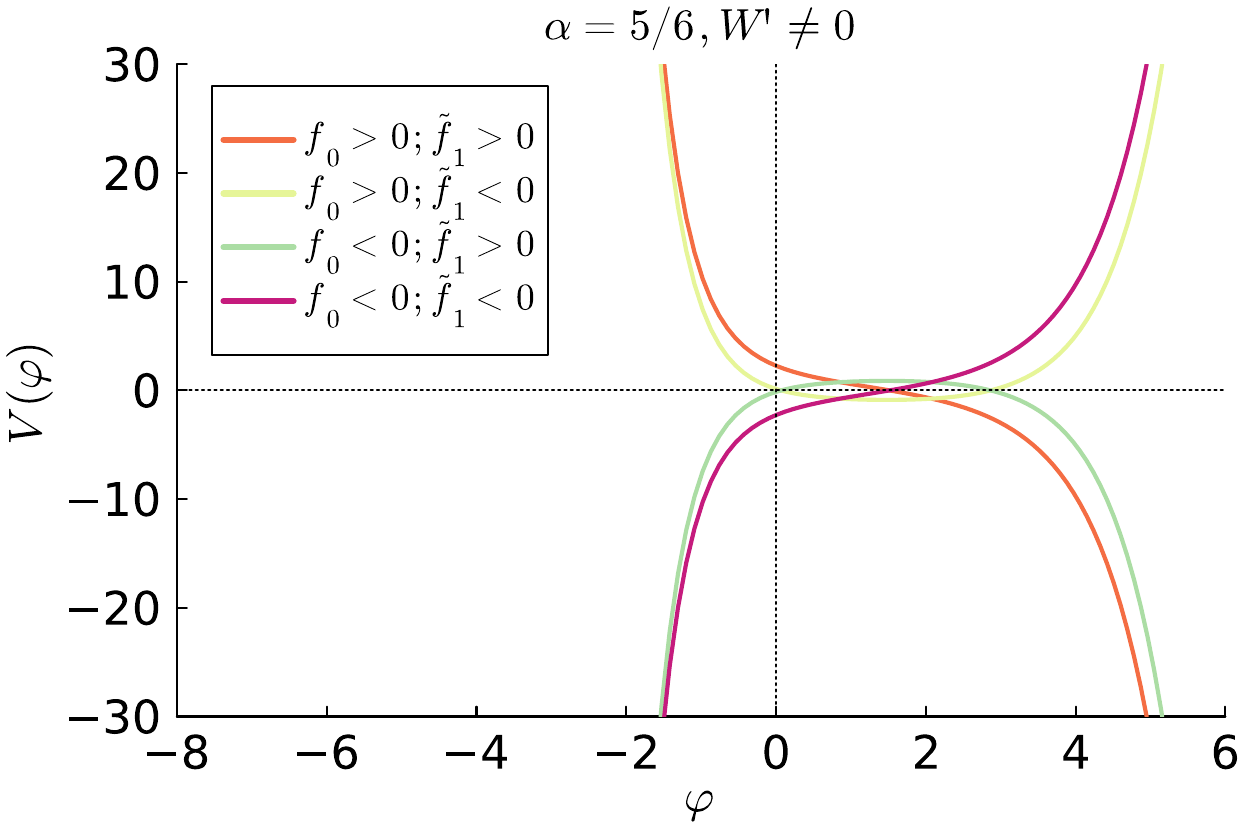}}}
	\caption{Potential $V$ given in Eq. \eqref{vz} for $d=4$ (see also Eq. \eqref{k11}), for $C_1=C_t=1$. We demand that $W$ has no extremum and set $C_2=-1$. We choose three qualitatively distinct values of $\a$ and vary the integration constants $f_0$ and $\tilde{f}_1$. The latter is defined in Eq. \eqref{ftild}.}
	\label{fig:0K2}
\end{figure}

In Figs.  \ref{fig:0K1} and \ref{fig:0K2} we show the form of the potential $V$ for $d=4$ and different choices of integration constants. In both figures, we set $C_1 = 1$ via rescalings of the radial coordinate \eqref{scaling2} and $C_t = 1$ by virtue of the scaling symmetry $(f,T,V)\to \lambda (f,T,V)$ of Eqs. \eqref{f10_1}, \eqref{w55} and \eqref{w55b}. In Fig. \ref{fig:0K1} we demand that the superpotential has an extremum, which we locate at $\f=0$ by setting $C_2 = \alpha/(1-\alpha)$, while in Fig. \ref{fig:0K2} we demand there is no extremum of the superpotential and set $C_2=-1$. In both cases we study three values of $\alpha$ that give qualitatively different behaviours of the potential, as we explain below, and vary the two integration constants $f_0$ and $\tilde{f}_1$, which we define as
\begin{equation}\label{ftild}
\tilde{f}_1 = f_1 C_1^2 C_2 - 72 \alpha (3\alpha - 1)C_t C_1\,.
\end{equation}
Generically, we observe that the potential $V$ can have up to three local extrema, and diverge to positive or negative values at the boundary of field space.

The coefficients $(h_i,a,c)$ are uniquely determined from the six constants of integration: ($C_1$,$C_2$,$\alpha$,$C_t$,$f_0$,$f_1$). Consequently, not all coefficients appearing in the potential $V$ are independent from one another. In general, for $d\neq 4$, the hypergeometric functions at hand are independent. However, at $d=4$ the Hypergeometric functions become elementary functions and the potential simplifies and depends on four independent parameters as shown below.
Generically for $d\not=4$  the potential has $6$ free coefficients instead of $15$ and, once those are determined, no free integration constant.

We now study further the case of $d=4$. Then $f$ and $V$ simplify significantly using the identity $ \, _2F_1(1,a;a;z)=(1-z)^{-1}\,$:
\begin{equation}\label{k10}
   f = f_0 + z^{-2+3\alpha} \left( \frac{1}{2}\frac{\alpha}{(-2+3\alpha)}C_1f_1-\frac{36  \alpha^2}{2(-2+3\alpha)}\frac{C_t}{C_2} + \frac{1}{2}\frac{-1+\alpha}{(-1+3\alpha)}C_2 f_1 z \right)\,,
\end{equation}
\begin{equation}\label{k11}
\begin{split}
V &= -f_0 C_1^2 z^{-2\alpha} \left(\frac{1 }{6}\frac{(-2+3\alpha)}{-1+\alpha} + \frac{C_2}{C_1} z + \frac{C_2^2 }{6 C_1^2} \frac{(-1+3\alpha)}{\alpha}z^2 \right) \\
	&+ (-2+3\alpha)^{-1}\left(\frac{f_1 C_1^2C_2}{6(-1+3\alpha)} - 12\alpha C_t C_1 \right) z^{-1+\alpha}\left(\frac{\alpha}{(-1+\alpha)}  + \frac{C_2(-1+\alpha)}{C_1\alpha}z\right)\,.
\end{split}
\end{equation}
The previous potential is a linear combination of five linearly independent terms, except for specific values $\alpha=1/3\,, 2/3\,$, where we have three linearly independent terms in each case. For generic $\alpha$ we may define
\begin{equation}
\beta_1=\alpha \,,\,\,\,\, \beta_2 = f_0 C_1^2\,,\,\,\,\, \beta_3 =(-2+3\alpha)^{-1} \left(\frac{f_1 C_1^2C_2}{6(-1+3\alpha)} - 12\alpha C_t C_1 \right) \,,\,\,\,\, \beta_4 = \frac{C_2}{C_1}
\end{equation}
\noindent
and the potential $V$ can be written as
\begin{equation}\label{k13}
\begin{split}
V &= -\beta_2 z^{-2\alpha} \left(\frac{1 }{6}\frac{(-2+3\alpha)}{-1+\alpha} + \beta_4 z + \beta_4^2 \frac{(-1+3\alpha)}{\alpha}z^2 \right) \\
	&+ \beta_3 z^{-1+\alpha}\left(\frac{\alpha}{(-1+\alpha)}  + \beta_4 \frac{(-1+\alpha)}{\alpha}z\right)\,.
\end{split}
\end{equation}
The potential is tuned, because four constants fix five independent terms. However, given such a potential, we can construct families of solutions, by varying the undetermined integration constants $(f_0,f_1)$. The case of $\alpha=1/3\,,2/3$ is analogous, except that now we have three constants to fix three linearly independent terms, so the potential $V$ is less tuned in that sense.

We now study under what conditions the flow constructed in this section feature the type 0, type I or type II asymptotic structure, uncovered in Appendix \ref{asymp}, as we approach the boundary of field space: $z\to 0$ and $z\to \infty$. We restrict this analysis to the case of $d=4$ dimensions. It follows from Appendix \ref{asymp} that type I or II asymptotics take place whenever $\frac{fW^2}{V}$ approaches a constant value at the boundary, and type 0 asymptotics otherwise. We find the following possibilities:
\begin{itemize}
\item[(a)] $z\to 0$ ($\f\to-\infty)$:
\begin{enumerate}
\item[(a.1)] $\a\in(0,1/3)$. For generic integration constants, combining Eqs. \eqref{eq:d42}, \eqref{k10} and \eqref{k11}, we find that
\begin{equation}\label{to}
\frac{fW^2}{V}\propto z^{-1}+ \dots
\end{equation}
which is not constant as $z\to0$ and therefore corresponds to the type 0 asymptotic solution. The relation \eqref{to} can be modified if the leading term of $f$ vanishes.

We find that, if the leading term of $f$ as $z\to 0$ in Eq. \eqref{k10} vanishes, i.e. for
\begin{equation}\label{ti}
f_1 = \dfrac{36 \alpha C_t}{C_1 C_2}
\end{equation}
then the discriminant $\frac{fW^2}{V}$ approaches a constant value and this corresponds to either type I or type II asymptotics. In this case, the function $f$ diverges because $\alpha\in(0,1/3)$ and we learn that the choice \eqref{ti} gives rise to type II asymptotics.

\item[(a.2)]$ \a \in (1/3,2/3) $.
 For generic integration constants, combining Eqs. \eqref{eq:d42}, \eqref{k10} and \eqref{k11}, we find that
\begin{equation}\label{to2}
\frac{fW^2}{V}\propto z^{3\a-2}+ \dots
\end{equation}
and we obtain type 0 asymptotics. The asymptotic structure can be modified if the leading term in $f$ vanishes, which is achieved by the choice \eqref{ti}. For this choice, the function $f$ given in \eqref{k10} approaches a constant value, $f = f_0 +  \dots$, and we identify this solution with the type I asymptotics if $f_0\neq 0$, and with the type II asymptotics if $f_0 = 0$.
Additionally, for the type I asymptotic solution to be acceptable, we require that the potential satisfies the Gubser bound (see Appendix \ref{app:l1}). Indeed, in the case where $f_0\neq 0$, the potential diverges as $z^{-2\a} = e^{-2 \sqrt{\a/(3\a-3)}\f}$, which satisfies the Gubser bound for $\a<2/3$.

\item[(a.3)]$ \a \in (2/3,1) $.
 For generic integration constants, combining Eqs. \eqref{eq:d42}, \eqref{k10} and \eqref{k11}, we find that both the discriminant and $f$ approach a constant value:
\begin{equation}\label{to3}
\frac{fW^2}{V}\propto z^{0}+ \dots\qquad f = f_0 + \dots
\end{equation}
and we identify this solution with the type I asymptotic solution. However, in this case the Gubser bound is violated, as we demonstrated in the previous item, and such a solution is not acceptable in the Gubser sense. The previous asymptotics can be modified if $f_0=0$, in which case the leading terms in $f$ and $V$ vanish. In this case, the discriminant diverges, $\frac{fW^2}{V}\propto z^{-1}$, and we have type 0 asymptotics. Finally, if both $f_0=0$ and \eqref{ti} is satisfied, $\frac{fW^2}{V}$ approaches a constant value and $f$ vanishes. We identify this latter choice with the type II asymptotic solution.

\end{enumerate}

\item[(b)] $z\to \infty$ ($\f\to +\infty)$:
\begin{enumerate}
\item[(b.1)] $\a\in (0,1/3)$.
For generic integration constants, combining Eqs. \eqref{eq:d42}, \eqref{k10} and \eqref{k11}, we find that both the discriminant and $f$ approach a constant value:
\begin{equation}\label{to4}
\frac{fW^2}{V}\propto z^{0}+ \dots\qquad f = f_0 + \dots
\end{equation}
and we identify this solution with the type I asymptotic solution. For the type I solution to be acceptable, the potential $V$ must satisfy the Gubser bound (see Appendix \ref{app:l1}). In this case, the potential diverges as $V\propto z^{2-2\a} = e^{-2 \sqrt{(1-\a)/(3\a)}\f}$, and the Gubser bound is violated for $\a<1/3$. Therefore, the type I asymptotic solution with $\a\in (0,1/3)$ is not acceptable.

The previous asymptotics can be modified if $f_0=0$, in which case the leading terms in $f$ and $V$ vanish. In this case, the discriminant diverges, $\frac{fW^2}{V}\propto z$, and we have type 0 asymptotics. Finally, if both $f_0=0$ and $f_1=0$, then $\frac{fW^2}{V}$ approaches a constant value and $f$ vanishes. We identify this latter choice with the type II asymptotic solution.

\item[(b.2)]$ \a \in (1/3,2/3) $.
 For generic integration constants, combining Eqs. \eqref{eq:d42}, \eqref{k10} and \eqref{k11}, we find that
\begin{equation}\label{to5}
\frac{fW^2}{V}\propto z^{3\a-1}+ \dots
\end{equation}
and we obtain type 0 asymptotics. The asymptotic structure can be modified if the leading term in $f$ vanishes, which is achieved with $f_1=0$. For this choice, the function $f$ given in \eqref{k10} approaches a constant value, $f = f_0 +  \dots$, and we identify this solution with the type I asymptotics if $f_0\neq 0$, and with the type II asymptotics if $f_0 = 0$.
Additionally, the potential $V$ respects the Gubser bound for the type I asymptotic solution with $\a \in (1/3,2/3) $ and such a solution is acceptable.

\item[(b.3)] $\a\in(2/3,1)$. For generic integration constants, combining Eqs. \eqref{eq:d42}, \eqref{k10} and \eqref{k11}, we find that
\begin{equation}\label{to6}
\frac{fW^2}{V}\propto z+ \dots
\end{equation}
which diverges as $z\to\infty$, and therefore corresponds to the type 0 asymptotic solution. The relation \eqref{to} can be modified if the leading term of $f$ vanishes. We find that, if the leading term of $f$ as $z\to \infty$ in Eq. \eqref{k10} vanishes, i.e. for $f_1=0$, then the discriminant $\frac{fW^2}{V}$ approaches a constant value and this corresponds to either type I or type II asymptotics. In this case, the function $f$ diverges because $\alpha\in(2/3,1)$ and we learn that the choice $f_1=0$ gives rise to type II asymptotics.

\end{enumerate}

\end{itemize}

We conclude that type II asymptotic solutions are possible for any value of $\a$, while acceptable type I solutions are only possible if $\a\in (1/3,2/3)$.

It is  possible to have here a flow from $-\infty$ to $+\infty$ without stopping at $z_*$. This happens  if $C_2 C_1<0$. In that case,  the superpotential has no local extremum and the flow does not stop at finite $\f$. An example of a potential where the flow does not stop is in figure \ref{fig:0K2}. Such solutions necessarily contain an $A$-bounce (because $W$ vanishes if $C_2 C_1<0$), and therefore there is a naked bad singularity from rule 23 on page \pageref{ru23}.

\vskip 1cm
\subsection{Particular examples: Flows with event horizons}\label{aehor}
\vskip 1cm

We shall now study particular examples of flows in $d=4$ dimensions that have one endpoint at finite $\f$. Accordingly, we demand that $W$ has a regular extremum at finite $\f$, where the flow stops according to the rules of Sec. \ref{rul}. It is useful to note that the endpoint of the flow \eqref{k3} can be set to $z_*=1$ ($\f_*=0$) via a shift in $\f$ without loss of generality. Additionally, we can set $W(z_*)=1$ with a rescaling of the radial coordinate $u$ \eqref{scaling}. Both conditions are equivalent to choosing the integration constants $C_1$ and $C_2$ in \eqref{eq:d42} as
\begin{equation}\label{k14}
C_1 = 1-\alpha\,,\qquad C_2 = \alpha\,.
\end{equation}
In order to understand the possible flows contained in the superpotential \eqref{eq:d42}, we restrict the following discussion to the particular value $\alpha=1/2$. Eventually, we provide more examples with different values of $\alpha$ that respect the Gubser bound.

The choices of integration constants in Eq. \eqref{k14} along with $\alpha=1/2$ simplify the relations \eqref{eq:d42}, \eqref{k5}, \eqref{k10} and \eqref{k11} as follows:

\begin{equation}\label{k15}
W = \cosh\left(\dfrac{\varphi}{\sqrt{3}}\right)\,, \qquad T = -C_t \sinh\left(\dfrac{\varphi}{\sqrt{3}}\right)\,,
\end{equation}
\begin{equation}\label{k16}
f= f_0 + 36 C_t e^{-\f/\sqrt{3}}-\frac{1}{2}f_1\cosh\left(\dfrac{\varphi}{\sqrt{3}}\right)\,,
\end{equation}
\begin{equation}\label{k17}
V=-\dfrac{1}{4}f_0+\left(\frac{1}{6}f_1 -12C_t\right)\cosh\left(\dfrac{\varphi}{\sqrt{3}}\right)-\dfrac{1}{12}f_0\cosh\left(\dfrac{2\varphi}{\sqrt{3}}\right)\,,
\end{equation}
where we have reinstated $\f$ using the definition of $z$ given in Eq. \eqref{k2}. The flow ranges from the endpoint at $\f_*=0$ to either positive or negative infinity. Since the superpotential \eqref{k15}, with the choice of integration constants as in \eqref{k14}, is invariant under $\f \to -\f$, we can take $\f>0$. Then the assumption that $T>0$ translates into $C_t<0$. Note that for generic $\a$, the superpotential \eqref{eq:d42} is invariant under the simultaneous transformation $\a\to 1-\a$ and $\f\to -\f$, and we can similarly restrict to $\f>0$ while scanning the possible values for $\a$. Note that the inverse scale factor in Eq. \eqref{k15} vanishes at the endpoint $\f_*=0$, while $W(\f_*)=1$. This is enough to identify the endpoint with a (dS$_{5}$, AdS$_{5}$ or Minkowski$_{5}$) boundary.

In order to classify possible boundaries, we expand the potential $V$ around the $\f_*=0$ endpoint, in order  to find
\begin{equation}\label{k22}
V= V_*\left(1+\frac{\f^2}{6} + O(\f^4)  \right) + C_t O(\f^4)\,,
\end{equation}
where we have defined $V_*\equiv V(0)$. Therefore, we find three inequivalent cases:
\begin{itemize}
\item $V_*>0$. The geometry of the endpoint corresponds to a dS$_{5}$ boundary with $\Delta_{\pm}=2$.
\item $V_*<0$. The geometry of the endpoint corresponds to a AdS$_{5}$ boundary with $\Delta_{\pm}=2$.
\item $V_*=0$. The potential vanishes up to $O(\f^4)$, while the superpotential satisfies $W_*''/W_*=1/3$. We identify this solution with the spatial boundary of Minkowski studied in Appendix \ref{exth} (see Eqs. \eqref{e2c}-\eqref{e58}) for $d=4$.
\end{itemize}

The flow hits a singularity as it reaches the boundary in field space $\f\to \infty$. We focus on solutions where the singularity is covered by a black-hole event horizon. Examples of solutions with naked singularities can be found in Sec. \ref{sec:7}. Therefore, we demand that $f(\f_h)=0$ for some value of $\f_h$. In order to construct explicit solutions, it is useful to redefine the constants $f_0$ and $f_1$ in terms of the value of the potential at the endpoint, $V_*\equiv V(0)$, and the location of the horizon $\f_h$:

\begin{equation}
f_0 = -\frac{3}{2} \left(24 C_t \coth \left(\frac{\f_h}{2
   \sqrt{3}}\right)+V_* \coth ^2\left(\frac{\f_h}{2
   \sqrt{3}}\right)+V_*\right)\,,
\end{equation}
\begin{equation}
f_1 = \frac{144 C_t-12 e^{\frac{\f_h}{\sqrt{3}}} (12
   C_t+V_*)}{\left(e^{\frac{\f_h}{\sqrt{3}}}-1\right)^2}\,.
\end{equation}
With these definitions, we can obtain the value of $f$ at the endpoint, which we denote by $f_*$, as well as the value of the potential at the horizon $V_h$:
\begin{equation}\label{k20}
f_* = -3V_*\,,
\end{equation}
\begin{equation}\label{k21}
V_h = \dfrac{1}{4}V_*\left(2 \cosh \left(\frac{\f_h}{\sqrt{3}}\right)+\cosh \left(\frac{2 \f_h}{\sqrt{3}}\right)+1\right)-3 C_t\left(-2 \sinh \left(\frac{\f_h}{\sqrt{3}}\right)+\sinh
   \left(\frac{2 \f_h}{\sqrt{3}}\right)\right)\,.
\end{equation}

We shall now discuss three qualitatively distinct flows, depending on the geometry at the endpoint: Flows from an AdS$_{5}$ boundary, flows from a dS$_{5}$ boundary and flows form a Minkowski$_{5}$ boundary.

\subsubsection*{Flow from an AdS$_{5}$ boundary to an event horizon.}

\begin{figure}[h!]
	\centering
	\subfloat{\includegraphics[scale=0.32]{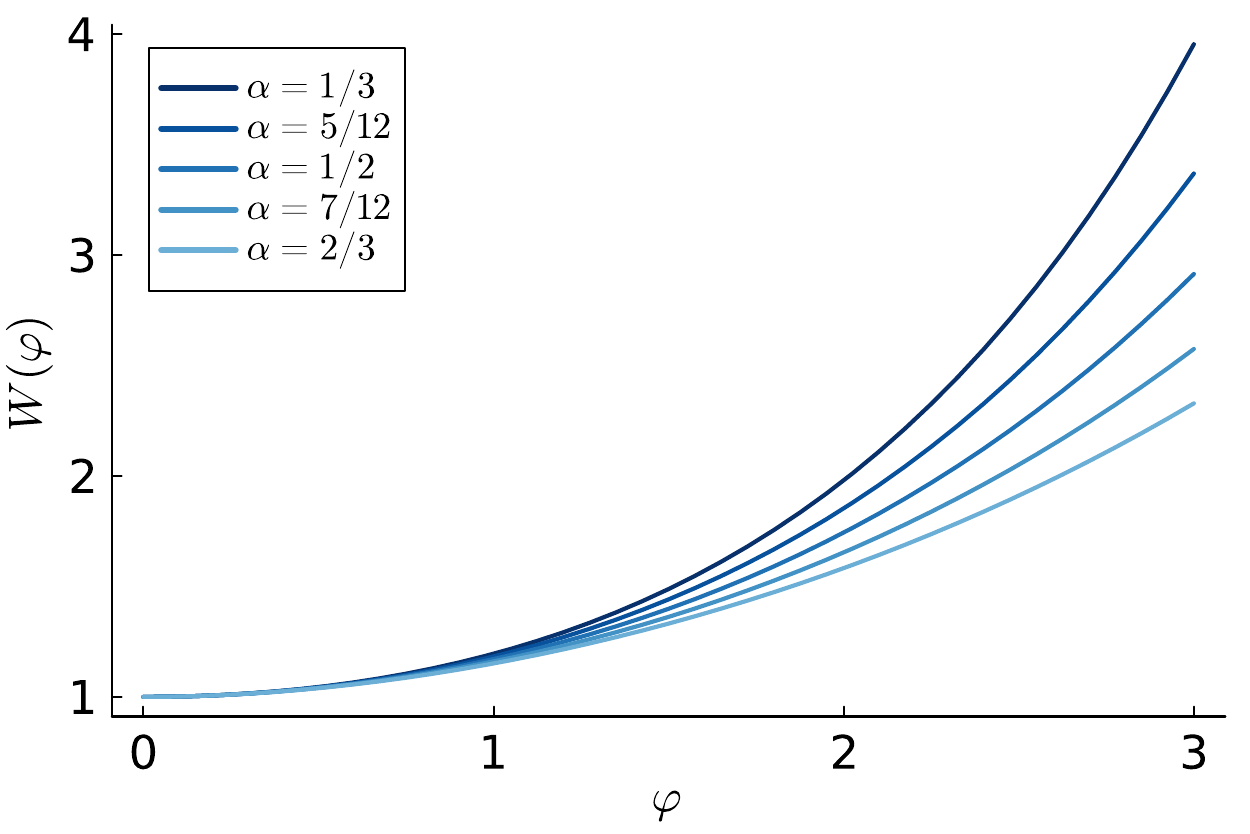}}
	\qquad
	\subfloat{\includegraphics[scale=0.32]{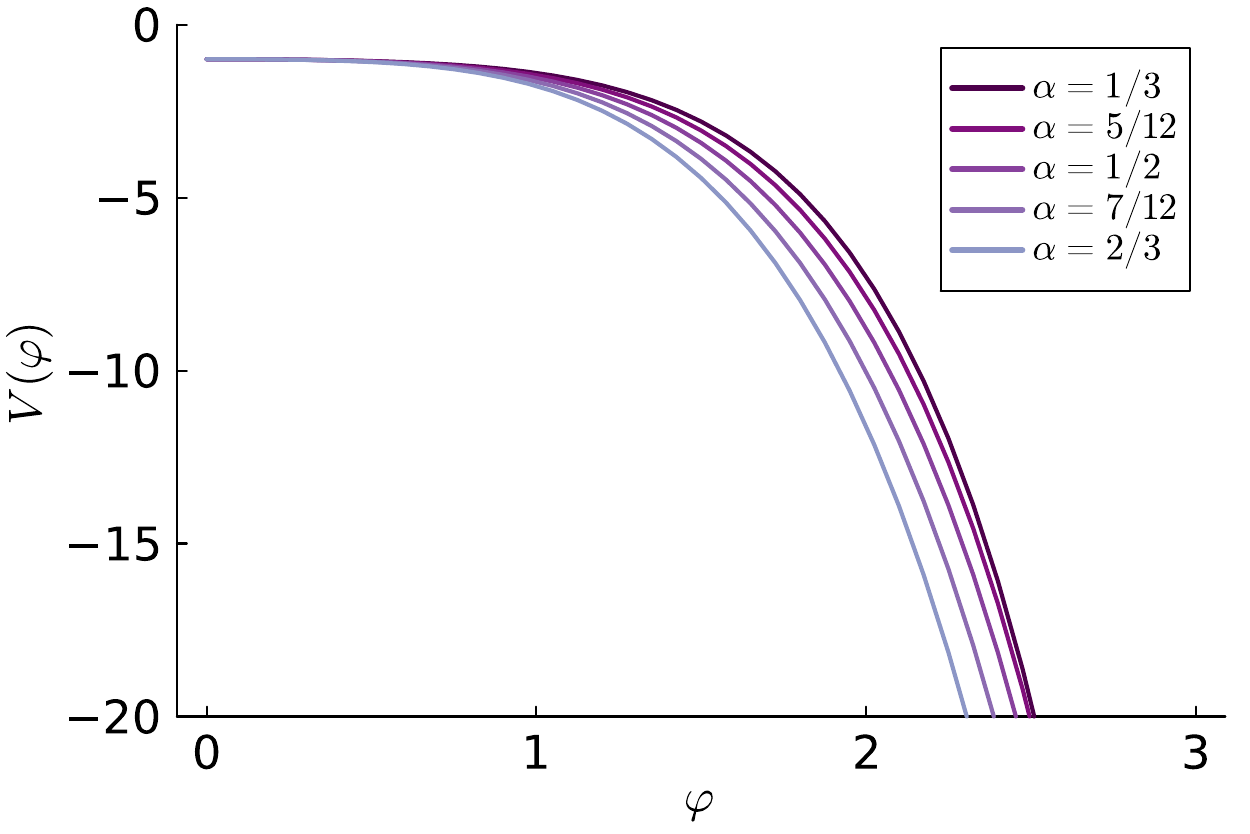}}
	\qquad
	\subfloat{{\includegraphics[scale=0.32]{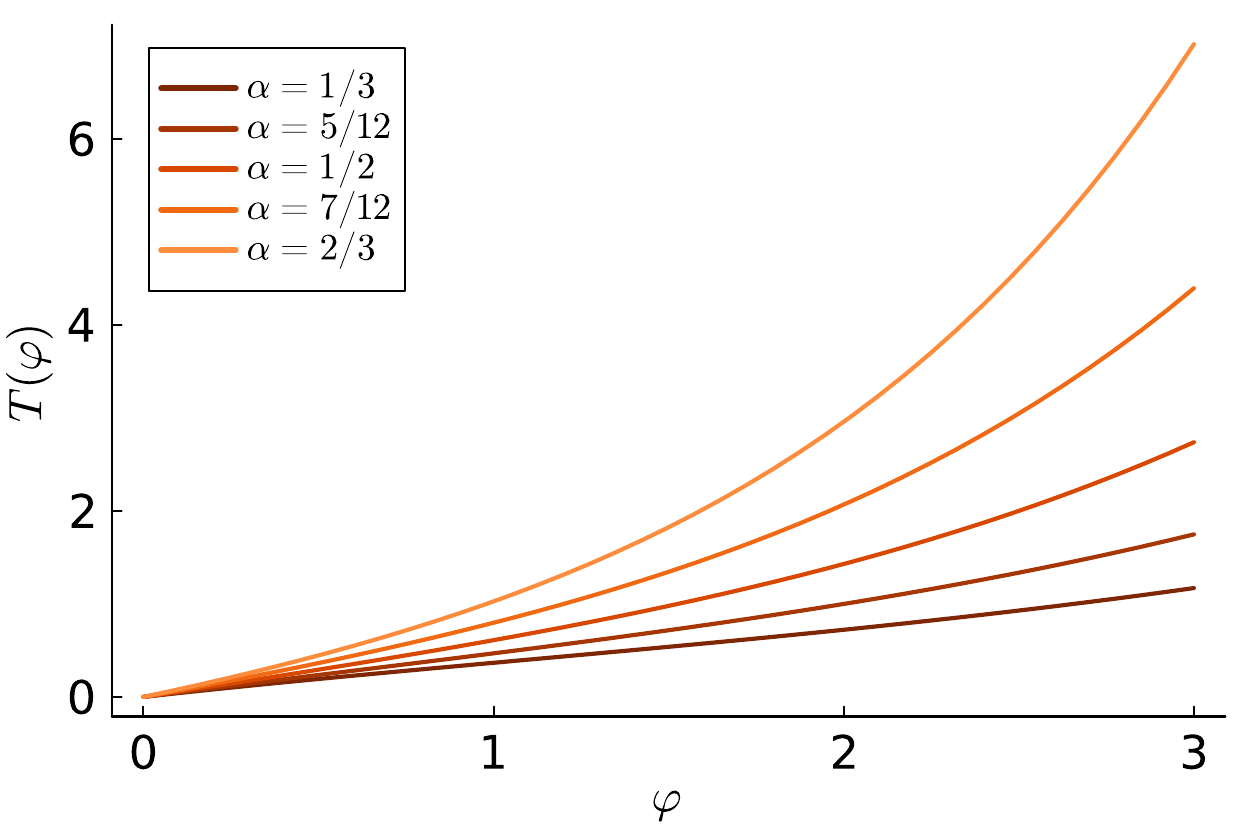}}}
	\qquad
	\subfloat{{\includegraphics[scale=0.32]{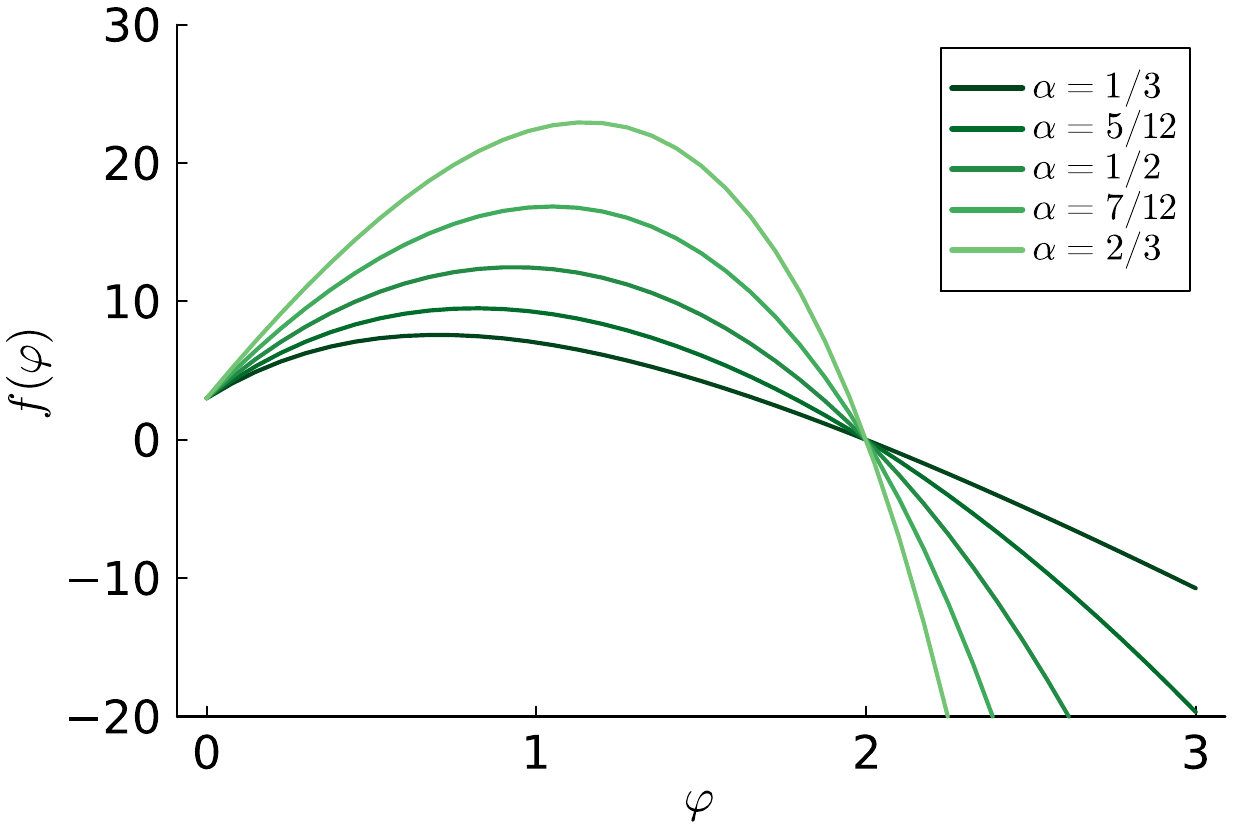}}}
	\caption{Graphical depiction of the flow described in Eqs. \eqref{k15}-\eqref{k17}. At the endpoint $\f=0$ the geometry is identified with an AdS$_{5}$ boundary, while at $\f_h=2$ there is a black-hole event horizon. At $\f\to\infty$ one encounters a (covered) singularity.}
	\label{fig:K1}
\end{figure}

In order for the endpoint to feature an AdS boundary we choose $V_*<0$. On the other hand, the assumption that $T>0$ requires that $C_t<0$. Therefore, from Eq. \eqref{k21} we conclude that the horizon is necessarily located in the AdS regime $V_h<0$. In Fig. \ref{fig:K1} we display several examples of the flow for different values of $\alpha$. We have chosen $V_*=C_t=-1$ and we locate the horizon at $\f_h=2$.

\subsubsection*{Flow from a dS$_{5}$ boundary to an event horizon.}

In this case, we choose $V_*>0$ to generate a dS boundary. From Eq. \eqref{k20} this implies that $f_*<0$ and, therefore, $f$ is negative in the near-boundary region. According to the discussion in Appendix \ref{app:J}, the outermost horizon is cosmological. We demand that there is a second horizon in order to avoid a naked singularity. The location of the second horizon can be controlled by $C_t$.

\begin{figure}[h!]
	\centering
	\subfloat{\includegraphics[scale=0.32]{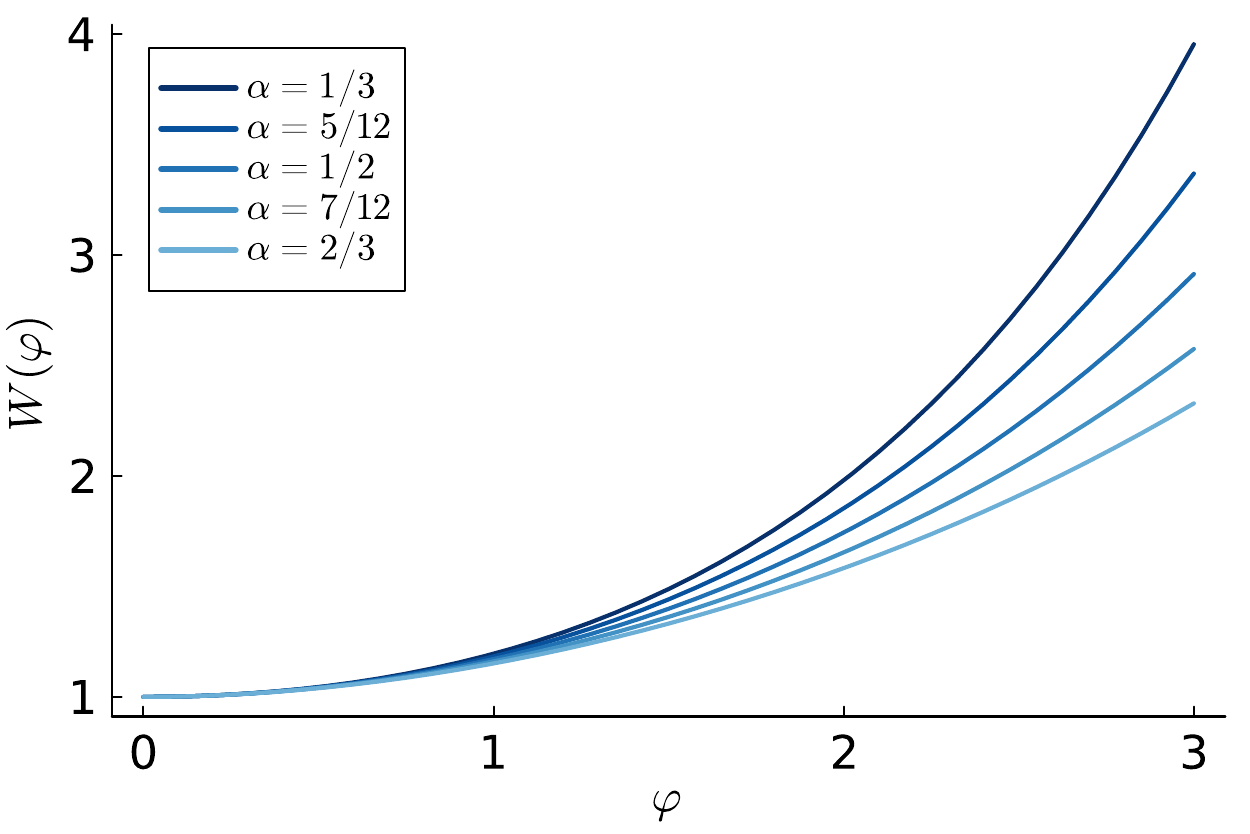}}
	\qquad
	\subfloat{\includegraphics[scale=0.32]{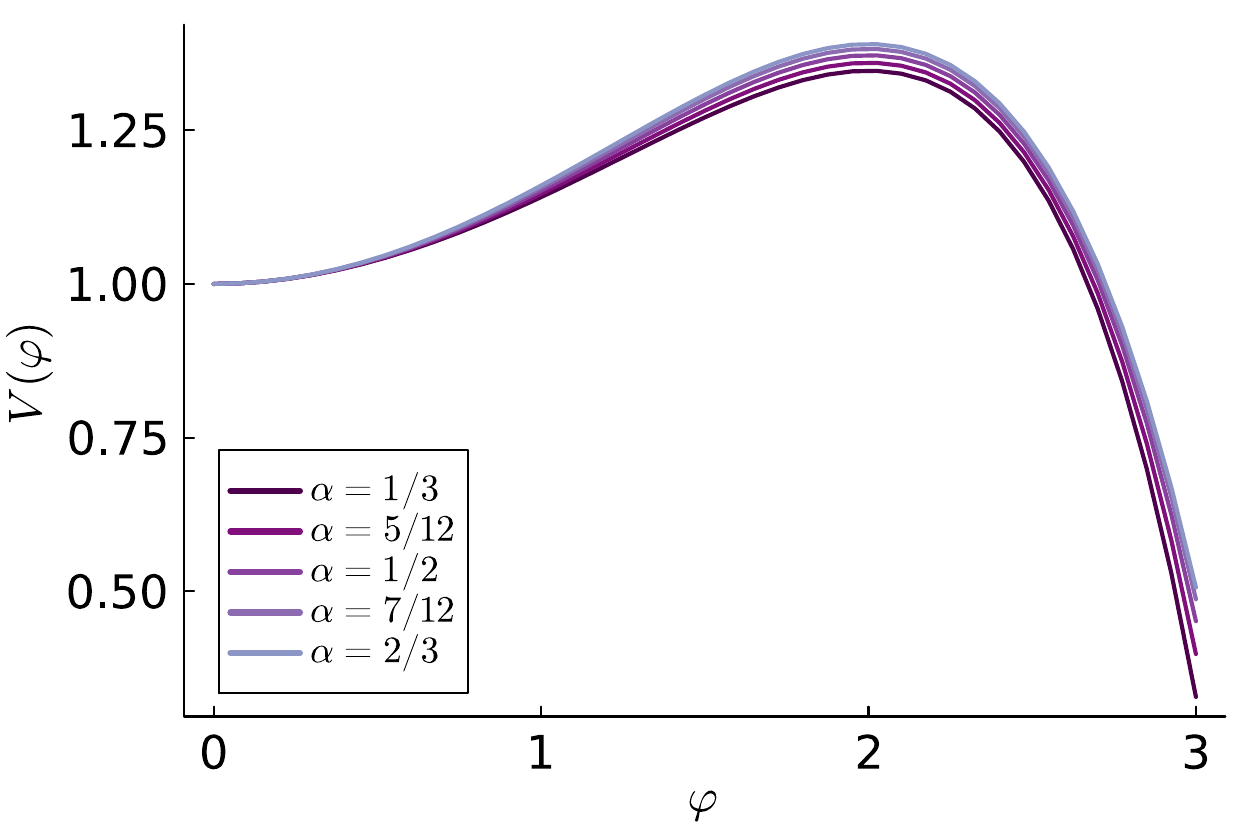}}
	\qquad
	\subfloat{{\includegraphics[scale=0.32]{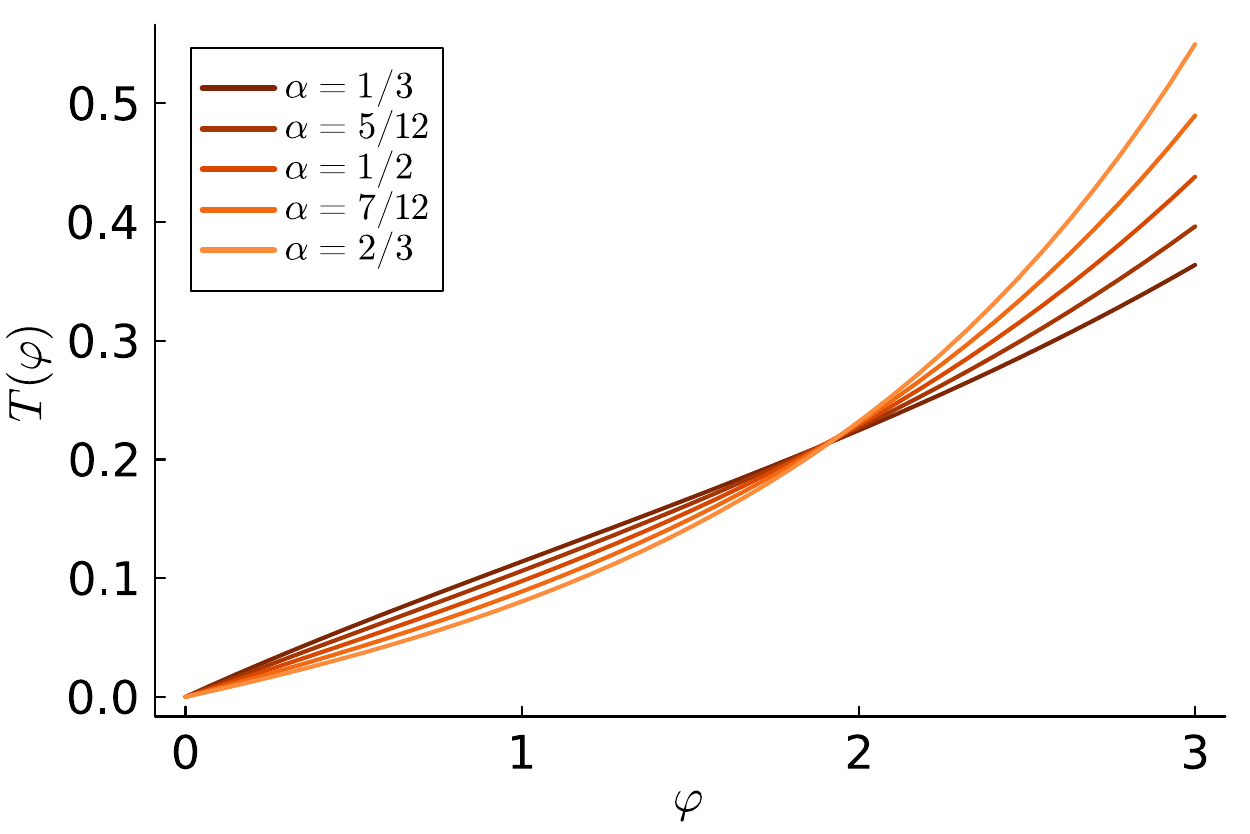}}}
	\qquad
	\subfloat{{\includegraphics[scale=0.32]{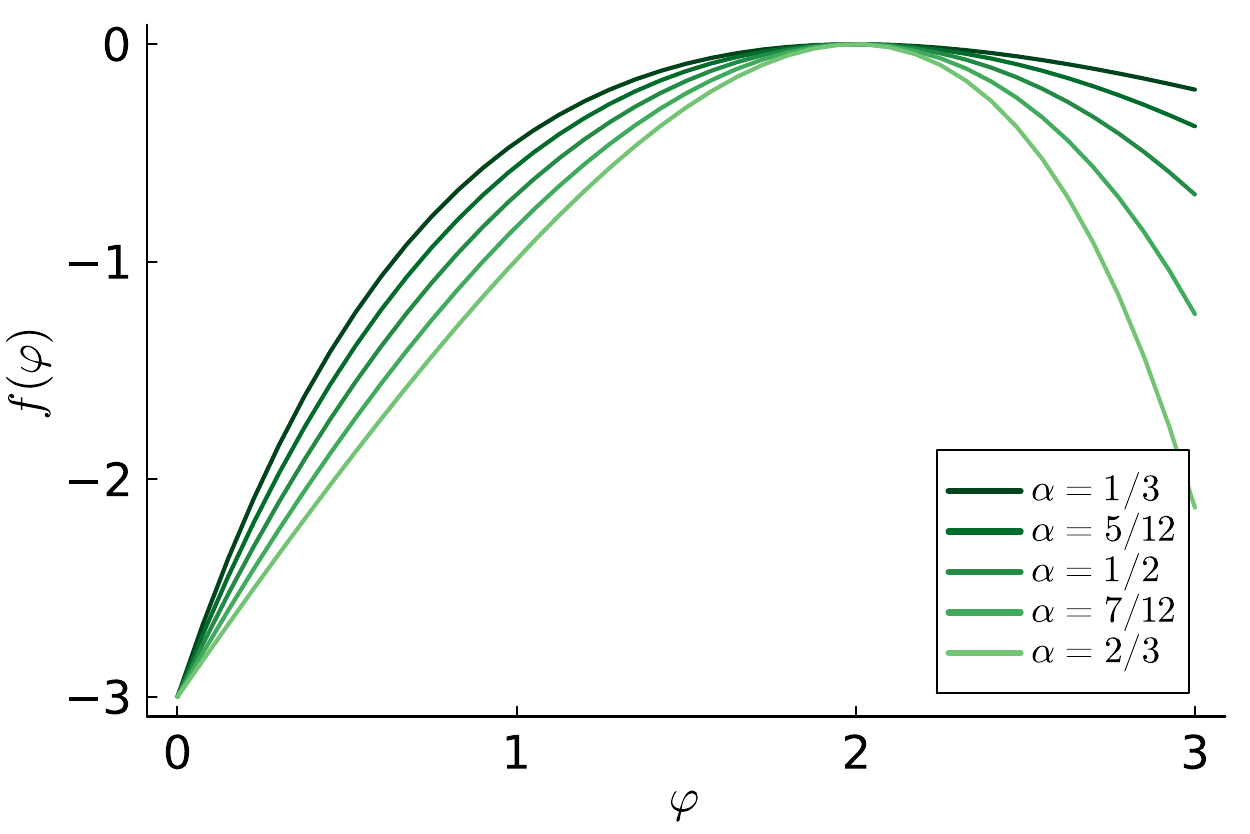}}}
	\caption{Graphical depiction of the flow described in Eqs. \eqref{k15}-\eqref{k17}. At the endpoint $\f=0$ the geometry is identified with a dS$_{5}$ boundary, while at $\f_h=2$ there is a Nariai black-hole event horizon. At $\f\to\infty$ one encounters a (covered) singularity.}
	\label{fig:K2}
\end{figure}

For concreteness, we  choose $V_*=1$ and $\f_h=2$, while the constant $C_t$ will be such that both horizon coincide and the solution features a Nariai horizon:

\begin{equation}
C_t = -\frac{1}{12}V_* \coth \left(\frac{\f_h}{2 \sqrt{3}}\right)\,.
\end{equation}
In Fig. \ref{fig:K2} we show examples of such flows for different values of $\alpha$. In all cases, the potential has a maximum at the location of the Nariai horizon, in agreement with the discussion in Appendix \ref{seho}. Note that the integration constants can be chosen such that the cosmological and event horizons do not coincide, as exemplified in Sec. \ref{fine2}.

\subsubsection*{Flow from an M$_{5}$ boundary to an event horizon.}

Finally, we study the case where $V_*=0$ where the geometry near the endpoint is identified with the spatial boundary of Minkowski. From Eq. \eqref{k20}, the function $f$ vanishes at the boundary $f_*=0$. However, the time component of the metric $g_{tt}\propto f/T$ remains finite as $\f\to0$. In particular, from Eqs. \eqref{k15} and \eqref{k16} we find
\begin{equation}
\lim_{\f\to0}\frac{f}{T} = 36\,,
\end{equation}
in agreement with the analysis in Appendix \ref{exth} (see Eq. \ref{51e}). In Fig. \ref{fig:K3} we show several examples of such flows. We have chosen to locate the horizon at $\f_h=2$, while we set $C_t=-1$ for definiteness.

\begin{figure}[h!]
	\centering
	\subfloat{\includegraphics[scale=0.32]{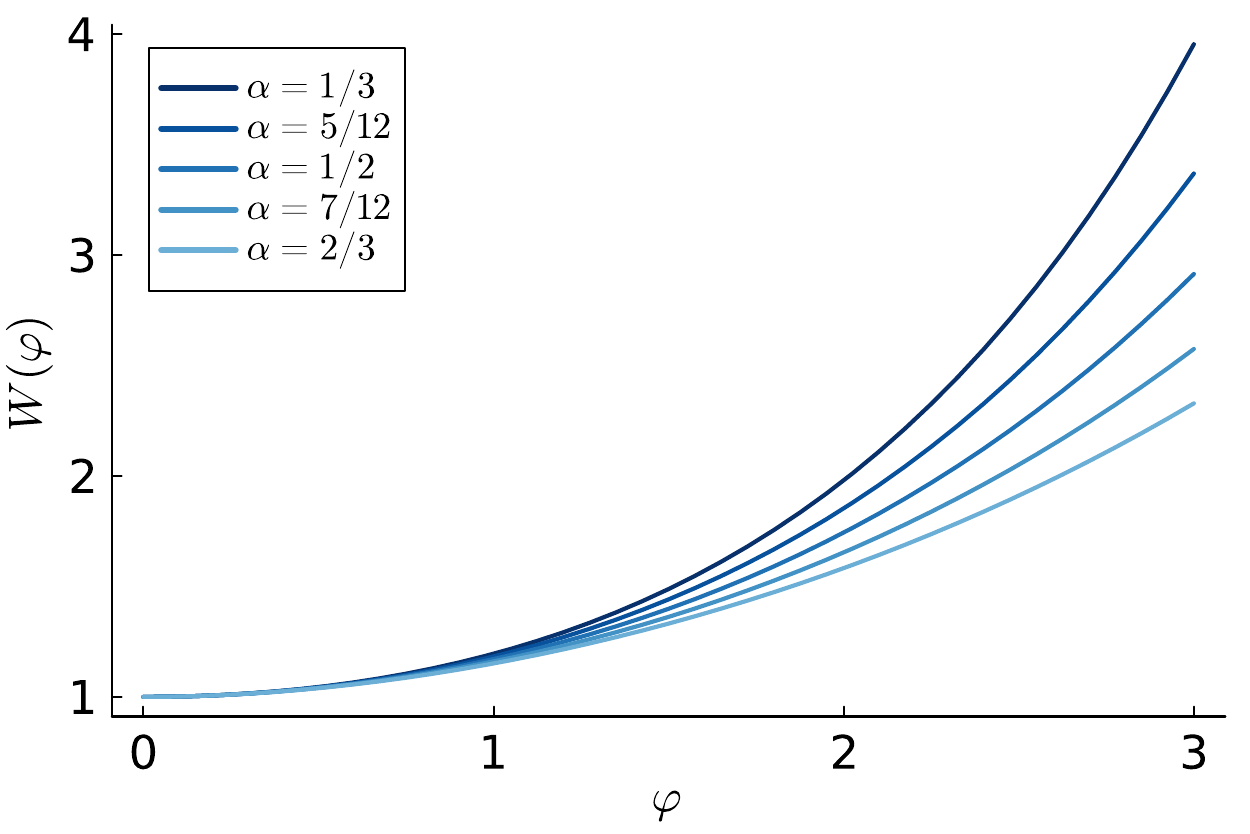}}
	\qquad
	\subfloat{\includegraphics[scale=0.32]{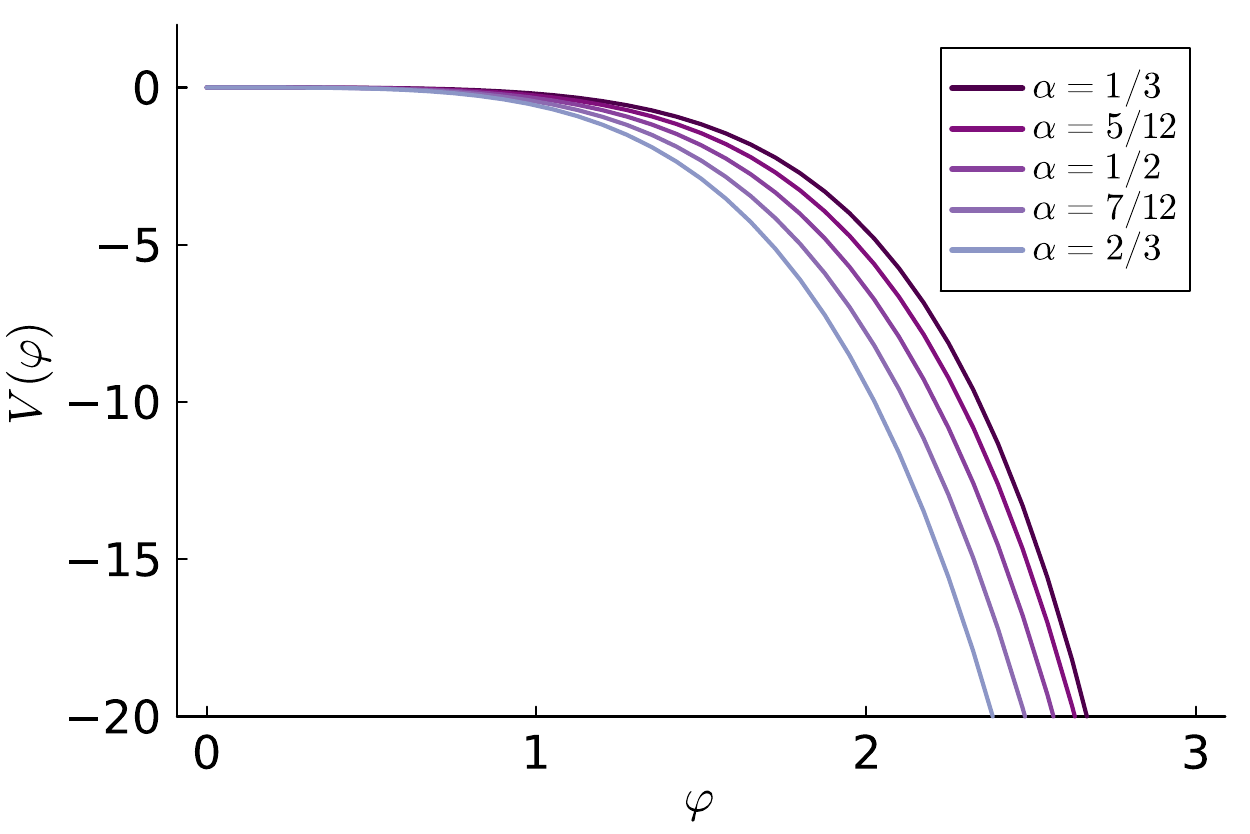}}
	\qquad
	\subfloat{{\includegraphics[scale=0.32]{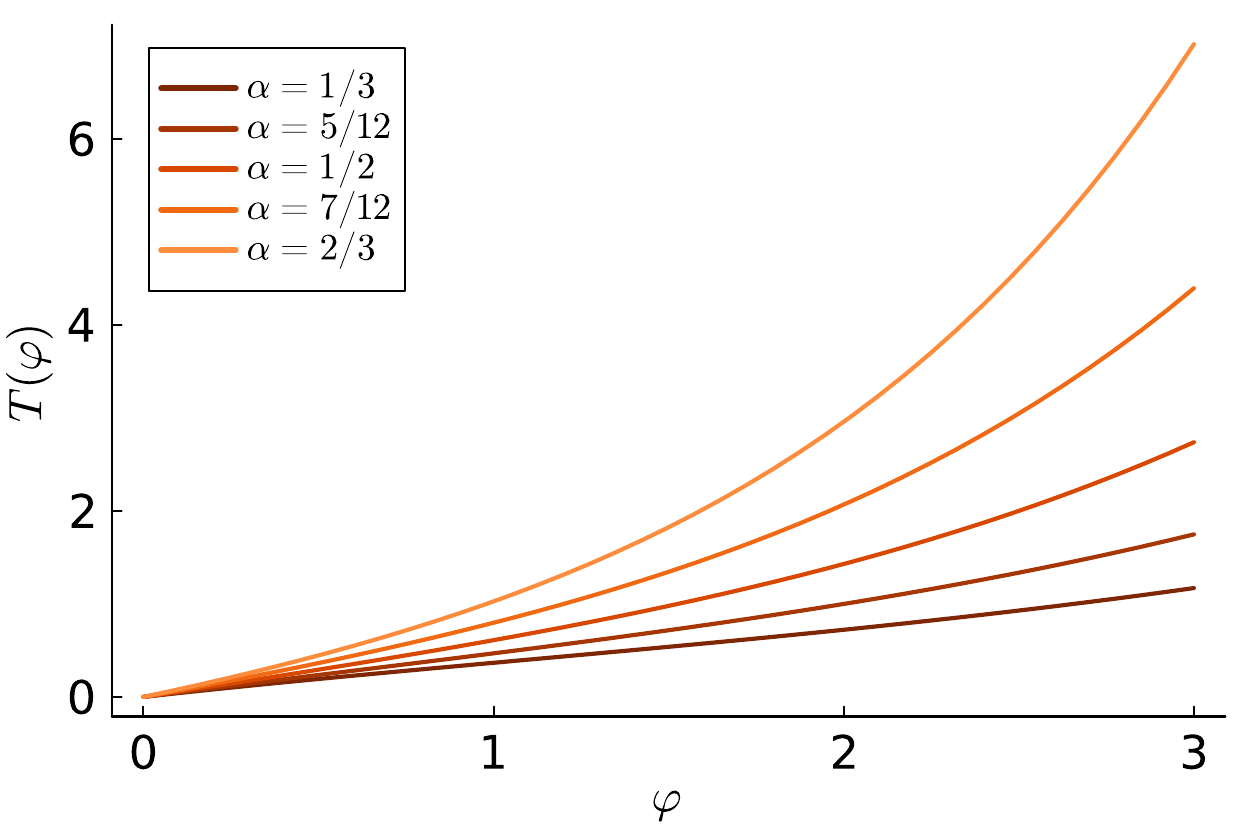}}}
	\qquad
	\subfloat{{\includegraphics[scale=0.32]{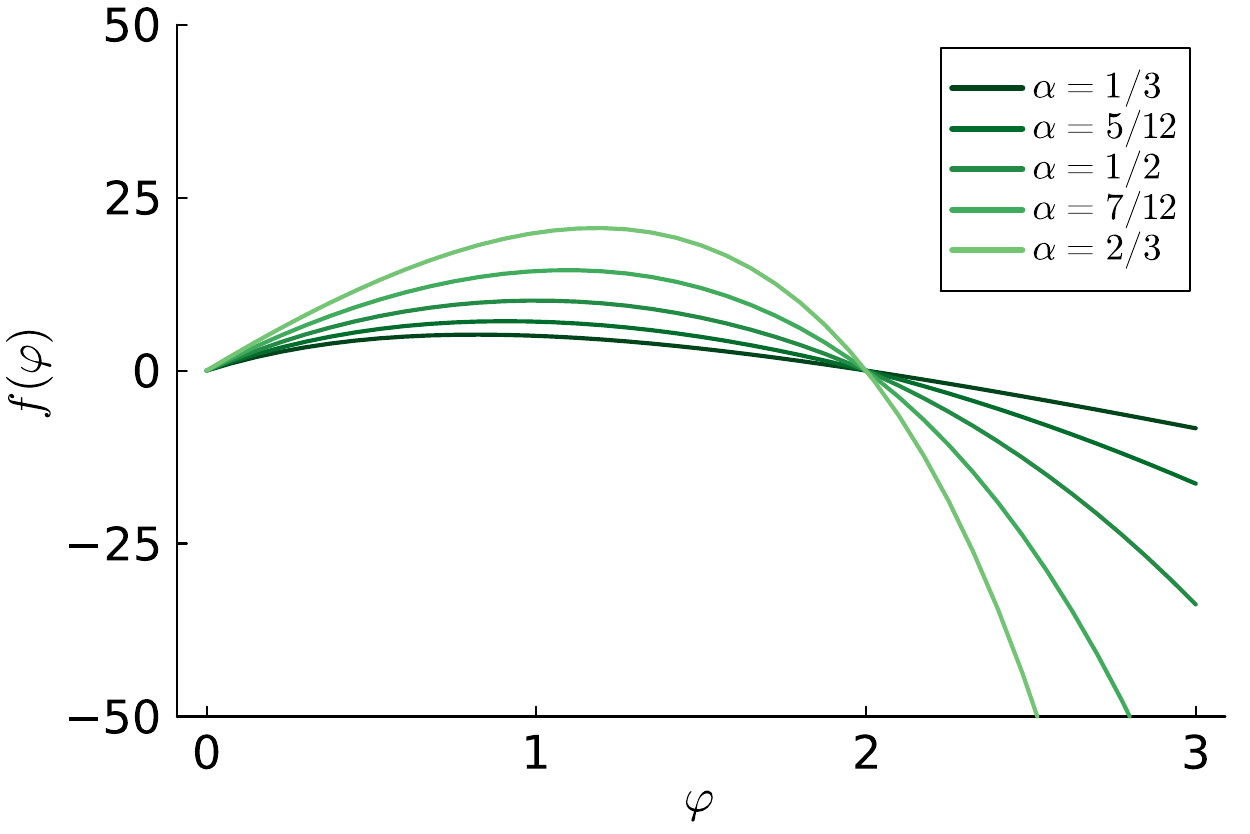}}}
	\caption{Graphical depiction of the flow described in Eqs. \eqref{k15}-\eqref{k17}. At the endpoint $\f=0$ the geometry is identified with the spatial boundary of Minkowski space-time, while at $\f_h=2$ there is a black-hole event horizon. At $\f\to\infty$ one encounters a (covered) singularity.}
	\label{fig:K3}
\end{figure}

\vskip 1cm
\section{On cosmological vs. event horizons}\label{app:J}
\vskip 1cm

In this appendix we shall review the properties of cosmological horizons and contrast them to those of event horizons.

We start with our ansatz with spherical slicing:

\begin{equation}
ds^2 = \dfrac{du^2}{f}-f e^{2A}dt^2 + e^{2A}R^2 d\Omega_{d-1}^2
\end{equation}

We  place the boundary at $u=+\infty$ and a horizon at some fixed position $u_h$. The function $f$ is positive (negative) near an AdS (dS) boundary. In order to distinguish cosmological from event horizons we study the motion of radial null geodesics across the horizon. The scale factor diverges or vanishes only at endpoints of the flow, so we focus in the case where the scale factor $e^A$ is finite at the horizon, in which case the horizon is identified by $f(u_h)=0$. This constitutes a coordinate singularity of the metric. In order to remove the apparent singularity we change to Eddington-Finklestein like coordinates:

\begin{equation}\label{eq:adv}
dv_{\pm}= dt \pm \dfrac{du}{e^A f}\,.
\end{equation}

\noindent
Choosing either sign results into a different extension of space-time as we shall shortly see. In Eddington-Finklestein coordinates the metric becomes

\begin{equation}
ds^2 = - f e^{2A} dv_{\pm}^2 \pm 2 e^A dudv_{\pm} + e^{2A}R^2 d\Omega_{d-1}^2\,,
\end{equation}

\noindent
or equivalently

\begin{equation}\label{eq:h}
2 e^A dudv_{\pm} = \mp (-f e^{2A} dv_{\pm}^2 -ds^2 + e^{2A}R^2 d\Omega_{d-1}^2)\,.
\end{equation}

In order to unveil the causal structure of space-time, we focus on time-like and null geodesics, which satisfy $ds^2\leq 0\,$. We first discuss some familiar examples:

\begin{itemize}
\item \textbf{Schwarzschild solution}. In the Schwarzschild solution, there is a single  horizon at some location $u_h$. For $u<u_h$ one has $f<0$. Using the $dv_+$ coordinate, and according to \eqref{eq:h}, we have $f<0 \Rightarrow dudv_+<0$ so that future directed geodesics ($dv_+>0$) are necessarily ingoing ($du<0$). For $u>u_h$ one has $f>0$, so that $dudv_+$ can be positive or negative, i.e. future-directed geodesics can be ingoing or outgoing. Therefore, we conclude that the $dv_+$ extension of the Schwarzschild solution covers the \textit{black-hole} region. Had we used the $dv_-$ extension we would have found that future-directed geodesics are necessarily outgoing for $u<u_h$. In other words, the $dv_-$ extension of the Schwarzschild solution covers the \textit{white hole} region.

\item \textbf{Exact de-Sitter solution in static coordinates} (see equation (\ref{a27})). In the dS solution, there is one horizon at some location $u_h$ while $f<0$ near the boundary at $u\to+\infty$. Therefore, if $u>u_h$ one has $f<0$.

    Using the $dv_-$ coordinate, and according to \eqref{eq:h}, we have $f<0 \Rightarrow dudv_->0$ so that future directed geodesics ($dv_->0$) are necessarily outgoing ($du>0$). For $u<u_h$ one has $f>0$, so that $dudv_-$ can be positive or negative, i.e. future-directed geodesics can be ingoing or outgoing. Therefore, we conclude that the $dv_-$ extension of the dS solution describes the region across a cosmological horizon. Had we used the $dv_+$ extension we would have found that future-directed geodesics are necessarily ingoing for $u>u_h$. In other words, the $dv_+$ extension of the Schwarzschild solution, covers the neighbourhood of the past cosmological horizon.
\end{itemize}

In subsequent examples we focus on the physically relevant extension, i.e. that containing black holes and future cosmological horizons rather than white holes and past cosmological horizons.

\begin{itemize}
\item \textbf{de-Sitter Schwarzschild solution} In this case, we have two horizons at some fixed locations $u_c$ and $u_h$ such that $\infty>u_c>u_h>0$. It is important to note that every time we find a horizon, we have the freedom to continue the metric with either of the $dv_{\pm}$ coordinates. At $u>u_c$ one has $f<0$, and it is easy to check that the $dv_-$ coordinates provides the extension across a cosmological horizon at $u_c$. The intermediate region $u_c>u>u_h$ has $f>0$ and supports both ingoing and outgoing geodesics. Finally, for $u<u_h$ we have again $f<0$ and one finds that the $dv_+$ extension describes extension to the interior of a black hole.

\item \textbf{Reissner-Nordstrom solution}. In this case, we have two horizons at some fixed locations $u_h$ and $u_C$ such that $\infty>u_h>u_C>0$. At $u>u_h$ one has $f>0$, and geodesics can be either outgoing or ingoing in both $dv_{\pm}$ extension. The intermediate region $u_h>u>u_C$ has $f<0$. Just like in the Schwarzschild solution, the $dv_+$ extension provides the extension across a black-hole horizon. In the interior we have again $f>0$ and both ingoing and outgoing geodesics are allowed. Contrary to the previous black holes, now the singularity is "timelike" and it lies in the causal past of observers that cross the second horizon at $u_C$. Therefore $u_C$ constitutes a Cauchy horizon.

\item \textbf{de Sitter-Reissner-Nordstrom solution}. We have three horizons located at $u_c$,$u_h$ and $u_C$ such that  $\infty>u_c>u_h>u_C>0$. The function $f$ is negative for $u>u_c$ and it changes sign as it crosses each horizon. Following a similar reasoning as in the examples above we find that the first horizon, at $u_c$ is a cosmological horizon using the $dv_-$ extension (the $dv_+$ extension gives a past cosmological horizon). The second horizon, at $u_h$, is a black-hole horizon using the $dv_+$ extension (white hole using the $dv_-$ extension). Finally, after the third horizon we have again $f>0$ and the singularity is timelike, so that $u_C$ is a Cauchy horizon.

\end{itemize}

Regarding our solutions in the main text, we encounter the following cases:

\begin{itemize}
\item In solutions that feature an AdS boundary and encounter a horizon at $u_h$, one has $f>0$ near the AdS boundary and $f<0$ inside the horizon. From \eqref{eq:h} with the $dv_+$ extension we conclude that we have a black-hole horizon.
\item In solutions that feature a dS boundary and encounter one horizon. We have $f<0$ near the dS boundary and $f$ positive inside of the horizon. Equation \eqref{eq:h} with the $dv_-$ extension reveal that it is a cosmological horizon. If such solutions have a singularity in the interior of a cosmological horizon, then it is a naked singularity.
\item  In solutions that feature a dS boundary and encounter two horizons. The situation is analogous to the dS black-hole solution. The outermost horizon is a cosmological horizon whereas the inner horizon is a black-hole event horizon.
\item In solutions that feature a Minkowski boundary and encounter one horizon at $u_h$, we have $e^{2A}f$ finite and positive (see e.g. Eqs. \eqref{5e5}, together with \eqref{e5e} and \eqref{e6e}) near the boundary, while it is negative inside the horizon. From \eqref{eq:h} with the $dv_+$ extension we conclude that we have a black-hole horizon.
\item In solutions that feature a Gubser-regular (type I or type II) endpoint with $V\to 0^-$ and encounter one horizon. At such a type I or type II  endpoint, the scale factor diverges, while $f$ is positive in the neighbourhood of that endpoint (see tables \ref{tab_IS} and \ref{tab_IIS}). Consequently, the $dv_+$ extension of Eq. \eqref{eq:h} describes a black-hole event horizon, with the bad singularity being away from the type I or type II endpoints.
\item In solutions that feature a Gubser-regular (type I or type II) endpoint with $V\to 0^+$ and encounter one horizon. At such a type I or type II  endpoint, the scale factor diverges, while $f$ is negative in the neighbourhood of that endpoint. Then, Eq. \eqref{eq:h} with the $dv_-$ extension reveals that the horizon is cosmological. If such solutions have a bad singularity in the interior of the cosmological horizon, then it is naked.

\item In solutions that feature a Gubser-regular endpoint (type I or type II) with $V\to 0^+$ and encounter two horizons. The situation is similar to that of a dS black-hole solution, namely the outermost horizon is cosmological while the inner one is a black-hole event horizon.
\end{itemize}

\vskip 1cm
\section{Solutions near asymptotic infinity in field space\label{asymp}}
\vskip 1cm

In this appendix, we study the asymptotic behaviour of the possible solutions as we approach the boundary in field space $|\f|\to \infty$. Without loss of generality,  we restrict the analysis to $\f>0$. The complementary case $\f<0$ is obtained by reversing the sign of the exponents $\a$, $\b$, $\g$ and $\delta$ defined in Eq. \eqref{z1} below. 
Specifically, we assume that the leading behaviour of the potential and metric functions is \footnote{The asymptotics of the potential are dictated by string theory paradigms that result from compactification. Although the parametrization of the solutions in (\ref{z1}) seems to be ad-hoc, it can be justified, using the tools of dynamical system theory, used recently in a similar problem in \cite{Raymond}.} \footnote{The variables $\a,\b,\g$ have also been used in other appendices. There is no relation between these variables in different appendices.}
\be
V\simeq -V_{\infty}e^{\a\f}\sp W\simeq W_{\infty}e^{\b\f} \sp f\simeq f_{\infty} e^{\g\f} \sp T\simeq T_{\infty}e^{\delta \f}\,.
\label{z1}\ee
Under the above assumption, the equations (\ref{reda1a}), (\ref{reda2a}) to leading order become

\begin{equation}
{ f_{\infty} W_{\infty}^2 g_1(\b,\g) e^{(2 \beta +\gamma-\alpha  )\varphi  }}+2(d-1)\alpha
   V_{\infty}=0\,,
\label{z2}\end{equation}
\be
f_{\infty} W_{\infty}^2 g_2(\b,\g) e^{ (2 \beta +\gamma-\alpha  )\varphi }-4(d-1)
   V_{\infty}=0\,,
\label{z4}\ee
respectively, where we have defined the prefactors $g_1$ and $g_2$ as

\be
 g_1\equiv 2(d-1) \beta \left(\beta^2 + \beta \gamma - \dfrac{d}{2(d-1)}\right)\,,
 \label{z3}  \ee
\be
g_2 \equiv -2(d-1)  \left(\beta^2 + \beta \gamma - \dfrac{d}{2(d-1)}\right)(1 + (d-1)\beta\gamma).
 \label{z5}  \ee
The solution for $T$ can be read off from Eq. \eqref{reda3a}:
\be
T=-{f_{\infty}W_{\infty}^2\over 2(d-1)(d-2)}\left(\b^2+\b\g-{d\over 2(d-1)}\right)e^{(2\b+\g)\f}
-{V_{\infty}\over (d-1)(d-2)}e^{\a\f}+\cdots
\label{z1a}\ee
In order to study the different solutions of Eqs. \eqref{z2} and \eqref{z4}, we first distinguish the cases where the exponential factor $e^{(2 \beta +\gamma-\alpha  )\varphi}$ is either leading or competing with the constant factor proportional to  $V_\infty$. The case where the exponential is subleading requires $V_\infty =0$, contradicting our initial assumption \eqref{z1}.
\vskip 1cm
{\bf The ``irregular" solutions (Type 0)}:  $ 2 \beta +\gamma-\alpha>0$
\vskip 1cm
In this case, we must solve the equations
\be
g_1=g_2=0
 \label{z6}\ee
and keep the solutions that satisfy the inequality $ 2 \beta +\gamma-\alpha>0$.
The solution of (\ref{z6}) is
\be
\g={d\over 2(d-1)\b}-\b\sp \b\not=0
  \label{z9} \ee
The inequality $ 2 \beta +\gamma-\alpha>0$ can be rewritten as
\begin{equation}\label{ms1}
\b + \dfrac{d}{2(d-1)\b}-\a>0~~~\to~~~ {1\over \b}\left(\b^2-\a\b+\dfrac{d}{2(d-1)}\right)>0
\end{equation}
and it is saturated for
\begin{equation}\label{ms2}
\beta = \a_\pm\equiv
{\a\over 2}\pm \sqrt{{\a^2\over 4}-{d\over 2(d-1)}}
\end{equation}
The $\a_\pm$ are degenerate for the value of $\a$ that saturates the Gubser bound:
\begin{equation}\label{ms3}
\a_G \equiv \sqrt{\dfrac{2d}{d-1}}\,.
\end{equation}
We now distinguish three subcases of solutions depending on the value of $\a$:

{\bf Type 0$_+$}: $\a>\a_G$.

In that case $\a_+>\a_->0$ and the inequality \eqref{ms1} is satisfied when
\be\label{ms4}
\b \in[0,\a_-]\cup [\a_+,\infty]\,,
\ee

{\bf Type 0$_m$ }: $|\a|<\a_G$

In this case, the binomial in (\ref{ms1}) is always positive and we must have
\be\label{ms5}
\b>0
\ee
to satisfy the inequality \eqref{ms1}.

{\bf Type 0$_-$}: $\a<-\a_G$

In that case $\a_-<\a_+<0$ and the inequality \eqref{ms1} is satisfied when
\be\label{ms6}
\b \in[\a_-,\a_+]\cup [0,+\infty]\,.
\ee

\noindent
Note that, from Eq. \eqref{z9}, $\gamma$ can be either positive or negative depending on the value of $\beta$. Specifically, we have
\begin{equation}\label{ms7}
\g<0 \Leftrightarrow \b\in\left(-\dfrac{\a_G}{2},0\right)\cup \left(\dfrac{\a_G}{2},\infty\right)
\end{equation}
while the complementary ranges of $\b$ lead to $\g>0$.

In all cases, the solution to Eqs. \eqref{z2} and \eqref{z4} is (including subleading terms proportional to $V_\infty$)
\be
f\simeq e^{\g\f}\left(f_{\infty} + f_{\infty}^{(1)} e^{-(2\b+\g-\a)\f}+\cdots\right) \sp W\simeq e^{\b\f}\left( W_{\infty}
+W_{\infty}^{(1)} e^{-(2\b+\g-\a)\f}+\cdots\right)
  \label{z13}
\ee
where the subleading coefficients are found to be
\be\label{ms8}
f_{\infty}^{(1)}={\a(d-2(d-1)\b^2)\over 2(d-1)\b\left(\b^2-\a\b+{d\over 2(d-1)}\right)^2 }~{V_{\infty}\over W_{\infty}^2}\,,
\ee
\be\label{ms9}
W_{\infty}^{(1)}={\b(2\b-\a) \over \left(\b^2-\a\b+{d\over 2(d-1)}\right)^2 }~  {V_{\infty}\over f_{\infty}W_{\infty}}\,.
\ee
Both $f_{\infty}$ and $W_{\infty}$, as well as the exponent $\b$ are free parameters. The function $T$ in (\ref{z1a}) is dominated in this case by the first term but the coefficient of this term vanishes.
The coefficient also of the subleading terms vanish and this solution has $T=0$ to all orders.
We look for deformations of the above solution by expanding
\be
W = W_\infty (e^{\b\f}+\dots)+e^{\b\f}\delta W\sp f = f_\infty (e^{\g\f}+\dots)+e^{\g\f}\delta f\;.
\ee
 Assuming that the deformations are small compared to the leading solution, we can linearise the equations \eqref{reda1a} and \eqref{reda2a}. Up to an exponential prefactor, the linearised equation governing the fluctuations $\delta W$ and $\delta f$ to leading order are
$$
\beta ^2 \delta f''-\beta  \left(\beta ^2+\frac{2-d}{2 (d-1)}\right)
   \delta f'-\beta  \left(\beta
   -\frac{d}{2 \beta  (d-1)}\right) \delta W''-
$$
\be
   -\left(\beta +\frac{d}{2 \beta  (d-1)}\right) \left(\beta
   ^2+\frac{2-d}{2 (d-1)}\right) \delta W'=0\,,
\label{ms11}\ee
\begin{equation}\label{ms12}
\beta ^2 \delta f'+\beta  \delta W''+\left(\beta
   ^2+\frac{d}{2 (d-1)}\right) \delta W'=0\,.
\end{equation}
The solution of the previous system of differential equations is given by
\begin{equation}
\delta W = C_0 + C_1 \f + C_T e^{ -\left(\beta + \frac{d-2}{2 \beta  (d-1)}\right)\f}\,,
\end{equation}
\begin{equation}
\delta f = C_2 + C_1 \frac{1}{2}  \left(-\frac{d}{\beta ^2 (d-1)}-2\right)\f  -\frac{C_T}{\beta ^2 (d-1)}e^{ -\left(\beta + \frac{d-2}{2 \beta  (d-1)}\right)\f}\,.
\end{equation}
The terms proportional to $C_0$, $C_1$ and $C_2$ are not subleading with respect to the unperturbed solution, and we set $C_0=C_1=C_2=0$ for consistency. The terms proportional to $C_T$ are subleading provided that $\b>0$. All in all, the full solution for $W$ and $f$ in this case is given by
\begin{equation}
W\simeq e^{\b\f}\left( W_{\infty}
+W_{\infty}^{(1)} e^{-(2\b+\g-\a)\f}+\cdots\right) + C_T e^{\b\f}\left( e^{ -\left(\beta + \frac{d-2}{2 \beta  (d-1)}\right)\f} + \dots\right)\,,(\b>0)
\end{equation}
\begin{equation}
f\simeq e^{\g\f}\left(f_{\infty} + f_{\infty}^{(1)} e^{-(2\b+\g-\a)\f}+\cdots\right)  - C_T e^{\g\f} \left(\frac{e^{ -\left(\beta + \frac{d-2}{2 \beta  (d-1)}\right)\f}}{\beta ^2 (d-1)}+\dots\right)\,,(\b>0)
\end{equation}
where $\gamma$, $W_\infty^{(1)}$ and $f_\infty^{(1)}$ are defined in equations \eqref{z9}, \eqref{ms8} and \eqref{ms9} respectively. There are four free parameters in this solution: $\beta$, $f_\infty$, $W_\infty$ and $C_T$.
We now obtain the function $T$ from the equation \eqref{reda3a}:
\begin{equation}\label{ms16}
T\simeq C_T\frac{ f_\infty V_\infty W_\infty^2 \left(2 \beta ^2
   (d-1)+d-2\right)^2 e^{\frac{\varphi }{\beta  (d-1)}}}{8 \beta ^2 (d-2) (d-1)^3}+\dots \ \ (\b>0)
\end{equation}
The case with $\b<0$ entails $C_T=0$, which can only happen with a flat slicing. {\em Therefore, in the spherical sliced ansatz \eqref{c39} we must have $\b>0$.} We find that the exponent $\delta$ defined in Eq. \eqref{z1} is $\delta = \frac{1}{(d-1)\beta} $ and the inverse scale factor diverges for this class of solutions.
Finally, the Kretschmann invariant for this class of solutions asymptotes to
\begin{equation}\label{l27}
K_2\simeq\frac{f_\infty^2 W_\infty^4 \left(4 \beta ^4 \left(d^2-1\right)-4 \beta ^2 (d-2)
   (d-1)+(d-2) d\right) }{16 (d-1)^2}e^{2 \varphi  (2 \beta +\gamma )}+\dots
\end{equation}
which , for  $\b>0$, diverges exponentially as $2\b+\g = \b+\frac{d}{2(d-1)\b}>0$.

The quantities $\rho$, $p$ and $\mathcal{I}$ controlling the curvature invariants (see appendix \ref{sect:inv_sphere}) also diverge exponentially:
\begin{equation}
\rho = \frac{1}{2} \beta ^2 f_\infty W_{\infty}^2 e^{\varphi  (2 \beta +\gamma )} + \dots \sp p = \frac{1}{2} \beta ^2 f_\infty W_{\infty}^2 e^{\varphi  (2 \beta +\gamma )} + \dots
\end{equation}
\begin{equation}
\mathcal{I} = \frac{f_\infty W_\infty^2 e^{\varphi  (2 \beta +\gamma )}}{4 (d-1)^2} + \dots
\end{equation}
Note that the temporal component of the metric behaves as

\begin{equation}
g_{tt}\sim f e^{2A}\sim f/T\sim e^{(\g-\delta)\varphi} = e^{-\frac{1}{\b}\left(\b^2-\frac{d-2}{2(d-1)}\right)\varphi}\,,
\end{equation}
which vanishes for $\b>\sqrt{\frac{d-2}{2(d-1)}}$ while it diverges for  $\b<\sqrt{\frac{d-2}{2(d-1)}}$. In both cases the sphere shrinks to zero size.

This is the generic class of solutions as $\f\to +\infty$ and they are badly singular. Their leading behavior is independent of the behavior of the scalar potential $V(\f)$, which affects these solutions to subleading orders. They therefore match the solutions given in Appendix \ref{nopot}, where $V=0$.

Overall, such solutions are not acceptable holographically and we therefore call them bad singularities in the sense of \cite{Gubser}.

 \vskip 1cm
{\bf The Gubser-regular solutions}: $ 2 \beta +\gamma-\alpha=0$
 \vskip 1cm
In this case, all terms of the equations \eqref{z2} and \eqref{z4} are of the same order, and we obtain two families of solutions:
\begin{equation}\label{ms18}
\textrm{I}: \beta = \dfrac{\a}{2} \sp \g=0 \sp f_\infty W_\infty^2 = \frac{8  V_\infty}{ \left(\a_G^2-\a^2\right)}
\end{equation}
\begin{equation}\label{ms19}
\textrm{II}: \beta = \dfrac{1}{(d-1)\a} \sp \g=\a\left(1-\dfrac{\a_C^2}{\a^2}\right) \sp f_\infty W_\infty^2 = \dfrac{2(d-1)^3\a^4V_\infty}{(d-1)(d-2)\a^2+2}
\end{equation}
where $\a_G$ is the value of $\a$ that saturates the Gubser bound defined in Eq. \eqref{ms3}, and we have defined $\a_C$ as the value of $\a$ that saturates the confinement bound\footnote{The names ``Gubser bound" and ``confinement bound" refer to the properties of holographic solutions with flat slicing in the AdS regime.
More information about the associated holographic physics can be found in \cite{Ahmad}.} :
\begin{equation}\label{ms019}
\a_C \equiv \sqrt{\dfrac{2}{d-1}} = \dfrac{1}{\sqrt{d}}\a_G\,.
\end{equation}

\vskip 0.7cm
{\bf Type I}: [Eq. \eqref{ms18}]
\vskip 0.7cm

The inverse scale factor $T$ can be obtained from Eq. \eqref{reda3a}. Under substitution of Eq. \eqref{ms18} we find that $T=0$. In order to fully characterize the solution, we first look for deformations about the given solution \eqref{ms18}. As usual, we parametrize the deformation as
\begin{equation}
W = W_\infty e^{\a\f/2} + e^{\a\f/2}\delta W\sp f = f_\infty  + \delta f\,,
\end{equation}
and assume that the deformations are small compared to the unperturbed solution. Now, we linearise the system of equations \eqref{reda1a} and \eqref{reda2a}. To leading order, we find that the fluctuations $\delta W$ and $\delta f$ satisfy the following system of equations:
\begin{multline}
\frac{1}{8} W_\infty^2 \left(\frac{2 d}{d-1}-\alpha ^2\right)\delta f + \frac{1}{16} \alpha  W_\infty^2 \left(\alpha ^2 (d-1)-2 (d+2)\right)\delta f' + \frac{1}{8} \alpha ^2 (d-1) W_\infty^2\delta f''  \\ +\frac{2 V_\infty}{W_\infty}\delta W + \frac{4 \alpha  (d-1) V_\infty}{\alpha ^2 (d-1) W_\infty-2 d W_\infty}\delta W' = 0\,,
\end{multline}
\begin{multline}
-\frac{2 \alpha  V_\infty}{W_\infty}\delta W + \frac{\left(4-6 \alpha ^2\right) d V_\infty+6 \alpha ^2 V_\infty}{W_\infty
   \left(\left(\alpha ^2-2\right) d-\alpha ^2\right)} \delta W'  -\frac{4 \alpha  (d-1) V_\infty}{W_\infty \left(\left(\alpha ^2-2\right) d-\alpha
   ^2\right)}\delta W'' \\ +\frac{\alpha  W_\infty^2 \left(\left(\alpha ^2-2\right) d-\alpha ^2\right)}{8 (d-1)} \delta f + \frac{\alpha ^2 W_\infty^2}{4}\delta f' = 0\,.
\end{multline}
The general solution to the previous equations is easily found to be:
\begin{equation}
\delta W = C_1+C_2 e^{-\a\left(1-\frac{\a_C^2}{\a^2}\right)\varphi}+C_3 e^{\frac{1}{2\a}(\a_G^2-\a^2)\varphi}\,,
\end{equation}
\begin{equation}
\delta f = C_1\frac{16  V_\infty}{W_\infty^3 \left(\alpha ^2 -\a_G^2\right)}+C_2\frac{32  V_\infty }{\alpha ^2(d-1) W_\infty^3 \left(\alpha ^2-\a_G^2\right)}e^{-\a\left(1-\frac{\a_C^2}{\a^2}\right)\varphi}+C_4
   e^{\frac{1}{2\a}(\a_G^2-\a^2)\varphi}\,.
\end{equation}
The solution proportional to $C_1$ corresponds to a deformation of the single integration constant of the leading solution and therefore we can set it to zero when we look for new integration constants.
 The deformations proportional to $C_{2,3,4}$ are allowed depending on the ranges of $\a$. Specifically, the consistency conditions are
\be\label{m39}
 \a\in(-\infty,-\a_G)\Rightarrow C_2=C_3=C_4=0
\ee
\be\label{m40}
\a\in\left(-\a_G,-\a_C\right) \Rightarrow C_2=0\,,
\ee
\be\label{m41}
\a\in\left(-\a_C,0\right) \Rightarrow C_2,C_3,C_4 \ \textrm{allowed}\;,
\ee
\be\label{m42}
\a\in\left(0,\a_C\right) \Rightarrow C_2=C_3=C_4=0
\ee
\be\label{m43}
 \a \in \left(\a_C,\a_G\right)\Rightarrow C_3=C_4=0\,.
\ee
\be\label{m44}
\a \in \left(\a_G,+\infty\right)\Rightarrow C_2,C_3,C_4 \ \textrm{allowed}\,.
\ee
We obtain now the inverse scale factor $T$ by algebraically solving Eq. \eqref{reda3a}:
\begin{equation}\label{m45}
T\simeq-\frac{2 C_2 V_\infty \left(\alpha ^2 -\a_C^2\right) \left(\alpha ^2 (d-1)+2
   (d-2)\right) }{\alpha ^2 (d-1) W_\infty
   \left(\alpha ^2 -\a_G^2\right)}e^{\frac{2 \varphi }{\alpha  (d-1)}} +\dots
\end{equation}
The inverse scale factor is proportional to $C_2$. The spherical slicing requires  a non-vanishing scale factor, which, because of (\ref{m39})-(\ref{m44})  is only achieved for
\be
\a \in \left(-\a_C,0\right)\cup\left(\a_C,+\infty\right)\;.
\ee

 In such a case, the exponent of the inverse scale factor is
\begin{equation}
\delta = \dfrac{2}{\a(d-1)}\,.
\end{equation}
The inverse scale factor diverges if the potential diverges, and vanishes if the potential vanishes.

All in all, the type I solutions for a spherical slicing can have up to four integration constants, $W_\infty$, $C_1$, $C_2$ and $C_3$. Explicitly, the solutions with a diverging potential ($\a>0$) that are Gubser-regular ($\a<\a_G$) with a spherical sliced ansatz ($T> 0$) we are using here, have only two integration constants, namely $W_\infty$ and $C_2$, and the asymptotic solution in this case can be written as
\begin{equation}
W = W_\infty e^{\a\varphi/2} + C_2 e^{-\frac{1}{2}\left(\a^2-a_C^2\right)\varphi} + \dots\hspace{1cm} \a \in \left(\a_C,\a_G\right)
\end{equation}
\begin{equation}\label{m49}
f = -\dfrac{8V_\infty}{\a^2-\a_G^2}  +C_2\frac{32  V_\infty }{\alpha ^2 (d-1) W_\infty^3 \left(\alpha ^2 -\a_G^2\right)}e^{-\a\left(1-\frac{\a_C^2}{\a^2}\right)\varphi}+ \dots
\end{equation}
\begin{equation}\label{m50}
T=-\frac{2 C_2 V_\infty \left(\alpha ^2 -\a_C^2\right) \left(\alpha ^2 (d-1)+2
   (d-2)\right) }{\alpha ^2 (d-1) W_\infty
   \left(\alpha ^2 -\a_G^2\right)}e^{\frac{2 \varphi }{\alpha  (d-1)}} +\dots
\end{equation}
Note that $f$ approaches a constant value that, for the given range in $\a$, is positive in the AdS regime ($V<0$, $V_\infty>0$) and negative in the dS regime. Additionally, the subleading term for $f$ in Eq. \eqref{m49} is anti-correlated with the leading term for $T$ in Eq. \eqref{m50}. Therefore, the requirement that $T>0$ implies that the blackening function $f$ decreases as we depart from the type I endpoints with a divergent potential.

Similarly, the solutions with a vanishing potential ($\a<0$) with a spherically sliced ansatz have $-\a_C<\a<0$ and have the four integration constants allowed. Finally, the Krestchmann invariant for the type I solution is, to leading order,
\begin{equation}
K_2 = \frac{4 d V_\infty^2  \left(\alpha ^4 (d-1)^2-4 \alpha ^2 (d-1)+2
   (d+1)\right)}{(d-1)^4 \left(\alpha ^2 -\a_G^2\right)^2}e^{2 \alpha  \varphi }\,,
\end{equation}
while the pressure, energy density and quantity $\mathcal{I}$ are given by
\begin{equation}\label{invregI1}
\rho = -\frac{\alpha_G^2 V_\infty }{\alpha ^2-\alpha_G^2}e^{\alpha  \varphi } + \dots \sp p = -\frac{ V_\infty(2 \alpha ^2-\alpha_G^2) }{\alpha ^2-\alpha_G^2}e^{\alpha  \varphi }+\dots
\end{equation}
\begin{equation}\label{invregI2}
\mathcal{I} =  -\frac{2 V_\infty
   }{(d-1)^2 \left(\alpha ^2-\alpha_G^2\right)}e^{\alpha  \varphi } + \dots
\end{equation}

The type-I singularity is milder that that of the type-0 solution because in type-0 we have $2\b+\g>\a$.
When $\a<\a_G$, the type I  solutions are holographically acceptable, as we discuss in the next section \ref{app:l1}. The general behaviour of the asymptotic solution at the type I endpoints is summarised in Table \ref{tab_IS} for the spherically sliced ansatz, and extended to the hyperbolic and flat ansatze in Tables \ref{tab_IH} and \ref{tab_IF} respectively.

\vskip 0.7cm
{\bf Type II}: [Eq. \eqref{ms19}]
\vskip 0.7cm

The leading inverse scale factor $T$ is obtained from Eq. \eqref{reda3a}, which, under substitution of Eq. \eqref{ms19} reads

\begin{equation}\label{m54}
T = \frac{V_\infty e^{\alpha  \varphi } }{2 (d-2)}\left(\alpha ^2 -\a_C^2\right) \,,
\end{equation}
with $\a_C$ defined in Eq. \eqref{ms019}. The coefficient $\delta$ defined in Eq. \eqref{z1} is $\delta = \a$. Therefore, the inverse scale factor diverges if the potential diverges, and vanishes if the potential vanishes.
Additionally, the spherically sliced ansatz \eqref{c39} requires $T>0$, which translates into $V_\infty (\a^2-\a_C^2)>0$. In the AdS regime ($V<0$, $V_\infty>0$), the previous condition translates into $|\a|>\a_C$, while in the dS regime ($V>0$, $V_\infty<0$) one requires that $|\a|<\a_C$.

In order to fully characterize the solution, we look for deformations around the given solution \eqref{ms19}:
\begin{equation}
W = W_\infty e^{\frac{1}{(d-1)\a}\varphi} + e^{\frac{1}{(d-1)\a}\varphi}\delta W \sp f = f_\infty e^{\a\left(1-\frac{\a_C^2}{\a^2}\right)\f} + e^{\a\left(1-\frac{\a_C^2}{\a^2}\right)\f}\delta f\,,
\end{equation}
where the perturbations $\delta f$ and $\delta W$ are assumed to be subleading with respect to $f_\infty$ and $W_\infty$ respectively. Linearising the equations \eqref{reda1a} and \eqref{reda2a}, we find that, to leading order, the perturbations obey the following system of equations:
\begin{multline}
\frac{\alpha ^2 (d-1)^2 V_\infty \left(\alpha ^2 (d-1)-2\right)}{W_\infty
   \left(\alpha ^2 (d-2) (d-1)+2\right)}\delta W''  + \frac{W_\infty^2}{2 \alpha ^2 (d-1)}\delta f''  -\frac{W_\infty^2 \left(\alpha ^2 (d-2) (d-1)+6\right)}{4 \alpha ^3 (d-1)^2}\delta f' \\+ \frac{\alpha  (d-1) V_\infty \left(\alpha ^4 (-(d-2)) (d-1)^2+2 \alpha ^2 (d-5)
   (d-1)+4\right)}{2 W_\infty \left(\alpha ^2 (d-2) (d-1)+2\right)}\delta W' +\frac{2 V_\infty}{W_\infty}\delta W \\+ \frac{W_\infty^2 \left(\alpha ^2 (d-2) (d-1)+2\right)}{2 \alpha ^4 (d-1)^3}\delta f = 0\,,
\end{multline}
\begin{multline}
\frac{2 \alpha ^3 (d-1)^2 V_\infty}{W_\infty \left(\alpha ^2 (d-2) (d-1)+2\right)}\delta W'' + \frac{\alpha ^2 (d-1) V_\infty \left(\alpha ^2 (-(d-4)) (d-1)-2\right)}{W_\infty
   \left(\alpha ^2 (d-2) (d-1)+2\right)}\delta W'  \\ -\frac{2 \alpha  V_\infty}{W_\infty}\delta W + \frac{W_\infty^2}{\alpha ^2 (d-1)^2}\delta f' + -\frac{W_\infty^2 \left(\alpha ^2 (d-2) (d-1)+2\right)}{2 \alpha ^3 (d-1)^3}\delta f =0 \,.
\end{multline}
The general solution to the previous system of equations is given by
\begin{equation}
\delta W = C_+ e^{\left(\a_1 + \sqrt{\a_1^2-\a_2}\right)\frac{\f}{4\a}} + C_- e^{\left(\a_1 - \sqrt{\a_1^2-\a_2}\right)\frac{\f}{4\a}} + C_3\,,
\end{equation}
\begin{equation}
\delta f = -\frac{2 \alpha ^6  (d-1)^3 V_\infty}{W_\infty^3 \a_1}\left(C_+e^{\left(\a_1 + \sqrt{\a_1^2-\a_2}\right)\frac{\f}{4\a}}+C_-e^{\left(\a_1 - \sqrt{\a_1^2-\a_2}\right)\frac{\f}{4\a}}+\frac{2C_3}{(d-1)\a^2}\right) +C_4 e^{\frac{\a_1}{2\a}\f}
\end{equation}
where we have defined
\begin{equation}
\a_1 = \alpha ^2 (d-2)+\frac{2}{d-1} \sp \a_2 = 8 (d-2) \left(\alpha ^2-\frac{2}{d-1}\right) \left(\alpha ^2+\frac{2}{(d-2)
   (d-1)}\right)\,.
\end{equation}
The solution proportional to $C_3$ is a deformation of the single integration constant of the leading solution and as such is not interesting, therefore we set it to zero.

 The solution proportional to $C_4$ is subleading only if $\a<0$ (assuming $d>2$).

 The situation for $C_{\pm}$ is more delicate, since the exponents can be complex. The consistency conditions, under the assumption $2<d<10$, are summarised in the following:
\begin{itemize}
\item If $\a<-\frac{3\sqrt{2}}{\sqrt{(10-d)(d-1)}}$, the exponents $\a_1 \pm\sqrt{\a_1^2-\a_2}$ are complex. A real solution is achieved by appropriately combining $C_+$ and $C_-$. Both deformations are allowed in this case as ${\a_1\over \a}<0$

\item If  $-\frac{3\sqrt{2}}{\sqrt{(10-d)(d-1)}}<\a<0$, the exponents are real. The deformation proportional to $C_+$ is always allowed while the deformation proportional to $C_-$ is allowed only for $\a<-\a_C$.
\item If  $0<\a<\frac{3\sqrt{2}}{\sqrt{(10-d)(d-1)}}$, then the exponents are real. The deformation proportional to $C_+$ is never allowed and we have to set $C_+=0$. Conversely, the deformation proportional to $C_-$ is allowed only if $\a<\a_C$.
\item If $\a>\frac{3\sqrt{2}}{\sqrt{(10-d)(d-1)}}$, then the exponents are again complex, but none of the deformations is allowed. Then we require $C_+=C_-=0$.
\end{itemize}
The induced deformations on the inverse scale factor $T\to T + \delta T$ are obtained again from Eq. \eqref{reda3a}, directly giving
\begin{equation}
\delta T = -\frac{\alpha ^2 V_\infty \left(\alpha ^2 (d-1)-2\right)}{2 (d-2) W_\infty}\left(C_+ e^{\frac{\varphi  \left(4 \alpha ^2+\sqrt{\a_1^2-\a_2}+\a_1\right)}{4 \alpha
   }}+C_- e^{\frac{\varphi  \left(4 \alpha ^2-\sqrt{\a_1^2-\a_2}+\a_1\right)}{4 \alpha
   }}\right)\,.
\end{equation}

In the particular case of $\a>0$ in the  AdS regime ($V_\infty>0$) for the spherically sliced ansatz ($\a>\a_C$), the full solution has a single integration constant, $W_\infty$. The inverse scale factor $T$, the superpotential $W$ and the blackening function $f$ are diverging. However, the temporal component of the metric vanishes exponentially: $g_{tt}\sim f/T\sim e^{-\a_G^2\f/(d\a)}\to 0$.
The Kretschmann invariant diverges for this class of solutions for $\a>0$:
\begin{equation}
K_2 = \frac{2V_\infty^2 e^{2 \alpha  \varphi } \left(\alpha ^4 (d-1)^2 (d ((d-5) d+11)-9)+4
   \alpha ^2 (d-1) ((d-6) d+7)+4 (3 d-5)\right)}{(d-2) (d-1)^2 \left(\alpha ^2 (d-2)
   (d-1)+2\right)^2}+\dots
\end{equation}
In this case, the pressure, energy density and quantity $\mathcal{I}$ are given by
\begin{equation}\label{invregII1}
\rho = V_\infty\dfrac{(d-1)\alpha^2 + \a_C^2}{(d-2)\a^2 + \a_C^2} e^{\alpha\varphi} + \dots \sp p = -V_\infty\dfrac{(d-3)\alpha^2 + \a_C^2}{(d-2)\a^2 + \a_C^2} e^{\alpha\varphi}+\dots
\end{equation}
\begin{equation}\label{invregII2}
\mathcal{I} =  \frac{1}{2}V_\infty\left(\frac{\a_C^2-\alpha ^2}{d-2}+\frac{\alpha ^4}{\a_C^2+\alpha
   ^2 (d-2)}\right)e^{\a \varphi}+ \dots
\end{equation}
Again this is less singular compared to the type 0 solutions and as singular as the type I solutions. The general behaviour of the asymptotic solution at the type II endpoints is summarised in Table \ref{tab_IIS} for the spherical slicing, and extended in Tables \ref{tab_IIH} and \ref{tab_IIF} for the hyperbolic and flat slicing respectively.

\begin{table}[]
\centering
\begin{tabular}{|c|c|c|c|c|c|c|}
\hline
 \multicolumn{7}{|c|}{Type I, $\qquad R_c>0$  } \\ \hline
 & $V$  &  $|W|$  &  $T$  &  $f$ & $\rho$ & Free parameters  \\ \hline
 \multirow{2}{*}{$\alpha<-\a_G$}  & \multirow{2}{*}{$0^\pm$}  & \multicolumn{5}{|c|}{ \multirow{2}{*}{Does not exist}}  \\
  &    &  \multicolumn{5}{|c|}{}   \\ \hline
  \multirow{2}{*}{$\alpha \in (-\a_G,-\a_C)$}  & \multirow{2}{*}{$0^\pm$}  & \multicolumn{5}{|c|}{ \multirow{2}{*}{Does not exist}}  \\
  &    &  \multicolumn{5}{|c|}{}    \\\hline
   \multirow{2}{*}{$\alpha \in (-\a_C,0)$}  & $0^-$  & \multirow{2}{*}{$0$}  & \multirow{2}{*}{$0^+$}   & $>0$  & $0^+$   & \multirow{2}{*}{$4$}  \\ \cline{2-2} \cline{5-6}
  & $0^+$  &   &    & $<0$  &$0^-$  &   \\ \hline
 \multirow{2}{*}{$\alpha\in(0, \a_C)$}  & \multirow{2}{*}{$\pm\infty$}  & \multicolumn{5}{|c|}{ \multirow{2}{*}{Does not exist}}  \\
  &    &  \multicolumn{5}{|c|}{}     \\ \hline
  \multirow{2}{*}{$\alpha \in (\a_C,\a_G)$}  & $-\infty$  & \multirow{2}{*}{$\infty$}  & \multirow{2}{*}{$+\infty$}   & $>0$  & $+\infty$   & \multirow{2}{*}{$2$}  \\ \cline{2-2} \cline{5-6}
  & $+\infty$  &   &    & $<0$  &$-\infty$  &   \\ \hline
   \multirow{2}{*}{$\alpha >\a_G$}  & $-\infty$  & \multirow{2}{*}{$\infty$}  & \multirow{2}{*}{$+\infty$}   & $<0$  & $-\infty$   & \multirow{2}{*}{$4$}  \\ \cline{2-2} \cline{5-6}
  & $+\infty$  &   &    & $>0$  &$+\infty$  &   \\ \hline
\end{tabular}
\caption{Properties of the type I solutions for an ansatz sliced with a positive constant curvature manifold, whose curvature is denoted $R_c$.}\label{tab_IS}
\end{table}

\begin{table}[]
\centering
\begin{tabular}{|c|c|c|c|c|c|c|}
\hline
 \multicolumn{7}{|c|}{Type I, $\qquad R_c<0$  } \\ \hline
 & $V$  &  $|W|$  &  $T$  &  $f$ & $\rho$ & Free parameters  \\ \hline
 \multirow{2}{*}{$\alpha<-\a_G$}  & \multirow{2}{*}{$0^\pm$}  & \multicolumn{5}{|c|}{ \multirow{2}{*}{Does not exist}}  \\
  &    &  \multicolumn{5}{|c|}{}   \\ \hline
  \multirow{2}{*}{$\alpha \in (-\a_G,-\a_C)$}  & \multirow{2}{*}{$0^\pm$}  & \multicolumn{5}{|c|}{ \multirow{2}{*}{Does not exist}}  \\
  &    &  \multicolumn{5}{|c|}{}    \\\hline
   \multirow{2}{*}{$\alpha \in (-\a_C,0)$}  & $0^-$  & \multirow{2}{*}{$0$}  & \multirow{2}{*}{$0^-$}   & $>0$  & $0^+$   & \multirow{2}{*}{$4$}  \\ \cline{2-2} \cline{5-6}
  & $0^+$  &   &    & $<0$  &$0^-$  &   \\ \hline
 \multirow{2}{*}{$\alpha\in(0, \a_C)$}  & \multirow{2}{*}{$\pm\infty$}  & \multicolumn{5}{|c|}{ \multirow{2}{*}{Does not exist}}  \\
  &    &  \multicolumn{5}{|c|}{}     \\ \hline
  \multirow{2}{*}{$\alpha \in (\a_C,\a_G)$}  & $-\infty$  & \multirow{2}{*}{$\infty$}  & \multirow{2}{*}{$-\infty$}   & $>0$  & $+\infty$   & \multirow{2}{*}{$2$}  \\ \cline{2-2} \cline{5-6}
  & $+\infty$  &   &    & $<0$  &$-\infty$  &   \\ \hline
   \multirow{2}{*}{$\alpha >\a_G$}  & $-\infty$  & \multirow{2}{*}{$\infty$}  & \multirow{2}{*}{$-\infty$}   & $<0$  & $-\infty$   & \multirow{2}{*}{$4$}  \\ \cline{2-2} \cline{5-6}
  & $+\infty$  &   &    & $>0$  &$+\infty$  &   \\ \hline
\end{tabular}
\caption{Properties of the type I solutions for an ansatz sliced with a negative constant curvature manifold, whose curvature is denoted $R_c$.}\label{tab_IH}
\end{table}

\begin{table}[]
\centering
\begin{tabular}{|c|c|c|c|c|c|c|}
\hline
 \multicolumn{7}{|c|}{Type I, $\qquad R_c=0$  } \\ \hline
 & $V$  &  $|W|$  &  $T$  &  $f$ & $\rho$ & Free parameters  \\ \hline
 \multirow{2}{*}{$\alpha <-\a_G$}  & $0^-$  & \multirow{2}{*}{$0$}  & \multirow{2}{*}{$=0$}   &  $<0$ & $0^-$  & \multirow{2}{*}{$1$}  \\ \cline{2-2} \cline{5-6}
  & $0^+$  &   &    & $>0$  & $0^+$ &   \\ \hline
 \multirow{2}{*}{$\alpha \in(-\a_G,-\a_C)$}  & $0^-$  & \multirow{2}{*}{$0$}  & \multirow{2}{*}{$=0$}   &  $>0$ & $0^+$   & \multirow{2}{*}{$3$}  \\ \cline{2-2} \cline{5-6}
  & $0^+$  &   &    &  $<0$ & $0^-$  &   \\ \hline
  \multirow{2}{*}{$\alpha \in (-\a_C,0)$}  & $0^-$  & \multirow{2}{*}{$0$}  & \multirow{2}{*}{$=0$}   & $>0$  & $0^+$   & \multirow{2}{*}{$3$}  \\ \cline{2-2} \cline{5-6}
  & $0^+$  &   &    & $<0$  &$0^-$  &   \\ \hline
 \multirow{2}{*}{$\alpha \in(0,\a_G)$}  & $-\infty$  & \multirow{2}{*}{$\infty$}  & \multirow{2}{*}{$=0$}   & $>0$  & $+\infty$  & \multirow{2}{*}{$1$}  \\ \cline{2-2} \cline{5-6}
  & $+\infty$  &   &    & $<0$  & $-\infty$  &   \\ \hline
  \multirow{2}{*}{$\alpha \in (\a_C,\a_G)$}  & $-\infty$  & \multirow{2}{*}{$\infty$}  & \multirow{2}{*}{$=0$}   & $>0$  & $+\infty$   & \multirow{2}{*}{$1$}  \\ \cline{2-2} \cline{5-6}
  & $+\infty$  &   &    & $<0$  &$-\infty$  &   \\ \hline
   \multirow{2}{*}{$\alpha >\a_G$}  & $-\infty$  & \multirow{2}{*}{$\infty$}  & \multirow{2}{*}{$=0$}   & $<0$  & $-\infty$   & \multirow{2}{*}{$3$}  \\ \cline{2-2} \cline{5-6}
  & $+\infty$  &   &    & $>0$  &$+\infty$  &   \\ \hline
\end{tabular}
\caption{Properties of the type I solutions for an ansatz sliced with a zero curvature manifold, whose curvature is denoted $R_c=0$. Note that the function $T$ is identically zero for the ansatz sliced with a zero curvature manifold. As a consequence, the constant $C_2$ in Eqs. \eqref{m39}-\eqref{m45} must be set to zero.}\label{tab_IF}
\end{table}

\begin{table}[]
\centering
\begin{tabular}{|c|c|c|c|c|c|c|}
\hline
 \multicolumn{7}{|c|}{Type II, $\qquad R_c>0$  } \\ \hline
 & $V$  &  $|W|$  &  $T$  &  $f$ & $\rho$ & Free parameters  \\ \hline
 \multirow{2}{*}{$\alpha <-\a_G$}  & $0^-$  & $0$  & $0^+$   &  $ 0^+$ & $0^+$  & $4$ \\ \cline{2-7}
  & $0^+$  &  \multicolumn{5}{c|}{Does not exist}  \\ \hline
 \multirow{2}{*}{$\alpha \in(-\a_G,-\a_C)$}  & $0^-$  & $0$  & $0^+$   &  $0^+$ & $0^+$  & $4$ \\ \cline{2-7}
  & $0^+$  &  \multicolumn{5}{c|}{Does not exist}  \\ \hline
 \multirow{2}{*}{$\alpha \in(-\a_C, 0 )$}  & $0^-$  & \multicolumn{5}{c|}{Does not exist} \\ \cline{2-7}
  & $0^+$  &   $0$  & $0^+$   &  $-\infty$ & $0^-$  & $3$ \\ \hline
 \multirow{2}{*}{$\alpha \in(0, \a_C )$}  & $-\infty$  & \multicolumn{5}{c|}{Does not exist} \\ \cline{2-7}
  & $+\infty$  &   $\infty$  & $+\infty$   &  $0^-$ & $-\infty$  & $2$ \\ \hline
 \multirow{2}{*}{$\alpha \in(\a_C, \a_G )$}  & $-\infty$  &    $\infty$  & $+\infty$   &  $+\infty$ & $+\infty$  & $1$\\ \cline{2-7}
  & $+\infty$  & \multicolumn{5}{c|}{Does not exist} \\ \hline
 \multirow{2}{*}{$\alpha >\a_G$}  & $-\infty$  &    $\infty$  & $+\infty$   &  $+\infty$ & $+\infty$  & $1$\\ \cline{2-7}
  & $+\infty$  & \multicolumn{5}{c|}{Does not exist} \\ \hline
\end{tabular}
\caption{Properties of the type II solutions for an ansatz sliced with a positive constant curvature manifold, whose curvature is denoted $R_c$.}\label{tab_IIS}
\end{table}

\begin{table}[]
\centering
\begin{tabular}{|c|c|c|c|c|c|c|}
\hline
 \multicolumn{7}{|c|}{Type II, $\qquad R_c<0$  } \\ \hline
 & $V$  &  $|W|$  &  $T$  &  $f$ & $\rho$ & Free parameters  \\ \hline
 \multirow{2}{*}{$\alpha <-\a_G$}  & $0^-$  &  \multicolumn{5}{c|}{Does not exist} \\ \cline{2-7}
  & $0^+$  & $0$  & $0^-$   &  $ 0^-$ & $0^-$  & $4$  \\ \hline
 \multirow{2}{*}{$\alpha \in(-\a_G,-\a_C)$}  & $0^-$  &  \multicolumn{5}{c|}{Does not exist} \\ \cline{2-7}
  & $0^+$  &  $0$  & $0^-$   &  $0^-$ & $0^-$  & $4$  \\ \hline
 \multirow{2}{*}{$\alpha \in(-\a_C, 0 )$}  & $0^-$  &  $0$  & $0^-$   &  $+\infty$ & $0^+$  & $3$   \\ \cline{2-7}
  & $0^+$  &  \multicolumn{5}{c|}{Does not exist}\\ \hline
\multirow{2}{*}{$\alpha \in(0, \a_C )$}  & $-\infty$  &   $\infty$  & $-\infty$   &  $0^+$ & $+\infty$  & $2$  \\ \cline{2-7}
  & $+\infty$  & \multicolumn{5}{c|}{Does not exist}\\ \hline
 \multirow{2}{*}{$\alpha \in(\a_C, \a_G )$}  & $-\infty$  & \multicolumn{5}{c|}{Does not exist}\\ \cline{2-7}
  & $+\infty$  &     $\infty$  & $-\infty$   &  $-\infty$ & $-\infty$  & $1$\\ \hline
\multirow{2}{*}{$\alpha >\a_G$}  & $-\infty$  &\multicolumn{5}{c|}{Does not exist}\\ \cline{2-7}
  & $+\infty$  &     $\infty$  & $-\infty$   &  $-\infty$ & $-\infty$  & $1$ \\ \hline
\end{tabular}
\caption{Properties of the type II solutions for an ansatz sliced with a negative constant curvature manifold, whose curvature is denoted $R_c$.}\label{tab_IIH}
\end{table}

\begin{table}[]
\centering
\begin{tabular}{|c|c|c|c|c|c|c|}
\hline
 \multicolumn{7}{|c|}{Type II, $\qquad R_c=0$  } \\ \hline
 & $V$  &  $|W|$  &  $T$  &  $f$ & $\rho$ & Free parameters  \\ \hline
$\forall \a$  &  \multicolumn{6}{c|}{Does not exist except for $\a=\pm \a_C$} \\ \hline
\end{tabular}
\caption{Type II solutions generically do not exist for an ansatz sliced with a zero curvature manifold, $R_c=0$. For the case of $\a=\pm \a_{C}$ it coincides with the type I solution.}\label{tab_IIF}
\end{table}

\subsection{Acceptable singularities}\label{app:l1}

In this section, we discuss a criterion for  the singularities can be acceptable in the holographic sense.
Gubser's criterion, \cite{Gubser} instructs us to accept a singularity if it can be cloaked by an infinitesimal  horizon.
Equivalently, a singularity is acceptable if the scalar potential is bounded above during the flow.
It is important to stress here that Gubser's analysis focused on Lorentz-invariant ($f=1$) flat-sliced solutions.
Moreover, he assumed that solutions start at an AdS boundary.
It is not {\it a priori} clear which of his analysis is valid when $f\not=1$, or when we have positive or negative curvature slices.

However, since then, several Gubser-regular asymptotics were shown explicitly to be resolvable by lifting to higher dimensions (see for example \cite{cgkkm1}).
Here, we give a more general definition of Gubser-regular solutions that goes beyond the one given in \cite{Gubser}.

There is a generalization of the aforementioned criterion: If solutions belong to the boundary of a manifold of otherwise acceptable solutions, then they are acceptable.
This is the case with the solutions that can be cloaked with an infinitesimal horizon. All solutions with regular horizons are acceptable in holography and therefore by continuity their extremal limits should be acceptable. There are other examples of this criterion that does not involve regular horizons.
As it was shown in \cite{Jani}, solutions that are singular even after uplifting in higher dimensions must be accepted as regular for consistency of the holographic approach.
The example discussed in \cite{Jani} involves, in higher dimensions, a conifold where two spheres shrink to zero at the same point.

\subsubsection{Gubser's criterion}\label{sec:gc}

In order to determine whether an asymptotic singularity can be covered by a horizon, we study the solutions to the equations of motion with a potential given by
\begin{equation}\label{l65}
V = - V_{\infty} e^{\a \f}\,,
\end{equation}
assume that $\a>0$, and focus on the singularity at $\f\to\infty$\footnote{We do not consider here, the case where as $\f\to\pm\infty$, $V\to 0^{\pm}$. These also correspond to naked singularities but their nature is different.}. We suppress subleading terms to the potential as $\f\to\infty$.

We study under what conditions a type 0, I or II asymptotic solution can be deformed such that it is covered by a horizon.

\vskip .5cm
\subsubsection*{The type 0 asymptotic solution}
\vskip 0.5cm

The type 0 asymptotic solution has all the integration constants allowed by the equations of motion, and therefore admits no further deformation.
Therefore, either type 0 singularities are already covered by a horizon (in which case they are deemed acceptable) or they are naked. In the second case, we cannot further deform the solution to create a horizon.  We conclude that type 0 singularities when not covered by a  horizon are not acceptable singularities.

\vskip .5cm
\subsubsection*{The type I asymptotic solution}

\vskip .5cm

One exact solution to Eqs. \eqref{f10_1}-\eqref{w55b}, is given by
\begin{equation}\label{l66}
W = W_0 e^{\frac{\a}{2} \f}\,,\quad T=0\,,\quad  f= \frac{8  V_{\infty}}{W_0^2 \left(\alpha_G ^2 -\a^2 \right)}+\frac{f_1}{W_0^2} e^{\frac{1}{2
   \alpha }(\a_G^2-\a^2)\f}\,, \qquad (\a \neq \a_G)
\end{equation}
which is a deformation of the type I asymptotic solution described above, with the deformation parameter being $f_1$.
Note that fluctuations about the above solution allow for $T\neq 0$, as required by the spherically sliced ansatz. $W_0$ and $f_1$ are two integration constants. We solve now the flow equation $W'=\dot{\f}$ to find
\begin{equation}\label{l67}
\f(u) = -\frac{2}{\a}\log\left(-\frac{1}{4}W_0\a^2(u-u_0)\right)\,,
\end{equation}
where the location of the singularity is denoted as $u_0$. We rewrite the function $f$ in the above solution \eqref{l66} as a function of $u$, and set the integration constant $f_1$ such that a  horizon is located at $u_h$. Additionally, we set $W_0^2 = 8$ without loss of generality by virtue of the scaling symmetry \eqref{scaling}. With this choice of integration constants, the function $f$ becomes
\begin{equation}\label{l68}
f= \frac{V_{\infty}}{\a_G^2-\a^2}\left[1- \left(\dfrac{u_h - u_0}{u - u_0}\right)^{-1+\a_G^2/\a^2}\right]\,.
\end{equation}
Generically, the singularity at $u_0$ is of type 0. However, if the location of the horizon is arbitrarily close to the location of the singularity ($u_h\to u_0$), and $\a<\a_G$, the singularity at $u_0$ has type I asymptotics. We conclude that, in a flat sliced ansatz ($T=0$), the type I asymptotic solution can be deformed such that the singularity is covered by a small horizon only if $\a < \a_G$. In such a case, the type I asymptotic solution is acceptable \`a la Gubser (i.e. Gubser-regular).

For the spherically sliced ansatz, we need to further deform the asymptotic solution \eqref{l66}. In particular, we assume that $T = C_T e^{\delta \f}$, where $\delta$ is determined by the equations of motion,  Eq. \eqref{eqtt},\begin{equation}
\delta = \dfrac{2}{(d-1)\a}
\end{equation}
Also $C_T\ll 1$ and we obtain the appropriate corrections to $W$ and $f$.  In particular, we write $W \to W + C_T \delta W$ and $f\to f + C_T \delta f$.

Subsequently, from Eqs. \eqref{w55} and \eqref{w55b}, we find that the perturbation for the superpotential is determined by the following equations:
$$
\frac{f_1 e^{\frac{d \varphi }{\alpha  (d-1)}} \left(4 \delta W ''-4 \alpha  \delta W ' +\alpha ^2 \delta W\right)}{8 W_0}-C_T (d-2) (d-1)
   e^{\frac{2 \varphi }{\alpha  (d-1)}}+
$$
\begin{equation}\label{l69}
+\frac{ V_{\infty} e^{\frac{\alpha  \varphi }{2}}
   \left(-4 \alpha  \delta W'' +2 \left(\alpha ^2 + \a_G^2 \right) \delta W' - \alpha \a_G^2
   \delta W \right)}{\alpha W_0(\a^2 - \a_G^2)}=0\,.
\end{equation}
We obtain only the leading solution to the previous equation. The corrections proportional to $C_T$ are obtained when the term proportional to $f_1$ or the term proportional to $V_\infty$ are of the same order to the term proportional to $C_T$. Whether the term proportional to $f_1$ dominates over the term proportional to $V_\infty$ is determined by the value of $\a$, and we obtain the following two possibilities:
\begin{itemize}
\item $\a<\a_G$.
In this case, the deformation to Eq. \eqref{l66} is given by
\begin{equation}
W = W_0 e^{\frac{\a}{2}\f} +  C_T\frac{8 \alpha ^2 W_0  (d-2) (d-1)^3}{f_1 \left(\alpha ^2( d-1)+2(d-2)\right)^2}e^{-\frac{d-2}{\a (d-1)}\f} + \dots
\end{equation}
\begin{equation}
T = C_T e^{\frac{2}{(d-1)\a}\f} + O(C_T^2)
\end{equation}
\begin{equation}
f = \frac{8  V_{\infty}}{W_0^2 \left(\alpha_G ^2 -\a^2 \right)}+\frac{f_1}{W_0^2} e^{\frac{1}{2
   \alpha }(\a_G^2-\a^2)\f}  -C_T \frac{32  (d-2) (d-1)^2 e^{-\frac{1}{\a}\varphi  \left(\a^2-\a_C^2
   \right)}}{W_0^2 \left(\alpha ^2-\left(\alpha ^2+2\right) d+4\right)^2} + \dots
\end{equation}
The previous deformation is only subleading to the unperturbed solution if $\a>\a_C$. Similarly to the flat sliced case, we set the integration constant $f_1$ such that there is a horizon at $\f_h$, and we set $W_0^2 = 8/(\a_G^2-\a^2)$ without loss of generality. Then, the function $f$ can be rewritten as
$$
f = V_{\infty}\left(1-e^{\frac{(\a_G^2-\a^2)}{2\a}(\f - \f_h)}\right) -
$$
\begin{equation}
 -C_T \frac{4 (d-2) (d-1) \left(\alpha_G ^2-\a^2\right) e^{-\frac{1}{\alpha }
   \left(\alpha ^2-\a_C^2\right)\f}}{\left(\alpha ^2 (d-1)+2
   (d-2)\right)^2}\left( 1 - e^{\frac{\left(\alpha ^2 (d-1)+2 (d-2)\right)}{2 \alpha  (d-1)} (\varphi -\f_h)}\right)
\end{equation}

Similarly to the flat sliced case, we recover the type I asymptotic solution in the limit where the horizon approaches the location of the singularity $\f_h\to\infty$, provided that
$$\a_C<\a<\a_G.$$ For finite $\f_h$, the singularity has type 0 asymptotics.

\item $\a>\a_G$.

In this case, the deformation to \eqref{l66} is given by
\begin{equation}
W = W_0 e^{\frac{\a}{2}\f} -  C_T\frac{\alpha ^2 W_0  (d-2) (d-1)^2 \left(\alpha ^2-\a_G^2\right)}{2 V_\infty
   \left(\alpha ^2 - \a_C^2 \right) \left(\alpha ^2 (d-1)+2 (d-2)\right)}e^{-\left(\frac{\a}{2}-\frac{2}{(d-1)\a}\right)\f} + O(C_T^2)
\end{equation}
\begin{equation}
T = C_T e^{\frac{2}{(d-1)\a}\f} + O(C_T^2)
\end{equation}
\begin{equation}
f = \frac{8  V_{\infty}}{W_0^2 \left(\alpha_G ^2 -\a^2 \right)}+\frac{f_1}{W_0^2} e^{\frac{1}{2
   \alpha }(\a_G^2-\a^2)\f}  -C_T\dfrac{16(d-1) (d-2)e^{-\frac{1}{\a}\varphi  \left(\a^2-\a_C^2
   \right)}}{W_0^2(\a^2-\a_C^2)(2(d-2)+\a^2(d-1))} + \dots
\end{equation}
We set the integration constant $f_1$ such that there is a horizon located at $\f_h$, and set $W_0^2 = 8/(\a^2-\a_G^2)$. In such a case, the function $f$ can be rewritten as

$$
f = -V_{\infty}\left(1-e^{\frac{(\a_G^2-\a^2)}{2\a}(\f - \f_h)}\right) -
$$
\begin{equation}\label{dfs2}
  -C_T\dfrac{2(d-1) (d-2)(\a^2-\a_G^2)e^{-\frac{1}{\a}\varphi  \left(\a^2-\a_C^2
   \right)}}{(\a^2-\a_C^2)(2(d-2)+\a^2(d-1))}\left(1- e^{\frac{\alpha ^2 (d-1)+2 (d-2)}{2 \alpha  (d-1)}(\f-\f_h)}\right)
\end{equation}
The location of the horizon cannot be brought arbitrarily close to the singularity, $\f_h\to \infty$, because $\a>\a_G$ and the first exponential term in \eqref{dfs2} diverges.

\end{itemize}
We conclude that type I singularities with a diverging potential can be covered by an infinitesimal regular horizon only if $\a<\a_G$, and this is valid for the spherical, flat or hyperbolically sliced ansatze.

\subsubsection*{The type II asymptotic solution}

We turn our attention now to an exact solution to the equations of motion, given the potential \eqref{l65}, that is a deformation of the type II asymptotic solution previously described. This exact solution is given by

$$
W = W_0 e^{\frac{\f}{(d-1)\a}}\,, \quad T= \frac{V_\infty \left(\a^2-\alpha_C ^2\right)}{2 (d-2) }  e^{\a \f}\,, \quad
$$
\begin{equation}\label{l78}
f = \frac{2 \alpha ^4 (d-1)^3 V_\infty e^{\a  \left(1 -\frac{\a_C^2}{\alpha^2}\right)\f}}{W_0^2 \left(\alpha ^2 (d-2) (d-1)+2\right)} + \dfrac{f_1}{W_0^2}e^{\left(\frac{d}{2\a}-\frac{1}{(d-1)\a} \right)\f}\,.
\end{equation}
Note that the function $T$ is non-trivial, and such a solution does not exist in the flat sliced ansatz except if $\a=\a_C$.
We first find the solution to the flow equation $\dot{\f}=W'$:
\begin{equation}
\f(u) = -(d-1)\a \log\left(-\frac{W_0}{(d-1)^2\a^2}(u-u_0)\right)\,,
\end{equation}
where we have denoted $u_0$ as the location of the singularity. Then, we can rewrite $f$ as a function of $u$. We set the integration constant $f_1$ such that there is a horizon located at $u_h$, and we find that
\begin{equation}\label{l80}
f = V_\infty (u-u_0)^{2-\alpha ^2 (d-1)}
   \left(1-\left(\frac{u_h-u_0}{u-u_0}\right)^{\frac{1}{2} \alpha ^2
   (d-2) (d-1)+1}\right)
\end{equation}
where the integration constant $W_0$ has been adjusted without loss of generality in order to normalize the prefactor in $f$:

$$W_0 = \alpha ^2 (d-1)^2 2^{\frac{1}{\alpha ^2 (d-1)}} \left((d-1) \left(\alpha ^2 \left(d^2-3
   d+2\right)+2\right)\right)^{\frac{1}{\alpha ^2-\alpha ^2 d}}\,.$$
We observe that the location of the horizon can be set arbitrarily close to the singularity, $u_h\to u_0$, for any value of $\a$. In the coincident limit, we recover the type II asymptotic solution uncovered in the previous section. Note that if $u_h\neq u_0$, the singularity has type 0 asymptotics. We conclude that type II singularities with a diverging potential can be always be cloaked by infinitesimal horizons, regardless of the value of $\a$.

There is a further condition that can characterise resolvable naked singularities and this is whether they can be resolved by uplifting the theory to a higher dimension, \cite{cgkkm1}. This has been studied recently in special cases in \cite{Ghodsi,GKN,Ahmad,Jani}.
The general case, however, requires further analysis and will not be pursued here.

\section{The solutions when $V(\f)=0$}\label{nopot}

There are periods in the evolution of solutions where the potential is negligible in the equations, see appendix \ref{forbid} for an example.
In such a case, the solutions becomes approximately equal to these with $V=0$.
In particular, the generic singular solutions  as $\f\to \pm\infty$ that were called type $0_{\pm,m}$ in the previous section, are in this class, as can be seen from the analysis in appendix \ref{asymp}.

In this appendix, we shall find the general solution to equations (\ref{reda1a}), (\ref{reda2a}) with $V=0$.

One obvious solution of  (\ref{reda2a}) is $W=W_0$ constant, and  (\ref{reda2a}) implies that $f=0$.
Therefore, this solution is not acceptable. The other solutions must satisfy
\be
 W'f' + f\left(W''-\frac{d }{2
   (d-1)}W\right)=0~~~\to~~~f(\f)={e^{{d\over 2(d-1)}\int_{\f_0}^{\f}dx{W(x)\over W'(x)}}\over W'(\f)}
\label{V1}   \ee
Since
\be
{W\over W'}=-2(d-1){dA\over d\f}
\ee
we obtain
\be
f(\f)={e^{{-d}\int_{\f_0}^{\f}dx{dA\over d\f}}\over W'(\f)}={e^{-d A+dA_0}\over W'}=Z_0{e^{-dA}\over W'}\,.
\label{V1a}   \ee

Substituting $f$ into (\ref{reda1a}) we obtain
\be
2(d-1)(W'''W'-(W'')^2)+(d-2)(WW''-(W')^2)=0
\label{V2} \ee
which can be equivalently rewritten as
\be
2(d-1)\left({W''\over W'}\right)'-(d-2)\left({W\over W'}\right)' =0
\label{V3} \ee
and which can be integrated once to obtain
\be
2(d-1)W''-(d-2)W=2CW'
\label{V4} \ee
where $C$ is an arbitrary real integration constant.
This equation is linear with constant coefficients and its general solution is
\be
W=C_+e^{\rho_+\f}+C_-e^{\rho_-\f}\sp \rho_{\pm}={C\pm\sqrt{C^2+2(d-1)(d-2)}\over 2(d-1)}
\label{V5} \ee
We can now determine $f$ from (\ref{V1}) to be
\be
f=f_0~e^{-{d(\r_++\r_-)\over (d-2)}\f}{|W'|^{{d\over (d-2)}}\over W'}
\label{V6} \ee
The scale factor  as a function of $\f$ is given by integrating ${dA\over d\f}=-{1\over 2(d-1)}{W\over W'}$. We obtain
\be
e^{A}={e^{A_0+{(\r_++\r_-)\over d-2}\f}\over  |W'|^{{1\over d-2}}}\sp fW' e^{dA}=f_0 e^{dA_0}={\rm constant}
\ee
To compute $T$ we must take derivatives of absolute values properly. We use $|g(x)|'={g(x)g'(x)\over |g(x)|}$ to obtain
\be
f'=\left[-{d(\r_++\r_-)\over (d-2)}+{2\over d-2}{W''\over W'}\right]f
\ee
as well as T from (\ref{reda3a})
$$
T={f ( WW''-(W')^2 )\over
 2 (d-2) (d-1)}=
$$
\be
={(\r_+-\r_-)^2C_+C_-\over 2(d-1)(d-2)}e^{{(\r_++\r_-)}\f}~f  ={1\over R^2}e^{-2A}
\ee
We must therefore have ${fC_+C_-}>0$ and
\be
f=e^{-2A}{ 2(d-1)(d-2)  \over  R^2(\r_+-\r_-)^2C_+C_-}e^{-{(\r_++\r_-)}\f}
\ee

From (\ref{w54a}) we obtain that
\be
R^2={{\rm sign}(W')\over  e^{2A_0}f_0}{2(d-1)(d-2)\over (\r_+-\r_-)^2C_+C_- }>0
\ee
which implies that
\be
 {\rm sign}(W')f_0C_+C_->0
 \ee

We can also calculate the monotonous quantities (\ref{ev1}), (\ref{ev2})
\be
f\dot A e^{dA}=-{fW e^{dA}\over 2(d-1)}=-{f_0 e^{dA_0}\over 2(d-1)}{W\over W'}
\ee
\be
\dot f e^{dA}=f'W' e^{dA}=f_0 e^{dA_0}{f'\over f}=f_0 e^{dA_0}\left[-{d(\r_++\r_-)\over (d-2)}+{2\over d-2}{W''\over W'}\right]=
\ee
$$
=f_0 e^{dA_0}\left[{1\over (d-1)}{W\over W'}-(\r_++\r_-)\right]
$$
Indeed, for $V=0$ their rate of change is the same and therefore they differ by a constant.
We have
\be
{d\over du}\left(f\dot A e^{dA}\right)=\dot \f{d\over d\f}\left(f\dot A e^{dA}\right)=
-{f_0 e^{dA_0}\over 2(d-1)}W'{d\over d\f}\left({W\over W'}\right)=-{f_0 e^{dA_0}\over 2(d-1)}{(W')^2-WW''\over W'}=
\ee
$$
=-{f_0 e^{dA_0}\over 2(d-1)}{C_+C_- \over (C_+\r_+e^{-\r_-\f}+C_-\r_-e^{-\r_+\f})}
$$

This is the most general solution as it has four arbitrary integration constants.
It is clear that $f$ has a correlated sign with  $W'$ and if $W'$ vanishes at a finite point then $f$ also vanishes and changes sign at that same point.
Note that if $f$ changes sign, then $T$ changes sign.

This solution agrees with the potential independent terms of the solutions of the appendix \ref{asymp}.

It is clear that $\r_+>0$ always and $\r_-<0$ always. Another way of parametrizing the two exponents is
\be
\r_+=\beta>0\sp \r_-=-{(d-2)\over 2(d-1)~\beta}
\label{V11} \ee
By varying $C$ through all real values, $\b$ takes all non-negative values.
The precise map is
\be
\beta={C+\sqrt{C^2+2(d-1)(d-2)}\over 2(d-1)}\sp C=(d-1)\beta-{d-2\over 2\b}
\ee

When $\f\to +\infty$, from (\ref{V5}), (\ref{V6})  the leading behavior is
\be
W\simeq C_+ e^{\beta \f}\to\infty \sp f\simeq {\rm sign}(C_+)~f_0~(\beta|C_+|)^{2\over d-2}~e^{\gamma\f}\ \,, \ \g=-(\r_++\r_-)={d\over 2(d-1)\b}-\b
\label{V12} \ee
\be
e^A\simeq {e^{A_0-{1\over 2(d-1)\b}\f}\over |\beta C_+|^{1\over d-2}}\to 0
\ee
\be
T\simeq {\rm sign}(C_+)~f_0~{C_+C_-(\r_+-\r_-)^2\over 2(d-1)(d-2)}(\beta|C_+|)^{2\over d-2}  e^{\delta\f}\to \infty\sp \delta ={1\over (d-1)\b}
\ee
Since $\delta>0$ always, as $\f\to +\infty$ $e^A\to 0$, always.
The radius of the time circle is
\be
fe^{2A}\simeq f_0 e^{2A_0 + \e\f}\sp \e\equiv -\beta+{(d-2)\over 2(d-1) \b}
\ee
We can compute $A(u)$ and $\f(u)$ as
\be
e^{\beta\f}={1\over \b^2 |C_+||u_0-u|}\sp e^A=e^{A_0}|u_0-u|^{1\over 2(d-1)\b^2}\sp u\to u_0
\ee

We also compute the monotonous quantities (\ref{ev1}), (\ref{ev2})
\be
f\dot A e^{dA}\sim {-fW T^{-{d\over 2}}\over 2(d-1)}\sim {W\over W'}\sp
\dot f e^{dA}\simeq f'W'T^{-{d\over 2}}\sim -{d\over d-2}(\r_++\r_-)+{2\over d-2}{W''\over W'}
\ee
Near the $\f\to + \infty$ boundary we have

\begin{equation}
f \dot{A}e^{dA}\simeq \dfrac{f_0 e^{dA_0}}{2(1-d)} + \dfrac{f_0 C_-(d-2)(\rho_+ - \rho_-)e^{dA_0}}{ 4 C_+ (d-1)^2\beta^2\rho_+\rho_-}e^{\eta \varphi} + \dots \qquad \eta \equiv -\beta - \dfrac{d-2}{2(d-1)\beta}
\end{equation}

\begin{equation}
\dot{f}e^{dA}\simeq f_0 \gamma e^{dA_0} - \dfrac{f_0 C_-(d-2)(\rho_+ - \rho_-)e^{dA_0}}{2  C_+ (d-1)^2\beta^2\rho_+\rho_-}e^{\eta \varphi} + \dots
\end{equation}
It is clear above that one is increasing and the other is decreasing.

\begin{figure}[h!]
	\centering
	\subfloat{\includegraphics[scale=0.28]{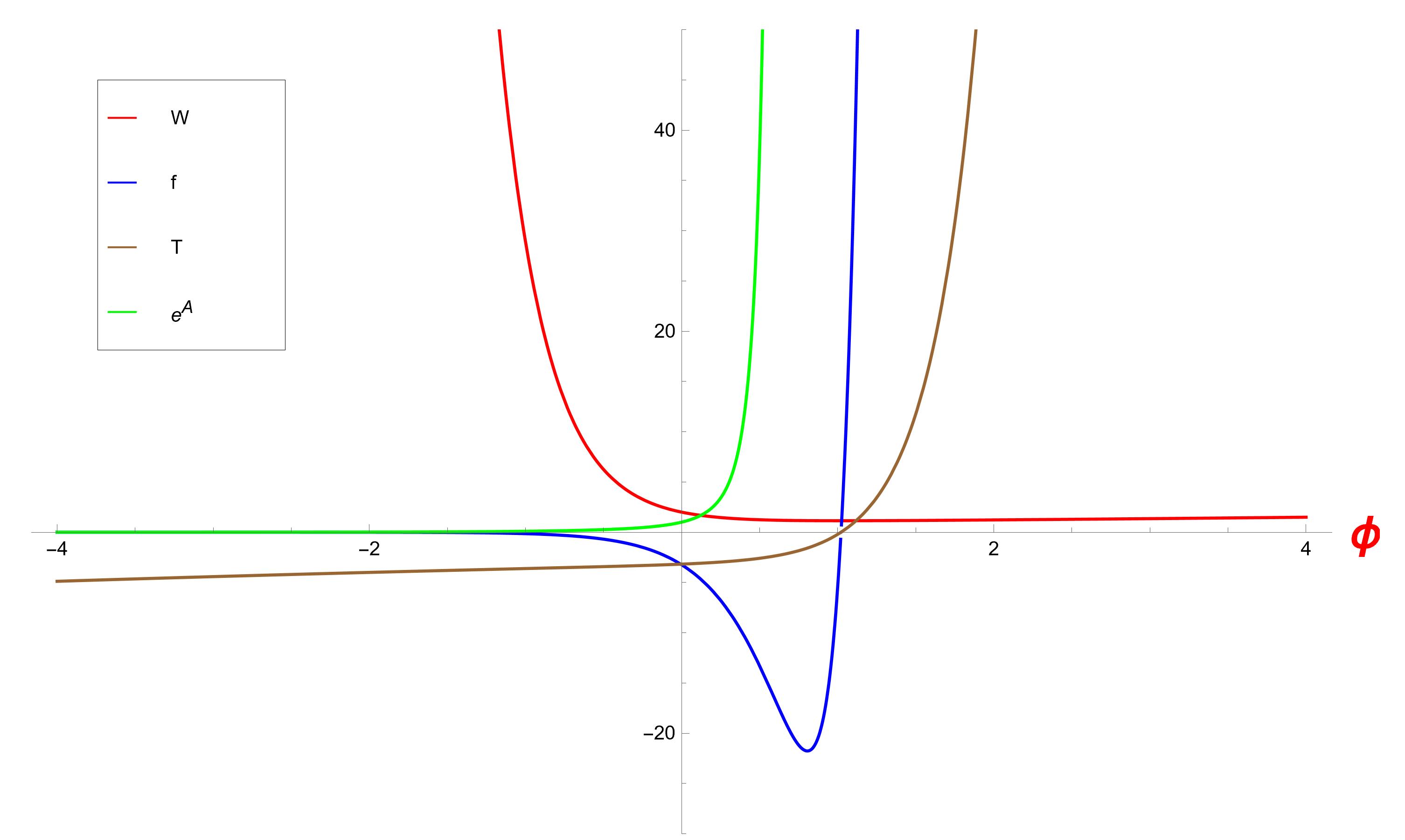}}
	\qquad
	\subfloat{\includegraphics[scale=0.28]{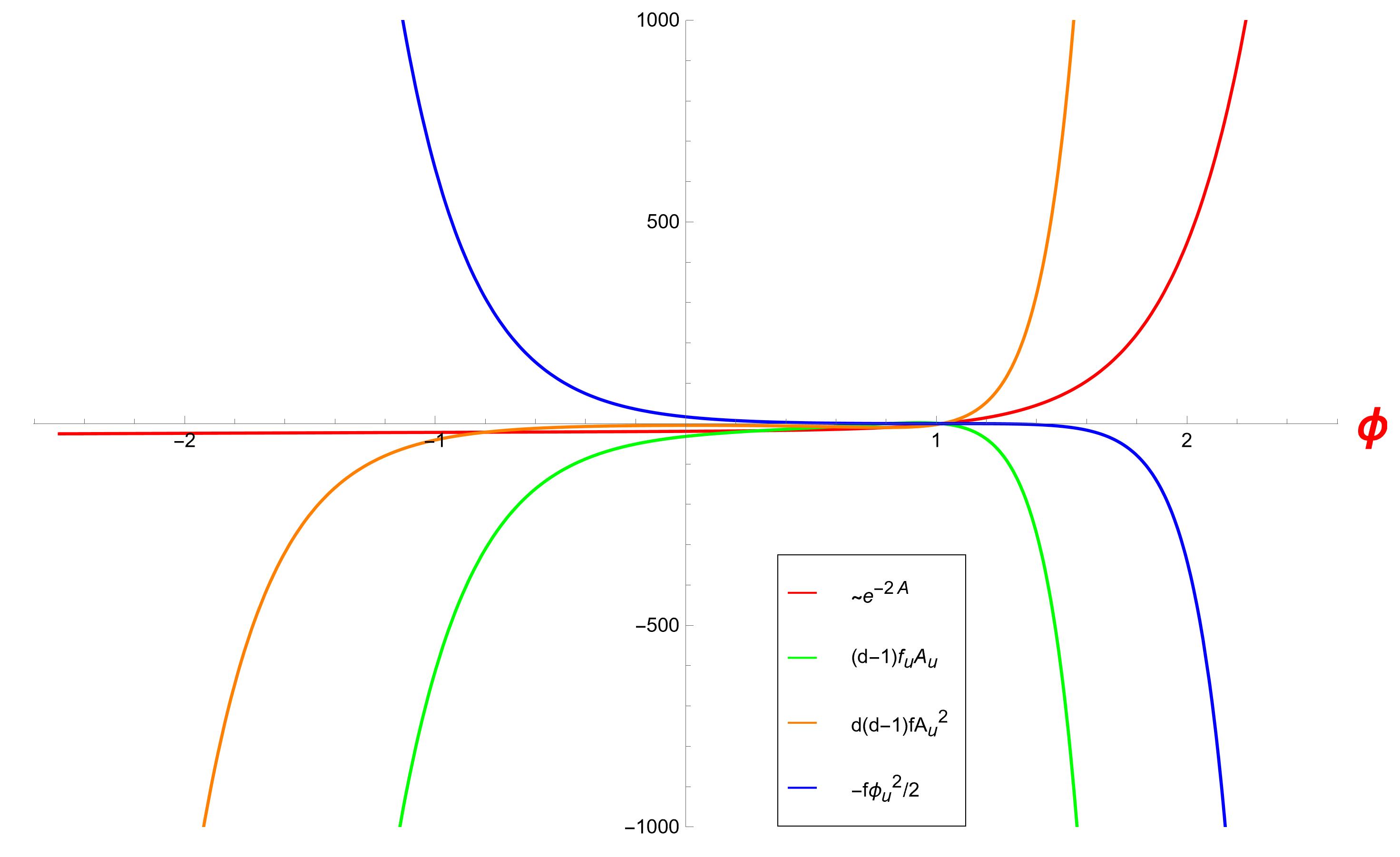}}
	\caption{The solutions with $d=4$, $\beta=0.1$, $C_+=C_-=f_0=1$.
  In this solution the curvature function $T$ changes sign. However, pieces of such solutions, near $\f=\pm\infty$ can be asymptotics  of the solutions of the system with $V\not=0$. In this case the
  $\f\to +\infty$ asymptotics have $T>0$ and are acceptable. On the right-hand figure we plot the different terms of the Hubble equation,  (\ref{f23c}). The subscript $u$ in the various quantities in the figures stands for a derivative with respect to $u$. }
	\label{fM1}
\end{figure}

\begin{figure}[h!]
	\centering
	\subfloat{\includegraphics[scale=0.28]{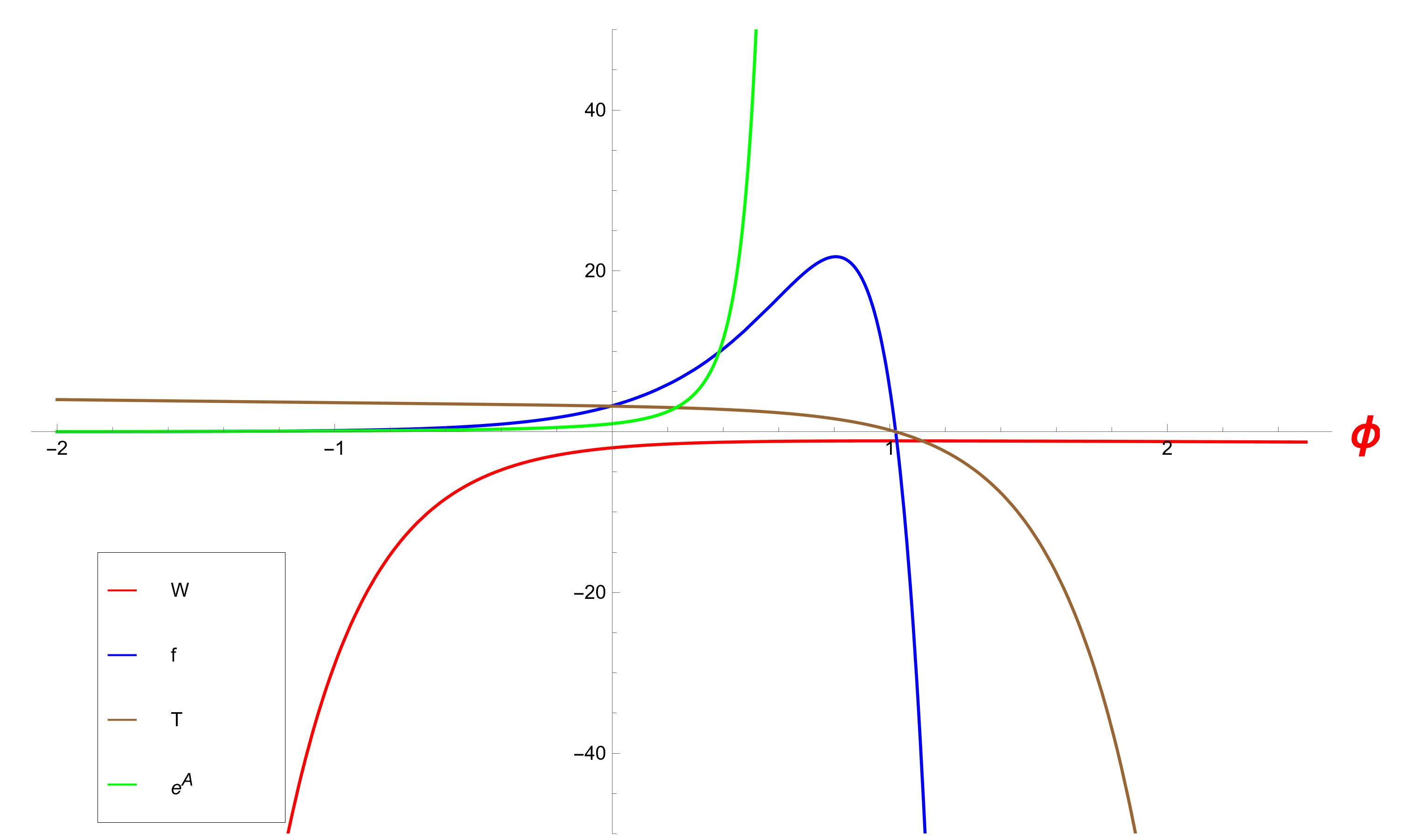}}
	\qquad
	\subfloat{\includegraphics[scale=0.28]{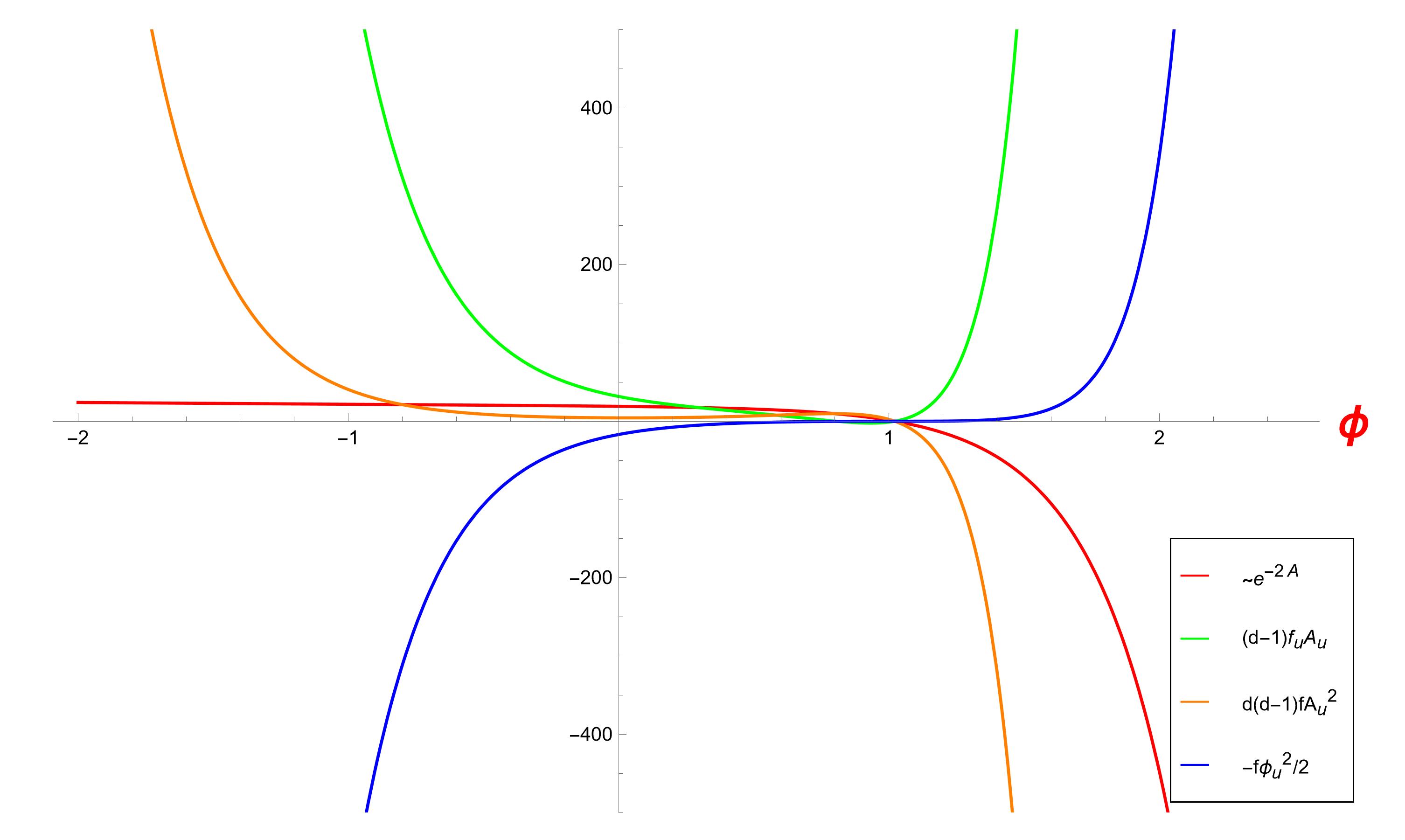}}
	\caption{The solutions with $d=4$, $\beta=0.1$, $C_+=C_-=-1$, $f_0=1$. The curvature function $T$ in this solution  changes sign.
However, pieces of such solutions, near $\f=\pm\infty$ can be asymptotics  of the solutions of the system with $V\not=0$.
In this case, the $\f\to -\infty$ asymptotics have $T>0$ and are acceptable. On the right-hand figure we plot the different terms of the Hubble equation,  (\ref{f23c}). The subscript $u$ in the various quantities in the figures stands for a derivative with respect to $u$.  }
	\label{fM2}
\end{figure}

If instead we analyze the $\f\to -\infty$ limit, then we obtain
\be
W\simeq C_-~e^{-{(d-2)\over 2(d-1)~\beta}\f}\to \infty\sp f\simeq -{\rm sign}(C_-)f_0|\r_- C_-|^{2\over d-2} e^{\bar \g\f}\ \,, \ \bar \g=-{d\b\over d-2}+{d-2\over 2(d-1)\b}
\ee
\be
e^A\simeq {e^{A_0+{\b\over d-2}\f}\over |\r_- C_-|^{1\over d-2}}\to 0
\ee
\be
T\simeq -{\rm sign}(C_-)f_0~{C_+C_-(\r_+-\r_-)^2\over 2(d-1)(d-2)}|\r_- C_-|^{2\over d-2} ~e^{\bar \delta \f}\to \infty\sp \bar \delta=-{2\b\over d-2}
\ee
The radius of the time circle is
\be
fe^{2A}\simeq e^{\bar \e\f}\sp \bar \e=\e=-\b+{d-2\over 2(d-1)\b}
\ee

As $\bar \delta<0$ always, as $\f\to -\infty$ $e^A\to 0$, always.

The qualitative nature of the generic solution is as follows. First we can always change the sign of W without loss of generality ($u\to -u$). We can always also interchange the role of $\rho_+$ and $\rho_-$ by $\f\to -\f$.

Therefore we have two distinct cases:

\vskip 0.5cm

$\bullet$ $C_+>0, C_->0$. In this case $W\to +\infty$ as $\f\to \pm \infty$ with a single positive minimum in-between where $W'=0$. At this place, $\dot \f=0$.  $f$ vanishes at the minimum of $W$ and therefore this point is a horizon.
However, $T$ changes sign at the horizon and therefore such solutions are not solutions of the second order system.

Despite this, the solution can appear a part of  solution of the $V\not=0$ system, if this solution eventually ends up at $\f\to \pm \infty$. In such only a piece of this solution is relevant, and if this part has $T>0$ then this is acceptable. We shall show such examples when we discuss the solutions that start from a shrinking endpoint in the dS region and end at  $\f\to -\infty$ in appendix \ref{forbid}.

In figures \ref{fM1} and \ref{fM2} we plot characteristic solutions in this class.

\vskip 0.5cm

\begin{figure}[h!]
	\centering
	\subfloat{\includegraphics[scale=0.28]{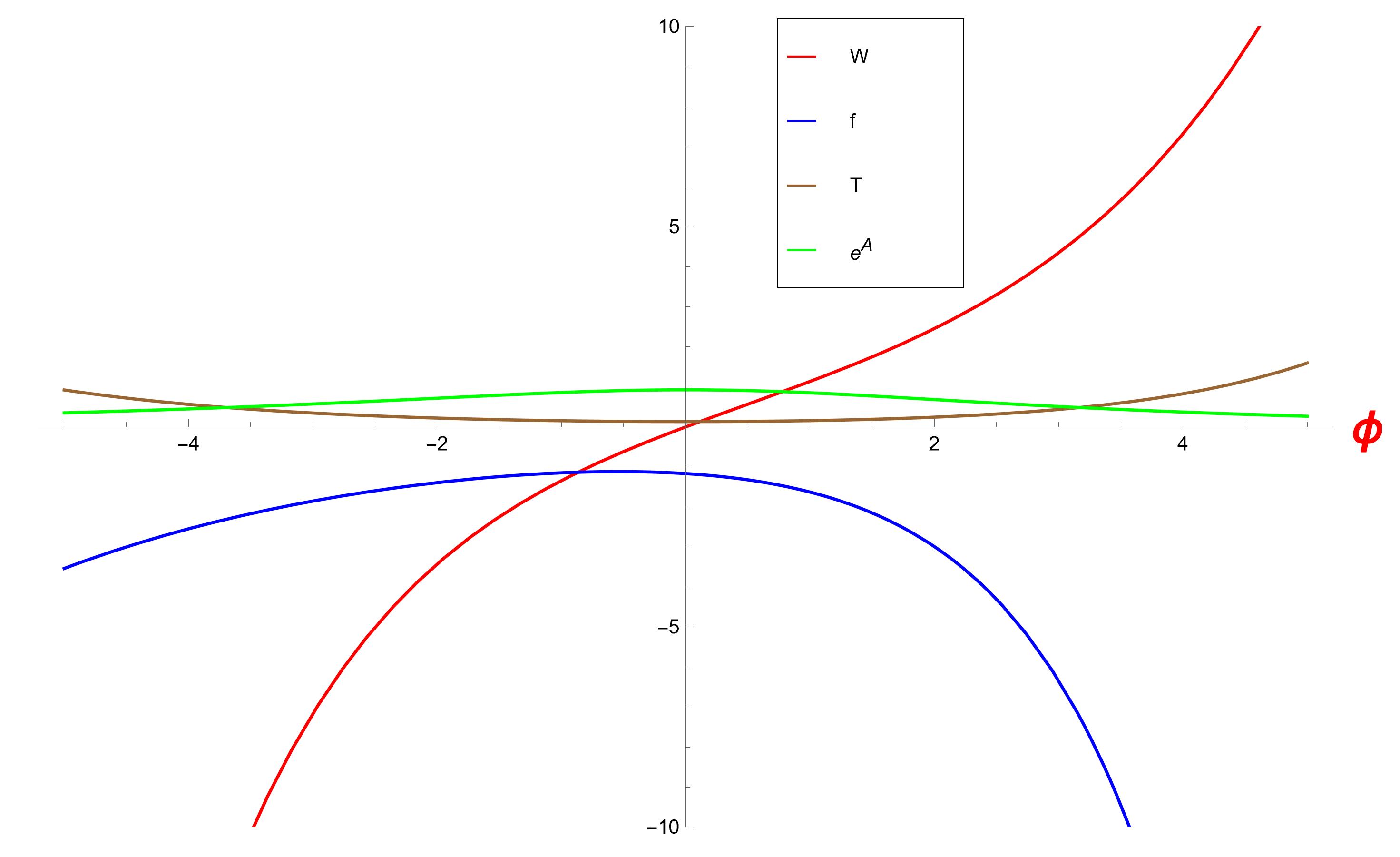}}
	\qquad
	\subfloat{\includegraphics[scale=0.28]{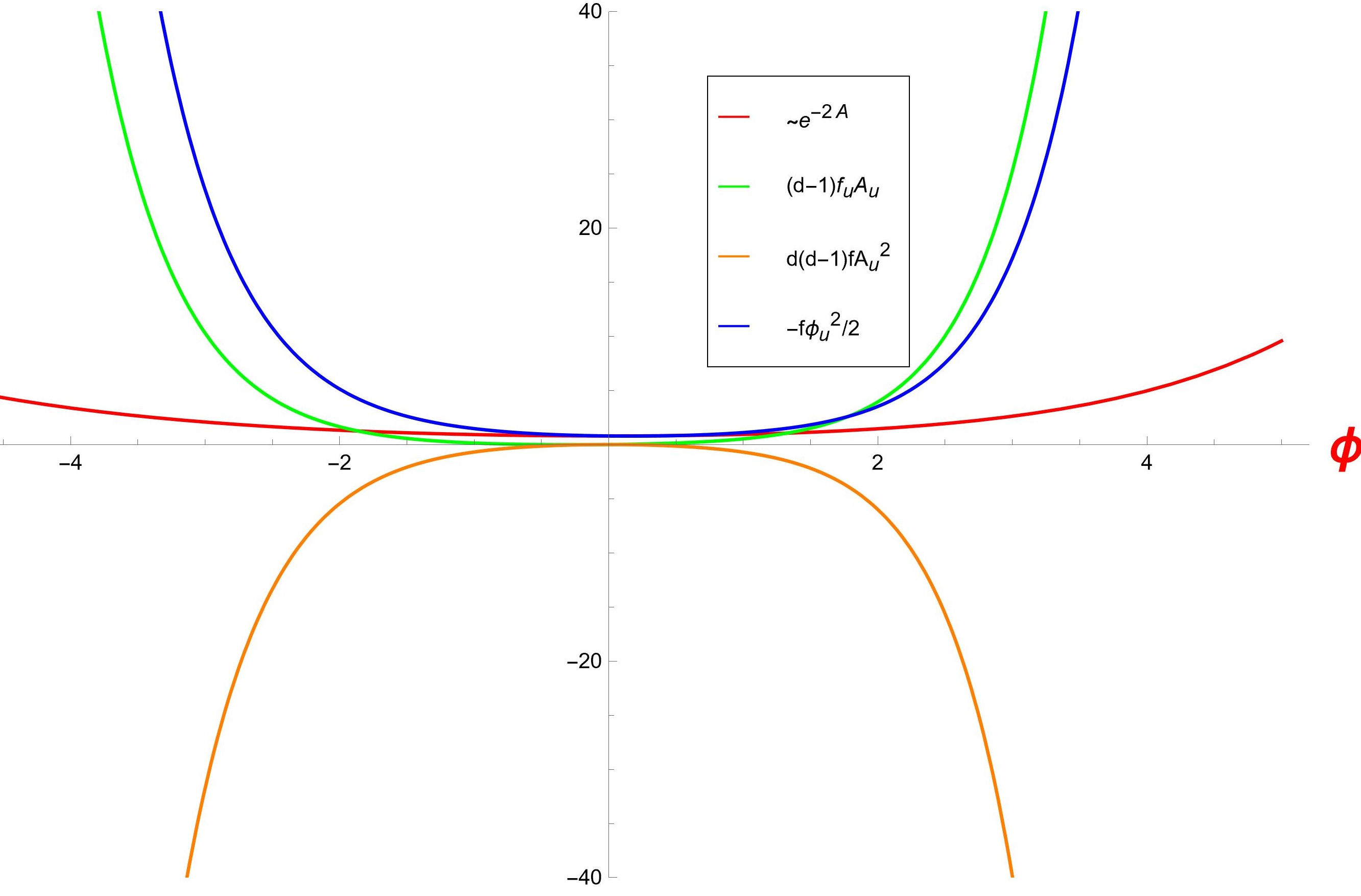}}
	\caption{The solutions with $d=4$, $\beta={1\over 2}$, $C_+=-C_-=1$, $f_0=-1$.  The sign of $f_0$ was chosen so that the curvature function $T>0$. Left: In green the scale factor $e^A$ is plotted. Right: Plot of the size of the four distinct terms in the Hubble equation,  (\ref{f23c}). The subscript $u$ in the various quantities in the figures stands for a derivative with respect to $u$.}
	\label{fM3}
\end{figure}

\begin{figure}[h!]
	\centering
	\subfloat{\includegraphics[scale=0.28]{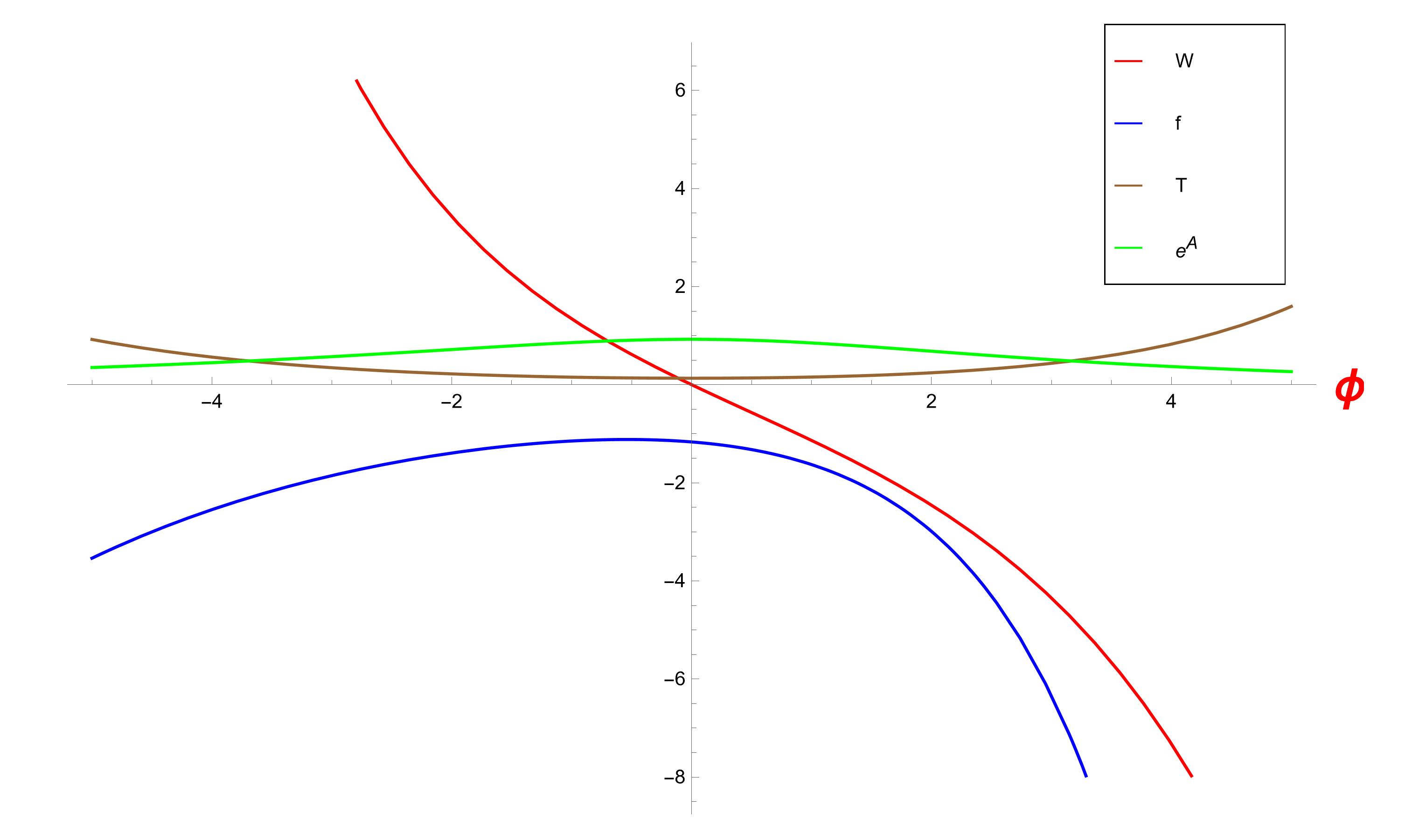}}
	\qquad
	\subfloat{\includegraphics[scale=0.28]{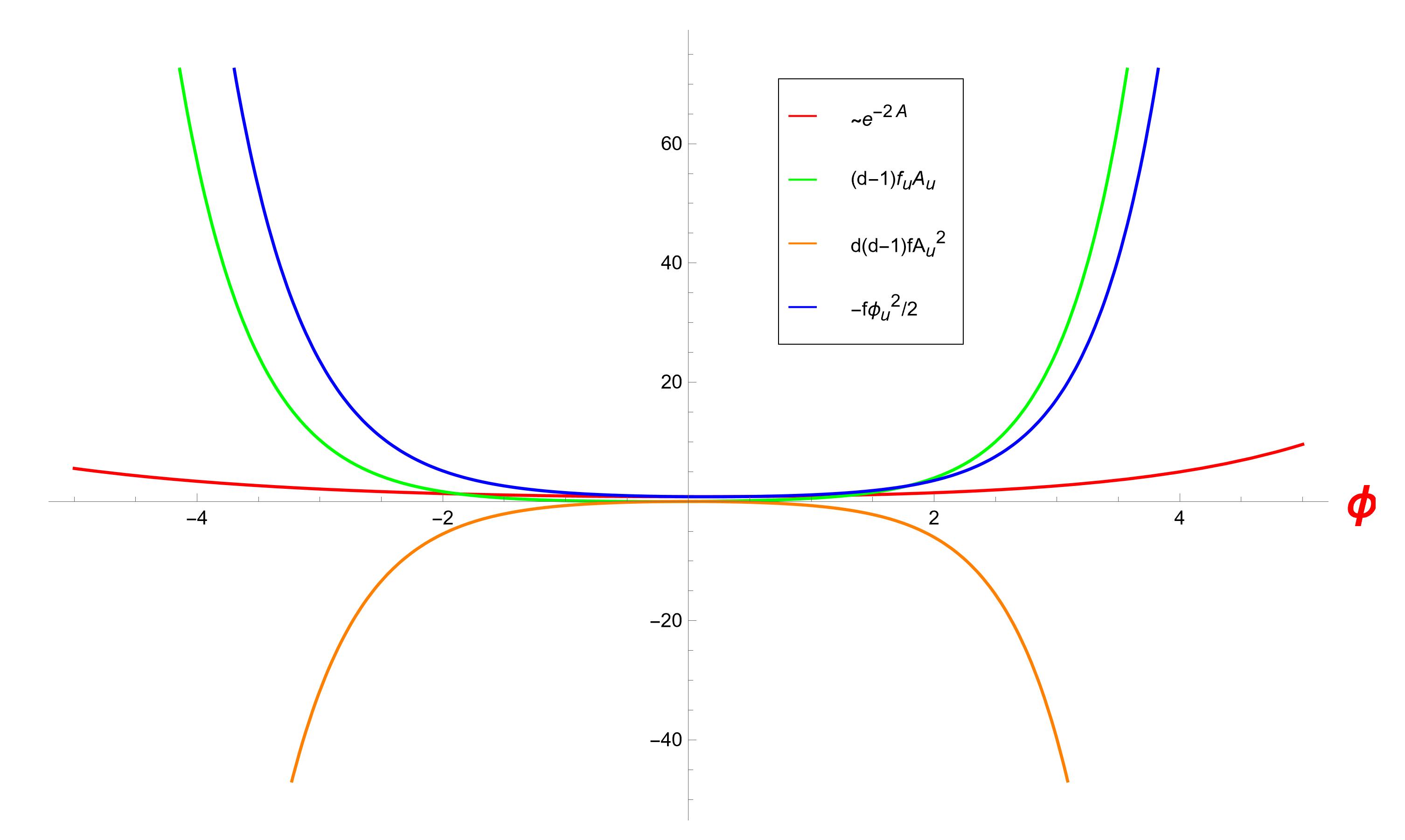}}
	\caption{The solutions with $d=4$, $\beta={1\over 2}$, $C_+=-C_-=-1$, $f_0=1$. In both cases the sign of $f_0$ was chosen so that the curvature function $T>0$. Left: In green the scale factor $e^A$ is plotted. Right: Plot of the size of the four distinct terms in the Hubble equation,  (\ref{f23c}). The subindex $u$ in the various quantities in the figures stands for a derivative with respect to $u$. The subscript $u$ in the various quantities in the figures stands for a derivative with respect to $u$.}
	\label{fM4}
\end{figure}

$\bullet$ $C_+>0, C_-<0$. In this case $W$ increases monotonically from $-\infty$ at $\f\to -\infty$ to $+\infty$ for $\f\to +\infty$. $W'>0$ and is nowhere vanishing. Therefore $f$ has always the same sign that is the same as  the sign of $f_0$. The sign of $T$ in this case is opposite to the sign of $f_0$ and therefore we must have $f_0<0$.
Therefore $f$ is negative over the whole domain. There is an $A$-bounce at the point where the superpotential vanishes.

  Generically speaking, in such solutions $T$ has the same sign if $W'$  always have the same sign over the whole solution. Otherwise, $f$ changes sign and then T changes sign.
  The conditions for $T>0$ in the whole domain $\f\in(-\infty,+\infty)$ are
  $$f_0C_->0~~~~{\rm and}~~~f_0 C_+<0.$$
Solutions in this class are plotted in \ref{fM3} and \ref{fM4}.

  As mentioned above, when this solution is part of solution with a potential, these properties must happen in the part of the solution where the potential becomes negligible.

As for the exponent $\b$, there are four possible regimes that are determined by the three values
\be
\b_1\equiv {d-2\over \sqrt{2d(d-1)}}\sp
\b_{2}\equiv \sqrt{d-2\over 2(d-1)}\sp \b_3\equiv \sqrt{d\over 2(d-1)}
\ee

\begin{itemize}

\item $0<\b<\b_1$. In this case,
\be
(f,fe^{2A})\to (\infty, \infty)\sp \f\to +\infty
\ee
\be
(f,fe^{2A})\to (0, 0)\sp \f\to -\infty
\ee
\begin{figure}[h!]
	\centering
	\subfloat{\includegraphics[scale=0.28]{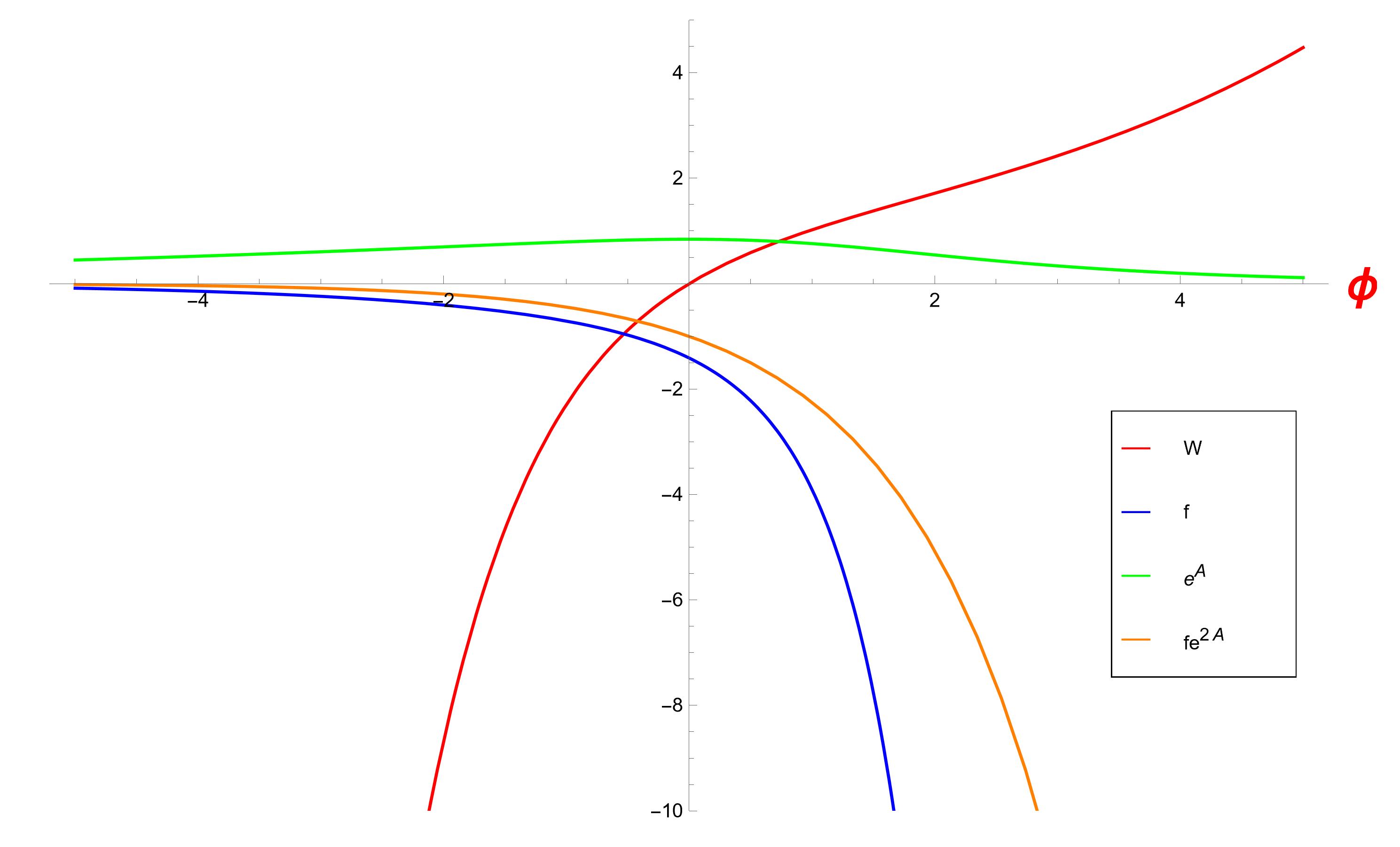}}	
\qquad
\subfloat{\includegraphics[scale=0.28]{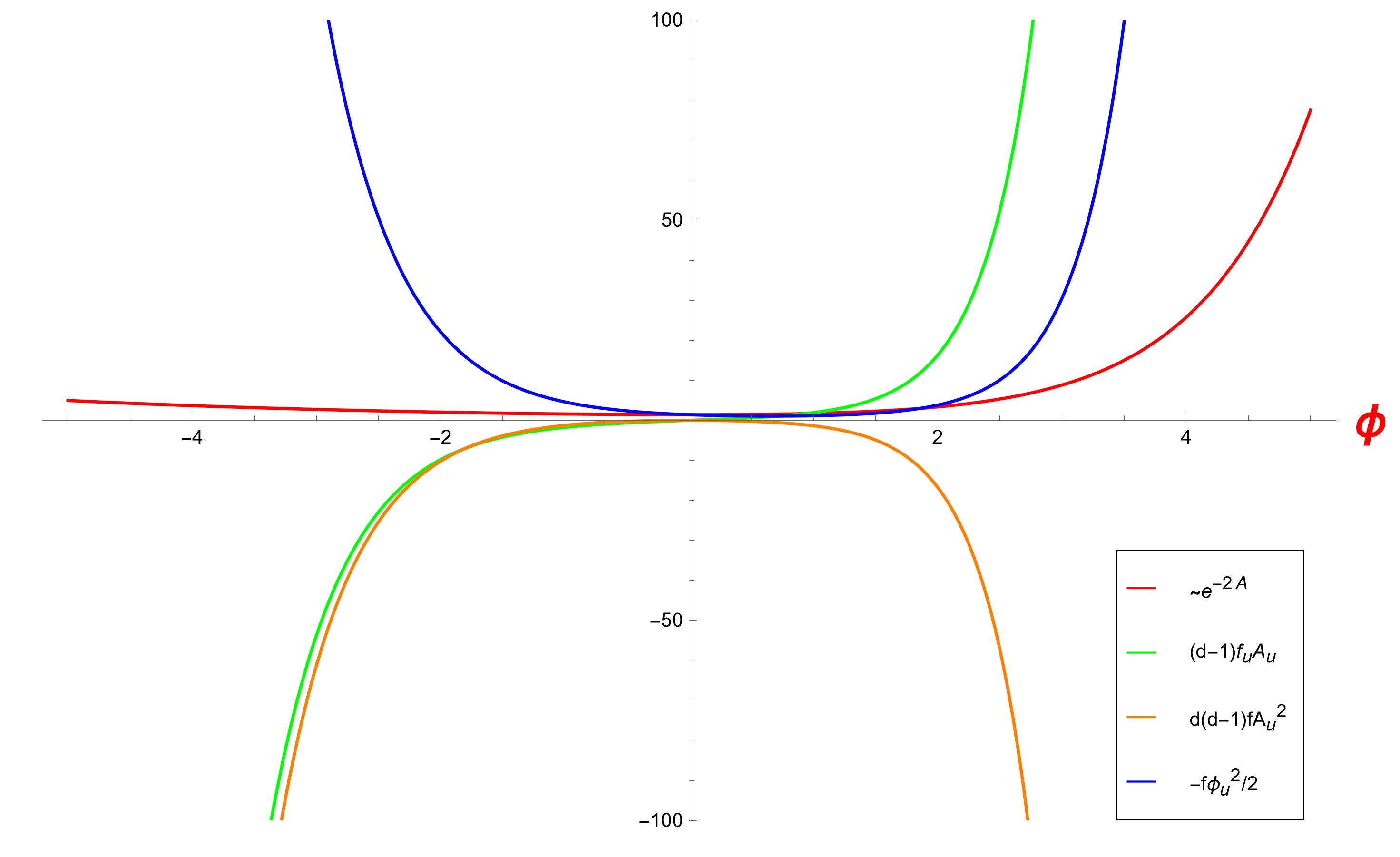}}	
	\caption{The solutions with $d=4$, $\beta=0.3<\b_1$, $C_+=-C_-=1$, $f_0=-1$. At the right the various terms of the Hubble equation,  (\ref{f23c}).}
	\label{fM5}
\end{figure}
An example of a solution in this class is shown in figure \ref{fM5}.

\item $\b_1<\b<\b_2$. In this case,
\be
(f,fe^{2A})\to (\infty, \infty)\sp \f\to +\infty
\ee
\be
(f,fe^{2A})\to (\infty, 0)\sp \f\to -\infty
\ee
An example of a solution in this class is shown in figure \ref{fM6}.

\begin{figure}[h!]
	\centering
	\subfloat{\includegraphics[scale=0.28]{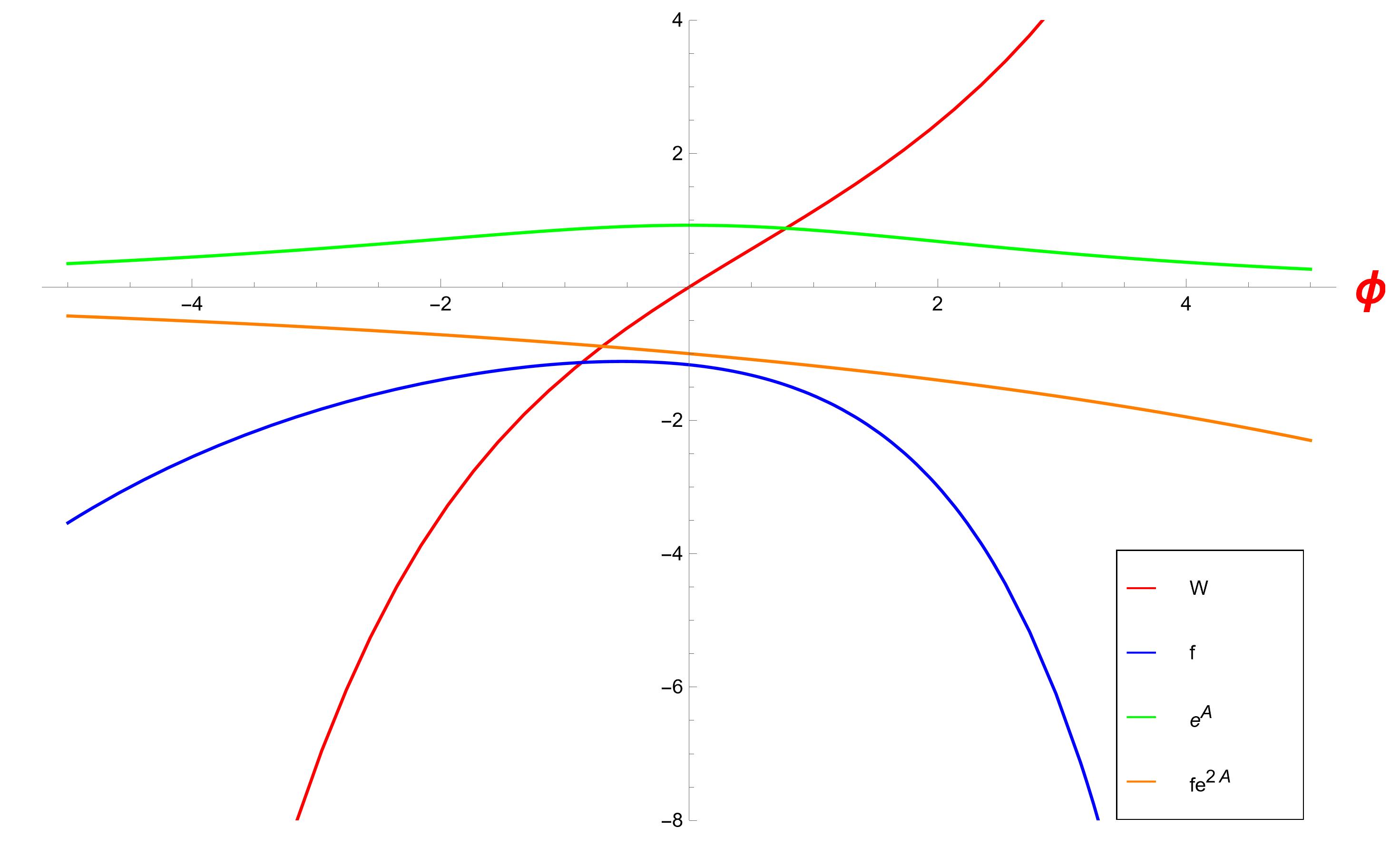}}	
\qquad
\subfloat{\includegraphics[scale=0.28]{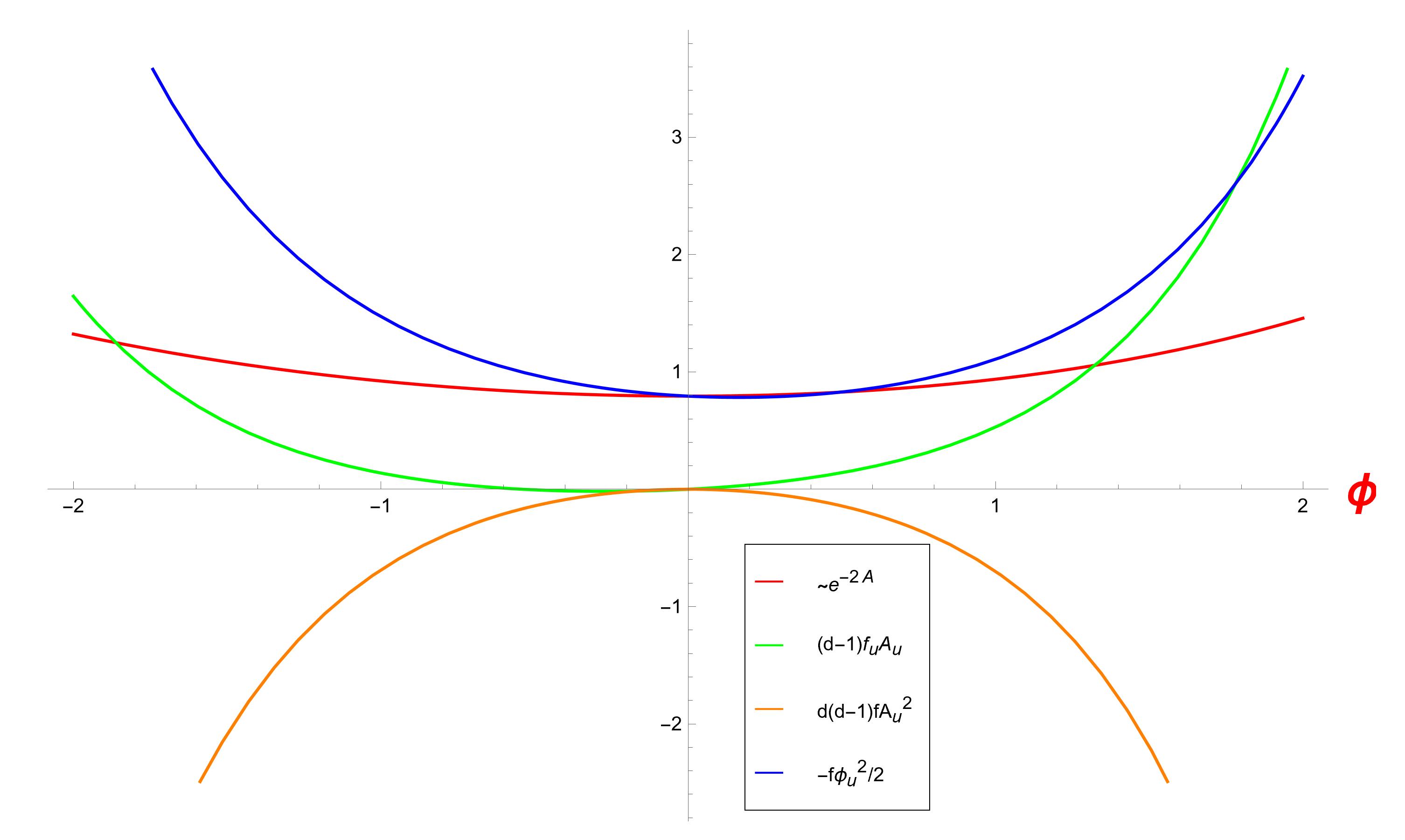}}	
	\caption{The solutions with $d=4$, $\b_1<\beta=0.5<\b_2$, $C_+=-C_-=1$, $f_0=-1$.At right the various terms of the Hubble equation,  (\ref{f23c}). The subscript $u$ in the various quantities in the figures stands for a derivative with respect to $u$.The subscript $u$ in the various quantities in the figures stands for a derivative with respect to $u$.}
	\label{fM6}
\end{figure}

\item $\b_2<\b<\b_3$. In this case,
\be
(f,fe^{2A})\to (\infty, 0)\sp \f\to +\infty
\ee
\be
(f,fe^{2A})\to (\infty, \infty)\sp \f\to -\infty
\ee
An example of a solution in this class is shown in figure \ref{fM7}.
\begin{figure}[h!]
	\centering
	\subfloat{\includegraphics[scale=0.28]{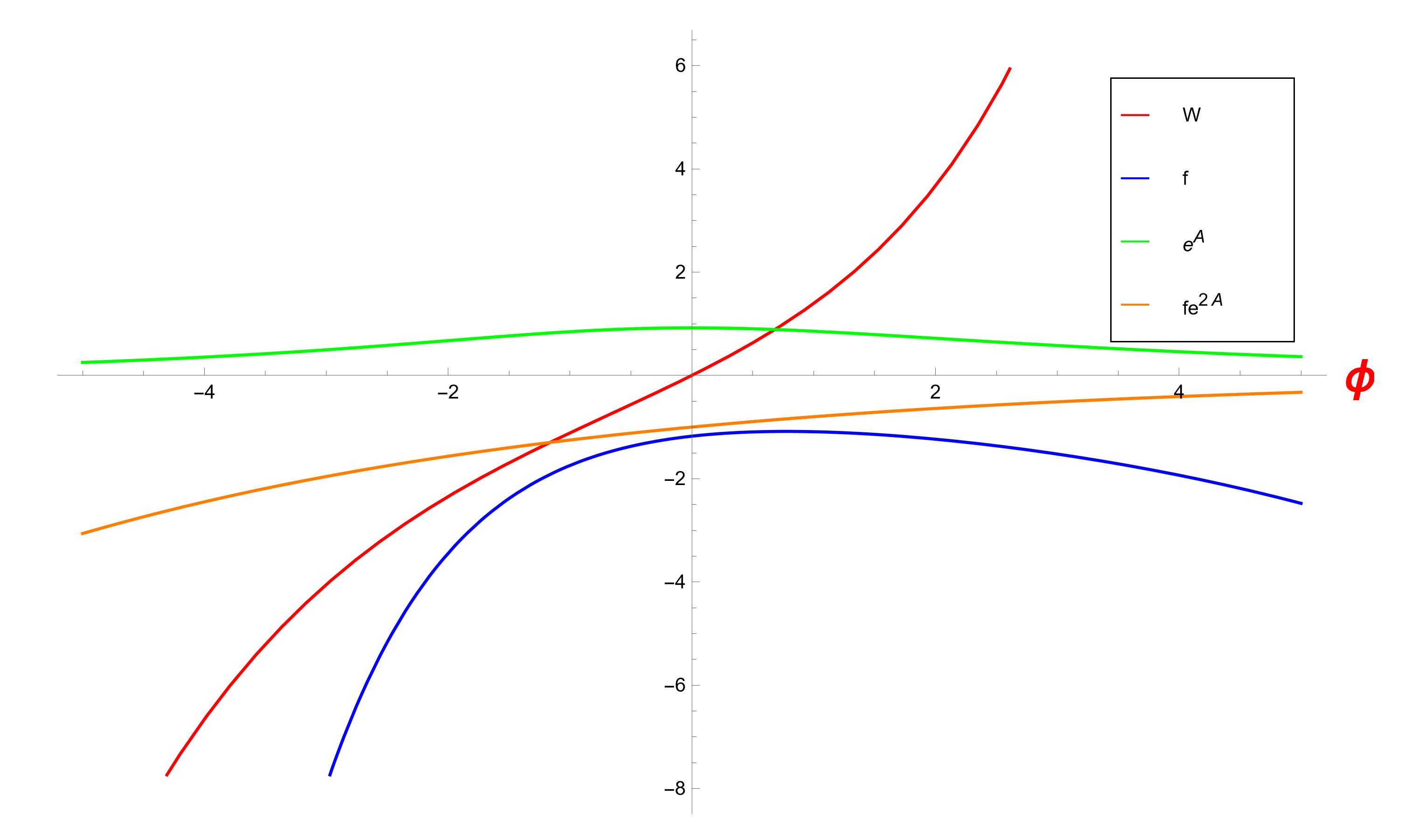}}	
\qquad
\subfloat{\includegraphics[scale=0.28]{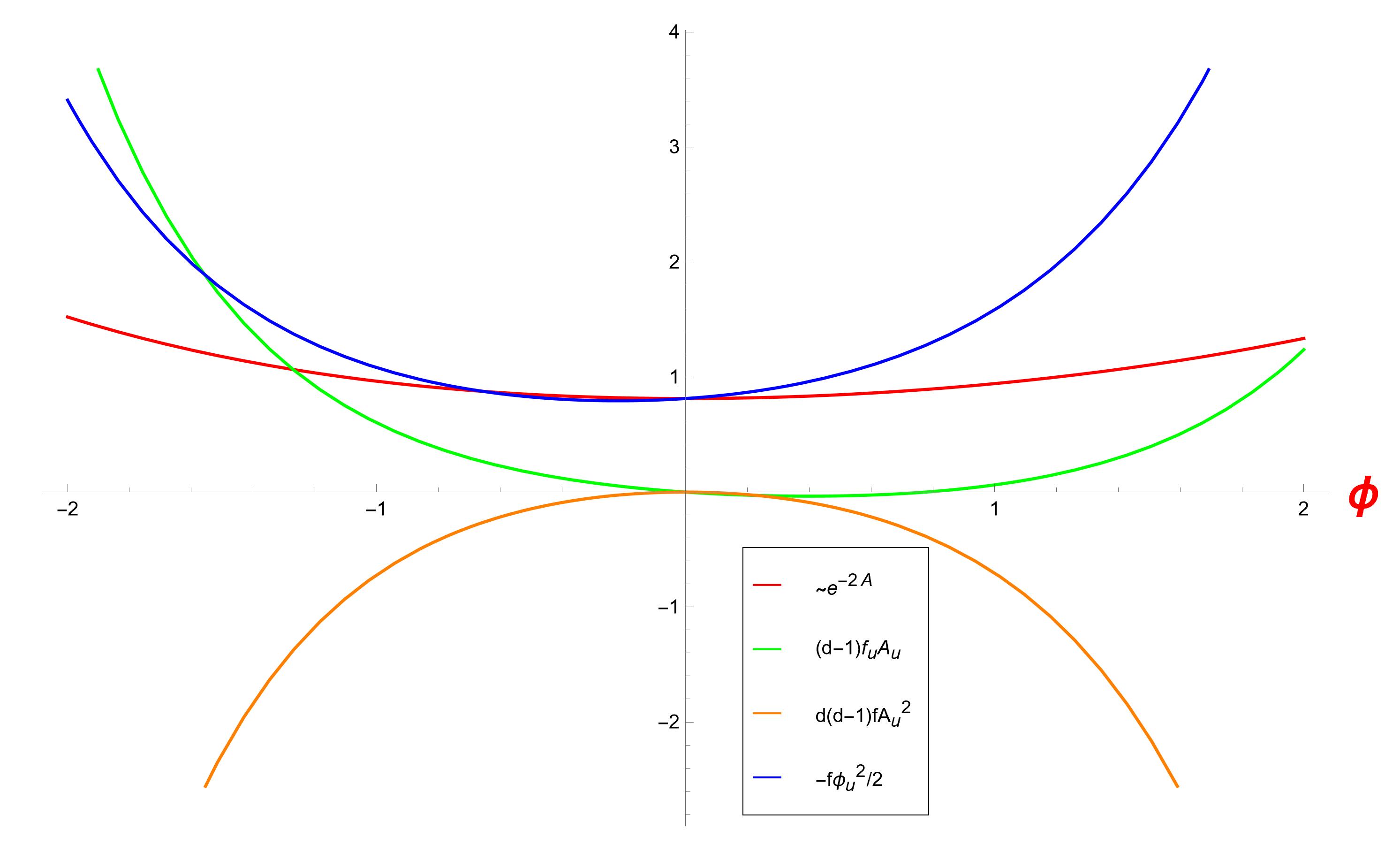}}
	\caption{The solutions with $d=4$, $\b_2<\beta=0.7<\b_3$, $C_+=-C_-=1$, $f_0=-1$.At right, the various terms of the Hubble equation,  (\ref{f23c}). The subscript $u$ in the various quantities in the figures stands for a derivative with respect to $u$.}
	\label{fM7}
\end{figure}

\item $\b>\b_3$. In this case,
\be
(f,fe^{2A})\to (0, 0)\sp \f\to +\infty
\ee
\be
(f,fe^{2A})\to (\infty, \infty)\sp \f\to -\infty
\ee
An example of a solution in this class is shown in figure \ref{fM8}.
\begin{figure}[h!]
	\centering
	\subfloat{\includegraphics[scale=0.28]{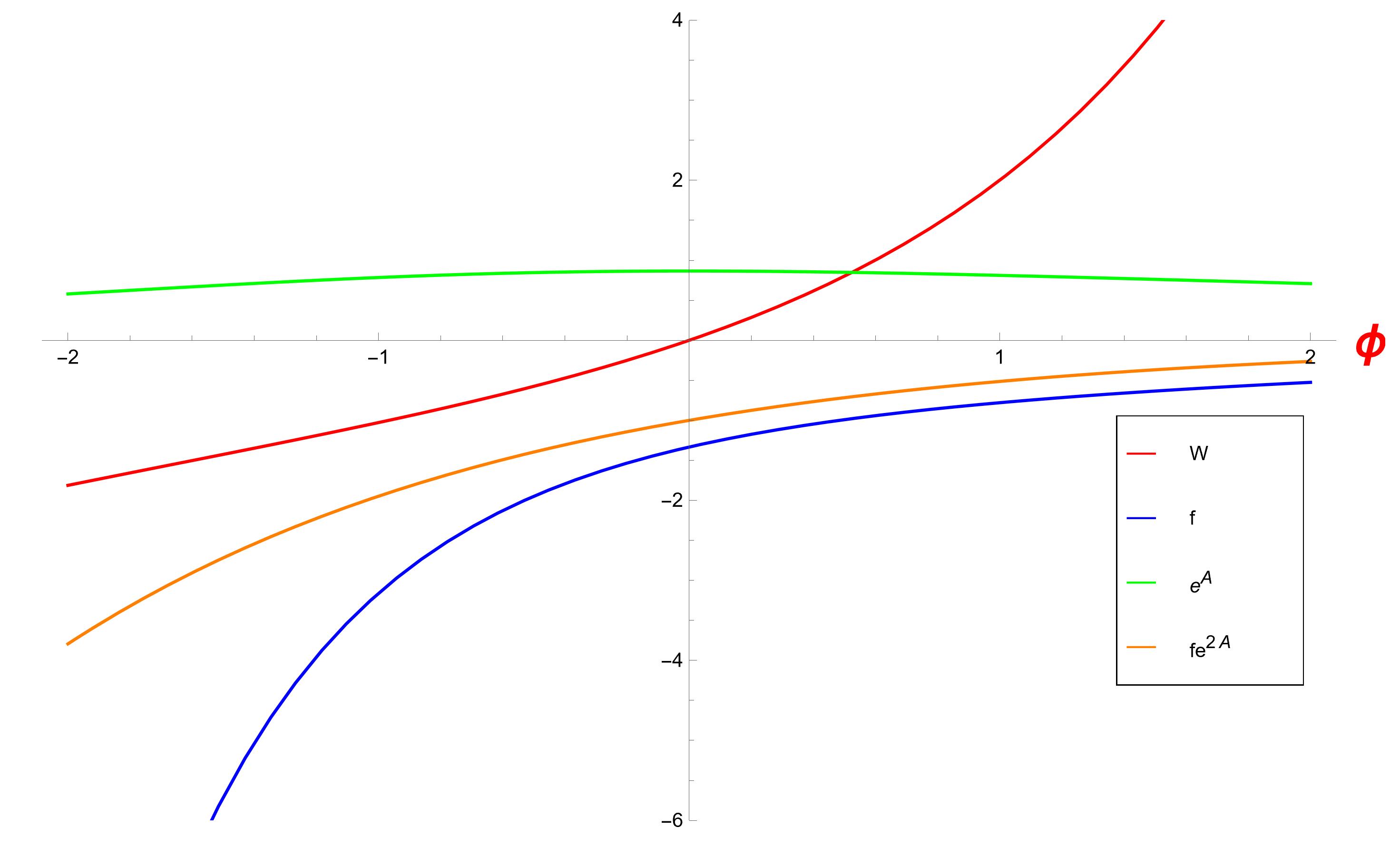}}	
\qquad
\subfloat{\includegraphics[scale=0.28]{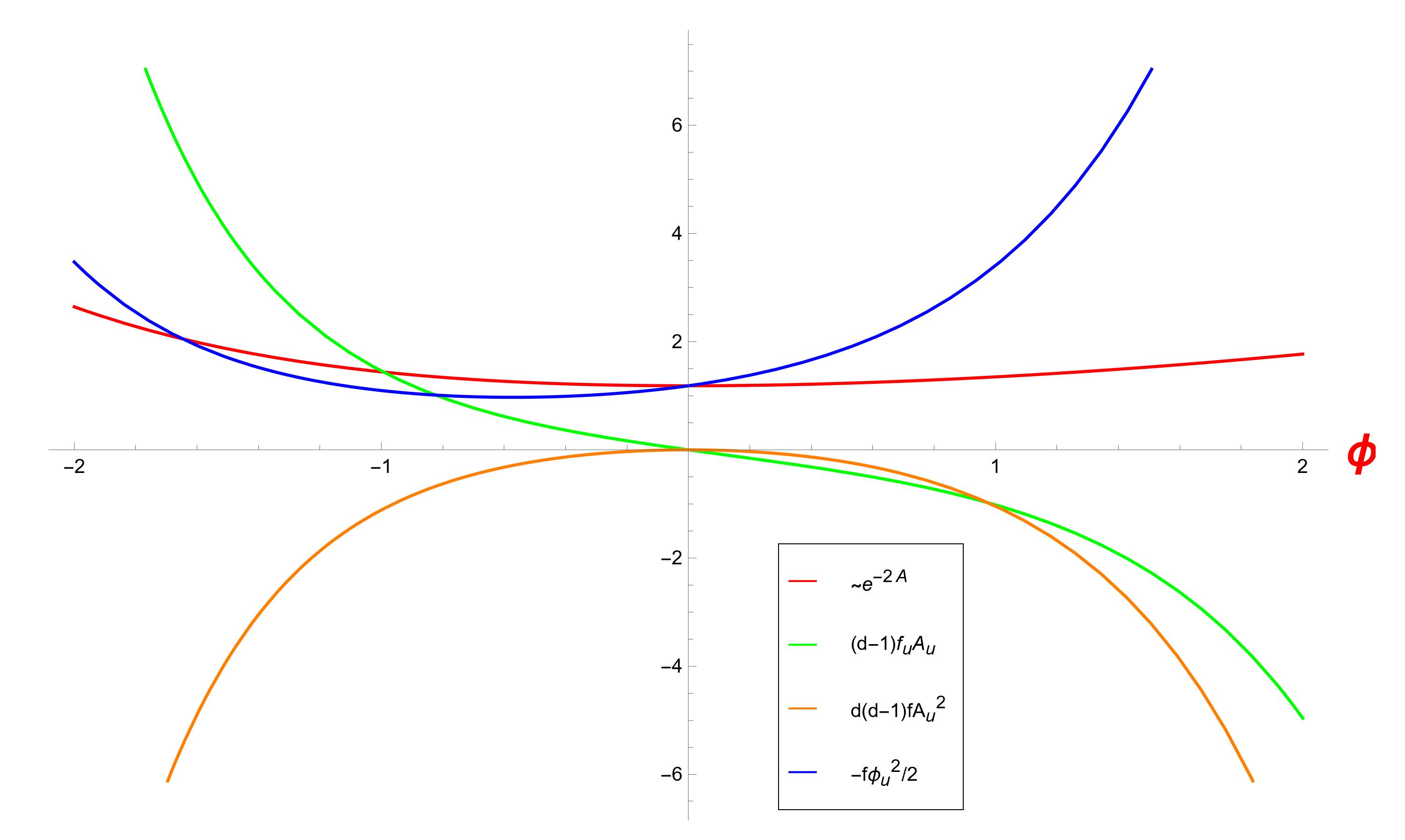}}	
	\caption{The solutions with $d=4$, $\b_3<\beta=1$, $C_+=-C_-=1$, $f_0=-1$. At right, the various terms of the Hubble equation,  (\ref{f23c}).The subscript $u$ in the various quantities in the figures stands for a derivative with respect to $u$.}
	\label{fM8}
\end{figure}

\end{itemize}

\begin{figure}[h!]
	\centering
	\subfloat{\includegraphics[scale=0.55]{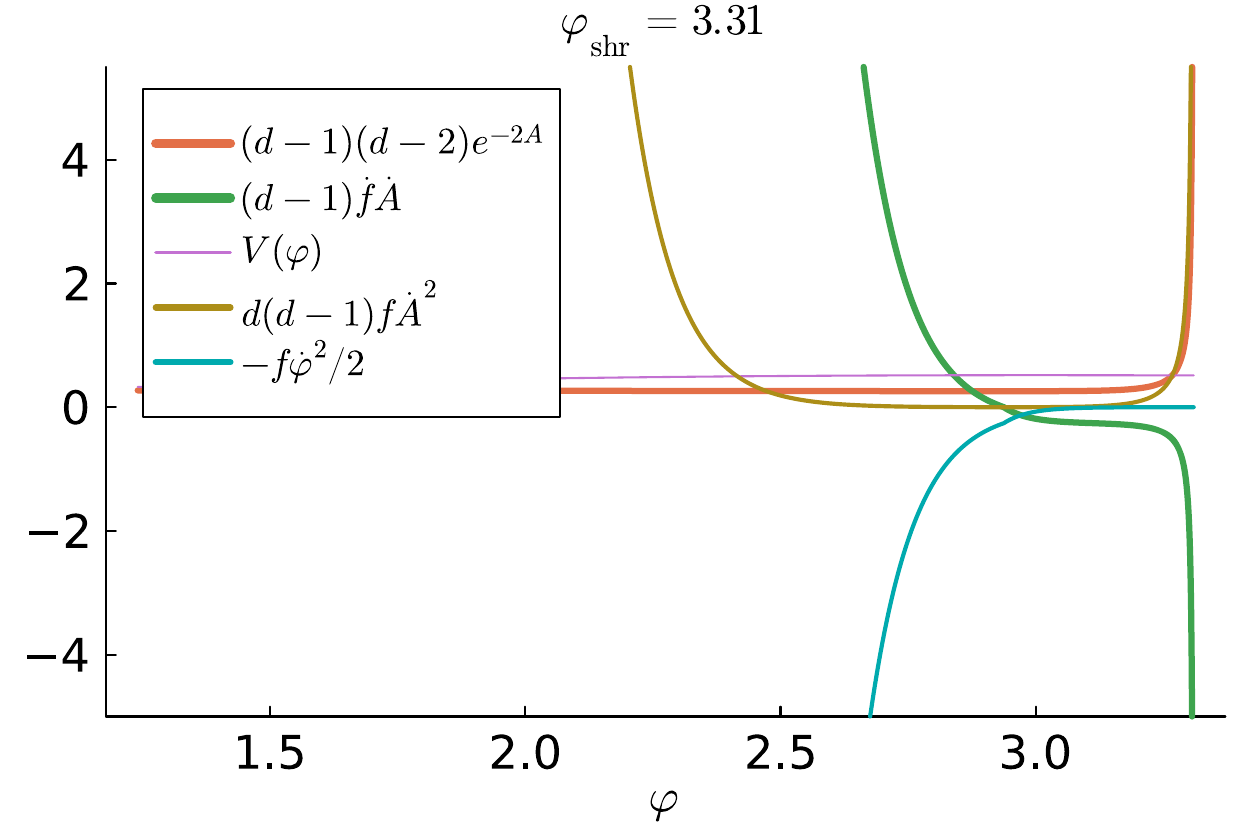}}	
	\caption{The various terms in Hubble equation for a solution where the sphere shrinks at $\f=3.31$ in the dS regime, and runs to a naked singularity.}
	\label{fM9}
\end{figure}

We can compare now our solutions in this appendix with an example of the solutions,shown in Fig. \ref{fM9}, constructed in Appendix \ref{forbid}, where the sphere shrinks to zero size in the dS regime. We observe that this solution matches the one in figure \ref{fM2} in the $\f\to -\infty$ regime.

\section{Constructing flows that end at $\f\to \pm\infty$}\label{app:7}

In this appendix, we discuss in detail how to construct the solutions presented in Sec. \ref{sec:7}, where at least one of the endpoints is of type I or type II. In all cases, we require that the potential behaves as $V\sim e^{\a\f}$ as $\f\to +\infty$, such that the exponent satisfies the Gubser bound:
\begin{equation}
\a<\a_G =\sqrt{\dfrac{2d}{d-1}}\,,
\end{equation}
where $\a_G$ is referred to as the Gubser bound. Whether or not a given solution can be acceptable \`a la Gubser is addressed in Appendix \ref{app:l1}. Depending on the sign of $\alpha$, the potential diverges or vanishes as we approach the $\f\to\pm\infty$  endpoints. Moreover, the potential can be either in the dS or AdS regimes. We work with $d=4$ space dimensions in this appendix.

\subsection{From $d+1$-dimensional boundaries to $V(\infty)=\pm\infty$}\label{app:71}

In Appendix \ref{app:I} we constructed a family of exact solutions to the equations of motion. These naturally contain examples of solutions where the potential diverges as $\f\to\infty$ with the type I or type II asymptotic structure. As a concrete example, we study the solution given in Eqs. \eqref{k15}, \eqref{k16} and \eqref{k17}:

\begin{equation}\label{a715}
W = \cosh\left(\dfrac{\varphi}{\sqrt{3}}\right)\,, \qquad T = -C_t \sinh\left(\dfrac{\varphi}{\sqrt{3}}\right)\,,
\end{equation}
\begin{equation}\label{a716}
f= f_0 + 36 C_t e^{-\f/\sqrt{3}}-\frac{1}{2}f_1\cosh\left(\dfrac{\varphi}{\sqrt{3}}\right)\,,
\end{equation}
\begin{equation}\label{a717}
V=-\dfrac{1}{4}f_0+\left(\frac{1}{6}f_1 -12C_t\right)\cosh\left(\dfrac{\varphi}{\sqrt{3}}\right)-\dfrac{1}{12}f_0\cosh\left(\dfrac{2\varphi}{\sqrt{3}}\right)\,,
\end{equation}
where $f_0$, $f_1$ and $C_t$ are integration constants, and the spherical slicing requires $C_t<0$ for solutions with $\f>0$.
The superpotential \eqref{a715} has a single extremum at $\f=0$, corresponding to a five dimensional boundary (dS$_5$, AdS$_5$ or M$_5$). Therefore, the flows contained in \eqref{a715} can only be from $\f=0$ to $\f\to\infty$.

If $f_1\neq 0$ in Eq. \eqref{a716}, the blackening function $f$ diverges as dictated by the irregular solutions, see Eq. \eqref{z9}. Alternatively, for $f_1=0$, the function $f$ approaches to a constant value at $\f\to\infty$, in agreement with the type I Gubser-regular asymptotics of Eq. \eqref{ms18}. Additionally, we set $C_t=-1/12$ without loss of generality\footnote{Note that $C_t<0$ in order to have a spherical slicing, and the equations of motion are invariant under $(f,T,V)\to \lambda (f,T,V)$ for some constant $\lambda$.}, and we redefine
\begin{equation}
f_0\equiv-3(V(0)-1)
\end{equation}
where $V(0)$ is the value of the potential at $\f=0$.

 Note that for $V(0)=1$, then $f_0=0$ and the function $f$ vanishes exponentially as dictated by the Gubser-regular type II asymptotics \eqref{ms19}. In summary, we consider Eqs. \eqref{a715}, \eqref{a716} and \eqref{a717} with $(f_0,f_1,C_t) = (3-3V(0),0,-1/12)$. With this choice of integration constants, the potential $V$ diverges at $\f\to \infty$ as
\begin{equation}
V = \dfrac{1}{8}(V(0) - 1) e^{2\varphi/\sqrt{3}} + O(e^{\varphi/\sqrt{3}})\,.
\label{acoonf}\end{equation}
The solutions are characterized by $V(0)$. The different possibilities are discussed in Sec. \ref{sec:71} and shown in Fig. \ref{fig:gubx}.

\subsection{From dS$_2$ boundaries to $V(\infty)=\pm\infty$}\label{app:74}

In this section, we construct solutions that interpolate between dS$_2$ boundaries and Gubser-regular endpoints, where the potential is necessarily divergent. We engineer such a solution by choosing some superpotential with the appropriate behaviour. At a dS$_2$ boundary, the superpotential vanishes as dictated by \eqref{wds22}, while at the boundary of field space $\f\to+\infty$ we assume that it diverges exponentially. These conditions are satisfied by the following superpotential:
\begin{equation}\label{awgds2}
W = c_1 \left[\cosh\left(\frac{1}{2}\b \f\right)-\cosh\left(\frac{1}{2}\b_2\f\right)\right]^2\,.
\end{equation}
We set $c_1=1$ without loss of generality by virtue of the scaling symmetry \eqref{scaling}. Additionally, we assume that $\b_2<\b$ without loss of generality. The superpotential \eqref{awgds2} vanishes at $\f=0$ as $\f^4$. This corresponds to the asymptotic solution \eqref{wds22} with $\delta_{\pm}=\frac{1}{2}$, and we can identify the point $\f=0$ with a dS$_2$ boundary. The superpotential \eqref{awgds2} has only one regular extremum, at $\f=0$, and, according to rule 1 on page \pageref{ru1}, the flow connects $\f=0$ with $\f\to +\infty$ or with $\f\to -\infty$. We shall restrict ourselves to a flow from a dS$_2$ boundary to a Gubser-regular endpoint at $\f\to+\infty$. An equivalent construction can be made demanding that the Gubser-regular endpoint is at $\f\to-\infty$.

By construction, the superpotential \eqref{awgds2} diverges as
\begin{equation}
W = e^{\b \f}+\dots\sp \f\to +\infty
\end{equation}
In order to set the value of $\b$, we assume (and later verify) that the asymptotic solution as $\f\to\infty$ is of type I. This is convenient because, according to the properties of the regular endpoints given in tables \ref{tab_IS} and \ref{tab_IIS}, the type I asymptotic solutions can have a positive or negative potential at the regular endpoint independently of the value of $\beta$, while type II solutions with a positive potential are mutually exclusive with type II solutions with a negative potential. Assuming type I asymptotics, \eqref{ms18} implies that the potential diverges as
\begin{equation}
V \sim e^{2\b \f}+\dots
\end{equation}
Additionally, Gubser-regular solutions with a spherical slicing restrict the exponent with which the potential diverges to lie between the confinement and Gubser bounds:
$$\a_C<2\b<\a_G\,.$$
As a particular example, we study the superpotential \eqref{awgds2} with
\begin{equation}
\b = \dfrac{2}{\sqrt{15}}\,,\qquad \b_2 = \dfrac{1}{\sqrt{15}}\,.
\end{equation}
We compute the inverse scale factor $T$ by solving the first relation of Eq. \eqref{eqtt}:
\begin{equation}\label{o11}
T = C_T \cosh ^4\left(\frac{\varphi }{4 \sqrt{15}}\right) \left(4 \cosh \left(\frac{\varphi
   }{2 \sqrt{15}}\right)-1\right)^3\,,
\end{equation}
where $C_T$ is an integration constant. We require $C_T>0$ for consistency with the spherically sliced ansatz. Note that $C_T$ can be scaled away by means of the symmetry $(f,V,T)\to\lambda(f,V,T)$ of the equations of motion. Momentarily, we keep $C_T$ generic. We can integrate once the equation \eqref{f10_1} to obtain the first derivative of the blackening function:
$$
f' = \frac{f_1 \cosh ^4\left(\frac{\varphi }{4 \sqrt{15}}\right) \left(1-4 \cosh
   \left(\frac{\varphi }{2 \sqrt{15}}\right)\right)^5 \coth ^3\left(\frac{\varphi }{4
   \sqrt{15}}\right)}{2 \cosh \left(\frac{\varphi }{2 \sqrt{15}}\right)+1}+\frac{C_T \left(1-4 \cosh \left(\frac{\varphi }{2
   \sqrt{15}}\right)\right)^5}{40500 \sqrt{15} \left(2 \cosh
   \left(\frac{\varphi }{2 \sqrt{15}}\right)+1\right)}\times
$$
$$
   \cosh ^4\left(\frac{\varphi }{4 \sqrt{15}}\right)  \coth ^3\left(\frac{\varphi }{4 \sqrt{15}}\right) \left\{-18750
   \text{csch}^2\left(\frac{\varphi }{4 \sqrt{15}}\right)+3645 \text{sech}^4\left(\frac{\varphi
   }{4 \sqrt{15}}\right)+\right.
$$
$$
  + \left.90396 \text{sech}^2\left(\frac{\varphi }{4 \sqrt{15}}\right)-4
   \left[\frac{724992}{4 \cosh \left(\frac{\varphi }{2
   \sqrt{15}}\right)-1}-\frac{368640}{\left(1-4 \cosh \left(\frac{\varphi }{2
   \sqrt{15}}\right)\right)^2}+\right. \right.
   $$
   $$
   \left. \left. +\frac{921600}{\left(4 \cosh \left(\frac{\varphi }{2
   \sqrt{15}}\right)-1\right)^3} +140625 \log \left(\sinh \left(\frac{\varphi }{4
   \sqrt{15}}\right)\right)+331695 \log \left(\cosh \left(\frac{\varphi }{4
   \sqrt{15}}\right)\right)\right. \right. $$
\begin{equation}\label{fpg}
   \left. \left. -400000 \log \left(2 \cosh \left(\frac{\varphi }{2
    \sqrt{15}}\right)+1\right) +163840 \log \left(4 \cosh \left(\frac{\varphi }{2
   \sqrt{15}}\right)-1\right)\right]\right\}\,,
\end{equation}
where the integration constant has been denoted as $f_1$. As we approach the dS$_2$ boundary, $f'$ diverges as
\begin{equation}\label{fpct}
f' = \dfrac{8640000}{\f^5}C_T + O(\f^4)\,.
\end{equation}
This is consistent with the perturbative expansion around a dS$_2$ boundary given in Eq. \eqref{fds2} with our chosen values of $\delta_{\pm}=1/2$. Conversely, as we approach the boundary of field space $\f\to\infty$, the function \eqref{fpg} diverges as
\begin{equation}
f'  = -\frac{2 e^{\sqrt{\frac{3}{5}} \varphi } \left(30375 f_1 + 61696 \sqrt{15} C_T
   \log 2\right)}{30375}+ \dots
\end{equation}
This corresponds to the irregular (type 0) asymptotic solution, where the function $f$ diverges according to Eq. \eqref{z9}, i.e. as $e^{\g \f}$ with $\g=2/(3\b)-\b = \sqrt{3/5}$. In order to obtain the Gubser-regular asymptotic solution, we choose $f_1$ such that the leading contribution to $f'$ vanishes, that is
\begin{equation}
f_1 = -\frac{61696 C_T \log 2}{2025 \sqrt{15}}\,.
\end{equation}
With this choice, now $f'$ vanishes as $\f\to\infty$:
\begin{equation}\label{o16}
f'=80 \sqrt{\frac{5}{3}} C_T e^{-\frac{1}{2} \sqrt{\frac{3}{5}} \varphi }+ \dots
\end{equation}
and, as a consequence, the function $f$ approaches a constant value, that we denote $f(\infty)$, in agreement now with the type I asymptotic solution. We shall integrate the differential equation \eqref{fpg} for different choices of $f(\infty)$. Once $f$ is known, we reconstruct the potential $V$ by algebraically solving Eq. \eqref{w55}. Upon substitution of $W$ and $T$, Eq. \eqref{w55} becomes
$$
V = 6 C_T \cosh ^4\left(\frac{\varphi }{4 \sqrt{15}}\right) \left(4 \cosh \left(\frac{\varphi
   }{2 \sqrt{15}}\right)-1\right)^3+\frac{4}{15}  \left(2 \cosh \left(\frac{\varphi }{2 \sqrt{15}}\right)+1\right)^2 \times
   $$
   $$
   \sinh ^6\left(\frac{\varphi }{4
   \sqrt{15}}\right)\left\{\sqrt{15} \left[2 \sinh \left(\frac{1}{2} \sqrt{\frac{3}{5}} \varphi \right)+\sinh
   \left(\frac{\varphi }{2 \sqrt{15}}\right)+\sinh \left(\frac{\varphi
   }{\sqrt{15}}\right)\right] f' \right. $$
   \begin{equation}
   \left. +\left[-6 \cosh \left(\frac{1}{2}
   \sqrt{\frac{3}{5}} \varphi \right)+5 \cosh \left(\frac{\varphi }{2 \sqrt{15}}\right)+4 \cosh
   \left(\frac{\varphi }{\sqrt{15}}\right)+15\right]f\right\}\,.
\end{equation}
From the previous expression, we can read off the behaviour of the potential $V$ as we approach the dS$_2$ boundary ($\f=0$) and as we approach the type I endpoint ($\f\to+\infty$):
\begin{equation}\label{o18}
V(\f\to 0) = 162 C_T \left(1 + \dfrac{1}{24}\f^2 + O(\f^4)\right)\,, \quad V(\f\to\infty) = -\frac{f(\infty)}{80}e^{\frac{4}{\sqrt{15}}\f} + O\left(e^{\frac{\sqrt{5}}{2\sqrt{3}}}\right)\,,
\end{equation}
again in agreement with the dS$_2$ boundary asymptotic solution of Sec. \eqref{g53} and with the type I asymptotic solution of Eq. \eqref{ms18}. Note that in the particular case where $f(\infty) =0$, the asymptotic behaviour becomes that of the type II solutions. Additionally, we observe that the sign of the potential as $\f\to\infty$ is anti-correlated with the sign of $f(\infty)$, while at the dS$_2$ boundary the potential is always positive. Different values of $f(\infty)$ can give rise to qualitatively different solutions. These are discussed in Sec. \ref{sec:74} and shown in Fig. \ref{fig:gubxs}.

\subsection{From shrinking endpoints to  $V(\infty)=0$}\label{app:72}

We intend to construct solutions where the potential vanishes exponentially $V\to 0^{\pm}$ as we approach the boundary in field space $\f\to \infty$, and such that they admit the regular asymptotic structure of Appendix \ref{asymp}. In this case, Eqs. \eqref{ms18} and \eqref{ms19} imply that both the potential $V$ and the superpotential $W$ vanish exponentially as $\f\to\infty$.

We assume that the same flow has another endpoint at a finite $\f$. According to rule 1 on page \pageref{ru1} of Sec. \ref{sec:global}, the endpoint should be an extremum of the superpotential. Additionally, the fact that $W(\f\to\infty)\to 0$ implies that the extremum is a maximum (minimum) if the superpotential is positive (negative). It follows from rule 2 on page \pageref{ru2} that these are shrinking end-points. This is the context of rule 17 on page \pageref{ru17} in section \ref{rul}.

Consider the following superpotential:
\begin{equation}\label{asugu}
W = c_1 e^{\beta \f} + c_2 e^{\beta_2 \f}\,.
\end{equation}
We assume that $\beta_2<\beta$ without loss of generality. Then, the fact that $W$ vanishes exponentially at $\f\to +\infty$ translates to $\beta<0$. We can set $c_1=1$ using the scaling symmetry \eqref{scaling}. Additionally, we assume that $W$ has an extremum at some finite value of $\f$, which is a shrinking endpoint and which can be set to zero by a shift in $\f$ . According to Eq. \eqref{eqdiv}, the presence of a shrinking endpoint further requires that $W''/W = 1/(d-1)$ at the extremum. For concreteness we work in $d=4$ dimensions. The previous conditions are satisfied for
\begin{equation}
c_1 = 1 \sp c_2 = - 3\beta^2 \sp \beta_2 = \dfrac{1}{3\beta}\sp -\dfrac{1}{\sqrt{3}}<\beta<0\,.
\end{equation}
Similarly to the construction in the previous appendix, \ref{app:74}, we choose $\b$ such that there exists a type I endpoint at the boundary of field space. As a concrete example, we choose $\beta=-1/3$.
We solve analytically Eqs. \eqref{eqtt} and \eqref{f10_1} to find the functional form of the inverse scale factor $T$ and the blackening function $f$:
\begin{equation}
T= \frac{C_T e^{-\f}}{1-e^{-2
   \varphi/3}} \,,\quad f = f_0+\frac{3}{8} f_1 \coth ^{-1}\left(e^{\varphi /3}\right) + \frac{1}{16\sinh  \frac{\varphi }{3}} \ \left(648 C_T+f_1 \coth
   \frac{\varphi }{3}-4 f_1\right)\,,
\end{equation}
where $f_0$, $f_1$ and $C_T$ are three integration constants. We require $C_T>0$ for consistency with the spherically sliced ansatz. Around the maximum of the superpotential, the two functions behave as
\begin{equation}
T = \dfrac{3 C_T}{2\f}+O(\f^0)\sp f = \dfrac{9 f_1}{16\f^2} + \dfrac{3(162 C_T - f_1)}{4\f} + O(\f^0)\,.,
\end{equation}
while the asymptotic solution around a shrinking endpoint \eqref{soch} and \eqref{sochu} require that the diverge as $\f^{-1}$. Therefore, the maximum of the superpotential is a shrinking endpoint only for $f_1=0$. Then, $f$ simplifies to
\begin{equation}\label{afgub}
f = f_0 + \frac{81 C_T}{2\sinh  \frac{\varphi }{3}}\,.
\end{equation}
We reconstruct the potential $V$ from Eq. \eqref{w55}:
\begin{equation}
V = -\dfrac{f_0e^{-2\f/3}}{54}\left(15 - 6 e^{-2\f/3}-e^{-4\f/3}\right)-12 C_T e^{-\f}\,.
\end{equation}
As we approach the boundary of field space $\f\to\infty$, the functions behave as
\begin{equation}\label{avgub}
V = -\frac{f_0}{54}e^{-2\f/3}+\dots \sp W = e^{-\f/3}+\dots\sp f= f_0 + \dots \sp T=C_T e^{-\f}+\dots
\end{equation}
If $f_0\neq 0$, the previous asymptotic behaviour match the type I asymptotic solution of Appendix \ref{asymp} that are Gubser-regular \eqref{ms18}\,, while for $f_0 =0$ it corresponds to the type II asymptotics. Therefore, such solutions connect a shrinking endpoint at $\f=0$ with an endpoint at $\f\to \infty$, with a regular asymptotic structure, where the potential vanishes. We parametrise $f_0$ in terms of the value of the potential at the shrinking endpoint $V(0)$:
\begin{equation}
f_0 = -27/4(12 C_T + V(0))
\end{equation}
 Note that the equations of motion are invariant under $(f,T,V)\to \lambda (f,T,V)$ for some constant $\lambda$. We exploit the previous symmetry to set $12 C_T = 1$. This leaves a single free parameter for these solutions: $V(0)$. The solutions for different choices of $V(0)$ are discussed in Sec. \ref{sec:72} and shown in Fig. \ref{fig:gubxv}.

\subsection{From $V(\infty)\to \pm \infty$ to $V(\infty)\to 0^{\pm}$}\label{app:73}

We proceed now to construct flow solutions that run between two Gubser-regular endpoints as $|\f|\to \infty$. From rules 17 on page \pageref{ru17} and 18 on page \pageref{ru18}, it is necessary that the potential vanishes at one of the endpoints, while it diverges at the second one.

Consider the following superpotential, also given in Eq. \eqref{sugu}:
\begin{equation}\label{asugu2}
W = c_1 e^{\beta \f} + c_2 e^{\b_2 \f}\,.
\end{equation}
The constant $c_1$ can be set to unity by means of the scaling symmetry \eqref{scaling}. The superpotential \eqref{asugu2} has no local extrema provided that  $c_2>0$ and $\b\b_2>0$. In such a case, the flow cannot stop at finite $\f$ (rule 1 on page \pageref{ru1}), and must run to the boundary of field space on both ends. We assume that $0<\beta_2<\beta$ without loss of generality. Then, the superpotential diverges as $W\sim e^{\b\f}$ for $\f\to +\infty$ and vanishes as $W\sim e^{\b_2\f}$ for $\f\to-\infty$.

In order to choose the values for $\b$ and $\b_2$, we assume (and later confirm) that the asymptotic solution is of type I at both endpoints. This is convenient because type I solutions can accommodate different behaviours of the potential with the same asymptotic behaviour of the potential $V$ and superpotential $W$, while in type II solutions, the asymptotic behaviour of the (super)potential depends on the sign of $V$ (see Table \ref{tab_IIS}). Given the superpotential \eqref{asugu2}, the potential behaves as $V\sim e^{2\b \f}$ for $\f\to \infty$ and as $V\sim e^{2\b_2\f}$ for $\f\to-\infty$. We can choose both $\b$ and $\b_2$ such that the Gubser bound is respected and the restrictions of type I asymptotics in table \ref{tab_IS} are satisfied:
\begin{equation}
\a_C<2\b<\a_G\,,\quad 0<2\b_2<\a_C\,.
\end{equation}
For concreteness, we choose
\begin{equation}\label{apam}
\b= \sqrt[3]{\dfrac{2}{3}}\,, \quad \b_2 = \dfrac{1}{4}\sqrt{\dfrac{3}{2}}\,,
\end{equation}
while we keep $c_2$ momentarily undetermined. We can now solve the first relation of Eq. \eqref{eqtt} to obtain the inverse scale factor:
\begin{equation}\label{tgg}
T = \frac{864 C_T e^{\frac{4}{3} \sqrt{\frac{2}{3}} \varphi }}{\left(9 c_2+16
   e^{\frac{7 \f }{12 \sqrt{6}}}\right)^2}\,,
\end{equation}
with $C_T$ an integration constant, which is positive in the spherically sliced ansatz. Subsequently, we integrate Eq. \eqref{f10_1} once to obtain
\begin{equation}\label{fpgub}
f' = \frac{35831808 \sqrt{6} C_T e^{\frac{7 \varphi }{6 \sqrt{6}}} \left(153 c_2+328
   e^{\frac{7 \varphi }{12 \sqrt{6}}}\right)+697 f_1 e^{\frac{55 \varphi }{12 \sqrt{6}}}}{697
   \left(9 c_2+16 e^{\frac{7 \varphi }{12 \sqrt{6}}}\right)^5}\,,
\end{equation}
with $f_1$ the integration constant. Note that as $\f\to+\infty$, the blackening function, and its derivative, diverge as $\sim f_1 e^{5\f/(3\sqrt{6})}$. This corresponds to the irregular (type 0) asymptotics, equation \eqref{z9}, found in appendix \ref{asymp}. Therefore, the regular asymptotic structure at $\f\to \infty$ requires that $f_1=0$.\footnote{Alternatively, one can keep $f_1\neq0$, such that there is a bad singularity at $\f\to\infty$. In such a scenario, one can choose $f_1$ such that the singularity is covered by a black-hole event horizon. This construction has been carried out in Appendix \ref{app:75}.} With this condition, we integrate Eq. \eqref{fpgub} to obtain
\begin{equation}\label{fgg}
f = f_0-\frac{1259712 C_T \left(2025 c_2^2+14400 c_2 e^{\frac{7 \varphi
   }{12 \sqrt{6}}}+20992 e^{\frac{7 \varphi }{6 \sqrt{6}}}\right)}{4879 \left(9 c_2+16
   e^{\frac{7 \varphi }{12 \sqrt{6}}}\right)^4}\,,
\end{equation}
where $f_0$ is an integration constant. For generic integration constants, the blackening function asymptotes to constant values as $\f\to\pm \infty$, as dictated by the type I asymptotic structure \eqref{ms18}. We can now solve algebraically Eq. \eqref{w55} for the potential $V$ to find
\begin{equation}\label{vgg}
V = \frac{e^{\frac{1}{2} \sqrt{\frac{3}{2}} \phi } }{8430912}\left(\frac{139968 C_T \left(111375
   c_2^2+590400 c_2 e^{\frac{7 \varphi }{12 \sqrt{6}}}+713728 e^{\frac{7 \varphi
   }{6 \sqrt{6}}}\right)}{\left(9 c_2+16 e^{\frac{7 \varphi }{12
   \sqrt{6}}}\right)^2}\right.
\end{equation}
$$
\left. -4879 f_0 \left(495 c_2^2+864 c_2 e^{\frac{7
   \varphi }{12 \sqrt{6}}}+320 e^{\frac{7 \varphi }{6 \sqrt{6}}}\right)\right)\,.
$$
As we approach the boundary of field space on both directions, the potential becomes, asymptotically,
\begin{equation}\label{vasym}
V|_{\f\to+\infty} = -\frac{5}{27}f_0 e^{2\sqrt[3]{\frac{2}{3}}\f} +\dots\,,\qquad V|_{\f\to-\infty} = \left(\frac{111375 C_T}{4879}-\frac{55 c_2^2 f_0}{192}\right)e^{\frac{2}{2}\sqrt{\frac{3}{2}}\f}+\dots
\end{equation}
both of which are compatible with the type I asymptotic solutions, see table \ref{tab_IS}.
We can set $C_T$ to any positive arbitrary value without loss of generality through the scaling symmetry $(f,T,V)\to \lambda(f,T,V)$. In addition, we choose a value of $c_2$ that simplifies the asymptotic equations above \eqref{vasym}, and reparametrize $f_0$ in terms of the leading coefficient of the potential at $\f\to -\infty$, denoted as $V_{-\infty}$. Specifically, we set
\begin{equation}
C_T = \frac{4879}{111375}\,,\quad c_2 = \sqrt{\frac{192}{55}}\,,\quad f_0 = 1-V_{-\infty}\,.
\end{equation}
With this choice, we rewrite the asymptotic behaviour of the potential \eqref{vasym}
\begin{equation}\label{o36}
V|_{\f\to+\infty} = -\frac{5}{27}(1-V_{-\infty}) e^{2\sqrt[3]{\frac{2}{3}}\f} +\dots\,,\qquad V|_{\f\to-\infty} = V_{-\infty}e^{\frac{2}{2}\sqrt{\frac{3}{2}}\f}+\dots
\end{equation}
while for $f$ we have
\begin{equation}\label{o37}
f|_{\f\to+\infty} = (1-V_{-\infty}) +\dots\,,\qquad f|_{\f\to-\infty} = -V_{-\infty}+\dots
\end{equation}
Different values of $V_{-\infty}$ give rise to qualitatively different solutions. All the possibilities are discussed in Sec. \ref{sec:73} and shown in Fig. \ref{fig:gubxi}.

\subsection{From $V(\infty)\to 0^\pm$ to a black hole}\label{app:75}

In this section, we construct examples of solutions that feature a Gubser-regular endpoint where $V\to 0^{\pm}$ together with a black-hole event horizon. According to the classification of horizons in Appendix \ref{app:J}, the presence of a black hole requires that $f$ vanishes once if $V\to 0^-$, or that $f$ vanishes twice if $V\to 0^+$. Inside the black hole, the flow runs again to the boundary of field space, where it shall encounter a bad singularity.

Similarly to the previous sections, we construct a superpotential that can accommodate such a solution. In particular, we demand that the superpotential $W$ vanishes exponentially as $\f\to-\infty$, with an asymptotic behaviour that is compatible with the Gubser-regular type of endpoints. Furthermore, the flow runs to the boundary of field space inside the black hole. This requires that $W$ does not have regular extrema at finite $\f$, otherwise the flow would stop there. A suitable superpotential with these characteristics has already been used in the previous section (Appendix \ref{app:73}), i.e. Eq. \eqref{asugu2} with the parameters given in \eqref{apam}:
\begin{equation}
W= e^{\sqrt[3]{\frac{2}{3}}\f} + e^{\frac{1}{4}\sqrt{\frac{3}{2}}}\,,
\end{equation}
where we have also set $c_2=1$ for concreteness. In this case, the inverse scale factor $T$ and $f'$ are directly given in Eqs. \eqref{tgg} and \eqref{fpgub}:

\begin{equation}\label{tgg2}
T = \frac{864  e^{\frac{4}{3} \sqrt{\frac{2}{3}} \varphi }}{\left(9 +16
   e^{\frac{7 \f }{12 \sqrt{6}}}\right)^2}\,, \quad f' = \frac{35831808 \sqrt{6}  e^{\frac{7 \varphi }{6 \sqrt{6}}} \left(153 +328
   e^{\frac{7 \varphi }{12 \sqrt{6}}}\right)+697 f_1 e^{\frac{55 \varphi }{12 \sqrt{6}}}}{697
   \left(9 +16 e^{\frac{7 \varphi }{12 \sqrt{6}}}\right)^5}\,,
\end{equation}
where we have set $C_T=1$ without loss of generality. In the previous section \ref{app:73}, we had set  $f_1=0$ in order to have Gubser-regular asymptotics as $\f\to\infty$. In this section, the integration constant $f_1$ shall control the location of the event horizon. Now, we integrate the second relation in Eq. \eqref{tgg2} to obtain

$$
f=f_0+\frac{1}{3649960683765760}\left\{ -69060178395 \sqrt{6} f_1 e^{\frac{1 \varphi
   }{2 \sqrt{6}}} \, _2F_1\left(\frac{6}{7},1;\frac{13}{7};-\frac{16}{9}
   e^{\frac{7 \varphi }{12 \sqrt{6}}}\right) +  \right.
$$
$$
+ \frac{1}{\left(16 e^{\frac{7
   \varphi }{12 \sqrt{6}}}+9\right)^4}\left[-942385585748705280 \left(14400 e^{\frac{7 \varphi
   }{12 \sqrt{6}}}+20992 e^{\frac{7 \varphi }{6 \sqrt{6}}}+2025\right) +  \right.
$$
$$
\left. 2091 \sqrt{6} f_1 e^{\frac{1 \varphi }{2\sqrt{6}}}  \left(32 \left(42597653355 e^{\frac{7 \phi }{12
   \sqrt{6}}}+95319966348 e^{\frac{7 \varphi }{6 \sqrt{6}}}+84029695488 e^{\frac{7 \varphi }{4
   \sqrt{6}}} \right. \right. \right.
$$
\begin{equation}\label{fgg2}
\left.  \left. \left. \left.+16330180608 e^{\frac{7 \varphi }{3 \sqrt{6}}}  -4248502272 e^{\frac{35 \varphi }{12
   \sqrt{6}}}+2045575168 e^{\frac{7 \varphi }{2
   \sqrt{6}}}\right)+216692410545\right)\right]\right\}\,.
\end{equation}
where $f_0$ is another integration constant. As $\f\to -\infty$, the function $f$ approaches a constant value $f(-\infty) = f_0-388800/4879$, in agreement with the type I asymptotics of Appendix \ref{asymp}. The potential $V$ is reconstructed from Eq. \eqref{w55}, and is given by
$$
V=\frac{e^{\frac{1}{2} \sqrt{\frac{3}{2}} \varphi }}{1204416}\left\{ -697 \left(864 e^{\frac{7 \varphi }{12 \sqrt{6}}}+320 e^{\frac{7 \varphi }{6
   \sqrt{6}}}+495\right) f(\varphi ) + \frac{12 e^{\frac{7 \varphi }{6 \sqrt{6}}}}{\left(16 e^{\frac{7 \varphi }{12
   \sqrt{6}}}+9\right)^4} \times  \right.
$$
\begin{equation}\label{vgg2}
 \left.\left[  697 f_1 \sqrt{6} e^{\frac{41 \varphi }{12 \sqrt{6}}} \left(e^{\frac{7 \varphi }{12
   \sqrt{6}}}+1\right) +746496 \left(339264 e^{\frac{7 \varphi }{12 \sqrt{6}}}+272896 e^{\frac{7 \varphi }{6
   \sqrt{6}}}+100521\right) \right]\right\}\,,
\end{equation}
where we have employed the second relation in \eqref{tgg2}. As we approach the Gubser-regular endpoint at $\f\to-\infty$, the potential vanishes asymptotically as
\begin{equation}
V(-\infty)= -\frac{55}{192} f(-\infty) e^{\frac{1}{2} \sqrt{\frac{3}{2}} \varphi }+\dots \equiv -V_{-\infty}e^{\frac{1}{2} \sqrt{\frac{3}{2}} \varphi }+\dots
\end{equation}
which is compatible with the  type I asymptotic solutions of Appendix \ref{asymp}.

So far, the solution has two integration constants: $f_0$ and $f_1$. In order to construct solutions running from a Gubser-regular endpoint to a horizon, we must demand that $f$ vanishes at least one, at a location $\f_h$. For concreteness, we set $\f_h=15$. This condition fixes one of the integration constants. The second integration constant can be fixed in terms of the value $V_{-\infty}$. Specifically, we have
\begin{equation}
f_0 = 79.6885\, -3.49091 V_{-\infty}\,, \qquad f_1 =-2360.5 + 103.488 V_{-\infty}.
\end{equation}

The value $V_{-\infty}$ distinguishes qualitatively different solutions. We discuss them in Sec. \ref{sec:75} and shown in Fig. \ref{fig:gubx5}.
\section{On the forbidden flows from an AdS boundary to dS\label{forbid}}

\begin{figure}[h]
\centering
\includegraphics[ height=10cm, width=15cm]{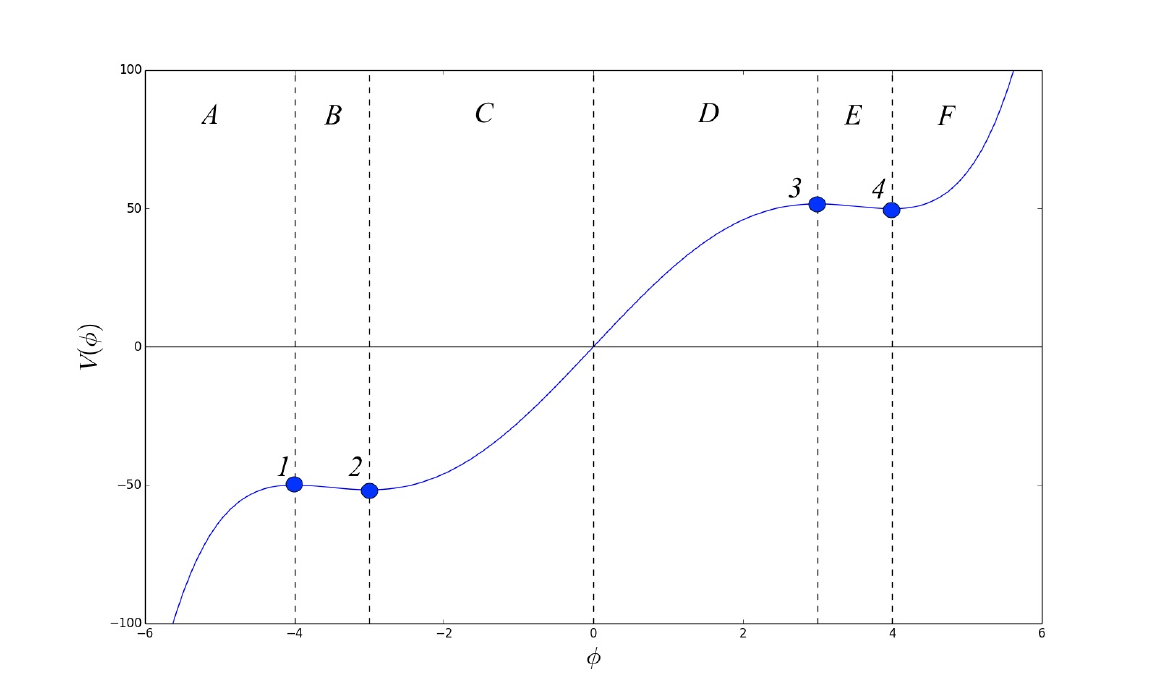}
\caption{A generic potential that contains minima and maxima in both the AdS and dS regimes. The four extrema divide our region of study into six subregions, ranging from $A$ to $F$, with respect to the placing of the extrema. The numbers $1$ to $4$ label each of the extrema.}
\label{plot1}
\end{figure}

In this appendix, we show the numerical attempts to construct a solution interpolating between an AdS$_{5}$ boundary to a shrinking endpoint in the dS regime.

We illustrate the numerical strategy that we follow based on a generic potential $V$ as shown in Fig. \ref{plot1}. In this case, there is an AdS$_5$ boundary at point $1$, while the dS shrinking endpoint can be in region $E$ where the potential is decreasing. In section \ref{shr}, we showed that the solution ``climbs up" the potential as it departs from the shrinking endpoint.
However,  if the solution features an even (respectively odd) number of $\f$-bounces, the shrinking endpoint should be in region E (respectively regions D or F).

The solution around a shrinking endpoint has less integrations constants than allowed by the system of differential equations (see appendix \ref{G.2.2}), and for this reason it is convenient to start the numerical integration from the shrinking endpoint. In particular, around a shrinking endpoint there is a single integration constant, denoted $W_0$, which can be set to $1$ with the scaling symmetry \eqref{scaling}. As a result, the only free parameter is the location in field space of the shrinking endpoint, which we denote $\f_{\rm shr}$. Given a potential $V(\f)$, we vary the location of the dS shrinking endpoint, and study where such solutions end.

\begin{figure}[h!]
	\centering
	\subfloat{\includegraphics[scale=0.33]{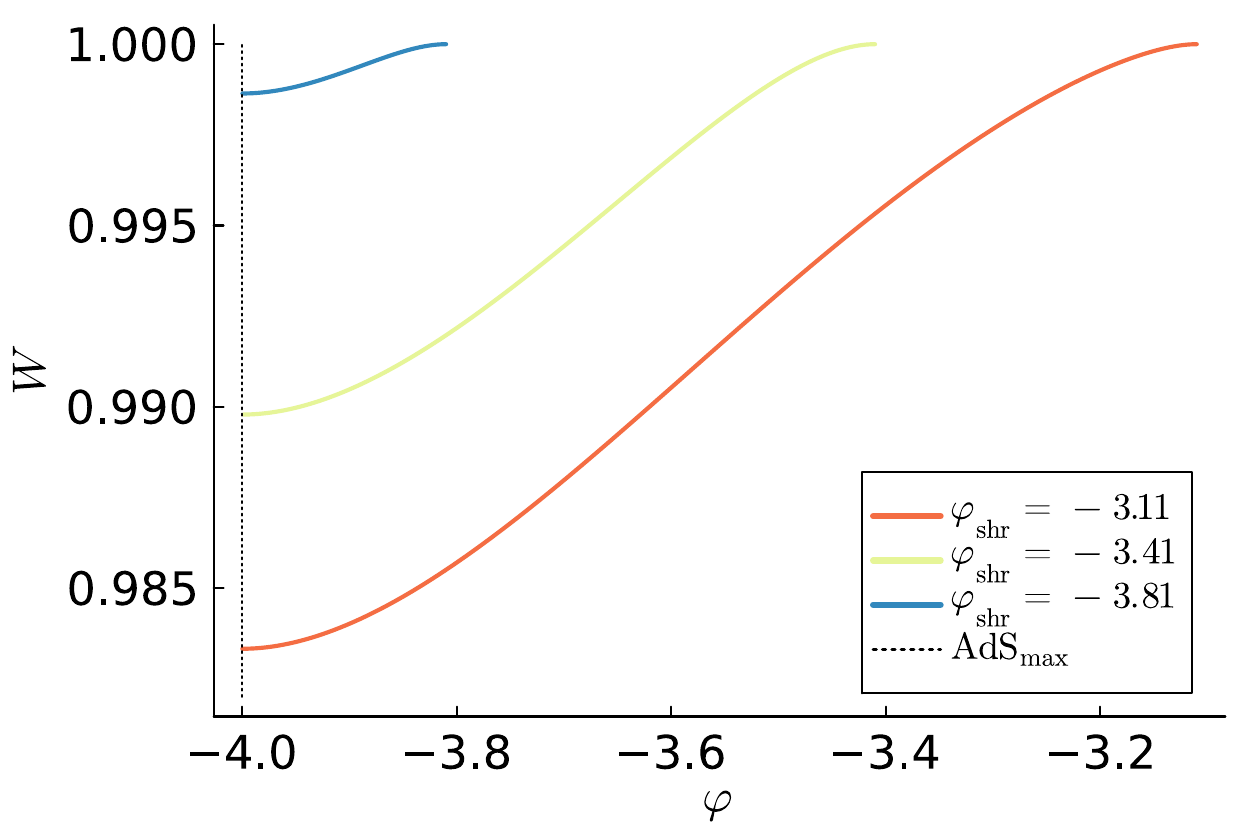}}
	\qquad
	\subfloat{{\includegraphics[scale=0.33]{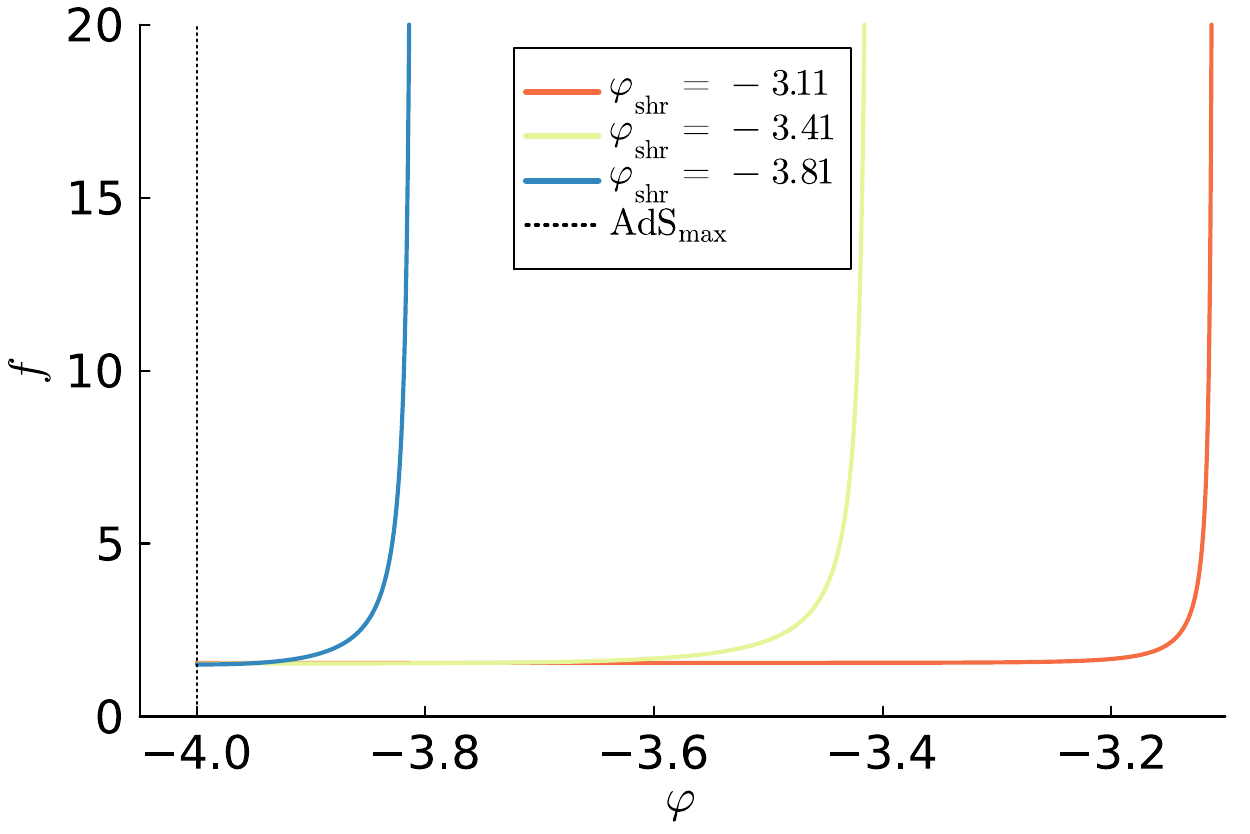}}}
	\caption{Solutions to the flow equations \eqref{f23a}-\eqref{f23c} for the potential $V_a$, Eq. \eqref{d1Wk} (see figure \ref{fig:O0}), starting from several shrinking endpoints in the AdS regime, in the region B of figure \ref{plot1}. The dotted black line marks the location of the maximum of the potential $V_a$ in the AdS regime. These solutions correspond to the standard holographic RG-flows between an AdS$_5$ boundary at $\f=-4$ and a shrinking endpoint at $\f_{\rm shr}$.}
	\label{fig:OA}
\end{figure}

\begin{figure}[h!]
	\centering
	\subfloat{\includegraphics[scale=0.33]{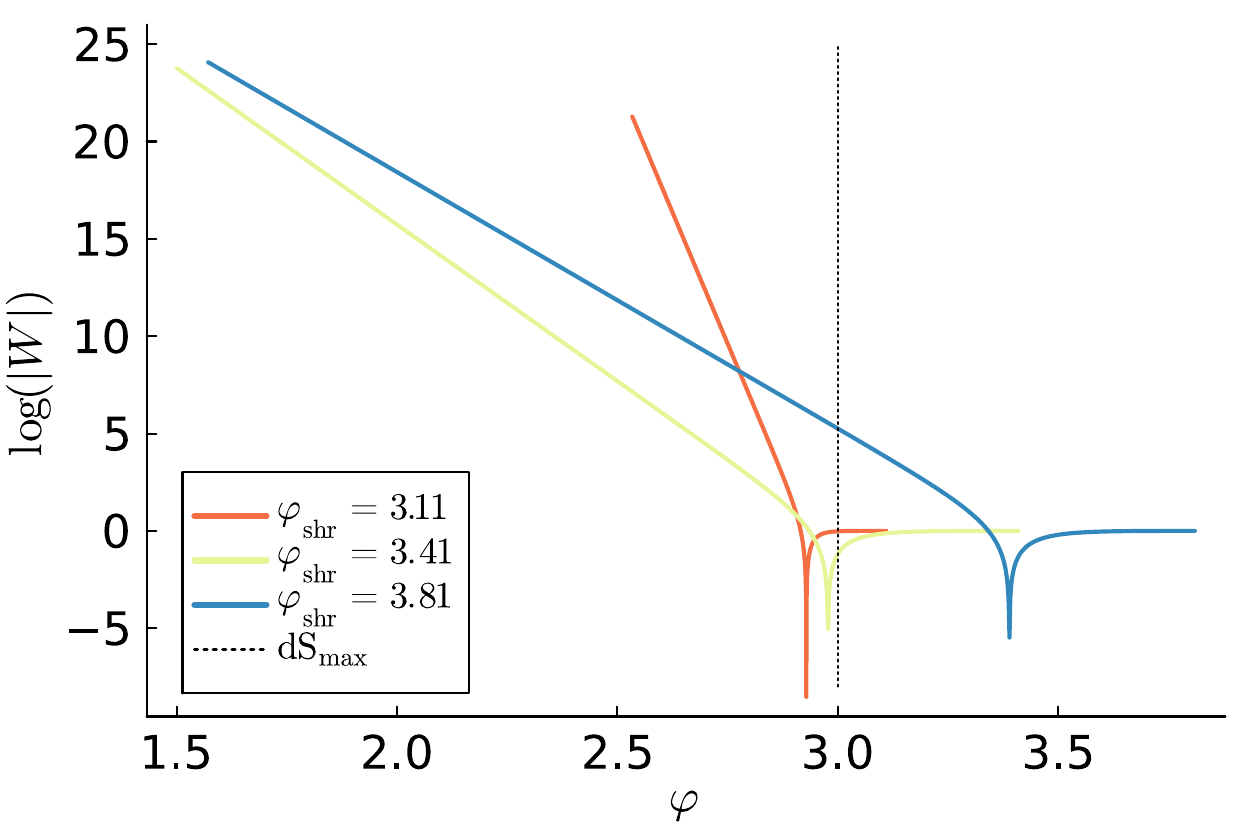}}
	\qquad
	\subfloat{{\includegraphics[scale=0.33]{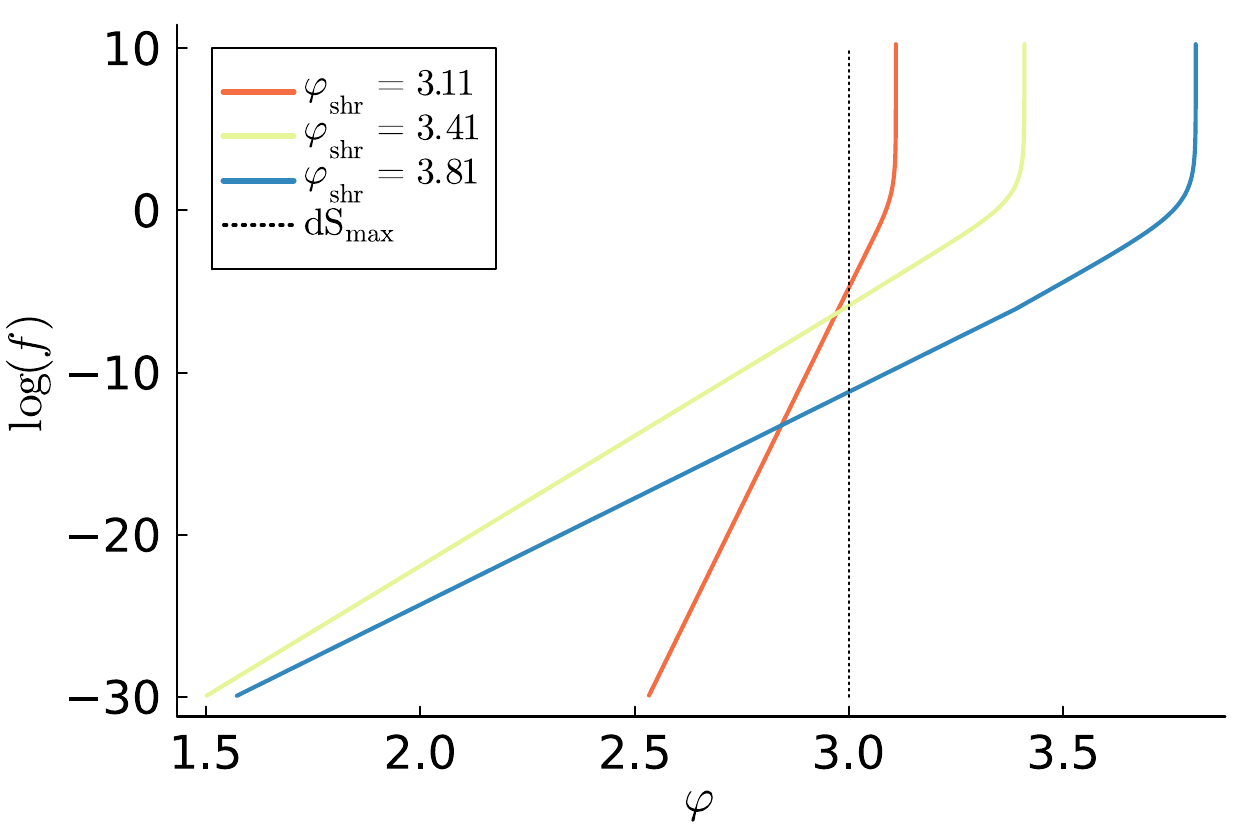}}}
	\caption{Solution to the flow equations \eqref{f23a}-\eqref{f23c} for the potential $V_a$, Eq. \eqref{d1Wk} (see figure \ref{fig:O0}), starting from several shrinking endpoints in the dS regime, in region E of the potential in \ref{plot1}. Note that the superpotential (left panel) vanishes at the point where $\log|W|$ features a spike. This marks a vanishing of $\dot A$ and it is therefore an A-bounce.   The dotted black line marks the location of the maximum of the potential $V_a$ in the dS regime. The dS minimum lies outside of these plots.}
	\label{fig:O1}
\end{figure}

\begin{figure}[h!]
	\centering
	\subfloat{\includegraphics[scale=0.55]{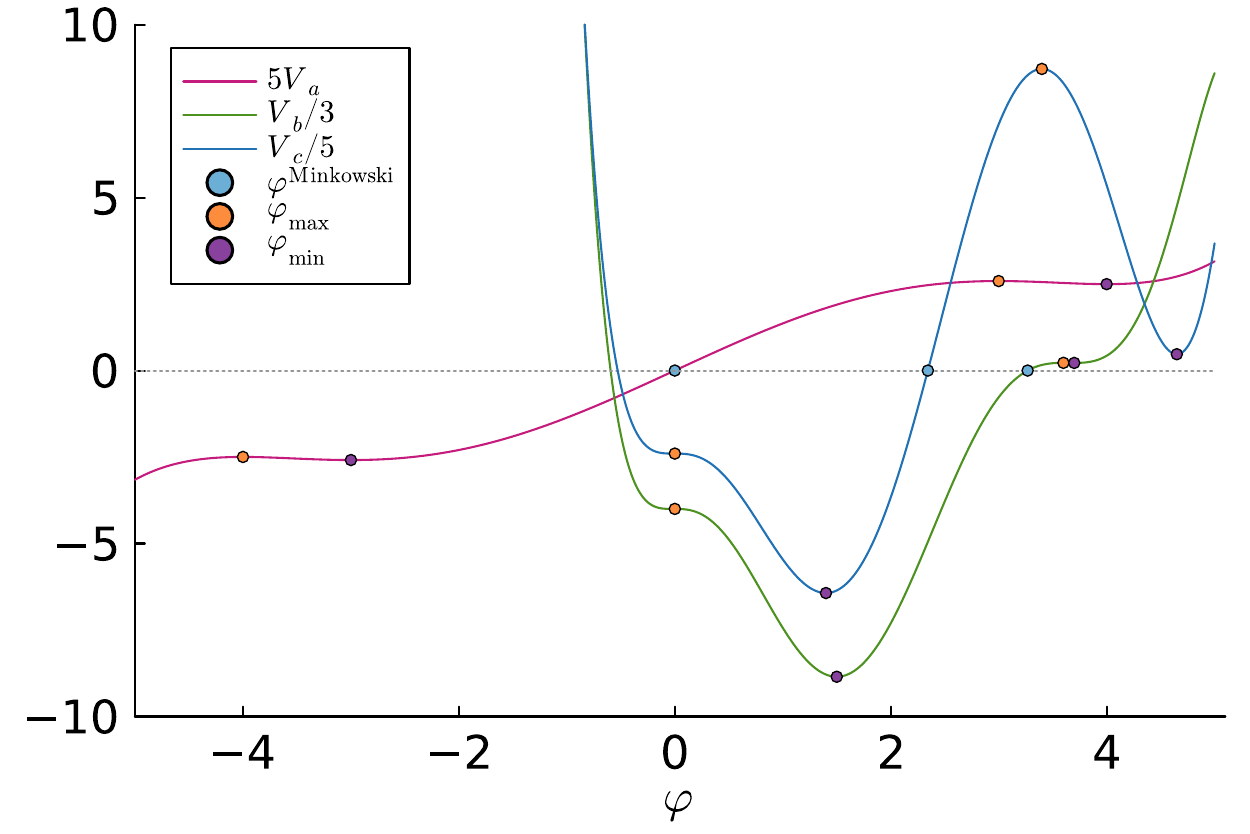}}
\caption{Three scalar potentials $V_{a,b,c}$ employed in the numerical attempts to construct flows from AdS$_5$ boundaries to dS shrinking endpoints. The overall magnitude of the potentials is rescaled as indicated in the labels for visual clarity. The colored dots indicate the extrema of each potential.}
	\label{fig:O0}
\end{figure}

As a particular example, we study the solutions in $4+1$ dimensions for the potential displayed in Fig.\ref{plot1} (see also Fig. \ref{fig:O0}), explicitly given by

\be
 V_a(\f)=\frac{4}{5} \left[e^{-{\f\over 8}} \left(-\f^4-32 \f^3-743 \f^2-11888 \f-95248\right)+\right.
\label{d1Wk}\ee
$$+\left.
e^{\f\over 8} \left(\f^4-32 \f^3+743
   \f^2-11888 \f+95248\right)\right].
$$
Generic AdS$_5$ boundaries are located at $\f=-4$, where the potential has an AdS maximum. We can construct the standard holographic RG-flows solving the equations of motion \eqref{f23a}-\eqref{f23c} from a shrinking endpoint in the AdS regime, $\f_{\rm shr}\in(-4,-3)$, connecting them to the AdS$_5$ boundary at the AdS maximum of the potential: $\f=-4$. In figure \ref{fig:OA} we show three examples of such solutions. The blackening function $f$ (right panel) diverges to $+\infty$ at the shrinking endpoint and decreases monotonically as it approaches the endpoint of the flow at $\f=-4$, where the boundary of AdS$_5$ is located. The superpotential $W$ (left-panel) is monotonic along the flow, and it features two extrema at the two endpoints of the flow.

We now solve the equations of motion \eqref{f23a}-\eqref{f23c} starting from shrinking endpoints in the dS regime for the same potential $V_a$. We locate the shrinking endpoint in the range $\f_{\rm shr}\in (3,4)$ (region E of Fig.\ref{plot1}). We use boundary conditions in agreement with the analysis of appendix \ref{G.2.2} with $W_0=1$.

\begin{figure}[h!]
	\centering
	\subfloat{\includegraphics[scale=0.55]{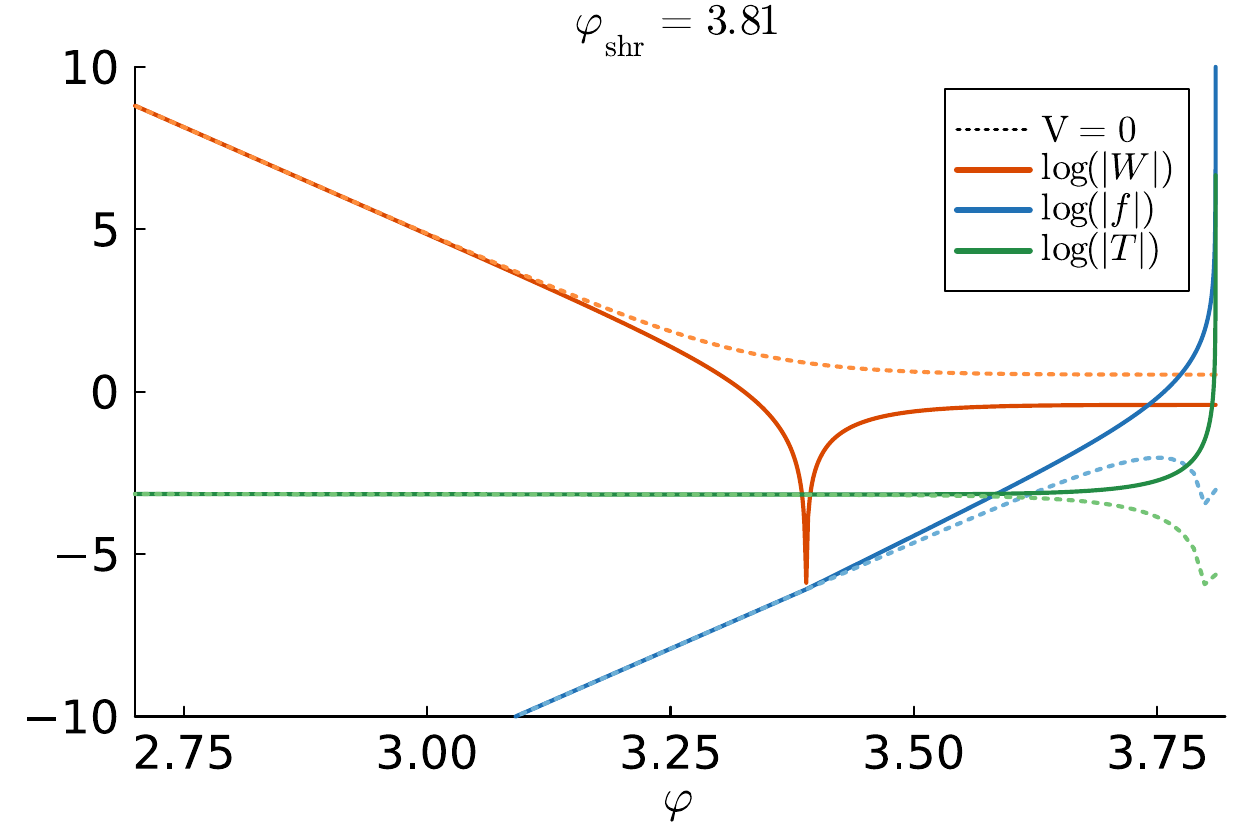}}
	\caption{Solution to the flow equations \eqref{f23a}-\eqref{f23c} for the potential \ref{d1Wk} starting at the shrinking endpoint $\f_{\rm shr}=3.81$ in the dS regime. The solid lines correspond to the numerical solution, while the dotted lines represent a solution without potential $V=0$, as studied in appendix \ref{nopot}, which has the same asymptotic behaviour at $\f\to-\infty$ as the numerical solution.}
	\label{fig:O2}
\end{figure}

In figure \ref{fig:O1} we show the behaviour of the superpotential $W$ and the blackening function $f$ (in logarithmic scale) for three representative choices of $\f_{\rm shr}$. These three choices correspond to shrinking endpoints near the two end-points of region E as well as in the middle.

 Strikingly, the superpotential vanishes once in these solutions, indicating the presence of an A bounce, where $\dot A=0$. This is one of the main differences with respects to the flows from an AdS shrinkpoint to an AdS boundary, where the superpotential (left panel of figure \ref{fig:OA}) remains positive along the flow. From the definition \ref{w53}, the vanishing of the superpotential is equivalent to the scale factor $e^A$  not being monotonic for these solutions. It then follows from rule 3 on page \pageref{ru3} of section \ref{rul} that such flows cannot be connected to an AdS boundary. Neither can they be connected to another shrinking endpoint (see section \ref{pib}). During this flow the scale factor vanishes at the shrinking endpoint, then grows to a maximum, and finally turns around and vanishes again in a singular way at $\f\to -\infty$.

In figure \ref{fig:O2} we verify explicitly that the solution for $\f_{\rm shr} = 3.81$ becomes, asymptotically, the irregular (type 0) solutions at the boundary of field space of appendix \ref{asymp}. We have verified that the same asymptotic behaviour is achieved for the other solutions with different $\f_{\rm shr}$. The potential $V$ is irrelevant for the irregular asymptotic solutions, and we are able to glue the asymptotic behaviour of the functions at $\f\to-\infty$ to the $V=0$ solutions of appendix \ref{nopot}. Note that the inverse scale factor $T$ for the solution with  $V=0$ (green dotted line) vanishes at some point along the flow. This corresponds to the solutions of appendix \ref{nopot} that are not well defined globally (see figure \ref{fM1}), but that can be acceptable when glued to a solution with non-trivial potential, as it happens in \ref{fig:O2} (solid lines).

\begin{figure}[h!]
	\centering
	\subfloat{\includegraphics[scale=0.33]{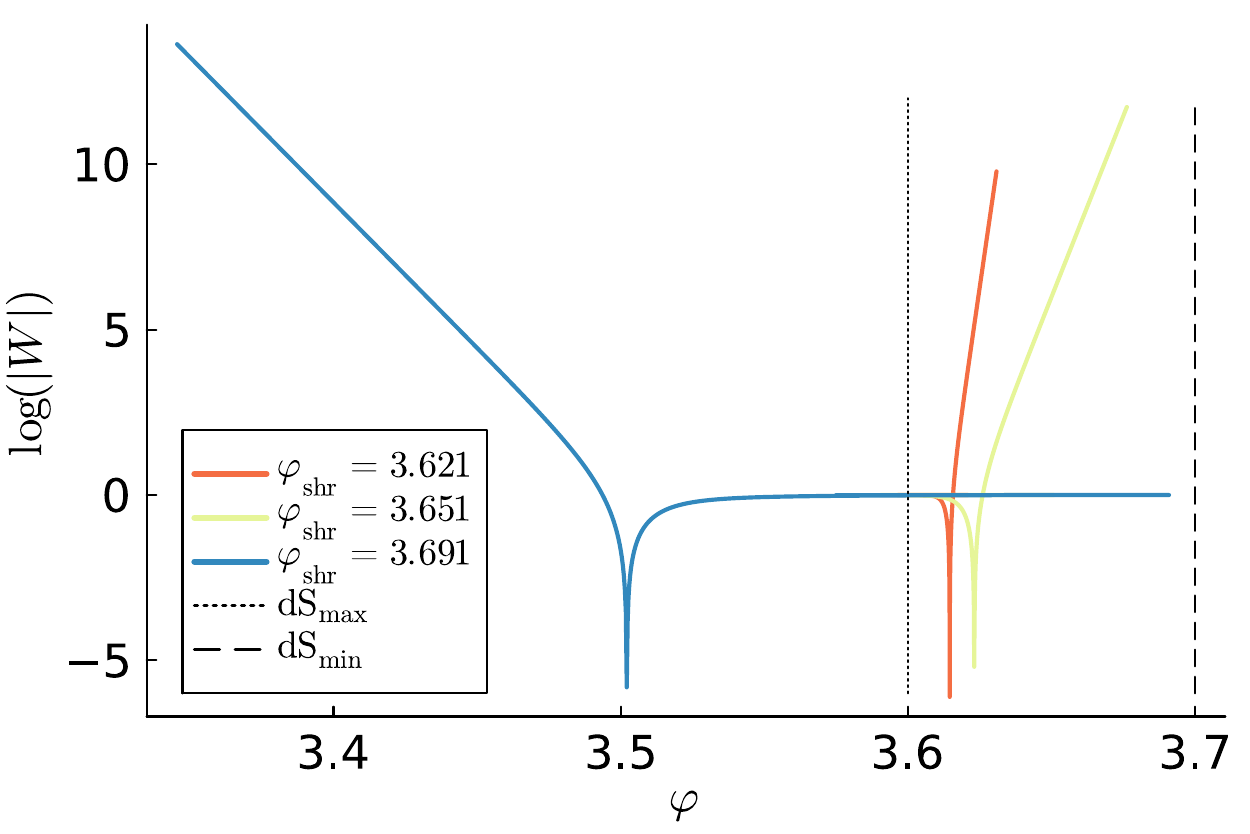}}
	\qquad
	\subfloat{{\includegraphics[scale=0.33]{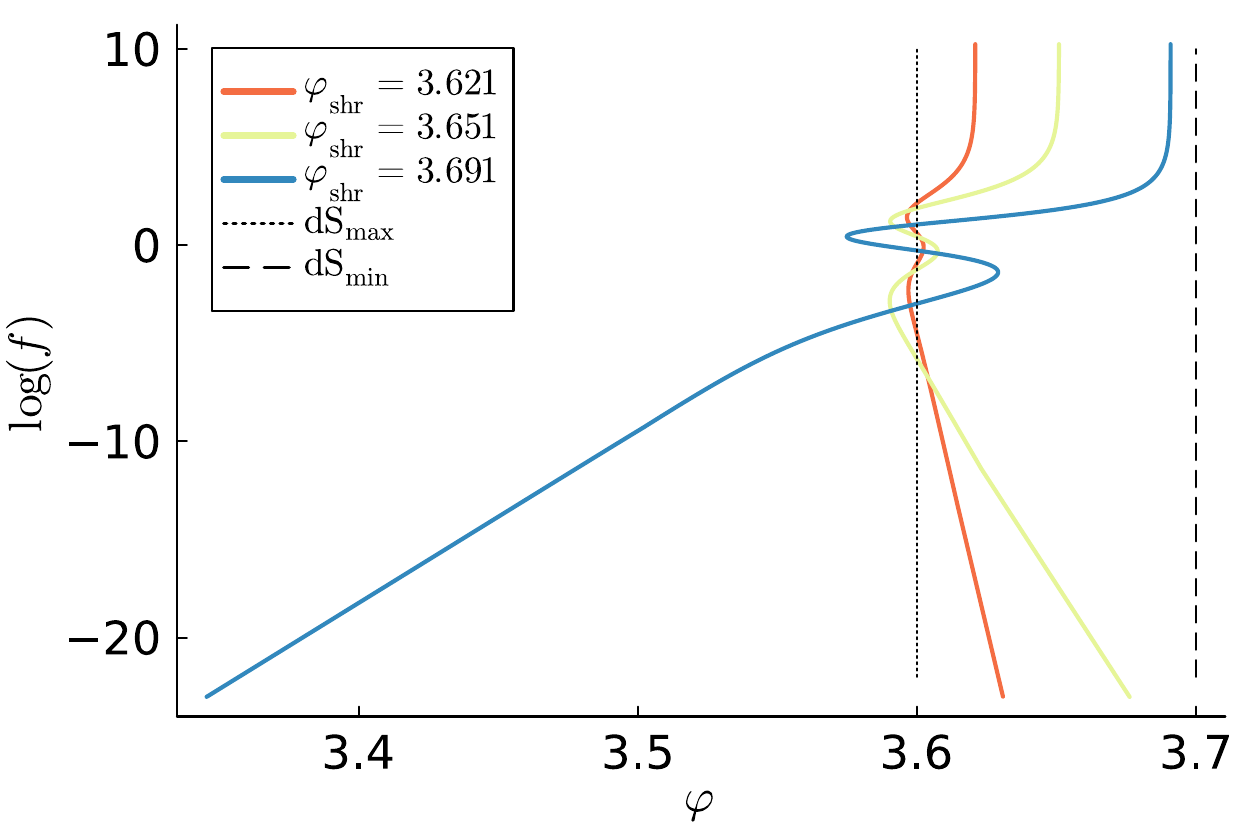}}}
	\caption{Solution to the flow equations \eqref{f23a}-\eqref{f23c} for the potential $V_b$, Eqs. \eqref{pots} and \eqref{vb} (see figure \ref{fig:O0}), starting from several shrinking endpoints in the dS regime. Note that the superpotential (left panel) vanishes at the point where $\log|W|$ features a spike. The dashed (dotted) black line marks the location of a minimum (maximum) of the potential $V_b$ in the dS regime. }
	\label{fig:O3}
\end{figure}

\begin{figure}[h!]
	\centering
	\subfloat{\includegraphics[scale=0.33]{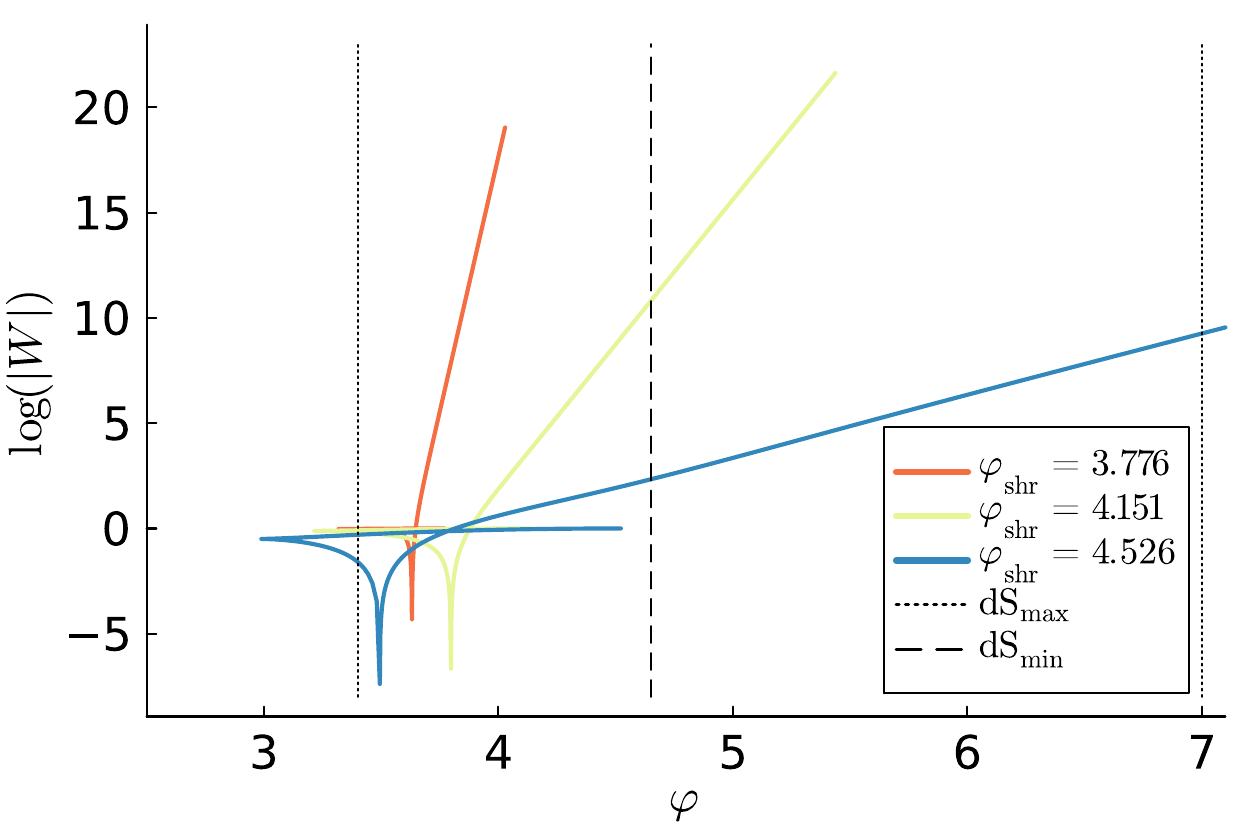}}
	\qquad
	\subfloat{{\includegraphics[scale=0.33]{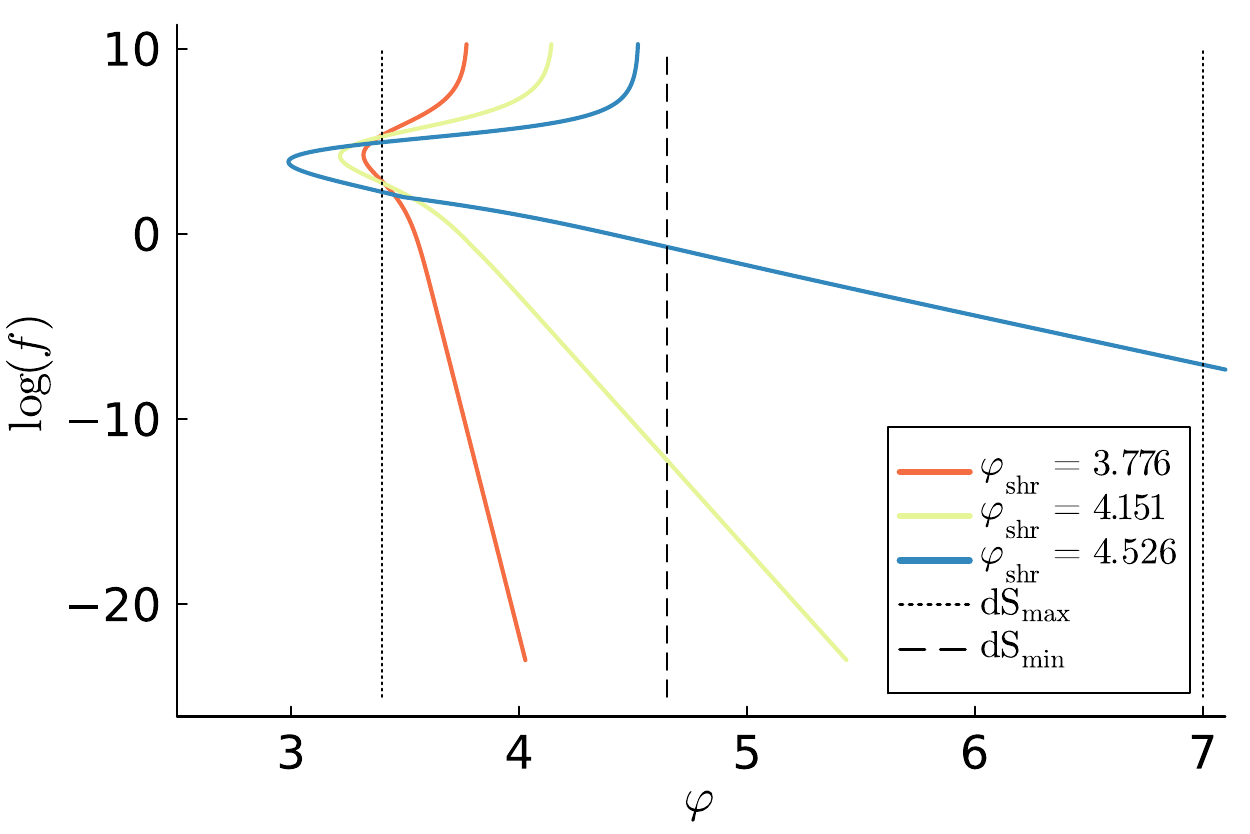}}}
	\caption{Solution to the flow equations \eqref{f23a}-\eqref{f23c} for the potential $V_c$, Eqs. \eqref{pots} and \eqref{vc} (see figure \ref{fig:O0}), starting from several shrinking endpoints in the dS regime. Note that the superpotential (left panel) vanishes at the point where $\log|W|$ features a spike. The dashed (dotted) black lines marks the location of a minimum (maximum) of the potential $V_c$ in the dS regime.}
	\label{fig:O4}
\end{figure}

In addition to the potential $V_a$, given in Eq. \eqref{d1Wk}, we have studied similar solutions for two more potentials, denoted $V_b$ and $V_c$. The two potentials are constructed as a solution of

\begin{equation}\label{pots}
V_{b,c}' = -\f  (\f -\f_0) (\f -\f_1) (\f -\f_2) (\f
   -\f_3) \left(\f -\frac{\Delta  (\Delta -d)}{\f_0 \f_1 \f_2 \f_3}\right)\,,
\end{equation}
which has extrema at $0$, $\f_1$, $\f_2$, $\f_3$, $\f_4$ and $\Delta (\Delta-d)/\Pi_i\f_i$. The integration constant is fixed such that $V(0) = -d(d-1)$. Therefore, the extemum of the potential at $\f=0$ is an AdS extremum and the mass of the scalar is $m^2 = \Delta(\Delta-d)$ at that point. We set $d=4$, while the rest of the parameters are
\begin{equation}\label{vb}
V_b: \quad \f_0 = 1.5\,,\ \f_1 = 3.6\,,\ \f_2 = 3.7\,,\ \f_3 = 5.15\,,\ \Delta = 3 \,.
\end{equation}
\begin{equation}\label{vc}
V_c: \quad \f_0 = 1.4\,,\ \f_1 = 3.4\,,\ \f_2 = 4.65\,,\ \f_3 = 7\,,\ \Delta = 3\,.
\end{equation}
The two potentials are also shown in Fig. \ref{fig:O0}. Qualitatively speaking, the potential $V_b$ is shallower in the neighbourhood of the extrema in the dS regime, while the potential $V_c$ is considerably steeper in the same region.

In figure \ref{fig:O3} we show the behaviour of the superpotential $W$ and the blackening function $f$, in logarithmic scale, for solutions that start at a shrinking endpoint in the dS regime for the potential $V_b$. Three representative choices of $\f_{\rm shr}$ are displayed. Note that the solutions feature two or three $\f$-bounces, where the flow reverses its direction, as is clearly seen in the behaviour of the function $f$. The $\f$-bounces happen in the vicinity of the dS maximum.

The superpotential (left panel of Fig. \ref{fig:O3}) vanishes at one point. Similarly to the solutions discussed above for the potential $V_a$, the vanishing of $W$ signals an $A$-bounce. After the $A$-bounce, the solution runs to the boundary of field space. The situation is completely analogous for the flows that start at a dS shrinking endpoint in the third potential, $V_c$. Again, the superpotential $W$ and the blackening function $f$ are shown in logarithmic scale for the potential $V_c$ in figure \ref{fig:O4}. The solutions feature two $\f$-bounces and one $A$-bounce, which eventually forces the flow to run to the boundary in field space.

\section{Thin Brane Solutions}\label{app:thinW}

In this appendix  we collect the details of the construction of the thin-brane domain-wall solutions. We begin by endowing our space-time with a metric written in coordinates of the form
\begin{equation}\label{eq:rfol}
\mathrm{d}s^2 = g_{\mu\nu}\mathrm{d}x^\mu\mathrm{d}x^\nu=\mathcal{N}(r)^2\mathrm{d}r^2 + \gamma_{ij}(r,x)\mathrm{d}x^i\mathrm{d}x^j.
\end{equation}
We first wish to determining under what conditions two space-times written in such coordinates can be ``smoothly glued together" along a surface of constant $r$. These conditions are the content of the Israel junction conditions.

The location of the gluing, $r_g$ identifies a hypersurface $\Sigma$ described by $\Phi_\Sigma(r,x) = r-r_g = 0$. A unit vector normal to this hypersurface, which points in the direction of increasing $\Phi_\Sigma$, is
\begin{equation}
n = \frac{1}{\mathcal{N}}\partial_r.
\end{equation}
The vectors $\theta^r = 0$ and $\theta^i = \partial_i$ are clearly tangent to the hypersurface, and are convenient for describing its intrinsic and extrinsic geometry. For example, the induced metric is simply
\begin{equation}
\mathrm{d}s^2_\Sigma = g_{\mu\nu}\theta^\mu_i\theta^\nu_j\mathrm{d}x^i\mathrm{d}x^j = \gamma_{ij}\mathrm{d}x^i\mathrm{d}x^j
\end{equation}
and the extrinsic curvature and its trace are given by
\begin{equation}
K_{ij} = \nabla_\nu n_\mu \theta^\mu_i \theta^\nu_j = \frac{1}{2\mathcal{N}}\partial_r\gamma_{ij} \qquad \mathrm{and} \qquad K = \gamma^{ij}K_{ij}
\end{equation}
respectively.

We now turn to the junction conditions. Imagine a space-time partitioned by the hypersurface into $\mathcal{M}^-$ and $\mathcal{M}^+$, and choose the convention in which the unit normal $n$ points towards $\mathcal{M}^+$. For any tensor $T$ defined on either side of the hypersurface, we introduce the notation
\begin{equation}
\left[T \right] = T\left(\mathcal{M}^+\right)\big |_\Sigma-T\left(\mathcal{M}^-\right)\big |_\Sigma.
\end{equation}
The junction conditions can then be written
\begin{equation}
\left[\gamma_{ij} \right] = 0 \qquad \mathrm{and} \qquad \Big(\left[K_{ij}\right]-\left[K\right]\gamma_{ij} \Big)=-S_{ij}
\end{equation}
where $S_{ij}$ is the surface stress energy tensor, which is proportional to the pull-back of the putative membrane stress tensor, like
\begin{equation}
T_D^{\mu\nu} =  \delta(s)S^{ij}\theta^\mu_i\theta^\nu_j
\end{equation}
where $s$ is taken to be the proper distance from the hypersurface.

We next consider a $d+1$ space-time with a metric of the form
\begin{equation}
\mathrm{d}s^2 = \frac{\mathrm{d}u^2}{f} + e^{2A}\left(-f\mathrm{d}t^2+R^2\mathrm{d}\Omega_{d-1}^2 \right).
\end{equation}
In these coordinates, AdS can be written in ``static patch" coordinates like
\begin{equation}\label{eq:AdSsp}
e^A = e^{-\frac{u}{\ell}}, \qquad f = 1+e^{2\frac{u}{\ell}}, \qquad R = \ell
\end{equation}
with the boundary obtained as $u\to-\infty$ and the center as $u\to \infty$. Similarly dS is given by
\begin{equation}\label{eq:dSsp}
e^{A}=e^{Hu}, \qquad f=-1+e^{-2Hu}, \qquad R = \frac{1}{H}.
\end{equation}
which is evidently the static patch of dS in which one could consider an ``observer'' at $u\to -\infty$, who is in causal contact with the region bounded by the cosmological horizon at $u = 0$. Continuing, the future boundary is achieved as $u\to \infty$.

In this ansatz, we compute
\begin{align}
K_{tt} & = -\frac{1}{2}\sqrt{f}\left(2fA'+f' \right)e^{2A} = \frac{1}{2}\sqrt{f}\left(2A'+\frac{f'}{f} \right)\gamma_{tt}\nonumber\\
K_{\rho\sigma} & = \sqrt{f}A'\gamma_{\rho\sigma}
\end{align}
where $\rho$, $\sigma$ are directions on the sphere.

The trace of the extrinsic curvature is given by
\begin{equation}
K = \sqrt{f}\left(d A'+\frac{1}{2}\frac{f'}{f} \right)
\end{equation}
and it follows that the first junction condition is given by
\begin{equation}
[\gamma_{ij}] = 0 \Longrightarrow \left[f e^{2A} \right] = \left[R^2 e^{2A} \right] =0.\label{eq:1jc}
\end{equation}

In passing, we observe that the trivial scenario in which we take the metric to be written in the same coordinates in $\mathcal{M}^\pm$ does not appear viable. This is because the first junction condition then requires $[A] = [ f] = 0$.

From the second junction equation, we further derive
\begin{align}
S_{tt} & = \sqrt{f}\left[(d-1)A' \right]\gamma_{tt}\\
S_{\rho\sigma} & = \sqrt{f}\left[(d-1)A'+\frac{1}{2}\frac{f'}{f} \right]\gamma_{\rho\sigma}.\label{eq:2jc2}
\end{align}
We next explore the possibility of satisfying the complete set of  junction conditions (\ref{eq:1jc})-(\ref{eq:2jc2}) for a domain wall separating AdS and dS regions in the coordinates of (\ref{eq:AdSsp}) and (\ref{eq:dSsp}).

As we are interested in solutions with an AdS boundary, we take the metric on $\mathcal{M}^-$ to be of the form
\begin{equation}
\mathrm{d}s^2_- = \frac{\mathrm{d}u^2}{\left(1+e^{2u}\right)} + e^{-2u}\left[-\left(1+e^{2u}\right)\mathrm{d}t^2 +\mathrm{d}\Omega_{d-1}^2\right]
\end{equation}
where we have arranged without loss of generality for the AdS scale $\ell$ to be set to one. In these coordinates, the boundary of AdS is located at $u\to-\infty$.

Similarly, we take the metric on $\mathcal{M}^+$ to be
\begin{equation}
\mathrm{d}s^2_+ = \frac{\mathrm{d}\mathfrak{u}^2}{\left(e^{2 H\mathfrak{u}}-1\right)} + e^{-2H\mathfrak{u}}\left[-\left(e^{2H\mathfrak{u}}-1\right)\mathrm{d}\mathfrak{t}^2 +\frac{1}{H^2}\mathrm{d}\Omega_{d-1}^2\right]
\end{equation}
On general grounds we might expect that were a domain wall with an AdS boundary to exist, the dS side would asymptote to the ``shrinking endpoint'' identified with the location of the observer above. We have written the metric $\mathcal{M}^+$ in coordinates such that this point is obtained as $\mathfrak{u}\to \infty$.

From the first junction condition in the sphere directions, we note that
\begin{equation}\label{eq:sJC1}
e^{-2u_g} = \frac{1}{H^2}e^{-2H\mathfrak{u}_g},
\end{equation}
where $u_g$ and $\mathfrak{u}_g$ are the locations of the brane in $\mathcal{M}^-$ and $\mathcal{M}^+$ respectively.

As a first pass,we assume that $\mathfrak{t}= t+t_s$ with $t_s$ an arbitrary constant. In this case, the first junction condition in the $tt$ direction yields the additional constraint
\begin{equation}
e^{-2u_g}\left(1+e^{2u_g} \right) = e^{-2H\mathfrak{u}_g}\left(e^{2H\mathfrak{u}_g}-1 \right)
\end{equation}
which, upon using (\ref{eq:sJC1}) reduces to
\begin{equation}
H^2 = -1.
\end{equation}
Accordingly, in this case we discover that no such domain wall is permitted.

We next exploit the fact that the ansatz is preserved by the simultaneous scaling $t\to\alpha t$ and $H\to H/\alpha$. One can use this to take $\mathfrak{t} = \alpha t$ such that the metric becomes
\begin{equation}
\mathrm{d}s^2_+ = \frac{\mathrm{d}\mathfrak{u}^2}{\left(e^{2 \frac{H}{\alpha}\mathfrak{u}}-1\right)} + \alpha^2\,e^{-2\frac{H}{\alpha}\mathfrak{u}}\left[-\left(e^{2\frac{H}{\alpha}\mathfrak{u}}-1\right)\mathrm{d}t^2 +\frac{1}{H^2}\mathrm{d}\Omega_{d-1}^2\right].
\end{equation}
Repeating the above exercise, we obtain
\begin{equation}\label{eq:jc1ss}
e^{-2u_g} = \frac{\alpha^2}{H^2}e^{-2\frac{H}{\alpha}\mathfrak{u}_g}\qquad\mathrm{and}\qquad e^{2\frac{H}{\alpha}\mathfrak{u}_g} = \frac{\alpha^2}{H^2}\left( \frac{1+H^2}{\alpha^2-1}\right).
\end{equation}
Before continuing we make several observations: first, the parameter $\alpha$ which scales the leaves of the radial foliation must satisfy $\alpha > 1$ if a solution is to exist. Next, at the location of the gluing, we observe that $f_\pm > 0$ for any finite allowed values of $\alpha$, $H$.

Turning to the second junction condition, we evaluate
\begin{align}
\sqrt{f}(d-1)A'\Big|^+ & = -\frac{H}{\alpha}(d-1)\sqrt{e^{2\frac{H}{\alpha}\mathfrak{u}_g}-1}\nonumber\\
\sqrt{f}(d-1)A'\Big|^- & = - \frac{H}{\alpha}(d-1)\sqrt{e^{2\frac{H}{\alpha}\mathfrak{u}_g}+\frac{\alpha^2}{H^2}}
\end{align}
as well as
\begin{align}
\frac{1}{2}\frac{f'}{\sqrt{f}}\Big|^+ & = \frac{H}{\alpha}\frac{e^{2\frac{H}{\alpha}\mathfrak{u}_g}}{\sqrt{e^{2\frac{H}{\alpha}\mathfrak{u}_g}-1}}\nonumber\\
\frac{1}{2}\frac{f'}{\sqrt{f}}\Big|^- & = \frac{e^{2u_g}}{\sqrt{e^{2u_g}+1}} = \frac{H}{\alpha}\frac{e^{2\frac{H}{\alpha}\mathfrak{u}_g}}{\sqrt{e^{2\frac{H}{\alpha}\mathfrak{u}_g}+\frac{\alpha^2}{H^2}}}
\end{align}
Note that a necessary condition for solutions with a tensionless brane is given by
\begin{equation}
\sqrt{e^{2\frac{H}{\alpha}\mathfrak{u}_g}-1} = \sqrt{e^{2\frac{H}{\alpha}\mathfrak{u}_g}+\frac{\alpha^2}{H^2}}
\end{equation}
or, using (\ref{eq:jc1ss}),
\begin{equation}
\frac{1}{H}\left(\frac{\alpha^2+H^2}{\alpha^2-1} \right)^{1/2} = \frac{\alpha}{H}\left(\frac{\alpha^2+H^2}{\alpha^2 -1} \right)^{1/2}
\end{equation}
which has no solution for $\alpha>1$. If, however, one is willing to entertain solutions with stress-energy on the brane, there appears to be no immediate obstruction. We now investigate this claim in more detail.

From (\ref{eq:jc1ss}), we can write the second junction condition as
\begin{equation}\label{eq:STc1A}
S_{tt} = \frac{1}{\alpha}(1-d)\left(\frac{\alpha^2+H^2}{\alpha^2-1} \right)^{1/2}\left(1-\alpha\right)\gamma_{tt}
\end{equation}
and
\begin{equation}\label{eq:STc2A}
S_{\rho\sigma} = \left(1-\alpha\right)\left[\frac{1}{\alpha}(1-d)-\left(\frac{1+H^2}{\alpha^2+H^2}\right) \right]\left(\frac{\alpha^2+H^2}{\alpha^2-1} \right)^{1/2}\gamma_{\rho\sigma}
\end{equation}
where we again keep in mind the fact that $\alpha>1$ for the solution to be sensible.

At this point, the state of affairs is that {\it if} one can arrange for a brane stress tensor of the form (\ref{eq:STc1A}-\ref{eq:STc2A}), then a domain-wall solution exists. The question then becomes what must the theory on the brane be, in order to give rise to such a stress tensor?

Locally, the desired stress energy is that of a perfect fluid. To determine this, one can adopt a local frame $\mathrm{e}^i$ such that
\begin{equation}
\gamma_{ij}\mathrm{e}^i_a\mathrm{e}^j_b = \eta_{ab} \qquad \mathrm{and} \qquad S_{ab} \propto \mathrm{diag}\left(\rho, p,p,\ldots,p \right).
\end{equation}
This motivates a simple action of the form
\begin{equation}
S_D = -\frac{1}{2\kappa_D^2} \int_\Sigma\mathrm{d}x^d \sqrt{-\gamma}\Big( R[\gamma]+\mu\Big).
\end{equation}

The contribution to the junction conditions from the non-trivial brane action is determined by requiring the stationarity of the full gravitational action, which can be written as a sum of a bulk gravitational action $S_B$, a Gibbons-Hawking boundary term $S_{GH}$ which contains contributions from either side of the domain wall, and the brane action $S_D$. The variation of the first two terms, evaluated on either side of the domain wall, yields (see e.g. \cite{reall})
\begin{equation}
\delta S_B + \delta S_{GH} = \int_{\Sigma^+}\mathrm{d}^dx\sqrt{-\gamma} \left(K^{ij}-K\gamma^{ij} \right)\delta g_{ij}-\int_{\Sigma^-}\mathrm{d}^dx\sqrt{-\gamma} \left(K^{ij}-K\gamma^{ij} \right)\delta g_{ij}
\end{equation}
in our conventions. Demanding that this variation is of equal magnitude but opposite sign to the variation $\delta S_D$, together with continuity of the induced metric across the brane, then gives
\begin{equation}
\Big(\left[K_{ij}\right]-\left[K\right]\gamma_{ij} \Big)=-S_{ij}
\end{equation}
with
\begin{equation}
S_{ij} = -\frac{1}{2\kappa_D^2}\left(\frac{1}{2}\mu \gamma_{ij}-R[\gamma]_{ij}+\frac{1}{2}R[\gamma]\gamma_{ij} \right).
\end{equation}

Noting that $\gamma_{ij}$ is a metric on $\mathbb{R}\times S^{d-1}$, we pause to collect some geometric facts about $\gamma_{ij}$. First write
\begin{equation}
\mathrm{d}s_\Sigma^2 = -\mathcal{T}^2_g \,\mathrm{d}t^2 + \mathcal{R}^2_g\, \mathrm{d}\Omega_{d-1}^2
\end{equation}
where we have defined the length scales
\begin{equation}
\mathcal{T}^2_g = f e^{2A}\Big |_{u_g} \qquad \mathrm{and} \qquad \mathcal{R}^2_g = e^{2A}\Big|_{u_g}.
\end{equation}
Their values are given in terms of the parameters $\alpha$ and $H$ via the first junction condition above, (\ref{eq:jc1ss}). We then introduce the obvious orthonormal frame
\begin{equation}
\mathrm{e}^0 = \mathcal{T}_g \mathrm{d}t \qquad \mathrm{and} \qquad \mathrm{e}^r = \mathcal{R}_g\overline{\mathrm{e}}^r
\end{equation}
where $\overline{\mathrm{e}}^r$ are $(d-1)$ one-forms on the sphere satisfying
\begin{equation}\label{eq:sbein}
\mathrm{d}\overline{\mathrm{e}}^r = -\overline{\omega}^r\,_s \wedge \overline{\mathrm{e}}^s \qquad \mathrm{and} \qquad \overline{\rho}^{rs} =  \overline{\mathrm{e}}^r \wedge  \overline{\mathrm{e}}^s.
\end{equation}
Because the metric on $\Sigma$ is the product space $\mathbb{R}\times S^{d-1}$ the curvature tensors are just those of the sphere factor. In detail,
\begin{align}
R^0\,_a & = 0 \nonumber\\
R^r\,_s & = \frac{(d-2)}{\mathcal{R}_g^2}\delta^r_s \nonumber\\
R  & = \frac{(d-1)(d-2)}{\mathcal{R}^2_g}.
\end{align}

Evaluating this on the brane metric yields
\begin{align}\label{eq:STd1}
S_{tt} & = -\frac{1}{4\kappa_D^2}\left(\mu+\frac{(d-1)(d-2)}{\mathcal{R}^2_g} \right)\gamma_{tt}\\
S_{\alpha\beta} & = -\frac{1}{4\kappa_D^2}\left(\mu+\frac{(d-3)(d-2)}{\mathcal{R}^2_g} \right)\gamma_{\alpha\beta}\label{eq:STd2}
\end{align}
which demonstrates that such a brane action is indeed capable of reproducing the stress-energy necessary to support the domain wall.

Explicitly, we determine the brane parameters $\mu$ and $\kappa_D$ by equating (\ref{eq:STc1}), (\ref{eq:STc2}) with (\ref{eq:STd1}), (\ref{eq:STd2}) to obtain
\begin{equation}
 \frac{1}{2\kappa_D^2} = \left(\frac{\alpha-1}{d-2}\right)\sqrt{\frac{\alpha^2-1}{\alpha^2+H^2}}
\end{equation}
and
\begin{equation}
\mu =(1-d)(d-2)\left(\frac{\alpha^2+H^2}{\alpha^2-1} \right)\left[\frac{2}{\alpha}+\frac{1+H^2}{\alpha^2+H^2} \right]
\end{equation}
as presented in the main text, (\ref{eq:kD}) and (\ref{eq:muD}). We note that $\mu$ is necessarily negative in this simple setup.

\end{document}